\documentclass[BCOR2.0cm,DIV14,twoside,12pt,
abstracton,smallheadings,cleardoubleempty,
openright,pagesize,abstractoff,bibtotoc]{scrreprt} 

\newif\ifcomment

\newif\iffinal
\newif\ifarxiv
\newif\ifprint
\newif\ifabstract
\newif\ifindex
\newif\ifappendix
\newif\iftocs
\newif\ifbib
\newif\ifacro
\newif\ifgerman
\newif\ifallpages

\finaltrue
\arxivtrue
\tocstrue
\bibtrue
\appendixtrue
\acrotrue

\ifprint
\germantrue
\fi

\usepackage[german,english]{babel}
\ifgerman
\fi

\ifindex
\usepackage{makeidx}
\fi

\usepackage{exscale} 
\usepackage[latin1]{inputenc} 
\usepackage{amsmath,amssymb}
\usepackage{bbm,mathrsfs}
\usepackage{floatflt}
\usepackage{tabularx}
\usepackage[square,numbers,sort&compress,nonamebreak]{natbib}
\usepackage{hypernat}
\usepackage{doc}
\usepackage{url}
\usepackage{ae}
\usepackage{array}
\usepackage{subeqnarray}
\usepackage[tight]{subfigure}
\usepackage{times}
\usepackage{color}
\usepackage[printonlyused,smaller]{acronym}

\newif\ifpdf
\ifx\pdfoutput\undefined
 \pdffalse
\else
 \pdftrue 
\fi

\ifpdf
\ifarxiv
.<<<Cannot have arxiv mode with pdflatex>>>
\fi
\usepackage[pdftex]{graphicx}
\iffinal
\else
 \usepackage{pdfdraftcopy}
 \draftstring{Draft \today}
\fi
\ifprint
 \usepackage[pdftex,draft,a4paper]{hyperref}
\else
 \usepackage[pdftex,backref,pagebackref,colorlinks,
             linkcolor=blue,citecolor=blue,anchorcolor=blue,
             pagecolor=blue,urlcolor=red,a4paper]{hyperref}
\fi
\pdfoutput=1        
\pdfcatalog{/PageMode(/UseOutlines)} 
\else 
\usepackage[dvips]{graphicx}
\usepackage{rotating}
\iffinal
\else
 \usepackage[dvips,english,all,light]{draftcopy}
\fi
\ifprint
 \usepackage[dvips,draft,a4paper]{hyperref}
\else
 \usepackage[ps2pdf,backref,pagebackref,colorlinks,
            linkcolor=blue,citecolor=blue,anchorcolor=blue,
            pagecolor=blue,urlcolor=red,a4paper]{hyperref}
\fi
\hypersetup{
  pdfauthor   = {Constantinos A. Loizides},
  pdftitle    = {Jet physics in ALICE},
  pdfcreator  = {},
  pdfproducer = {},
  pdfsubject  = {},
  pdfkeywords = {}
 }
\fi

\ifarxiv
\germanfalse
\fi

\graphicspath{ {./pics/} 
               {./pics/chap2/} 
               {./pics/chap3/} 
               {./pics/chap4/} 
               {./pics/chap5/} 
               {./pics/chap6/} 
             }
\def\eos    {equation of state}
\def\cms    {centre-of-mass}
\def\stn    {signal-to-noise}
\newcommand {\snn}{\sqrt{s_{\scriptscriptstyle{{\rm NN}}}}}
\newcommand {\dndy}{\ensuremath{\mathrm{d}N/\mathrm{d}y}}

\newcommand {\dncdy}{\ensuremath{\mathrm{d}N_{\mathrm{ch}}/\mathrm{d}y}}
\newcommand {\dncde}{\ensuremath{\mathrm{d}N_{\mathrm{ch}}/\mathrm{d}\eta}}
\newcommand {\av}[1]{\ensuremath{\left< #1 \right>}}
\newcommand {\abs}[1]{\ensuremath{\left| #1 \right|}}
\newcommand {\lsim}{\,{\buildrel < \over {_\sim}}\,}
\newcommand {\gsim}{\,{\buildrel > \over {_\sim}}\,}
\newcommand {\hrefurl}[1]{\href{#1}{\url{#1}}}

\newcommand {\Ref}[1]{Ref.~\cite{#1}}
\newcommand {\Refs}[1]{Refs.~\cite{#1}}
\newcommand {\eq}[1]{eq.~(\ref{#1})}
\newcommand {\Eq}[1]{Equation~(\ref{#1})}
\newcommand {\sect}[1]{section~\ref{#1}}

\newcommand {\chap}[1]{chapter~\ref{#1}}
\newcommand {\Chap}[1]{Chapter~\ref{#1}}
\newcommand {\fig}[1]{fig.~\ref{#1}}
\newcommand {\Fig}[1]{Figure~\ref{#1}}
\newcommand {\tab}[1]{table~\ref{#1}}
\newcommand {\Tab}[1]{Table~\ref{#1}}
\newcommand {\peq}[1]{eq.~(\ref{#1}) on page~\pageref{#1}}

\newcommand {\psect}[1]{section~\ref{#1} on page~\pageref{#1}}

\newcommand {\pfig}[1]{fig.~\ref{#1} on page~\pageref{#1}}

\newcommand {\ptab}[1]{tab.~\ref{#1} on page~\pageref{#1}}

\newcommand {\page}[1]{page~\pageref{#1}}

\newcommand {\hide}[1]{\color{white}#1\color{black}}
\newcommand {\pgfig}[1]{Abb.~\ref{#1} auf Seite~\pageref{#1}}

\newcommand {\Lumi} {\ensuremath{{\cal{L}}}}
\newcommand {\mrm}  {\mathrm}
\newcommand {\pt}   {\ensuremath{p_{\mathrm{T}}}}
\newcommand {\px}   {\ensuremath{p_{\mathrm{x}}}}
\newcommand {\py}   {\ensuremath{p_{\mathrm{y}}}}
\newcommand {\pz}   {\ensuremath{p_{\mathrm{z}}}}
\newcommand {\et}   {\ensuremath{E_{\mathrm{T}}}}
\newcommand {\xt}   {\ensuremath{x_{\mathrm{T}}}}

\newcommand {\kt}   {\ensuremath{k_{\mathrm{T}}}}
\newcommand {\Pt}   {\ensuremath{P^{\mathrm{max}}_{\mathrm{T}}}}
\newcommand {\Ptj}  {\ensuremath{P^{\mathrm{jet}}_{\mathrm{T}}}}
\newcommand {\Etj}  {\ensuremath{E^{\mathrm{jet}}_{\mathrm{T}}}}

\newcommand {\dd}   {\mrm{d}}
\newcommand {\eg}   {e.g.}
\newcommand {\ie}   {i.e.}
\newcommand {\cf}   {cf.}
\newcommand {\etc}  {etc.}

\newcommand {\RAA}  {R_{\rm AA}}
\newcommand {\RAB}  {R_{\rm AB}}

\newcommand {\const}{\ensuremath{\mathrm{const}}}
\newcommand {\as}   {\ensuremath{\alpha_{\rm S}}}
\newcommand {\mub}  {\ensuremath{\mu_{\rm B}}}
\newcommand {\tc}   {\ensuremath{T_{\rm C}}}
\newcommand {\Ncoll}{\ensuremath{N_{\rm coll}}}
\newcommand {\Npart}{\ensuremath{N_{\rm part}}}
\newcommand {\Nhard}{\ensuremath{N_{\rm hard}}}
\newcommand {\Nch}  {\ensuremath{N_{\rm ch}}}
\newcommand {\lQCD} {\ensuremath{\Lambda_{\rm QCD}}}
\newcommand {\lum}   {\mbox{${\rm cm}^{-2} {\rm s}^{-1}$}}

\newcommand {\tev}   {\mbox{${\rm TeV}$}}
\newcommand {\gev}   {\mbox{${\rm GeV}$}}
\newcommand {\mev}   {\mbox{${\rm MeV}$}}

\newcommand {\mom}   {\mbox{\rm GeV$\kern-0.15em /\kern-0.12em c$}}
\newcommand {\gmom}  {\mbox{\rm GeV$\kern-0.15em /\kern-0.12em c$}}
\newcommand {\mass}  {\mbox{\rm GeV$\kern-0.15em /\kern-0.12em c^2$}}
\newcommand {\mmass} {\mbox{\rm MeV$\kern-0.15em /\kern-0.12em c^2$}}
\newcommand {\mmom}  {\mbox{\rm MeV$\kern-0.15em /\kern-0.12em c$}}
\newcommand {\barn}  {\mbox{${\rm b}$}}
\newcommand {\mbarn} {\mbox{${\rm mb}$}}

\newcommand {\nb}    {\mbox{\rm nb}}
\newcommand {\m}     {\mbox{${\rm m}$}}
\newcommand {\cm}    {\mbox{${\rm cm}$}}
\newcommand {\mm}    {\mbox{${\rm mm}$}}
\newcommand {\mum}   {\mbox{$\mu {\rm m}$}}

\newcommand {\fm}    {\mbox{${\rm fm}$}}

\newcommand {\hz}  {\mbox{${\rm Hz}$}}
\newcommand {\khz} {\mbox{${\rm kHz}$}}
\newcommand {\mhz} {\mbox{${\rm MHz}$}}

\newcommand {\s}     {\mbox{${\rm s}$}}
\newcommand {\ms}    {\mbox{${\rm ms}$}}
\newcommand {\musec} {\mbox{$\mu {\rm s}$}}
\newcommand {\ns}    {\mbox{${\rm ns}$}}
\newcommand {\ps}    {\mbox{${\rm ps}$}}
\newcommand {\bit}   {\mbox{${\rm bit}$}}

\newcommand {\kbyte} {\mbox{${\rm kB}$}}

\newcommand {\mbyte} {\mbox{${\rm MB}$}}

\newcommand {\mbyteps}{\mbox{${\rm MB/s}$}}

\newcommand {\gbyteps}{\mbox{${\rm GB/s}$}}
\newcommand {\pp}    {\mbox{pp}}
\newcommand {\NN}    {\mbox{NN}}
\newcommand {\PbPb}  {\mbox{Pb--Pb}}
\newcommand {\AuAu}  {\mbox{Au--Au}}
\newcommand {\AaAa}  {\mbox{A--A}}
\newcommand {\pA}    {\mbox{p--A}}
\newcommand {\dAu}   {\mbox{d--Au}}
\newcommand {\pPb}   {\mbox{p--Pb}}

\newcommand {\NNex}  {\mbox{nucleon--nucleon}}
\newcommand {\ppex}  {\mbox{proton--proton}}
\newcommand {\pAex}  {\mbox{proton--nucleus}}
\newcommand {\AAex}  {\mbox{nucleus--nucleus}}
\newcommand {\pizero}{\mbox{$\mathrm {\pi^0}$}}

\newcommand {\Jpsi}  {\mbox{J\kern-0.05em /\kern-0.05em$\psi$}}

\newcommand {\ppbar} {\mbox{$\mathrm {p\overline{p}}$}}
\newcommand {\epem}  {\mbox{$\mathrm {e^{+}e^{-}}$}}
\newcommand {\ccbar} {\mbox{$\mathrm {c\overline{c}}$}}

\newcommand {\quark}[1] {\mbox{$\mathrm {#1}$}}
\newcommand {\qubar}[1] {\mbox{$\mathrm {\overline{#1}}$}}

\ifindex
\makeindex
\fi

\iffinal
\else
\fi

\begin{document}
\setcounter{page}{1}
\pagenumbering{roman}
\pagestyle{empty}
\thispagestyle{empty}

\ifprint
\else 
\title{\vspace*{1.5cm}Jet physics in ALICE}
\author{ Phd.~thesis\\  
       {\bf Constantinos A. Loizides}\\
       \vspace*{1.0cm}\\
       Johann Wolfgang von Goethe -- Universität\\
       Frankfurt am Main, Germany}
\iffinal
\date{ \large February 4, 2005\\
       \vspace*{0.5cm}
}
\else
\date{ \large \today}
\fi
\fi

\iffinal
\maketitle
\thispagestyle{empty}
\ifprint
 \thispagestyle{empty}
 {\large
 \vspace*{9cm}\noindent 
 Vom Fachbereich Physik der Johann Wolfgang Goethe-Universität\\
 als Dissertation angenommen.\\

 \vspace*{5cm}\noindent
 Dekan: \hspace{5cm}Prof.~Dr.~Wolf Assmus\\

 \vspace*{0.4cm}\noindent
 Gutachter: \hspace{4.35cm}Prof.~Dr.~Reinhard Stock\\

            \hspace{6.25cm}Prof.~Dr.~Harald Appelshäuser\\

 \vspace*{0.8cm}\noindent
 Datum der Disputation: \hspace{1.75cm}4.~April 2005}
 \pagebreak
\fi
\ifabstract
 \begin{abstract}
%

This works at the performance of the \acs{ALICE} detector for the measurement 
of high-energy jets at mid-pseudo-rapidity in ultra-relativistic nucleus--nucleus 
collisions at \acs{LHC} and their potential for the characterization of the partonic 
matter created in these collisions. 

In our approach, jets at energies of $30~\gev$ and higher are reconstructed with a cone 
jet finder, as typically done for jet measurements in hadronic collisions. The goal is 
to study its capabilities of measuring jets, to quantify obtainable rates and the quality 
of jet reconstruction, both, in proton--proton and in lead--lead collisions. 

In particular, we will address, whether modification of the jet fragmentation 
in the charged-particle sector can be detected within the high particle-multiplicity 
environment expected in central lead--lead collisions. We will comparatively treat 
these topics in view of an \acs{EMCAL} proposed to complete the central \acs{ALICE} 
tracking detectors.

The topcis covered in the thesis are the following:\\
Determination of the potential for exclusive jet measurements in \acs{ALICE}.\\
Determination of jet rates that can be acquired with the \acs{ALICE} setup.\\
Development of a parton-energy loss model.\\
Simulation and study of the energy-loss effect on jet properties.

 \end{abstract}
 \thispagestyle{empty}
\fi

\ifgerman
 \setcounter{page}{1}
 \selectlanguage{german}
%

\chapter*{Zusammenfassung}
\pagestyle{headings}
\pagenumbering{roman}
Diese Arbeit bestimmt die experimentellen Möglichkeiten des\acs{ALICE}
Detektors, Jets mit einer transversalen Energie von mehr als $50~\gev$  
zu messen. \acs{ALICE} ist eines der vier vorgesehenen 
Detektorsysteme am derzeit in der Bauphase befindlichen \ac{LHC}.
Das System ist speziell für Bleikernkollisionen bei einer Schwerpunktsenergie 
von $\snn=5.5~\tev$ pro Nuk\-leonpaar optimiert.

Die Untersuchung von  Kernstößen bei relativistischen 
und ultrarelativistischen Ener\-gien ermöglicht es, die Eigenschaften
von stark wechselwirkender Materie unter verschiedenen Bedingungen
von ---im Vergleich zu normaler Kernmaterie--- hoher Dichte und 
Temperatur zu studieren. 
Bei den Experimenten am Bevatron/Bevalac und am \ac{AGS} in den 70er 
und 80er Jahren, lag das Interesse vornehmlich der Zustandsbeschreibung 
von angeregter Kernmaterie~\cite{stock2004}.

Mit wachsendem Verständnis fundamentaler Prinzipien der Natur, 
vor allem verbunden mit \acs{QCD} als grundlegender Theorie zur dyna\-mischen 
Beschreibung von Quarks und Gluonen (Partonen), wurden die ursprünglichen 
Ziele selbstverständlich verfeinert und gleichermaßen erweitert. 
Man vermutete, daß die bei der Kollision erzeugte Energie ausreicht, 
um die normalerweise in Protonen und Neutronen eingeschlossenen Partonen 
zumindest kurzzeitig freizusetzen, und auf diese Weise ihre ungebundene 
Wechselwirkung experimentell zugänglich wird. 

Über viele Jahre motivierte die Suche nach diesem freien Zustand, dem \ac{QGP}, 
sowie die Lokalisation und Natur des Phasenübergangs von hadro\-nischer 
(eingeschlossen) zu partonischer (freier) Materie die Forschung auf dem Gebiet 
der Schwerionenphysik. 
Die ersten Hinweise auf Existenz des Plasmas stammten aus einer Serie von 
experimentellen Beobachtungen am \ac{SPS}~\cite{satz2002}. Seit kurzem sind diese
durch eine Reihe von Experimenten am \ac{RHIC} bestätigt worden. Die Natur des 
Phasenübergangs und damit auch die Zustandsgleichung des \ac{QGP} sind jedoch 
weiterhin unverstanden~\cite{gyulassy2004}.

Das Verständnis des Phasendiagramms der \acs{QCD} und insbesondere des 
Phasenübergangs sind von außerordentlichem Interesse für die Hochenergiephysik. 
Der Übergang von gebundener hadronischer zu freier partonischer Materie ist der 
einzige im Standardmodell vorhergesagte Phasenübergang, der unter Laborbedingungen 
zugänglich ist (und der daher Fluktuationen fundamentaler Quantenfelder 
beinhaltet). Der Ursprung hadronischer Massen, beispielsweise des Protons 
und Neutrons, sind eng mit ihm verbunden. Numerische Lösungsverfahren 
der \acs{QCD} formuliert in diskretisierter Raumzeit (Gitterrechnungen) sagen 
sowohl die Art des Übergangs, als auch die Wiederherstellung von chiraler Symmetrie 
voraus~\cite{karsch2001,karsch2003}. Sie stellen damit den theoretischen 
Zusammenhang dieser Fragen zu Eigenschaften der \acs{QCD} im thermodynamischen 
Gleichgewicht her.

Die am \ac{LHC} dreißig mal höhere Schwerpunktsenergie erzeugt im Vergleich zu 
\ac{RHIC} weitaus idealere Bedingungen; insbesondere entsteht eine heißere, räumlich 
ausgedehntere und länger lebendende partonische Phase. 
Die Aufgabe des Schwer\-ionenprogrammes am \ac{LHC} wird es sein, die Eigenschaften dieses 
\ac{QGP} genau zu untersuchen. Wir sprechen zentrale Aspekte der Schwer\-ionenphysik am 
\ac{LHC} in Kapitel~\ref{chap2} an.

Jedoch ändert sich das in der Kollision erzeugte System rapide: vom extrem dichten, 
partonischen Anfangszustand zum ausgedünnten, hadronischen und damit meßbaren Endzustand. 
Das Verständnis dieser auf sehr kurzen Zeitskalen agierenden Prozesse erfordert andere 
Konzepte als die der bloßen Beschreibung von \acs{QCD} im Gleichgewicht. Stattdessen bedarf 
es einer Kombination von Methoden aus unterschiedlichsten Gebieten der Physik, von der 
Elementarteilchen- und Kernphysik, über die Beschreibung von Gleichgewichts- 
und Nichtgleichgewichtsprozessen bis hin zur Hydrodynamik.

Eine direkte Verbindung zwischen störungstheoretischen Vorhersagen der \acs{QCD} und 
experimentellen Beo\-bachtungsgrößen wird durch eine begrenzte Anzahl von Meßgrößen 
hergestellt. Diese sogenannten \glqq harten\grqq\ Proben enstehen am Anfang in der 
Nichtgleichgewichtsphase der Reaktion, bei der die Dynamik noch vorwiegend durch harte 
Stöße innerhalb des partonischen Sys\-tems dominiert wird. Die harten Prozeße prüfen
das Medium: Veränderungen an ihren bekannten Eigenschaften lassen sich auf Eigenschaften 
des erzeugten Mediums zurückführen.

Bei den Experimenten am \ac{RHIC} sind harte Prozesse zum ersten Mal in Schwer\-ionenreaktionen
mit ausreichender Statistik zugänglich. Sie äußern sich in \AuAu\ Kollisionen bei
$\snn=200~\gev$ beispielsweise in Hadronen mit einem Transversal\-impuls von mehr als $5~\gev$.
Die gemessene Anzahl der Teilchen mit solchen Impulsen in zentralen Kollisionen ist um 
einen Faktor vier bis fünf kleiner als man nach Messungen in weniger zentralen oder in \pp\ 
(oder $\dAu$) Kollisionen bei gleicher Schwerpunktsenergie erwarten würde~\cite{arsene2003,
adams2003,adler2003,adler2003b,back2003b}.
Da Hadronen mit hohen Transversalimpulsen bevorzugt in der Fragmentation von partonischen Jets 
entstehen, wird ihre Unterdrückung in der Literatur \glqq \mbox{Jet Quenching}\grqq\ 
genannt~\cite{wang2003}. 
Der Effekt wird dem Energieverlust von harten Partonen zugeschrieben, der durch die zusätzliche 
Wechselwirkung mit dem erzeugten, partonischen Medium verursacht wird. Die Quantifizierung 
der Unterdrückung harter Partonen, die im Anfangsstadium der Reaktion erzeugt werden, ist
von großem Interesse, da sie die Charakterisierung des, vornehmlich partonischen, Mediums bei 
hohen Dichten erlaubt.

Am \ac{LHC} in Kollisionen bei $\snn=5.5~\tev$ werden energetische Proben, leichte Quarks und 
Gluonen, in großen Mengen produziert, sogar mit Energien, die mehr als eine Größenordnung 
über denen am \ac{RHIC} liegen. Auch schwere Quarks und elektromagnetische Prozesse werden durch
relativ hohen Produktionsraten zugänglich. Die vorliegende Arbeit handelt von Partonen, die durch
ihre Fragmentation in hadronische Jets hoher Energie ($\Etj\ge50~\gev$) identifiziert werden.
Im Unterschied zu \ac{RHIC} wird erwartet, daß diese Jets experimentell zugänglich sind
und trotz der Komplexität der Schwerionen\-ereignisse exklusiv rekonstruiert werden können.

Jets sind in zahlreichen Streuversuchen in elementaren (\epem) oder hadronischen (\pp/\ppbar) 
Prozessen an verschiedenen Beschleunigern mit unterschiedlichen Energien untersucht worden. 
Mit Hilfe von Konzepten aus der \acs{pQCD}, sowie nicht-perturbativen Ergänzungen 
sind sowohl ihr Produktionswechselwirkungsquerschnitt, als auch ihre Fragmentationsfunktion 
beschreibbar. Diese hadronischen Eigenschaften sind universell, also unabhängig von den 
spezifischen Kollisionsprozessen. Wir stellen in Abschnitt~\ref{chap3:jetpp} die wesentlichen 
Konzepte der Jetphysik zusammen.

Jetfragmentierung im Vakuum ereignet sich vorzugsweise innerhalb eines Radius von $R=0.7$ 
in der Ebene aufgespannt von Pseudorapidität ($\eta$) und Azimuthalwinkel ($\phi$),
$R^2=\eta^2+\phi^2$. Für Jets mit einer Energie von $\Etj\gsim 50~\gev$ befinden sich 
davon bereits mehr als $80$\% innerhalb von $R=0.3$. Das Auffinden eines hadronischen Jets 
erfolgt typischerweise mit einem iterativen Verfahren, \glqq Cone Finder\grqq, mit 
festgewähltem Radius $R$, indem Teilchenenergien oder akkumulierte Energien von calorimetrischen 
Messungen summiert werden. 

Man erwartet, daß das in Schwerionenkollisionen erzeugte Medium die bekannten 
Fragmentationseigenschaften verändert. Die Beschreibung von Jeteigenschaften unter dem 
Einfluß der Wechselwirkung mit dem Medium ist Gegenstand aktueller Forschung. 
In Abschnitt~\ref{chap3:partoneloss} erläutern wir den \acs{BDMPS-Z-SW} Formalismus.
Dieses erlaubt die statistische Berechnung des Energieverlusts, $\Delta E$, von energetischen 
Quarks und Gluonen der Energie $E$ (wobei $E\rightarrow\infty$ in der Eikonalapproximation), 
der durch die zusätzliche Gluonstrahlung im Medium der Länge $L$ induziert wird. 
Das Medium wird dabei durch den Transportkoeffizienten, $\hat{q}$, charakterisiert, 
der den durchschnittlichen Impulsübertrag auf das partonische Projektil pro freier Weglänge 
quantifiziert. 

Abschnitt~\ref{chap3:expresultsrhic} fasst den experimentellen Befund für harte Proben 
in Kollisionen \ac{RHIC} Bedingungen zusammen. Der Vergleich von Messungen in zentralen \AuAu\ 
Kollisionen mit periphären und mit \pp\ Kollisionen zeigt, daß Hadronen mit hohen Transversalimpulsen
($\pt>5~\gev$) signifikant unterdrückt werden. Desweiteren verschwinden Teilchenkorrelationen, 
die auf Jetfragmentation hinweisen, bei mittleren Impulsen ($\pt>2~\gev$) und werden erst bei 
niedrigen Impulsen ($\pt<2~\gev$) wieder sichtbar. In allen Fällen wurde durch eine 
Kontrollmessung in $\dAu$ bestätigt, daß diese Effekte vorwiegend durch den Einfluß der 
partonischen Phase im Endzustand und nicht durch veränderte Anfangsbedingungen in den 
Partonverteilungen der einfallenden Kernen zustandekommt.

Die vorgestellten Effekte werden mittlerweile durch verschiedene Modellrechnungen erklärt, 
von denen der überwiegende Teil die Beschreibung von Partonenergieverlust durch zusätzliche 
Gluonstrahlung im Medium beinhaltet. Abschnitt~\ref{chap3:pqm} behandelt das \ac{PQM}, unser 
Monte Carlo Verfahren, mit dem für zentrale Rapiditäten alle Observablen (außer dem elliptischen 
Fluß) im harten Sektor von \ac{RHIC} beschrieben werden können. Dazu wird die \acs{BDMPS-Z-SW} 
Wahrscheinlichkeitsverteilung $P(\Delta E)$ derart erweitert, daß der abgestrahlte Energieverlust
auch für kleine Partonenergien immer $\Delta E\le E$ erfüllt. Durch eine konsistente Integration 
der Kollisionsgeometrie in den Formalismus wird zudem sichergestellt, daß nur ein einziger 
Parameter im Modell verbleibt, der die Skala des Energieverlusts bestimmt. 
Für zentrale \AuAu\ Kollisionen bei $\snn=200~\gev$ erhalten wir mit \ac{PQM} einen 
durchschnittlichen Transportkoeffizienten von $\av{\hat{q}}=14~\gev^2/\fm$.

Die Tatsache, daß der Transportkoeffizient proportional mit der Gluondichte wächst, 
$\hat{q}\propto n^{\rm gluons}$, kann man für die Extrapolation zu den Bedingungen 
am \ac{LHC} ausnutzen. Im \acs{EKRT} Modell, das sehr erfolgreich Teilchenmultiplizitäten
für \ac{RHIC} vorausgesagt hat, skaliert diese mit 
$n^{\rm gluons}\propto{\rm A}^{0.383}\, \left(\snn\right)^{0.574}$. Für zentrale \PbPb\ 
Kollisionen bei $\snn=5.5~\tev$ erwarten wir daher einen Transportkoeffizienten 
von $\av{\hat{q}}\simeq100~\gev^2/\fm$ (siehe \pgfig{chap3:fig:RAAlhc}). 
Dieselbe Methode führt zur erfolgreichen Beschreibung der Daten für \AuAu\ bei 
$\snn=62.4~\gev$ am \ac{RHIC}.

Die absoluten Werte für den Transportkoeffizienten beinhalten  relativ große, systematischen 
Unsicherheiten. Zum einen ist die Erweiterung zu endlichen (und sehr kleinen, $E\le15~\gev$) 
Partonenergien theoretisch nicht fundiert, und zum anderen hängt die Berechnung 
linear von dem gewählten, festen Wert von $\as$ ($1/3$) ab. 
Zudem saturiert seine Bestimmung durch Verhältnisse von Teilchenspektren alleine 
dadurch, daß die Elternpartonen bevorzugt von der Oberfläche des Feuerballs emittiert 
werden (siehe \pgfig{chap3:fig:raavsq}). 
In jedem Fall, auch für vierfach kleinere Dichten, sagt \ac{PQM} nach heutigem 
Stand der Theorie vorraus, daß der Energieverlust für Partonen mit einer Anfangsenergie von 
mehr als $50~\gev$ signifikant ist, und daß auch am \ac{LHC} Teilchenspektren vornehmlich
durch den Oberflächeneffekt dominiert sein werden.

In diesem Licht stellen sich bezüglich des Jetphysikprogramms bei \acs{ALICE} vor allem die Fragen, 
in welcher Form der Einfluß des Mediums auf Jetproduktion und Jetfragmentation erfolgen wird, 
und ob man ---im Unterschied zu Teilchenspektren--- die komplette Jetenergie, und damit die Energie 
des zugrundeliegenden Partons, in Schwerionenreaktionen am \ac{LHC} rekonstruieren kann. 
Zur Zeit vermutet man auf Grundlage einer Rechnung auf partonischer Ebene~\cite{salgado2003b}, 
daß die durch die Gluonstrahlung emittierte Energie für diese Jets zu $85$\% innerhalb
von $R=0.3$ verbleibt, die zusätzlich abgestrahlten Gluonen also innerhalb des Kegels 
hadronisieren. Danach besteht die Modifikation der Fragmentationsfunktion vor allem darin, 
daß die longitudinalen und transversalen Impulse der assoziierten Teilchen im Jet in 
Bezug auf die Jetachse vermindert bzw.~erhöht werden.

Es wird erwartet, daß gerade Veränderungen an diesen Verteilungen mit \acs{ALICE} meßbar sind, 
denn \acs{ALICE} zeichnet sich von den anderen Experimenten am \ac{LHC} dadurch aus, daß es von Beginn 
an für die Bestimmung von Impulsen geladener Teilchen von $\pt\simeq 100~\mev$ bis zu $100~\gev$ 
bei gleichzeitig hohen Erwartungen an die Teilchenmultiplizität von maximal $\dncdy\approx8000$ 
geladenden Teilchen pro Rapiditätseinheit in zentraler Akzeptanz geplant wurde.
Das Detektorsystem wird in Kapitel \ref{chap4} beschrieben. Für die Jetrekonstruktion sind 
insbesondere die Spurfindungsdetektoren wichtig. Sie bestehen aus \ac{ITS}, \ac{TPC} und \ac{TRD}
und befinden sich in dem  zentralen Akzeptanzbereich, $-0.9 \leq \eta \leq 0.9$, innerhalb eines
Magnetfeldes von $0.5$~T. Seit kurzem gibt es den Vorschlag, den zentralen Bereich um ein 
\ac{EMCAL} mit einer Akzeptanz von $\abs{\phi}\le120^\circ$ und $\abs{\eta}\le0.7$ zu erweitern. 
Zudem ist vorgesehen, die begrenzte Datennahme des Experimentes durch Echtzeitdatenanalyse mit dem 
\ac{HLT} zu unterstützen. Dieses System wird in der Lage sein, Kollisionsereignisse vor der endgültigen 
Speicherung auf ihren physikalischen Inhalt zu untersuchen, um vorwiegend seltene Ereignisse zu 
selektieren. Beispielsweise ist der  Produktionswechselwirkungsquerschnitt von Jets mit $\Etj=100~\gev$ 
bereits zwei Größenordnungen niedriger als der von $50~\gev$ Jets (see \pgfig{chap5:fig:spectrapythia}). 
Durch den \ac{HLT} kann die Statistik für Jets hoher Energie bei gleichbleibender Bandbreite 
zum Massenspeichersystem signifikant vergrößert werden.

In Kapitel \ref{chap5} werden die Rekonstruktionsmöglichkeiten von \acs{ALICE}
unter Beteiligung verschiedener Detektorsysteme für Jets mit $\Etj\gsim 50~\gev$
in zentralen \PbPb\ Kollisionen im Vergleich zu reinen \pp\ Ereignissen untersucht. 
Dabei beschränken wir uns auf Jets, deren Achse innerhalb $\abs{\eta}\le0.5$ liegt.
Die Jetidentifikation erfolgt in beiden Fällen mit dem gleichen Algorithmus (\acs{ILCA}). 
Jedoch entstehen in zentralen \PbPb\ Ereignissen starke Korrelationen, die vorwiegend durch die 
zahlreiche Produktion von Minijets bei kleinen Energien ($\Etj\lsim10~\gev$) verursacht werden.
Ihre Anwesenheit macht es notwendig, den typischen Radius von $R=0.7$ auf $R=0.3$ zu verringern.
Zusätzlich erfordert die \mbox{hohe} Teilchendichte in zentralen Kollisionen (in der Simulation mit 
\acs{HIJING} sind  es um die $6000$ geladene Teilchen pro Rapiditätseinheit vorwiegend bei kleinen 
Transversalimpulsen), geladene Teilchen mit $\pt<2~\gev$ nicht bei der Jetrekonstruktion zu 
berücksichtigen.

Mit diesen Einschränkungen sind Jets mit $50~\gev$ und höher auch in zentralen \PbPb\ Kollisionen 
identifizierbar. Doch ist ihre Energieauflösung limitiert durch intrinsische Fluktuationen in der 
Jetfragmentierung, Teilchenproduktion außerhalb des limitierten Jetkegels, sowie Fluktuationen 
im verbleibenden Minijetuntergrund. Das gilt auch für einen idealen Detektor und für Jets mit 
weitaus höherer Energie. Ohne \ac{EMCAL} ist die Auflösung vorwiegend dominiert durch Fluktuationen 
im Verhältnis von geladenen Teilchen zu neutralen Teilchen in der Jetfragmentation. Im Durchschnitt 
werden in diesem Fall $50$\% der Energie rekonstruiert, mit einer Breite von $50$\%. 
Mit dem \ac{EMCAL} erhöht sich das Verhältnis zu $60$\% bei einer Breite von $30$\%.
In beiden Fällen ist die räumliche Auflösung der Jetachse (in $\eta$ und $\phi$) besser 
als $0.01$. Die räumliche Auflösung steigt moderat mit höherer Jetenergie, wohingegen
die Energieauflösung dominiert durch intrinsische Fluktuationen relativ unabhängig von 
der Jetenergie ist. In allen Fällen ist der durchschnittliche Beitrag des Untergrundes 
innerhalb des reduzierten Kegels ist beträchtlich ($15$\% bei $50~\gev$).

Bei Jets mit weniger als $50~\gev$ wird es zunehmend schwieriger, daß Signal vom Untergrund
zu trennen. In diesem Energiebereich, in dem Jetraten relativ hoch sind, werden inklusive
Meßmethoden zur Anwendung kommen. In \PbPb\ Kollisionen fällt die Rate produzierter Jets 
innerhalb der zentralen Akzeptanz von einem Jet pro Ereignis mit \mbox{$\Etj\ge20~\gev$}
auf ungefähr einen Jet in 1000 Ereignissen mit \mbox{$\Etj\ge100~\gev$}. Auf Jets mit mehr 
als $50~\gev$ wird man daher mittels des \acp{HLT} triggern.

Unseren Studien zur Triggersimulation verwenden denselben Jetrekonstruktionsalgorithmus 
wie für die bisherige Analyse, auch mit denselben Parametern, da die Laufzeit für zentrale
\PbPb\ Ereignisse auf einem Pentium III mit $800~\mhz$ unter $50~\ms$ liegt. Diese Laufzeit
ist vernachläßigbar gegenüber der Rekonstruktionsdauer des Ereignisses ($\propto \s$). 
Es zeigt sich, daß ein solcher Trigger die gespeicherte Datenmenge für \PbPb\ um einen Faktor 
$50$ reduziert, und dabei $10$\% der Jets mit $\Etj\ge50~\gev$ und  $50$\% mit $\Etj\ge100~\gev$ 
selektiert.
Ohne weitere Unterstützung durch Triggerdetektoren würden in einem \acs{ALICE} Jahr
($10^{6}~\s$) bei einer durchschnittlichen Luminosität von $\Lumi=0.5\,\mbarn^{-1}\s^{-1}$ 
insgesamt $10^{7}$ Ereignisse aufgezeichnet, in denen $4\cdot10^{5}$ Jets mit
$\Etj>50~\gev$ und $10^{5}$ mit $\Etj>100~\gev$ vorhanden sind.

In Kapitel \ref{chap6} untersuchen wir die Meßbarkeit von Mediumeffekten in zentralen 
\PbPb\ Kollisionen anhand von Eigenschaften identifizierter Jets. Dazu werden modizierte 
(hadronische) Jets mit einer speziellen Version von \acs{PYTHIA} prepariert. 
Diese erzeugt zunächst die partonischen Jets und berechnet dann für jedes Parton im 
Endzustand den Energieverlust mittels \ac{PQM}. Die abgestrahlte Energie wird mit bis 
zu sechs Gluonen pro Parton ausgeglichen, bevor die endgültige Fragmentierung in den 
hadronischen Endzustand erfolgt. Die Jets werden für Mediumsdichten von 
$\av{\hat{q}}=1.2$, $12$ und $24~\gev^2/\fm$ erzeugt und vor der Jetrekonstruktion 
mit zentralen \acs{HIJING} Ereignissen vermischt.

\enlargethispage{1cm}
Unsere Analyse beschränkt sich auf Jets mit $\Etj\gsim50~\gev$, die mit den oben beschriebenen
Einschränkungen exklusiv gemessen werden können. Im Vergleich zu der Referenzmessung in \pp\ 
zeigt sich, daß die Jeteigenschaften deutlich durch den Untergrund der zentralen 
Schwerionenkollision beeinflußt werden, sogar bis zu Energien von $100~\gev$. 
Allerdings verschwindet der Einfluß mit steigendem Energieschwellenwert für die 
rekonstruierten Jets in der Analyse, und die meisten Observablen zeigen zumindest 
moderate Abhängigkeit von den gewählten Transportkoeffizienten.

Innerhalb der limitierten Möglichkeiten des Models zeigt sich, daß Messungen mit
Spurdetektoren alleine, ohne Unterstützung des \acp{EMCAL}, ausreichen, um zwischen 
einer Mediumsdichte von $\av{\hat{q}}=1.2$  und $24~\gev^2/\fm$ unterscheiden zu können. 
Basierend auf dieser Untersuchung könnte daher die Veränderung der Jetfragmentationsfunktion 
alleine durch die Messung von geladenen Teilchen festgestellt werden. 

Wenn allerdings, wie mit \ac{PQM} ohne assoziierte Teilchenproduktion (also ohne \acs{PYTHIA})
vorrausgesagt, auch Jets mit $\Etj\gsim 150~\gev$ stark unterdrückt werden (beispielsweise durch 
näherungsweise Thermalisierung der Elternpartonen im Medium),
wird man \glqq \mbox{Jet Quenching}\grqq\ vorzugsweise durch kalorimetrische
Bestimmung der Jetproduktionswahrscheinlichkeit im Vergleich mit Messungen in periphären 
und \pp\ Kollisionen nachweisen. In diesem Fall wäre das \ac{EMCAL} außerordentlich
wichtig für die Verbesserung der Bestimmung der Jetenergie.

 \selectlanguage{english}
 \acresetall
\fi
\fi

\iftocs
\addtocontents{toc}{\relax \protect \pagestyle{empty}}
\addtocontents{toc}{\relax \protect \thispagestyle{empty}}
\addtocontents{toc}{\relax \protect \enlargethispage{0.5cm}}
\addtocontents{lof}{\relax \protect \pagestyle{empty}}
\addtocontents{lof}{\relax \protect \thispagestyle{empty}}
\addtocontents{lot}{\relax \protect \pagestyle{empty}}
\addtocontents{lot}{\relax \protect \thispagestyle{empty}}
\tableofcontents
\thispagestyle{empty}
\fi


\allpagestrue
\cleardoublepage
\pagestyle{headings}
\thispagestyle{headings}
\pagenumbering{arabic}

\iffinal
%

\chapter{Introduction}
\label{chap1}
\setcounter{page}{1}
Research on \AAex\ collisions at relativistic and ultra-relativistic 
energies addresses the properties of strongly interacting matter under 
varying conditions of high density and temperature; that is high with 
respect to normal nuclear matter constituting our known world.
During recent decades a large number of experiments have been carried out.
Starting at the Bevatron/Bevalac facility in the early 70's and followed 
by the \ac{AGS} programme in the 80's, the original focus was the nuclear 
\eos\ away from the ground state.

With increasing understanding of fundamental principles of nature, and with 
the advent of \ac{QCD}, naturally, the physics goals have been refined, 
culminating in the location and characterization of the hadron--parton
deconfinement phase transition. Over many years, the search for the \ac{QGP} 
and determination of its \eos, motivated research in the field at ever
increasing \cms\ energies. First evidence of its existence stemmed from a series 
of experimental observations at the \ac{SPS}. Recently the \ac{RHIC} experiments 
confirmed that the \ac{QCD} phase transition exists, however, the \eos\ 
still remains unknown. 

The \ac{QCD} phase transition is the only one predicted by the Standard Model 
(and thus involving fundamental quantum fields) that can be reached with laboratory 
experiments. The transition to deconfined matter and its inverse process into confinement, 
collective hadronization, are intrinsically linked to the origin of hadronic mass.
Lattice \ac{QCD} calculations, predicting the nature and phase boundary of the 
transition, as well as the approximate restoration of chiral symmetry in the 
deconfined phase, provide the connection with properties of the \ac{QCD} Lagrangian
in thermodynamical equilibrium.

At the \ac{LHC} lead ions are foreseen to collide at energies about 30 times 
higher than at \ac{RHIC}. It is expected that these collisions will provide rather 
ideal conditions: In particular, hotter, larger and longer-living \ac{QGP} matter will
be created that allows significant qualitative improvement with respect to the 
previous studies. The task of the \ac{LHC} heavy-ion programme, therefore, will 
be to investigate the properties of deconfined matter. 

However, the partonic system created in \AAex\ collisions rapidly changes 
from extreme initial conditions into dilute final hadronic states. 
The understanding of these fast evolving processes 
goes far beyond the exploration of equilibrium \ac{QCD} described by
lattice methods. Instead a combination of concepts from elementary-particle
physics, nuclear physics, equilibrium and nonequilibrium thermodynamics, 
as well as hydrodynamics is needed for the theoretical description.

A direct link between \ac{QCD} predictions and experimental observables is provided 
by a limited number of observables, classified as hard probes. They are produced 
during the initial, non-equilibrated stage of the collision, when the collision 
dynamics is dominated by hard scatters within the interacting partonic system. 
Modification of their known properties gives information about the properties of 
the medium.

At \ac{RHIC}, the yield of high-momentum particles is significantly reduced in 
central collisions compared to peripheral ones. This effect, somewhat misleadingly 
called jet quenching, is commonly attributed to an apparent energy loss of 
energetic partons propagating through partonic matter created in the collision. 
The attenuation of hard partons, created during the primordial, dynamical evolution 
of ultra-relativistic \AAex\ collisions, is an observable of interest in view of 
the characterization of the `fireball' matter at high energy density, supposedly 
constituting of deconfined partonic compositions. The initial parton created by
hard scattering acts as a test particle for the \ac{QCD} structure of the medium 
it has traversed.

At the \ac{LHC}, energetic probes, light quarks and gluons, will be abundantly 
produced, even at energies of more than one order of magnitude higher than at \ac{RHIC}. 
Also heavy quarks and other types of probes will become available with fairly high 
rates. The probes, which are the principal topic of this thesis, might be identified
by their fragmentation into hadronic jets of high energy. In contrast with \ac{RHIC},
their initial energy should be high enough to allow the full reconstruction of the 
hadronic jet, even in the heavy-ion environment.

Understanding the dependence of high-energy jet production and fragmentation on the 
created medium is an open field of active research. 
Generally, the energy loss of the primary parton 
is attributed to medium-induced gluon radiation. 
It is suggested that hadronization products of these, rather soft gluons may be 
contained within the jet emission cone, resulting in a modification of the characteristic 
jet fragmentation, as observed via longitudinal and transverse momentum distributions
with respect to the direction of the initial parton, as well as of the multiplicity 
distributions arising from the jet fragmentation.

\bigskip
In the present work, we focus on jet physics with \acs{ALICE}, the dedicated heavy-ion  
experiment at the \ac{LHC}. The goal is to study its capabilities of measuring high-energy 
jets (with $\Etj\gsim50~\gev$), to quantify obtainable rates and the quality of reconstruction, 
in both proton--proton and lead--lead collisions. In particular, we shall address 
whether modification of the charged-particle jet fragmentation can be detected 
within the high-particle-multiplicity environment of central lead--lead collisions. 
We shall consider the benefits of an \ac{EMCAL} proposed to complete the central 
\acs{ALICE} tracking detectors. 

\bigskip
After the introduction in \chap{chap1},
we start in \chap{chap2} with an outline of qualitative and quantitative new conditions 
for heavy-ion collisions at the \ac{LHC}. In \chap{chap3}, we describe in detail the physics 
framework used within the thesis. Since we are aiming at full jet reconstruction in heavy-ion 
collisions, we first reproduce in \sect{chap3:jetpp} the concepts of jet physics in 
hadron--hadron collisions.
Then in \sect{chap3:partoneloss}, we outline a state-of-the-art \acs{pQCD} framework for
the calculation of partonic energy loss in partonic matter. In \sect{chap3:expresultsrhic}, 
we summarize the recent high-transverse-momentum measurements at \ac{RHIC}. 
Finally in \sect{chap3:pqm}, we describe our Monte Carlo \ac{PQM} and its current application 
to \acs{RHIC} data, as well as predictions for \ac{LHC} conditions and limitations of the approach.  
\Chap{chap4} deals with the experimental setup of \acs{ALICE}, mainly addressing 
the description of the \acs{ALICE} detector system, as well as the Trigger, 
\ac{DAQ} and \ac{HLT} complex. In \chap{chap5}, we first introduce obtainable jet rates 
within the central \acs{ALICE} detectors. Then we discuss the jet reconstruction capabilities 
in \ppex\ and lead--lead collisions, as well as, the corresponding trigger rates, both
with respect of the possible integration of the \ac{EMCAL}. We close the chapter with a short 
summary on back-to-back jet or photon--jet correlation measurements.
In \chap{chap6}, we focus on measurements of jet quenching within identified jets in central 
lead--lead collisions at \ac{LHC} conditions. 
In this context, we introduce simple, model-independent observables and discuss their sensitivity 
to the density of medium. \Chap{chap7} attempts a summary concerning the prospects of jet 
spectroscopy and, in particular, the quantification of jet attenuation in the medium.

\bigskip
The main topics of the thesis are the following:
\begin{itemize}
\item {\em Determination of the potential for exclusive jet measurements in \acs{ALICE}:}\\
We introduce a simulation strategy of realistic jet spectra in \pp\ and in \PbPb\ to
compare jet-reconstruction methods for online and offline usage. Of particular 
interest are systematic errors introduced in central \PbPb\ collisions by the underlying soft 
event and qualitative improvements by the proposed \ac{EMCAL}. A preliminary study of 
reconstructed jets in \pp\ based on charged tracks is performed to compare different 
track-reconstruction algorithms, both, for online and offline usage. These topics are 
covered in \chap{chap5}.

\item {\em Determination of jet rates that can be acquired with the \acs{ALICE} setup:}\\
Originally, \acs{ALICE} was designed to measure soft, hadronic properties of the bulk,
with the \ac{DAQ} system comfortably well able to cope with the expected background rates 
without need for higher-level triggering. However, jets with very high energy of more than 
$100~\gev$ are rare, also at the \ac{LHC}, and require to be identified online. We set up 
a complete simulation of the \ac{HLT} system, to obtain the trigger rates in \pp\ and \PbPb\
based on the information of the charged-track content in the event. The topic is covered in
\chap{chap5}; the results are reported in \sect{chap5:recontriggerrates}. 

\item {\em Development of a parton-energy loss model:}\\
We develop a Monte Carlo model, \ac{PQM}, where the collision geometry is incorporated 
into the framework of the \acs{BDMPS-Z-SW} quenching weights and uses mid-rapidity 
data from \ac{RHIC} to tune its single parameter. Extrapolating the medium density 
found at \ac{RHIC} to the expectations at the \ac{LHC}, we address the possible quenching 
scenario at the \ac{LHC}. The model, its results and the derived expectations for 
leading-hadron spectroscopy, as well as its limitations 
are discussed in \sect{chap3:pqm}. This part was carried out in close collaboration 
with A.~Dainese; published in~\Ref{dainese2004}.

\item {\em Simulation and study of the energy-loss effect on jet properties:}\\
For simulation of medium-modified jets, we combine the quenching model 
with the \acs{PYTHIA} generator. The aim is to study modified jets for different medium 
densities in central \PbPb\ collisions and evaluate the sensitivity of several jet observables, 
as measured by their hadronic content with respect to their values obtained in reference 
measurements from \pp\ collisions. The results are discussed in \chap{chap6}.
\end{itemize}

%

\newif\ifexpcond
\expcondtrue
\newif\ifpartmult
\partmulttrue
\newif\ifdeepdeconfinement
\deepdeconfinementtrue
\newif\ifnovelaspects
\novelaspectstrue

\chapter{Heavy-ion physics at the Large Hadron Collider}
\label{chap2}
The \acf{LHC} scheduled to start operation in
2007 will accelerate protons, light and heavy nuclei up to
\cms\ energies of several \tev\ per \NNex\ pair. 
For \AAex\ collisions at energies about 30 times higher 
than at the \acf{RHIC} and 300 times higher than at 
the \acf{SPS} one expects that particle production will 
mostly be determined by saturated parton densities 
and hard processes will significantly contribute 
to the total \AAex\ cross section. In addition to
the long life-time of the \ac{QGP} state and its 
high (initial) temperature and density, these qualitatively new
features will allow one to address the task of the \ac{LHC} heavy-ion 
programme: the systematic study of the properties of deconfined matter.

\section{Experimental running conditions}
\label{chap2:expcond}
\ifexpcond
Like the former \ac{SPS} and current \ac{RHIC} programme,
the heavy-ion programme at the \ac{LHC} will be based on two components:
use of the largest available nuclei at the highest possible energy and
the variation of system sizes (\pp, \pA, \AaAa) and beam energies.
The ion beams will be accelerated up to a momentum of 7~\tev\ per unit 
of $Z/A$, where $A$ and $Z$ are the mass and the atomic numbers of the ions.
Thus, an ion ($A$,~$Z$) will acquire a fraction $p(A,Z)=Z/A \, p^{\rm p}$
of the momentum, $p^{\rm p}=7~\tev$, for a proton beam. Neglecting masses,
the \cms\ energy per \NNex\ pair in the collision of two ions 
($A_1$,~$Z_1$) and ($A_2$,~$Z_2$) is given by 
\begin{equation*}
\snn=\sqrt{(E_1+E_2)^2 - (\vec{p}_1+\vec{p}_2)^2} \simeq \sqrt{4\,p_1\,p_2} = 
2\,\sqrt{\frac{Z_1\,Z_2}{A_1\,A_2}} p^{\rm p}\;.
\end{equation*}

The running programme~\cite{pprvol1} of \ac{ALICE}, which is dedicated 
to heavy-ion collisions at the \ac{LHC}, initially foresees:
\begin{itemize}
\item Regular \pp\ runs at $\sqrt{s}=14~\tev$;
\item $1$--$2$ years with \PbPb~runs at $\snn=5.5~\tev$;
\item $1$ year with \pPb~runs at $\snn=8.8~\tev$ 
(or \mbox{d--Pb} or \mbox{$\alpha$--Pb});
\item $1$--$2$ years with \mbox{Ar--Ar} at $\snn=6.3~\tev$.
\end{itemize}

The \NNex\ and \pAex\ runs are required to establish a basis
for the comparison of the results obtained in \PbPb\ collisions.
This point is detailed during the discussion on hard probes
in \sect{chap2:hardprobes}. The runs with lighter ions facilitate the change 
of the energy density and the volume of the produced system. Concerning 
the hard sector, running at different \cms~energies for different systems 
is not expected to introduce large uncertainties in the comparisons 
since \ac{pQCD} calculations are quite safely applicable for the extrapolation 
to different energies, \ie~to scale the jet cross section and shapes measured in 
\pp\ at $14~\tev$ to the energy of \PbPb, $5.5~\tev$ as mentioned in 
\psect{chap5:jetrates}. Further \ac{ALICE}-specific details are given in \chap{chap4}.
\fi

\section{Expected particle multiplicity}
\label{chap2:partmult}
\ifpartmult
The average charged-particle multiplicity per rapidity unit ($\dndy$) 
is one of the most fundamental, global observables in heavy-ion collisions.
On the theoretical side, it enters the calculation of most other 
observables, as it is related to the attained energy density 
of the medium produced in the collision.
It can be estimated at the time of local thermal equilibration 
using the Bjorken estimate~\cite{bjorken1983}
\begin{equation}
\label{chap2:eq:bjorkendensity}
\varepsilon = \left(\dndy\right)_{y=0}\, 
\frac{\av{\et}}{ A\, \tau_0}\;,
\end{equation}
where $(\dndy)_{y=0}$ specifies the number of emitted particles (or partons) 
per unit of rapidity at mid-rapidity having the average transverse 
energy $\av{\et}$.~\footnote{The longitudinal rapidity  of a 
particle with four-momentum $(E,\vec{p})$ is defined as 
\mbox{$y=\frac{1}{2}\,\ln\left(\frac{E+p_z}{E-p_z}\right)$}, where 
$z$ is the direction along the beams.} 
The effective initial volume is characterized by the area 
$A = \pi R_{\rm A}^2$ with the nuclear radius $R_{\rm A}$ and 
longitudinally by the formation time $\tau_0$ of the thermal medium.
It is about $1~\fm$ at \ac{SPS}, $0.2~\fm$ at \ac{RHIC} 
and expected to be $0.1~\fm$ at the \ac{LHC}.
On the experimental side, the average charged-particle multiplicity 
per unit rapidity largely determines the accuracy with which many 
observables can be measured and, thus, constitutes the main unknown 
in the detector performance. 
Another important---closely related---observable is the 
total transverse energy per rapidity unit at mid-rapidity. 
It quantifies how much of the total initial longitudinal 
energy is converted into the transverse plane. 
Up to now, there is no first-principles calculation of these observables 
starting from the \ac{QCD} Lagrangian, since particle production is dominated 
by soft, non-perturbative and long-range \ac{QCD}
on the large (nuclear) scale of $R_{\rm A} \approx A^{1/3}~\fm$.

Understanding the multiplicity in \pp\ collisions is a
prerequisite for the study of multiplicity in \AaAa, but already 
here, at the \NNex\ level, the difficulties in the 
theoretical description arise. The inclusive hadron 
rapidity density for ${\rm pp}\to{\rm hX}$ is defined as
\begin{equation*}
\rho_{\rm h}(y)=\frac{1}{\sigma^{\rm in}_{\rm pp}} \;
\int_0^{\pt^{\rm max}} \, \dd^2\pt\,\frac{\dd\sigma_{{\rm pp}\to{\rm hX}}}  
{\dd y \, \dd^2\pt}\;,
\end{equation*}
where $\sigma^{\rm in}_{\rm pp}$ is the inelastic \pp\
cross section. 
Its energy dependence and especially the slow rise
above $\sqrt{s}=20$~GeV is poorly understood by 
first-principles \ac{QCD} calculations, because 
for scattering processes with large \cms\ energies, 
but without large virtualities in the intermediate 
states, both, perturbation theory and numerical 
Euclidian lattice methods, fail~\cite{hebecker2001}.
For high energy the dependence roughly follows 
a power law $s^\alpha$ or a logarithm $\ln s$ or $\ln^2 s$.
By general arguments such as unitarity and analyticity 
the cross section is asymptotically bounded by $\const \ln^2 (s/s_0)$, 
the Froissart bound~\cite{froissart1961,martin1965}.
Recently there has been evidence from ${\rm \gamma p}$ and 
${\rm \pi p}$ reactions for its saturation~\cite{block2004}. 

The hadron rapidity density at mid-rapidity $\rho_{\rm h}(y=0)$,
or equivalently the total multiplicity $N$, grows as well with energy.
It can be parametrized for charged particles by 
\begin{equation}
\label{chap2:eq:rhoparampp}
\Nch \equiv \rho_{\rm ch}(y=0) \approx a \ln^2\sqrt{s} + b \ln\sqrt{s} + c\;,
\end{equation}
plotted for $a=0.049$, $b=0.046$ and $c=0.96$ in 
\fig{chap2:fig:ncharge} (solid line). Thus, the total
charged multiplicity in \pp\ is about 2 at \ac{SPS} energies, 
about 2.5 at \ac{RHIC} energies and extrapolates to about 5 
at \ac{LHC} energies.~\footnote{Model calculations typically compute 
the total multiplicity $N$ and assume $\Nch=2/3 N$ because of 
iso-spin conservation. If resonance decays are included, 
the ratio drops from $0.67$ to about $0.6$.}

The high multiplicities in central \AAex\ collisions typically arise
from the large number of independent and successive \NNex\ collisions,
occurring when many nucleons interact several times on their path
through the oncoming nucleus. Studies of \pAex\ collisions 
have revealed that the total multiplicity 
does not scale with the number of binary collisions ($\Ncoll$)
in the reaction, but rather with the number of `wounded nucleons'
($\Npart$), which participate inelastically~\cite{elias1978}.~\footnote{The 
`wounded nucleon' scaling is approximately correct at \ac{SPS} energies, 
at \ac{RHIC} energies processes violating the $\Npart$ scaling become 
available, thus one there assumes $N \propto (1-x)\,\Npart + x\, \Ncoll$,
or $N \propto \Npart\,\log \Npart$ because of saturation effects.}
The number of participants is $\Npart=2$ for \pp\ and $\Npart=\Ncoll+1$ for 
\pA\ and about $2A$ for central \AaAa\ collisions. In general, both quantities 
depend on the impact parameter $b$ of the collision and can be related 
through simple phenomenological (Glauber) models~\cite{glauber1970,bialas1976}.
As the aim of studying heavy-ion collisions is to discover
qualitatively new effects at the scale $R_{\rm A}$,
not observed in \pp\ collisions, one typically
scales the particle yields measured in \AaAa\
collisions by $\Npart/2$ to directly compare
with similar yields in elementary collisions. 
At \ac{RHIC} energies and at top \ac{SPS} energies the charged-particle 
multiplicity in central collisions normalized by $\Npart/2$ scales 
with $\sqrt{s}$ in the same way as elementary $\epem$ into hadrons 
data at the same \cms\ energy. Also in \pp\ or \ppbar\ collisions the
scaling agrees, however at the effective \cms\ energy given by the \pp\ 
or \ppbar\ \cms\ energy minus the energy of leading particles~\cite{back2003}, 
indicating a common particle productions mechanism for the different systems 
at high energies. A further hint to an universal mechanism is the suggestion 
of the limiting fragmentation hypothesis~\cite{back2002,nouicer2002}.

\begin{figure}[htb]
\begin{center}
\includegraphics[width=13cm]{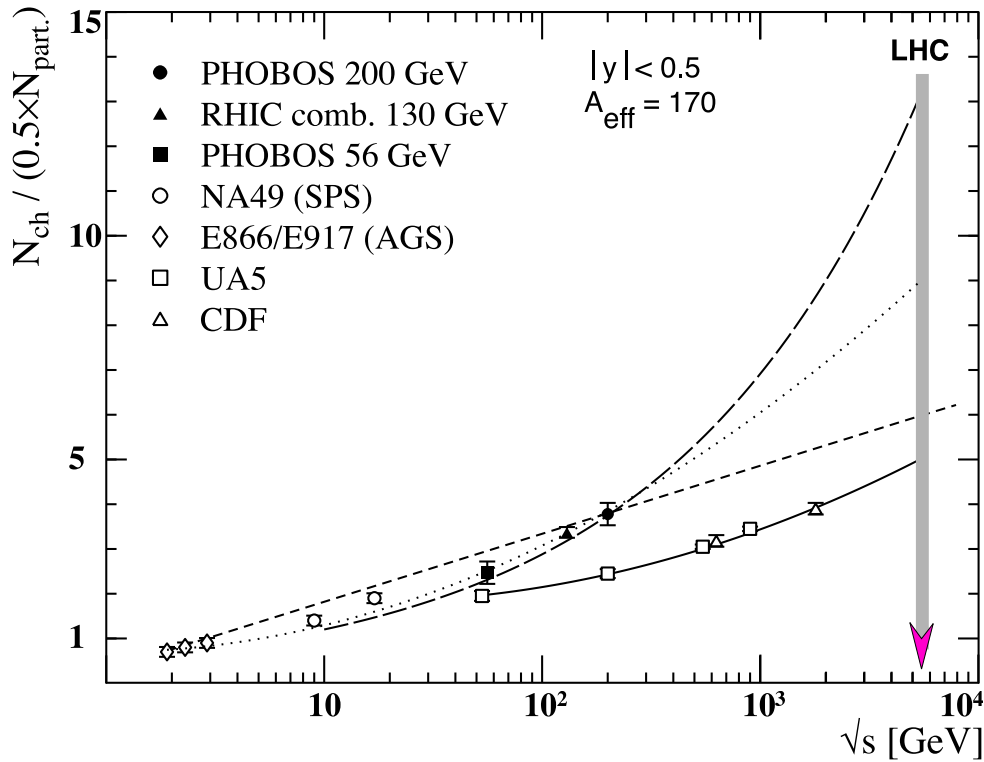}
\end{center}
\vspace{-0.3cm}
\caption[xxx]{Charged-particle multiplicity per participant pair at 
mid-rapidity as  a function of \cms\ energy for \AuAu\ collisions 
at \acs{RHIC} (closed symbols) and \pp\ collisions (open symbols) 
measured at various accelerators. The long-dashed line is the 
extrapolation to \acs{LHC} energies using the saturation model 
(\acs{EKRT})~\cite{eskola1999}. The other lines are different fits 
of the nuclear and \pp\ data to \eq{chap2:eq:rhoparampp}. 
Further details are given in the text.} 
\label{chap2:fig:ncharge}
\end{figure}

In \fig{chap2:fig:ncharge} the charged multiplicity normalized 
to the number of participant pairs 
as a function of the \cms\ energy is shown for \AuAu\ 
data at \ac{RHIC} (closed symbols) and a variety of \pp\ data 
(open symbols). Assuming universality, a fit for the extrapolation 
to the \ac{LHC} energy of all nuclear data to \eq{chap2:eq:rhoparampp}, 
gives $b=0.68$ and $c=0.26$ for $a=0$ fixed (dashed line), 
and $a=0.028$ and $c=0.7$ for $b=0$ fixed (dotted line). 
The long-dashed line is the extrapolation given by the saturation 
model (\acs{EKRT})~\cite{eskola1999}. Like most models in that
context (see \Ref{kharzeev2000} and references therein) 
it assumes that the phase space available for quarks and gluons 
saturates at some dynamical energy scale $Q_{\rm sat, A}(\sqrt{s})$,
the saturation scale, at which, by the uncertainty, principle 
the parton wave functions start to overlap in the transverse 
plane.~\footnote{Although there is one remarkable 
difference: typically, parton-saturation models assume the saturation 
of the {\em incident} partons, whereas the \acs{EKRT} model assumes 
the saturation of the {\em produced} partons.} 
In such a scenario the total multiplicity basically is 
determined by the transverse energy density per unit rapidity. 
\begin{equation}
\label{chap2:eq:Qsaturation}
N(\sqrt{s})  \sim Q^2_{\rm sat, A}(\sqrt{s}) \, R^2_{\rm A} 
             \sim A^\alpha\,\sqrt{s}^\beta\;,
\end{equation}
The original \acs{EKRT} result~\cite{eskola1999} ---refined in \Ref{eskola2001}---
for $\alpha=0.922$, $\beta=0.383$  and a proportionality constant of $1.383$ 
quite successfully predicted the \ac{RHIC} multiplicities. 
A very recent estimation~\cite{armesto2004}, using the argument of geometrical
scaling found at small-x lepton-proton data from \acs{HERA} extended to nuclear 
photo-absorption cross sections, finds \eq{chap2:eq:Qsaturation}
with $\alpha=0.089$, $\beta=0.288$ and a proportionality constant of $0.5$.

For the extrapolation to \ac{LHC} energies, the crucial point is, 
whether the total multiplicity as a function of $\sqrt{s}$ 
has a power-law behaviour like \eq{chap2:eq:Qsaturation} 
or rather grows with the power of the logarithm like 
\eq{chap2:eq:rhoparampp}. The range within \ac{RHIC} is small and
that from \ac{RHIC} to \ac{LHC} is large (see \fig{chap2:fig:ncharge}). 
There is a lot of room for error, as within \ac{RHIC} one cannot 
reliably distinguish between the different parametrical descriptions.

\begin{table}[htb]
\begin{center}
\begin{tabular}{lcl}
\hline\hline
{ Model} & {$\Nch$} & {Comments} \\
\hline
$\ln s$ fit   & $\simeq$ 1000 & \eq{chap2:eq:rhoparampp};  $a=0$, $b=0.68$, $c=0.26$ \\
$\ln^2s$ fit  & $\simeq$ 1500 & \eq{chap2:eq:rhoparampp};  $a=0.028$, $b=0$, $c=0.7$\\
\acs{EKRT}    & $\simeq$ 2200 & \eq{chap2:eq:Qsaturation}; $\alpha=0.922$, $\beta=0.383$, $\const=1.383$\\
Geom.~scaling & $\simeq$ 1700 & \eq{chap2:eq:Qsaturation}; $\alpha=0.089$, $\beta=0.288$, $\const=0.5$\\
Initial parton saturation 
              & $\simeq$ 1900 & \eq{chap2:eq:Qsaturation}, but calculated from~\acs{CGC}~\cite{kharzeev2004}\\
\acs{HIJING}~1.36  &   $\simeq$ 6200 & with quenching     \\
                   &   $\simeq$ 2900 & without quenching  \\
\acs{DPMJET}-II.5  &   $\simeq$ 2300 & with baryon stopping    \\
                   &   $\simeq$ 2000 & without baryon stopping \\
\acs{SFM}          &   $\simeq$ 2700 & with fusion    \\
                   &   $\simeq$ 3100 & without fusion \\
\hline\hline
\end{tabular}
\end{center}
\vspace{-0.4cm}
\caption[xxx]{Charged-particle multiplicity predictions of different models at $\eta\approx 0$.}
\label{chap2:tab:models}
\end{table}

In \tab{chap2:tab:models} we summarize the expectations of the 
charged-particle multiplicity at mid-pseudo-rapidity for the different 
models.~\footnote{The pseudo-rapidity is defined as 
$\eta = -\ln \left [ \tan ( \theta/2) \right ]$,
where $\theta$ is the polar angle with respect to the beam direction.
It is $\eta = y$ for a massless particle and $\eta \approx y$ 
if the particles' velocity approaches unity.}

In addition to estimates already mentioned, we quote the 
predictions for various Monte Carlo event generators.
At the time of the \acs{ALICE} technical proposal~\cite{alicetp1} 
and before the start-up of \ac{RHIC}, 
the predictions for \PbPb\ collisions at $\snn = 5.5~\tev$ 
ranged between $1500$--$8000$ charged particles at central 
rapidity~\cite{eijndhoven1995}. Now most generators have been updated,
of which we mention \acs{HIJING},
\acs{DPMJET} and \acs{SFM}. \acs{HIJING} is a \acs{QCD}-inspired 
model of jet production~\cite{mchijing1991,mchijing1994} with the Lund 
model~\cite{mclund1983} for jet fragmentation.
The multiplicity in central events with and without jet 
quenching differs by more than a factor of 2. Including jet
quenching it predicts the highest multiplicities of all models.
The \acs{DPMJET} model~\cite{mcdpmjet1999} is an implementation 
of the two-component \ac{DPM}~\cite{mcdpm1994} 
based on the Glauber--Gribov approach. It treats 
soft and hard scattering processes in an unified way
and uses the Lund model~\cite{mclund1983} for fragmentation.
Predictions with and without the baryon stopping mechanism are shown,
and baryon stopping increases the multiplicity by about 15\%. 
The \acf{SFM}~\cite{amelin2001} includes in its initial stage 
both soft and semi-hard components leading to the formation of 
colour strings. Collectivity is taken into account by means of string 
fusion and string breaking leads to the production of secondaries. 
Predictions with string fusion reduce the multiplicity by about 10\%
compared to calculations without fusion. 

The large variety of available models of heavy-ion collisions 
gives a wide range of predicted multiplicities from $1000$ to $6200$ charged
particles at mid-rapidity for central \PbPb\ collisions. The multiplicity 
measured at \ac{RHIC}, $\dncde\simeq 750$ ($\dncdy\simeq 650$), at 
$\snn=200~\gev$, was found to be about a factor 2 lower than 
what was predicted by most models~\cite{eskola2001b}. 
In view of this fact, the multiplicity at the \ac{LHC} is probably
between $1500$ and $3000$ charged particles per unit of rapidity. 
Though, as we will briefly touch on in \chap{chap4} the \ac{ALICE} detectors are 
designed to cope with multiplicities up to 8000 charged particles per
rapidity unit, a value which ensures a comfortable safety margin. 
\fi

\section{Deconfinement region}
\label{chap2:deepdeconfinement}
\ifdeepdeconfinement
Starting from the estimates of the charged multiplicity or average
transverse energy, most parameters of the medium produced in the collision 
can be inferred by assuming (local) thermodynamical equilibrium with a certain
\eos. Because of its simplicity, one often considers the energy density 
of an equilibrated ideal gas of particles with $n_{\rm dof}$~degrees of 
freedom~\cite{safarik1999}
\begin{equation}
\label{chap2:eq:epsidealgas}
\varepsilon = n_{\rm dof}\,\frac{\pi^2}{30}\,T^4
\end{equation}
according to the Stefan-Boltzmann law. For a pion gas 
the degrees of freedom are only the three values of the 
iso-spin for $\pi^+,\,\pi^0,\,\pi^-$. For a QGP with two 
quark flavours the degrees of freedom are $n_g+7/8\,(n_q+n_{\bar{q}})
= N_g(8)\,N_{\rm pol}(2)+7/8 \, \times 2\times N_{\rm flav}(2) \, 
N_{\rm col}(3)\,N_{\rm spin}(2)=37$. The factor 7/8 accounts 
for the difference between Bose-Einstein for gluons and 
Fermi-Dirac statistics for quarks.

\begin{table}[htb]
\begin{center}
\begin{tabular}{ll|ccc}
\hline
\hline
{Parameter} & & {SPS} & {RHIC} & {LHC} \\
\hline
$\snn$                 & [$\gev$]       & $17$       & $200$       & $5500$ \\
d$N_{\rm gluons}/$d$y$ &                & $450$      & $1200$      & $4700$ \\
$\dncdy$               &                & $350$      & $800$       & $3000$ \\
$Q_{\rm sat,A}$        & [$\gev$]       & $0.71$     & $1.13$      & $2.13$ \\
Initial temperature    & [$\gev$]       & $0.38$     & $0.6$       & $>1$   \\
Initial energy density & [$\gev/\fm^3$] & $\sim 5$   & $\sim 25$ & $\sim 250$  \\
Freeze-out volume      & [$\fm^3$]      & $\sim 10^3$& $\sim 10^4$ & $\sim 10^5$ \\
Life time              & [$\fm$]        & $<2$       & $2$-$4$     & $>10$ \\ 
\hline
\hline
\end{tabular}
\end{center}
\vspace{-0.4cm}
\caption[xxx]{Comparison of the most relevant---model-dependent---parameters characterizing 
central \AAex\ collisions at different energy scales~\cite{eskola1999}.}
\label{chap2:tab:spsrhiclhc}
\end{table}

In \tab{chap2:tab:spsrhiclhc} we present a comparison of the most 
relevant ---model-dependent---parameters for \ac{SPS}, \ac{RHIC} and \ac{LHC} 
energies, where the \eos\ describes a free gas of gluons [$n_{\rm dof}=16$ 
in~\eq{chap2:eq:epsidealgas}] and adiabatic longitudinal expansion 
is included in the hydrodynamical calculation~\cite{eskola1999}. A slightly 
refined calculation~\cite{eskola2001} using an \eos\ with quark and gluonic 
degrees of freedom and including transverse expansion in the hydrodynamical 
phase as well as hadronic resonances (and decays) at freeze-out (see 
\Ref{sollfrank1996}) gives only slightly different results of the order of 
$10$--$15$\%.

The high energy in the collision \cms\ at the \ac{LHC} determines 
a very large energy density 
and an initial temperature at least a factor 2 higher than at \ac{RHIC}. 
The high initial temperature extends the life time and the volume of the 
deconfined medium, since the \ac{QGP} has to expand while cooling down to the 
critical (or freeze-out) temperature, which is about $175\pm 15~\mev$ and 
relatively independent of $\sqrt{s}$ above the \ac{SPS} energy 
(see \fig{chap2:fig:phasediag}). In addition, the large number of 
gluons favours energy and momentum exchanges, 
thus considerably reducing the time needed for the thermal equilibration 
of the medium. Thus, the \ac{LHC} will create a hotter, larger and longer-living 
\ac{QGP} state than the present heavy-ion facilities. The main advantage is
due to the fact that in the deconfinement scenario the \ac{QGP} is more 
similar to the thermodynamical equilibrated \ac{QGP} theoretically investigated  
by means of (Euclidian) lattice \ac{QCD}~\cite{ejiri2003,fodor2004,karsch2000,
karsch2003,karsch2003b,laermann2003} and of statistical hadronization 
models~\cite{becattini2004,braun-munzinger2003}.
Both methods---and further phenomenological models (see \Ref{stephanov2004} and
references therein)--- map out the different phase boundaries of strongly 
interacting matter described by \ac{QCD}.

\subsection{Phase diagram of strongly interacting matter}
\label{chap2:phasediagram}
The current knowledge of the phase diagram~\cite{hands2001}
is displayed in~\fig{chap2:fig:phasediag} as a function of the
temperature, $T$, and of the baryo-chemical potential, $\mub$,
as a measure of the baryonic density.~\footnote{The baryo-chemical 
potential $\mub$ of a strongly interacting system (in thermodynamical equilibrium) 
is defined as the change in the energy $E$ of the system, when the total
 baryonic number $N_{\rm B}$ (baryons minus anti-baryons) is increased by 
one unit: $\mub=\partial E/\partial N_{\rm B}$.} 

At low temperatures and for $\mub\simeq m_{\rm p}\simeq 1~\gev$, 
there is the region of the ordinary matter of protons and neutrons. 
Increasing the energy density of the system, by `compression' 
(towards the right) or by `heating'  (upward), the hadronic gas phase 
is reached in which nucleons interact and form pions, excited states of 
the proton and of the neutron such as $\Delta$ resonances and other hadrons. 
If the energy density is further increased, the transition to the deconfined 
\ac{QGP} phase is predicted~\cite{cabibbo1975}: the density of the partons, 
quarks and gluons, becomes high enough that the confinement of quarks in hadrons 
vanishes~(deconfinement). 

The phase transition can be reached along different `paths' on the ($\mub$, $T$) 
plane. In heavy-ion collisions, both, temperature and density increase, possibly 
bringing the system beyond the phase boundary. In \fig{chap2:fig:phasediag} 
the regions of the fixed-target (\acs{AGS}, \acs{SPS}) and collider 
(\acs{RHIC}) experiments are shown, as well as the freeze-out 
temperatures and densities from $\chi^2$-minimization fits of the measured 
particle yields versus the yields calculated from the statistical partition 
function of an ideal hadron-resonance gas for two different approaches: 
the \acf{SHM}~\cite{becattini2004} and the \acf{THM}~\cite{braun-munzinger2003}). 
\acs{THM} assumes full thermodynamical equilibrium using the grand canonical
ensemble, whereas \acs{SHM} allows the non-equilibrium fluctuation 
of the total strangeness content by introducing one additional parameter
($\gamma_{\rm S}$) to account for the suppression of hadrons 
containing valence strange-quarks.

A more fundamental, complementary method to explore the qualitative 
features of the \ac{QGP} and to quantify its properties is the numerical 
evaluation of expectation values from path-integrals in discrete 
space-time on a lattice~\cite{wilson1974}.~\footnote{The effect of discrete 
space-time is to regularize the ultra-violet divergences, since distances smaller 
than the lattice spacing corresponding to large momentum exchanges are neglected. 
Depending on the discrete approximation of the continuum action and 
realization of the fermionic degrees of freedom on the lattice 
(Wilson or Kogut-Susskind fermions), quite significant systematic errors may
be introduced. Variation of the lattice parameters and bare couplings 
(and in principle also masses) in accordance with the renormalization group 
equations pave the way for a proper normalization scheme and 
allow the extrapolation of the continuum and chiral limit 
(see, for example, \Refs{karsch2001,loizides2001} and references therein).}
As phase transitions are related to large-distance phenomena, implying
correlations over a large volume, and because of the increasing strength 
of \ac{QCD} interactions with distance, such phenomena cannot be treated 
using perturbative methods. 

\begin{figure}[htb]
\begin{center}
\includegraphics[width=13cm]{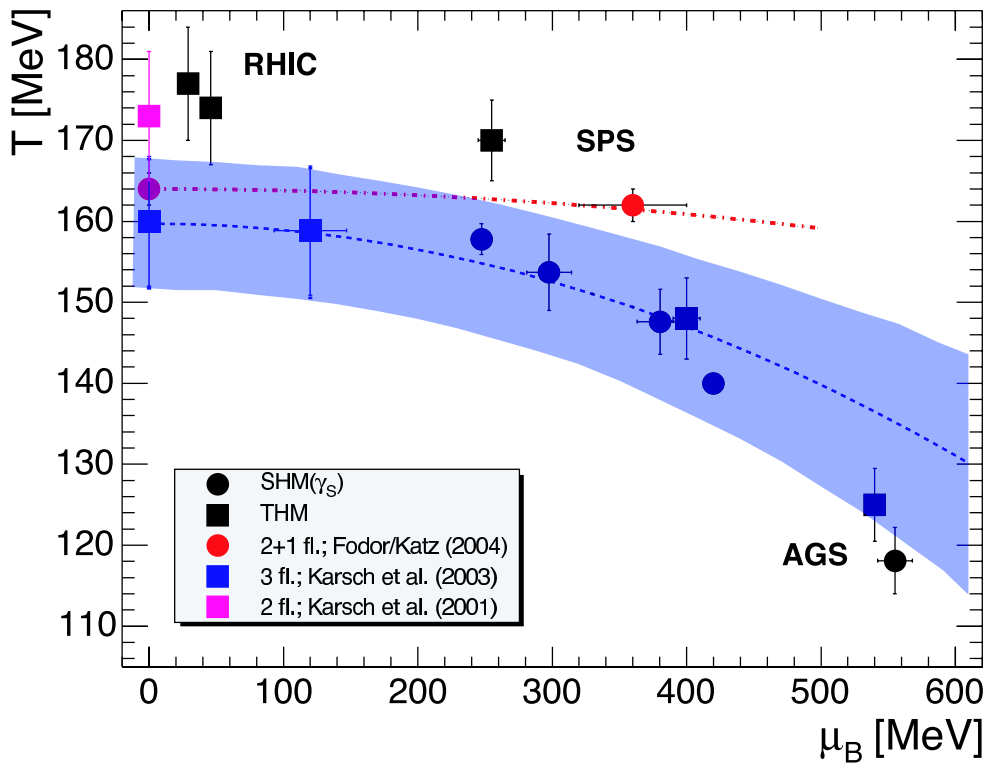}
\end{center}
\vspace{-0.3cm}
\caption[xxx]{The phase diagram of strongly interacting \acs{QCD} matter as 
a function of the temperature $T$ and of the baryonic chemical potential $\mub$. 
The freeze-out points for \acs{SHM} are taken from \Ref{becattini2004} and for 
\acs{THM} from \Ref{braun-munzinger2003}; the lattice point in the chiral limit 
at $\mub=0$ for two flavours~from~\Ref{karsch2000} and the phase-boundary curves 
for $2+1$~flavours (red curve) with physical quark masses from~\Ref{fodor2004} 
and three flavours (blue curve) from~\Ref{karsch2003b} in the chiral limit. 
When available, statistical errors are shown.}
\label{chap2:fig:phasediag}
\end{figure} 

Lattice calculations are most reliably performed for a baryon-free  
system, as the introduction of a finite potential $\mub\ne 0$
in the Wick-rotated Euclidian path-integrals imposes severe problems
for the numerical Monte Carlo evaluation.~\footnote{The reason is that
for non-vanishing potential the functional measure, 
the determinant of the Euclidian Dirac operator, becomes complex, 
thus spoiling the Monte Carlo technique based on `importance sampling'.}
Recently, several methods have been introduced allowing one to 
address moderate chemical potentials on the lattice~\cite{fodor2002}.
The results obtained from different lattice calculations are shown 
in~\fig{chap2:fig:phasediag} for a $2$-flavours calculation in the chiral 
limit~\cite{karsch2000} 
and two very recent calculations 
of the phase boundary: $2+1$ flavours with physical quark masses~\cite{fodor2004} 
and three flavours in the chiral limit~\cite{karsch2003b}.
The precise location of the various phase points and the nature of the phase 
transition vary quantitatively but also qualitatively depending on the number 
of flavours, their (bare) masses 
and the extrapolation to the chiral and continuum limit (if done at all). 
These recent results support the following picture:
The phase transition is of first order starting from 
($T$,~$\mub$) at low temperature and high density till the critical 
end-point ($\tc^{\rm E}$,~$\mub^{\rm E}$) at low baryo-chemical 
potentials of about $100$--$500~\mev$ followed by a cross-over 
region until (\tc,~$0$).~\footnote{The transition is of second 
order in the chiral limit of $2$-flavours \acs{QCD} and of first order 
in the chiral limit of $3$-flavours \acs{QCD}. For physical quark masses 
it is expected to be a (rapid) cross-over~\cite{stephanov2004}.}
The exact location of the end-point is of great interest for the 
heavy-ion community. Its precise determination, however,
highly depends on the values for the quark masses used on the lattice.

\subsection{Towards the Stefan-Boltzmann limit}
\label{chap2:towards}
The cross-over, however, is expected to take place in a narrow
temperature interval, which makes the transition between the hadronic 
and partonic phases quite well localized. This is reflected in a rapid rise 
of the energy density in the vicinity of the cross-over temperature 
in \fig{chap2:fig:qcdeos} for lattice calculations at $\mub=0$.
Shown is the normalized energy density $\varepsilon/T^4$ as a
function of $T$ for the pure gauge sector of \ac{QCD} alone, 
for $2$- and $3$-flavours \ac{QCD} in the chiral limit, as well as
the expected form for the case of two degenerate light and one heavier 
(strange) quark with $m_{\rm s} \sim \tc$ (indicated by the stars)~\cite{karsch2004}.

The number of flavours and the masses of the quarks constitute the main 
uncertainties in the determination of the critical temperature 
and critical energy density. They are estimated to be
$\tc=175\pm 15~\mev$ and 
$\varepsilon_{\rm C}/\tc^4 \simeq 6\pm 2$ leading to 
$\varepsilon_{\rm C}\simeq 0.3$--$1.3~\gev/\fm^3$.
Clearly, the transition is not of first order, which would be characterized 
by a discontinuity of $\varepsilon$ at $T=\tc$. 
However, a large increase of $\Delta\varepsilon/\tc^4\simeq 8$ in the 
energy density is observed in a small temperature 
interval of only about $20$--$40~\mev$ for the 
$2$-flavours calculation.~\footnote{This fact is sometimes interpreted as the
latent heat of the transition.}
The dramatic increase of $\varepsilon/T^4$ is related 
to the change of $n_{\rm dof}$ in~\eq{chap2:eq:epsidealgas} from 3 in the 
pion-gas phase to $37$~for two flavours and $47.5$ for three flavours in the 
deconfined phase, as soon as the additional colour and quark flavour degrees 
of freedom become available. The transition temperature for the physically realized
quark-mass spectrum ($2+1$-flavours \ac{QCD}) is expected to be close to 
the value for two~flavours, since the strange quarks have a mass of 
$m_{\rm s} \sim \tc$ and therefore do not contribute 
to the physics at a temperature close to $\tc$, but will 
do so at higher temperature.

In general, $\mub=0$ is not valid for heavy-ion collisions, 
since the two colliding nuclei carry a total baryon number equal 
to twice their mass number. But, the baryon content of the system 
after the collision is expected to be concentrated rather near the rapidity of 
the two colliding nuclei. Therefore, the larger the rapidity of the 
beams, with respect to their centre of mass, the lower the 
baryo-chemical potential in the central rapidity region. The rapidities 
of the beams at \ac{SPS}, \ac{RHIC} and \ac{LHC} are 2.9, 5.3 and 8.6, 
respectively. Thus, the \ac{LHC} at mid-rapidity is expected to be much more 
baryon-free than \ac{RHIC} and closer to the conditions simulated in lattice \ac{QCD} 
for $\mub=0$.

\begin{figure}[htb]
\begin{center}
\includegraphics[width=14cm]{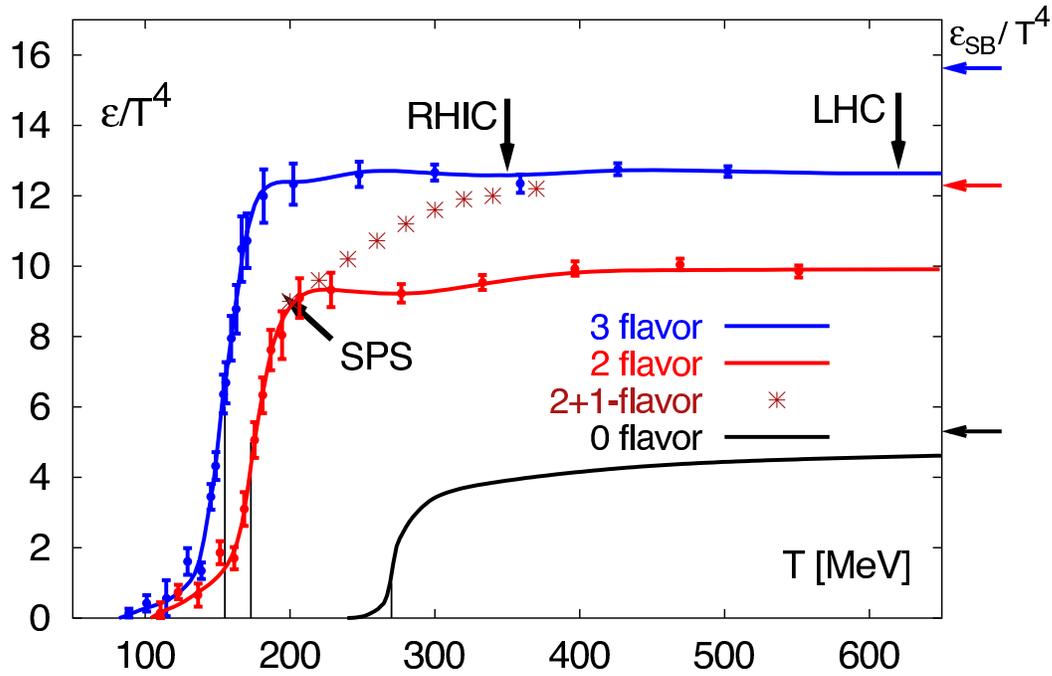}
\end{center}
\vspace{-0.3cm}
\caption[xxx]{The (normalized) energy density in \ac{QCD} with different 
(degenerate) quark flavours at $\mub=0$ in the chiral limit 
and a sketch of the expected form of the energy density for 
\acs{QCD} with two degenerate light quarks and a fixed strange 
quark mass $m_{\rm s} \sim \tc$~\cite{karsch2004}. 
The arrows indicating the energy densities reached in the initial stage 
of heavy-ion collisions at the \acs{SPS}, \acs{RHIC} and \ac{LHC} 
are based on the Bjorken estimate, \eq{chap2:eq:bjorkendensity}. The arrows
on the right-side ordinates show the value of the Stefan-Boltzmann
limit for an ideal quark-gluon gas, \eq{chap2:eq:epsidealgas}.}
\label{chap2:fig:qcdeos}
\end{figure} 

The difference of $\varepsilon$ computed on the lattice compared
to the Stefan-Boltzmann limit calculated from \eq{chap2:eq:epsidealgas}
with $n_{\rm dof}=16$ (gluons only), $37$ (two flavours) and $47.5$ (three flavours)
(see \fig{chap2:fig:qcdeos}) indicates that significant non-perturbative effects
are to be expected at least up to temperatures $T\simeq (2$-$3)\,\tc$. 
The strong coupling constant in the range $T\ge \tc$ is 
estimated~\cite{kajantie2003} as
\begin{equation*}
 \alpha_{\rm S}(T) = \frac{4\pi}{18\,\ln(5\,T/\lQCD)}=\left\{
\begin{array}{ll}
0.43 & {\rm for}~T=\tc \\
0.3  & {\rm for}~T=2\,\tc \\
0.23 & {\rm for}~T=4\,\tc \\
\end{array}\right.
\end{equation*}
where the numbers are obtained by using the fact that 
the QCD scaling constant $\lQCD\approx 200~\mev$ 
is of the same order of magnitude as $\tc$. 
The values for \as\ confirm that non-perturbative 
effects are still sizeable in the range $T<2\,\tc$, where 
the \ac{QCD} recently has been called \acs{sQGP}~\cite{shuryak2004}.
With an initial temperature of about $4$--$6~\tc$ predicted for 
central \PbPb~collisions at $\snn=5.5~\tev$ (see \tab{chap2:tab:spsrhiclhc}), 
the \ac{LHC} will provide quite ideal conditions (with smaller non-perturbative 
effects) possibly allowing a direct comparison to perturbative calculations:
\fi

\section{Novel kinematical range}
\label{chap2:novelaspects}
\ifnovelaspects
Heavy-ion collisions at the \ac{LHC} access not only a quantitatively 
different regime of much higher energy density providing
ideal initial conditions, but also a qualitatively 
new regime of parton kinematics, mainly because:

\begin{itemize}
\item Saturated parton distributions dominate particle production;
\item Hard processes contribute significantly to the total \AaAa\ cross section.
\end{itemize}

\ifallpages
\pagebreak
\fi
\subsection{Low-x parton distribution functions}
\label{chap2:lowx}
In the inelastic, hard collision of an elementary particle
with an hadron, the Bjorken-$x$ variable (in the infinite-momentum frame) 
is essentially determined by the fraction of the hadron momentum 
carried by the parton that enters the hard scattering process. 
The hard scatter is characterized by the momentum transfer 
squared, $Q^2=-q^2$, between the elementary particle and the 
participating parton in the inelastic scattering process.
$Q$ is called the virtuality and typically represents the 
hard-scattering scale of \ac{pQCD}~\cite{soper2000}. 

\begin{figure}[b!]
\begin{center}
\subfigure[Structure function, $F_2$]{
\label{chap2:fig:f2data}
\includegraphics[width=7.5cm]{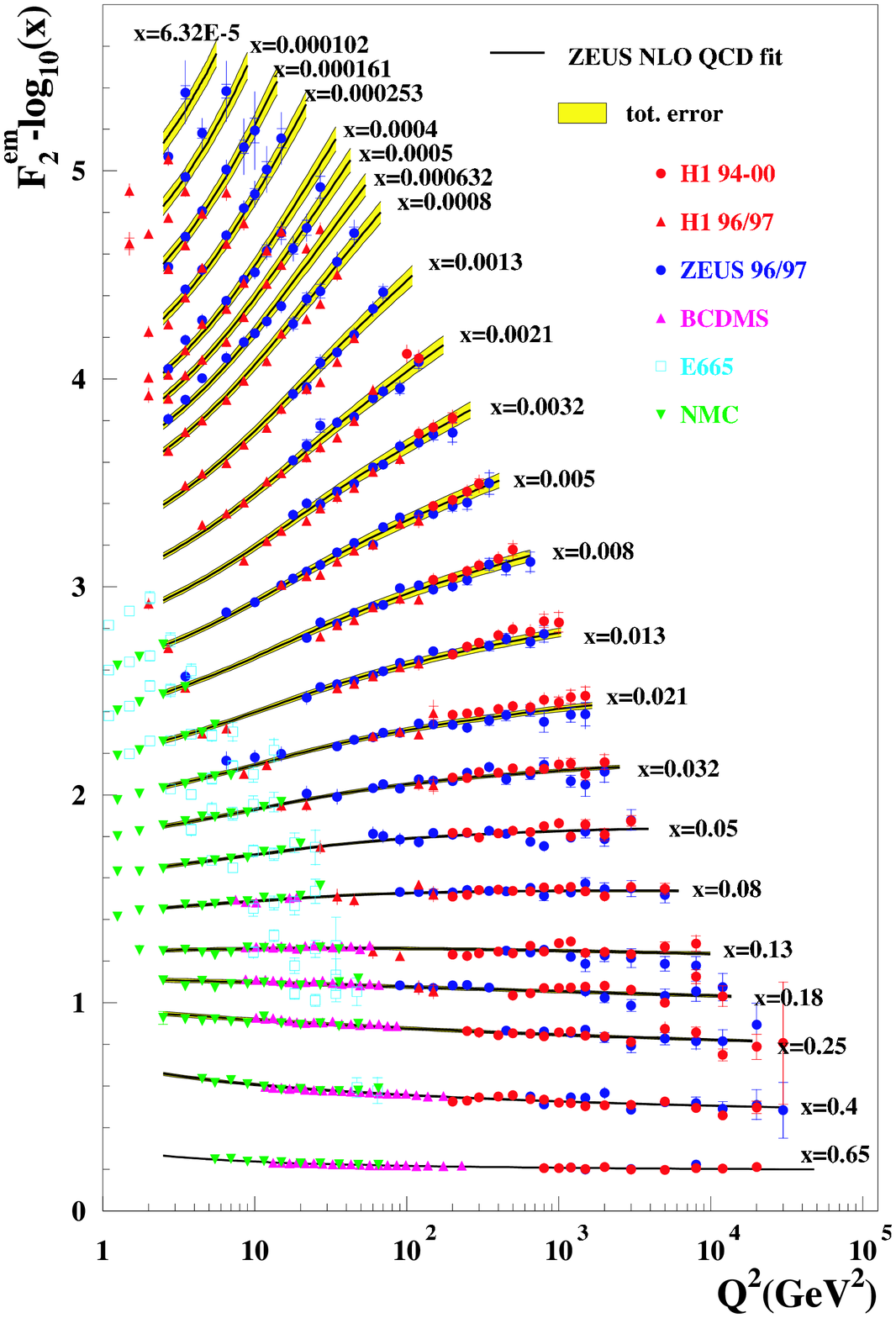}}
\hspace{0.3cm}
\subfigure[Comparison of global fits]{
\label{chap2:fig:pdfglobal}
\includegraphics[width=6.5cm,height=8cm]{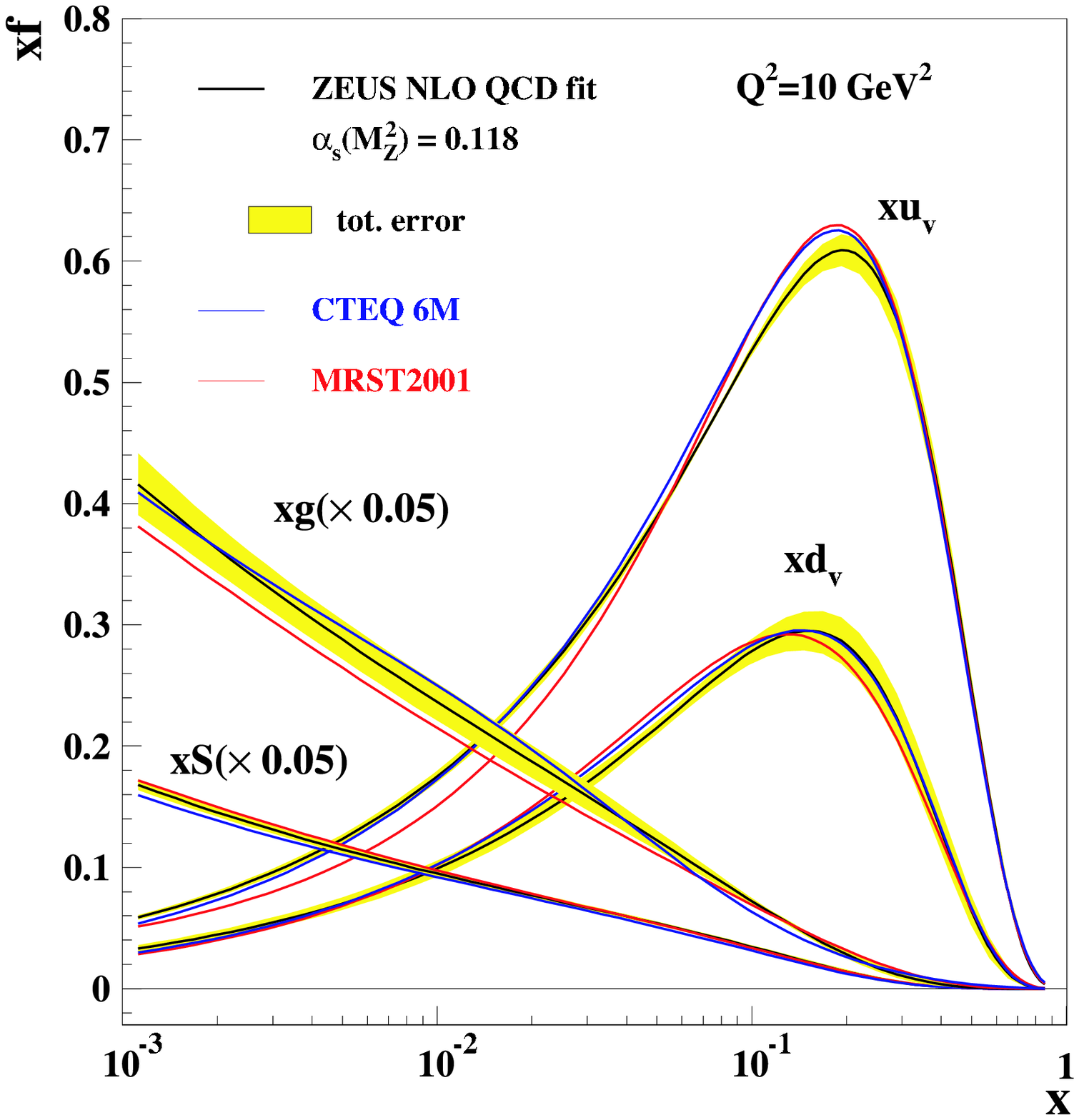}}
\vspace{-0.1cm}
\caption[xxx]{\subref{chap2:fig:f2data} $F_2$ (from pure $\gamma$ exchange) from 
\acs{HERA} and fixed target experiments compared with \acs{ZEUS} 
\acs{NLO} \acs{QCD} fit. 
\subref{chap2:fig:pdfglobal} Comparison of the \acp{PDF} from the 
\acs{ZEUS} fit~\cite{werner2003} to global fits by 
\acs{MRST}2001~\cite{martin2001} and \acs{CTEQ}~6M~\cite{pumplin2002}. 
Both figures are taken from \Ref{werner2003}.} 
\label{chap2:fig:f2pdf}
\end{center}
\end{figure}

The momentum-fraction distribution for a given parton type 
(\eg~gluon, valence quark, sea quark), $f_i(x)$, is called 
\ac{PDF}. It gives the probability that a parton of type $i$ 
carries a fraction $x$ of the hadron's (longitudinal) momentum. 
The \ac{PDF} cannot be computed by perturbative methods 
and, so far, it has not been possible to compute them with lattice 
methods either. Thus, non-perturbative input from data on 
various hard processes must be taken for its extraction. 
The momentum distributions of partons within a hadron are 
assumed to be universal, which is one of the essential features of 
\ac{QCD}. In other words, the \acp{PDF} derived from any process 
can be applied to other processes. Uncertainties from the \acp{PDF} 
result from uncertainty in the input data.
The main experimental knowledge on the proton \acp{PDF} comes 
from \ac{DIS} measuring the proton structure functions, 
in particular from \acs{HERA} for the 
small-$x$ region~\cite{olsson2001,werner2003}. 

\Fig{chap2:fig:f2data} shows the proton structure 
function $F_2(x,Q^2)$  measured at \acs{H1} and \acs{ZEUS} 
together with data from fixed-target experiments. 
The steep rise of $F_2$ at small $x$ is driven by the gluons.
The data are spread over four orders of magnitude in $x$ and $Q^2$ 
and are well described by the \acf{DGLAP} parton 
evolution~\cite{gribov1972,gribov1972b,altarelli1977,dokshitzer1977}.

Several groups (\acs{MRST}~\cite{martin1998,martin2001}, 
\acs{CTEQ}~\cite{lai1996,lai1999,pumplin2002}, \acs{GRV}~\cite{gluck1998}) 
have developed parametrizations for the \acp{PDF} by global 
fits to most of the available \ac{DIS} data, where typically the 
\acp{PDF} are parametrized at a fixed starting scale $Q^2=Q^2_0$ 
and determined by a \ac{NLO} \ac{QCD} fitting procedure.~\footnote{\acf{LO} 
means that the perturbative calculation only contains Feynman diagrams of lowest 
(non-zero) order in $\as$; \acf{NLO} calculations include also diagrams 
of the next oder in $\as$ (see \psect{chap3:inclusivexsec}).} 
Using the framework of \ac{DGLAP} for parton evolution in \ac{pQCD}
one then can extrapolate the \acp{PDF} at different kinematical
ranges. The extracted \acp{PDF} derived by \acs{ZEUS}~\cite{werner2003} at the 
scale of $Q^2=10~\gev^2$ compared to the global analyses of 
\acs{MRST}2001~\cite{martin2001} and \acs{CTEQ}~6M~\cite{pumplin2002} 
are shown in~\fig{chap2:fig:pdfglobal}. Within the estimated total error 
on the \acp{PDF} the different sets are consistent.

\begin{figure}[htb]
\begin{center}
\subfigure[\mbox{$Q=10~\gev$}]{
\label{chap2:fig:cteqpdfsa}
\includegraphics[width=7cm]{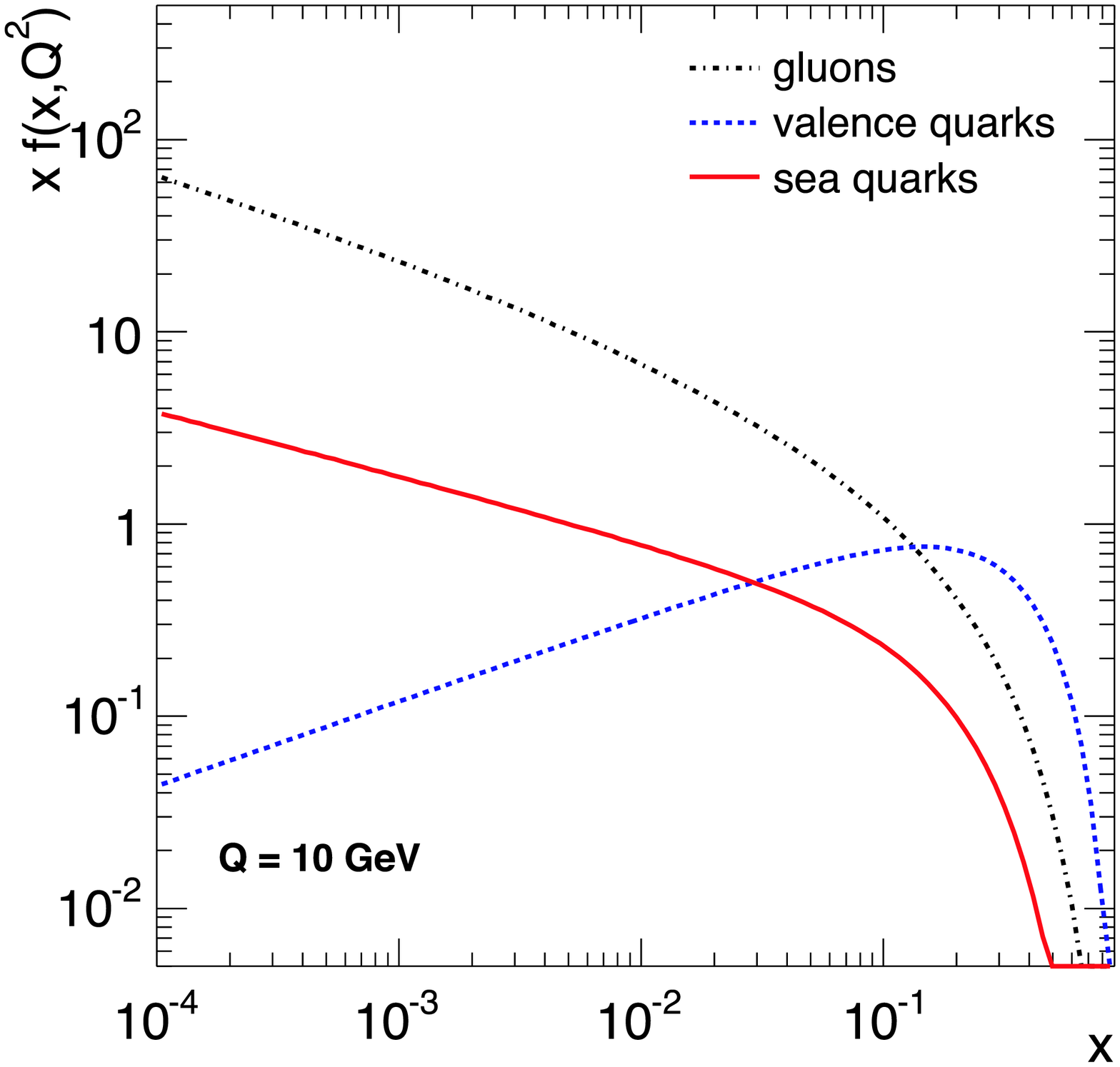}}
\hspace{0.5cm}
\subfigure[\mbox{$Q=100~\gev$}]{
\label{chap2:fig:cteqpdfsb}
\includegraphics[width=7cm]{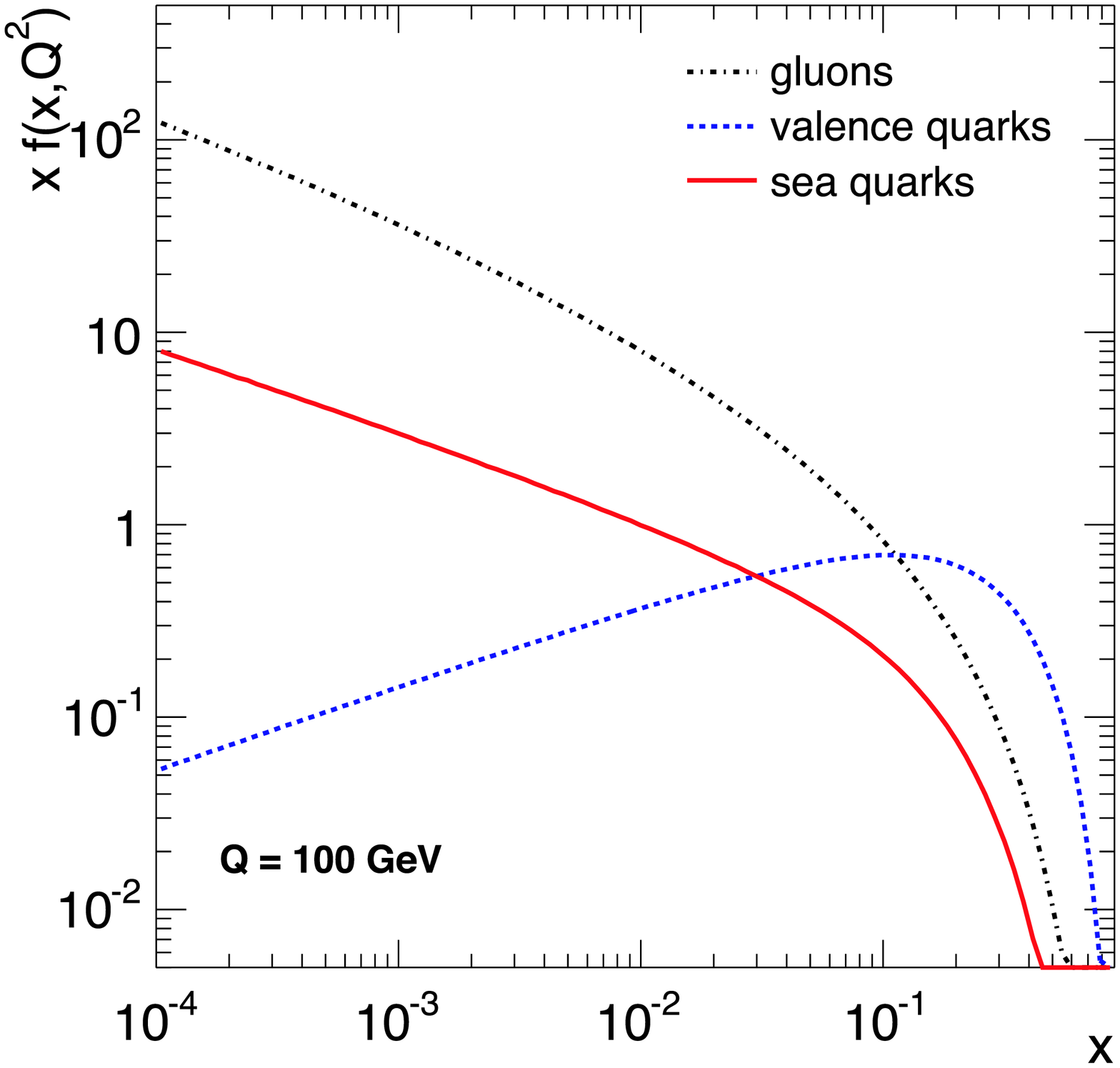}}
\end{center}
\vspace{-0.5cm}
\caption[xxx]{The \acs{CTEQ}~4L parametrization of the proton \acsp{PDF} for 
valence quarks, sea quarks and gluons inside 
the proton at the scale of \mbox{$Q=10~\gev$} and \mbox{$Q=100~\gev$}.} 
\label{chap2:fig:cteqpdfs}
\end{figure}

\ifallpages
\enlargethispage{0.5cm}
\fi
\Fig{chap2:fig:cteqpdfs} shows the \acp{PDF} of valence quarks,
sea quarks and gluons inside the proton at two scales, $Q=10~\gev$ and 
$Q=100~\gev$, in the \acs{CTEQ}~4L parametrization~\cite{lai1996}. Note that 
the \acsp{PDF} are weighted by $x$ to indicate the differences between the 
parton types, which somewhat hides the strong growth of the gluon and sea 
quark contribution at small~$x$. We shall see in the next section that at 
\acs{LHC} values of $x\gsim 0.005$ contribute to the production of jets at 
mid-rapidity. 

\subsection{Accessible x-range at the \acs{LHC}}
\label{chap2:accessablex}
At the \ac{LHC} the \acp{PDF} of the nucleon and, in the case of \pA~and 
\AaAa\ collisions, their modifications in the nucleus, will be probed down 
to unprecedented low values of $x$. 
In the following, we consider the case of the production of a dijet 
through \ac{LO} two-parton kinematics $p_1+p_2=p_3+p_4$ (\eg~gluon--gluon, 
quark--qluon or quark--quark scattering) in the collision of 
two ions ($A_1$,~$Z_1$) and ($A_2$,~$Z_2$).~\footnote{The derivation is done 
along the lines of \Ref{thesisdainese}. A similar calculation can be found 
in \Ref{ellisqcd}.}
The $x$ range actually probed depends on the value of the \cms\ energy per 
nucleon pair $\snn$, on the invariant mass $M_{\rm jj}$ of the dijet produced 
in the hard scattering representing the virtuality of the process and on its 
rapidity $y_{\rm jj}$.\footnote{The invariant mass for two particles with 
four-momenta $(E_1,\vec{p}_1)$ and $(E_2,\vec{p}_2)$ is defined as the modulus 
of the total four-momentum $M=\sqrt{(E_1+E_2)^2-(\vec{p}_1+\vec{p}_2)^2}$.}
Neglecting the intrinsic transverse momenta of the partons in the 
nucleon, we can approximate the four-momenta of the two incoming partons
by \mbox{$p_1=(x_1,0,0,x_1) \, Z_1/A_1 \,\sqrt{s_{\rm pp}}/2$} and 
\mbox{$p_2=(x_2,0,0,-x_2) \, Z_2/A_2 \,\sqrt{s_{\rm pp}}/2$},
where $x_1$ and $x_2$ are the momentum fractions carried by the partons, 
and $\sqrt{s_{\rm pp}}$ is the \cms\ energy for \pp\ collisions 
($14~\tev$ at the \ac{LHC}).
Thus, we derive the square of the invariant mass of the dijet
\begin{equation*}
\label{chap2:eq:sx1x2M2}
M^2_{\rm jj} = \hat{s} = x_1\,x_2\,s_{\rm NN} = 
x_1\,\frac{Z_1}{A_1} \, x_2\,\frac{Z_2}{A_2}\,s_{\rm pp}
\end{equation*}
and its longitudinal rapidity in the laboratory system
\begin{equation*}
\label{chap2:eq:rapidityx1x2}
y_{\rm jj} = \frac{1}{2}\ln \left[\frac{E+p_z}{E-p_z}\right] 
= \frac{1}{2}\ln\left[\frac{x_1}{x_2} \, 
\frac{Z_1\,A_2} {Z_2\,A_1}\right]\;.
\end{equation*}
From the two relations we get the dependence of $x_1$ and $x_2$ on the properties
of the colliding system, $M_{\rm jj}$ and $y_{\rm jj}$, as
\begin{equation*}
x_1 = \frac{A_1}{Z_1}\,\frac{M_{\rm jj}}{\sqrt{s_{\rm pp}}}
\exp\left({+y_{\rm jj}}\right) \hspace{0.5cm} \text{and} \hspace{0.5cm}
x_2 = \frac{A_2}{Z_2}\,\frac{M_{\rm jj}}{\sqrt{s_{\rm pp}}}
\exp\left({-y_{\rm jj}}\right)\;,
\end{equation*}
which for a symmetric colliding system ($A_1=A_2$, $Z_1=Z_2$) simplifies to
\begin{equation*}
\label{chap2:eq:yx1x2}
x_1 = \frac{M_{\rm jj}}{\snn}\exp\left({+y_{\rm jj}}\right)
\hspace{0.5cm} \text{and} \hspace{0.5cm}
x_2 = \frac{M_{\rm jj}}{\snn}\exp\left({-y_{\rm jj}}\right)\;.
\end{equation*}

In the case of asymmetric collisions, as \mbox{p--Pb} and \mbox{Pb--p}, 
the centre of mass moves with a longitudinal rapidity
\begin{equation*}
y_{\rm c.m.} = \frac{1}{2}\ln\left(\frac{\rm Z_1 A_2}{\rm Z_2 A_1}\right)
\end{equation*}
and the rapidity window covered by the experiment is consequently shifted by
\begin{equation*}
\Delta y = y_{\rm lab.~system}-y_{\rm c.m.~system} = y_{\rm c.m.}
\end{equation*}
corresponding to $+0.47$ ($-0.47$) for \mbox{p--Pb} (\mbox{Pb--p}) collisions.
Therefore, running with both \mbox{p--Pb} and \mbox{Pb--p} will allow one
to cover the largest interval in Bjorken-$x$. 

\begin{figure}[htb]
\begin{center}
\includegraphics[width=10cm]{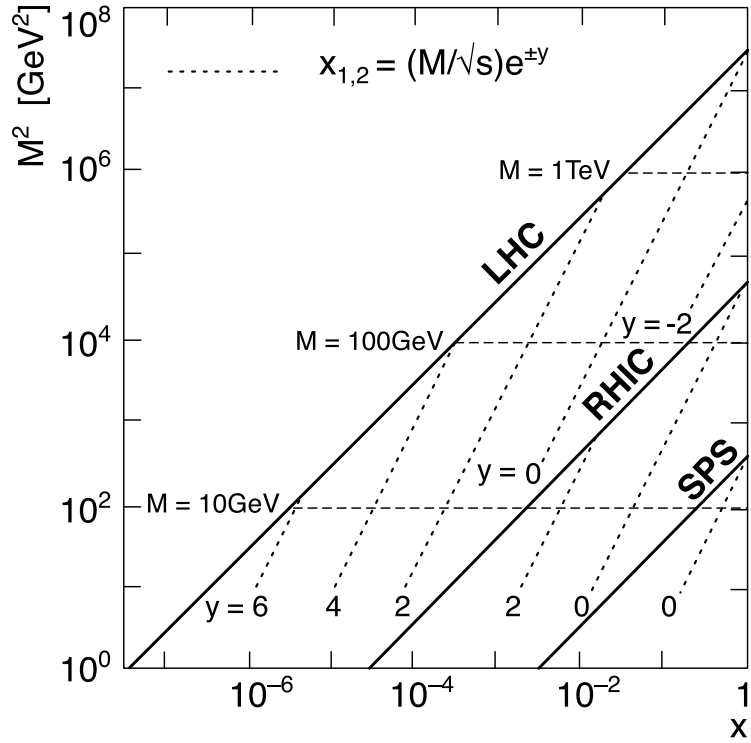}
\end{center}
\vspace{-0.3cm}
\caption[xxx]{The range of Bjorken-$x$ and $M^2$ relevant for particle 
production in \AaAa\ collisions at the top \acs{SPS}, \acs{RHIC}
and \acs{LHC} energies. Lines of constant rapidity are indicated.} 
\label{chap2:fig:partonkinematics}
\end{figure}

\begin{table}[htb!]
\vspace{0.5cm}
\begin{center}
\begin{tabular}{lcccc}
\hline\hline
Machine    & \acs{SPS} & \acs{RHIC} & \acs{LHC} & \acs{LHC} \\
System     & \PbPb     & \AuAu      & \PbPb     & \pp       \\
$\snn$     & 17~\gev   & 200~\gev   & 5.5~\tev  & 14~\tev   \\
\hline
$\et=10\hide{0}~\gev$  & $x\simeq1.18$  & $x\simeq0.10$ & $x\simeq0.004$ & $x\simeq0.001$\\ 
$\et=25\hide{0}~\gev$  & -              & $x\simeq0.25$ & $x\simeq0.009$ & $x\simeq0.004$\\ 
$\et=50\hide{0}~\gev$  & -              & $x\simeq0.50$ & $x\simeq0.018$ & $x\simeq0.007$\\
$\et=100~\gev$         & -              & -             & $x\simeq0.036$ & $x\simeq0.014$\\
$\et=200~\gev$         & -              & -             & $x\simeq0.072$ & $x\simeq0.028$\\
\hline
\hline
\end{tabular}
\end{center}
\vspace{-0.4cm}
\caption[xxx]{Bjorken-$x$ values probed by jet production at various transverse 
jet energies for different systems at mid-rapidity.}
\label{chap2:tab:xtable}
\end{table}

\Fig{chap2:fig:partonkinematics} shows the range of accessible
values of Bjorken-$x$ in \AAex\ collisions at the \acs{SPS}, 
\acs{RHIC} and \acs{LHC} energies. Clearly, the
\ac{LHC} will open a novel regime of $x$-values as low as $10^{-5}$, 
where strong gluon shadowing is expected and the initial gluon density
is close to saturation, such that the time evolution of the system
might be described by classical chromodynamics (see \sect{chap2:npdf}).

At central rapidities for $y_{\rm jj}\approx 0$ we have $x_1\simeq x_2$, 
and their magnitude is determined by the ratio of the invariant 
mass to the \cms\ energy. The invariant dijet mass is given by the 
transverse jet energy, $M_{\rm jj} = 2\,\et \simeq 2\,\pt$;
therefore with $\xt=2\,\pt/\snn$ we get $x_1\simeq x_2 \simeq \xt$. 
In terms of the outgoing parton momenta, $p_1+p_2=p_3+p_4$, 
we find by applying momentum conservation (at \ac{LO})
\begin{equation}
\label{chap2:eq:yx1x2out}
x_1 = \frac{1}{2}\,\xt\,\left( e^{y_3} + e^{y_4} \right)
\hspace{0.5cm} \text{and} \hspace{0.5cm}
x_2 = \frac{1}{2}\,\xt\,\left( e^{-y_3} + e^{-y_4} \right)\;.
\end{equation}

\Tab{chap2:tab:xtable} reports the Bjorken-$x$  values for jets with 
transverse energy between  \mbox{$\et=10~\gev$} and \mbox{$\et=200~\gev$} 
for a variety of systems. The $x$-regime relevant for jet production of 
$50$--$200$~\gev\ at \ac{LHC} ($0.005\lsim x\lsim0.1$) is between one and 
two orders of magnitude smaller than at \ac{RHIC}, where the cross 
section for the hard jets ($\et\ge50~\gev$) is essentially zero 
(see \tab{chap2:tab:ettable}).

\subsection{Nuclear-modified parton distribution functions}
\label{chap2:npdf}
So far, we have looked at the \acp{PDF} extracted from the 
structure function $F_{2}$ of the free proton. Experimentally
for various nuclei, the ratios of $F_2^A$ to the structure function 
of deuterium, $F_2^A(x,Q^2)/F_2^D(x,Q^2)$, reveal clear deviations
from unity. 
This indicates that the parton distributions of bound protons 
are different from those of the free protons, 
$f_{i/\rm A}(x,Q^2)\neq f_i(x,Q^2)$.
The nuclear effects in the ratio $F_2^A/F_2^D$ are usually 
divided into the following regions in Bjorken $x$:
\begin{itemize}
\item Fermi motion, an excess for $x\rightarrow1$ and beyond;
\item \acs{EMC} effect, a depletion at $0.3 \lsim x\lsim0.7$;
\item anti-shadowing, an excess at $0.1 \lsim x \lsim 0.3$;
\item nuclear shadowing, a depletion  at $x \lsim 0.1$;
\item saturation effect, saturation of the depletion at $x \lsim 0.001$.
\end{itemize}

\begin{figure}[htb]
\begin{center}
\includegraphics[width=9cm]{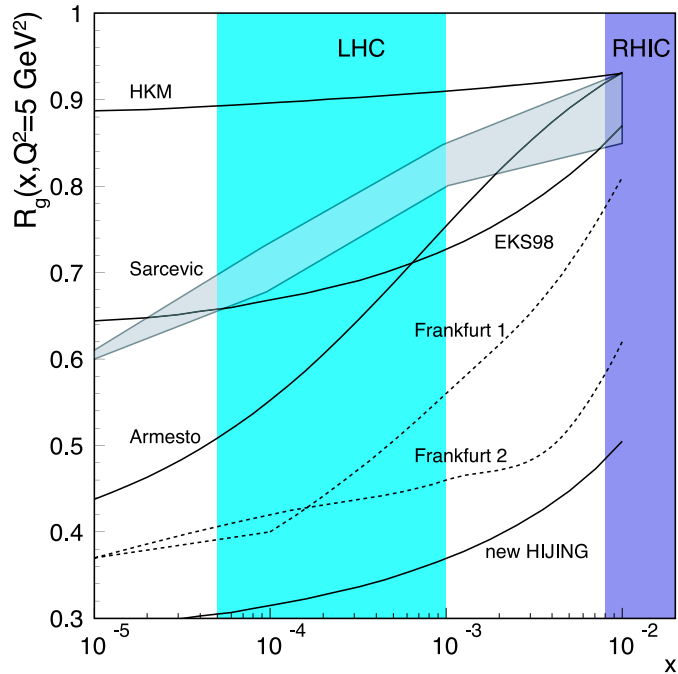}
\vspace{-0.2cm}
\caption[xxx]{Ratio of the gluon distribution in the lead nucleus 
over the gluon distribution in the proton for different models 
at $Q^2=5~\gev^2$, corresponding to $\ccbar$ production at threshold. 
The $x$-region for the production at \ac{RHIC} and \ac{LHC} are
highlighted. The figure is taken from \Ref{armesto2003}, which 
lists the original references to the model calculations.}
\label{chap2:fig:shadowingmodels}
\end{center}
\end{figure}

Currently, there is no unique theoretical description of
these effects. It is believed that different mechanisms are
responsible for them in different kinematic regions~\cite{arneodo1992}.
In a very simplified picture, the extension of the Bjorken-$x$ range 
down to about \mbox{$10^{-3}$--$10^{-5}$} at the \acs{LHC} means 
that a large-$x$ parton in one of the two colliding lead nuclei 
resolves the other incoming nucleus as a superposition of about 
\mbox{${\rm A} \times 10^{3}$--$10^{5}$ $\approx 10^5$--$10^7$} gluons. 
Thus, there are many incoming small-$x$ gluons, which are densely packed 
and have a large wave-length (via the Heisenberg uncertainty) so that the  
low-momentum gluons tend to merge together: two gluons with momentum 
fractions $x_1$ and $x_2$ combine into a single gluon with momentum fraction 
\mbox{$x_1+x_2$ ($g_{x_1}g_{x_2}\to g_{x_1+x_2}$}). As a consequence of the 
combination process towards larger $x$, affecting not only gluons, 
but all partons, the nuclear parton densities are depleted in the small-$x$  
region ($x\lsim 0.1$) and slightly enhanced in the large-$x$ region 
\mbox{($0.1 \lsim x \lsim 0.3$)} with respect to the parton densities of the 
proton. Eventually at certain small-$x$ values, the nuclear gluon densities 
saturate as a result of non-linear corrections to the \ac{DGLAP} 
evolution equations~\cite{gribov1984,mueller1985,iancu2003}.
The saturation scale, which is proportional to the gluon density
per unit area and grows as \mbox{$Q^2_{\rm sat, A}\sim A^{1/3}/x^{\delta}$} 
(\mbox{$\delta\approx 0.2$--$0.3$} at \acs{HERA}), 
determines the critical values of the momentum transfer, at which 
the parton systems becomes dense and recombination frequently 
happens. At \ac{LHC} the expected saturation scale, $Q_{\rm sat, A}>2~\gev$,
is in the perturbative regime and heavy-ion collisions are
depicted as a collision of dense gluon walls.~\footnote{For a short overview of 
the saturation physics see the QM~2004 talk by U.A.~Wiedemann~\cite{wiedemann2004}.}

\begin{figure}[htb]
\begin{center}
\includegraphics[width=9cm]{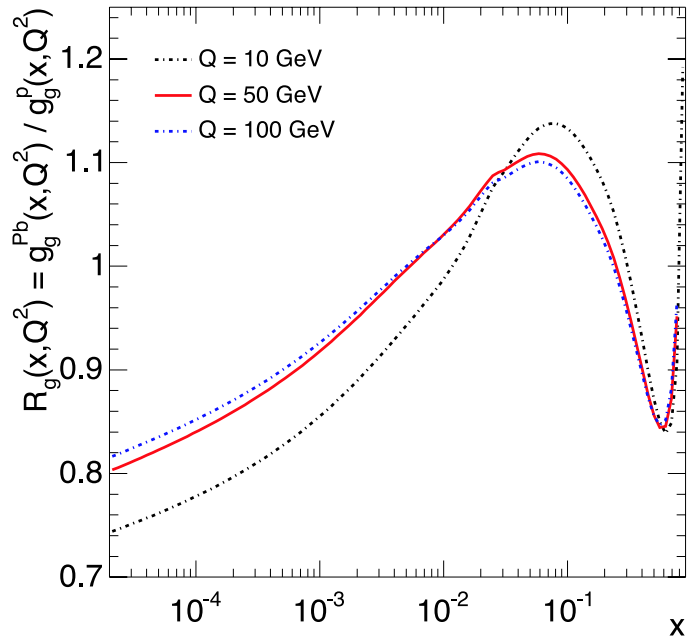}
\vspace{-0.2cm}
\caption[xxx]{Ratio of the gluon distribution in the lead nucleus
over the gluon distribution in the proton for \acs{EKS98}
parametrization for different scales.}
\label{chap2:fig:gluonmodpdf}
\end{center}
\end{figure}

Recently, the nuclear shadowing effect has been analysed 
in the \ac{DGLAP} framework using data from electron--nucleus \ac{DIS} 
in the range $0.005<x<1$~\cite{eskola2001c}. However, no data are available 
in the complete $x$-range covered by the \acs{LHC} and the existing data provide only
weak constraints for the gluon \acp{PDF}, which enter the measured structure 
functions at \ac{NLO}. Two groups, \acs{EKRS}~\cite{eskola1998,eskola1998b} 
and \acs{HKM}~\cite{hirai2001}, applied the same strategies as in the case of the 
proton \acp{PDF}, in order to obtain a parametrization (and extrapolation to 
low-$x$ values) of the \acp{nPDF}.~\footnote{The parametrization of \acs{EKRS} 
is known as \acs{EKS98}.} 
There are a couple of other models, which try to describe \acp{nPDF}. 
Though they all tend to disagree, where no experimental constraints 
are available. The present situation is summarized~\cite{armesto2003} 
in~\fig{chap2:fig:shadowingmodels} representing the results of the 
different models as the ratio of the gluon distribution in the lead nucleus 
over the gluon distribution in the proton,
\begin{equation}
\label{chap2:eq:shadowing}
 R_g(x,Q^2)=\frac{g^{\rm Pb}(x,Q^2)}{g^{\rm p}(x,Q^2)}\;.
\end{equation}

The predictions for the gluon shadowing, $g(x\sim 10^{-3}$--$10^{-5})$, 
at the \ac{LHC} range between 30\% and 90\%. The large uncertainty might 
be reduced in the future by more data from nuclear \ac{DIS}, 
\pA\ data collected at \ac{RHIC} and by the measurements of charm and 
beauty production in \pPb\ at the \acs{LHC}~\cite{thesisdainese}.

In~\fig{chap2:fig:gluonmodpdf} we show the ratio~\eq{chap2:eq:shadowing}
for the \acs{EKS98} parametrization~\cite{eskola1998b} as a function of
$x$ for different scales. The deviation from the proton \ac{PDF} 
in relevant region for jet production at mid-pseudo-rapidity 
($0.005\lsim x\lsim 0.1$) is about 10\%. Therefore, if not otherwise 
indicated, we often neglect the nuclear modification of the \acp{PDF} 
in the present work.

\subsection{Hard scattering processes}
\label{chap2:highQ}
Hard processes are expected to be abundant at \ac{LHC} energies.
Practically in every minimum-bias event at the \ac{LHC} high-$\pt$ partons 
are expected to be produced in scattering processes at 
$Q^2\gg\Lambda^2_{\rm QCD}$.~\footnote{Minimum-bias events are events where 
no (or, at least, as few as possible) selection cuts are applied.}
At the scale much larger than $\lQCD\simeq 200~\mev$, these hard processes can be calculated 
using \ac{pQCD} and are expected to be under reasonable theoretical control. Since high-$\pt$ 
partons tend to fragment hard, the measured transverse-momentum spectrum is expected to be 
harder than at \ac{RHIC} or \ac{SPS}.

\begin{figure}[htb]
\begin{center}
\includegraphics[width=10cm]{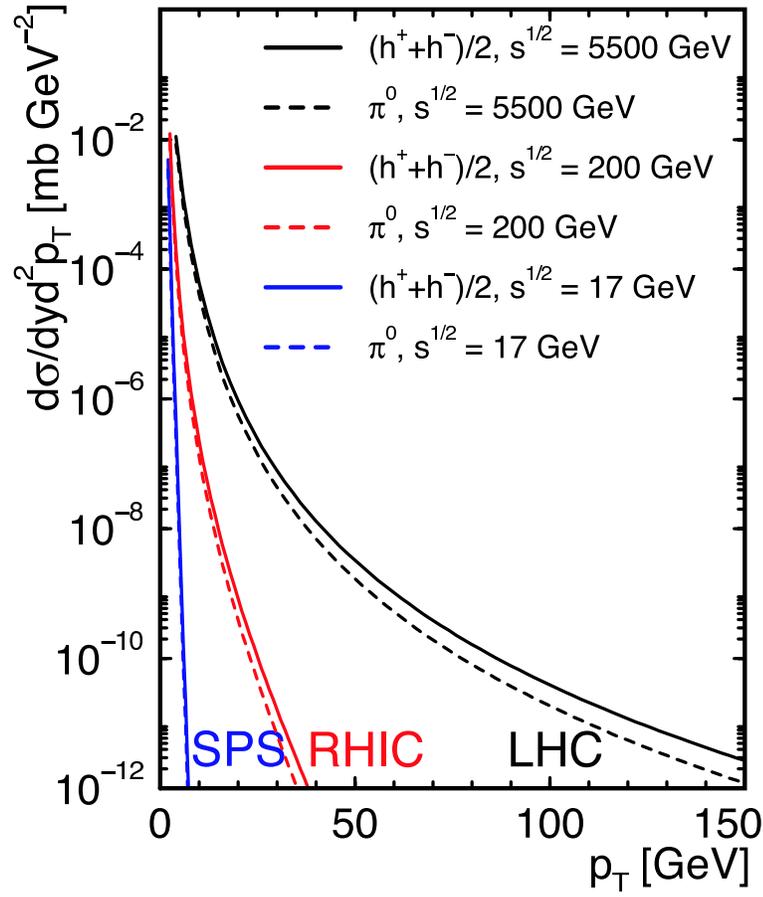}
\vspace{-0.2cm}
\caption[xxx]{The predicted, \acs{LO}, differential cross section 
for neutral pion and inclusive charged hadron production in 
\pp\ collisions at mid-rapidity ($y=0$) for $\snn=17,~200,~5500~\gev$. 
The figure is taken from~\Ref{vitev2002}. (All scales are set to $\pt^2$
and $\av{{k}_{\rm T}^2}=1.8~\gev^2$.)}
\label{chap2:fig:hardptvitev}
\end{center}
\end{figure}

\begin{table}[htb!]
\begin{center}
\begin{tabular}{lcccc}
\hline
\hline
Machine    & \acs{SPS} & \acs{RHIC} & \acs{LHC}  & \acs{LHC} \\
System     & \PbPb     & \AuAu      & \PbPb      & \pp       \\
$\snn$     & $17~\gev$ & $200~\gev$ & $5.5~\tev$ & $14~\tev$ \\
\hline
$\et^0\ge10\hide{0}~\gev$  & - & $1.5\cdot10^{-3}$ & $3.2\cdot10^{-1}$ & $1.4\cdot10^{-0}$\\ 
$\et^0\ge25\hide{0}~\gev$  & - & $1.6\cdot10^{-5}$ & $4.0\cdot10^{-2}$ & $1.6\cdot10^{-1}$\\ 
$\et^0\ge50\hide{0}~\gev$  & - & $1.8\cdot10^{-8}$ & $2.8\cdot10^{-3}$ & $1.3\cdot10^{-2}$\\
$\et^0\ge100~\gev$         & - & -              & $1.5\cdot10^{-4}$ & $8.2\cdot10^{-4}$\\
$\et^0\ge200~\gev$         & - & -              & $5.8\cdot10^{-6}$ & $4.3\cdot10^{-5}$\\
\hline
\hline
\end{tabular}
\end{center}
\vspace{-0.4cm}
\label{chap2:tab:ettable}
\caption[xxx]{The inclusive, accumulated jet cross section, $1/(T_{\rm AB}\sigma^{\rm geo}_{\rm AB}) 
\, \sigma(\et\ge\et^0)$, in units of $\mbarn$ for jets with $\et\ge\et^0$ in $\abs{\eta}<0.5$.}
\end{table}

The qualitative statement is clearly confirmed by \fig{chap2:fig:hardptvitev}, 
which shows the transverse-momentum distribution of neutral pions 
and inclusive charged hadrons predicted by a recent \ac{LO} \ac{pQCD} calculation 
invoking the standard factorized \ac{pQCD} hadron-production formalism~\cite{vitev2002}
(see \psect{chap3:hadronprod}).
The significant hardening of the spectra with $\sqrt{s}$ leads
to two important consequences for \pA\ and \AaAa\ collisions (see below):
a notably reduced sensitivity to initial state (kinematic) effects,
smaller Cronin effect, and larger variation of the final-state 
effects, such as parton energy loss, with $\pt$.

Another impressive example of the expected hardness of the \ac{LHC}
events is given in~\tab{chap2:tab:ettable} reporting the inclusive,
accumulated jet cross section per participant pair, 
$1/(T_{\rm AB}\sigma^{\rm geo}_{\rm AB}) \, \sigma(\et\ge\et^0)$,
for jets with $\et\ge\et^0$ at central pseudo-rapidity ($\abs{\eta}<0.5$)
for various transverse minimum jet energies $\et^0$. At \ac{LHC} energies,
high-$\et$ jets will be copiously produced in heavy-ion collisions 
and therefore for the first time experimentally accessible.
We shall see in the next section and throughout the next chapter that they 
are a very profound tool to probe the partonic medium created in such 
collisions.

\section{Hard processes as probes of QGP}
\label{chap2:hardprobes}
Assuming the absence of nuclear and \ac{QCD} medium effects, 
a \AAex\ collision can be considered as a superposition of 
independent \NN~collisions. Thus, the cross section for hard 
processes should scale from \pp\ to \AaAa\ with 
the number of inelastic \NNex\ collisions according to binary 
scaling~\cite{broniowski2001}. The effects modifying the 
simple scaling with $\Ncoll$ are usually divided in two classes:
\begin{itemize}
\item {\em Initial}-state effects are effects acting on the hard 
cross section in a way that depends on the size and energy of the 
colliding nuclei, but not on the medium formed in the collision,
such as Cronin enhancing~\cite{cronin1974}, nuclear shadowing and 
gluon saturation (described in~\sect{chap2:novelaspects}).
\item {\em Final}-state effects are effects induced by the 
created medium that influence the yields and the kinematic 
distributions of the produced hard partons, such as partonic energy 
loss. Final-state effects are not correlated to initial-state effects;
they depend strongly on the properties of the medium (gluon density, 
temperature and volume). Therefore, they provide information on these 
properties.
\end{itemize}

In order to distinguish the influence of the different effects on the
various observables and to draw conclusions, a systematic study 
of the effects in \pp\, \pA\ and \AaAa\ is required, such as has recently
been undertaken at \ac{RHIC}~\cite{gyulassy2004b}. Initial state effects 
can be studied in \pp\ and \pA\ collisions and then reliably 
extrapolated to \AaAa. If the \ac{QGP} is formed in \AaAa\ collisions, 
the final state effects will be significantly stronger than what is
expected by the extrapolation from \pA.
In this context, hard scattering processes are an excellent experimental 
probe in heavy-ion collisions inasmuch as they posses the following 
interesting properties:

\begin{itemize}
\item They are produced in the early stage of the collision in the
primary, short-distance, partonic scattering 
with large virtuality $Q^2$. Thus, owing to the uncertainty relation,
their production happens on temporal and spatial scales, 
\mbox{$\Delta\tau\sim 1/Q$} and \mbox{$\Delta r\sim 1/Q$}, which are 
sufficiently small to be unaffected by the properties 
of the medium (\ie~by final-state effects) and therefore
they directly probe the partonic phase of the reaction.
\item Because of the large virtuality, their production cross 
section can be reliably calculated with \ac{pQCD} (collinear factorization 
plus Glauber multi-scattering, see \psect{chap3:hadronprod}) or 
via the \ac{CGC} framework~\cite{iancu2003}.
In fact, since \ac{QCD} is asymptotically free~\cite{gross1973,politzer1973},
the running coupling constant (calculated up to two internal loops)
\begin{equation}
\label{chap2:eq:runningalphas}
\as(Q^2) = \frac{1}{b\,L}\, (1-\frac{b^\prime}{b}\,\frac{\ln L}{L})
\hspace{0.2cm}\text{where}\hspace{0.2cm}
L=\ln \frac{Q^2}{\Lambda^2}
\end{equation}
becomes small for large values of $Q^2\gg\Lambda\simeq\lQCD$.~\footnote{The 
scale $\Lambda$ is a fundamental scale and depends on the renormalization 
scheme and the number of active flavours. Its value is of the order of $\lQCD$. 
$b$ and $b^\prime$ are positive constants determined
by the perturbative expansion of the renormalization group equation. Their 
values are independent of the renormalization scheme.}
Hence, the higher-order terms (in general, higher than \ac{NLO}) can be 
neglected in an expansion of the cross section in powers of $\as$.
\item In the absence of medium effects, their cross section in \AaAa\
reactions is expected to simply scale with that measured in \NN\ 
collisions times the number of available point source
scattering centres (binary scaling).
\item They are expected to be significantly attenuated through the 
special \ac{QCD} type energy loss mechanisms, when they propagate 
inside the medium. The current theoretical understanding of these 
mechanisms and of the magnitude of the energy loss are extensively 
covered in the next chapter, with particular focus on the 
suppression of high-energy jets.
\end{itemize}
In short, hard probes are perturbative processes testing non-perturbative 
physics. The input (yields and $\pt$ distributions) is known from 
the measurements carried out in \pp\ (and \pA) interpolated to the \AaAa\ 
energy by means of \ac{pQCD} (and typically scaled according to $\Ncoll$).
The comparison of the measured outcome, after being influenced by the medium, 
to the known input allows to extract information of the medium properties.
Typical probes include the production of quarkonia and heavy flavours~\cite{berardi2004}, 
direct photons and photon tagged jets~\cite{arleo2004}, and ---as we will see
in the next chapter--- jet and di-jet production~\cite{accardi2003} and within 
limited scope leading-particle $\pt$-spectra. 
\fi

%

\newif\ifconcepts
\conceptstrue
\newif\ifpeloss
\pelosstrue
\newif\ifexpresults
\expresultstrue
\newif\ifpqm
\pqmtrue

\chapter{Jets in heavy-ion collisions}
\label{chap3}
High-energy jets are sensitive probes of the partonic medium 
produced in \NNex\ collisions. In fact, they possess all properties
listed at the end of the previous chapter: 

\begin{itemize}
\item Their initial production is not affected by final state effects, 
because the large value of the virtuality $Q=2\,\et$ 
for $\et\ge10~\gev$ implies production space-time 
scales of 
$\sim 1/(2\,\et)\lsim 0.01~\fm$, which are much smaller than the expected 
life-time of the partonic phase at the \ac{LHC}, \mbox{$\tau \gsim 10~\fm$}.
Thus, the early-produced partons (from which the jets eventually emerge) 
will experience the partonic evolution of the collision. 
\item As a consequence of the large virtuality compared to 
$\lQCD$, their production cross section measured in \pp\ (or \ppbar) 
and \pA\ collisions is calculable within the framework of \ac{pQCD}.
If $A^{\frac{1}{3}} \lQCD \ll Q$ and assuming that initial state effects
are known and under control the cross section can be safely scaled to \AAex\ 
collisions. We review the general ideas behind jets physics at hadron colliders 
in~\sect{chap3:jetpp}.
\item Strong final state effects are expected to influence the propagation of 
high-energy partons through the medium formed in \AAex\ collisions. 
Of particular interest is the predicted medium-induced energy loss of the 
hard partons via gluon radiation in a dense partonic medium.
Depending on the hadronization and thermalization lengths of the penetrating 
probes, see \chap{chap6}, jet tomography will be useful 
to investigate such phenomena. 
We summarize the theoretical framework of various `jet quenching' models in 
\sect{chap3:partoneloss}.
\end{itemize}
The experimental situation at \ac{RHIC}, where for the first time hard
processes are experimentally accessible in heavy-ion collisions
with sufficiently high rates, is reported in \sect{chap3:expresultsrhic}
In \sect{chap3:pqm} we describe a final state quenching model, which 
describes most of the high-$\pt$ observables at \ac{RHIC}.

\section{Concepts of jet physics}
\label{chap3:jetpp}
\ifconcepts
In the collision of high-energy hadrons one of four different 
types of scattering reactions can occur: elastic, diffractive, 
soft-inelastic and hard. Elastic collisions are interactions
where the initial and final particles are of the same type and energy. 
They can be regarded as diffractive processes, but involving the exchange 
of quantum numbers of the vacuum only. Inelastic diffractive 
processes are similar, here one or both of the incident hadrons break 
apart. Soft-inelastic collisions also induce the breakup of the 
incident hadrons but at rather low momentum transfers. They are best 
described by exchanges of virtual hadrons (Regge theory) 
and comprise the largest part of the total cross section. 

\begin{figure}[htb]
\begin{center}
\includegraphics[width=10cm]{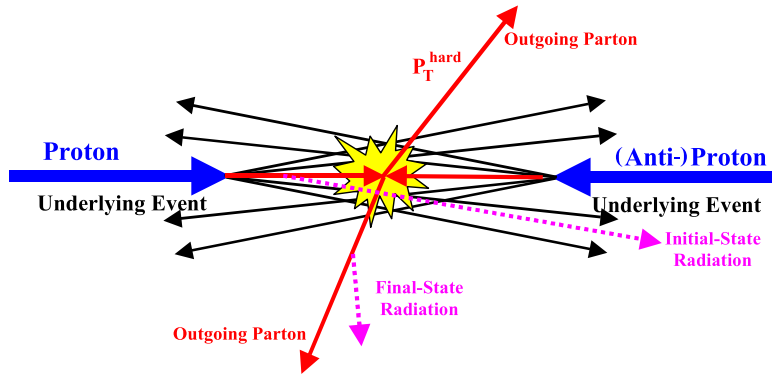}
\end{center}
\vspace{-0.3cm}
\caption[Illustration of the hard $2$-to-$2$ parton scattering.]
{Illustration of $2$-to-$2$ parton scattering in the hard collision 
of two incident hadrons. 
The figure is adapted from~\Ref{field2002}.} 
\label{chap3:fig:hardscatters}
\end{figure}

Of particular interest are the hard collisions, visualized 
in \fig{chap3:fig:hardscatters}, in which the partons within 
the hadrons (\eg~proton or anti-proton) interact directly. 
The incident hadrons break apart and many new particles are created. 
The outgoing partons from the hard sub-process fragment into jets of 
particles. The rest of the particles in the event are rather soft
particles, which mostly arise due to the break up of the remnants of 
the incident hadrons, and together form the 
underlying event. The hard-scattering component of the event consists 
of the outgoing two jets including \ac{ISR} and \ac{FSR}. \ac{ISR} and 
\ac{FSR} introduce corrections to the basic $2$-to-$2$ \ac{QCD} processes, 
which mimic \ac{NLO} topologies in Monte Carlo event generators.

\begin{figure}[b!]
\begin{center}
\subfigure[Experiment]{
\label{chap3:fig:evolution1}
\includegraphics[width=6cm]{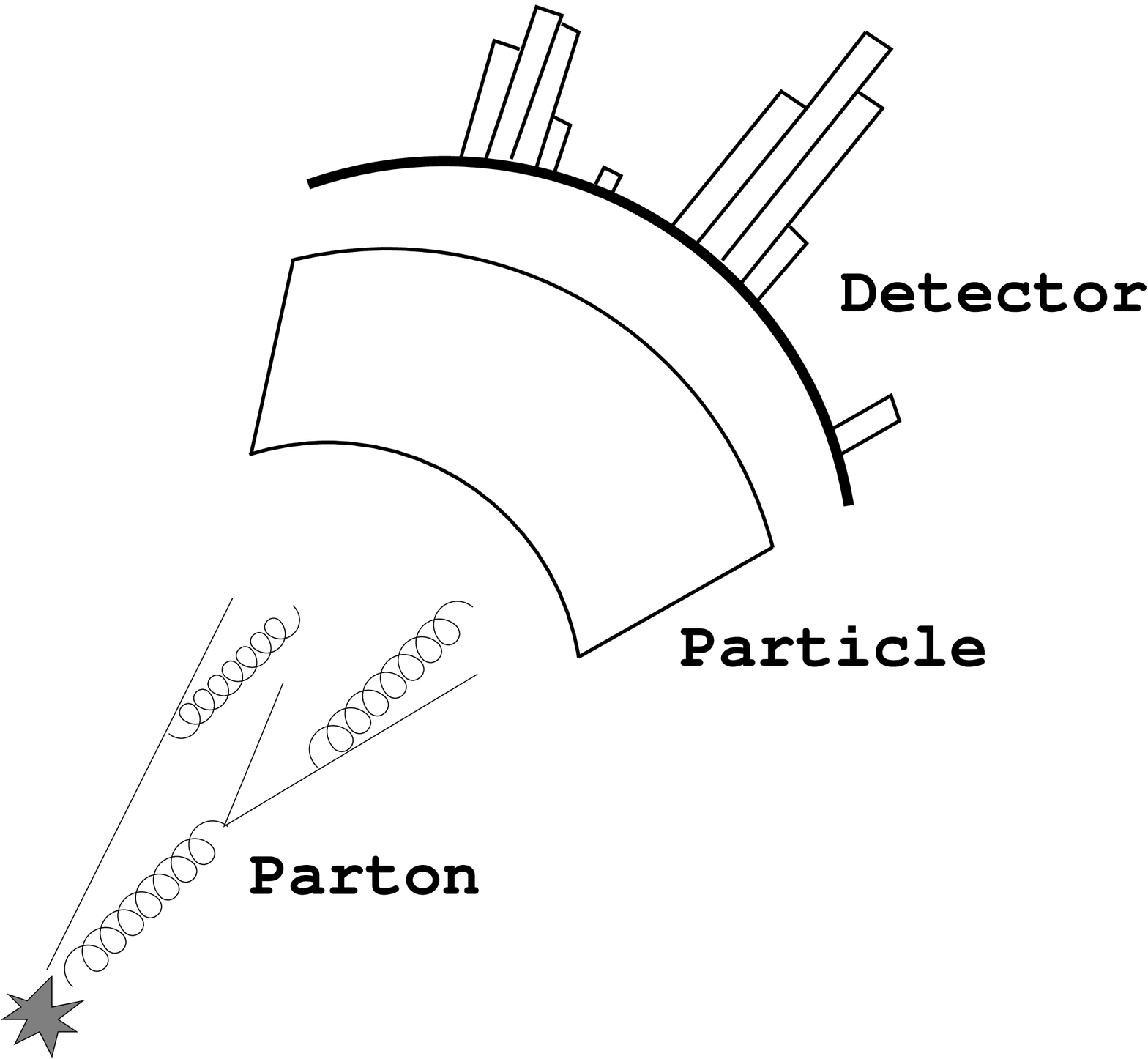}}
\hspace{0.5cm}
\subfigure[Monte Carlo]{
\label{chap3:fig:evolution2}
\includegraphics[width=6cm]{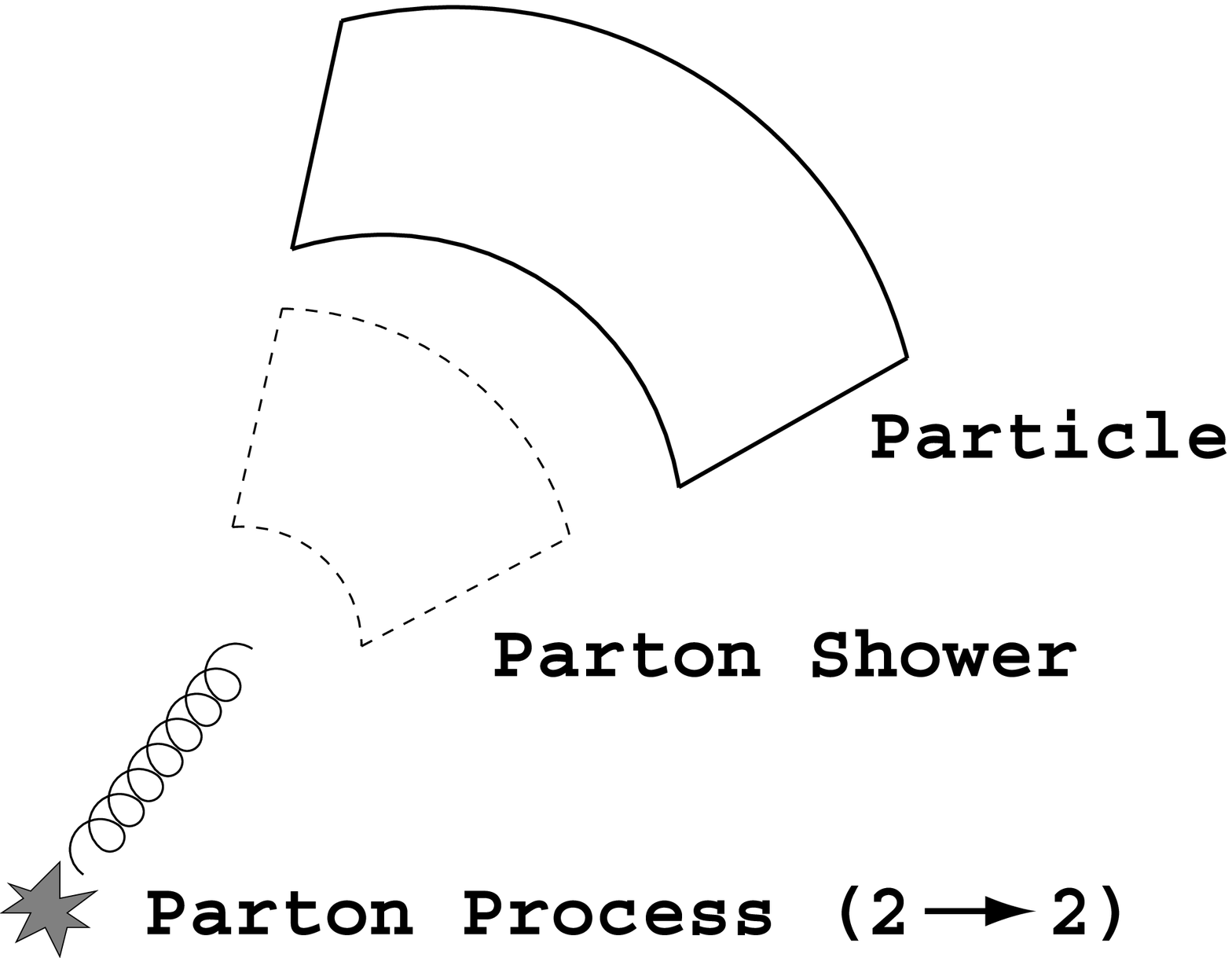}}
\end{center}
\vspace{-0.5cm}
\caption[Graphical representation of the evolution of a jet]
{Graphical representation of the evolution of a jet:
\ref{chap3:fig:evolution1}~at parton, particle and detector levels 
in experiments; \ref{chap3:fig:evolution2}~at parton, 
parton shower and particle levels in Monte Carlo simulations.} 
\label{chap3:fig:evolution}
\end{figure}

By definition, hard collisions involve very large momentum transfers, $Q$, 
and probe the structure of the hadrons at short distances. As a consequence 
of asymptotic freedom, the \ac{QCD} running coupling constant becomes small 
at this scale ($\as\lsim0.3$, see \peq{chap2:eq:runningalphas}) and 
perturbative methods become applicable. 

\pagebreak
Thus, the measurement of inclusive 
jet and dijet cross sections, as well as various other jet properties can 
be used to test the predictions of \ac{pQCD}, improve the knowledge on 
$\as$ and \acsp{PDF} at large $x$ and look for quark 
compositeness~\cite{affolder2001c}.

In order to draw comparisons between the data and the theoretical 
descriptions, jet finding (and defining) algorithms are used at 
the detector level, as clusters of towers in calorimeters,
see \fig{chap3:fig:evolution1}, and in Monte Carlo simulations, 
as final-state hadrons or outgoing partons from the hard sub-process, 
see \fig{chap3:fig:evolution2}.
A well-defined jet algorithm must not be sensitive to the level of 
the input it is applied to, as we will outline in~\sect{chap3:jetdefintion}.
Because perturbative calculations deal only with gluons and quarks, 
the subsequent jet evolution from partons and in particular
the generation of the underlying event typically is performed with 
Monte Carlo generators. The event generators mimic non-perturbative 
fragmentation and  hadronization processes converting partons into 
color-confined hadrons. Although, at hadron level, they fail to predict the 
shape of the measured (differential) jet cross-sections, presented in 
\sect{chap3:inclusivexsec}, they allow to study the performance of the jet 
finding at the Monte Carlo or ---including a realistic detector response 
simulation--- at the (simulated) detector level. 

\subsection{Jet production in pQCD}
\label{chap3:jetprodinpqcd}
The perturbative component of the hard-scattering cross section, 
the parton--parton cross section, can be analytically expanded 
in orders of $\as$, which becomes relatively small for large $Q^2$.
The contribution of each order to the scattering amplitude conveniently 
is expressed in the framework of Feynman diagrams.

\begin{figure}[b!]
\begin{center}
\subfigure[LO]{
\includegraphics[width=3cm]{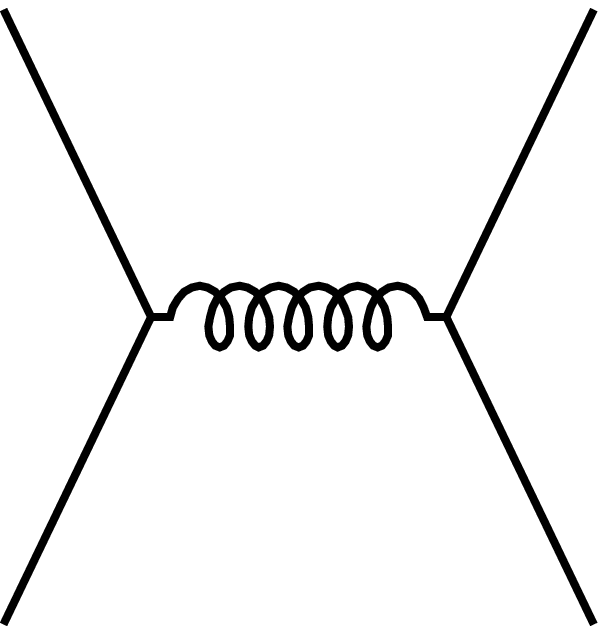}}
\hspace{0.8cm}
\subfigure[NLO (UV)]{
\includegraphics[width=3cm]{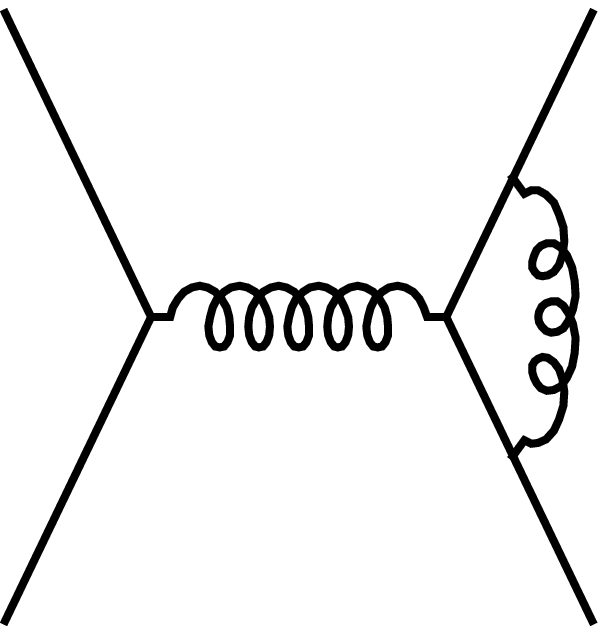}}
\hspace{0.8cm}
\subfigure[NLO (IR)]{
\includegraphics[width=3.3cm]{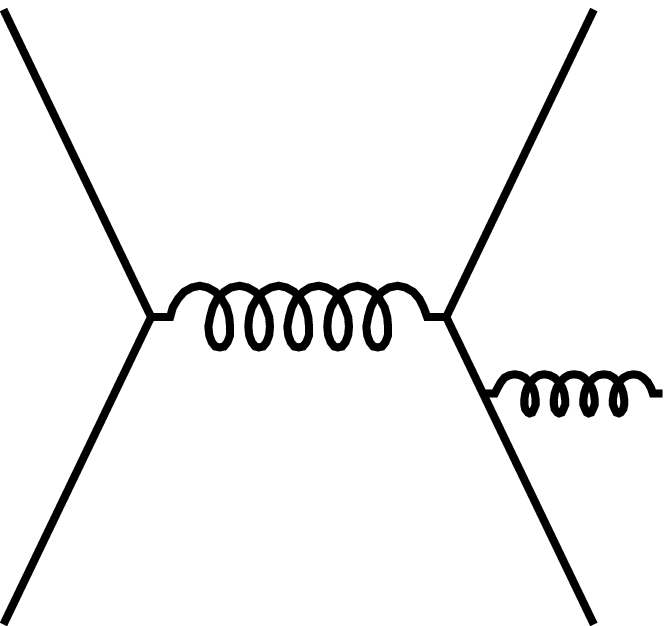}}
\end{center}
\vspace{-0.5cm}
\caption[xxx]{Example of some \acs{QCD} processes at \acs{LO}
and \acs{NLO}.} 
\label{chap3:fig:feynman}
\end{figure}


\Fig{chap3:fig:feynman} shows a few $2$-to-$2$
and $2$-to-$3$ processes contributing to \ac{LO} ($\as^2$) and 
\ac{NLO} ($\as^3$). The \ac{LO} diagrams consist of all ways connecting
the two incoming partons with the two outgoing partons using
the basic \ac{QCD} interaction vertices and do not include any 
internal loops. 
The \ac{NLO} processes are much more complicated 
because the diagrams with 2 or 3 partons in the final state have 
infra-red and ultra-violet divergences\label{defqcddivs}. 
The processes with 3 partons in the final state diverge, 
if two of the partons become collinear or one of them soft
(infra-red divergency). The processes with 2 partons 
in the final state must have one internal loop 
introducing another kind of divergent integral. 
These ultra-violet divergencies are isolated with well-defined 
regularization schemes (\eg~cut-off or dimensional 
regularization methods). Introducing the renormalization scale, 
$\mu_{\rm R}$, the singularities eventually are absorbed into the 
(bare) parameters of the theory (\eg~coupling, quark masses and 
vertices), which in turn become dependent on the momentum transfer 
and renormalization scales (and, at higher orders, also on the 
regularization scheme). 
In the massless limit and for suitably defined, inclusive observables
the collinear and soft contributions from the real and virtual 
gluon diagrams cancel after regularization~\cite{ellisqcd}.

Since the statistical momentum distributions of the initial hard-scattering 
partons are known, any cross section involving partons in the initial state 
is given by the convolution of the \acp{PDF} and the partonic 
cross section summed over all contributing partons 
and all Bjorken-$x$ values. The factorization of the cross section 
allows the separation of the long-distance and short-distance 
physics~\cite{collins1989} (see below).
The scale, $\mu_{\rm F}$, introduced by the factorization
distinguishes the two domains. Both, the \acp{PDF} and the 
partonic cross section, therefore depend on it. 
Typically, one takes the same value for the factorization scale as for 
renormalization scale ($\mu_{\rm F} \simeq \mu_{\rm R}$)~\cite{soper1996}.

\begin{figure}[htb]
\begin{center}
\includegraphics[width=10cm]{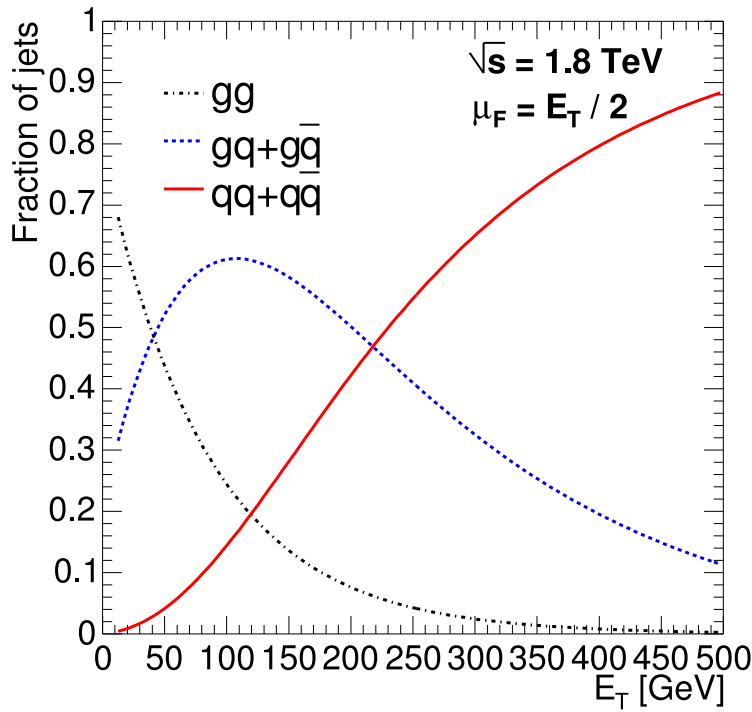}
\end{center}
\vspace{-0.8cm}
\caption[xxx]{The relative contribution of parton--parton sub-processes 
contributing to the inclusive single-jet cross section at 
mid-pseudo-rapidity ($\eta_1\simeq\eta_2\simeq 0$) at 
$\snn=1.8~\tev$ as a function of the transverse jet energy. 
The calculation uses $\mu_{\rm F}=\et/2$
and the \acs{CTEQ}~4L parameterization for the \acp{PDF}.} 
\label{chap3:fig:relativecontrib}
\end{figure}

The \acp{PDF}, $f_{i}(x,\mu^{2}_{\rm F})$, describe the initial parton 
momentum of the flavour $i$ (\quark{u}, \qubar{u}, \quark{d}, 
\qubar{d}, \quark{g}, \etc ) as the fraction $x$ of the incident 
hadron momentum (explained in~\psect{chap2:lowx}). 
To compute the relative contribution of the sub-processes 
to the partonic cross section we use the \acs{CTEQ}~4L parameterization. 
According to~\peq{chap2:eq:yx1x2out} for jets at mid-pseudo-rapidity 
$x\simeq \xt = 2\,\et/\snn$ holds. For the factorization scale we 
take $\mu_{\rm F}=\et/2$. The resulting contribution for $\snn=1.8~\tev$ 
based on the type of the incoming partons as a function of the transverse 
jet energy at mid-pseudo-rapidity is shown in \fig{chap3:fig:relativecontrib}. 
At low $\et$, jet production is dominated by gluon--gluon 
($\quark{g}\quark{g}$) and \mbox{gluon--(anti-)quark} 
($\quark{g}\quark{q}+\quark{g}\qubar{q}$) scattering. At high $E_T$ 
it is largely \mbox{quark--(anti-)quark} ($\quark{q}\quark{q}+\quark{q}\qubar{q}$) 
scattering. The \mbox{gluon--(anti-)quark} scattering is still about $40$\% 
at $E_T=250~\gev$ because of the large color factor associated 
with the gluon, and significantly contributes to the cross section 
at all $\et$ values.

\subsection{Jet defining and finding procedures}
\label{chap3:jetdefintion}
The distribution of final state quarks and gluons 
cannot be measured directly as, due to confinement, 
the final state objects of the hard-scattering reaction 
are colorless particles (mostly hadrons). 
For studies of parton-level interactions, event properties,
which only are weakly affected by long distance 
processes and which closely relate the partonic and
hadronic final states, are desirable.
The concept of jets and the jet identification algorithms  
allows to associate the partons and the hadrons 
observed in final states of high energy collisions,
a correspondence referred to as 
\ac{LPHD}~\cite{azimov1984}.
If \ac{LPHD} is satisfied, the study of jets may be regarded
as a tool for mapping the observed long-distance hadronic final 
states onto underlying short-distance partonic states.

\begin{figure}[htb]
\begin{center}
\subfigure[Calorimeter]{
\label{chap3:fig:cdfjetsa}
\includegraphics[width=8cm]{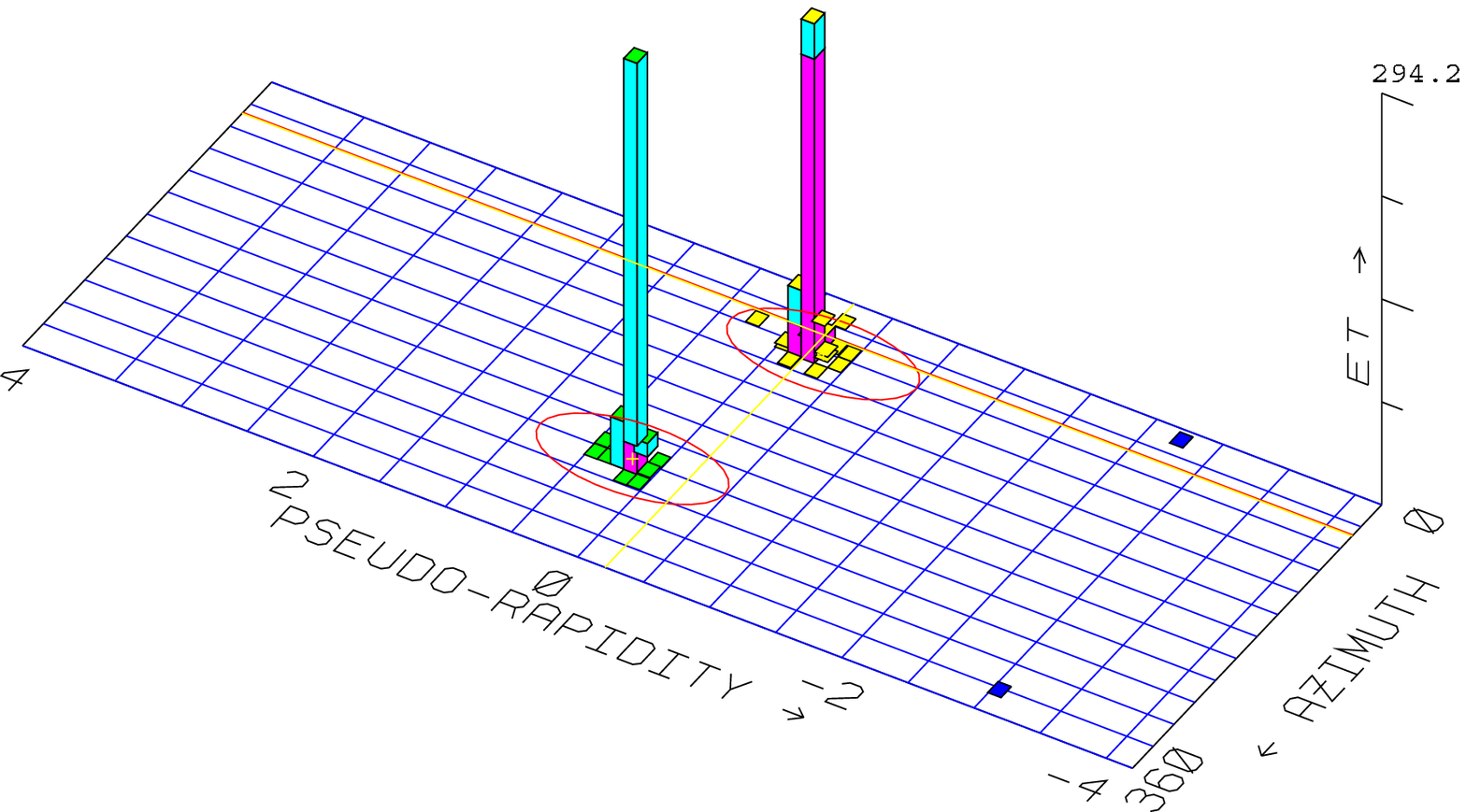}}
\hspace{0.8cm}
\subfigure[Tracking chamber]{
\label{chap3:fig:cdfjetsb}
\includegraphics[width=5cm]{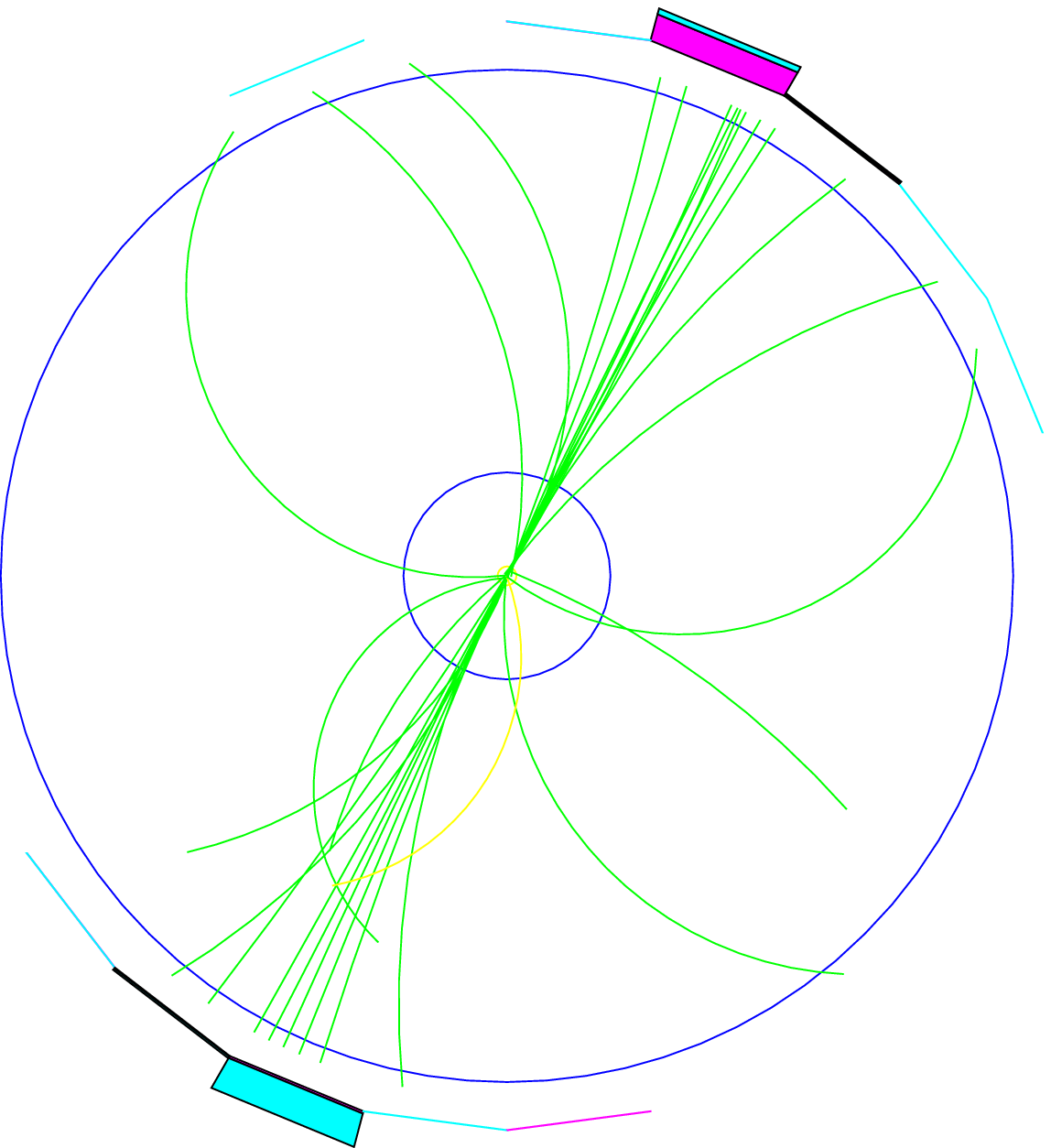}}
\end{center}
\vspace{-0.5cm}
\caption[xxx]{Jet event in the \acs{CDF} calorimeter~\subref{chap3:fig:cdfjetsa} 
and central tracking chamber~\subref{chap3:fig:cdfjetsb} identified by the cone jet
finder, \acs{JetClu}, with $R=0.7$ (see below). The figure is taken from~\Ref{affolder2001}.} 
\label{chap3:fig:cdfjets}
\end{figure}

Although `everyone knows a jet, when they see it', because
they stand out by their nature (\fig{chap3:fig:cdfjets}), 
precise definitions are elusive and detailed. 
Jet finding algorithms define a functional mapping 
\begin{equation*}
\label{chap3:eq:mapping}
\mathbf{particles}
\stackrel{\mathbf{jet\;algorithm}}
{-\!\!\!-\!\!\!-\!\!\!-\!\!\!-\!\!\!-\!\!\!-\!\!\!-\!\!\!-\!\!\!
\longrightarrow}
\mathbf{jets}
\end{equation*}
between the particles in the 
event, given by their kinematical description 
(\eg~momenta) and the configuration of jets, represented 
by suitable jet variables. 

\pagebreak
Ideally, jet defining algorithms 
must be~\cite{blazey2000}
\label{chap3:featurelist}
\begin{itemize}
\item {\em fully specified}: the jet finding procedure, the kinematic 
variables of the jet and the various corrections should be uniquely 
and completely defined.
\item {\em theoretically well behaved}: the algorithm should be 
infrared and collinear safe, without the need for 
ad-hoc parameters. 
\item {\em detector independent}: there should be no dependence on
detector type, segmentation or size.
\item {\em consistent}: the algorithm should be equally applicable at 
the theoretical and experimental levels.
\end{itemize}

The first two criteria must be fulfilled by every algorithm as 
\ac{LPHD} can only be satisfied if the applied jet algorithm is 
infrared safe. This ensures that its outcome is insensitive to the 
emission of soft or collinear partons. Therefore, the jet
observable must not change by adding an additional particle
with $E\rightarrow 0$ to the final state or when replacing
a pair of particles by a single particle with the summed momentum.
The last two criteria, however, can probably never be totally true,
since it is not possible to completely remove dependencies
on the experimental apparatus.

\subsubsection{Jet kinematics}
\label{chap3:jetkinematics}
The interacting partons are not generally in the \cms\ frame of the
colliding system, because the fraction of the hadron momentum carried by 
each parton varies from event to event. 
Therefore, the \cms\ system of the partons is randomly boosted along 
the direction of the colliding hadrons, so that jets 
are conveniently described by longitudinally boost-invariant 
variables:
\begin{eqnarray*}
\rm{mass}\hspace{3.5cm} & m & = \sqrt{E^2-\px^2-\py^2-\pz^2} \\ 
\rm{transverse~momentum}\hspace{0.4cm} & \pt & = \sqrt{\px^2+\py^2} \\
\rm{azimuthal~angle}\hspace{1.5cm} & \phi & = \arctan \left( \py / \px \right) \\ 
\rm{rapidity}\hspace{2.9cm} & y & = {\rm arctanh} \left( \pz / E \right) 
                                  = \frac{1}{2} \ln \left(\frac{E+\pz}{E-\pz} \right )\;.
\end{eqnarray*}
In the high energy limit, when $p>>m$, the directly measured quantities conveniently
are: energy ($E$) or transverse energy ($\et=E\,\sin\theta \simeq \pt$), 
the azimuth ($\phi$) and the pseudo-rapidity
\begin{equation*}
\eta = -\ln \left[ \tan \left( \theta/2 \right) \right ]\;,
\end{equation*}
where the polar angle is given by $\theta=\arctan(\pt/\pz)$.

\subsubsection{Jet algorithms}
\label{chap3:jetalgorithms}
Even though the criteria listed in~\sect{chap3:jetdefintion} 
lead to restrictions on possible algorithms, a variety of jet 
definitions emerged over time (see \Ref{blazey2000} for an overview). 
They can be grouped into two fundamental classes: 
recombination (clustering) algorithms~\cite{sterman1977,banner1982,
bethke1988,bethke1991,catani1991,dokshitzer1997,grigoriev2003} 
and cone algorithms~\cite{arnison1983,albajar1988,huth1990,abe1991,ellis1992,
abbott1997,seymour1997}.
Both are based on the assumption that hadrons associated 
with a jet will be `nearby' each other. 
The definition of cone jets 
is based on vicinity in real space (angles), whereas 
the recombination algorithms make use of vicinity in 
momentum space and, nowadays, go by the name of $\kt$ algorithms.

\begin{itemize}
\item The $\kt$ algorithms~\cite{catani1991,dokshitzer1997}
are inspired by \ac{QCD} parton showering. The algorithms 
try to mimic the hadronization processes backwards and
successively merge pairs of particles (or rather `vectors') 
in order of increasing transverse momentum. Typically they 
contain a parameter, $D$, which controls the termination of 
merging. By design, they are infra-red and collinear safe 
to all orders and were developed for precise 
$\epem\rightarrow \rm{jets}$ studies (see \Ref{moretti1998}
for a recent comparison). However, problems arise 
when the $\kt$ algorithm is applied at hadron--hadron colliders. 
This is mostly due to difficulties with the substraction 
of energy from spectator fragments and from the pile-up of 
the multiple hadron--hadron interactions. Only recently solutions 
to these problems have been developed~\cite{grinstein2003,blazey2000} 
\item The cone algorithms~\cite{arnison1983,abe1991,abbott1997} 
historically developed for jet definition in hadron--hadron
collisions group all particles within a cone of radius $R$ 
in $\eta\times\phi$ space into a single jet. 
The radius is defined as $R = \sqrt{\Delta\eta^2 + \Delta\phi^2}$, 
where $\Delta\eta$ and $\Delta\phi$ are the separation of the particles 
(or partons) in pseudo-rapidity and azimuthal angle (in radians)
to the jet axis.
The way the grouping procedure operates is such that the 
center of the cone is aligned with the jet direction.
Typically, the algorithm starts with a number of (high energy) 
seeds, but also seedless implementations exist.
As cones may overlap, a single particle could belong to 
two or more cones. Thus, a procedure is introduced to specify 
how to split or merge overlapping jets.  At the parton level
\ac{NLO} calculations require the addition of an ad-hoc separation
parameter, $R_{\rm sep}$, to regulate the clustering of the partons 
and simulate the role of seeds. 
\end{itemize}

\subsection{Improved Legacy Cone Algorithm}
\label{chap3:ilca}
The decision to use a cone finder for the present work
is based on the fact that the anticipated huge background 
(or underlying event) for $\rm{Pb}+\rm{Pb}\rightarrow\rm{jet}+\rm{X}$ 
at the \ac{LHC} will spoil the recombination scheme of the $\kt$ 
algorithms. Furthermore, the $\mathcal{O}(n^3)$ run-time of the
$\kt$ algorithms might be too time-consuming for the online version 
(trigger) of the jet finder.~\footnote{Here, $n$ symbolically denotes
the size of the input, \eg~number of particles or towers.}
Instead, we decided to implement the \acf{ILCA}~\cite{blazey2000},
which has been developed jointly by \acs{D0} and \acs{CDF} 
before the start of Run~II and which is supposed to possess the
required features listed on \page{chap3:featurelist}.

The basic idea of the algorithm is to find all of the circles 
in the $\eta \times \phi$ space (cones in three dimensional space) 
of a preselected, fixed radius $R$ that contain stable jets.
The algorithm starts with an input list of particles, 
partons or pre-towers, which are grouped into towers according to a 
simple pre-clustering procedure . 
Each tower in the event is assigned a massless four-vector 
$(E_k=\abs{\vec{p}_k},\,\vec{p}_k)\equiv(E_k,\phi_k,\eta_k)$
pointing into the direction of the tower.

\begin{figure}[htb]
\begin{center}
\subfigure[Clustering]{
\label{chap3:fig:seedless}
\includegraphics[width=7.0cm]{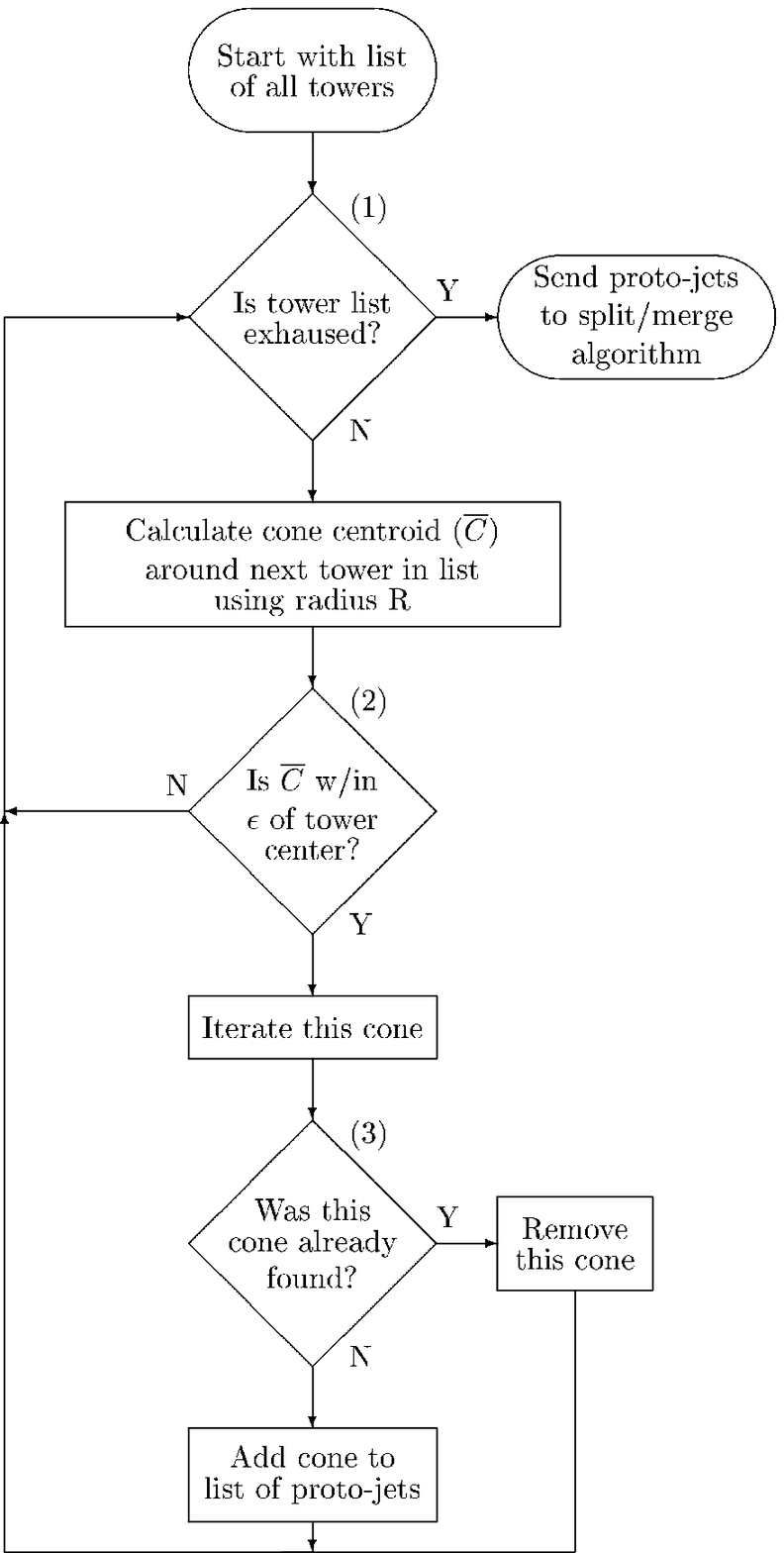}}
\hspace{0.5cm}
\subfigure[Splitting and merging]{
\label{chap3:fig:splitmerge}
\includegraphics[width=7.0cm]{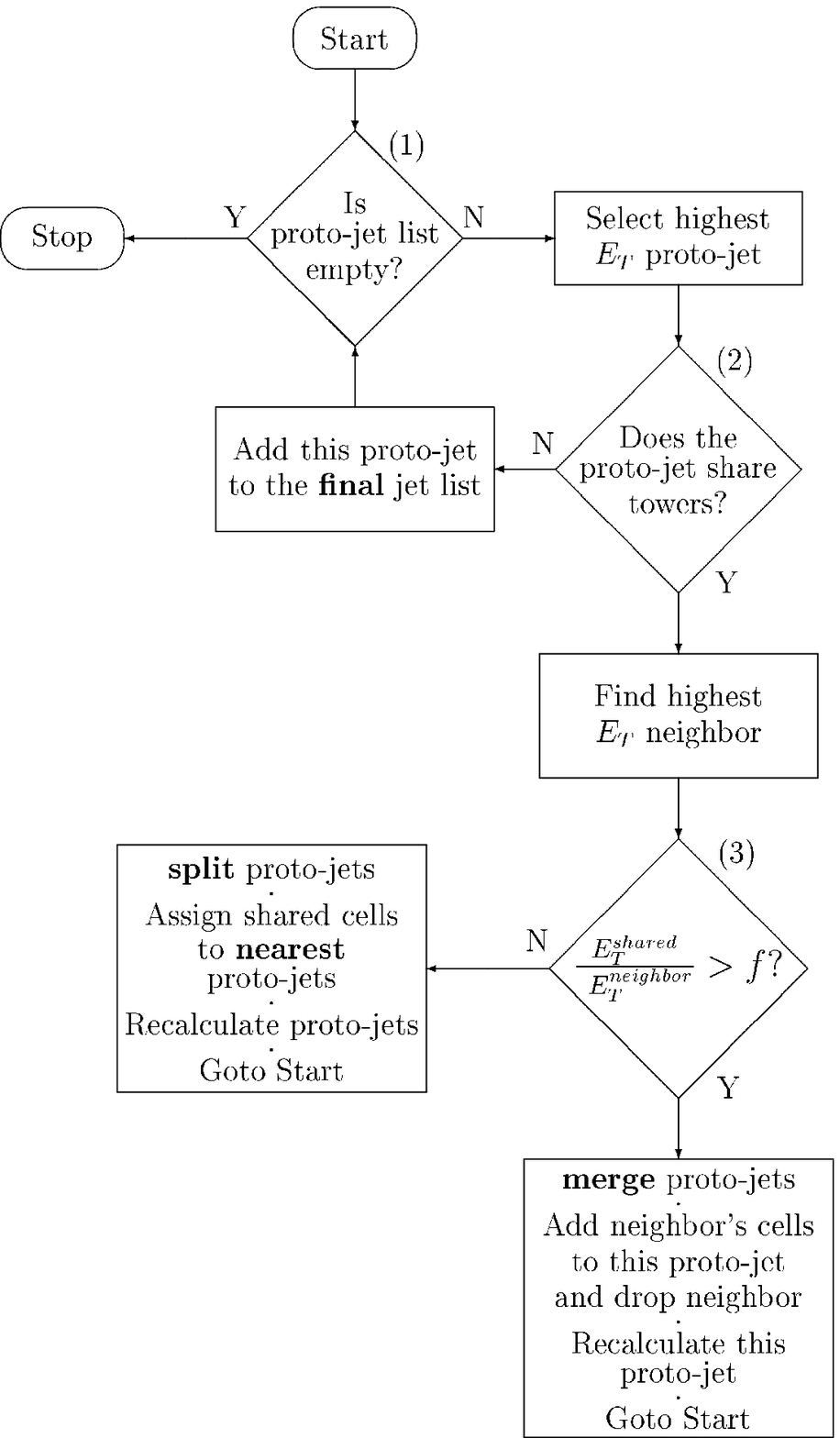}}
\end{center}
\vspace{-0.5cm}
\caption[The (seedless) cone jet finder algorithm]
{\subref{chap3:fig:seedless} The seedless clustering procedure and 
\subref{chap3:fig:splitmerge} the splitting-and-merging procedure of
the \acs{ILCA} algorithm. For details see the text.}
\label{chap3:fig:ilca}
\end{figure}

\pagebreak
The jets are defined in three sequential steps:
\begin{enumerate}
\item  In the {\em clustering} procedure, displayed in \fig{chap3:fig:seedless}, 
towers belonging to a jet are iteratively accumulated until stable proto-jets
are found. 
\item In the {\em splitting-and-merging} procedure, displayed in \fig{chap3:fig:splitmerge}, 
overlapping jets are split or merged depending on the fraction of 
energy they share. 
\item In the {\em recombination} procedure, the kinematical variables 
of the jets are computed according to a given recombination scheme 
(\eg~Snowmass, modified Run~I or energy scheme, see~\Ref{blazey2000}).
\end{enumerate}

\pagebreak
The clustering method displayed in \fig{chap3:fig:seedless} starts by looping 
over all towers. For each tower $k$, with center $\overrightarrow{k} 
= \left( \eta_{k},\phi_{k}\right)$, we define a cone of size 
$R$ centered on the tower
\begin{eqnarray*}
\overrightarrow{C_{k}} &=&\left( \eta_{C_{k}}=\eta_{k},\phi_{C_{k}}=\phi_{k}\right) \\
i &\in & C_{k}\; : \; {\left(\eta_{i}-\eta_{C_{k}}\right)^{2} + 
                       \left(\phi_{i}-\phi_{C_{k}}\right)^{2}} \leq R^2\;,
\end{eqnarray*}
which contains all towers falling into its circumference.~\footnote{The 
proposed clustering method is seedless. An alternative speeding up the 
algorithm, is to loop over a set of seeds instead, \eg~to loop over towers 
with $\et^{\rm tower}\ge\et^{\rm seed}$. In order to ensure infra-red 
insensitivity, points in between the seeds (`midpoints') have to be 
added~\cite{seymour1997}. The corresponding algorithms is called \acs{MidPoint} 
algorithm.} For each cone we, then, evaluate the $\et$-weighted average 
centroid $\overrightarrow{\bar{C}_{k}} = \left(\bar{\eta}_{C_{k}},
\bar{\phi}_{C_{k}}\right)$, where
\begin{equation*}
\bar{\eta}_{C_{k}} = \frac{\sum_{i\in C_{k}}E_{{\mrm T}_{i}}\eta_{i}}{E_{{\mrm T}_{C_{k}}}}
\hspace{0.5cm} \text{and} \hspace{0.5cm}
\bar{\phi}_{C_{k}} = \frac{\sum_{i\in C_{k}}E_{{\mrm T}_{i}}\phi_{i}}{E_{{\mrm T}_{C_{k}}}}
\end{equation*}
with its transverse energy content
\begin{equation*}
E_{{\mrm T}_{C_{k}}} = \sum_{i\in C_{k}}E_{{\mrm T}_{i}}\;.
\end{equation*}

In general, the centroid $\overrightarrow{\bar{C}_{k}}$
is not identical to the geometric center $\overrightarrow{C_{k}}$ 
and, thus, the cone is not stable. If the calculated centroid of the cone 
lies outside of the initial tower, further processing of that cone is 
skipped and the cone is discarded.~\footnote{The specific exclusion 
distance, $\epsilon$, used in this cut is an arbitrary parameter.
It is adjusted to maximize jet finding efficiency and minimize 
the run-time of the algorithm.}
All the cones, which yield a centroid within the original tower, 
are kept for re-iteration. For these cones the process of calculating 
a new centroid about the previous centroid is repeated. Thus, 
the cones are allowed to `flow' away from the original towers.
The iteration continues until either a stable cone center is found
or the centroid moves out of the fiducial volume. 
All the surviving stable cones constitute the list of 
proto-jets.~\footnote{For \PbPb\ collisions with large anticipated background, 
it may also be useful to apply some minimum $\et$-threshold to the list of proto-jets. 
In \pp\, the threshold could be set near the noise level of the detector.}

Typically, a number of overlapping proto-jets, for which towers are shared 
by more than one cone, will be found after applying the clustering 
procedure. These are subject of the splitting-and-merging procedure
sketched in \fig{chap3:fig:splitmerge}. The suggested algorithm starts with the 
list of all proto-jets and always works with the highest $\et$ proto-jet 
remaining on the list. 
After a merging or splitting occurred, the $\et$ ordering on the list of
remaining proto-jets can change, since the survivor of merged jets
may move up while split jets may move down. Once a proto-jet shares
no tower with any of the other proto-jets, it becomes a jet stored
on the list of final jets, which is not affected by the subsequent 
merging and splitting of the remaining proto-jets. 
The decision to split or merge a pair of overlapping proto-jets is based 
on the percentage of transverse energy shared by the lower $\et$ proto-jet. 
Proto-jets, which share a fraction greater than $f$ (typically $f=50\%$),
will be merged; others will be split with the shared towers individually 
assigned to the proto-jet, which is closest in $\eta \times \phi$ space.
The method will perform predictably even in the case of multiply split 
and merged jets, but there is no requirement that the centroid 
of the split or merged proto-jet still coincides precisely with 
its geometric center.

To complete the jet finding process the jet variables have to be
computed according to a suitable recombination prescription. 
Typically, we follow the original Snowmass ($\et$-) scheme~\cite{huth1990}
\begin{eqnarray}
\label{chap3:eq:snowmass} 
E_{{\rm T}_{J}}  &=&\sum_{i\in J=C}E_{{\rm T}_{i}} 
\nonumber\\
\eta_{J} &=&\frac{1}{E_{{\rm T}_{J}}}\sum_{i\in J=C}E_{{\rm T}_{i}}\eta_{i}
\nonumber \\
\phi_{J} &=&\frac{1}{E_{{\rm T}_{J}}}\sum_{i\in J=C}E_{{\rm T}_{i}}\phi_{i}\;,
\end{eqnarray}
which simply uses the stable cone variables. That way computing time 
is reduced, because there is no need to loop over the associated particles 
(or towers) in the jet, as one would need to do in the energy ($E$-) scheme 
in order to calculate the jet variables by adding four-vectors of 
the associated particles (or towers).~\footnote{In most cases we
write $\et$ ($P_{\rm T}$) for the transverse jet energy (momentum).
Only when we emphasize the compositeness of jets,
we denote the resulting transverse energy of the jet 
as $\Etj$ or using rapidity instead of pseudo-rapidity in the 
$E$-scheme also as $\Ptj$.} As reported by \acs{CDF}~\cite{affolder2001}, 
in practice the difference between the two representations is negligible.

\subsection{Inclusive single-jet cross section}
\label{chap3:inclusivexsec}
The inclusive single-jet cross section can be determined from the
process 
\begin{equation}
\label{chap3:eq:reaction}
\rm{h}_1 + \rm{h}_2 \rightarrow \rm{jet} + \rm{X}
\end{equation}
visualized in \fig{chap3:fig:visualcross}.

\begin{figure}[htb]
\begin{center}
\includegraphics[width=6cm]{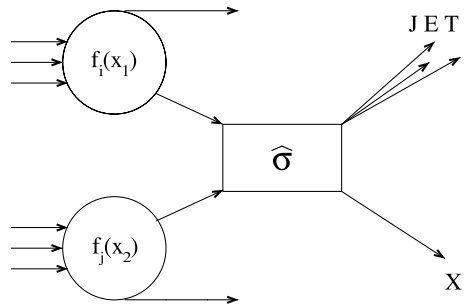}
\end{center}
\vspace{-0.3cm}
\caption[xxx]{Visualization of the jet production process in a hadron--hadron 
collision. The \acsp{PDF} give the probability to pick up a certain
parton with momentum fraction~$x$ of the hadron momentum. The elementary 
parton--parton interaction can be calculated in \acs{pQCD}.}
\label{chap3:fig:visualcross}
\end{figure}

\pagebreak
In the framework of \ac{QCD} improved parton model, which we partially
have outlined above, the corresponding cross section writes 
as~\footnote{See, for example, \Ref{ellisqcd} for a 
detailed discussion.}
\begin{equation}
\label{chap3:eq:ppjetcrossection}
\sigma_{\rm jet}
= \sum_{i,j} \int \dd x_{1} \dd x_{2} 
\,f_{i}(x_{1},\mu^{2}_{F})\,f_{j}(x_{2},\mu^{2}_{\rm F})\, \hat{\sigma}_{ij} 
[ x_{1}P, x_{2}P, \as(\mu_{\rm R}^2), \frac{Q^2}{\mu_{\rm R}^2}, 
\frac{Q^2}{\mu_{\rm F}^2}]
\end{equation} 
where, for simplicity, we omit the notation of parton evolution and 
fragmentation processes, as well as the jet finding procedure at the 
hadron (or parton) level. 
The short-distance, two-body parton-level cross section, 
$\hat{\sigma}_{i j}$, is a function of the momentum carried by 
each of the incident partons ($x_1 P$ and $x_2 P$), the strong coupling 
$\as(\mu^2_{\rm R})$, and the ratio of the renormalization and factorization 
scales, $\mu^2_{\rm R}$ and $\mu^2_{\rm F}$, 
to the characteristic scale of the hard interaction, $Q^2$.
The \ac{LO} calculation includes only the contribution of tree-level diagrams 
for the 2$\rightarrow$2 scattering processes 
given in~\Refs{combridge1977,ellisqcd}. 
The \ac{NLO} calculation adds the diagrams which describe the emission 
of a gluon  as an internal loop and as a final state 
parton~\cite{aversa1988,aversa1988b,aversa1988c,ellis1988,ellis1989,ellis1990}.  
The scales $\mu_{\rm R}$ and $\mu_{\rm F}$ are intrinsic parameters 
in a fixed order perturbation theory. In what follows, they are set 
equal, $\mu_{\rm F}=\mu_{\rm R}\equiv\mu$. Although the choice of the 
$\mu$ scale is arbitrary, a reasonable value is related to a 
physical observable, such as the $\et$ of the jets.~\footnote{After
fixing the scale, the predictions for the inclusive jet cross section 
depend on the choice of the scale, $\mu$. No such dependence 
would exist if the perturbation theory were calculated to all orders. 
The addition of higher order terms in the calculation 
reduces the $\mu$ dependence. Fixing $\mu$ is to a constant 
between $0.5\et$ and $2\et$ results in roughly a factor of two variation 
in the calculated cross section at \ac{LO} and $20$--$30$\% at \ac{NLO} 
in the range of $20~\gev\le\et\le500~\gev$~\cite{ellis1993}. The variation
can be used to estimate the systematic error from the fixed 
order calculation. However, a subtlety in the choice of scale arises at 
\ac{NLO}. At \ac{LO} there are only two partons of equal $\et$, whereas 
at \ac{NLO} the partons might be grouped together to form (parton-level) 
jets, not necessarily with equal $\et$. 
In order to avoid  more than one scale per event in the 
\ac{NLO} calculations, one typically chooses the $\et$ of the 
leading parton (leading jet) for the choice of scale.}

However, since it is impossible to measure the total jet cross section, 
one obtains the predictions for the jet cross section as a function of $\et$ 
from the general expression, \eq{chap3:eq:ppjetcrossection}, using
\begin{equation}
\label{chap3:eq:diffppjetcrossection}
\frac{ E \dd^3 \sigma}{\dd p^3} \equiv \frac{\dd^3 \sigma}{\dd \pt^2 \dd y} = 
\frac{1}{2\pi E_T}\frac{\dd^2 \sigma} {\dd \et d\eta}\;, 
\end{equation}
where (as often done in jet calculations) the mass of the partons 
has been assumed to be zero ($\pt=\et$).
Experimentally, the inclusive (differential) jet cross section is 
defined as the number of jets in a bin of $\et$ normalized by acceptance 
and integrated luminosity. All the jets in each event falling within the acceptance 
region contribute to the cross section measurement as appropriate for an inclusive 
quantity. Usually, measurements are performed in the central pseudo-rapidity interval 
($\abs{\eta}<1$) and results are averaged in the $\eta$-interval.

\begin{figure}[htb]
\begin{center}
\includegraphics[width=10cm]{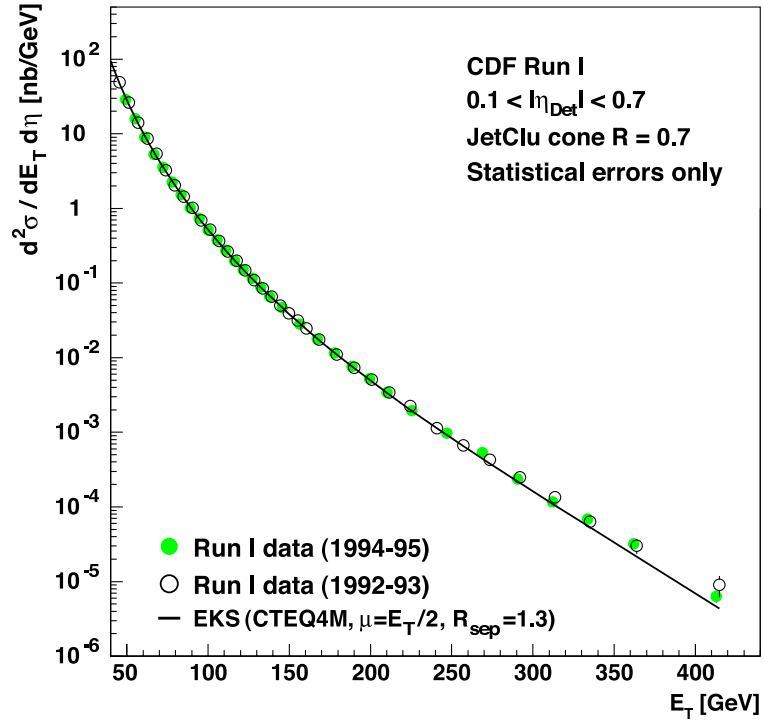}
\end{center}
\vspace{-0.3cm}
\caption[Inclusive single-jet cross section Run~I]{Inclusive 
single-jet cross section from Run~I at $\sqrt{s}=1.8~\tev$ for 
$0.1<|\eta|<0.7$ as published by \acs{CDF}. The jets are identified 
with \acs{JetClu} at $R=0.7$. 
The data from Run~1A~(92-93) and Run~1B~(94-95) are compared 
to the \acs{NLO} calculation of \acs{EKS} used with the \acs{CTEQ}~4M 
at $\mu=\et/2$ and a parton-separation value of $R_{\rm sep}=1.3$. 
The figure is adapted from~\Ref{affolder2001}.} 
\label{chap3:fig:cdfcross}
\end{figure}

\begin{figure}[htb]
\begin{center}
\subfigure[\acs{CDF} measurement]{
\label{chap3:fig:cdftevatroncross}
\includegraphics[width=7.5cm,height=5.4cm]{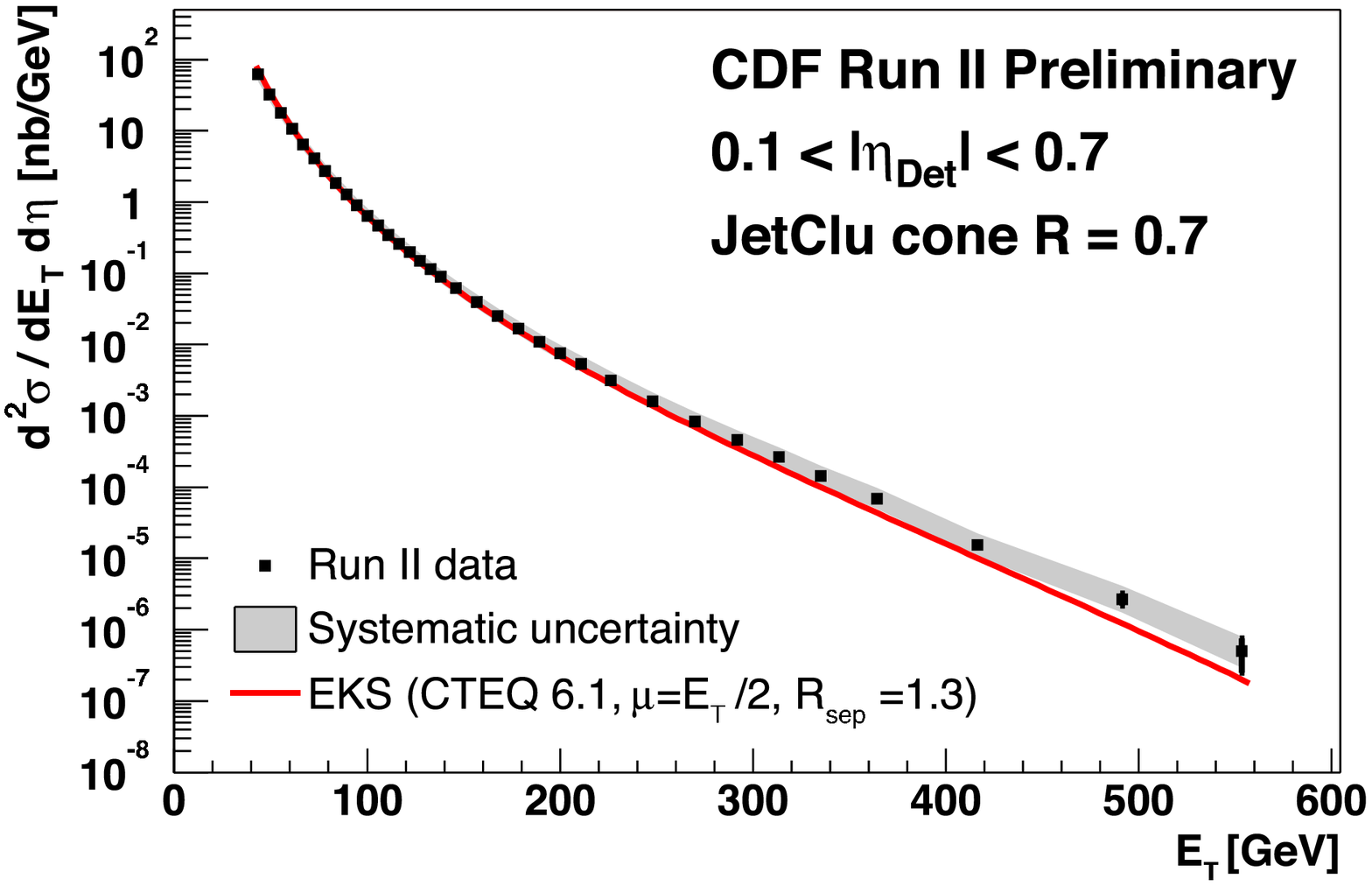}}
\subfigure[\acs{D0} measurement]{
\label{chap3:fig:d0tevatroncross}
\includegraphics[width=6.5cm,height=5.4cm]{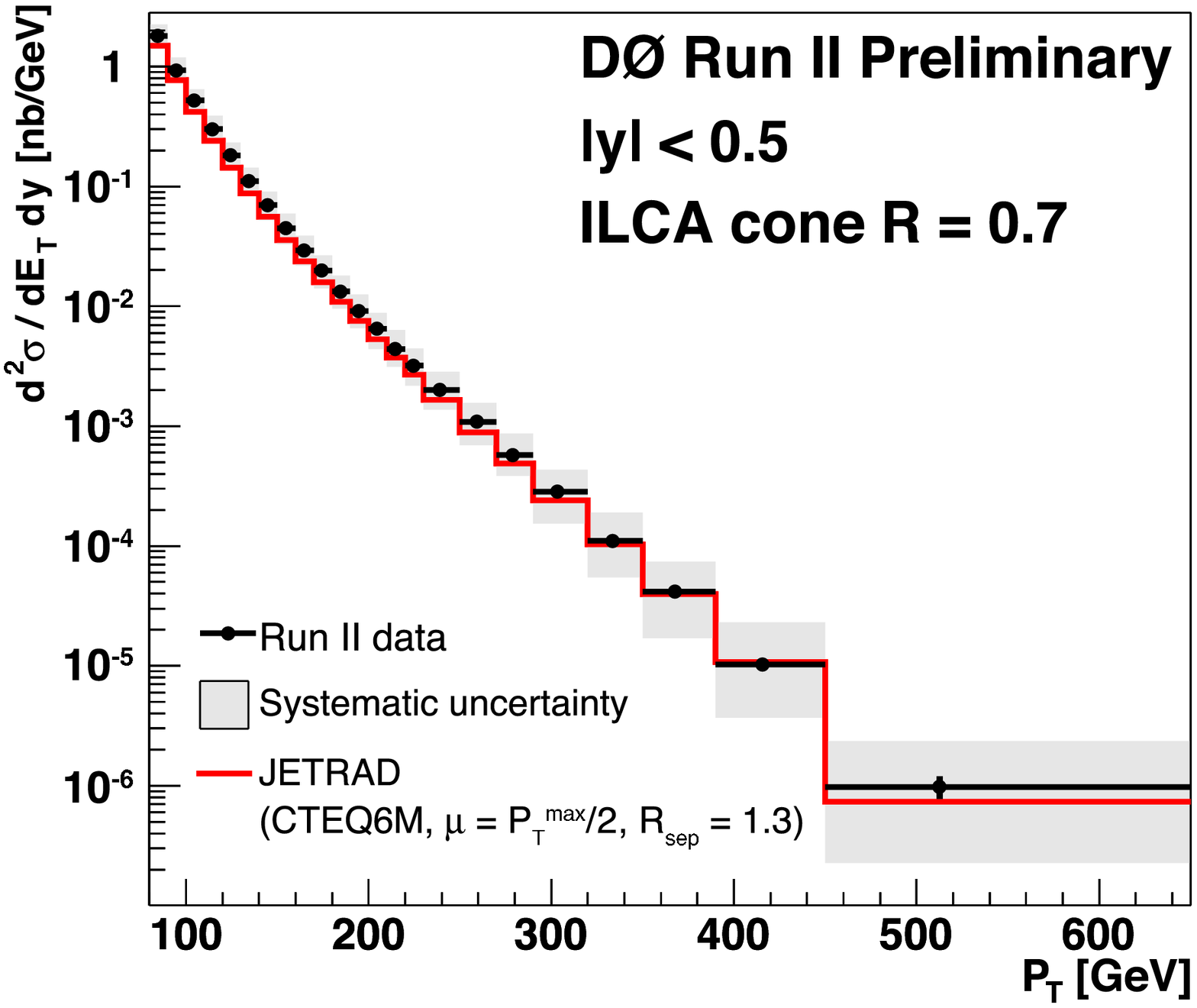}}
\end{center}
\vspace{-0.5cm}
\caption[Inclusive single-jet cross section Run~II]{Inclusive 
single-jet cross section from Run~II at $\sqrt{s}=1.96~\tev$.
\subref{chap3:fig:cdftevatroncross} The \acs{CDF} measurement 
(preliminary) using \acs{JetClu} with $R=0.7$ is compared to \acs{EKS} 
with \acs{CTEQ}~6.1 at $\mu=\et/2$ and $R_{\rm sep}=1.3$. 
The figure is adapted from~\Ref{latino2004}.
\subref{chap3:fig:d0tevatroncross} The \acs{D0} measurement 
(preliminary) using \acs{ILCA} with $R=0.7$ is compared to 
\acs{JETRAD} with \acs{CTEQ}~6M at $\mu=\Pt/2$ and $R_{\rm sep}=1.3$. 
The figure is adapted from~\Ref{padley2004}.} 
\label{chap3:fig:tevatroncross}
\end{figure}

\Fig{chap3:fig:cdfcross} shows the measurement of the inclusive 
single-jet cross section at $\sqrt{s}=1.8~\tev$ for $0.1<|\eta|<0.7$ 
as a function of $\et$ from \acs{CDF} at the \acs{Tevatron} $\ppbar$ 
collider~\cite{affolder2001}. The jets are identified with a 
cone finder, called \acs{JetClu}~\cite{abe1991}, using a radius
of $R=0.7$. The measured and corrected differential cross section is compared 
to a \ac{NLO} \ac{pQCD} calculation of the \acs{EKS} program~\cite{ellis1992}.
The calculation computes the spectrum at the parton level and uses the \ac{NLO} 
\acs{CTEQ}~4M \acp{PDF} at $\mu=\et/2$ and a parton separation value 
of $R_{\rm sep}=1.3$. The systematic errors not shown in the figure range between 
$20$--$50\%$ at high energy. Experimentally the uncertainty is dominated by
the uncertainty associated with the Monte Carlo production of realistic jets and 
underlying events for the derivation of corrections needed to compare the
measured cross section at hadron level with calculations at parton level.
The theoretical uncertainty is dominated by uncertainty in the
\acp{PDF}, mainly at high-$x$. The experimental and theoretical developments,
thus, are correlated, since the corrections to the raw data depend on 
detailed modeling of the events, which in turn depends on data quality
and size.

Very recently, preliminary results from Run~II at \acs{Tevatron} 
became available confirming and extending the precise measurements 
of Run~I to higher $\et$. The preliminary results for the inclusive 
(differential) single-jet cross section at $\sqrt{s}=1.96~\tev$ by
\acs{CDF}~\cite{padley2004} and \acs{D0}~\cite{latino2004} are shown
in \fig{chap3:fig:tevatroncross}. The \acs{CDF} measurement is performed
using the Run~I cone finder \acs{JetClu} with $R=0.7$ for the definition of jets 
at the hadron level, whereas the \acs{D0} result has been obtained with the 
improved Run~II cone finder (\acs{ILCA}) with $R=0.7$. Both are compared 
to \acs{NLO} \acs{pQCD} calculations, \acs{EKS} with \acs{CTEQ}~6.1
and \acs{JETRAD} with \acs{CTEQ}~6M, respectively. 
The consistency between the data and the theoretical prediction over many orders 
of magnitude is remarkable 
Still, the main source for errors is attributed to the uncertainty in the gluon 
\ac{PDF} arising at about $x\gsim 0.1$, where the new data will provide new constraints 
for the gluon \ac{PDF}~\cite{martin2004}.

\subsection{Jet fragmentation}
\label{chap3:frag}
The single-jet cross section presented in the last section in 
\fig{chap3:fig:cdfcross} and \fig{chap3:fig:tevatroncross} is
consistent with theoretical predictions at parton level.
This is due to \ac{LPHD} and the way jet finding algorithms are 
constructed (see \sect{chap3:jetdefintion}). However, the fixed order 
calculations cannot predict details of the jet structure observed 
in experiments. Monte Carlo programs implement the parton 
shower approach approximating higher order \ac{QCD} processes 
followed by hadronization.
General purpose generators, like 
\acs{HERWIG}~\cite{mcherwig1992,mcherwig2000,mcherwig2002} and 
\acs{PYTHIA}~\cite{mcpythia1987,mcpythia1994,mcpythia2001}, 
provide a variety of elementary $2$-to-$2$ processes. 
After the leading order calculation the primary hard partons develop
into multi-parton cascades or showers by multiple gluon bremsstrahlung.
These cascades are based on soft and collinear approximations to the
\ac{QCD} matrix elements and distribute the energy fractions according 
to the \ac{DGLAP} parton-evolution equations.
The parton shower stops, when the virtuality of the initial parton
falls below a cut-off parameter, $Q_0\simeq 1\gev>\lQCD$. 
The non-perturbative evolution is then phenomenologically 
described by hadronization models like the cluster or string model, 
which turn the final state partons into hadrons locally distributed
in phase space~\cite{webber1994,webber1999}. Due to the cut-off the
hadronization process is independent of the hard scattering and
the development of the parton shower.

Opposed to hadronization, for which at present only models exist, 
the evolution of parton showers and the scaling of inclusive fragmentation 
into hadrons can be described by \ac{pQCD}. The total fragmentation
function for hadrons of type $\rm h$ in a certain process,
typically $\epem$ annihilation, is defined by
\begin{equation*}
F_{\rm h}(x,s) = \frac{1}{\sigma_{\rm tot}} \frac{\dd \sigma}{\dd x}
(\epem\rightarrow \rm{h X}) \;,
\end{equation*}
where $x=2E_{\rm h}/\sqrt{s}$ is the scaled hadron energy in the \cms\ frame.
The total fragmentation function can be decomposed into a sum of contributions 
arising from the different primary partons
\begin{equation*}
F_{\rm h}(x,s) = \sum_{i} \int_{x}^{1} \frac{\dd z}{z} \, C_i(z,\as(s)) \, D_{{\rm h}/i}(x/z,s)\;,
\end{equation*}
where $C_i$ are the coefficients for the particular process and 
$D_{{\rm h}/i}(x,s=Q^2)$ are the \acp{FF} for turning the parton $i$ into 
the hadron $\rm h$. Like the \acp{PDF}, to leading order,
the \acp{FF} have an intuitive probabilistic interpretation.
Namely, they quantify the probability that the parton $i$ produced
at short distance of $1/Q$ forms a jet that includes the hadron $\rm h$ with
(longitudinal) momentum fraction $x$ of $i$. Furthermore, they 
are universal in a sense that they are believed not to depend on the
particular processes from which they are derived~\cite{kniehl2001}. 

\begin{figure}[htb]
\begin{center}
\subfigure[Charged hadrons at $Q=5~\gev$]{
\label{chap3:fig:kkpa}
\includegraphics[width=7.0cm]{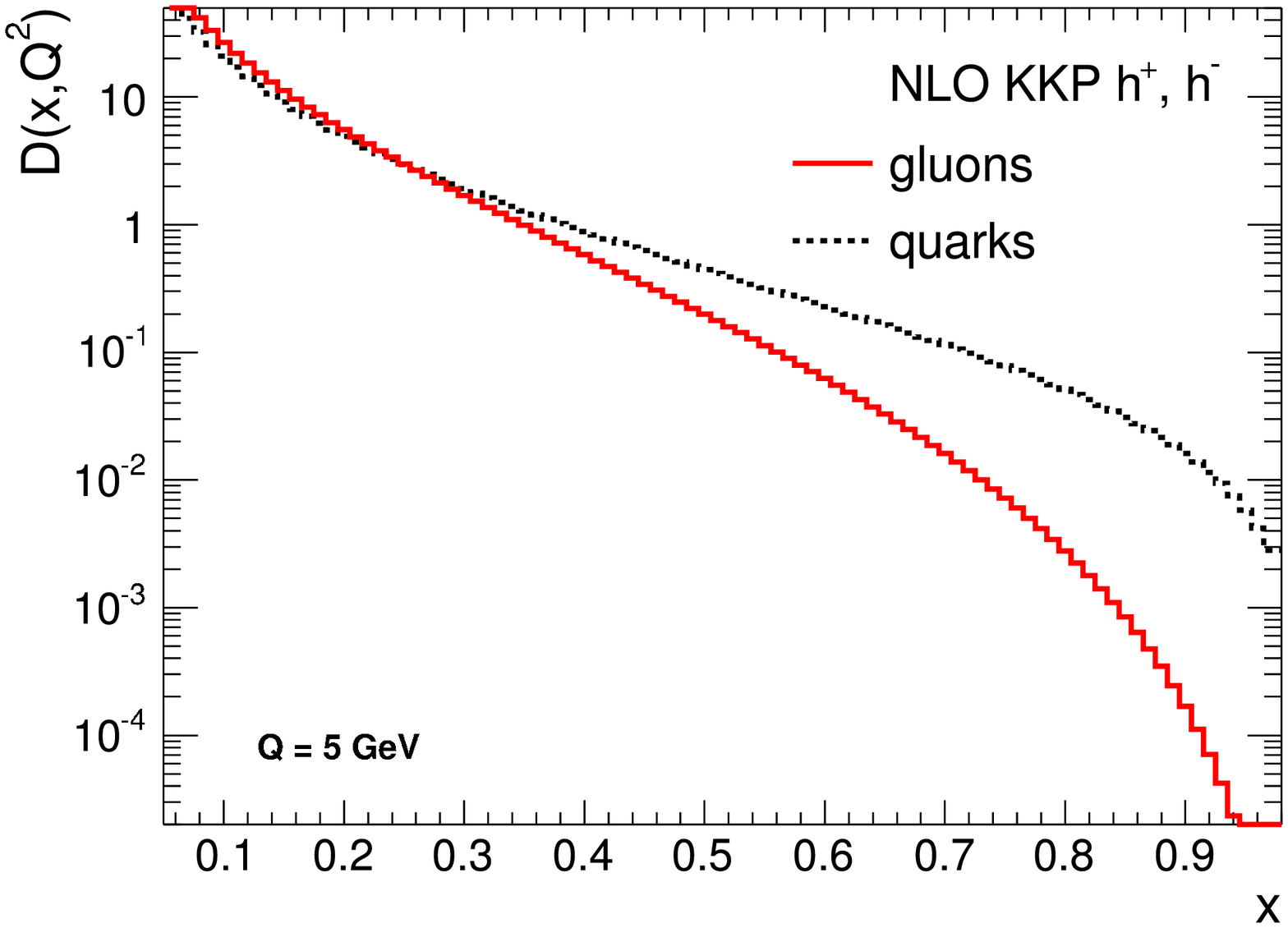}}
\hspace{0.4cm}
\subfigure[$D(x, 100^2~\gev^2) / D(x, 5^2~\gev^2)$]{
\label{chap3:fig:kkpb}
\includegraphics[width=7.0cm]{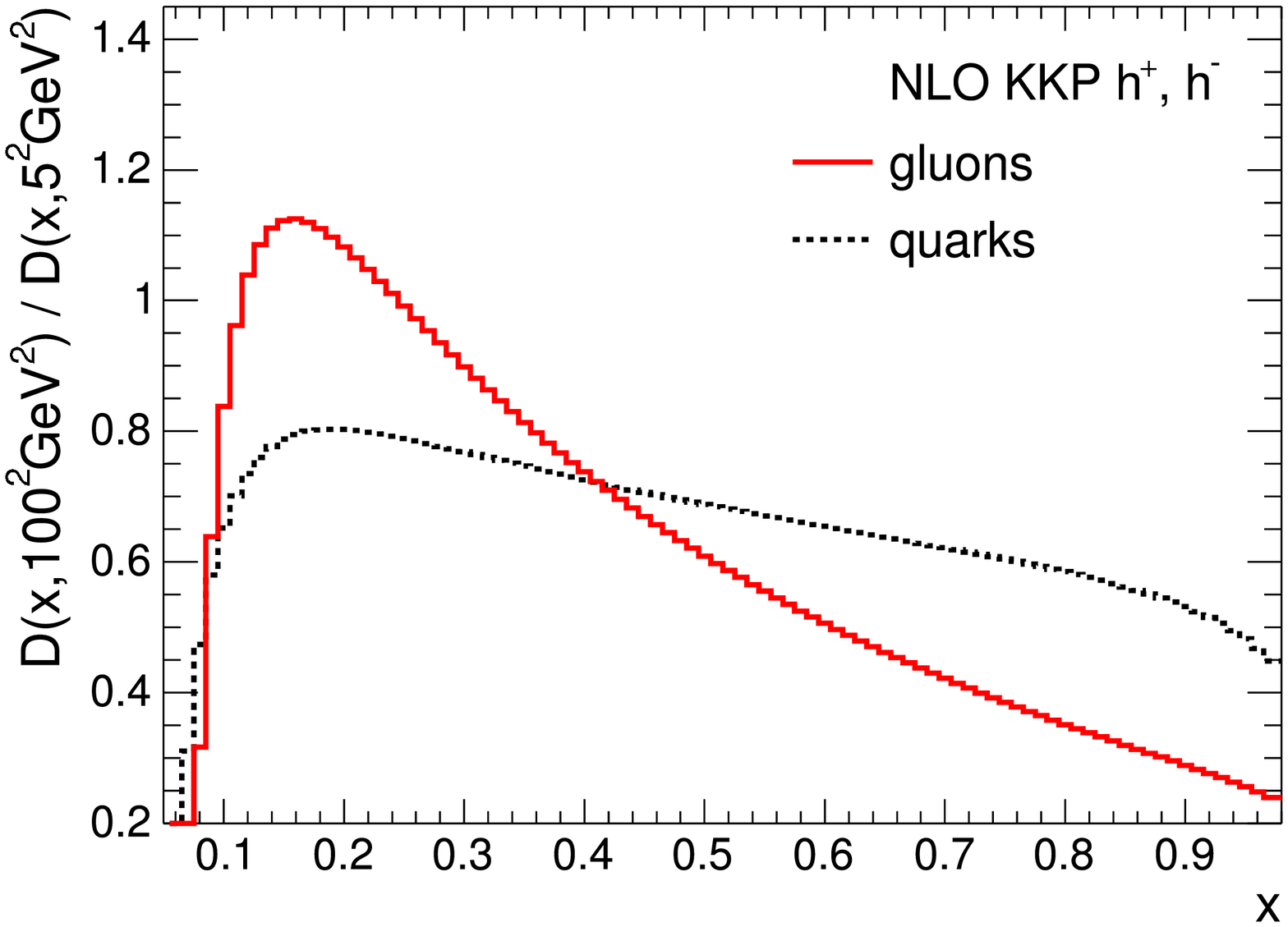}}
\vfill
\subfigure[Neutral pions at $Q=5~\gev$]{
\label{chap3:fig:kkpc}
\includegraphics[width=7.0cm]{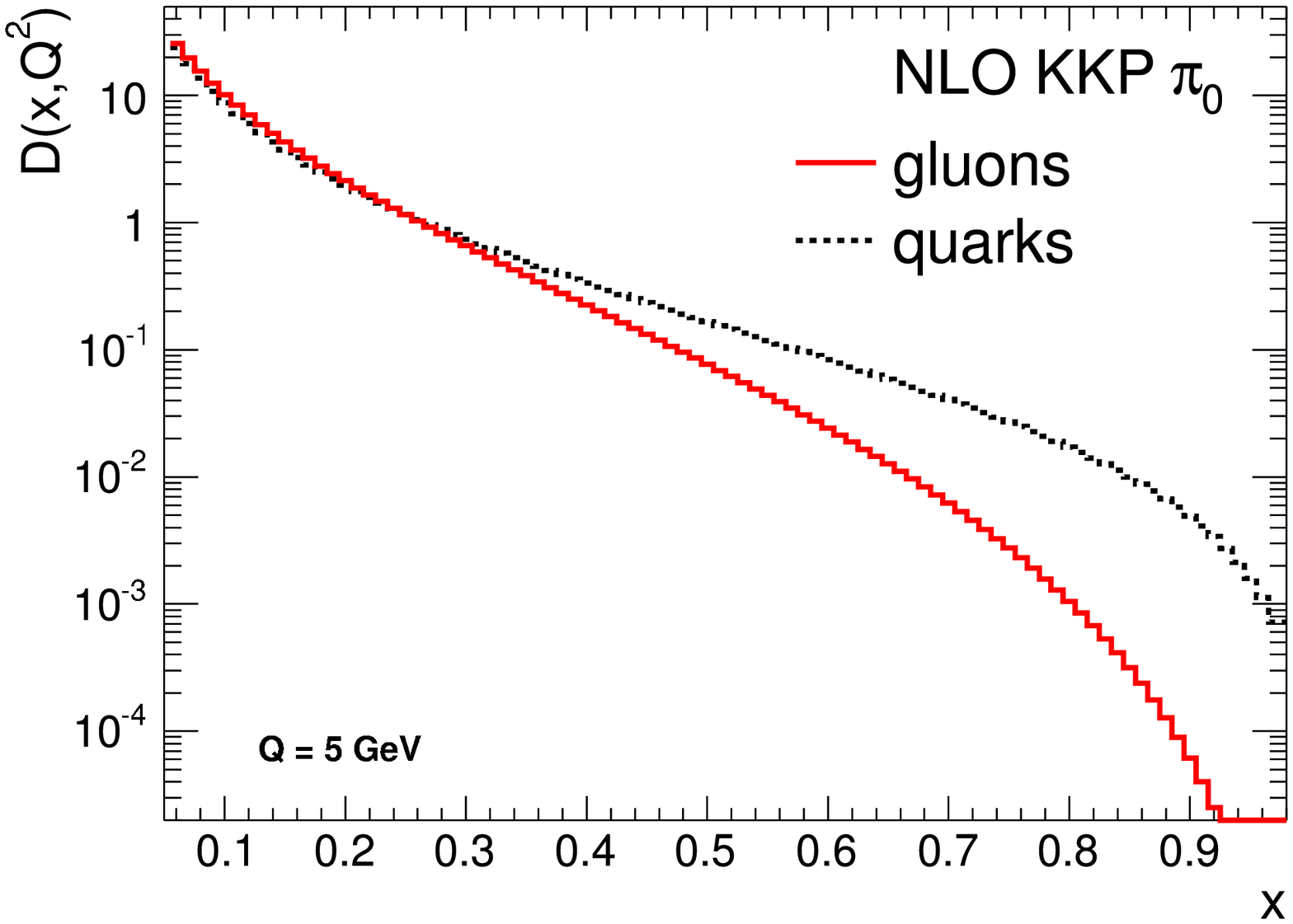}}
\hspace{0.4cm}
\subfigure[$D(x, 100^2~\gev^2) / D(x, 5^2~\gev^2)$]{
\label{chap3:fig:kkpd}
\includegraphics[width=7.0cm]{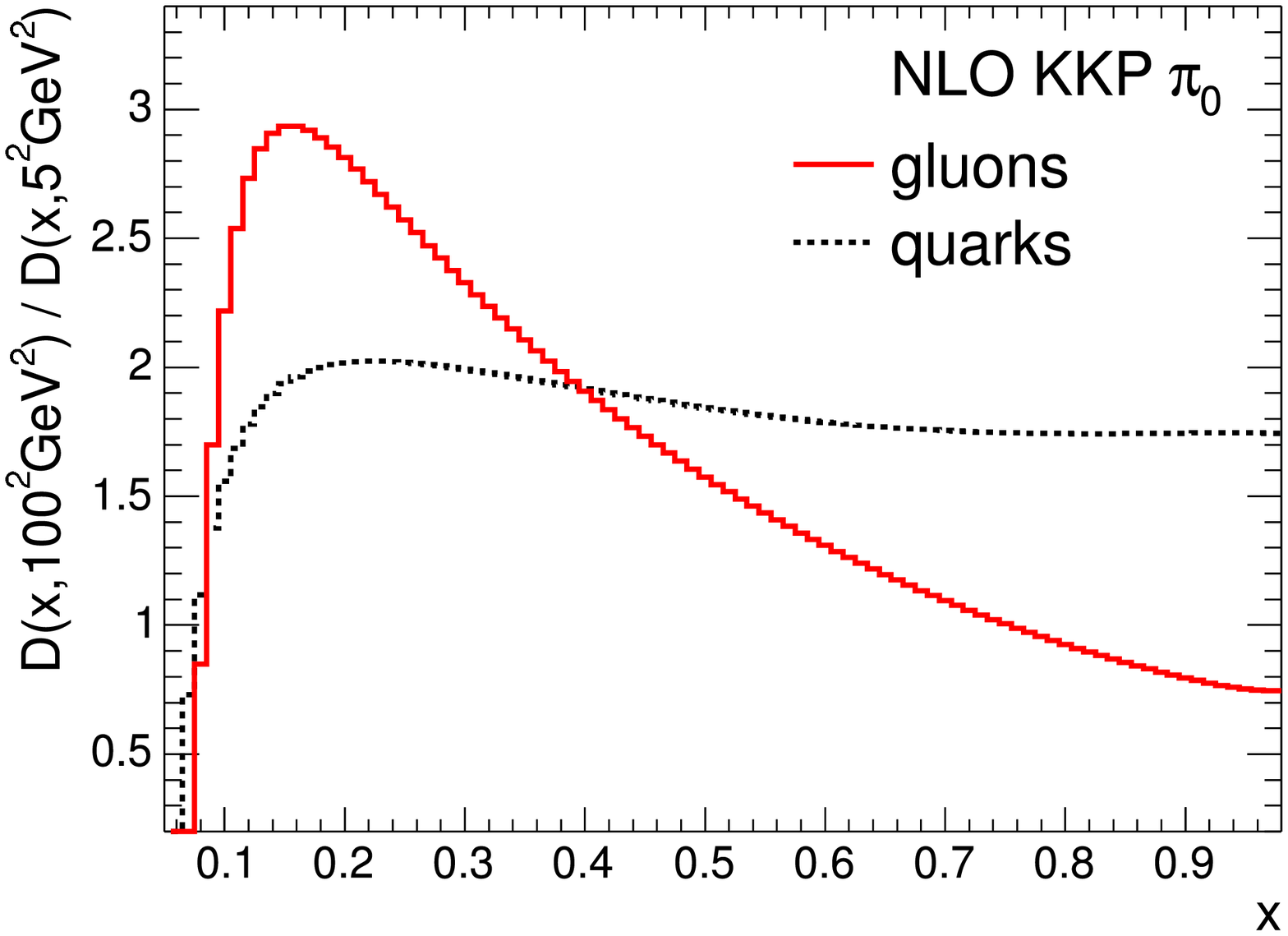}}
\end{center}
\vspace{-0.5cm}
\caption[xxx]{\acs{KKP} \acp{FF} for gluons and light quarks.
\subref{chap3:fig:kkpa} Fragmentation into charged hadrons at \mbox{$Q=5~\gev$} and
\subref{chap3:fig:kkpb} the ratio $D(x, 100^2~\gev^2) / D(x, 5^2~\gev^2)$.
\subref{chap3:fig:kkpc} Fragmentation into neutral pions at $Q=5~\gev$ and
\subref{chap3:fig:kkpd} the ratio $D(x, 100^2~\gev^2) / D(x, 5^2~\gev^2)$.}
\label{chap3:fig:kkp}
\end{figure}

\ifallpages
\pagebreak
\fi
The \acp{FF} themselves cannot be calculated by \acs{pQCD}, but as for the \acp{PDF} 
their scaling violation in $s=Q^2$ is described in the \ac{DGLAP} framework according to
\begin{equation*}
 s \frac{\partial}{\partial s}\, D_{{\rm h}/i}(z,s) = \sum_{j} \int _{x}^{1}
\frac{\dd z}{z} \, P_{j \, i} (z,\as(s)) \, D_{{\rm h}/j}(x/z,s)\;,
\end{equation*}
where the perturbatively calculable splitting functions $P_{j \, i}$ give
the evolution of parton $i$ into $j$~\cite{ellisqcd}.
Therefore, the \acp{FF} can be parameterized at some fixed scale, typically of the 
order of a few $\gev$, and then predicted at other scales. Several parameterizations 
of the \acp{FF} have been developed performing global \ac{NLO} fits to the 
available $\epem$ annihilation data from \acs{LEP} and $ep$ data from
\acs{HERA}~\cite{bourhis2000,kniehl2000,kretzer2000}.

In \fig{chap3:fig:kkp} we show the \acs{KKP} parameterization~\cite{kniehl2000}
at \acs{NLO} evaluated at the scale of $Q=5~\gev$ and the ratio 
\mbox{$D(x, 100^2~\gev^2) / D(x, 5^2~\gev^2)$} as a function of $x$
for the fragmentation of light quarks and gluons into charged 
hadrons and neutral pions. The \acs{KKP} parameterization is obtained by fits to 
available $\epem$ annihilation data performed within 
$0.1\le x\le 1$ in order to avoid mass and non-perturbative effects.
As can be seen, on average, quarks tend to fragment harder than gluons, an 
effect which increases with increasing fragmentation scale.

\begin{figure}[htb]
\begin{center}
\subfigure[Jet fragmentation functions]{
\label{chap3:fig:cdfjetfrag}
\includegraphics[width=7.2cm]{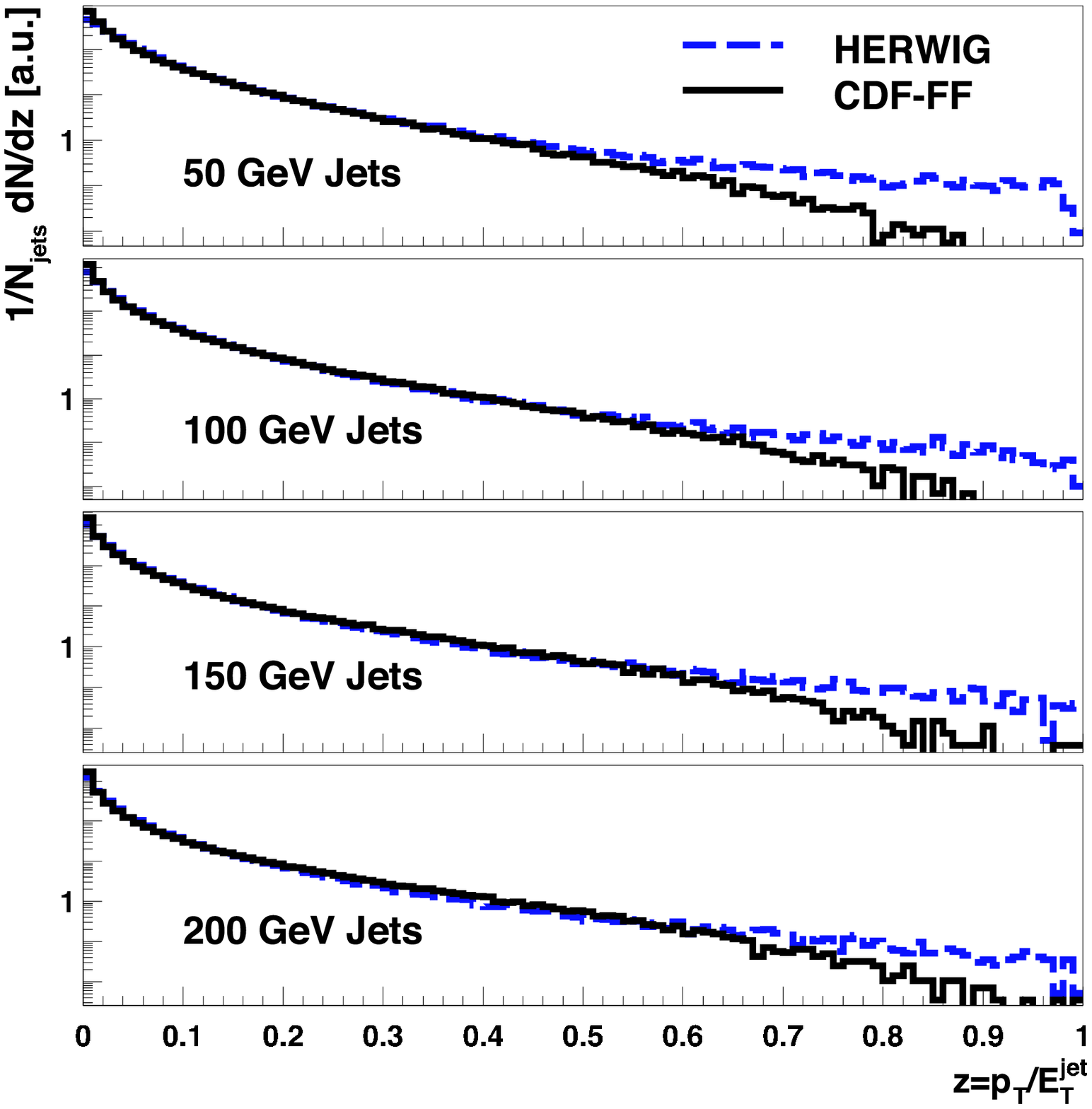}}
\hspace{0.1cm}
\subfigure[Fraction of jet energy in particles]{
\label{chap3:fig:cdfpartfrac}
\includegraphics[width=7.2cm, height=6.8cm]{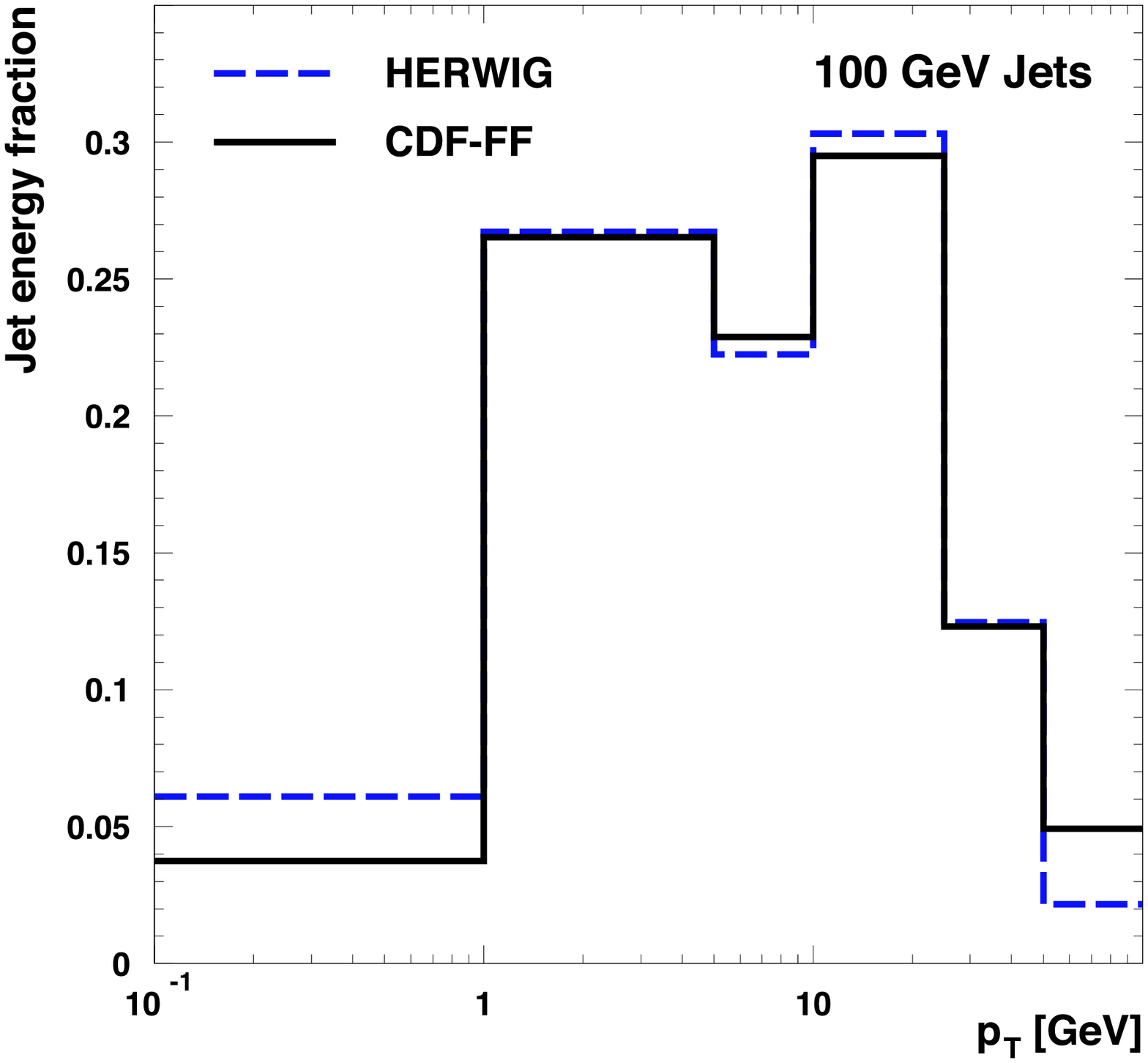}}
\end{center}
\vspace{-0.5cm}
\caption[xxx]{Jet fragmentation properties measured at mid-pseudo-rapidity comparing \acs{CDF-FF} 
(tuned \acs{ISAJET} to \acs{CDF} data) and \acs{HERWIG} fragmentation functions. 
\subref{chap3:fig:cdfjetfrag}~The associated, charged, particle $\pt$-spectrum normalized 
by the jet energy for different jet energies. \subref{chap3:fig:cdfpartfrac}~Fraction of
jet energy in associated particles of different $\pt$ for $\et=100~\gev$ jets.
Both figures are taken from~\Ref{affolder2001}.}
\label{chap3:fig:fragprop}
\end{figure}

The $\pt$-spectrum of charged particles in a jet has been obtained by \acs{CDF} using 
tracking information. The (normalized) distribution, $1/N_{\rm jets} \dd N/\dd z$,
where $z=\pt/\et^{\rm jet}$, can be regarded as an estimator for the jet \ac{FF}. The
distribution is shown in \fig{chap3:fig:cdfjetfrag} for different jet energies. 
The fragmentation function of \acs{ISAJET} (Feynman-Field fragmentation) tuned to give 
good agreement with data is called \acs{CDF-FF}. It is compared to \acs{HERWIG}, 
which uses cluster/string fragmentation adjusted to \acs{LEP} data. The change in 
the cross section, \eq{chap3:eq:diffppjetcrossection}, when \acs{HERWIG} \acp{FF}
were used instead of \acs{CDF-FF}, is smaller than the uncertainty attributed
to fragmentation functions in general, about $30$~\%~\cite{affolder2001}. For jets 
with $\et=100~\gev$ we reproduce in \fig{chap3:fig:cdfpartfrac} the fraction of jet 
energy carried by associated particles in the jet as a function of $\pt$; again both 
models are agree. Most of the jet energy is contained in particles of about $1$ to $30~\gev$. 
On average one third of the jet energy is manifested within the leading particle; 
a measured fact which is confirmed up to jet energies of $\et=400~\gev$~\cite{affolder2001}.

\subsection{Jet properties}
\label{chap3:jetproperties}
The internal structure of jets is dominated by multi-gluon emissions from the 
primary final-state partons. It is sensitive to the relative quark- and gluon-jet 
fraction and receives contributions from soft-gluon initial-state radiation and 
beam--beam remnant interactions. The structure is characterized by jet-shape
observables, which must be collinear and infra-red safe.
The study of jet shapes provides a stringent test of \ac{QCD} predictions and  
validates the models for parton cascades and soft-gluon emissions in 
hadron--hadron collisions. 

Jet shapes may be characterized in differential, $\rho(r)$, and integrated form,
$\Psi(r)$, where $r$ is a particle's radial distance from the jet axis.
They are defined as the average fraction of the jet transverse energy 
(or momentum) that lies inside an annulus or cone concentric to the jet axis
in the plane defined by the pseudo-rapidity ($\eta$) and azimuth ($\phi)$.
For an annulus of thickness $\Delta r$ and a cone of radius $R$ the differential 
distribution containing $N_{\rm jets}$ is defined as
\begin{equation}
\label{chap3:eq:diffjetshape}
\rho(r) = \frac{1}{\rm N_{\rm jets}} \sum_{\rm jets} 
\frac{\Etj(r-\frac{\Delta r}{2},r+\frac{\Delta r}{2}) }{\Etj(0,R)} 
\hspace{0.3cm}\text{for}\hspace{0.3cm} 0\le r \le R\;.
\end{equation}
The corresponding integrated distribution reads
\begin{equation}
\label{chap3:eq:intjetshape}
\Psi(r) = \frac{1}{\rm N_{\rm jets}} \sum_{\rm jets} \frac{\Etj(0,r) }{\Etj(0,R)} 
\hspace{0.3cm}\text{for}\hspace{0.3cm} 0\le r \le R\;.
\end{equation}

\begin{figure}[htb]
\begin{center}
\subfigure[Differential shape]{
\label{chap3:fig:diffjetshape}
\includegraphics[width=7cm, height=6cm]{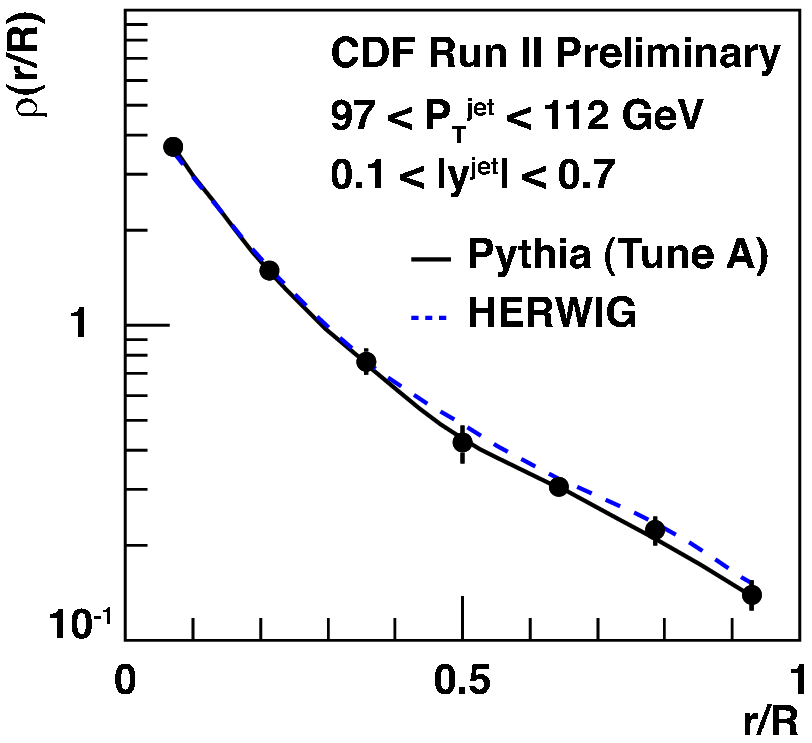}}
\subfigure[Integrated shape]{
\label{chap3:fig:intjetshape}
\includegraphics[width=7cm, height=6cm]{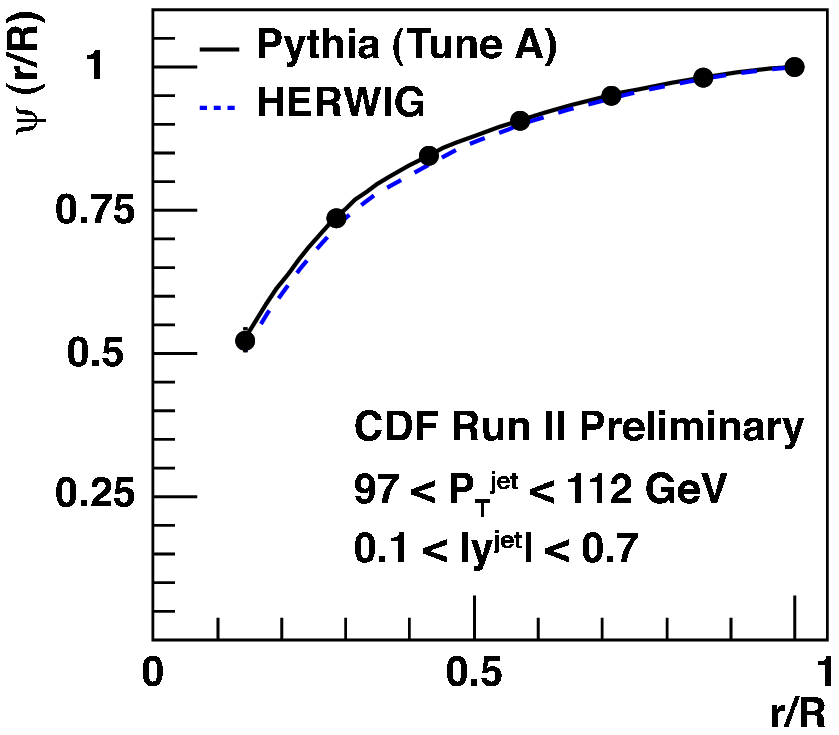}}
\end{center}
\vspace{-0.5cm}
\caption[xxx]{Differential~\subref{chap3:fig:diffjetshape} and integrated 
\subref{chap3:fig:intjetshape} shapes measured by \acs{CDF} 
(preliminary) compared to \acs{PYTHIA} (Tune A) and \acs{HERWIG}. 
The jets are defined using the \acs{MidPoint} algorithm with a cone size of $R=0.7$. 
The figure is taken from~\Ref{cdfpub}.}
\label{chap3:fig:jetshape}
\end{figure}

\begin{figure}[htb!]
\vspace{0.5cm}
\begin{center}
\subfigure[Data versus Monte Carlo generators]{
\label{chap3:fig:intjetshapea}
\includegraphics[width=7.0cm]{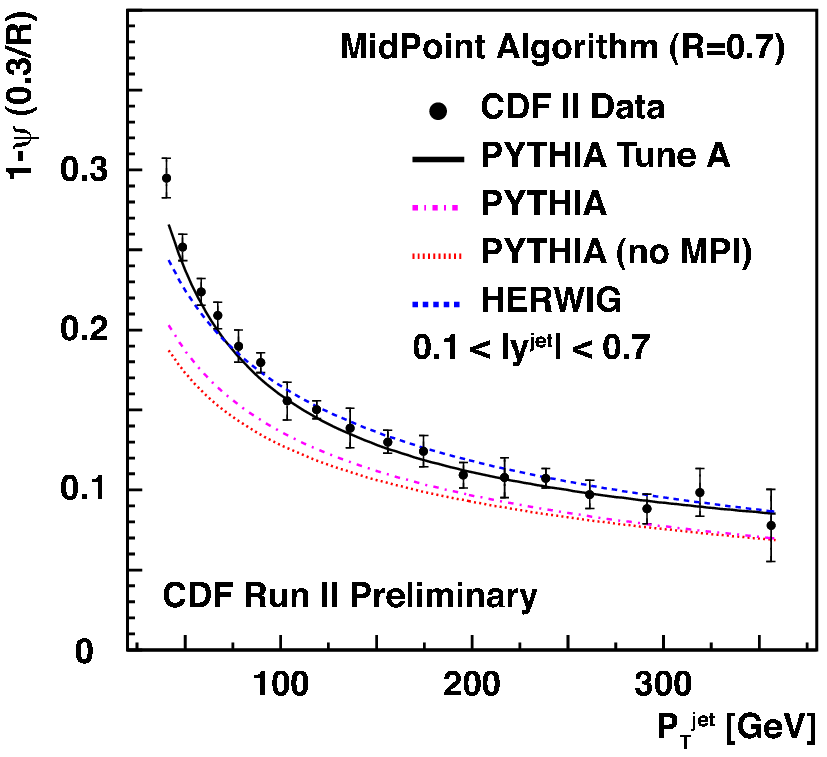}}
\hspace{0.5cm}
\subfigure[Data versus parton contribution (Tune A)]{
\label{chap3:fig:intjetshapeb}
\includegraphics[width=7.0cm]{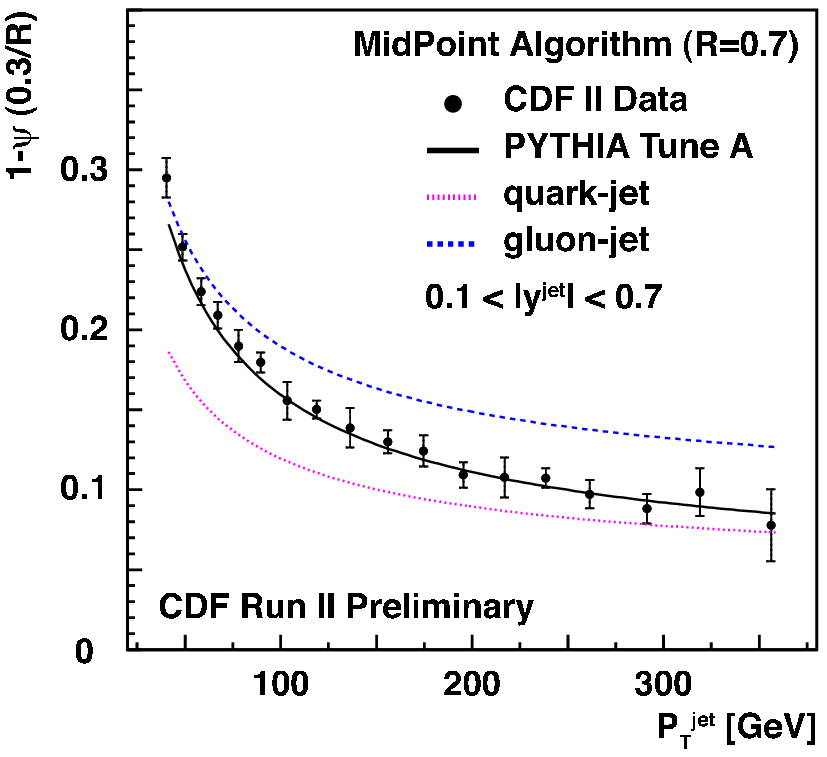}}
\end{center}
\vspace{-0.5cm}
\caption[xxx]{The fraction of the jet transverse momentum 
outside $r=0.3$ as a function of $\Ptj$ measured by 
\acs{CDF} (preliminary). The jets are defined using the 
\acs{MidPoint} algorithm with a cone size of $R=0.7$. 
\subref{chap3:fig:intjetshapea} The measured shape compared 
to simulations of \acs{HERWIG} and \acs{PYTHIA} 
(default, Tune A and without multiple parton interactions). 
\subref{chap3:fig:intjetshapea} The measured shape compared 
to calculations of \acs{PYTHIA} (Tune A) and the separated 
contributions from quark-  and gluon-jets.
Both figures are taken from~\Ref{martinez2004}.}
\label{chap3:fig:intjetshapes}
\end{figure}

The \acs{CDF} experiment has presented preliminary results on jet shapes 
for central jets ($0.1\le y \le 0.7$) with transverse momentum $37~\gev\le \Ptj\le 380~\gev$, 
where jets are searched using the \acs{MidPoint} algorithm with a cone size of 
$R=0.7$~\cite{cdfpub}.~\footnote{A $f=0.75$ merging fraction has been used instead of 
the default $0.5$.} The measured jet shapes of which we show $97\le \Ptj \le 112~\gev$
in \fig{chap3:fig:jetshape}, have been compared to calculations from \acs{PYTHIA} 
(with special parameters, Tune A~\cite{pythiatunea}, see below) and \acs{HERWIG}. 
Above $\Ptj\ge55~\gev$, both the tuned \acs{PYTHIA} and \acs{HERWIG} describe the data well, 
while below \acs{HERWIG} slightly deviates from the data.

\Fig{chap3:fig:intjetshapes} presents, for a fixed radius $r_0 = 0.3$, the average
fraction of the jet transverse momentum outside $r=r_0$, given by $1-\Psi(r_0/R)$, as 
a function of $\Ptj$. The preliminary measurements show that the fraction 
of jet transverse momentum at a given fixed $r_0/R$ increases ($1-\Psi(r_0/R)$ decreases) 
with $\Ptj$. This indicates that the jets become narrower as  $\Ptj$ increases. 
As can be seen from \fig{chap3:fig:intjetshapea} the tuned version of \acs{PYTHIA} 
describes all of the data well, whereas \acs{HERWIG} produces jets that are too narrow at 
$P_{T}^{\rm jet}\le55~\gev$. In order to illustrate the importance of proper modeling of 
soft-gluon radiation in describing measured jet shapes, in addition to the tuned \acs{PYTHIA}, 
two other \acs{PYTHIA} parameter sets have been compared using default options with and 
without the contribution from multiple parton interactions between the remnants.
\acs{PYTHIA} with default parameters produces jets systematically narrower than the data 
throughout the whole $P_{T}^{\rm jet}$ region. Looking at the difference between \acs{PYTHIA} and 
\acs{PYTHIA} without MPI the contribution from secondary parton interactions between 
remnants to the predicted jet shapes is relatively small and decreases as $P_{T}^{\rm jet}$ 
increases. \Fig{chap3:fig:intjetshapeb} shows the contributions of gluon and quark jets 
in \acs{PYTHIA} (Tune~A) compared to the measurement of $1-\Psi(0.3/R)$ as a function of 
$\Ptj$.~\footnote{Each jet at particle level from \acs{PYTHIA} is classified as a quark- or 
gluon-jet by matching its direction with that of one of the outgoing partons from the hard 
interaction.} The Monte Carlo predictions indicate that the measured jet shapes are dominated by
contributions from gluon-initiated jets at low $\Ptj$, while contributions from
quark-initiated jets become important at high $\Ptj$. This can be explained in terms of
the different partonic contents in the incident hadrons (proton and anti-proton) 
in the low- and high-$\Ptj$ regions, since the mixture of gluon- and quark-jet 
in the final state partially reflects the nature of the incoming partons 
that participate in the hard interaction (see \fig{chap3:fig:relativecontrib}).
\fi

\section{Medium-induced parton energy loss}
\label{chap3:partoneloss}
\ifpeloss
In the first formulation of parton energy loss
Bjorken predicted that high energy partons propagating 
through the \ac{QGP} suffer differential energy loss 
due to elastic scattering with the quarks and gluons in 
the plasma~\cite{bjorken1982}. 
He further pointed out that as an interesting signature 
events may be observed, in which the hard collisions 
may occur such that one jet is escaping without absorption, 
whereas the other is fully absorbed in the medium.
The resulting, collisional, loss was estimated to be 
$\dd E/\dd x\simeq \as^2\sqrt{\varepsilon}$, 
where $\varepsilon$ is the energy density of the (ideal) 
\ac{QGP}. The loss turns out to be quite low, of the order 
of $0.1~\gev/\fm$~\cite{thoma1995}.

However, \ac{QCD} gluon bremsstrahlung, as in \acs{QED} bremsstrahlung, 
is another important ---if not the dominant--- 
source of energy loss~\cite{gyulassy1993,wang1994}. 
Due to multiple inelastic scattering and induced gluon 
radiation high-energy jets and high-$\pt$ leading hadrons become 
depleted, quenched~\cite{gyulassy1990} or even extinct.
The radiative loss, as we will report in the following, 
is considerably larger than the collisional loss.
A number of theoretical studies have dealt with the 
subject~\cite{baier1994,baier1996,baier1996b,zakharov1996,zakharov1997,baier1998,
zakharov1998,zakharov1999,zakharov2000,wiedemann1999,wiedemann2000,wiedemann2000b,
gyulassy1999,gyulassy2000,gyulassy2000b}.~\footnote{See~\Refs{baier2000,gyulassy2003,kovner2003} 
for recent reviews.}

In the next section we briefly present the general ideas of the model 
proposed by \ac{BDMPS-Z}~\cite{baier1996,baier1996b,baier1998} and 
its evaluation in the framework of quenching probabilities 
(weights)~\cite{baier2001} for light quarks and gluons, as calculated 
by Salgado and Wiedemann~\cite{salgado2003}.

\subsection{Medium-induced radiative energy loss}
\label{chap3:radenloss}
It has been shown~\cite{baier1994} that the genuine \ac{pQCD} process, 
depicted in \fig{chap3:fig:gluonsstrahlung}, dominantly determines
the energy loss of an energetic parton traversing through
dense \ac{QCD} matter. 

\begin{figure}[htb]
\begin{center}
\includegraphics[width=10cm]{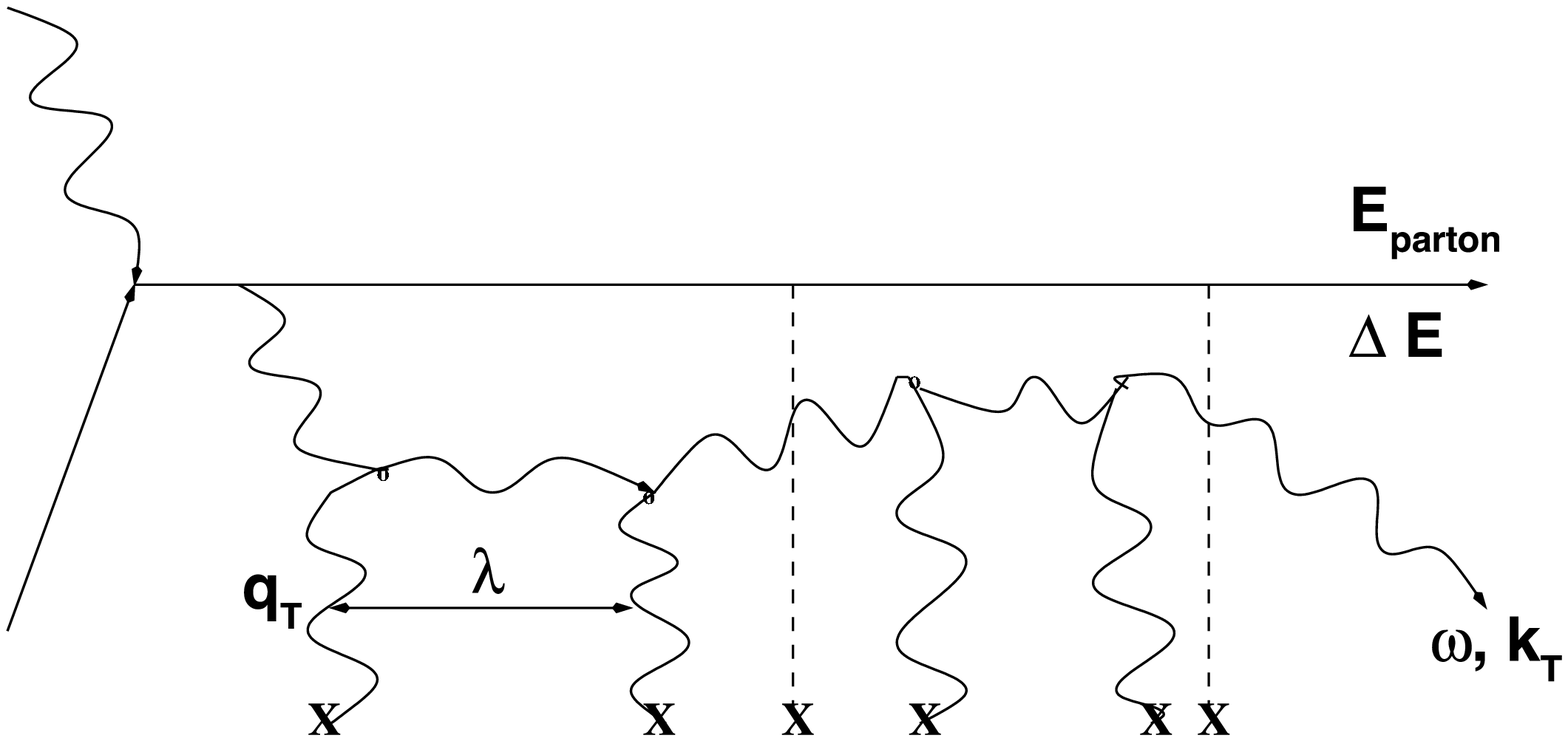}
\end{center}
\vspace{-0.3cm}
\caption[xxx]{Typical gluon-radiation diagram, adapted from~\Ref{baier2002}.}
\label{chap3:fig:gluonsstrahlung}
\end{figure}

After its initial production in a hard collision the energetic 
parton with energy $E$ radiates off a gluon with energy $\omega$ 
with a probability proportional to $L$, the size of the dense medium it 
traverses. Due to its non-abelian nature the radiated gluon,
itself, subsequently suffers multiple scattering due to the
interactions in the medium 
with a mean free path $\lambda$, 
which decreases as the density of the medium increases. 
The number of scatterings with momentum transfer $q^2_{\rm T}$ 
that the radiated gluon undergoes until it eventually decoheres, 
is proportional to $L$, too. Thus, the average energy loss
of the parton must be proportional to $L^2$.~\footnote{See 
\Ref{baier1996b} for an elaborated qualitative and quantitative 
derivation.}
This is the most distinctive feature of the \ac{QCD} energy loss compared 
to the \acs{QED} case being only proportional to $L$. It arises due 
to the non-abelian character of \ac{QCD}: the fact 
that gluons interact with each other, while photons do not. 
Note that we concentrate on the coherent regime of the \ac{LPM} effect 
valid for intermediate values of $\omega$, 
\begin{equation*}
\label{chap3:eq:regimes}
\omega_{\rm BH} \sim \lambda \, q^2_{\rm T}\ll \omega \ll 
\omega_{\rm fact} \sim L^2 \, q^2_{\rm T}/\lambda \le E\rightarrow\infty\;,
\end{equation*}
in-between the Bethe-Heitler 
and factorization regimes~\cite{baier1994,baier1996}.

Conveniently, the properties of the medium are collectively encoded in
the transport coefficient, $\hat{q}$, of the medium, which is defined as 
the average medium-induced transverse momentum squared transferred to the projectile 
per mean free path,
\begin{equation}
\label{chap3:eq:hatq}
\hat{q} = \frac{\av{q_{\rm T}^2}}{\lambda}\;.
\end{equation}

The scale of the radiated energy distribution $\omega\,\dd I/\dd\omega$ 
(or $\omega\,\dd^2I/\dd\omega\dd \kt$) is set by the characteristic energy 
\begin{equation}
\label{chap3:eq:wc}
\omega_{\rm c} = \frac{1}{2}\,\hat{q}\,L^2
\end{equation} 
of the emitted gluon,
which depends on the in-medium path-length of the projectile 
and on the transport coefficient of the medium.

In the original \ac{BDMPS-Z} calculation~\cite{baier1996,baier1996b,baier1998}
the radiated gluon is allowed to exploit the full transverse 
phase space regardless of its energy. However, physically 
the transverse momentum $k_{\rm T}$ of the radiated gluon 
is kinematically bound to be smaller than its energy $\omega$. 
This imposes a constraint on the emission probability 
via the dimensionless quantity
\begin{equation}
\label{chap3:eq:kinematicr}
R = \omega_{\rm c}\,L\,= \frac{1}{2}\,\hat{q}\,L^3
\end{equation}
first introduced as `density parameter' in~\Ref{salgado2002}.

\begin{figure}[htb]
\begin{center}
\includegraphics[width=10cm]{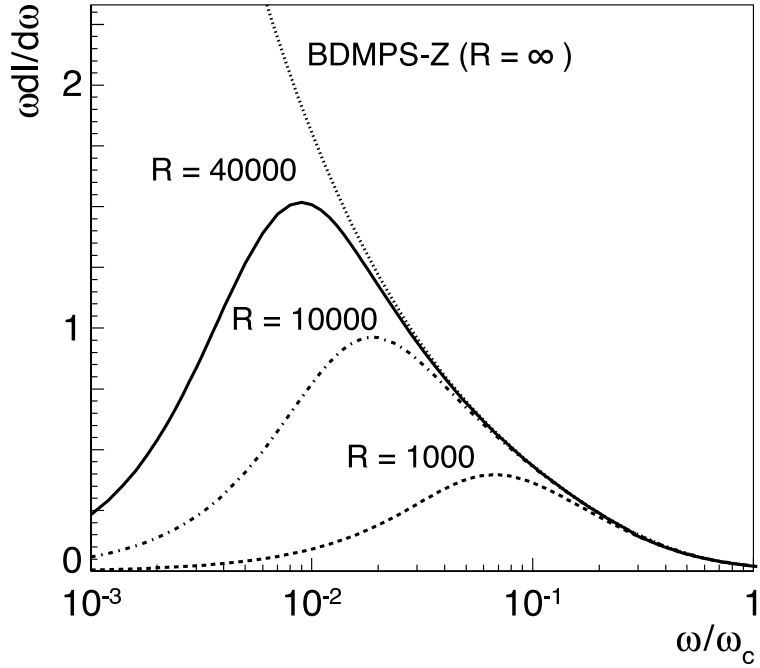}
\end{center}
\vspace{-0.8cm}
\caption[xxx]{The medium-induced radiated gluon energy distribution 
$\omega\,\dd I/\dd\omega$ for different values of the kinematical 
constraint $R=\omega_{\rm c}\,L$ and in the original form 
of \acs{BDMPS-Z} (for $R\rightarrow\infty$). The figure
is taken from~\Ref{salgado2003}.}
\label{chap3:fig:bdmpscomp}
\end{figure}

In the multiple scattering approximation of the inclusive
radiation spectrum~\Ref{salgado2003}, referred to 
as \acs{BDMPS-Z-SW}, the two parameters $\omega_{\rm c}$ 
and $R$ determine the energy distribution of radiated gluons, 
$\omega\,\dd I/\dd\omega$, reproduced in \fig{chap3:fig:bdmpscomp}.
While $\omega_{\rm c}$ sets the scale of the distribution, 
$R$ controls its shape in the region $\omega\ll \omega_{\rm c}$
suppressing soft-gluon emission through the kinematical 
bound on $k_{\rm T}$ ($k_{\rm T}<\omega$).
The \ac{BDMPS-Z} case corresponds to $R\to\infty$. It 
can be recovered by considering an infinitely-extended medium:
taking $L\to\infty$ for fixed finite $\omega_{\rm c}$.
In the limit $R\to\infty$ the distribution is of the form
\begin{equation} 
\label{chap3:eq:wdIdw}
\lim_{R\to\infty} \omega\frac{\dd I} {\dd \omega} \simeq 
\frac{2\,\as\,C_{\rm R}}{\pi}
\left\{
\begin{array}{lll}
\sqrt{\frac{\omega_{\rm c}}{2\omega}} & {\rm for} & \omega<\omega_{\rm c} \\
\frac{1}{12}\,\left(\frac{\omega_{\rm c}}{\omega}\right)^2 & {\rm for} & 
 \omega\ge\omega_{\rm c} \\
\end{array}
\right.
\end{equation}
where $C_{\rm R}$ is the \ac{QCD} coupling factor or Casimir factor 
between the  considered hard parton and the gluons in the medium; it is  
$C_{\rm F}=4/3$ if the parton is a quark and $C_{\rm A}=3$ if the parton 
is a gluon. 

\pagebreak
In the eikonal limit of very large parton initial energy $E$ 
($E\gg\omega_{\rm c}$), the integral of the radiated-gluon energy 
distribution estimates the average energy loss of the parton
\begin{equation}
\label{chap3:eq:avdE}
\av{\Delta E}_{R\to\infty} = \lim_{R\to\infty} \int_0^{\infty} 
\omega \frac{\dd I}{\dd \omega}\,\dd \omega
\propto \as\,C_{\rm R}\,\omega_{\rm c} \propto \as\,C_{\rm R}\,\hat{q}\,L^2\;.
\end{equation}
Note that, due to the steep fall-off at large $\omega$ in 
\eq{chap3:eq:wdIdw}, the integral is dominated by the 
region $\omega<\omega_{\rm c}$. The average energy loss is:
\begin{itemize}
\item proportional to $\as$ (typically we take $0.3$ instead of $0.5$);
\item proportional to $C_{\rm R}$ and, thus, larger by a factor 
$9/4$ for gluons than for quarks; 
\item proportional to the transport coefficient of the medium;
\item proportional to $L^2$; 
\item independent of the parton initial energy $E$.  
\end{itemize}

It is a general peculiarity of all calculations~\cite{baier1994,baier1996,baier1996b,
zakharov1996,zakharov1997,baier1998,zakharov1998,zakharov1999,zakharov2000,wiedemann1999,
wiedemann2000,wiedemann2000b,gyulassy1999,gyulassy2000,gyulassy2000b}
that the radiated energy distribution~\eq{chap3:eq:wdIdw} in the \ac{LPM} regime
does not depend on  energy $E$ of the initial parton. However, depending on the way 
various kinematic bounds are taken into account, 
the resulting $\Delta E$ is $E$-independent~\cite{baier1996,baier1996b,baier1998} 
or depends logarithmically on $E$~\cite{gyulassy1999,gyulassy2000,gyulassy2000b}.
In any case, there is always strong intrinsic dependence of the radiated energy on 
the initial energy, simply determined by the fact that the former cannot 
be larger than the latter, $\Delta E\leq E$, as we will further discuss 
in~\sect{chap3:step3}.

The transport coefficient can be related to the density $\rho$ of the 
scattering centers and to the typical momentum transfer in the 
gluon scattering off these centers, 
\mbox{$\hat{q}= \rho\,\int q^2\dd\sigma/\dd q^2$}. 
For cold nuclear matter
\begin{equation*}
\hat{q}_{\rm cold} \simeq 0.05~\gev^2/\fm \simeq 8\,\rho_0
\end{equation*}
has been obtained using the nuclear density $\rho_0=0.16~\fm^{-3}$,
the gluon \ac{PDF} of the nucleon and $\as=0.5$~\cite{baier1996,baier1996b}.
The value agrees with the extracted value $\hat{q}=(9.4\pm0.7)\,\rho_0$
resulting from the analysis of gluon $k_{\rm T}$-broadening in experimental 
data of $\Jpsi$ transverse-momentum distributions~\cite{kharzeev1997}.
The estimation for a hot medium~\cite{baier1996,baier1996b} 
\begin{equation}
\hat{q}_{\rm hot}\simeq 1~\gev^2/\fm\simeq 20\,\hat{q}_{\rm cold}
\end{equation} 
based on perturbative treatment ($\as=0.3$) of gluon scattering 
in an ideal \ac{QGP} with a temperature of $T\simeq 250~\mev$ 
resulted in the value of the transport coefficient of about a 
factor twenty larger than for cold matter.
The average energy loss of the cold and hot medium according to
\eq{chap3:eq:avdE} amounts to 
\begin{equation*}
\Delta E_{\rm cold} \approx 0.02\,\gev\, (L/\fm)^2 
\hspace{0.5cm}\text{and}\hspace{0.5cm}
\Delta E_{\rm hot} \approx 0.3\,\gev\,(L/\fm)^2\;.
\end{equation*}
The precise values must not be taken too serious. However, the large
difference suggests that the hot matter is  may be rather effective
in stimulating the energy loss. 
The reason is due to
\begin{itemize}
\item the higher density of color charges and the correspondingly 
shorter mean free path of the probe in the \ac{QGP};
\item the fact that deconfined gluons have harder momenta than confined gluons
and, therefore, the typical momentum transfers are larger. 
\end{itemize}

\begin{figure}[htb]
\begin{center}
\includegraphics[width=8cm]{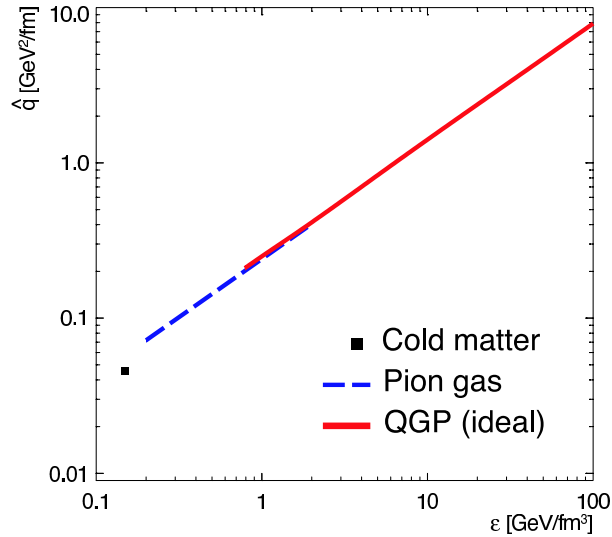}
\end{center}
\vspace{-0.3cm}
\caption[xxx]{Transport coefficient as a function of energy density for 
different media: cold matter, massless hot pion gas 
and ideal \acs{QGP}. The figure is adapted from~\Ref{baier2002}.}
\label{chap3:fig:qtranspvseps}
\end{figure}

\Fig{chap3:fig:qtranspvseps} shows the estimated dependence 
of the transport coefficient, $\hat{q}$, on the energy 
density, $\varepsilon$, for different equilibrated media 
(cold matter, hot pion gas and ideal \ac{QGP})~\cite{baier2002}. 
Assuming that the \ac{QGP} is formed (and sustained) at the \ac{LHC} 
around $\varepsilon\simeq 100~\gev/\fm^3$, one expects a transport 
coefficient of the order of $\hat{q}\simeq 10~\gev^2/\fm$. 

So far, we have assumed a static medium (with a constant transport 
coefficient). Though, due to the expansion of the system, the density 
of the medium decreases strongly in time. Hence, the transport
coefficient should be time-dependent.
However, it turns out~\cite{baier1998b,salgado2003} that a simple scaling 
law exists, which translates the gluon energy distribution for an expanding 
medium with a time-decreasing $\hat{q}(\xi)$ into an equivalent distribution 
for a static medium, with $\av{\hat{q}}=\const$, via
\begin{equation}
\label{chap3:eq:qscale}
\av{\hat{q}}= \frac{2}{L^2} \, \int_{\xi_0}^{L+\xi_0}  
\left( \xi - \xi_0 \right) \, \hat{q}(\xi) \, \dd \xi\;,
\end{equation}
where $\xi_0\sim 0.1~\fm\ll L$ is the time, at which the parton is
produced. Thus, depending on how one implements the expansion 
dynamics (\eg~ one-dimensional Bjorken expansion) one can translate 
the values for time-averaged into time-dependent transport coefficients.

\subsection{Quenching weights}
\label{chap3:qw}
The spectrum of the additional medium-induced energy loss 
due to scattering in spatially-extended \ac{QCD} matter can be 
characterized by the probability $P(\Delta E)$ that the radiated
gluons carry altogether the energy $\Delta E$. Assuming the
independent emission of soft gluons from the hard parton, 
the corresponding ansatz in the soft limit ($\Delta E\ll E$) 
reads
\begin{equation*}
P(\Delta E) = \sum_{n=0}^{\infty} \left[ \prod_{i=1}^{n} \int \dd
 \omega_i \,  \frac{\dd I(\omega_i)}{\dd \omega} \right ] 
\, \delta(\Delta E - \sum _{i=0}^{n}\omega_i) 
\, \exp{\left[ - \int \dd\omega\, \frac{\dd I}{\dd \omega}\right]} \;.
\end{equation*}
The expression can be explicitly evaluated,
because the summation over arbitrarily many gluon emissions can be 
factorized via Mellin and Lapclace transformations~\cite{baier2001}.
In general, the probability distribution $P(\Delta E)$,
also known as quenching weight, has a discrete and 
a continuous part~\cite{salgado2002},
\begin{equation}
\label{chap3:eq:pdeltae}
P(\Delta E;\,R,\omega_{\rm c})=p_0(R)\,\delta(\Delta E) + p(\Delta E;\,R,\omega_{\rm c})\;,
\end{equation}
which have recently been calculated in two different approximations~\cite{salgado2003}. 
The approximations differ in treating the medium as a source of many soft or a 
few hard momentum transfers. For the purpose of the thesis, the small numerical
differences between both are negligible, and we stay with the multiple scattering 
approximation of the \acs{BDMPS-Z-SW} model.
In addition to the indicated input parameters, the scale $\omega_{\rm c}$ and 
the kinematic constraint $R$, the discrete and continuous parts of the weight
and therefore also $P(\Delta E) \equiv  P(\Delta E;\,R,\omega_{\rm c})$
depend on the parton species (quark or gluon) of the projectile. 
The discrete weight $p_0\equiv p_0(R)$ gives the probability to have no 
medium-induced gluon radiation. For finite in-medium path length, 
there is always a finite probability $p_0\neq 0$ that the projectile
is not affected by the medium, but in the kinematic limit one finds
$\lim_{R\rightarrow \infty} p_0=0$.
The continuous weight $p(\Delta E)\equiv p(\Delta E;\,R,\omega_{\rm c})$ 
gives the probability to radiate an energy $\Delta E$, if at least one gluon 
is radiated. Due to the \ac{LPM} coherence effect, $P(\Delta E)$ is a
generalized probability, which might take negative values for some range
in $\Delta E$ as long as the normalization is unity, 
\begin{equation*}
\int \dd \epsilon \, P(\epsilon) = p_0 + \int \dd\epsilon \, p(\epsilon) = 1\;.
\end{equation*}

\begin{figure}[htb]
\begin{center}
\includegraphics[width=10cm]{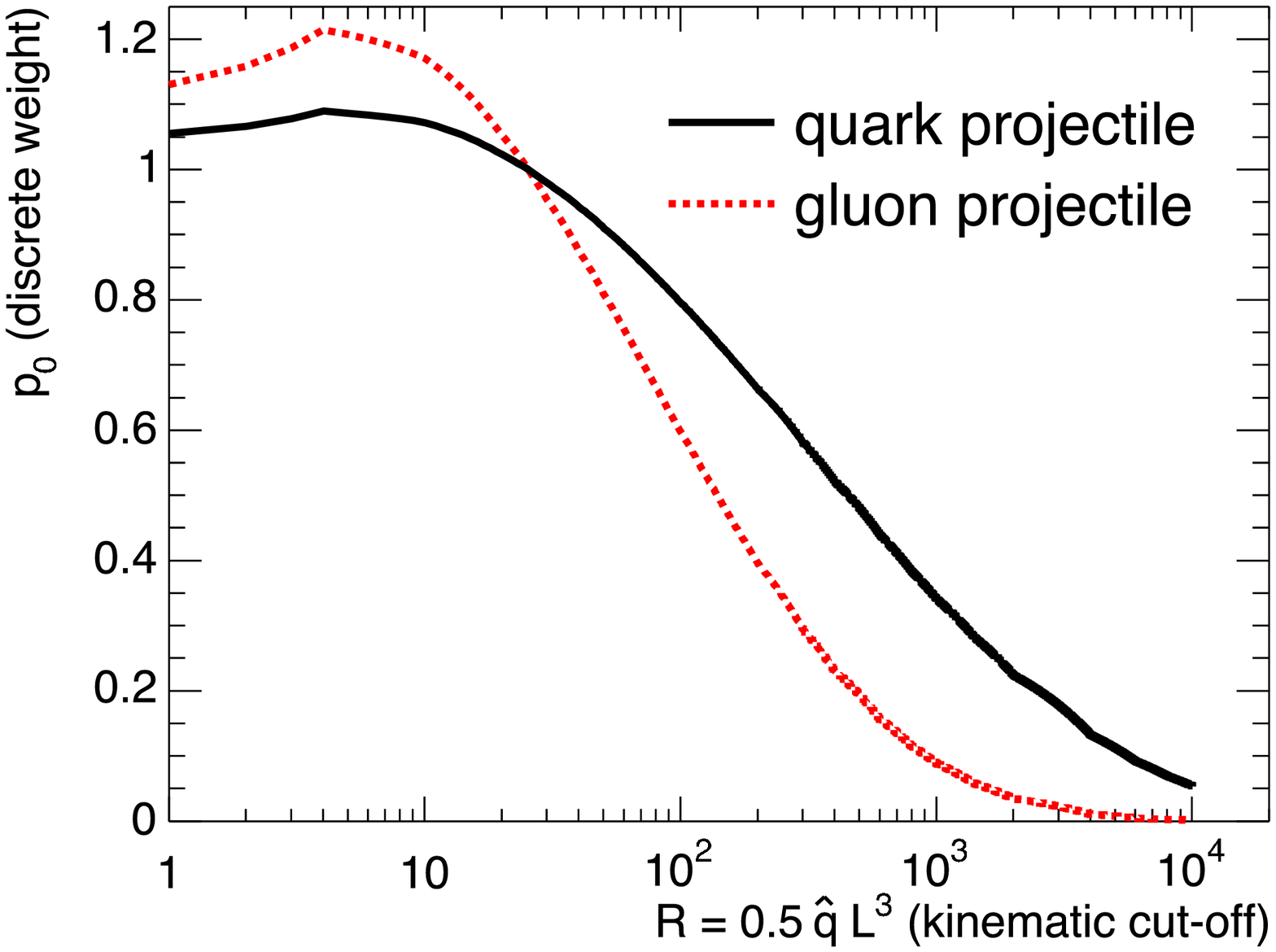}
\vspace{0.5cm}
\vfill
\includegraphics[width=4.9cm, height=3.9cm]{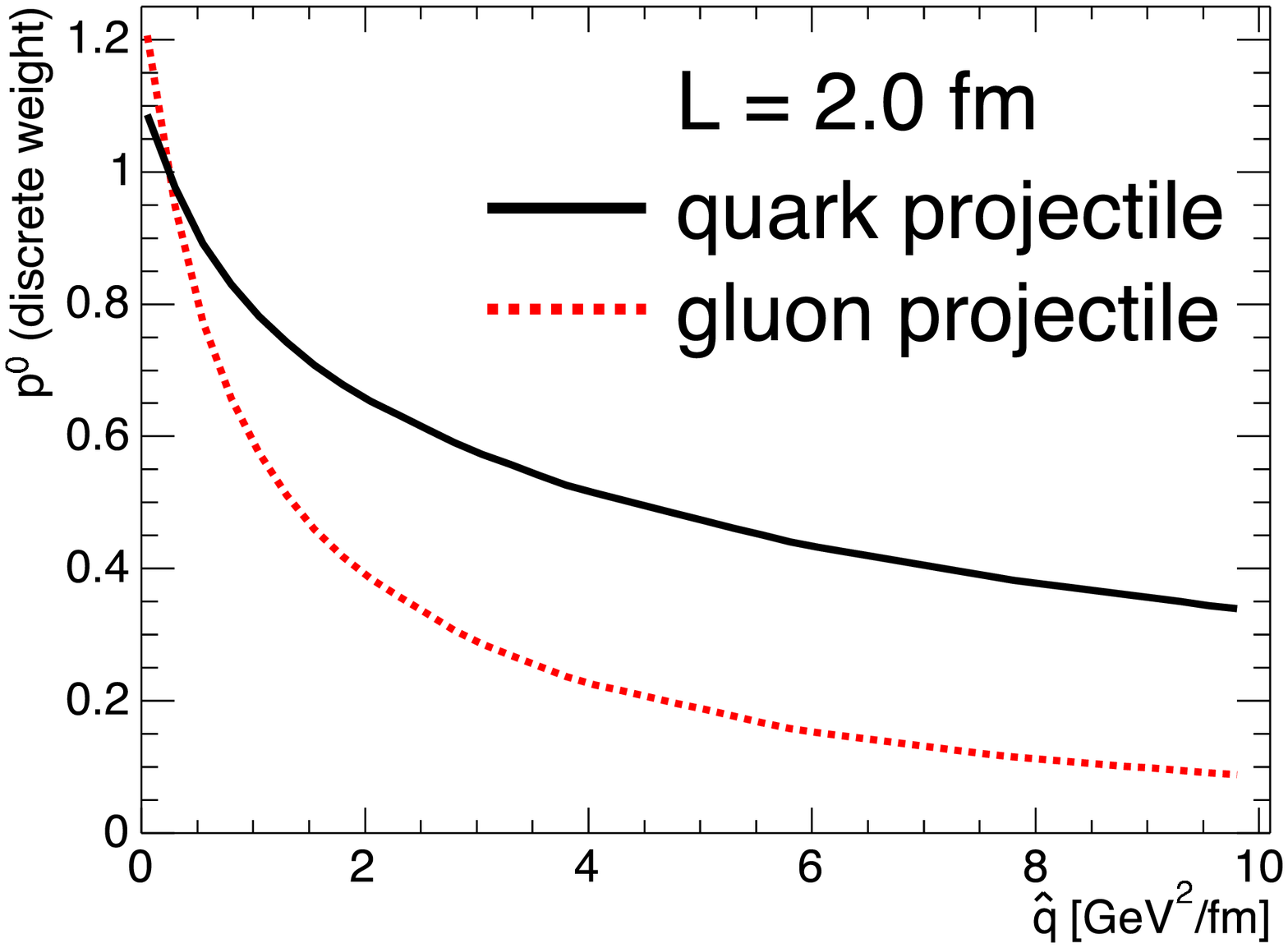}
\includegraphics[width=4.9cm, height=3.9cm]{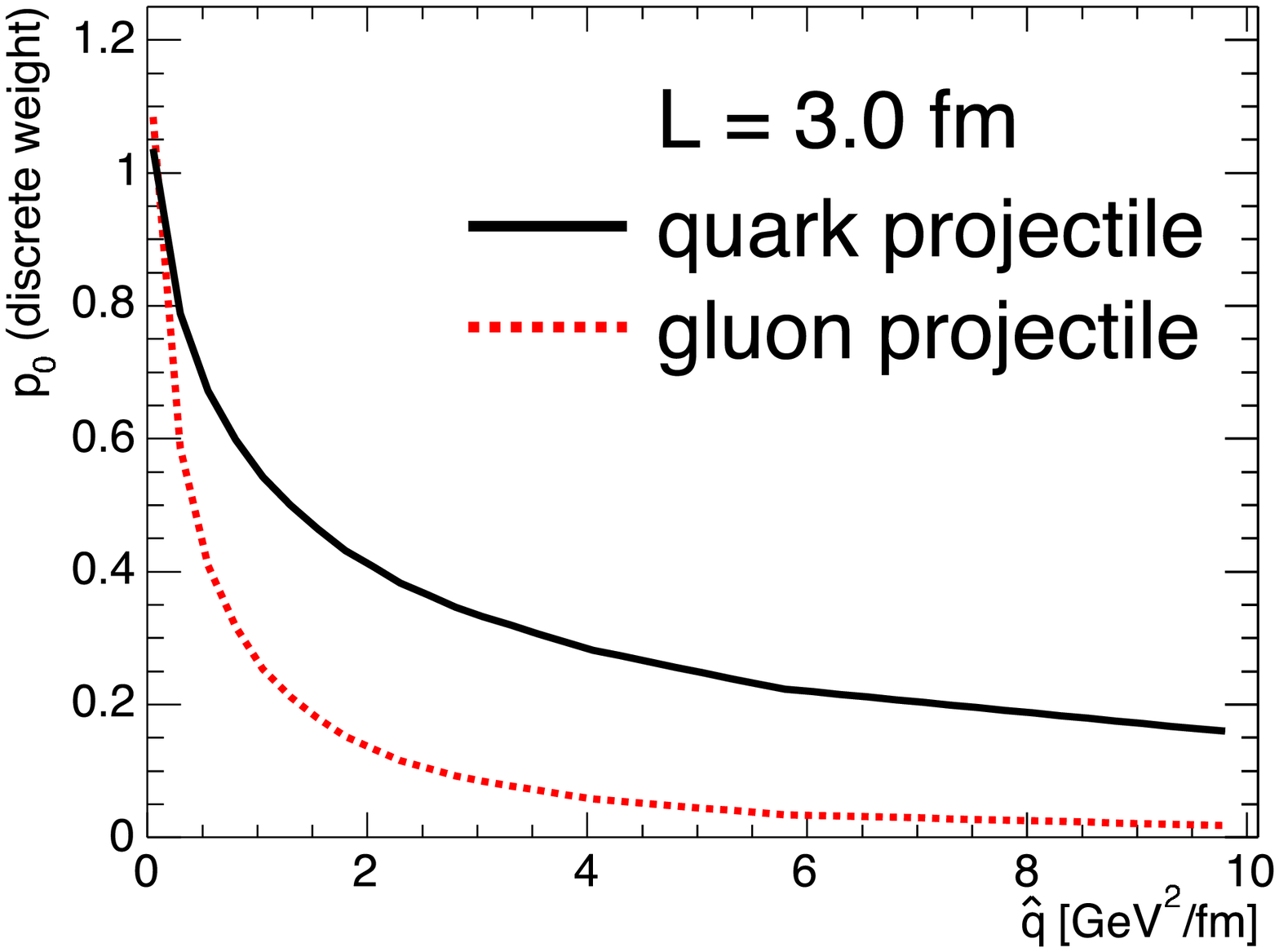}
\includegraphics[width=4.9cm, height=3.9cm]{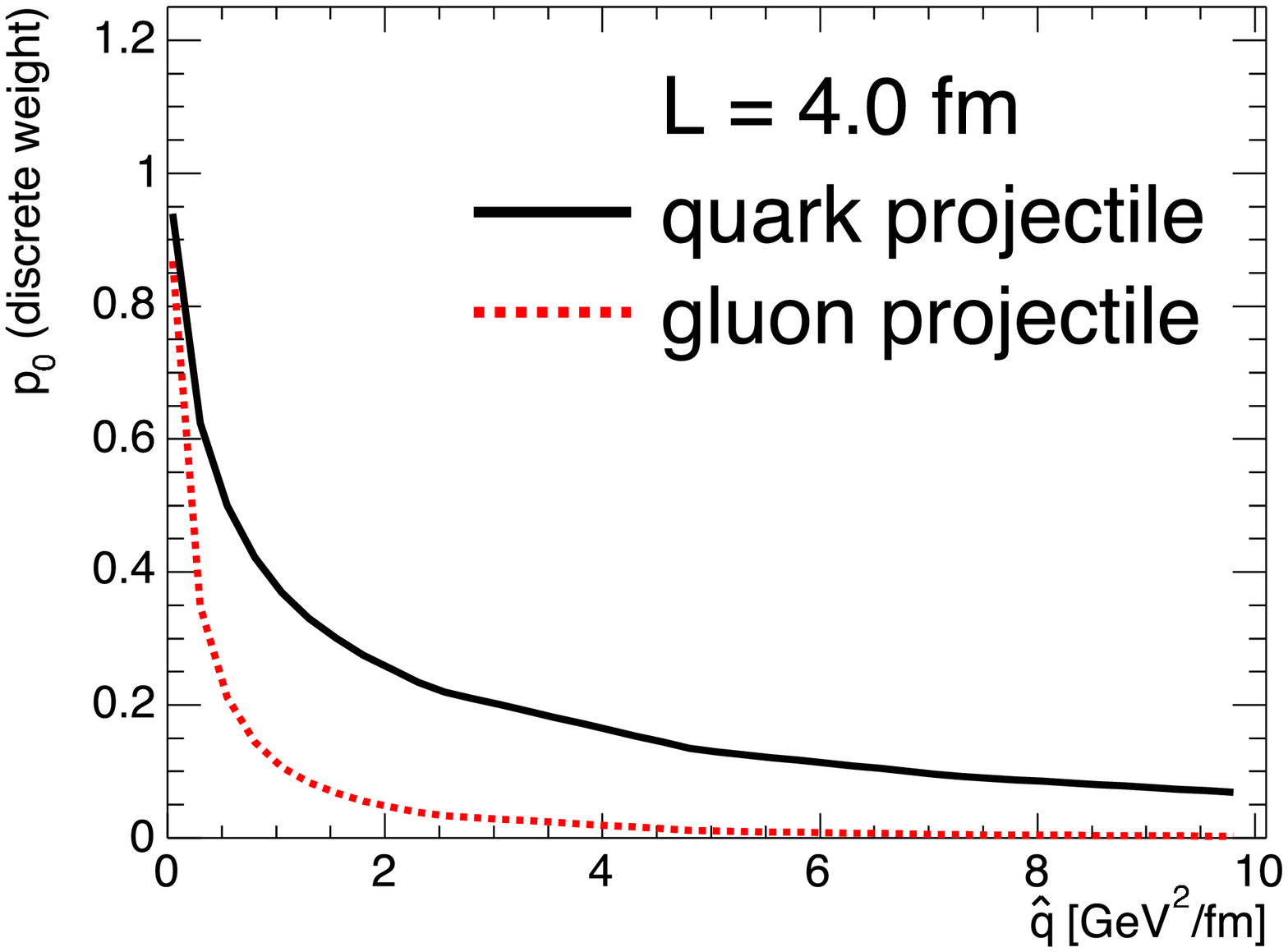}
\end{center}
\vspace{-0.5cm}
\caption[xxx]{The discrete weight of the quenching probability, 
$p_0$, as a function of the kinematic constraint $R$ 
(top) and as a function of the transport coefficient 
$\hat{q}$ for $L=2~\fm$ (left), $L=3~\fm$ (middle) and  
$L=4~\fm$ (right).}
\label{chap3:fig:p0weight}
\end{figure}

In \fig{chap3:fig:p0weight}, top panel, we reproduce the discrete weight $p_0$ 
as a function of the kinematic constraint $R$, whereas in \fig{chap3:fig:p0weight},
bottom panels, we show $p_0$ as a function of the transport coefficient $\hat{q}$ 
for a fixed in-medium path of $L=2~\fm$ (left), $L=3~\fm$ (middle) and  
$L=4~\fm$ (right).~\footnote{Here, 
and in the following, we use the Fortran subroutine 
of the quenching weights, provided by the authors 
of~\Ref{salgado2003} 
under \mbox{\hrefurl{http://cslagado.home.cern.ch}}.
The weights are evaluated at fixed value of \mbox{$\as=1/3$}.} 
The probability of no medium-induced radiation decreases with
increasing density of the medium. It is significantly 
larger for quarks than for gluons due to their lower \ac{QCD} 
coupling. The probability to radiate at least one gluon, $1-p_0$,
for $L=3~\fm$ at $\hat{q}=1~\gev^2/\fm$  is about $50$\% for quarks 
and about $80$\% for gluons; whereas at $\hat{q}=10~\gev^2/\fm$ 
it is about $80$\% quarks and almost $100$\% for gluons.

\begin{figure}[htb]
\begin{center}
\includegraphics[width=7cm, height=5.5cm]{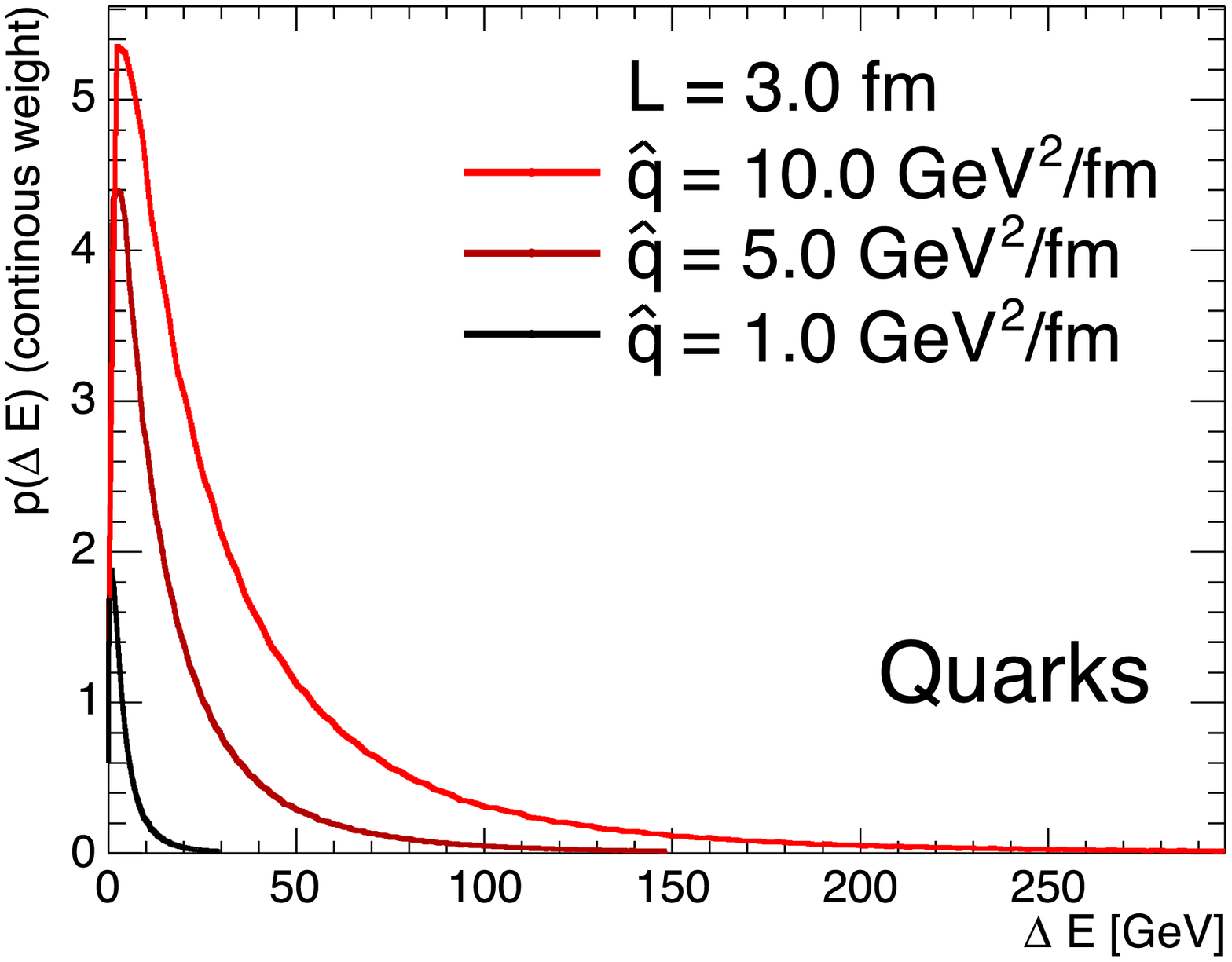}
\hspace{0.5cm}
\includegraphics[width=7cm, height=5.5cm]{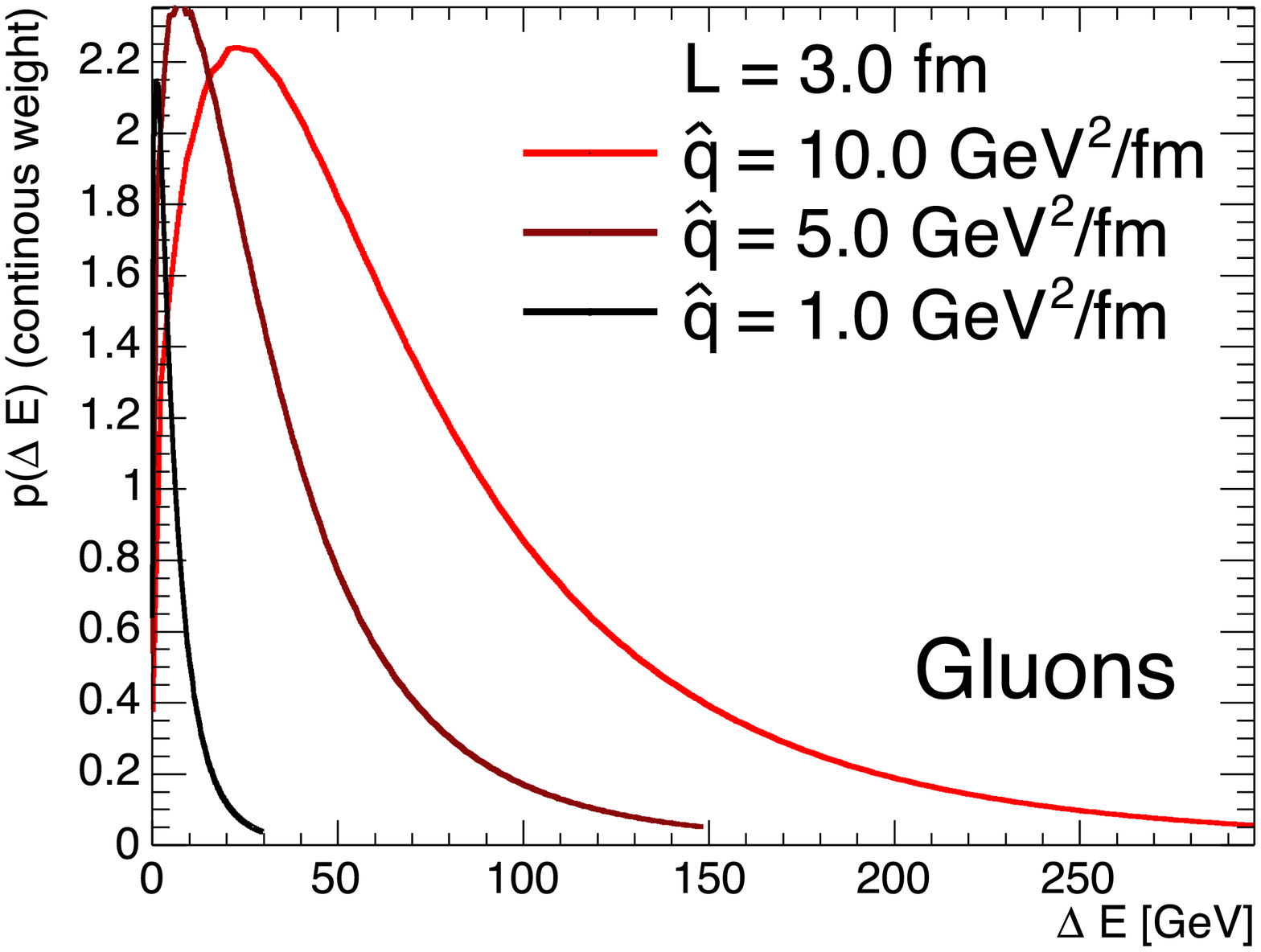}
\end{center}
\vspace{-0.4cm}
\caption[xxx]{Distribution of the continuous part of the quenching weight 
for quarks (left) and gluons (right) at fixed $L=3~\fm$ and for 
different values of $\hat{q}$ .}
\label{chap3:fig:qwcont}
\end{figure}

\begin{figure}[htb]
\begin{center}
\includegraphics[width=4.9cm, height=3.9cm]{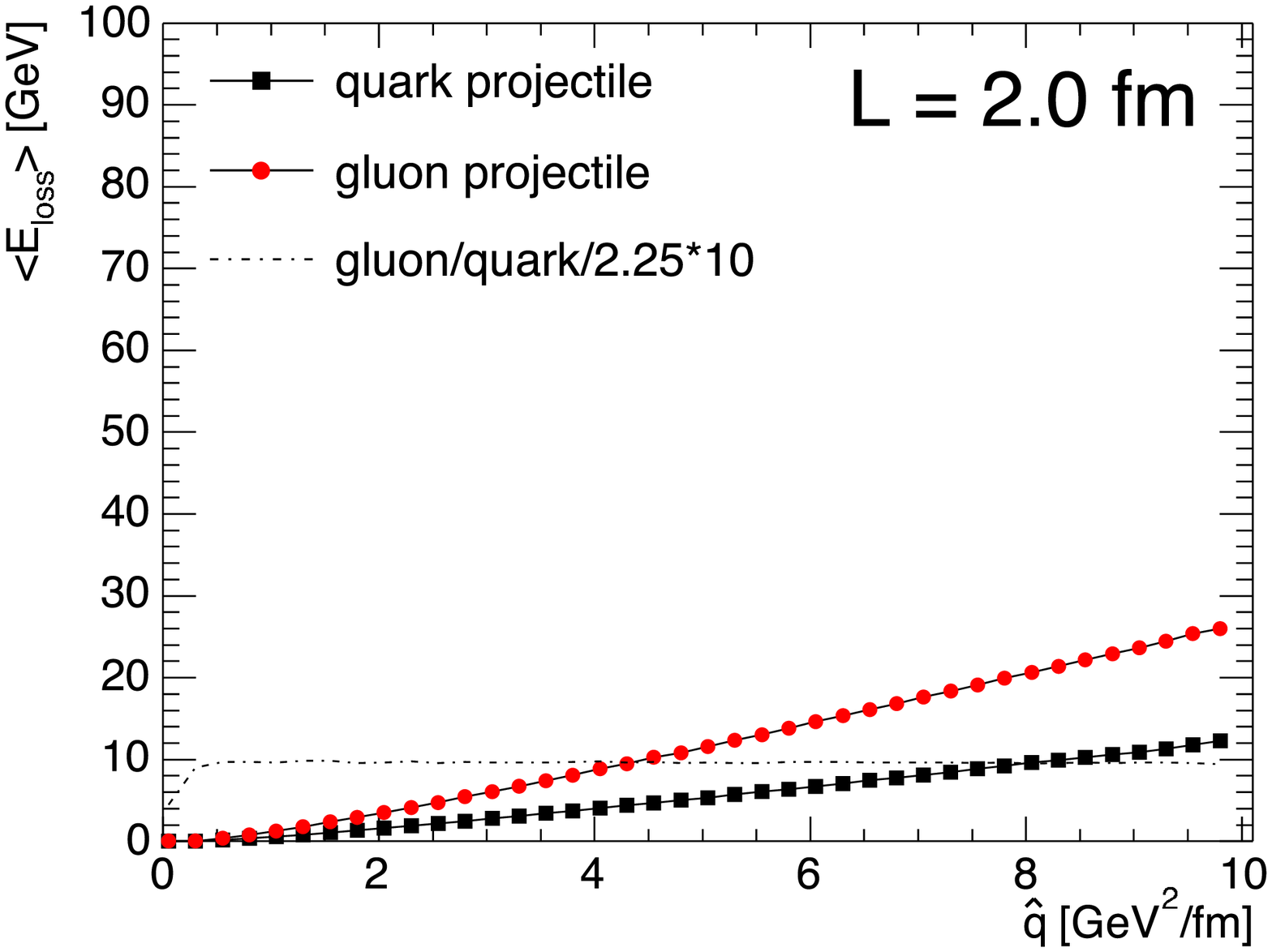}
\includegraphics[width=4.9cm, height=3.9cm]{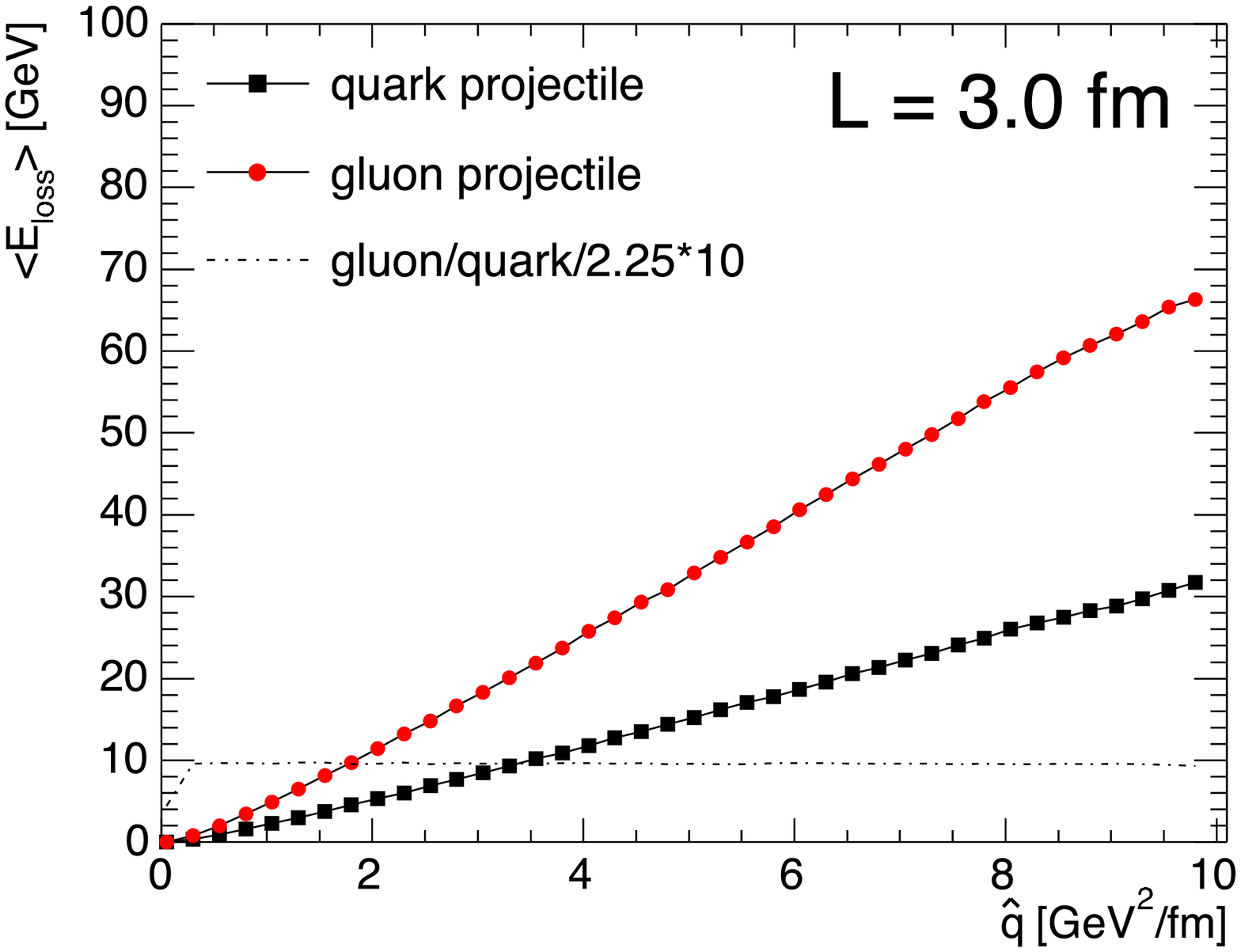}
\includegraphics[width=4.9cm, height=3.9cm]{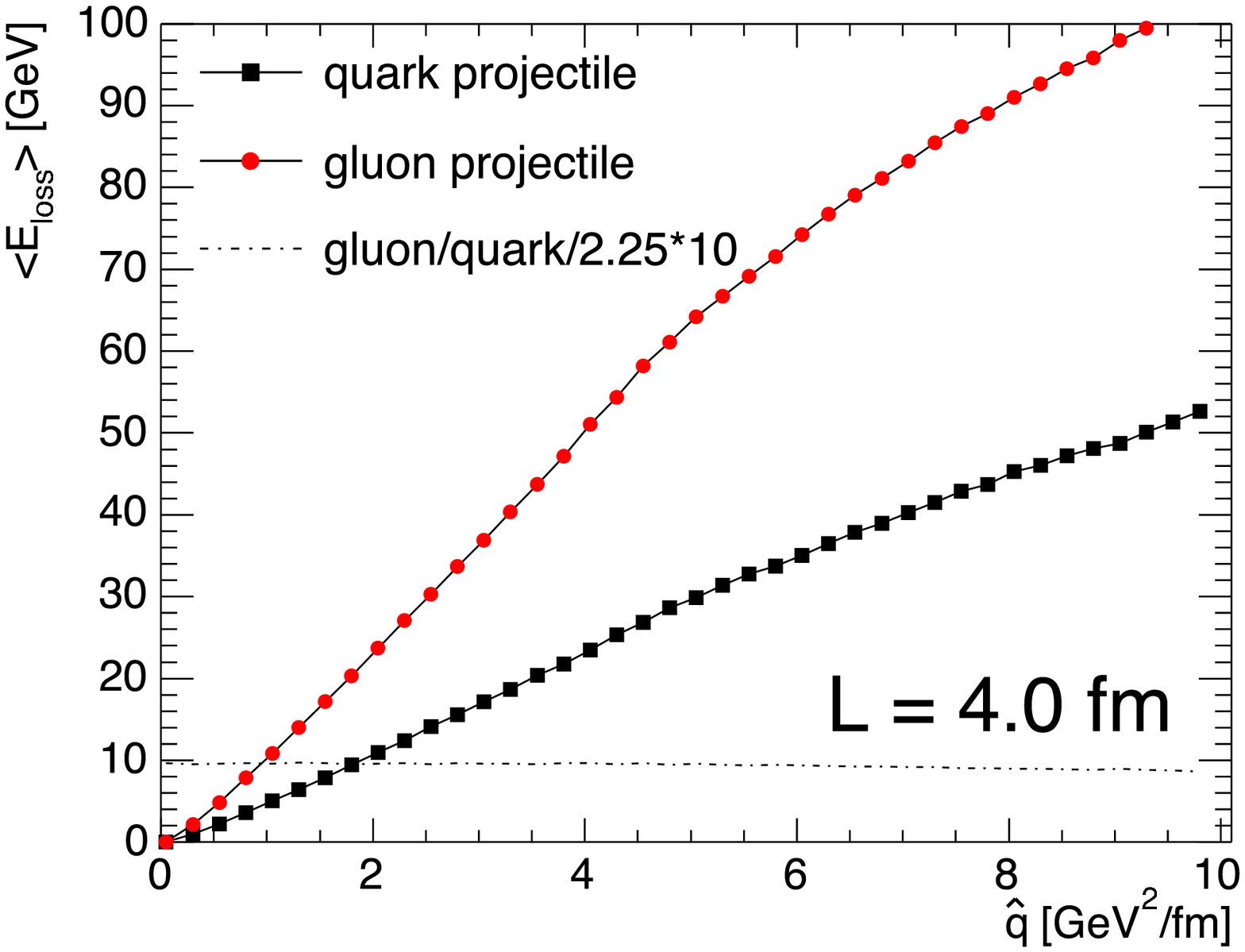}
\end{center}
\vspace{-0.4cm}
\caption[xxx]{Average energy loss of quarks and gluons as a 
function of the transport coefficient for $L=2~\fm$ (left), 
$L=3~\fm$ (middle) and $L=4~\fm$ (right).}
\label{chap3:fig:avgeloss}
\end{figure}

\Fig{chap3:fig:qwcont} reports the distribution of the continuous weight 
$p(\Delta E)$ for quark and gluon projectiles at fixed $L=3~\fm$ and for 
different values of $\hat{q}$. One observes that the gluon distribution 
is significantly broader compared to the quark distribution. This difference
resulting from the different \ac{QCD} coupling values of quarks and gluons
is most obvious reflected in the average energy loss, $\av{\Delta E}$, 
which we plot in \fig{chap3:fig:avgeloss} for quarks and gluons as a function 
of the transport coefficient for various values of the in-medium path length.
The calculation takes into account both the discrete and the continuous parts 
of the quenching weight. As expected from \eq{chap3:eq:avdE}, the gluon-to-quark 
ratio is exactly $9/4=2.25$ and $\av{\Delta E}$ grows approximately 
linearly with the transport coefficient and, thus, with the characteristic 
gluon energy $\omega_{\rm c}$. We find $\av{\Delta E}_{\rm quarks} \approx 0.1
\times \omega_{\rm c}$ for a quark and $\av{\Delta E}_{\rm gluons} \approx 
0.25\times \omega_{\rm c}$ for a gluon projectile.

The spatially integrated energy losses in the hot medium 
can be translated into losses per unit path length, 
$\dd E/\dd x_{\rm quarks}\approx 0.05\,\hat{q}\,L$ and
$\dd E/\dd x_{\rm gluons}\approx 0.125\,\hat{q}\,L$.
Even for conservatively chosen values of $\hat{q}=2~\gev^2/\fm$ and $L=3~\fm$
the resulting differential losses are one order of magnitude larger than 
those estimated by means of the Bjorken model for the collisional energy 
loss. 
Given the $L^2$-dependence of the effect, the differential 
energy loss should be quoted per unit path length squared, leading to
$\dd^2E/\dd x^2_{\rm quarks}\approx 0.05\,\hat{q}$. and
$\dd^2E/\dd x^2_{\rm gluons}\approx 0.125\,\hat{q}$.

As mentioned, the evaluation of the quenching weights in the \acs{BDMPS-Z-SW} 
model is done in the eikonal approximation (soft limit), where 
$\omega_{\rm c}\ll E \rightarrow \infty$ holds (\cf~\page{chap3:eq:regimes}). 
Since in practice calculations need to be performed at finite parton energies, 
we will in~\sect{chap3:step3} propose several possibilities to extrapolate to finite 
parton energies.
\fi

\section{Hard sector at RHIC}
\label{chap3:expresultsrhic}
\ifexpresults
At \ac{RHIC} the available \cms\ energy is for the first time in the history
of heavy-ion collisions high enough to allow hard scatters at the scale 
of $10$--$20~\gev$. Since this is too low to produce outstanding, 
high-energy jets to be identified on top of the heavy-ion background, 
the research focuses on inclusive (or leading-) particle spectra and (two-) 
particle correlations and their modification in \NNex\ with respect to 
\ppex\ collisions.

\enlargethispage{0.8cm}
The suppression of high-momentum leading particles is regarded 
as one of the major discoveries at \ac{RHIC}~\cite{gyulassy2004,gyulassy2004b}. 
In \AuAu\ collisions at various \cms\ energies (but mainly 
at $\snn=200~\gev$) the two experiments with high 
transverse-momentum capabilities, \acs{PHENIX} and \acs{STAR}, 
but also \acs{PHOBOS} and \acs{BRAHMS}, have measured: 
\begin{itemize}
\item the suppression of single-particle yields at high $\pt$ 
($\gsim 4~\gev$)~\cite{arsene2003,adams2003,adler2003,adler2003b,back2003b};
\item the disappearance, in central collisions, of jet-like 
correlations in the azimuthally-opposite side (away-side) of a 
high-$\pt$ leading particle~\cite{adler2002,adams2004} and, quite recently, 
the reappearance of the particles on the away-side manifested in 
low-momentum hadrons~\cite{wang2004,adams2005};
\item the absence of these (final-state) effects in \dAu\ collisions at the same 
\cms\ energy~\cite{adler2003c,adams2003b,arsene2003,back2003c}.
\end{itemize}

\subsection{Leading-hadron production in factorized pQCD}
\label{chap3:hadronprod}
Within the \ac{pQCD} collinear factorization framework~\cite{collins1989}, 
the inclusive cross section at \ac{LO} for the production of an high-$\pt$ hadron 
at central rapidity in the nuclear reaction of $\rm{A}+\rm{B} \rightarrow 
\rm{h}+\rm{X}$ can be expressed by~\cite{wang1996}
\begin{eqnarray} 
\label{chap3:eq:hcrossec}
\left.\frac{\dd^3\sigma_{{\rm AB}\rightarrow{\rm h X}}}{\dd^2\pt\,\dd y} 
\, \right|_{y=0} &=& K_{\rm NLO} \, \sum_{abc}\, \int \dd\vec{r} \,
\dd x_a \, \dd x_b \, \dd z_c  
\nonumber \\ & & 
\times F_{a/\rm A}(x_a,Q^2,\vec{r}) \, F_{b/\rm B}(x_b,Q^2,\,\vec{b}-\vec{r}) 
\nonumber \\ & &
\times \left.\frac{\dd^3\hat{s}_{ab\rightarrow c}}{\dd^3 p_{{\rm T},c}\,
\dd y_c}\,(x_a,x_b,Q^2)\right|_{y_c=0} \; \frac{D^{\rm mod}_{{\rm h}/c}(z_c,Q_{\rm f}^2)}{z_c^2} \;.
\end{eqnarray}
It gives the differential cross section as a convolution of 
generalized \acp{PDF} $F_{a/\rm A}$ for the interacting partons 
with generalized \acp{FF} $D^{\rm mod}_{{\rm h}/c}$ for the leading scattered 
parton into the final hadron and the parton--parton differential cross 
sections for the contributing, elementary sub-processes.~\footnote{Compare 
with~\sect{chap3:inclusivexsec}, mainly \eq{chap3:eq:ppjetcrossection}.} 
In this context $z_c=\pt/p_{{\rm T},c}$ is the momentum 
fraction of the hard parton, which is carried by the produced hadron. 
$K_{\rm NLO}$ is a phenomenological factor that is introduced to account for \ac{NLO} 
corrections. Like the hard cross section it is $\sqrt{s}$ and scale dependent. Usually
it takes values $\simeq 1-4$~\cite{vitev2002,eskola2002}.~\footnote{It can be omitted, 
if \ac{NLO} diagrams are included in the calculation of $\hat{\sigma}$~\cite{jeon2002}.} 
The various fragmentation, factorization and renormalization scales are fixed to the same
value, $Q=\alpha\,Q_{\rm f}=\kappa\,\pt$, where $1\le\alpha\le3$ and $0.5\le\kappa\le2$.

\Eq{chap3:eq:hcrossec} is applicable for hadron--hadron, 
hadron--nucleus and nucleus--nucleus interactions, and includes 
initial and final state effects.
The generalized \acp{PDF}
\begin{equation}
\label{chap3:eq:genpart}
F_{a/\rm A}(x_a,Q^2,\vec{b}) = T_{\rm A}(\vec{b})\, \int\dd^2k_{{\rm T}_a}\, 
g_{\rm A}(\vec{k}_{{\rm T}_a},Q^2,\vec{b})\, f_{a/\rm A}(x_a,Q^2)
\end{equation}
include the nuclear thickness function 
$T_{\rm A}$~\cite{denterria2003}~\footnote{In case A is a nucleon, the
thickness function reduces to $T_{\rm A}=\delta(\vec{b})$ in
units of $\fm^{-2}$.} and $\int\dd^2k_{\rm T}\, g_{\rm A}$ describing 
intrinsic-$k_{\rm T}$ (or $k_{\rm T}$-smearing)~\cite{owens1986} 
and $k_{\rm T}$-broadening for nuclei~\cite{wang1998}, as well as
the nuclear modification of the \ac{PDF} $f_{a/\rm A}$
(see \psect{chap2:npdf}).~\footnote{For the corresponding 
modification of the parton kinematics in~\eq{chap3:eq:hcrossec} 
in addition to the integration over $k_{\rm T}$, see~\Ref{owens1986}. 
For $g_{\rm A}(\vec{k}_{{\rm T}_a},Q^2,\vec{b})=\delta(\vec{k}_{{\rm T}_a})$
no intrinsic momentum is considered.}
The introduction of $k_{\rm T}$ is motivated by the \ac{pQCD} initial state 
radiation to correct the computation up to transverse momenta of $\pt \leq 4~\gev$. 
It is typically approximated by a Gaussian
\begin{equation}
\label{chap3:eq:gassm}
g_{\rm A}(k_{\rm T}) = \frac{e^{-k_{\rm T}^2/\langle {k}_{\rm T}^2 \rangle}}
{\pi\langle {k}_{\rm T}^2 \rangle } \; ,
\end{equation}
where the width  $\langle {k}_{\rm T}^2 \rangle$ 
enters as a phenomenological parameter and is typically set to
a value of the order of $1~\gev^2$.~\footnote{The width 
$\langle {k}_{\rm T}^2 \rangle$ is scale dependent and for nuclear 
broadening it is assumed to be proportional to the number of scatterings 
$\nu(\vec{b})$ the projectile suffers inside the nucleus~\cite{wang1998}.} 

The generalized \acp{FF}
\begin{equation*}
\label{chap3:eq:modff}
D^{\rm mod}_{{\rm h}/c}(z_c,Q_f^2)=\int \dd\epsilon\, P(\epsilon) \, 
\frac{1}{1-\epsilon} \, D_{{\rm h}/c}(\frac{z_c}{1-\epsilon},Q_f^2)
\end{equation*}
include the non-perturbative mechanism $D_{{\rm h}/c}$ of how the parton turns 
into the leading hadron (see \sect{chap3:frag}). The quenching probability, 
$P(\epsilon)$, denotes the possible in-medium modification in the final 
state~\cite{wang2002}. Prior to hadronization the parent parton of the hadron 
loses an energy fraction $\epsilon=\Delta E_c/E_c$ with probability $P(\epsilon)$. 
Therefore, the leading hadron is a fragment of a parton with lower energy 
\mbox{$(1-\epsilon)\,E_z$} and accordingly must carry a larger fraction of the 
parton energy \mbox{$z_c/(1-\epsilon)$}. In general, the quenching probability 
$P(\epsilon)$ ---given by a model of parton energy loss in dense deconfined matter 
(see \sect{chap3:partoneloss})--- is a function of the medium properties, 
the parton energy and type and the collision geometry. If no final state 
quenching is considered (\ie~for a nucleon) it reduces to 
$P(\epsilon)=\delta(\epsilon)$.

\Fig{chap3:fig:pizeroyieldsphenix} shows the application of \eq{chap3:eq:hcrossec}
for the production of neutral pions in \pp\ and \AuAu\ collisions measured by 
\acs{PHENIX}~\cite{adler2003c,adler2003b}. Whereas the binary scaled \pp\ and peripheral 
\AuAu\ data points are consistent and, both, described by \acs{NLO} \acs{pQCD}~\cite{jeon2002},
the central yields cannot be calculated by the standard \ac{pQCD} formalism 
(\ie~\eq{chap3:eq:hcrossec} without modified fragmentation functions). It is precisely
the breakdown of the expected incoherent parton-scattering assumption for high-$\pt$ 
production in non-peripheral \AuAu~collisions at \acs{RHIC} energies, which recently
has created much excitement. It is now attributed to strong final state effects
as we outline in the next section.

\begin{figure}[htb]
\begin{center}
\subfigure[Peripheral $\pi_0$ yields ($80$--$92$\%)]{
\label{chap3:fig:pizeroyieldsphenixa}
\includegraphics[width=7cm]{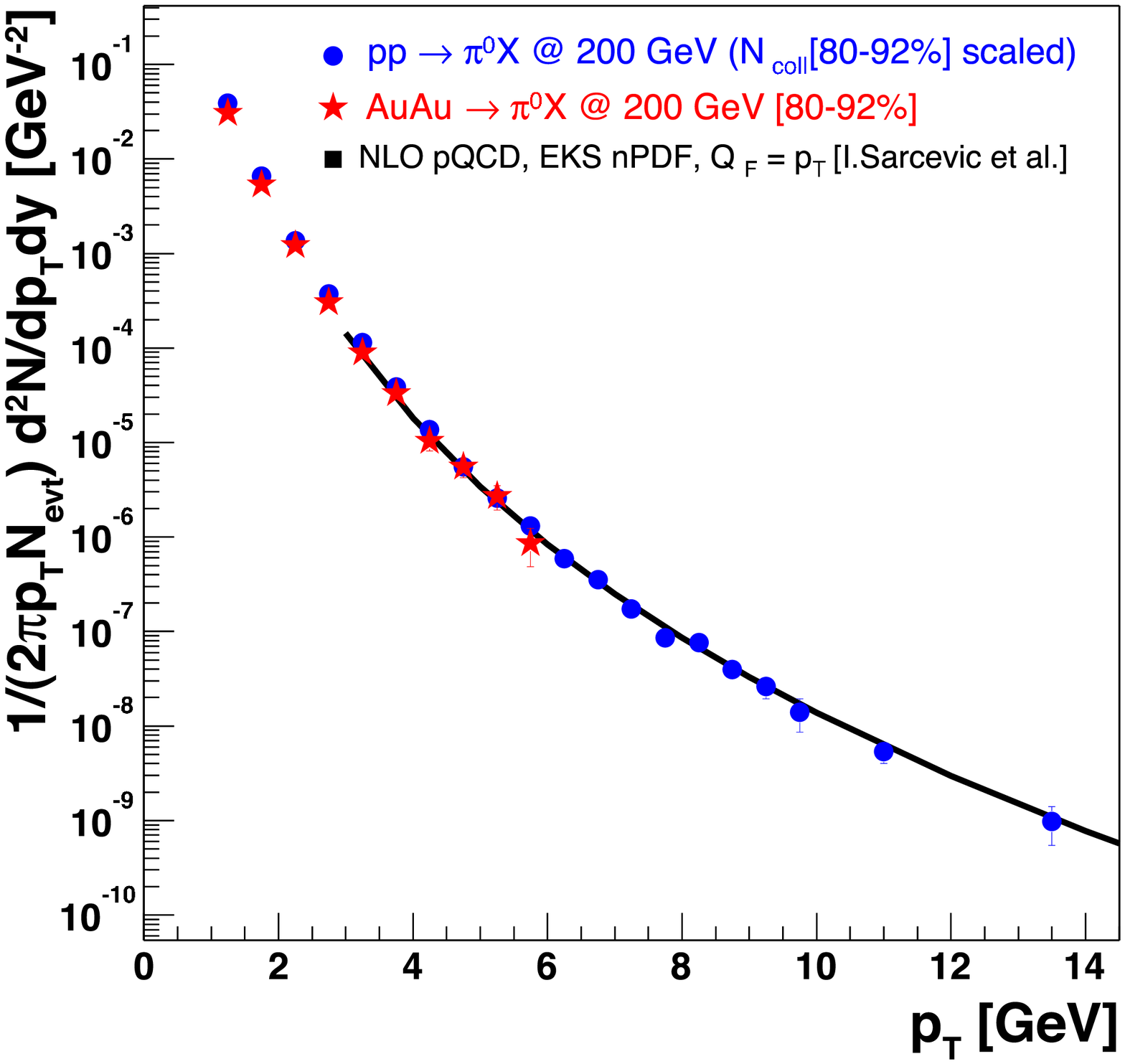}}
\hspace{0.5cm}
\subfigure[Central $\pi_0$ yields ($0$--$10$\%)]{
\label{chap3:fig:pizeroyieldsphenixb}
\includegraphics[width=7cm]{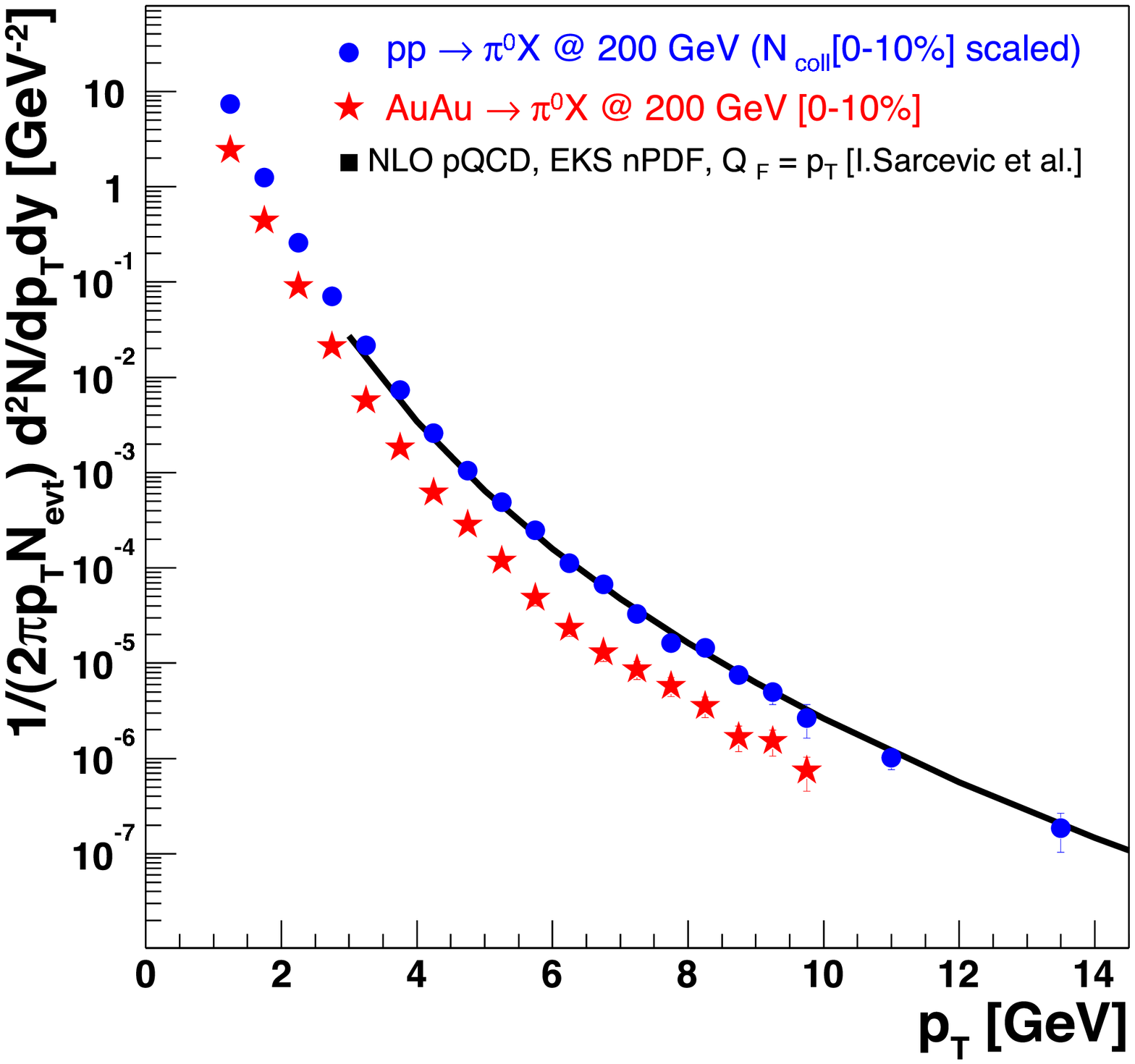}}
\end{center}
\vspace{-0.5cm}
\caption[xxx]{Invariant $\pi^0$ yields measured by \acs{PHENIX} in peripheral 
\subref{chap3:fig:pizeroyieldsphenixa} and in central 
\subref{chap3:fig:pizeroyieldsphenixb} \AuAu\ collisions~\cite{adler2003b} 
compared to the binary scaled \pp\ cross section~\cite{adler2003d} and to a 
standard \acs{NLO} \acs{pQCD} calculation~\cite{jeon2002b}. The overall normalization 
uncertainties in the scaled \pp\ yield is of the order of the symbol sizes.
The figures are adapted from~\cite{denterria2004}.}
\label{chap3:fig:pizeroyieldsphenix}
\end{figure}

\subsection{Leading-particle suppression as strong final state effect}
\label{chap3:leadpartsuppression}
The effect of the medium on the production of a hard probe
is typically quantified via the ratio of the shown spectra, the
nuclear modification factor,
\begin{equation}      
\label{chap3:eq:rab}
\RAB(\pt,\eta;\,b)= \frac{1}{\av{\Ncoll(b)}} \times \frac
{\dd^2N^{\rm hard}_{\rm AB}/\dd\pt\,\dd\eta}{\dd^2N^{\rm hard}_{\rm pp}/\dd\pt\,\dd\eta}\;,
\end{equation}
which measures the deviation of the \AAex\ from the superposition 
of independent \NNex\ collisions. In absence of strong nuclear 
initial state effects it should be unity, if binary collision scaling 
holds according to
\begin{equation}      
\label{chap3:eq:ncollscaling}
\dd^2N^{\rm hard}_{\rm AB}/\dd\pt\,\dd\eta = \av{\Ncoll(b)} \times 
\dd^2N^{\rm hard}_{\rm pp}/\dd\pt\,\dd\eta\;.
\end{equation}

However, in accordance with \fig{chap3:fig:pizeroyieldsphenixb}, at \ac{RHIC} in \AuAu\ 
collisions at $\snn=200~\gev$ strong suppression 
effects are observed~\cite{arsene2003,adams2003,adler2003,adler2003b,back2003b}, 
visible in \fig{chap3:fig:raavsptcent}, where we reproduce $\RAA$ as a function of 
\pt\ for charged hadron and neutral pions measured at mid-pseudo-rapidity in 
central events by \acs{STAR} and \acs{PHENIX}. 
The magnitude of suppression is the same for charged hadrons and neutral pions beyond 
$\pt\gsim 5~\gev$.~\footnote{The explanation of the physics behind the difference in the 
intermediate-\pt\ region is subject to an ongoing debate and far from 
understood~\cite{lamont2004,kotchetkov2004}.}
In \fig{chap3:fig:raavsnpart} we show the centrality dependence of the average
$\RAA$ as a function of \Npart\ for the same data sets. The average suppression for 
$\pt>4.5~\gev$ increases from peripheral to central events, up to about a factor of
$5$ in head-on collisions. Strong suppression exists also at 
$\snn=130~\gev$~\cite{adcox2001,adler2002b}. Recent measurements at $\snn=62.4~\gev$
show very moderate suppression for charged hadrons at intermediate 
$\pt\le4~\gev$~\cite{back2004b}.~\footnote{Though, 
preliminary \acs{PHENIX} results for neutral pions up to $7~\gev$ indicate 
the same tendency as for the $200~\gev$ data, with a suppression factor of 2--3 
at the highest \pt\ values.} Quite recently, suppression of high-$\pt$
particles already at the highest \ac{SPS} energy has been suggested~\cite{denterria2004b}.

\begin{figure}[htb]
\begin{center}
\includegraphics[width=11.9cm]{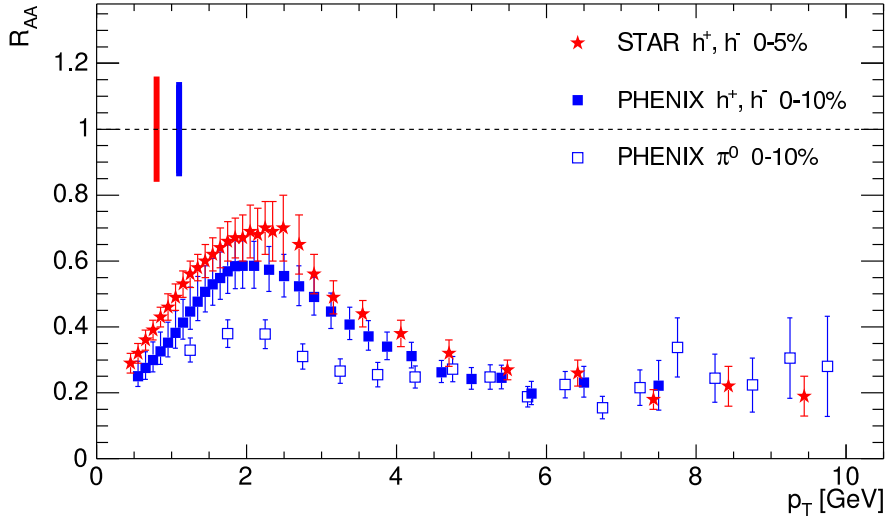}
\end{center}
\vspace{-0.3cm}
\caption[xxx]{Nuclear modification factor $\RAA(\pt)$ at $\eta\approx 0$ in central \AuAu\ 
collisions at $\snn=200~\gev$ for charged hadrons~\cite{adams2003,adler2003} 
and neutral pions~\cite{adler2003b}. The data are reported with statistical and 
$\pt$-dependent systematic errors (bars on the data points) and $\pt$-independent
systematic errors (bars at $\RAA=1$).}
\label{chap3:fig:raavsptcent}
\end{figure}

\begin{figure}[htb!]
\ifprint
\else
\vspace{0.4cm}
\fi
\begin{center}
\includegraphics[width=11.9cm]{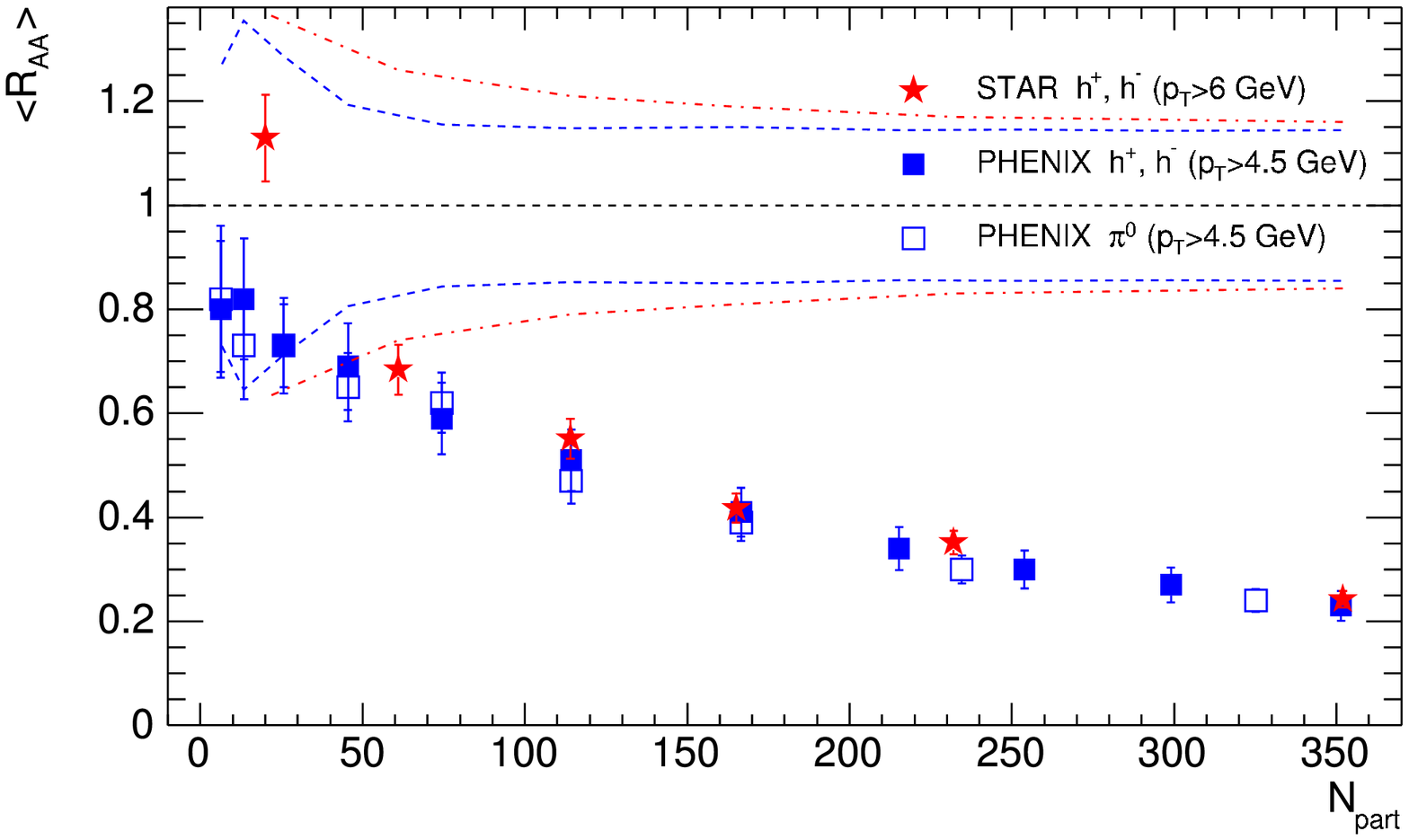}
\end{center}
\vspace{-0.3cm}
\caption[xxx]{Average nuclear modification factor $\av{\RAA}$ at 
$\eta\approx0$ in the range $4.5\le \pt \le 10~\gev$
as a function of collision centrality (expressed by the number of participants, 
$\Npart$) in central \AuAu\ collisions at $\snn=200~\gev$ for charged 
hadrons~\cite{adams2003,adler2003} and neutral pions~\cite{adler2003b}. 
The error bars are the sum of statistical and $\pt$-dependent systematic 
errors and the bands centered at $\RAA=1$ are the $\pt$-independent normalization 
errors for \acs{STAR} (dot-dashed) and \acs{PHENIX} (dashed).}
\label{chap3:fig:raavsnpart}
\end{figure}

The discrepancy of the expected scaling from \eq{chap3:eq:ncollscaling} by the
large factor of up to about $5$ at $\snn=200~\gev$ could be addressed by
the following pictures:
\begin{enumerate}
\item The breakdown of the leading-twist \ac{QCD} collinear factorization itself. 
In that case the incoherence between long- and short-distance effects, on which the
factorized product \eq{chap3:eq:hcrossec} relies upon, would not hold for \AaAa\ 
collisions. In addition, due to strong initial effects, the \acp{nPDF} might be 
modified, such that $f_{a/{\rm A}}\ll A\, f_{a/{\rm p}}$ in the relevant ($x$,~$Q^2$) 
range, reducing the number of effective partonic scattering centers.
\item Strong final-state effects in the deconfined medium, such as medium-induced 
parton energy loss (see~ sect{chap3:partoneloss}), modify the parton fragmentation 
functions 
compared to collisions in cold matter of free space.
\end{enumerate}

The first explanation is invoked by means of the \ac{CGC} model~\cite{iancu2003}; 
the latter by models employing parton energy loss; in most cases due to medium-induced 
gluon radiation in the \ac{QGP}, as we will outline in~\sect{chap3:pqm}.

\begin{figure}[htb]
\begin{center}
\subfigure[Control measurement ($\AuAu$~vs.~$\dAu$)]{
\label{chap3:fig:dauvsauau}
\includegraphics[width=7.3cm, height=6cm]{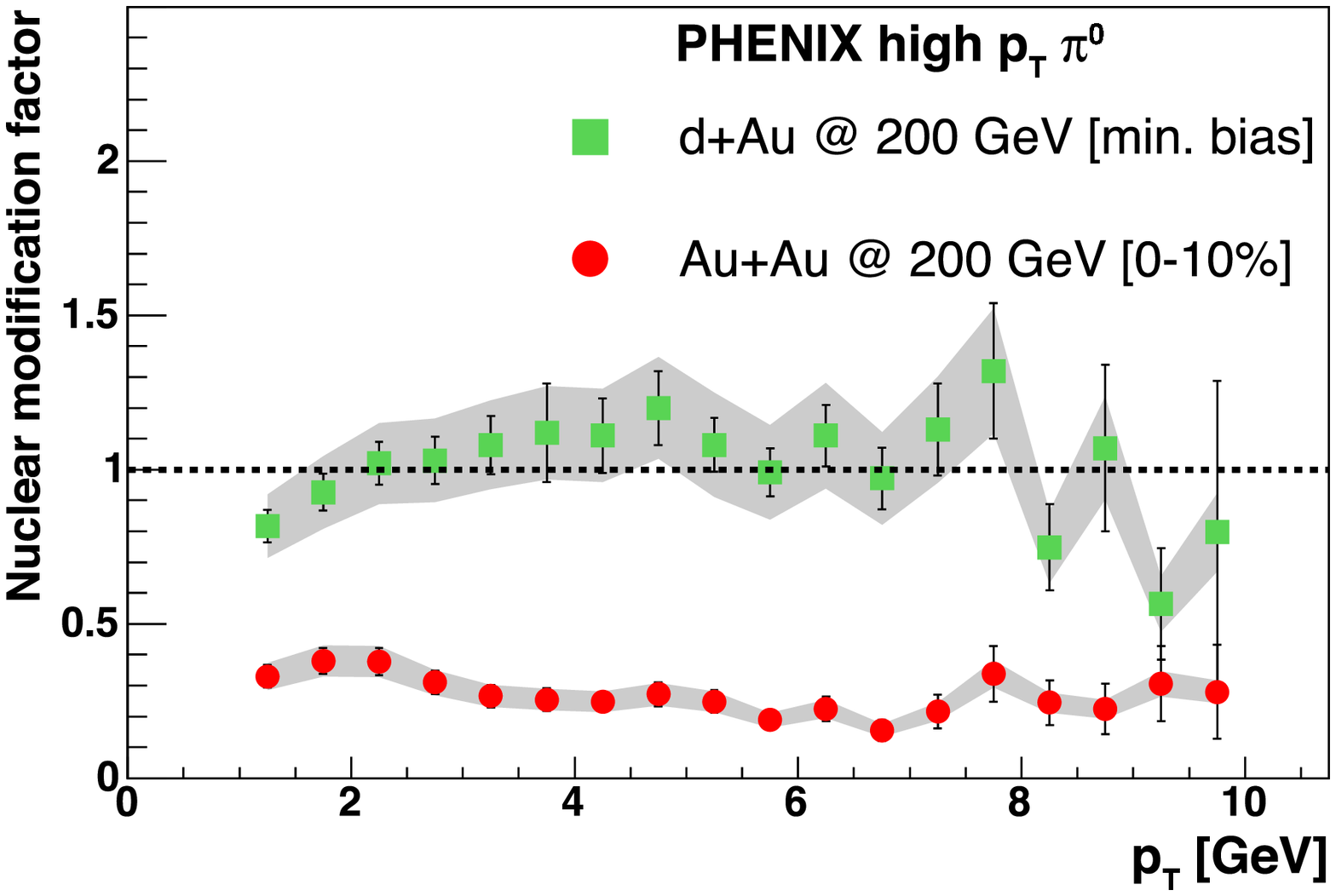}}
\subfigure[Direct photons excess]{
\label{chap3:fig:directphotons}
\includegraphics[width=7.3cm, height=6cm]{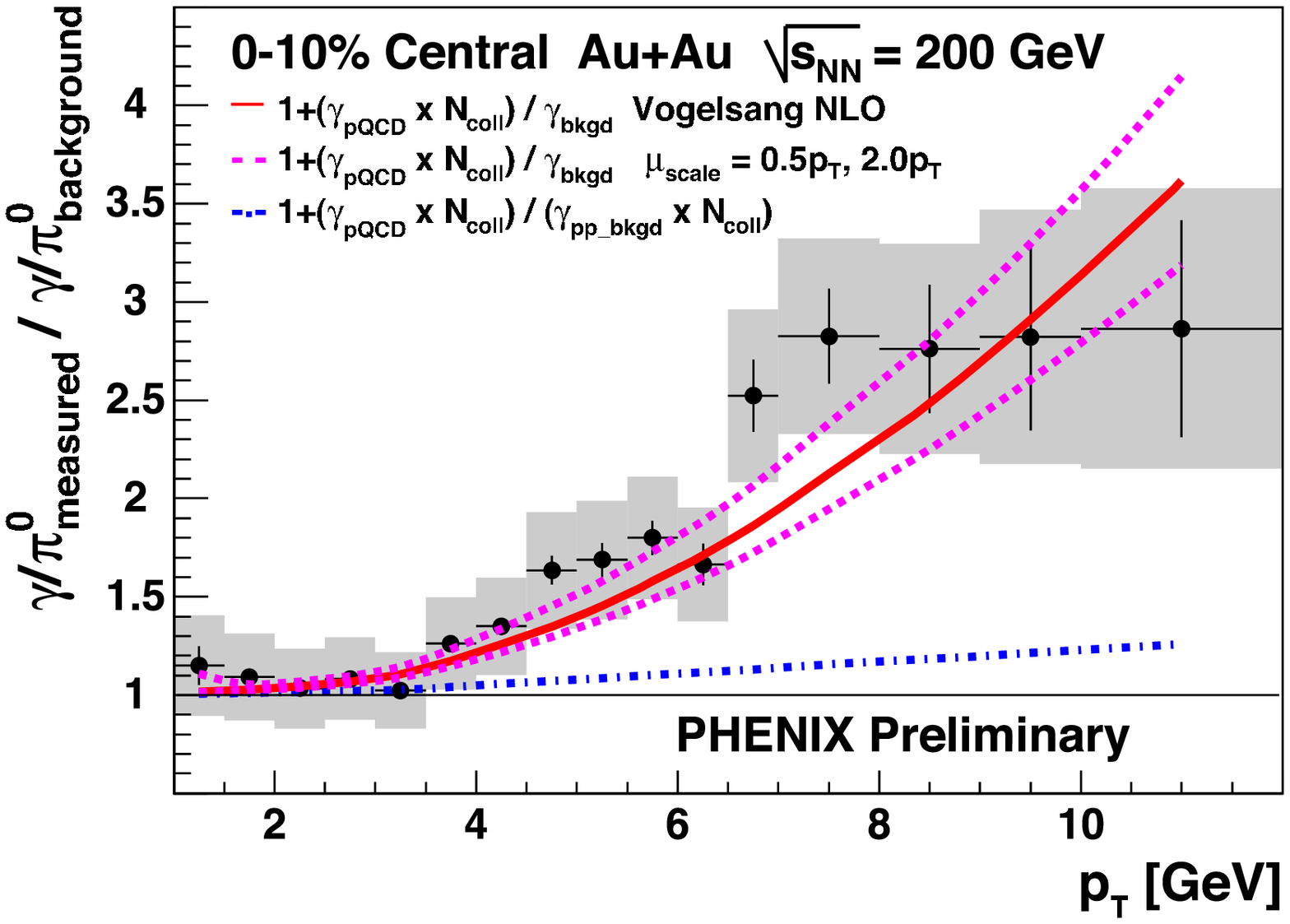}}
\end{center}
\vspace{-0.5cm}
\caption[xxx]{\subref{chap3:fig:dauvsauau} Nuclear modification factor, 
$R_{\rm dAu}(\pt)$ and $\RAA(\pt)$, for neutral pions measured by 
\acs{PHENIX} at mid-rapidity in minimum-bias \dAu~\cite{adler2003c} 
and central \AuAu~\cite{adler2003b} collisions, 
both at $\snn=200~\gev$. The figure is adapted from~\Ref{denterria2004}. 
\subref{chap3:fig:directphotons} Direct photon excess
for central \AuAu\ collisions at $\snn=200~\gev$ by \acs{PHENIX} 
compared to \ac{NLO} \ac{pQCD} yields scaled by $\Ncoll$. The dot-dash 
curve represents the expected excess, if there were no suppression 
of the background produced by meson decay. The figure is taken 
from~\Ref{frantz2004}.}
\label{chap3:fig:evidence}
\end{figure}

In order to disentangle between the two scenarios experimentally, it is vital
to compare to measurements in \dAu\ collisions at the same \cms\ energy,
where the nuclear modification factor $R_{\rm dAu}$ is determined by initial-state 
effects alone and no medium is expected to influence the final state. 
As shown in \fig{chap3:fig:dauvsauau} the high-$\pt$ production of
neutral pions at $\snn=200~\gev$ in \dAu\ is due to $\kt$-broadening 
(Cronin effect) even slightly enhanced ($R_{\rm dAu}\sim1.1$)~\cite{adler2003c}. 
This confirms that the suppression in \AuAu for central collisions cannot
be explained by initial-state effects. Further experimental evidence supporting 
the suppression driven by the formation of the dense partonic matter
is the measurement of direct photon excess from \acs{PHENIX}~\cite{frantz2004}. 
\Fig{chap3:fig:directphotons} shows the double-ratio of 
$\dd N^{\rm total}_{\gamma}/\dd\pt$ over $\dd N^{\rm decay}_{\gamma}/\dd\pt$ 
normalized by the $\pi_0$ spectra. It is consistent with $\Ncoll$ 
scaling as expected for an electro-magnetic (hard) probe since it is 
by its nature not sensitive to final-state medium effects.

\subsection{Disappearance and reappearance of the away-side correlations}
\label{chap3:jetcorrelations}
Another prominent result from \ac{RHIC} measured by \acs{STAR} at $\snn=200~\gev$
is the disappearance of jet-like azimuthal correlations in the opposite direction 
(away-side) of high-$\pt$ particles~\cite{adler2002,adams2004}. 

\begin{figure}[htb]
\begin{center}
\subfigure[Sketch of in/out-plane]{
\label{chap3:fig:inoutplane}
\ifarxiv
\includegraphics[width=4.5cm]{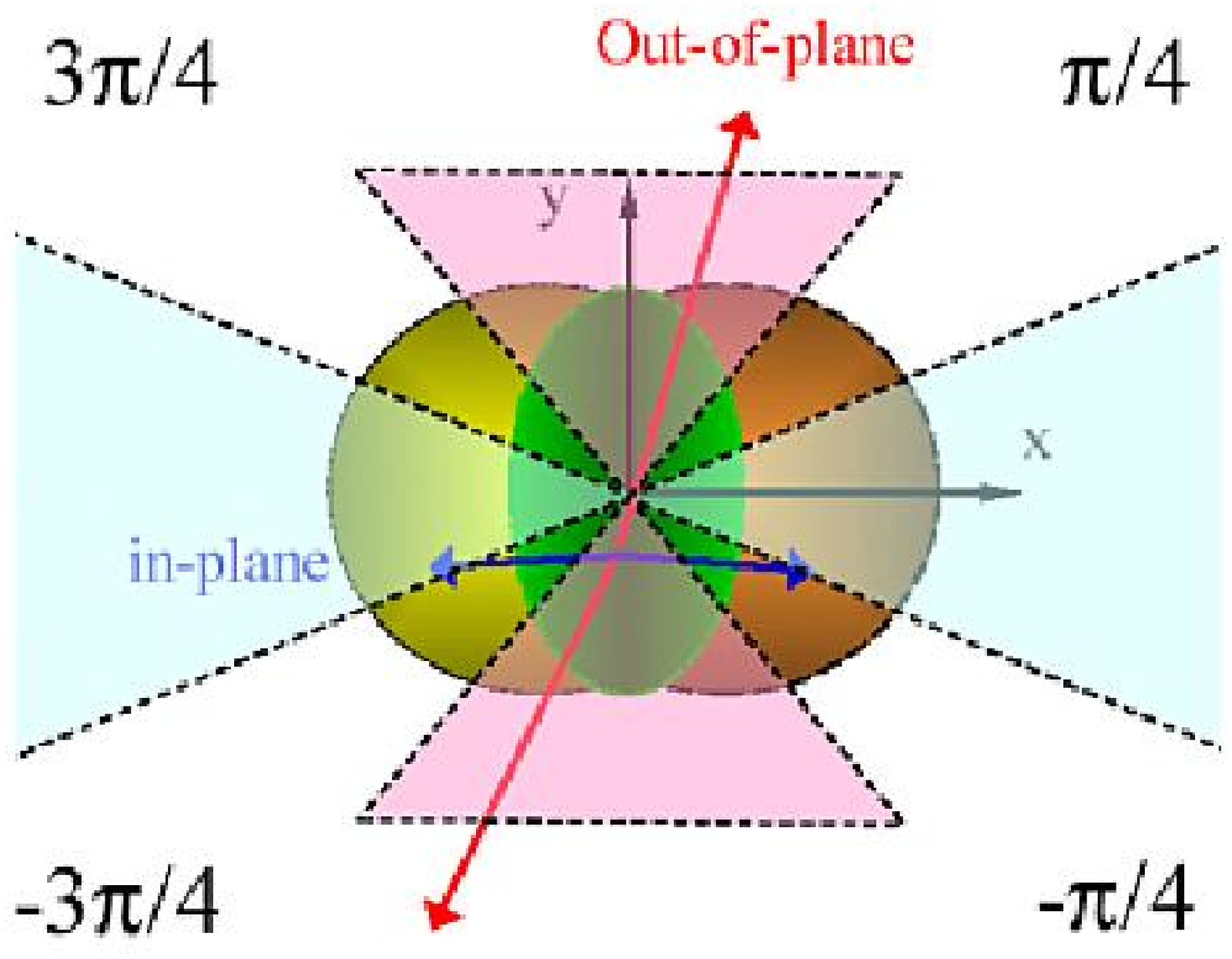}
\else
\includegraphics[width=4.5cm]{InOutPlane}
\fi
}
\hspace{0.5cm}
\subfigure[Associated particle correlation in azimuth]{
\label{chap3:fig:starcorrelation}
\includegraphics[width=8.5cm]{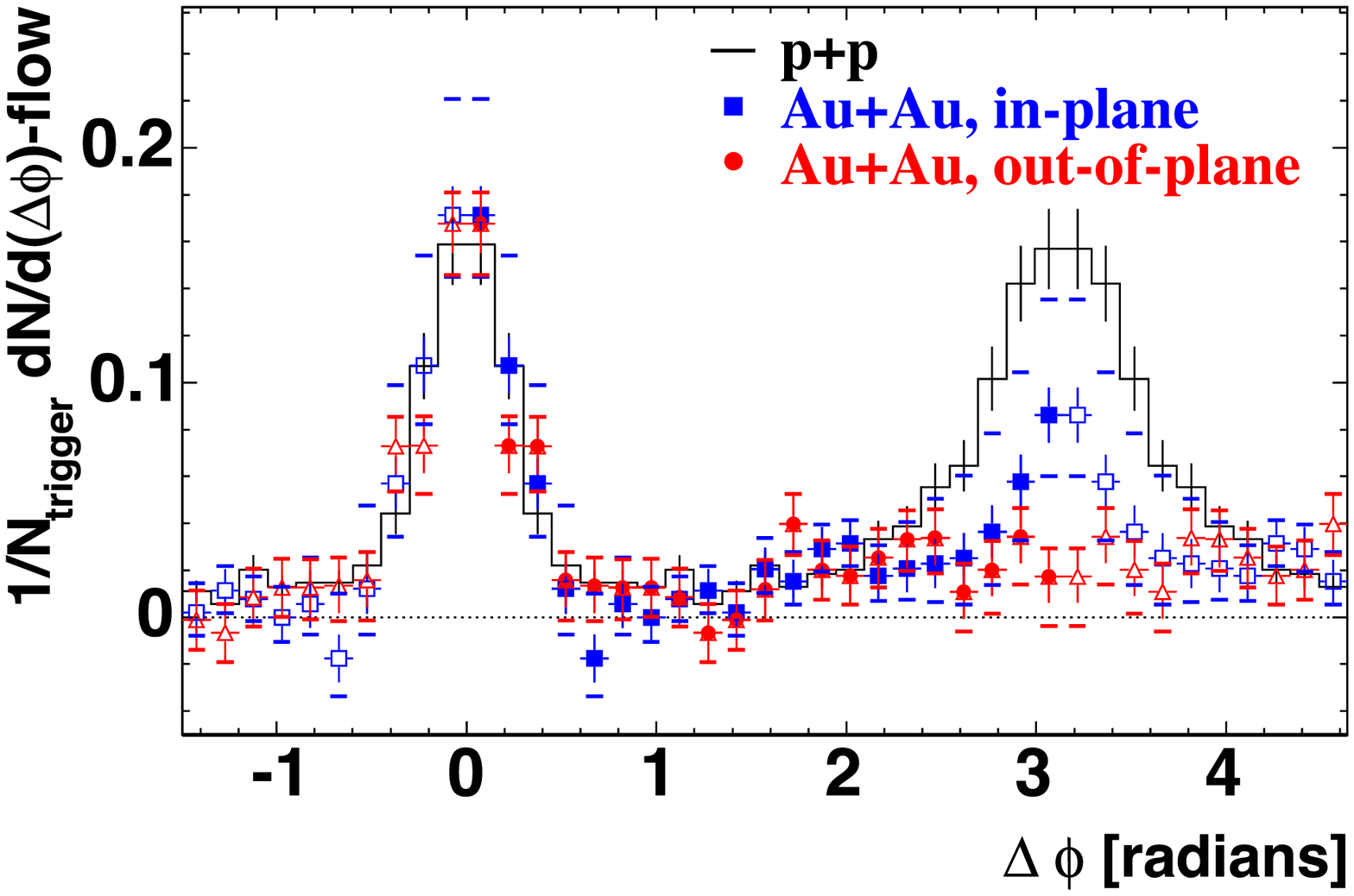}}
\end{center}
\vspace{-0.5cm}
\caption[xxx]{\subref{chap3:fig:inoutplane} Definition of in-plane and out-of-plane direction. 
Particle emission into the azimuthal cone of 45 degrees around the reaction plane given by the 
impact parameter (parallel to the $x$-axis in the sketch) and the $z$-axis is called in-plane.
Emission into the vertical cone of 45 degrees is out-of-plane. Every other direction is 
(in-) between-plane. \subref{chap3:fig:starcorrelation}~Azimuthal distributions 
of associated particles ($2 \le \pt \le \pt^{\rm trig}$) assigned to high-$\pt$ trigger 
particles ($4\le\pt^{\rm trig}\le6~\gev$), emitted in-plane and out-of-plane in \AuAu\ collisions 
at $\snn=200~\gev$ for $20$--$60$\% centrality, compared to the \pp\ reference at the same energy, 
measured by \acs{STAR}~\cite{adams2004}. The contribution of the elliptic flow ($v_2$) is 
subtracted. Further details are in the text.}
\label{chap3:fig:correlation}
\end{figure}

The effect is usually quantified using the correlation 
strength~\cite{wang2003}
\begin{equation}
\label{chap3:eq:daa}
D_{\rm AA} = \int^{p_{\rm t,1}}_{\pt^{\rm min}}\dd p_{\rm t,2}
\int_{\Delta\phi > \Delta\phi^{\rm min}}\dd\Delta\phi\,
\frac{\dd^3\sigma_{\rm AA}^{\rm{h}_1 \rm{h}_2}/\dd p_{\rm t,1} \, \dd p_{\rm t,2}
\, \dd\Delta\phi}{\dd\sigma_{\rm AA}^{h_1}/\dd p_{\rm t,1}}
\end{equation} 
for an associated particle $\rm{h}_2$ with transverse momentum 
$p_{\rm t,2}$ in the opposite azimuthal direction of a `trigger'
particle $\rm{h}_1$ with transverse momentum $p_{\rm t,1}$. 
The existing data allows trigger particles with $4<p_{\rm t,1}<6~\gev$ 
and associated particles with $p_{\rm t,2}>\pt^{\rm min} = 2~\gev$ and 
$p_{\rm t,2}<p_{\rm t,1}$, with $\Delta\phi\equiv |\phi_1-\phi_2| > 
\Delta\phi^{\rm min}=130^\circ$.~\footnote{New data where electromagnetic
calorimetry was used to trigger on high-$\pt$ neutral pions is been
analyzed and will extend the $\pt$-spectrum on the near side up to
$15~\gev$~\cite{thesistom}.}
The correlation strength is then corrected for combinatorial
background and azimuthal anisotropy of particle production in 
non-central collisions. The picture which emerged is that as central 
collisions are approached for increasing participants, the away-side 
correlations are gradually disappearing, until for most central collisions 
no correlation is observed (see \fig{chap3:fig:RAAIAAvsNpartb}).

Recently, the correlation has been measured depending on the emission 
direction of the trigger particle in non-central collisions~\cite{adams2004}. 
Since in non-central reactions the overlap \AAex\ region has an almond-like 
shape with shorter length in the in-plane than in the out-of-plane direction,
energy loss of partons which depends on the distance traveled through the
medium should differ for the two directions. The definition of the direction for 
particle emission into in-plane, out-of-plane and between-plane in a semi-peripheral 
collision is illustrated in \fig{chap3:fig:inoutplane}. In \fig{chap3:fig:starcorrelation} 
we show the azimuthal distribution of associated particles, 
defined by $2 \le \pt \le \pt^{\rm trig}$, in \AuAu\ collisions at $20$--$60$\% centrality 
and in \pp\ reference data. Depending on the direction of the trigger particle
(with $4\le\pt^{\rm trig}\le6~\gev$) ,
the associated particles are classified into different classes,
in-plane or out-of-plane. The \AuAu\ data are corrected for collective effects, 
by subtracting the elliptic flow component ($v_2$). 
The near-side ($|\Delta\phi|\lsim 0.5$) correlations measured in \AuAu\ 
are clearly jet-like and very similar to those in \pp\ collisions (and to
those in \dAu~\cite{adams2003b}, not shown here). The back-to-back correlations 
($|\Delta\phi-2\pi|\lsim 0.7$) in \AuAu\ collisions for trigger particles 
emitted in-plane are suppressed compared to \pp\ (and to \dAu~\cite{adams2003b},
not shown here), and even more suppressed for the out-of-plane trigger particles. 

\begin{figure}[htb]
\begin{center}
\includegraphics[width=8cm]{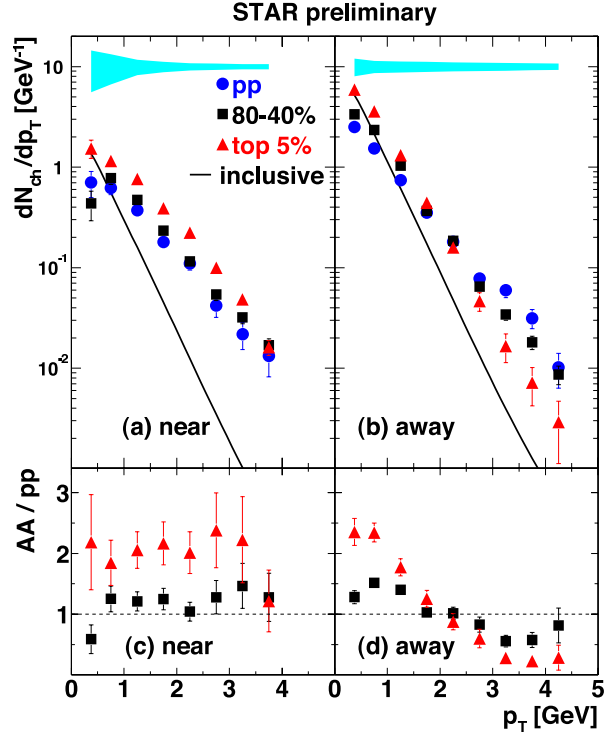}
\end{center}
\vspace{-0.3cm}
\caption[xxx]{Transverse-momentum distributions of near-side~(a) and away-side~(b) 
associated particles for \pp, peripheral and central \AuAu\ collisions 
at $\snn=200~\gev$ measured by \acs{STAR}. The bands show the systematic errors 
for central collisions. Ratios of the \AuAu\ to \pp\ distributions for the near-side~(c) 
and the away-side~(d). The figure is taken from~\Ref{wang2004}.}
\label{chap3:fig:reapstarcorrelation}
\end{figure}

Since energy must be conserved, it is expected that particles on the away-side 
should be rather soft, originating from thermalized (former hard) partons. 
Indeed, the reappearance of the particles on the away-side recently has been
confirmed in low-momentum hadrons~\cite{wang2004}. \Fig{chap3:fig:reapstarcorrelation}
shows the $\pt$-distribution of associated particles in \pp, peripheral ($40$--$80$\%)
and central ($0$--$5$\%) \AuAu\ collisions at $\snn=200~\gev$. While agreement 
is found between \pp\ and the peripheral \AuAu\ data on both sides,
the central \AuAu\ results differ from \pp, most significantly for the 
away-side. The ratios of the \AuAu\ to \pp\ distributions indicate that the leading
partons are modified in the medium created in the central \AuAu\ collisions. The
modifications lead to more associated particles on the near side, and shifts energy
from high to low momentum on the away side.

\enlargethispage{1cm}
In summary, the present, high-$\pt$ observations at \acs{RHIC} lead to the 
`model-independent' conclusion that partons traversing the dense medium in the core 
of the collision lose the majority of their energy, and the observed jet fragments 
are primarily those created from partons produced near the surface and directed outwards. 
In the next section we will try to give a quantitative description of these 
experimental findings.
\fi

\section{The Parton Quenching Model}
\label{chap3:pqm}
\ifpqm
The experimental observations at \acs{RHIC} have been explained in terms 
of attenuation or quenching models, where the energetic partons produced in 
the initial hard scattering processes as a consequence of the 
interaction with the dense QCD medium `loose' energy.
Several works exist on the subject~\cite{wang1998,wang1998b,wang2002,
wang2003,vitev2002b,adil2004,vitev2004,dainese2004,eskola2004}. 
Most models implement the parton energy loss due to medium-induced gluon 
radiation~(see \sect{chap3:partoneloss}). Also hadronic 
interactions~\cite{gallmeister2004,hardtke2004} have been investigated and 
found to contribute to the observed depletion of the hadron spectra.

In the following we describe the \ac{PQM}~\cite{dainese2004}, which combines 
the parton energy loss in the \acs{BDMPS-Z-SW} framework with a realistic 
collision geometry given by the Glauber model. The leading-particle suppression 
in \AaAa~collisions is obtained evaluating \eq{chap3:eq:hcrossec} in a Monte Carlo 
approach. The transverse momentum distributions for charged hadrons are 
generated by means of the following steps:
\begin{enumerate}
\item Generation of a quark or gluon with $\pt>3~\gev$, 
using \acs{PYTHIA}~\cite{mcpythia1987,mcpythia1994,mcpythia2001} 
in \pp\ mode with the \acs{CTEQ}~4L \ac{PDF}~\cite{lai1996};
\item determination of the parameters, $\omega_{\rm c}$ 
and $R$, \eq{chap3:eq:wc} and \eq{chap3:eq:kinematicr}, 
for the calculation of the quenching weights and
the energy-loss probability distribution $P(\Delta E)$
\eq{chap3:eq:pdeltae};
\item Monte Carlo sampling of the energy loss $\Delta E$ 
according to $P(\Delta E)$ to assign the quenched parton
transverse momentum, $\pt'=\pt-\Delta E$;
\item hadronization of the quenched parton using the \acs{KKP} 
fragmentation functions~\cite{kniehl2000}. 
\end{enumerate}
Steps 2 and 3 will be explained in more detail in the following paragraphs.
The quenched and unquenched $\pt$-distributions are obtained including or 
excluding the third step. The nuclear modification factor $R_{\rm AA}(\pt)$,
\eq{chap3:eq:rab}, is simply given by their ratio. 

\begin{figure}[htb]
\begin{center}
\includegraphics[width=10cm]{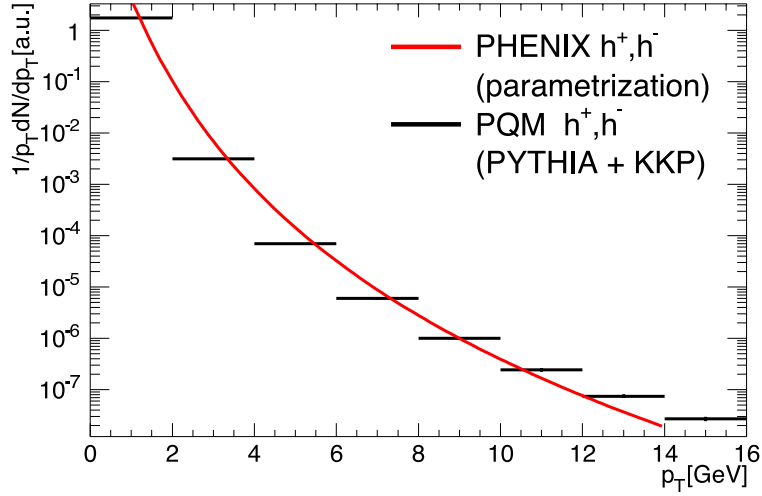}
\end{center}
\vspace{-0.3cm}
\caption[xxx]{Comparison between the charged hadron yield in \pp\ collisions 
measured and parameterized by \acs{PHENIX}~\cite{adler2003d} and the 
calculation by \acs{PQM}.}
\label{chap3:fig:compPPtoPQM}
\end{figure}

In \fig{chap3:fig:compPPtoPQM} we show that the hadron $\pt$-distribution 
at $\sqrt{s}=200~\gev$ agrees in shape with that measured for neutral pions 
in \pp\ collisions by \acs{PHENIX}~\cite{adler2003d}. The \pizero\ data
has been parameterized according to
\begin{equation*}
\frac{1}{\pt}\,\frac{\dd^2 N}{\dd\pt\dd\eta} 
= \const\,(1+\frac{p_0}{\pt})^{-n}\,r(\pt)\;,
\end{equation*}
where $p_0=1.22~\gev$ and $n=10$. For the correction from neutral
pions to charged hadrons we use a constant value of $r=1.6$.

\subsection{Parton-by-parton approach}
\label{chap3:step2}
The probability distribution $P(\Delta E)$, 
\eq{chap3:eq:pdeltae}, depends on the kinematical cutoff $R$ 
and on the characteristic gluon frequency $\omega_{\rm c}$. 
Due to the fact that $\hat{q}$ and $L$ are 
intuitively and physically more meaningful parameters, 
in most applications of the quenching 
weights~\cite{salgado2003,dainese2003,eskola2004}
the ($R$,~$\omega_{\rm c}$)-dependence of the quenching weights 
has been turned into a ($\hat{q}$,~$L$)-dependence,  
via \eq{chap3:eq:wc} and \eq{chap3:eq:kinematicr}.

The standard approach has been to fix a value for the transport coefficient, 
the same for all produced partons, and either to use a constant (effective)
in-medium length~\cite{salgado2003} or to calculate a different length for 
each parton according to a description of the collision 
geometry~\cite{dainese2003,eskola2004}.
However, that approach is not optimal, because there is no unique 
and exact definition of the in-medium path length when a realistic 
nuclear density profile is considered and the medium density is not 
constant over the whole nucleus--nucleus overlap region, but rather 
decreasing from the inner to the outer layers. 

In order to overcome these limitations, in \ac{PQM} we introduce an alternative 
approach. Namely, we determine the two parameters $\omega_{\rm c}$ and $R$ on a 
parton-by-parton basis: For a given centrality, the (transverse) density 
profile of the matter is computed and for each produced parton in the collision 
its path (azimuthal direction and length) through the matter determined. 
We, thus,  need to replace the fixed values of $\hat{q}$ and $L$ with their 
respective distributions as `seen' by the partons on the way out. 
Starting from \eq{chap3:eq:wc} and using \eq{chap3:eq:qscale} 
(for $\xi_0=0)$ with a space-point dependent transport coefficient $\hat{q}(\xi)$ and a 
path-averaged $\av{\hat{q}}$, we define the effective quantity
\begin{equation}
\label{chap3:eq:omegaceff}
\omega_{\rm c} \left |_{\rm effective}\right. 
\equiv \frac{1}{2} \av{\hat{q}}\,L^2
= \int_0^{\infty} \xi \, \hat{q}(\xi) \, \dd \xi\;,
\end{equation}
which on the r.h.s.~does not explicitly depend on $L$ anymore. 
Similarly, we define 
\begin{equation}
\label{chap3:eq:lqhateff}
\av{\hat{q}}L\left|_{\rm effective}\right. 
\equiv \int_0^{\infty} \hat{q}(\xi) \, \dd \xi
\end{equation}
and
\begin{equation}
\label{chap3:eq:kinematicreff}
R \left |_{\rm effective}\right. \equiv 
\frac{2\, \left(\omega_{\rm c}\left |_{\rm effective}\right)^2\right.}
{\av{\hat{q}}L\left |_{\rm effective}\right.}\;.
\end{equation}
For a step-function density distribution $\hat{q}(\xi)=\hat{q}_0\,\theta(L -\xi)$, 
\eq{chap3:eq:omegaceff} and \eq{chap3:eq:kinematicreff}
coincide with \eq{chap3:eq:wc} and \eq{chap3:eq:kinematicr}.

Using the definitions given by \eq{chap3:eq:omegaceff} and 
\eq{chap3:eq:kinematicreff} we incorporate the geometry of the collision 
via the local, space-point dependent, transport coefficient $\hat{q}(\xi)$.
The geometry is defined in the ($x$,~$y$) plane transverse to the beam direction $z$, 
in which the centers of two nuclei A and B colliding with an
impact parameter $b$ have the coordinates ($-b/2$,~$0$) and ($b/2$,~$0$), respectively.
We use the Glauber model (see~\sect{app:glauber}) to describe the geometry of the 
collision assuming the distribution of parton production points in the transverse plane 
and the transverse density of the medium both to be proportional to the $b$-dependent 
product $T_{\rm A}\,T_{\rm B}(x,y;\,b)\equiv T_{\rm A}(x,y)\times T_{\rm B}(x,y)$;
given by the thickness functions of the two nuclei. The nuclear thickness function 
is defined as the $z$-integrated Wood-Saxon nuclear density 
profile: $T_{\rm i}(x,y)\equiv\int \dd z\,\rho_{\rm i}^{\rm WS}(x,y,z)$. 
The parameters of the Wood-Saxon profile for different nuclei are tabulated 
from data~\cite{atomdata}. Note that $T_{\rm A}\,T_{\rm B}(x,y;\,b)$
estimates the transverse density of binary \NN\ collisions, 
$\rho_{\rm coll}(x,y;b)$, modulo the inelastic \NN\ cross section.

Since we only consider partons produced at very central rapidities, we 
assume that they propagate in the transverse plane ($E\approx p\approx \pt$). 
For a parton with production point ($x_0$,~$y_0$) and azimuthal propagation 
direction ($\cos\phi_0$,~$\sin\phi_0$) ($\phi_0$ is sampled uniformly), 
we define the local transport coefficient along the path of the parton 
inside the overlap region of the nuclei as
\begin{equation}
\label{chap3:eq:qofxi}
\hat{q}(\xi;b)=k\times 
T_{\rm A}\,T_{\rm B}(x_0+\xi \cos\phi_0,y_0+\xi \sin\phi_0;\,b)\;.
\end{equation}
The parameter $k$ (in $\fm$) is used to set the scale of the transport 
coefficient (in $\gev^2/\fm$). It is the only parameter
in \ac{PQM}. The idea is to keep $k$ fixed, once it is determined by
a fit to the data. For every parton (or every pair of partons in opposite directions) 
we compute the two integrals $I_0$ and $I_1$ given by \eq{chap3:eq:omegaceff} 
and \eq{chap3:eq:lqhateff}
\begin{equation*}
\label{chap3:eq:In}
I_n \equiv \int_0^\infty \xi^n\,\hat{q}(\xi;b)\,\dd\xi\hspace{1cm}n=0,\,1\;,
\end{equation*}
which determine the energy-loss probability distribution $P(\Delta E)$ 
according to 
\begin{equation}
\label{chap3:eq:wcRfromI0I1}
\omega_{\rm c}\left|_{\rm effective}\right.
=I_1\hspace{0.5cm}
{\rm and}\hspace{0.5cm}
R\left|_{\rm effective}\right.=2\,I_1^2/I_0\;.
\end{equation}

And for the corresponding effective path length and transport coefficient
we get 
\begin{equation}
\label{chap3:eq:L}
L\left|_{\rm effective}\right.=2\,I_1/I_0
\hspace{0.5cm}{\rm and}\hspace{0.5cm}
\hat{q}\left|_{\rm effective}\right.=I_0^2/(2\,I_1)\;.
\end{equation}
We point out that the resulting definition of 
$L$~\footnote{For simplicity, we drop the subscript 
`effective' from now on.} is, as necessary, independent of $k$. Furthermore,
it is the same used in~\Ref{dainese2003}. Note that the effective
$\hat{q}$ is proportional to $k$.

\begin{figure}[htb]
\begin{center}
\ifarxiv
\includegraphics[width=15cm]{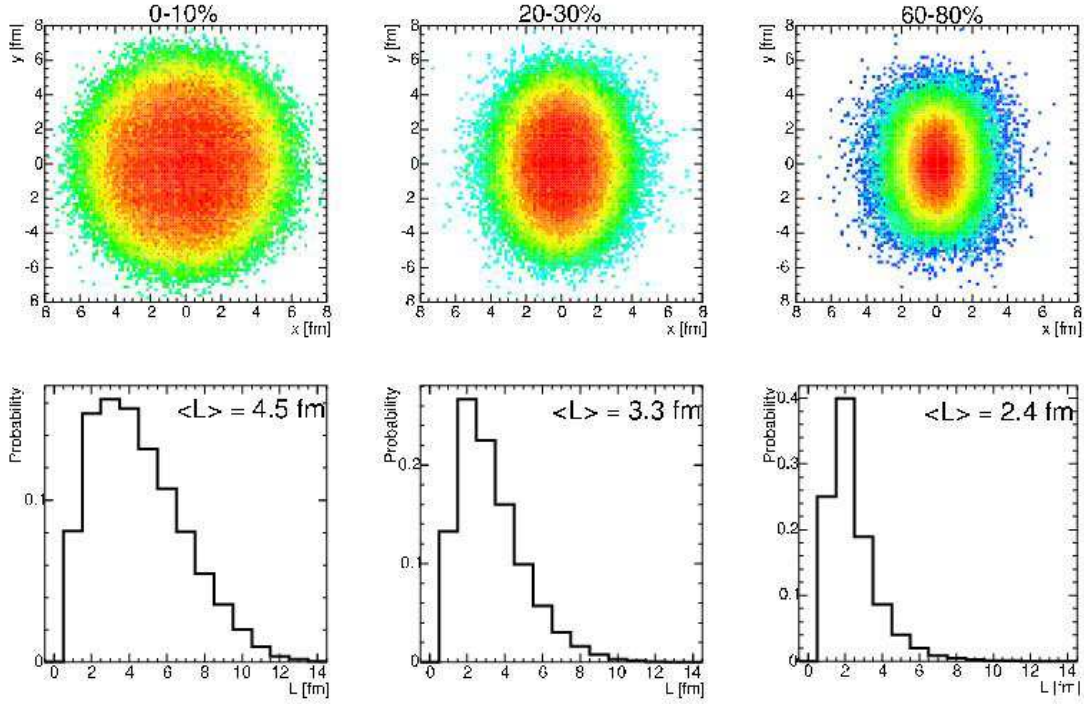}
\else
\includegraphics[width=15cm]{cXYandL-AuAu}
\fi
\end{center}
\vspace{-0.3cm}
\caption[xxx]{Distributions of parton production points in the transverse plane 
(upper row) and effective (\eq{chap3:eq:L}) in-medium path length (lower row) 
in central, semi-central and peripheral \AuAu\ collisions. The quantity $\av{L}$ 
is the average of the path-length distribution.} 
\label{chap3:fig:geo}
\end{figure}

The parton-by-parton approach allows a natural extension from central to 
peripheral nucleus--nucleus collisions. After the only free parameter, $k$, 
is determined to describe the measured nuclear modification factor in central 
collisions at $\snn=200~\gev$, the results for other centrality classes simply 
depend on the impact parameter dependence of the product $T_{\rm A}\,T_{\rm B}(x,y;\,b)$.
By means of the Glauber model, we translate the experimental definition 
of the centrality classes in terms of fractions of the geometrical cross 
section to a range in $b$. Within such range, we sample, for every parton
(or every parton pair) a value of $b$ according to the $b$-dependence 
of the average number of binary 
collisions, $\dd \av{N_{\rm AB}^{\rm coll}}/\dd b$.  
In \fig{chap3:fig:geo} we report the distributions of the parton production 
points ($x_0$,~$y_0$) in the transverse plane and of the effective 
in-medium path lengths, \eq{chap3:eq:L}, in central ($0$--$10$\%), 
semi-central ($20$--$30$\%) and peripheral (60--80\%) \AuAu\ collisions obtained 
with \ac{PQM}. The average length decreases from $\av{L}=4.4~\fm$ for most central, 
to $3.3~\fm$ for semi-central, down to $2.4~\fm$ for peripheral collisions.

\begin{figure}[htb]
\vspace{0.3cm}
\begin{center}
\includegraphics[width=12cm]{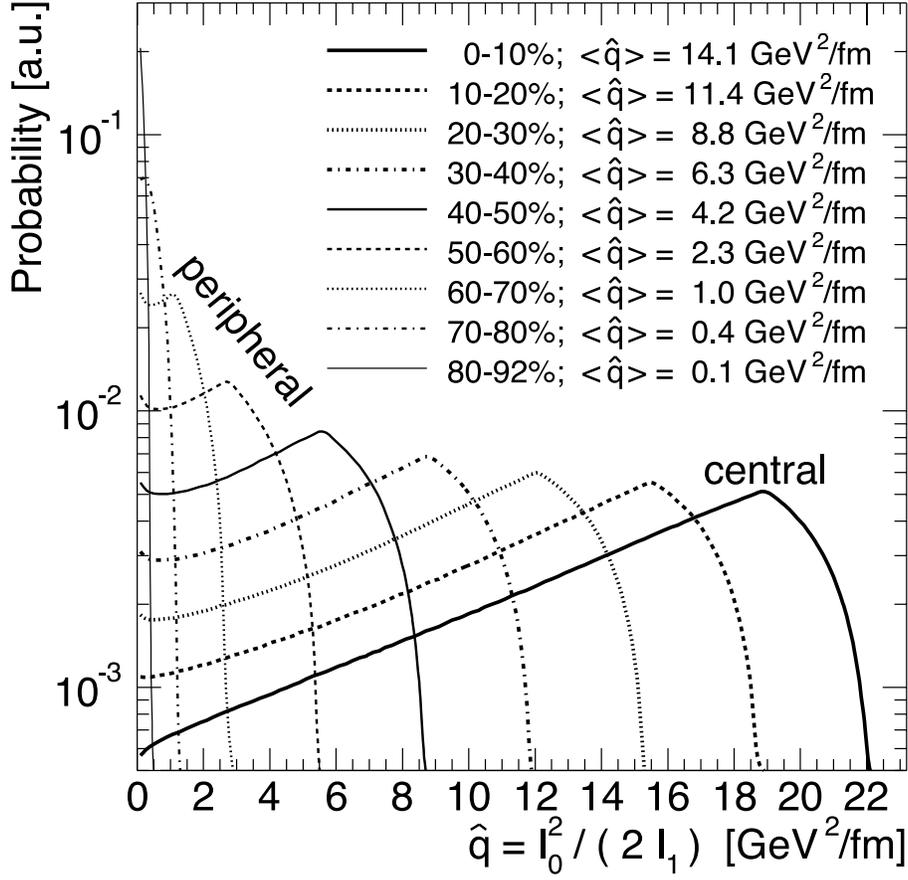}
\end{center}
\vspace{-0.3cm}
\caption[xxx]{Distributions of effective $\hat{q}$ (\eq{chap3:eq:L}), 
for different centralities. The $k$ parameter, setting the
scale, is fixed to the value  $k=5\cdot 10^6~\fm$ that allows to 
describe $\RAA$ for the most central \AuAu\ collisions 
at $\snn=200~\gev$.}
\label{chap3:fig:QhatAllCentralities}
\end{figure}

In \fig{chap3:fig:QhatAllCentralities} we show the distributions 
of effective $\hat{q}$, \eq{chap3:eq:L}, for different centralities.
The scale $k$ is fixed to the value  $k=5\cdot 10^6~\fm$ that allows to 
describe $\RAA$ for the most central \AuAu\ collisions at $\snn=200~\gev$
(see below). The $\hat{q}$ variation within a given centrality class 
reflects the different parton production points and, hence, the 
different medium densities encountered by the partons on their way out
of the interaction region. The rightmost (highest) value refers 
to partons originating from the border of almond region and 
traversing through the complete interior of the dense collision 
center.

\subsection{Finite energy constraints}
\label{chap3:step3}
For the calculation of the energy loss in \ac{PQM} 
we use the quenching weights in the 
\acs{BDMPS-Z-SW} framework, introduced
in \sect{chap3:qw}. 
According to the $P(\Delta E)$ distribution, 
obtained by the parameters $\omega_{\rm c}$ and $R$ as 
explained in the previous section, we sample the energy 
loss $\Delta E$, 
to get the reduced parton transverse momentum, $\pt'=\pt-\Delta E$.

The quenching weights are calculated in the eikonal approximation,
in which the energy of the primary parton is infinite
($E=\pt=\infty$). Therefore, when the realistic case of finite-energy 
partons is considered, a significant part of the energy-loss probability 
distribution $P(\Delta E)$ lies above the parton energy $E$, 
in particular for large values of $\omega_{\rm c}$ and $R$, 
or equivalently, of $\hat{q}$ and $L$. Since for a parton with
energy $E$ the energy loss must be constrained to $\Delta E\le E$, 
we define the constrained weights according to the following two 
prescriptions:
\begin{itemize}
\item {\em Reweighted}: Truncate $P(\Delta E)$ at $\Delta E=E$,
and renormalize it to unity  
\begin{equation}
\label{chap3:eq:prw}
P^{\rm rw}(\Delta E,\,E)=\frac{P(\Delta E) \, \Theta(E - \Delta E)}
{\int_0^E\,\dd\epsilon \, P(\epsilon)}\;.
\end{equation}
The Monte Carlo implementation of this approach is: sample $\Delta E$
from the original $P(\Delta E)$; if $\Delta E>E$, sample another $\Delta E$;
iterate until a $\Delta E\leq E$ is sampled.
\item {\em Non-reweighted}: Truncate $P(\Delta E)$ at $\Delta E=E$ 
and add
$\delta(E-\Delta E)\int_E^\infty\dd\epsilon\,P(\epsilon)$
\begin{equation}
\label{chap3:eq:nonprw}
P^{\rm non-rw}(\Delta E,\,E)=P(\Delta E) \, \Theta(E -\Delta E)
+\delta(E-\Delta E)\int_E^\infty\dd\epsilon\,P(\epsilon)\;.
\end{equation}
The integral of $P$ is, in this way, maintained equal to one.
The corresponding Monte Carlo implementation reads: sample an energy loss 
$\Delta E$ from the original $P(\Delta E)$; set $\Delta E=E$ 
if $\Delta E\geq E$.
\end{itemize}

Like the unconstrained weights the constrained quenching probabilities
depend also on the kinematical parameters, $\omega_{\rm c}$ and $R$.
Note, as necessary, in the eikonal limit $E\rightarrow\infty$ 
the constrained weights approach the unconstrained weights
\begin{equation*}
\lim_{E\rightarrow\infty} P^{\rm rw}(\Delta E,\,E) = 
\lim_{E\rightarrow\infty} P^{\rm non-rw}(\Delta E,\,E) =
P(\Delta E)\;.
\end{equation*}
Whereas, in the limit $E\rightarrow 0$ 
the constrained weights reduce to 
\begin{equation*}
\lim_{E\rightarrow\, 0} P^{\rm rw}(\Delta E,\,E) = 
\lim_{E\rightarrow\, 0} P^{\rm non-rw}(\Delta E,\,E) =
\delta(\Delta E)\;.
\end{equation*}

\begin{figure}[htb]
\begin{center}
\subfigure[Non-reweighted energy-loss distribution]{
\label{chap3:fig:elossfixednonrw}
\includegraphics[width=7cm]{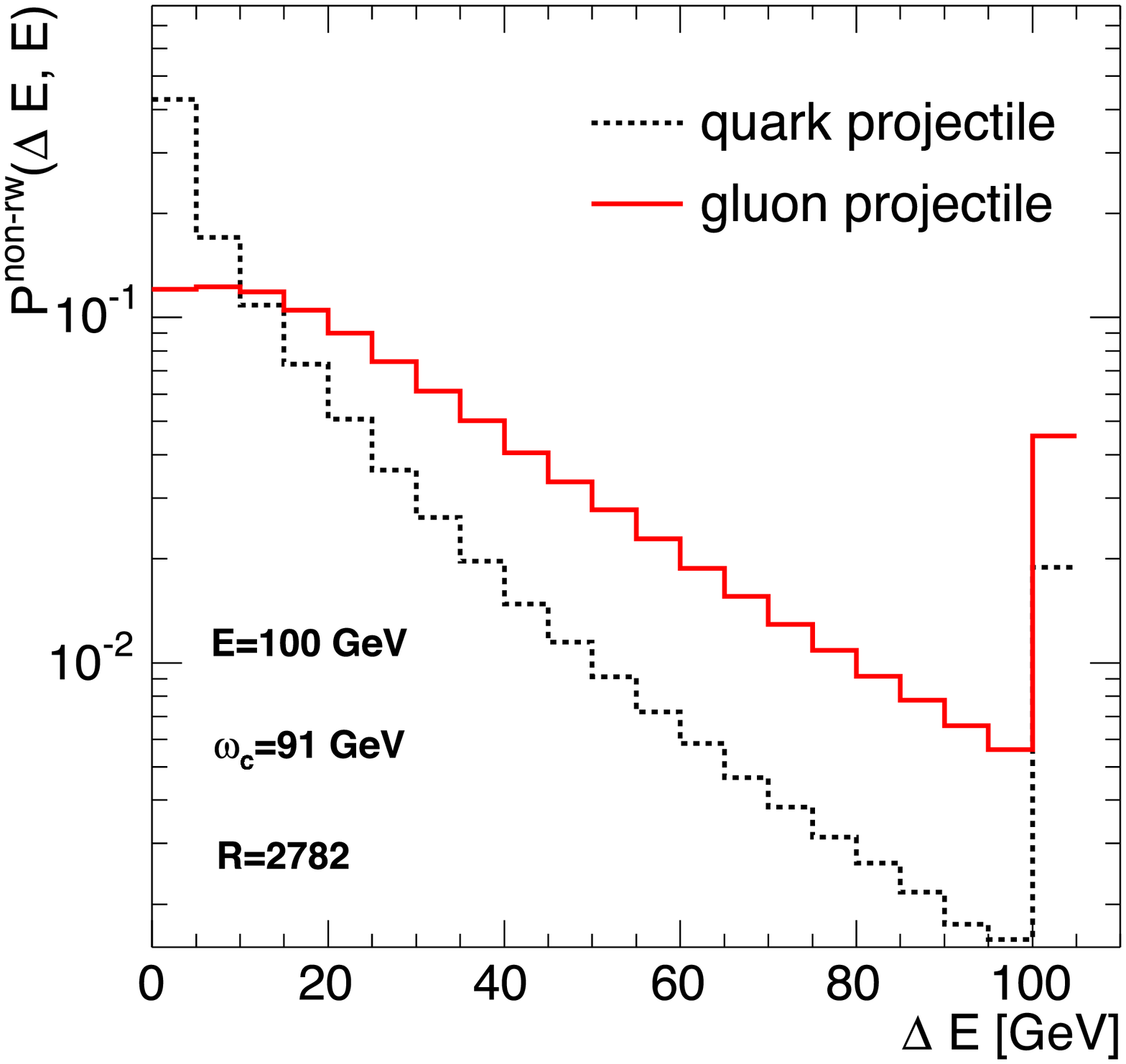}}
\hspace{0.5cm}
\subfigure[Reweighted energy-loss distribution]{
\label{chap3:fig:elossfixedrw}
\includegraphics[width=7cm]{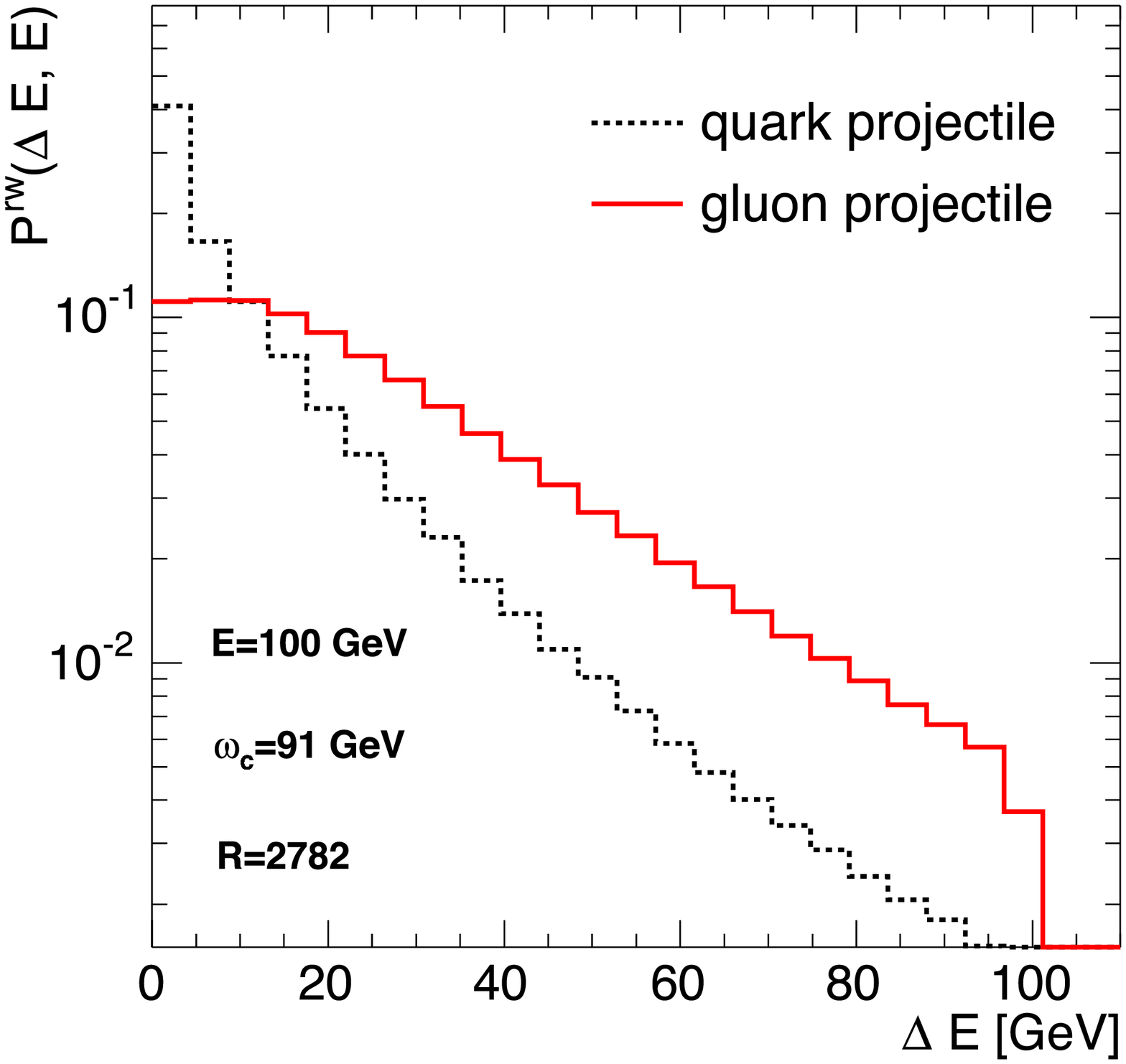}}
\end{center}
\vspace{-0.5cm}
\caption[Energy loss distribution]
{Energy-loss distribution $P^{\rm non-rw}(\Delta E,\,E)$~\subref{chap3:fig:elossfixednonrw} 
and $P^{\rm rw}(\Delta E,\,E)$~\subref{chap3:fig:elossfixedrw} for $E=100~\gev$ 
partons. The calculation uses fixed values of $\hat{q}=1~\gev^2/\fm$ and $L=6~\fm$, 
resulting in $\omega_{\rm c}=91~\gev$ and $R=2782$.}
\label{chap3:fig:elossfixed}
\end{figure}

\Fig{chap3:fig:elossfixed} shows the energy-loss distribution of quarks and gluons
with primary energy of $E=100~\gev$ for the non-reweighted and reweighted case.
The calculation is done using fixed values of $\hat{q}=1~\gev^2/\fm$ and $L=6~\fm$, 
which results in $\omega_{\rm c}=91~\gev$ and $R=2782$. The average loss
in the non-reweighted case is $14~\gev$ for quarks and $30~\gev$ for gluons;
in the reweighted case it is $12~\gev$ for quarks and $26~\gev$for gluons. 
In the non-reweighted case the accounted energy loss is generally larger, since the 
medium with a probability $\int_E^\infty\dd\epsilon\,P(\epsilon)$ may fully absorb 
the primary partons. 

It has been argued~\cite{salgado2003,eskola2004} that the difference in the 
values of the observables for the two different constraints illustrates the 
theoretical uncertainties of the \acs{BDMPS-Z-SW} framework evaluated
at finite parton energies. Therefore, along the lines of what has been 
done in~\Refs{eskola2004,dainese2004} we display the \ac{PQM} results as a 
band delimited by the reweighted case (smaller quenching) 
and the non-reweighted case (larger quenching). 

\subsection{PQM results at RHIC}
\label{chap3:pqmresults}
We will present the \ac{PQM} results on the nuclear modification factor;
further calculations concerning azimuthally-differential observables are reported 
in~\Ref{dainese2004}.

Before moving to the parton-by-parton approach of \ac{PQM} outlined 
in the previous sections, it is very instructive to perform the calculation 
of $R_{\rm AA}(\pt)$ in \AuAu\ collisions at $\snn=200~\gev$ defined 
by \eq{chap3:eq:rab} using a constant transport coefficient and the 
Glauber-based path-length distributions. The model results as well as the 
data ranging in centrality from $0$--$10$\% to $80$--$92$\% are shown in 
\fig{chap3:fig:QAllCentralities}.
The data on charged hadrons from \acs{STAR}~\cite{adams2003} and 
\acs{PHENIX}~\cite{adler2003b} and neutral pions from \acs{PHENIX}~\cite{adler2003} 
are reported with combined statistical and $\pt$-dependent systematic errors 
given by the bars on the data points and $\pt$-independent normalization 
errors given by the bars centered at $R_{\rm AA}=1$. We fix the transport coefficient
to $\hat{q}\simeq 15~\gev^2/\fm$ such that the data for the most central collisions 
are within the model band delimited by the reweighted and 
non-reweighted cases. For the most central case, using the constant transport 
coefficient of $15~\gev^2/\fm$ and the realistic length distribution, 
the measured hadron suppression can be fairly well described for $\pt \gsim 5~\gev$. 
At lower $\pt$ we do not apply the model, since initial-state effects and in-medium 
hadronization, that we do not include, might play an important role.
However, as clearly observed in \fig{chap3:fig:QAllCentralities}, using the same 
constant value for $\hat{q}$ with corresponding length distribution 
for the different centrality classes fails to reproduce the data for  
semi-central and peripheral collisions. We note that between the non-reweighted
and reweighted constraint there is a difference of about a factor 2 in the 
magnitude of $\RAA$. In addition, there is a change in the slope, which is slightly 
positive in the non-reweighted and slightly negative in reweighted case.

\begin{figure}[htb]
\begin{center}
\includegraphics[width=11.5cm]{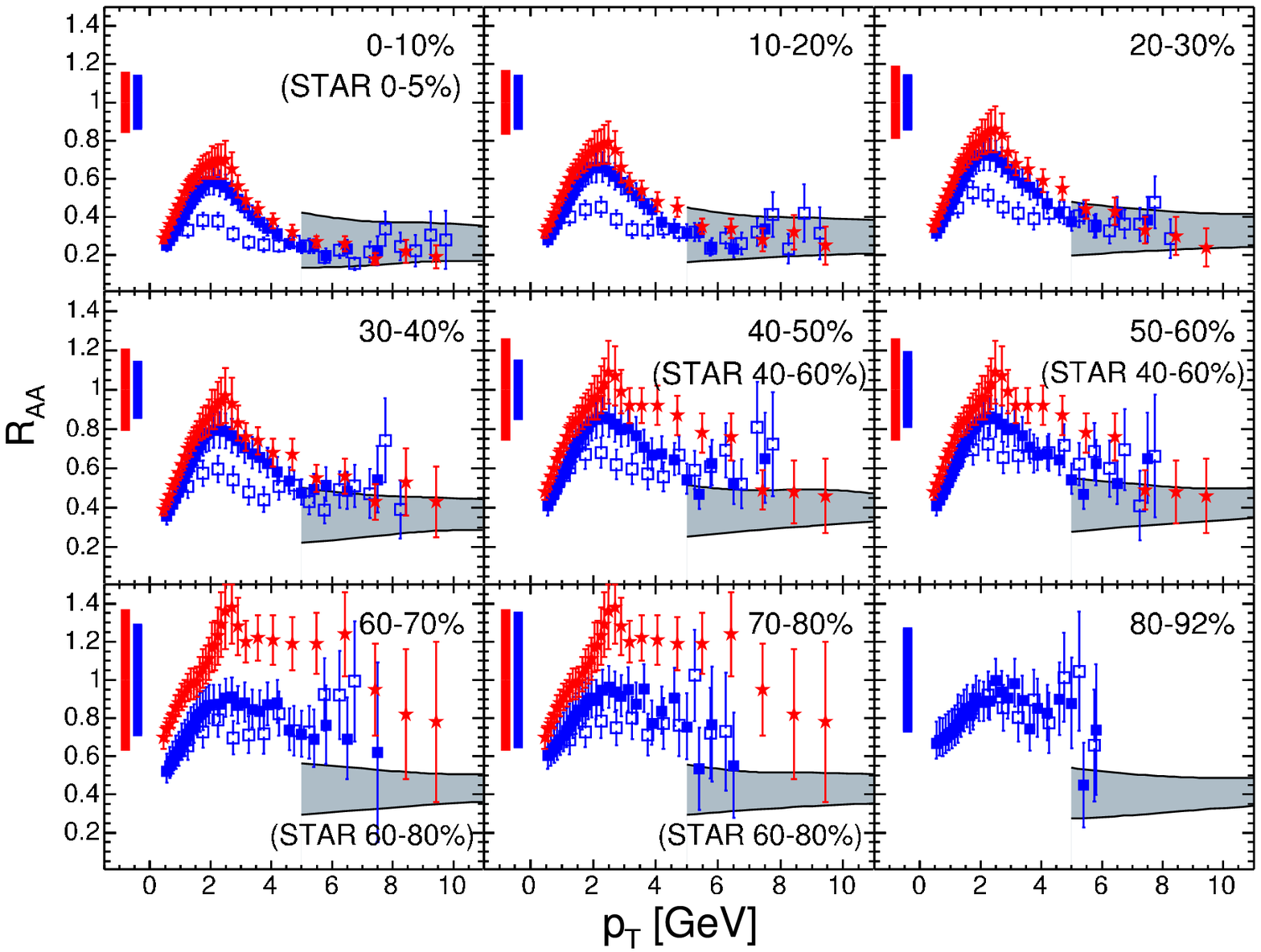}
\end{center}
\vspace{-0.4cm}
\caption[xxx]{$\RAA(\pt)$ for fixed $\hat{q}=15~\gev^2/\fm$ 
See caption of \fig{chap3:fig:KAllCentralities}.}
\label{chap3:fig:QAllCentralities}
\end{figure}

\begin{figure}[htb!]
\begin{center}
\includegraphics[width=11.5cm]{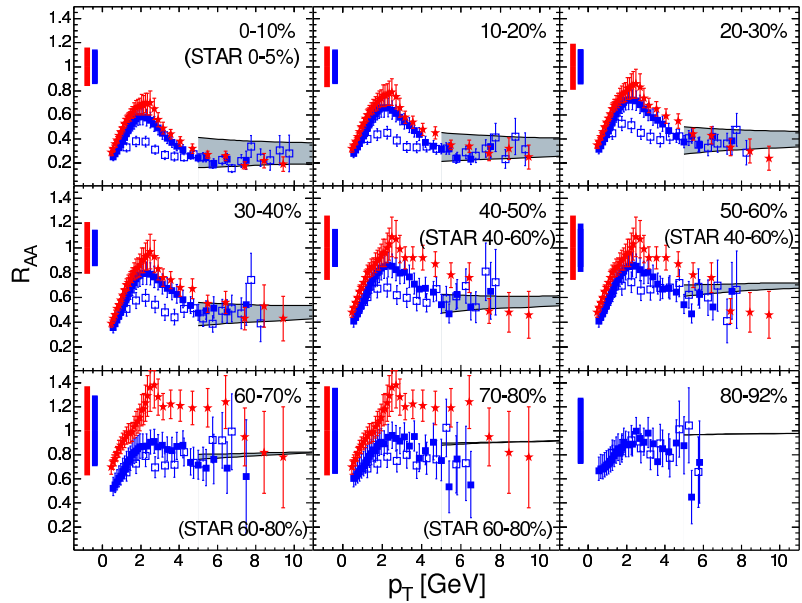}
\end{center}
\vspace{-0.4cm}
\caption[xxx]{$\RAA(\pt)$ for fixed $k=5\cdot 10^6~\fm$ in the \acs{PQM} parton-by-parton 
approach at mid-pseudo-rapidity for different centralities in \AuAu\ collisions at $\snn=200~\gev$. 
The measured points are for charged hadrons (stars, closed squares)~\cite{adams2003,adler2003} 
and neutral pions (open squares)~\cite{adler2003b}. The data are reported with statistical 
and $\pt$-dependent systematic errors (bars on the data points) and $\pt$-independent 
systematic errors (bars at $\RAA=1$).}
\label{chap3:fig:KAllCentralities}
\end{figure}

In order to address the centrality dependence of the high-$\pt$ suppression, 
we move to the parton-by-parton approach by \ac{PQM}. For the most central collisions, 
the result obtained with the scale parameter $k=5\cdot 10^6~\fm$  matches the data. 
From now on, we keep the same value of $k$ and merely vary the centrality 
by using dependence of the local transport coefficient via \eq{chap3:eq:qofxi} 
as outlined above. The \ac{PQM} parton-by-parton calculation is shown in 
\fig{chap3:fig:KAllCentralities}. The results nicely follow the increase of the 
measured $\RAA$ with decreasing centrality. The theoretical uncertainty band for 
the most central cases is very similar to that reported in \fig{chap3:fig:QAllCentralities}.
It is narrower for semi-central and peripheral collisions. As we will argue below the reason 
is that due to smaller size and density of the medium the probability to have $\Delta E>E$ in 
the quenching weights becomes less likely, and, therefore the differences introduced by the 
two constraints reduces. Note that the value of  $k=5\cdot 10^6~\fm$
corresponds to $\av{\hat{q}}\simeq 14~\gev^2/\fm$ in most central collisions 
(see \fig{chap3:fig:QhatAllCentralities}). Numerically, our value of 
$\av{\hat{q}}\simeq 14~\gev^2/\fm$ is smaller than $\hat{q} \simeq 10~\gev^2/\fm$ 
found in~\Ref{eskola2004} for central collisions. However, this is not an inconsistency. 
The value of $\as$ used in the calculation of the quenching weights 
is $\as=1/3$ here and $\as=1/2$ there. Since the scale of the energy loss is 
set by the product $\as\,\av{\hat{q}}$ (see \eq{chap3:eq:avdE}), the product is 
about the same for both calculations. Our results for the nuclear modification
factor at \ac{RHIC} are summarized in \fig{chap3:fig:RAAIAAvsNparta}, where we 
show the average $\RAA$ in the range $4.5<\pt<10~\gev$ compared to 
data~\cite{adams2003,adler2003,adler2003b} plotted as a function of the number of 
participant nucleons, $N_{\rm part}$, obtained from the Glauber model. 

\begin{figure}[htb]
\begin{center}
\subfigure[Nuclear modification factor]{
\label{chap3:fig:RAAIAAvsNparta}
\includegraphics[width=7cm]{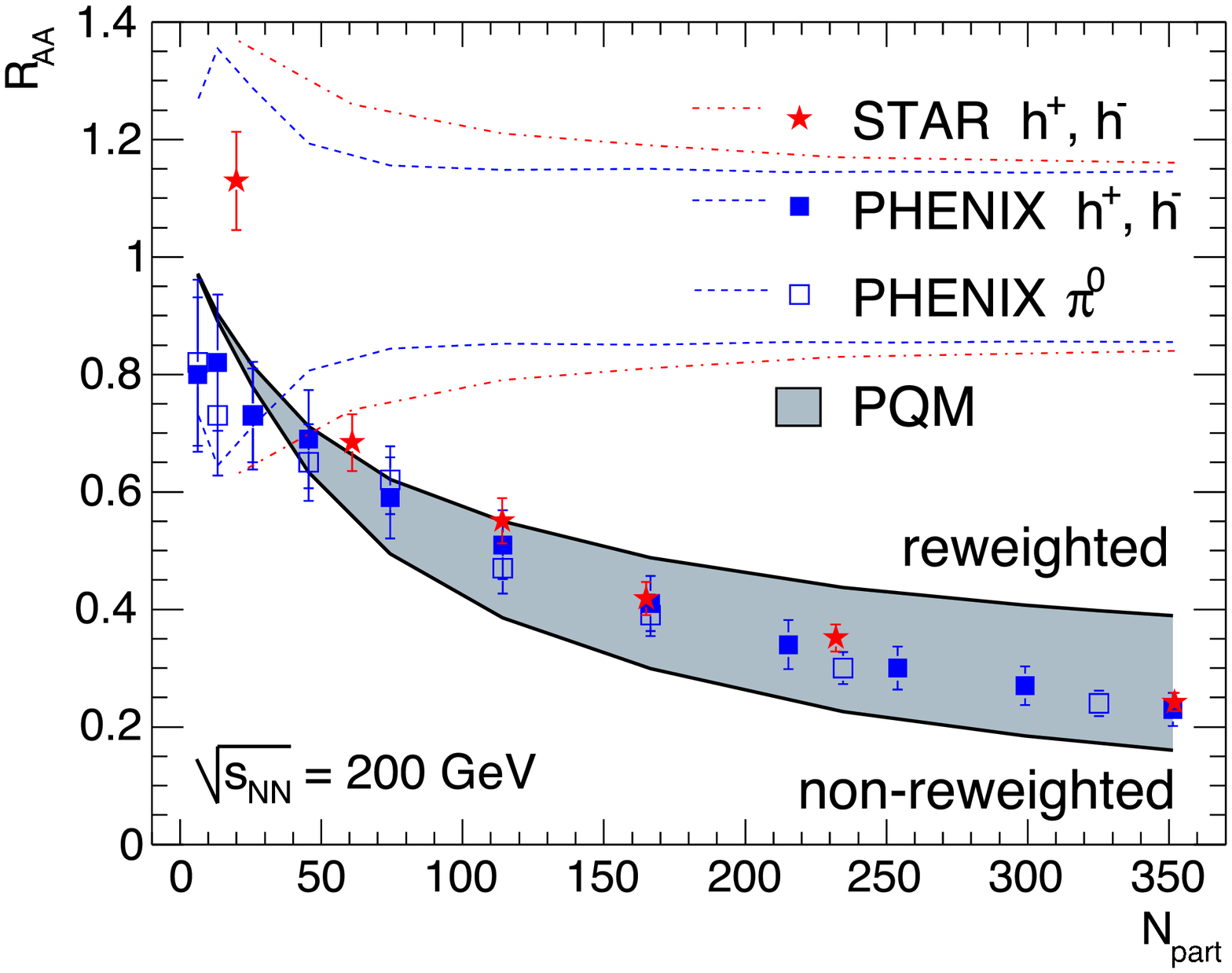}}
\hspace{0.2cm}
\subfigure[Suppression factor]{
\label{chap3:fig:RAAIAAvsNpartb}
\includegraphics[width=7cm]{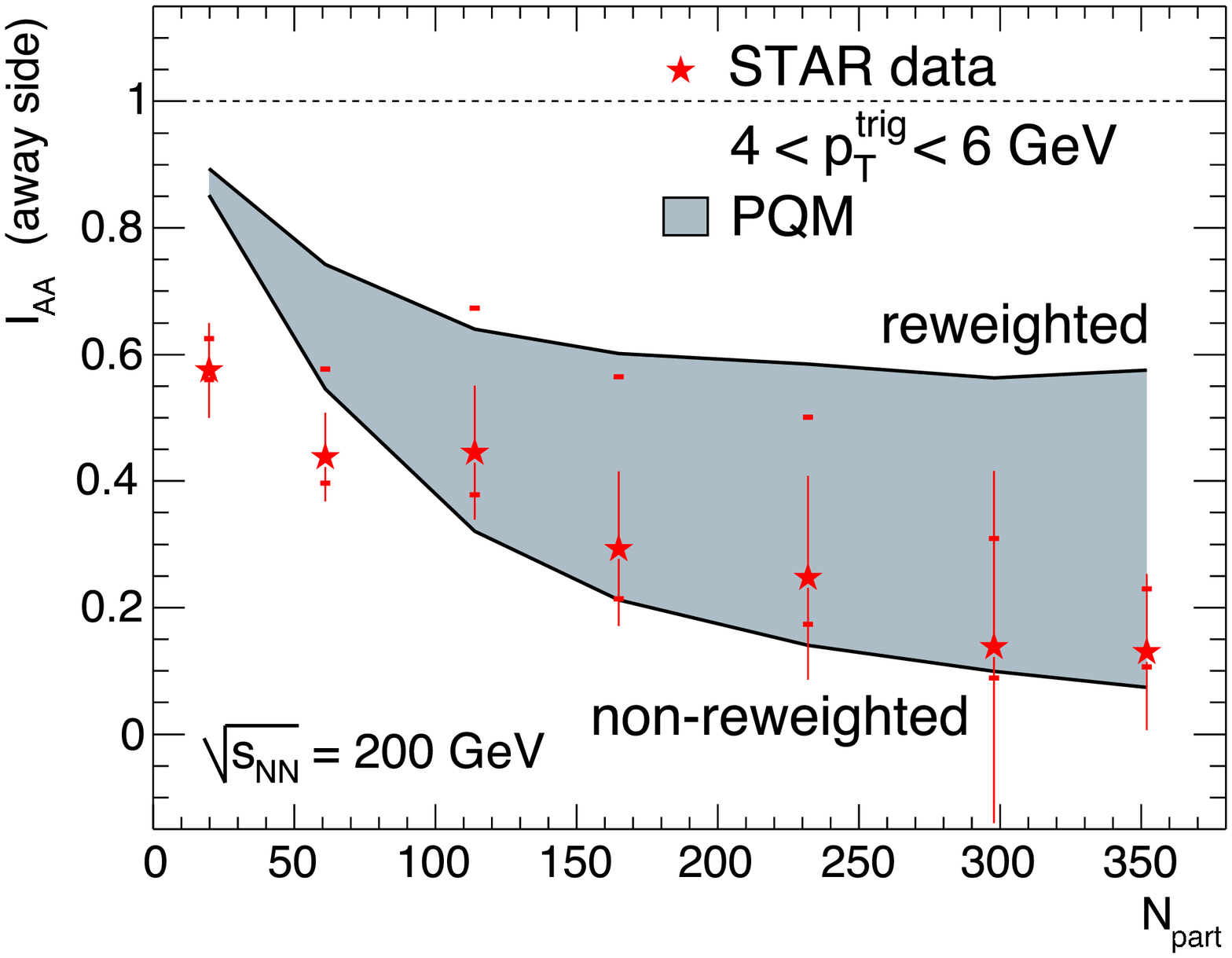}}
\end{center}
\vspace{-0.5cm}
\caption[xxx]{\acs{PQM} for $k=5\cdot 10^6~\fm$ compared to \acs{RHIC} data as a function of the 
number of participants, $N_{\rm part}$. \subref{chap3:fig:RAAIAAvsNparta}~Average $\RAA$ in 
the range $4.5 \le \pt \le 10~\gev$. Data are from~\Refs{adams2003,adler2003,adler2003b}.
The error bars are combined statistical and $\pt$-dependent systematic errors and the bands 
centered at $\RAA=1$ are the $\pt$-independent normalization errors for \acs{PHENIX} (dashed) 
and \acs{STAR} (dot-dashed). 
\subref{chap3:fig:RAAIAAvsNpartb} $I_{\rm AA}$ for the away-side jet. Data are from 
\acs{STAR}~\cite{adler2002}. The statistical (bars) and systematic (ticks) errors are shown.}
\label{chap3:fig:RAAIAAvsNpart}
\end{figure}

By generating pairs of back-to-back partons, we can study the centrality dependence of 
the disappearance of the away-side jet within the \ac{PQM} framework. Using the correlation 
strength, \eq{chap3:eq:daa}, in \NN\ relative to \pp\ collisions one conveniently defines 
the suppression factor
\begin{equation}
\label{chap3:eq:iaa}
I_{\rm AA}=\frac{D_{\rm AA}}{D_{\rm pp}}\;.
\end{equation}
For each parton we calculate $\omega_{\rm c}$ and $R$ and apply energy loss and fragmentation
as outlined above. We count as trigger particle every hadron $\rm{h}_1$ with $4<p_{\rm t,1}<6~\gev$ 
and as associated away-side particle the other hadron $\rm{h}_2$ of the pair, 
if its transverse momentum is in the range $2~\gev<p_{\rm t,2}<p_{\rm t,1}$. 
\begin{equation}
\label{eq:iaaus}
I_{\rm AA}=\left(\frac{N^{\rm associated}}{N^{\rm trigger}}\right)_{\rm with~energy~loss} \bigg/ \left(\frac{N^{\rm associated}}{N^{\rm trigger}}\right)_{\rm w/o~energy~loss}\,.
\end{equation}

\Fig{chap3:fig:RAAIAAvsNpartb} shows the \ac{PQM} result for $I_{\rm AA}$ versus $N_{\rm part}$,
compared to \acs{STAR} measurements~\cite{adler2002} in \AuAu\ collisions at $\snn=200~\gev$
with statistical (bars) and systematic (ticks) errors. The magnitude and centrality dependence 
of the suppression are described without changing the scale parameter value we extracted from 
$\RAA$ in central collisions.

\subsection{Extrapolation to the LHC energy}
\label{chap3:extrapollhc}
Within the \ac{PQM} model the choice of the scale parameter $k$ and,
hence, of the (average) transport coefficient is the main unknown for the 
extrapolation from \ac{RHIC} to \ac{LHC} energies. 
It is expected that the transport coefficient grows linearly with the 
initial (gluon) number-density of the medium~\cite{baier2001,baier2002},
$\hat{q}\propto n^{\rm gluons}\propto\epsilon^{3/4}$. 
For collisions of two nuclei with mass number $A$ at energy $\snn$
the initial gluon density in the \acs{EKRT} saturation 
model~\cite{eskola1999} is estimated to scale as 
\begin{equation}
\label{eq:ekrt}
n^{\rm gluons}\propto {\rm A}^{0.383}\, \left(\snn\right)^{0.574}. 
\end{equation}
Thus, we get for central \PbPb\ collisions at \acs{LHC} energies
\begin{equation*}
n^{\rm gluons}_{\rm Pb-Pb,\,5.5\,TeV}\simeq 7\times n^{\rm gluons}_{\rm Au-Au,\,200\,GeV}\;, 
\end{equation*}
\ie~$\av{\hat{q}}^{\rm LHC}\simeq 100~\gev^2/\fm$. 
According to \eq{chap3:eq:qofxi} and \eq{chap3:eq:L} the scaling can
be carried over to the $k$ parameter, $k^{\rm LHC}\simeq 3.5\cdot 10^7~\fm$.
To compute the expected nuclear modification factor in \PbPb~collisions at 
the \acs{LHC} we use \ac{PQM} as before, but generate the parton
momenta with \acs{PYTHIA} at $\sqrt{s}=5.5~\tev$. 

\begin{figure}[b!]
\begin{center}
\ifarxiv
\includegraphics[width=13cm]{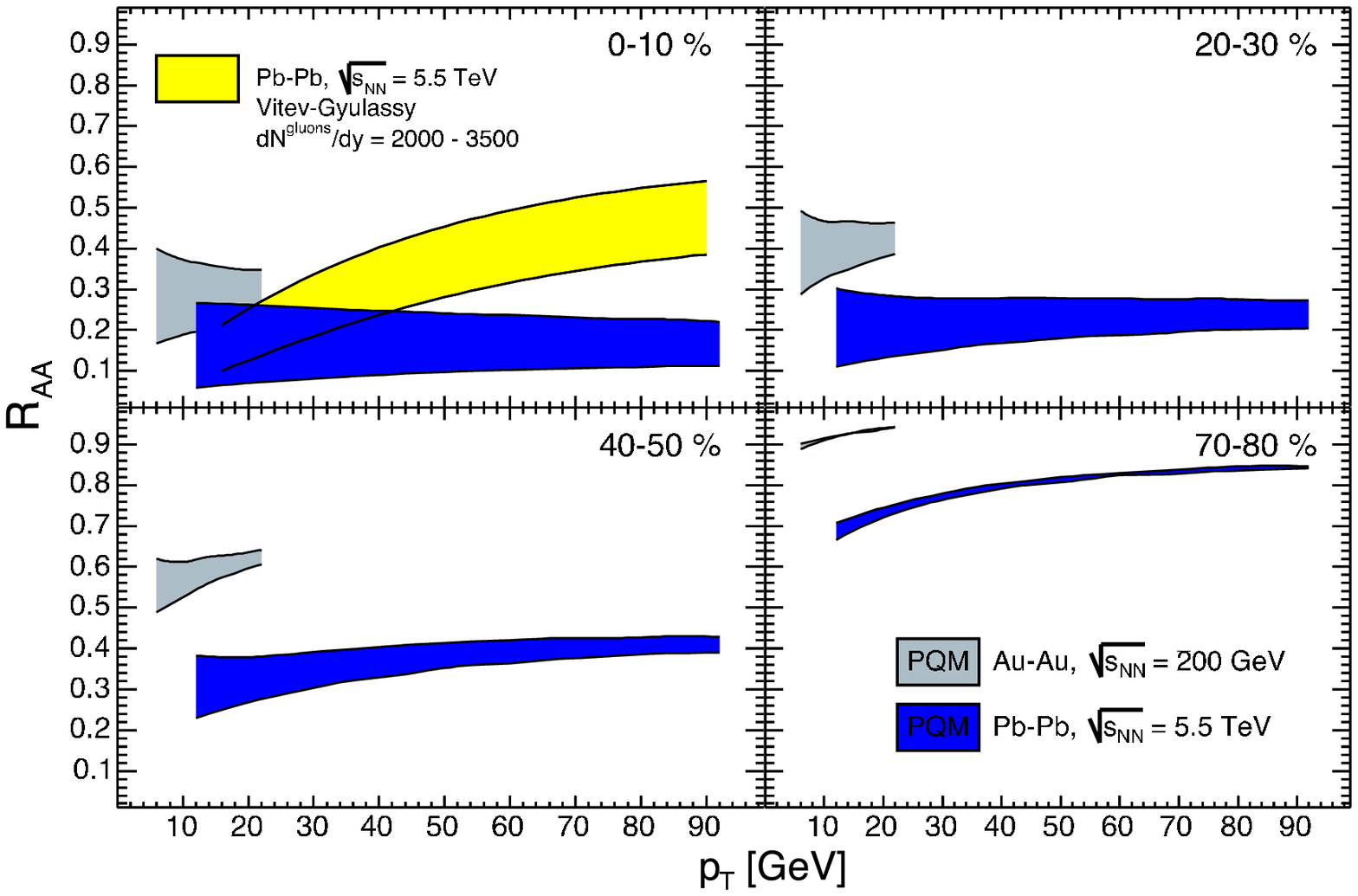}
\else
\includegraphics[width=14cm]{cRAAvsPt-LHCcentsPQMvsVitev}
\fi
\end{center}
\vspace{-0.3cm}
\caption[xxx]{\acs{PQM} predictions of $\RAA(\pt)$ in \PbPb\ collisions 
at $\snn=5.5~\tev$ for different centrality classes, as well as
the \AuAu\ results at $\snn=200~\gev$. The prediction for 
by Vitev and Gyulassy is taken from~\Ref{vitev2002}.} 
\label{chap3:fig:RAAlhc}
\end{figure}

In \fig{chap3:fig:RAAlhc} we report the expected transverse-momentum 
dependence of $\RAA$ at the \ac{LHC} in the range $10<\pt<90~\gev$ for 
different centralities; the results at $\snn=200~\gev$ are shown as well. 
In the most central collisions $\RAA$ is of $\approx 0.15$,
independent of $\pt$. Our prediction is almost a factor of 2 smaller 
than the measured value at $\snn=200~\gev$. 
It is in agreement, 
both in the numerical value and in the $\pt$-dependence, 
with~\cite{eskola2004} using the same quenching weights and the 
same $\as\av{\hat{q}}$, while it is quite different from the 
calculation by Vitev and Gyulassy~\cite{vitev2002b}.
For comparison, we report in the same figure their prediction,
which assumes an initial gluon rapidity-density $\dd N^{\rm gluons}/\dd y$ 
in the range 2000--3500. They expect $\RAA$ to rise significantly at 
large transverse momenta, from 0.1--0.2 at $20~\gev$ to 0.4--0.6 at 
$90~\gev$. Note that the difference between the two results 
is not due to the fact that we do not include nuclear (anti-)shadowing 
effects, since these are expected to determine a rather $\pt$-independent 
increase of $\RAA$ of about $10$\% in the range $25<\pt<100~\gev$~\cite{eskola2004}.

\subsection{Extrapolation to the intermediate RHIC energy}
\label{chap3:extrapolintrhic}
To verify the predictive power of \ac{PQM}, we recall the recent measurement 
of the nuclear modification factor for charged hadrons and neutral pions up to 
transverse momenta of $7$--$8~\gev$ at \ac{RHIC} for \AuAu\ collisions at 
$\snn=62.4~\gev$.
The scaling with the initial gluon number-density gives 
\begin{equation}
n^{\rm gluons}_{\rm Au-Au,\,62.4\,GeV} \simeq 0.5\times 
n^{\rm gluons}_{\rm Au-Au,\,200\,GeV}\;,
\end{equation}
which leads to $\av{\hat{q}}^{\rm 62.4 GeV}\simeq 7~\gev^2/\fm$ 
and $k^{\rm 62.4 GeV}=2.5\cdot 10^6~\fm$. We use \ac{PQM} as before, 
but generate the parton $\pt$ with \acs{PYTHIA} at $\sqrt{s}=62.4~\gev$.

\begin{figure}[b!]
\begin{center}
\includegraphics[width=8cm]{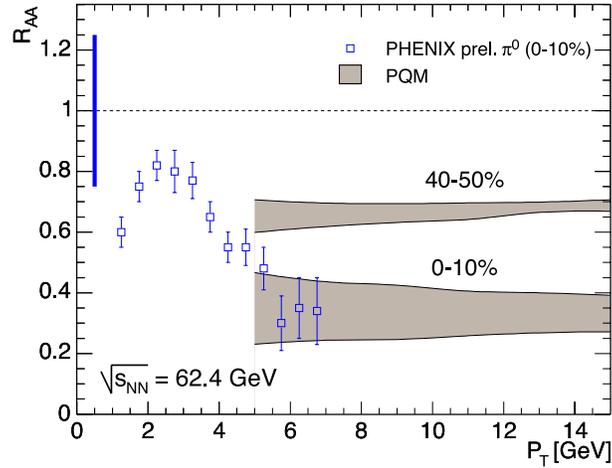}
\end{center}
\vspace{-0.3cm}
\caption[xxx]{\acs{PQM} results for $\RAA(\pt)$ in central and semi-peripheral 
\AuAu\ collisions at $\snn=62.4~\gev$. The preliminary $\pi^0$ data 
($0$--$10$\% centrality class) from \acs{PHENIX}~\cite{data62} are added for 
comparison. The \pp\ reference is the \acs{PHENIX} \mbox{${\rm pp}\to\pi^0+X$} 
parameterization, the error bars on the data points are the combined statistical 
and $\pt$-dependent systematic errors. The bar centered at \mbox{$\RAA=1$} 
gives the systematic error on the normalization.}
\label{chap3:fig:RAA62}
\end{figure}

The results are shown in \fig{chap3:fig:RAA62}, 
along with preliminary data from PHENIX~\cite{data62} for neutral pions 
up to $\pt\approx 7~\gev$ in $0$--$10$\% central collisions. 
For $\pt\gsim 5~\gev$, we find $\RAA\simeq 0.3$, in very good agreement 
with the data, in central ($0$--$10$\%) collisions. For semi-peripheral ($40$--$50$\%) 
collisions we predict a value of  $\simeq 0.7$.
\subsection{Parton emission from the surface}
\label{chap3:partonsfromsurface}
\ac{PQM} describes the centrality dependence of leading-hadron suppression 
(see \fig{chap3:fig:KAllCentralities}) and back-to-back di-hadron 
correlations (see~\Ref{dainese2004}) purely by the evolution with
the collision geometry.
This suggests that the high-opacity medium formed in \AuAu\ 
collisions at $\snn=200~\gev$ initially has a size and density, which 
decrease from central to peripheral events according to the 
the overlap profile of the colliding nuclei, $T_{\rm A}\,T_{\rm B}(x,y;\,b)$.
At the center of the medium the density is highest and partons crossing 
the central region are likely to be completely absorbed. Only partons produced in 
the vicinity of the surface and propagating outward can escape from the medium 
with sufficiently-high energy to fragment into hadrons with more than few 
$\gev$ in $\pt$. Such an ``emission from the surface'' scenario has also been pictured 
in~\Ref{drees2003}, where the centrality dependence of $\RAA$ and the back-to-back
correlation strength has been reproduced by a simple model of parton absorption 
whose only physical ingredient was a Glauber-based \NNex\ overlap profile. 

\begin{figure}[b!]
\begin{center}
\includegraphics[width=15cm]{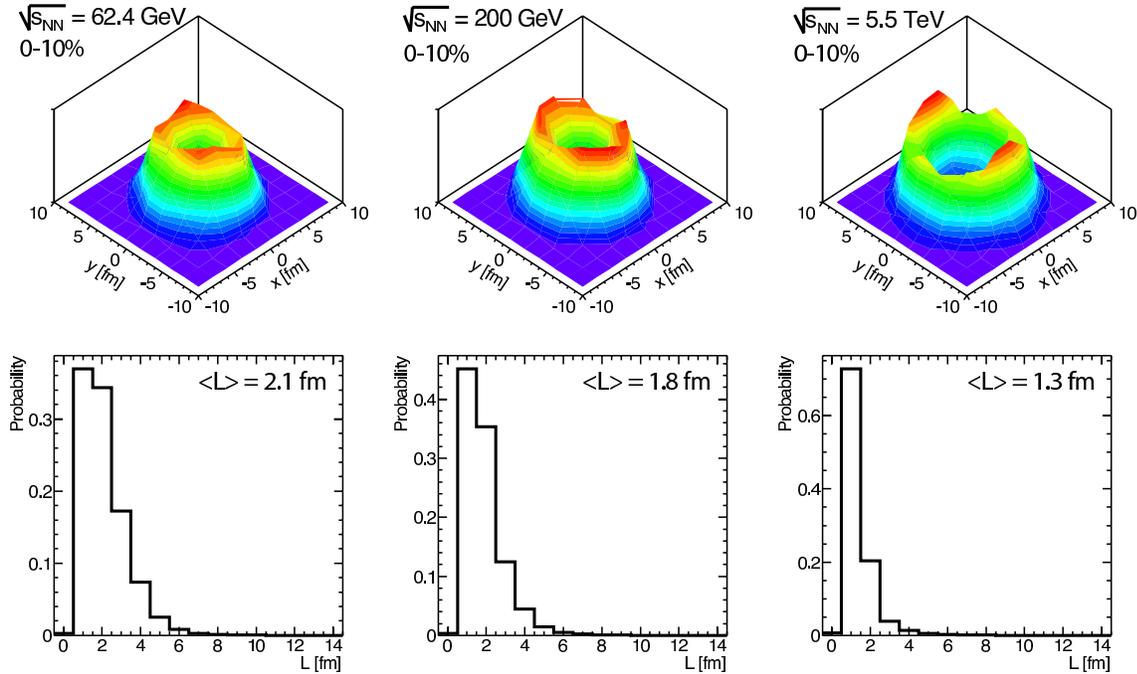}
\end{center}
\vspace{-0.3cm}
\caption[xxx]{Distributions of parton production points in the transverse plane (upper row)
and in-medium path length (lower row) for partons that escape the medium and produce 
hadrons with $\pt>5~\gev$ in central \AuAu\ collisions at 62.4 and 200~$\gev$ and in 
central \PbPb\ collisions at 5.5~$\tev$. The quantity $\av{L}$ denotes the average of the 
path-length distribution. All plots are in the non-reweighted case.} 
\label{chap3:fig:survivors}
\end{figure}

To quantify the effect of surface emission we visualize the region from which 
partons escape from the medium by plotting the distribution of production points 
($x_0$,~$y_0$) for partons fragmenting into high-energy hadrons ($\pt^{\rm hadron}>5~\gev$). 
The distributions for central \AuAu\ collisions at $62.4$ and $200~\gev$
and \PbPb\ collisions at $5.5~\tev$ are shown in \fig{chap3:fig:survivors}, 
along with the corresponding in-medium path-length distributions. The ``thickness'' 
of the surface is of order $2$--$3~\fm$ and it decreases as $\snn$ increases.~\footnote{Note
that for $0$--$10$\% a non-uniformity remains visable due to smaller losses in-plane
than out-of-plane.}  
The average depth decreases from $\av{L}=2.1~\fm$ at intermediate \acs{RHIC} energy,
to $1.8~\fm$ at $\snn=200~\gev$ and $1.3~\fm$ at \acs{LHC} energy. 
The reported values are for the non-reweighted case. It is interesting to note 
that in the reweighted case the `surface' region is much thicker; 
partons may escape almost from all depths and $\av{L}=3~\fm$ 
for all systems. This corresponds to our general statement~\cite{dainese2004}.
The reweighted approach generally simulates a softer transport coefficient,
which is not in accordance with the suppression of away-side correlations.

\begin{figure}[htb]
\begin{center}
\includegraphics[width=14cm]{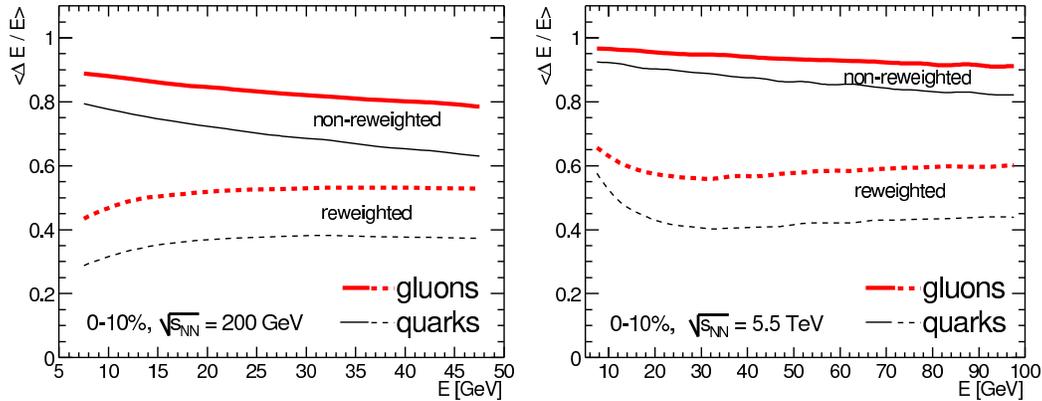}
\end{center}
\vspace{-0.3cm}
\caption[xxx]{Average relative energy loss versus parton energy for
quarks and gluons in central collisions at \acs{RHIC} (left) 
and \acs{LHC} (right) energies for the non-reweighted 
and reweighted cases.} 
\label{chap3:fig:avgreleloss}
\end{figure}

The strong parton absorption in central collisions at \ac{RHIC} 
suggests that the saturation regime of the energy loss, $\Delta E/E\to 1$, 
has been reached. Almost all hard partons produced in the inner core are 
thermalized ($\Delta E/E=1$) and, thus, cannot escape from the 
medium. 
Indeed, the average relative energy loss, $\av{\Delta E/E}$,
from the Monte Carlo calculation, plotted in \fig{chap3:fig:avgreleloss} 
as a function of the parton energy $E$ for central collisions at $\snn = 200$ 
and $5500~\gev$, is close to saturation. In the non-reweighted case it reaches 
$70$--$90$\% for gluons and $60$--$80$\% for quarks at \ac{RHIC} and is expected to 
be even higher at the \ac{LHC}.~\footnote{Note that the non-reweighted case for 
\ac{RHIC} and \ac{LHC} differ in the slope at low parton energy. This is
due to the different scale of the transport coefficients. For lower values
of the parton energy the result at \acs{RHIC} would turn around in slope and
approach the limit of $\Delta E/E=1$ for $E\rightarrow0$.} 
Due to the finite energy constraint and the fact that gluons are closer to energy-loss 
saturation than quarks, the ratio of gluon to quark $\av{\Delta E/E}$ is much smaller 
than the Casimir ratio $C_A/C_F=2.25$ expected from \eq{chap3:eq:avdE}. 
Furthermore, since absorption and, hence, saturation is more significant 
for low-energy partons, or, in other words, since high-energy partons can exploit 
larger energy losses, the genuine \ac{BDMPS-Z} $\Delta E/E\propto 1/E$ is replaced 
by a rather energy-independent effective $\Delta E/E$. 

\subsection{Relating the transport coefficient to energy density}
\label{chap3:pqmqtoen}
The static as well as the time-dependent transport coefficient
scale with the energy density of the medium like~\cite{baier2002}
\begin{equation}
\hat q(\tau)= c\, \epsilon^{3/4}(\tau)\;.
\label{chap3:eq:qepsilon}
\end{equation}
In this expression $c$ is a proportionality constant which is calculated for 
specific models of the medium. In particular, for ideal \acs{QGP} interacting 
perturbatively with the hard parton, the proportionality constant can be
extracted from \fig{chap3:fig:qtranspvseps} to the value of
\begin{equation*}
c_{\rm QGP}^{\rm ideal} \approx 2\;.
\label{chap3:eq:c}
\end{equation*}
The time-dependent coefficient for an expanding medium can be 
written as~\cite{baier1998b,salgado2003}
\begin{equation*}
\hat q(\tau) = \hat q_0\, \left(\tau_0/\tau \right)^\alpha\;.
\end{equation*}
The expansion parameter $\alpha$ is unity for a one-dimensional 
Bjorken expansion and is expected to stay close to unity for realistic 
expansion scenarios. From the measured time-averaged (equivalent static)
transport coefficient $\av{\hat q}$, we find via the dynamical 
scaling law \eq{chap3:eq:qscale} the transport coefficient for an 
initial time $\hat{q}_0=q(\tau_0)$
\begin{equation}
\label{chap3:eq:intitialqhat}
\hat{q}_0 = \hat{q} \, \frac{2-\alpha}{2}
\left( \frac{L}{\tau_0} \right)^{\alpha}\;,
\end{equation}
where we have assumed $\tau_0 \ll L$. Using this we can calculate the initial 
energy density, $\epsilon_0=\epsilon(\tau_0)$, according to \eq{chap3:eq:qepsilon}.
$L\equiv \tau$ denotes the time at which we extracted the transport coefficient 
given by the $\RAA$ data. Since the measurement is mainly determined by the partons
which escape from the medium, we take for $L$ the average value of the path-length 
distributions depicted in \fig{chap3:fig:survivors}.
Realistic expansion scenarios lie in the parameter range $0.75 < \alpha < 1.5$.
We chose one-dimensional Bjorken expansion ($\alpha=1$), for which we find 
the energy densities reported in \tab{chap3:tab:epsilonfromq}.

\begin{table}[htb]
\begin{center}
\begin{tabular}{l|ccccc}
\hline
\hline
System  & $\tau_0$ [$\fm$] & $L$ [$\fm$] & $\av{\hat{q}}$ [$\gev^2/\fm$] 
& $\hat{q}_0$ [$\gev^2/\fm$] & $\epsilon_0$ [$\gev/\fm^3$] \\
\hline
\AuAu\, 62.4 & 0.2 & 2.1 & \hide{00}7 & \hide{0}37 & \hide{00}715 \\
\AuAu\, 200  & 0.2 & 1.8 & \hide{0}14 & \hide{0}63 & \hide{0}1454 \\
\PbPb\, 5500 & 0.1 & 1.3 & 100        & 650        & 32670 \\ 
\hline
\hline
\end{tabular}
\end{center}
\vspace{-0.4cm}
\caption[xxx]{Initial energy density extracted from \eq{chap3:eq:qepsilon} and 
\eq{chap3:eq:intitialqhat} for $c_{\rm QGP}^{\rm ideal}=2$ and $\alpha=1$
(Bjorken expansion). The values for $L$ are taken from \fig{chap3:fig:survivors}.}
\label{chap3:tab:epsilonfromq}
\end{table}

The reported values yield to much larger initial densities compared to the values
given by the \acs{EKRT} model (see~\ptab{chap2:tab:spsrhiclhc}). Note that the 
calculation of $\hat{q}$ depends on $\as$, for which we chose a rather small value of 
$1/3$. However, using $\hat{q}=68~\gev^2/\fm$ (at $\as=0.5$) and an assumed 
energy density of $\epsilon=100~\gev/\fm^3$ the value of $c$ at the \ac{LHC} energy is 
found to be about $5$--$10$ times larger than perturbatively expected~\cite{eskola2004}.
This is a known and still outstanding problem: The hard probe seems to interact
much stronger as expected according to the perturbative estimate.
Although this perturbative estimate as well as the theoretical description of parton energy 
loss at finite (small) parton energies themselves bare significant uncertainties, the 
authors do not expect them to account for the large discrepancy alone.

\subsection{Limitations due to finite parton energies}
\label{chap3:pqmlimits}
The difference between the constrained distributions becomes quite large 
for low parton energies and sufficiently-large (effective) transport coefficients
and (effective) in-medium path lengths. It is controlled by the ratio of the 
maximum allowed energy loss, $\Delta E^{\rm max}=E$, to the characteristic 
emission frequency of the gluons, $\omega_{\rm c}$. Mathematically
$P^{\rm rw}(\Delta E,\,E)-P^{\rm non-rw}(\Delta E,\,E)$
is largely determined by $(1-\phi(E))/\phi(E)$, 
where $\phi(E)=\int_0^E\dd\epsilon\,P(\epsilon)$. 
We evaluate its dependence on $\omega_{\rm c}$ by defining
\begin{equation}
\label{chap3:eq:intweight}
\Phi(z) = \av{\int_0^z\,\dd\xi\,\omega_{\rm c} \, P(\xi\,\omega_{\rm c})}
\end{equation}
for $z=E/\omega_{\rm c}$. The brackets $\av{\cdot}$ in \eq{chap3:eq:intweight}
denote the average over the input parameters of the quenching weights, \ie~the
average over all parton paths and encountered local transport coefficients
determining $\omega_{\rm c}$ via \eq{chap3:eq:wcRfromI0I1}.

\begin{figure}[htb]
\begin{center}
\includegraphics[width=10cm]{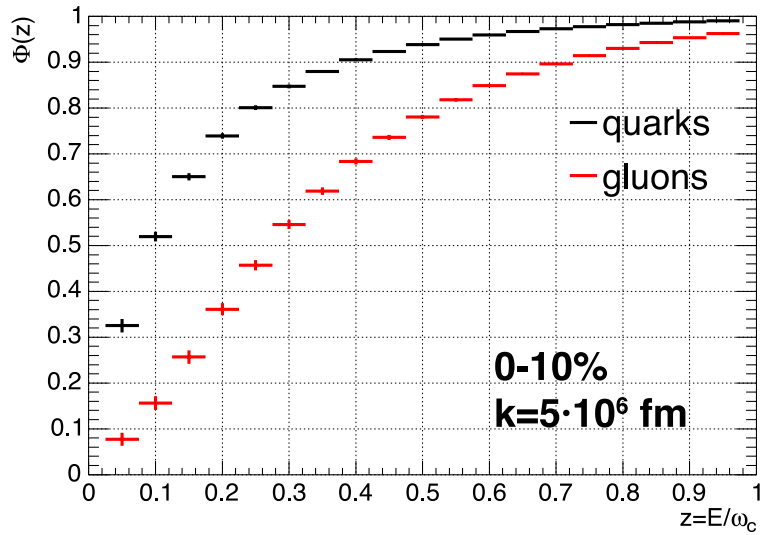}
\end{center}
\vspace{-0.3cm}
\caption[xxx]{The accumulated energy-loss probability distribution,
\eq{chap3:eq:intweight}, averaged over the input parameters of the
quenching weights, pairs of $\omega_{\rm c}$ and $R$.}
\label{chap3:fig:IntWeights}
\end{figure}

\begin{figure}[htb]
\begin{center}
\subfigure[Reweighted]{
\label{chap3:fig:IntWeightsUseda}
\includegraphics[width=7cm]{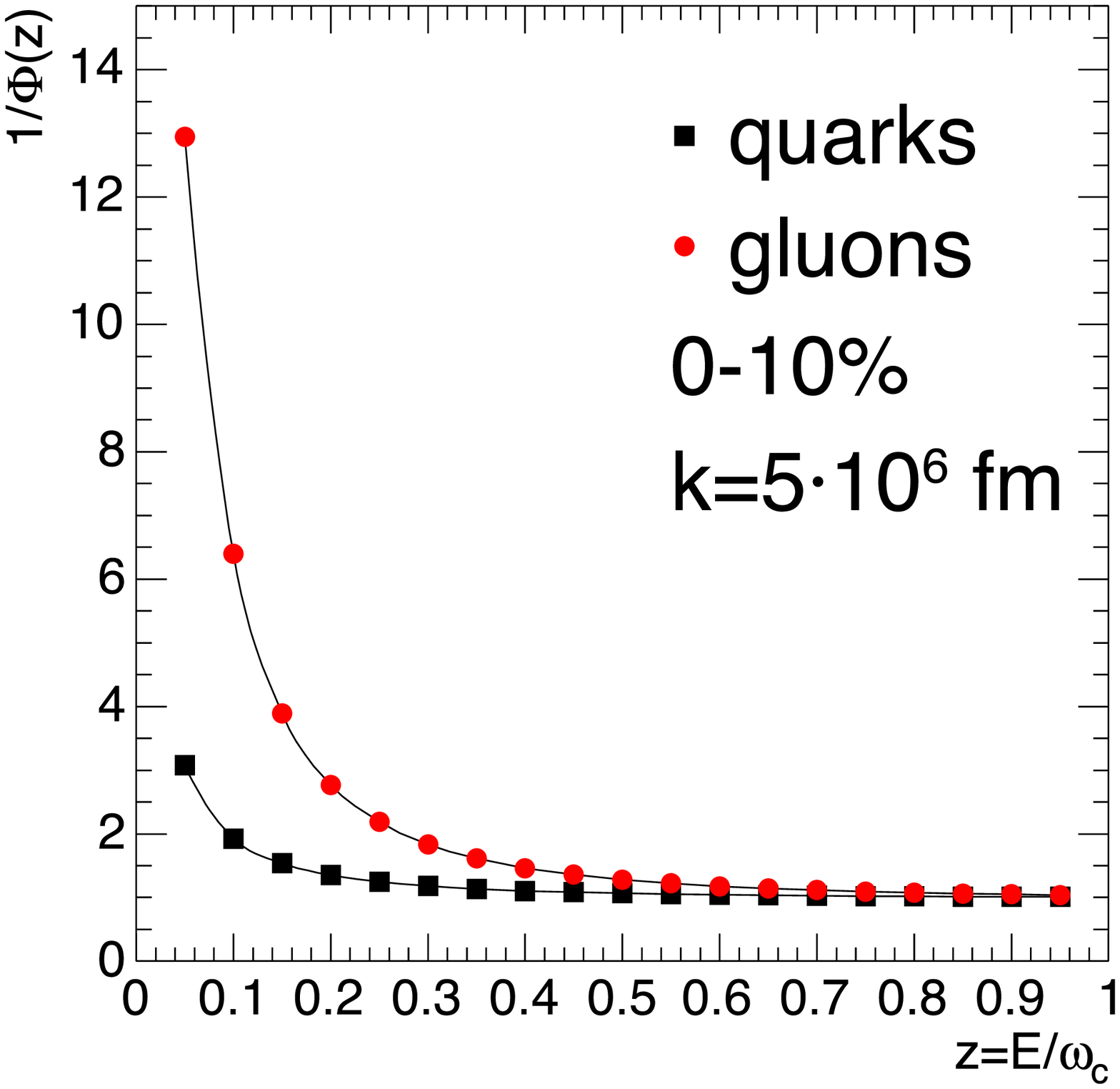}}
\hspace{0.5cm}
\subfigure[Non-reweighted]{
\label{chap3:fig:IntWeightsUsedb}
\includegraphics[width=7cm]{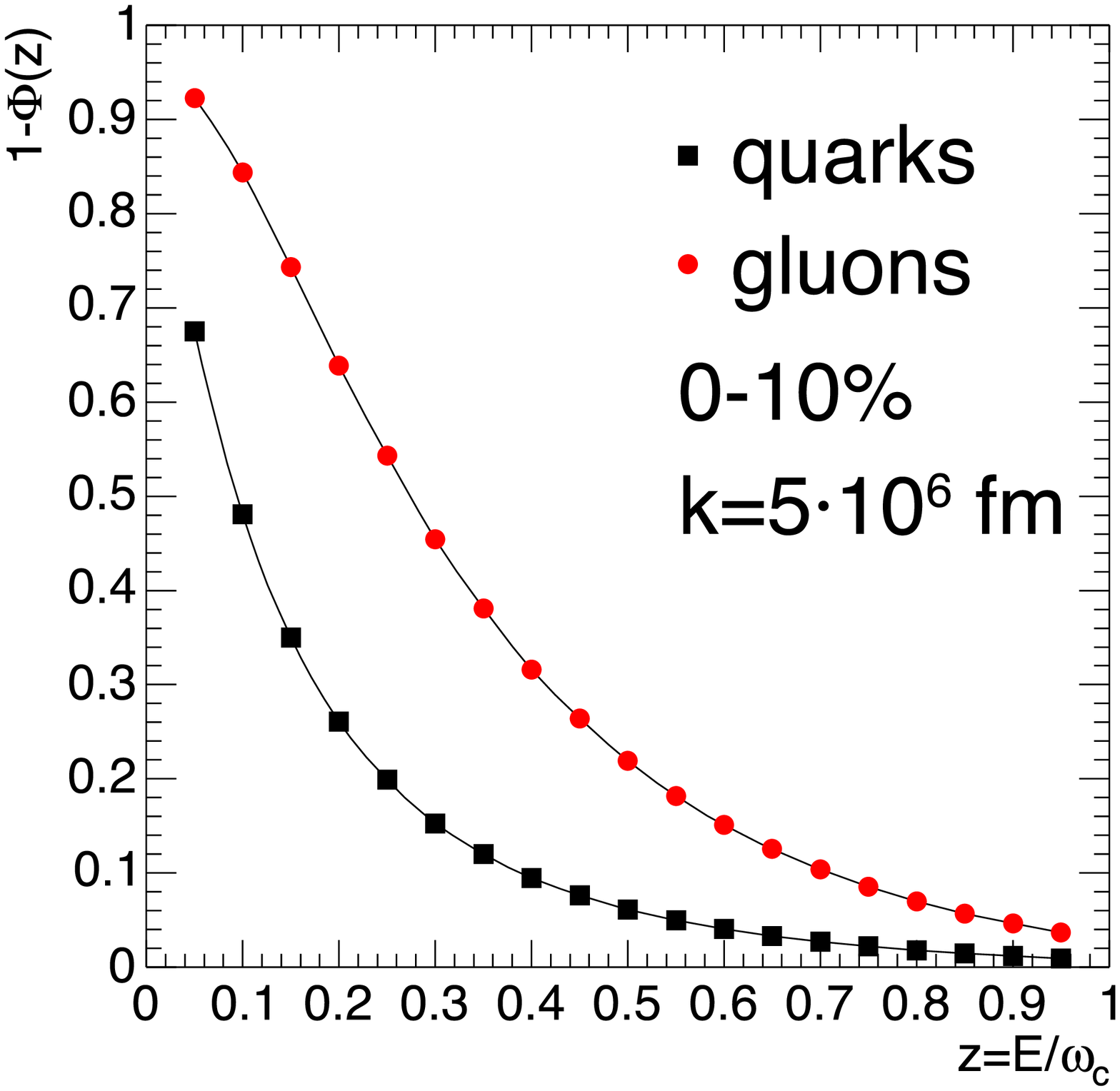}}
\end{center}
\vspace{-0.5cm}
\caption[xxx]{The deviation from the unconstrained quenching weights, 
$1/\Phi(z)$~\subref{chap3:fig:IntWeightsUseda} for reweighted and
$1-\Phi(z)$~\subref{chap3:fig:IntWeightsUsedb} for non-reweighted,
as a function of $z=E/\omega_{\rm c}$.}
\label{chap3:fig:IntWeightsUsed}
\end{figure}

The numerical result of $\Phi(z)$ as a function of $z$ as given 
by \eq{chap3:eq:intweight} is reported in \fig{chap3:fig:IntWeights}. 
The remarkable difference between quarks and gluons is due to the 
broader gluon-radiation distribution (see \eq{chap3:eq:wdIdw}) 
for a gluon projectile. 
Thus, for the same value of $\omega_{\rm c}$ the difference in the two approaches 
grows for smaller parton energies.~\footnote{The evaluation of the average in \eq{chap3:eq:intweight} 
has been obtained for the scale of $k=5\cdot10^{6}~\fm$ and the nuclear overlap function 
for \AuAu\ at $0$--$10$\%. In practice, this does not restrict the conclusion since 
empirically we find about the same curves starting from $k=1\cdot10^{6}~\fm$. Only if the scale
is reduced by one order of magnitude the picture changes. The difference on the nuclear
overlap for the change of the geometry to central \PbPb\ does not have an detectable influence.}

Furthermore, looking at \fig{chap3:fig:IntWeightsUsed} values of $\Phi(z)\ge 0.5$ are desirable, since 
$1/\phi(E)$ for the reweighted and $1-\phi(E)$ for the non-reweighted case control to which extent 
the quenching probabilities calculated in the eikonal approximation are modified by the finite energy 
constraint.  Thus, the parton energy should fulfill $E> 0.25\,\omega_{\rm c}$ for quarks and 
$E> 0.5\,\omega_{\rm c}$ for gluons, in order to keep the modification on the weights introduced 
by the constraints to the $20$\% level. For lower values of the initial parton energy relative to 
the characteristic frequency of the emitted gluons it might not longer be justified to consider
multiple successive and independent scatterings of the primary parton in the medium. A similar
remark can be found in~\Ref{arleo2002}.
 
The current difficulties of the theory to account for finite (small)
parton energies become obvious if we compute the distribution
of $\omega_{\rm c}$ for central \AuAu\ collisions shown in
\fig{chap3:fig:omegadist}. The mean value of the characteristic
emission frequency is very high, $\av{\omega_{\rm c}}=870~\gev$
(with large rms of $921~\gev$). The scale is huge compared to 
the parton energies encountered (surely below $100~\gev$). Quantitatively
we can estimate the effect on the constrained weights computing the 
distribution of $z=E/\omega_{c}$  shown in \fig{chap3:fig:omegaeratio}. 
The mean value is $\av{z}=0.05$ (with rms of $0.24$), which is below the
fraction we quoted above. Thus, for the scale $k=5\cdot10^{6}~\fm$, 
which fits the nuclear modification factor at \ac{RHIC} (and further observables 
as mentioned), we find in view of \fig{chap3:fig:IntWeightsUsed} that basically
every $z$-value used is from the $z$-region where the effect of the finite parton 
energies via the constrained weights is significant:
about factor of 2 (quarks) to 6 (gluons) for the reweighted, which is alarmingly high, 
and about $50$\% (quarks) to $90$\% (gluons) for the non-reweighted case (taking the 
value at $z=0.1$).

\begin{figure}[htb]
\begin{center}
\subfigure[Distribution of $\omega_{\rm c}$]{
\label{chap3:fig:omegadist}
\ifarxiv
\includegraphics[width=6cm]{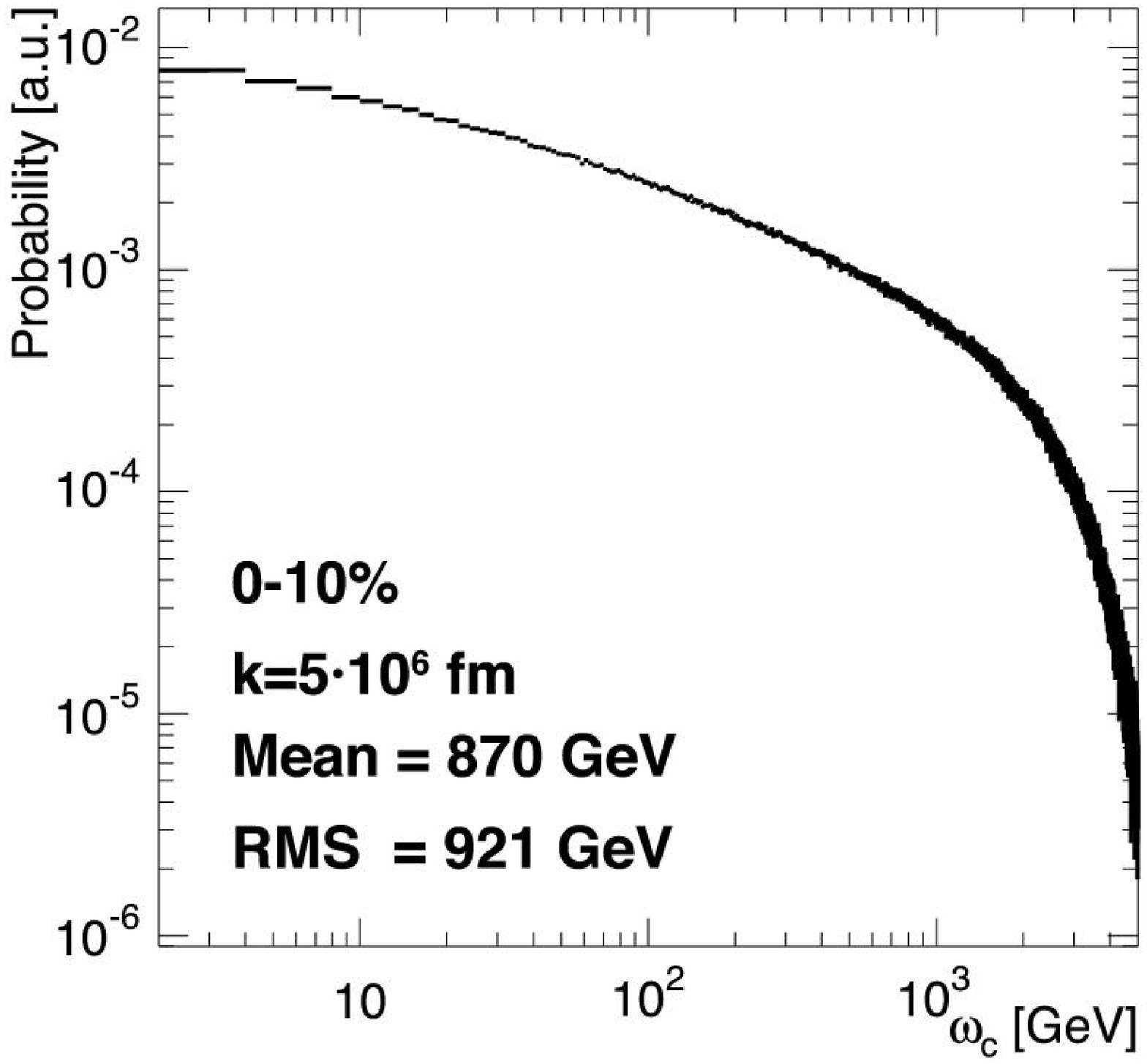}
\else
\includegraphics[width=6.6cm]{cOmegaDist}
\fi
}
\hspace{0.5cm}
\subfigure[Distribution of $E/\omega_{\rm c}$]{
\label{chap3:fig:omegaeratio}
\includegraphics[width=6.6cm]{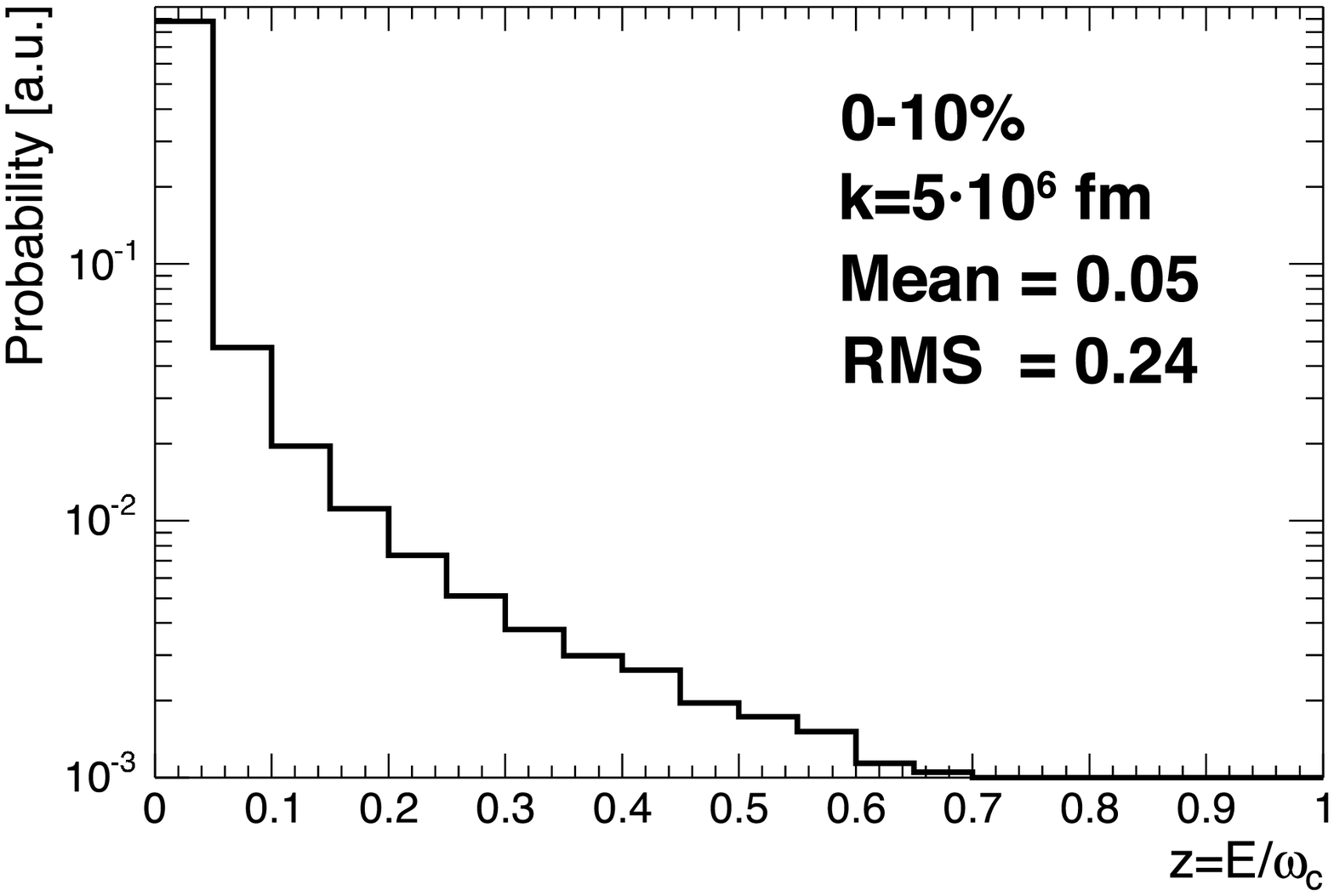}}
\end{center}
\vspace{-0.5cm}
\caption[xxx]{The distribution of $\omega_{\rm c}$~\subref{chap3:fig:omegadist} 
and $z=E/\omega_{\rm c}$~\subref{chap3:fig:omegaeratio} evaluated with
\acs{PQM} for central \AuAu\ collisions at $\snn=200~\gev$. The calculation
for $k=5\cdot 10^6~\fm$ results in $\av{\omega_{\rm c}}=870~\gev$ 
and $\av{z}=0.05$.}
\label{chap3:fig:omegarhic}
\end{figure}

\begin{figure}[htb!]
\vspace{0.5cm}
\begin{center}
\subfigure[Non-reweighted energy-loss distribution]{
\label{chap3:fig:elosskrhicnonrw}
\includegraphics[width=7cm]{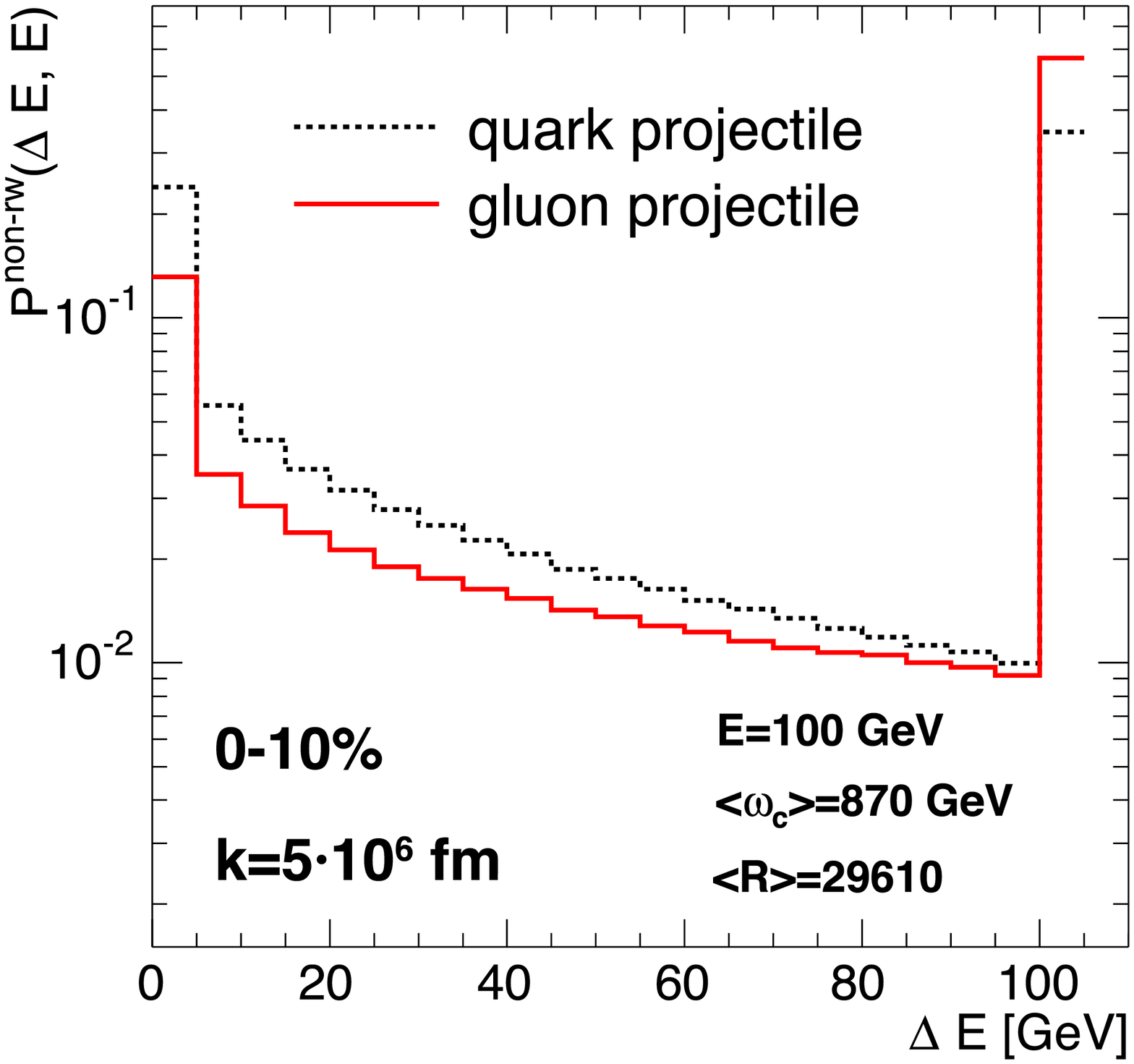}}
\hspace{0.5cm}
\subfigure[Reweighted energy loss distribution]{
\label{chap3:fig:elosskrhicrw}
\includegraphics[width=7cm]{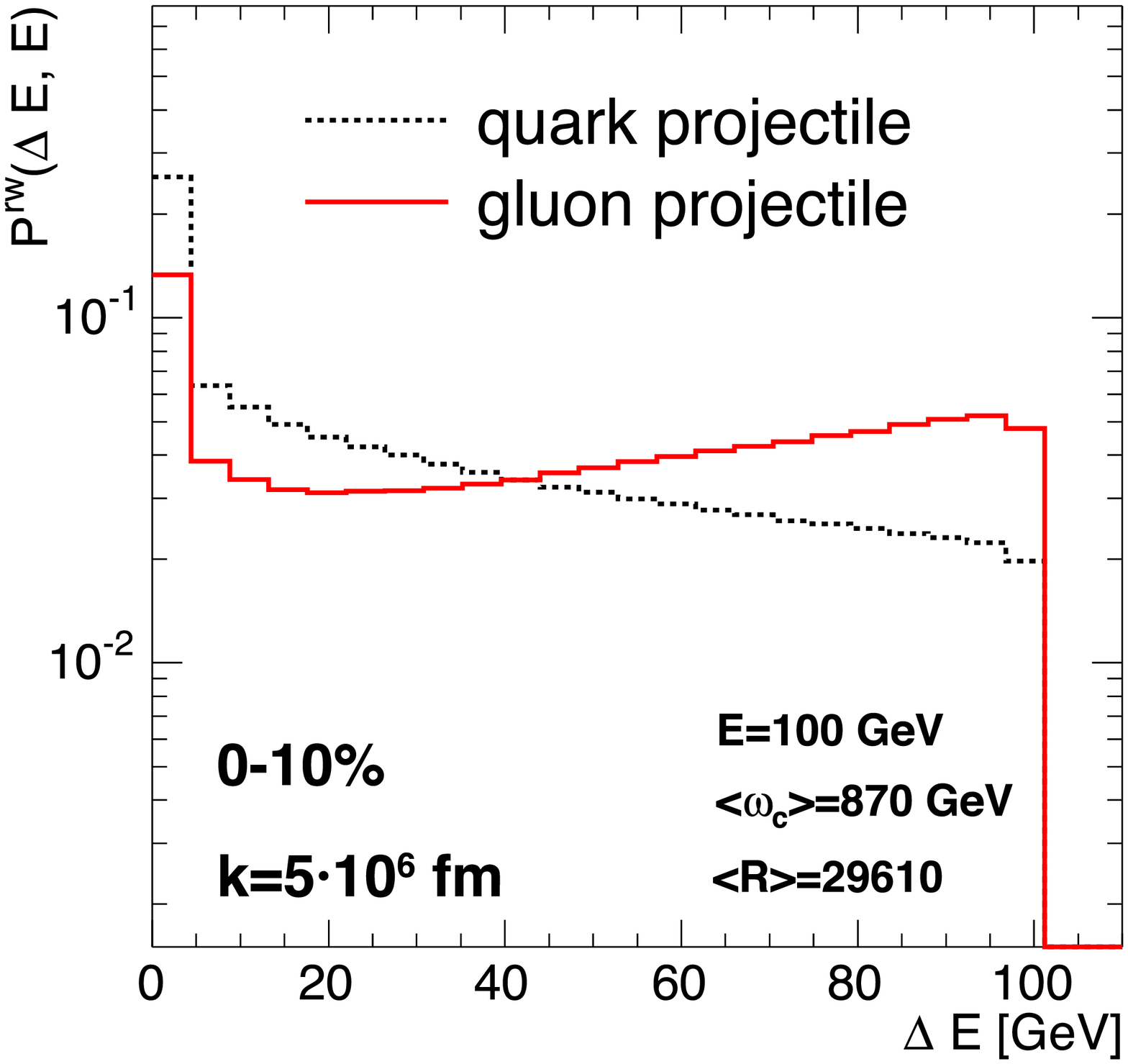}}
\end{center}
\vspace{-0.5cm}
\caption[Energy-loss distribution for central RHIC]
{Energy loss distribution $P^{\rm non-rw}(\Delta E,\,E)$~\subref{chap3:fig:elosskrhicnonrw} 
and $P^{\rm rw}(\Delta E,\,E)$~\subref{chap3:fig:elosskrhicrw} for $E=100~\gev$ partons
in central \AuAu\ collisions at $\snn=200~\gev$. The calculation for $k=5\cdot 10^6~\fm$ 
results in $\av{\omega_{\rm c}}=870~\gev$ and $\av{R}\approx 30000$. See the distribution
for lower (fixed) scale in \fig{chap3:fig:elossfixed}.}
\label{chap3:fig:elosskrhic}
\end{figure}

The effect of the high-$\omega_{\rm c}$, \ie~high-$k$ scale, 
can also be manifested in the energy-loss distribution for fixed parton energies. 
In \fig{chap3:fig:elosskrhic} it is plotted for $100~\gev$ partons. 
The average energy loss is $50~\gev$ for quarks and $70~\gev$ for gluons 
in the non-reweighted; in the reweighted case it is $33~\gev$ for quarks and $50~\gev$ for gluons. 
This is about $2$--$3$ times larger compared to the distribution obtained for a fixed length 
(shown in \fig{chap3:fig:elossfixed}). 
However, what concerns here, is the change of slope in the reweighted case, apparent
for gluons relative to quarks in \fig{chap3:fig:elosskrhicrw}. The non-reweighted distribution 
has a peak at zero and at maximal energy loss, as expected by the way it is constructed. Similarly
the reweighted distribution has the peak at zero, and for quarks does not possess the peak at 
maximum possible energy loss, whereas unexpectedly for gluons it does. Due to the high scale needed to describe 
the data at \acs{RHIC} the quenching weights are truncated, \peq{chap3:eq:prw}, in the rising part of 
the distribution and then reweighted, thus  amplifying the rise. Since  the (unconstrained) 
energy-loss probability distribution for the gluons (shown in \fig{chap3:fig:avgeloss}) is much broader 
for gluons than for quarks, this occurs for gluons at a lower value of the scale than for quarks.

We, therefore, come to the conclusion to abandon the reweighted approach from further discussion,
since at the extrapolated medium-density for \acs{RHIC} (and even more so for \acs{LHC}) it
introduces substantial deviation from the quenching weights leading to unphysical properties. 
In addition, it is not in accordance with the observed away-side suppression measured at 
\ac{RHIC}.~\footnote{See also the discussion in~\Ref{dainese2004}.}

\subsection{Limitations of leading-hadron spectroscopy}
\label{chap3:leadhadlimits}
The expected $\RAA$ calculated with \acs{PQM} in \sect{chap3:extrapollhc} for central 
\PbPb\ collisions at \ac{LHC} as a function of $\pt$ is rather flat, \ie~almost 
$\pt$-independent. As mentioned, this is in contrast to the estimation by Vitev and 
Gyulassy~\cite{vitev2002b}. The difference between the two predictions cannot to be 
attributed to nuclear (anti-)shadowing effects since they are included in a similar 
calculation~\cite{eskola2004} coming to the same observation.

In \fig{chap3:fig:raalhcomp} we present the $\RAA$ as a function of $\pt$ obtained with \ac{PQM} 
in central \PbPb\ at \ac{LHC} for different settings in the non-reweighted case. Shown are the 
calculations in the parton-by-parton approach for $k=5\cdot10^6~\fm$ (the value found at \ac{RHIC}),
$k=1\cdot10^7~\fm$ and $k=4\cdot10^7~\fm$ (the value obtained by the scaling from \ac{RHIC} 
to \ac{LHC}), as well as the result of a calculation with fixed $\hat{q}=10~\gev^2/\fm$ and 
fixed length of $4.4~\fm$ (the mean of the path-length distribution obtained with Glauber).
Clearly, the latter shows a stronger $\pt$-dependence than the other cases. 

The flatness obtained in the parton-by-parton approach is explained by
\begin{itemize}
\item Steeply falling production cross-section, $\propto \left(\frac{1}{p^{\rm hadron}_{\rm T}}
\right)^{n(\rm \pt)}$, where $n(\pt)$ is rising from about $7$ to $12$ (\ac{RHIC}) and from $6$ 
to $7$ (\ac{LHC}) in the relevant $\pt$ regime; 
\item emission from the surface, which for large medium densities dominates~\cite{mueller2002}.
\end{itemize}

The computation of $\RAA$, \eq{chap3:eq:rab}, at mid-rapidity, can be approximated 
by \cite{baier2001}
\begin{equation}
\label{chap3:eq:raaapprox}
\RAA(\pt) = \left.\int \dd\Delta E P(\Delta E,\, \pt+\Delta E) \, \frac{\dd N^{\rm pp}
(\pt+\Delta E)}{\dd\pt} \; \right/ \; \frac{\dd N^{\rm pp}(\pt)}{\dd\pt}\;,
\end{equation}
where, $\dd N^{\rm pp}/\dd\pt$ is the spectrum of hadrons (or partons) in the case of no medium 
(\ie~\pp\ neglecting initial state effects). The suppression computed with \eq{chap3:eq:raaapprox} 
is found to give a rather good approximation to the one computed with \ac{PQM} or with the full 
formula, \eq{chap3:eq:hcrossec}. 
In the case the production spectrum is (approximately) exponential the $\pt$-dependence cancels 
in the ratio and  we find $\RAA$ to be (approximately) independent of $\pt$. At \ac{RHIC} this is 
the case at about $\pt\ge 30~\gev$. Below that value and at the measured values of
$5$--$12~\gev$, as well as at the \acs{LHC} ($n(\pt)\le7$) the spectrum is given by a power-law
and it is expected~\cite{baier2001} that $\RAA\propto(1+c\,/\sqrt{n(\pt)\,\pt})^{-n(\pt)}$, 
\ie~reaching unity in the limit of large $\pt$. 

\begin{figure}[htb]
\begin{center}
\includegraphics[width=12cm]{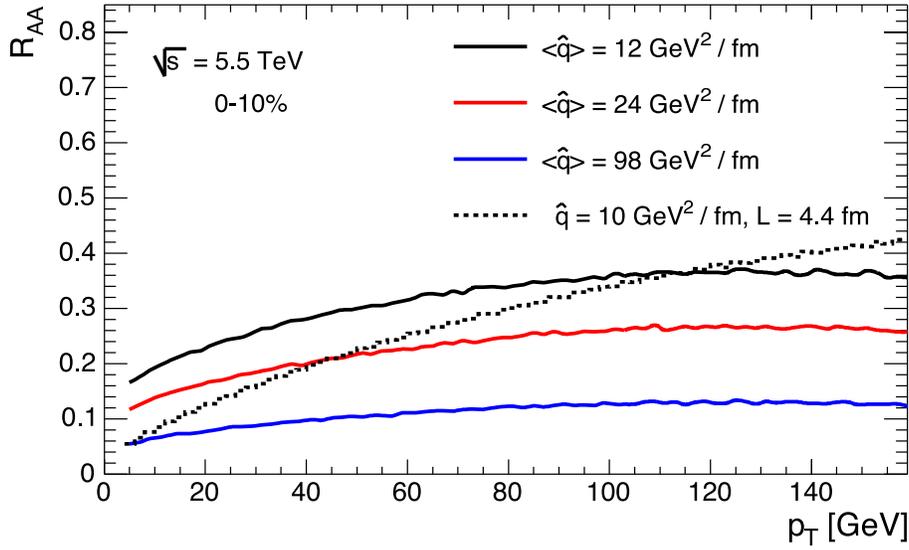}
\end{center}
\vspace{-0.3cm}
\caption[xxx]{$\RAA$ as a function of $\pt$ for $0$--$10$\% most central collisions at \acs{LHC} 
energy obtained by \acs{PQM}. The calculations in the parton-by-parton approach (solid lines) 
are compared to a calculation for fixed transport coefficient and length (dashed). 
All graphs are in the non-reweighted case.} 
\label{chap3:fig:raalhcomp}
\end{figure}

However, this neglects the fact that for dense media surface emission or, more generally, the 
probability to have no energy-loss, $P(\Delta E=0,\, E)$, plays a significant role, an effect 
which is even more pronounced at low $\pt$ (compared to $\omega_{\rm c}$).
To simplify our argumentation we allow either no loss ($\Delta E=0$) or complete loss ($\Delta E=E$)
in the non-reweighted case, $P(\Delta E,\, E)= p_0 \, \delta(\Delta E) + (1-p_0) \, \delta(\Delta E 
- E)$.~\footnote{For a dense medium the constrained weights at low parton energy indeed do 
have a sharp peak at zero and at maximum possible energy loss, whereas the values in between 
are negligible.}$^{\rm ,}$\footnote{A similar ansatz has been made in~\Ref{dainese2004}, 
but there it has been exactly computed in \acs{PQM} based on Monte Carlo and proven to 
describe the $\RAA$ and $I_{\rm AA}$ at \ac{RHIC}.}
Inserting the constrained weight into \eq{chap3:eq:raaapprox} we obtain
\begin{equation}
\label{chap3:eq:raaapproxtoymod}
\RAA(\pt) = p^{*} + (1-p^{*})\, \left. \frac{\dd N^{\rm pp}(2\,\pt)}{\dd\pt} \,
\right/ \, \frac{\dd N^{\rm pp}(\pt)}{\dd\pt}\;.
\end{equation}

\begin{figure}[htb]
\begin{center}
\includegraphics[width=10cm]{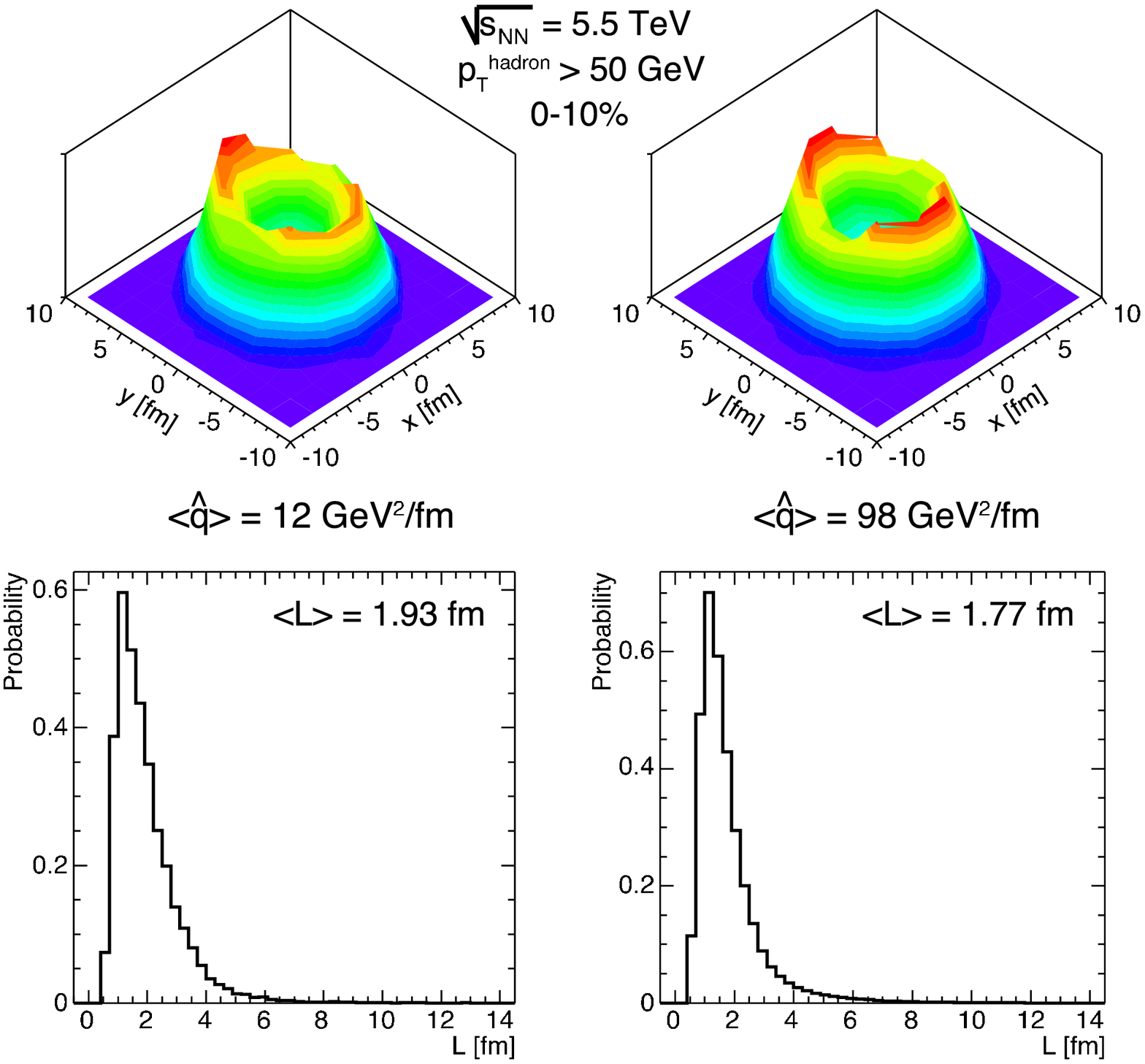}
\end{center}
\vspace{-0.3cm}
\caption[xxx]{Distributions of parton production points in the transverse plane (upper row)
and in-medium path length (lower row) for partons that escape the medium and produce 
hadrons with $\pt>50~\gev$ in central \PbPb\ collisions at 5.5~$\tev$ for
$k=5\cdot10^{6}~\fm$ (left) and $k=4\cdot10^{7}~\fm$ (right). The quantity $\av{L}$ denotes 
the average of the path-length distribution. All plots are in the non-reweighted case.} 
\label{chap3:fig:survivors50}
\end{figure}

\begin{figure}[htb]
\begin{center}
\includegraphics[width=10cm]{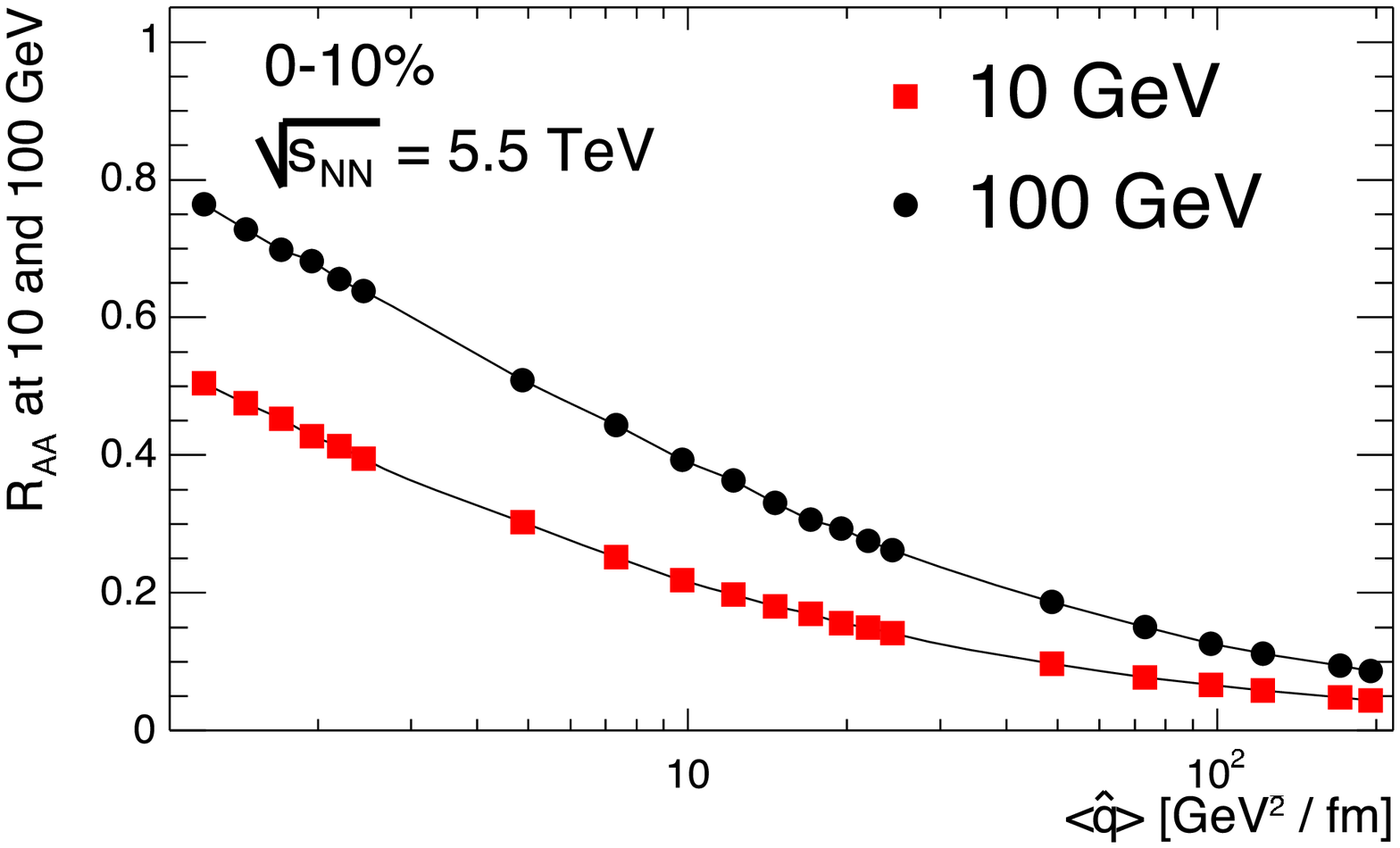}
\end{center}
\vspace{-0.3cm}
\caption[xxx]{$\RAA$ as a function of $\av{\hat{q}}$ for $10$ and $100~\gev$ hadrons
in $0$--$10$\% central \PbPb\ collisions at $\snn=5.5~\tev$. The calculation is
done with \acs{PQM} in the parton-by-parton approach for the non-reweighted case.}
\label{chap3:fig:raavsq}
\end{figure}

It is obvious that \eq{chap3:eq:raaapproxtoymod} is just a crude approximation, but it demonstrates
that the value of $\RAA$ is dominated by the fraction of hadrons (or partons), which 
escape without losing much of their energy. For the simple power-law production spectrum the 
contribution from higher $\pt$ is suppressed by about $(1+\Delta E/\pt)^{n(\pt)}$.
Taking into account only fixed values of $\hat{q}$ and $L$ the probability $p^{*}$ is given 
by the discrete weight, $p_0$, at $R=0.5\,\hat{q}\,L^3$. Note that for a proper calculation 
one must take into account the right production ratio of quarks-to-gluons.~\footnote{At 
$\sqrt{s}=200~\gev$, with \acs{CTEQ}~4L \acp{PDF}, gluons dominate the parton $\pt$-distribution 
up to about $20~\gev$. However, since quarks fragment harder than gluons, high-$\pt$ hadrons are 
mostly produced from quark fragmentation. Using \acs{KKP} \acp{FF} 75\% of the 
hadrons with $\pt>5~\gev$ come from quark fragmentation and 25\% from gluon fragmentation.}
For realistic path-length distributions $p^{*}$ is dominated by partons, which are emitted
close to the surface and, thus, enhanced relative to $p_0$ obtained at fixed scale.
It turns out that $p^{*}$ evaluated at \ac{LHC} central conditions, on average over path-lengths
and parton types, is independent of $\pt$ in the range shown above, and takes values of about 
$0.14$, $0.1$ and $0.05$ for the scales used in \fig{chap3:fig:raalhcomp}. 

To visualize the surface-emission scenario for high-$\pt$ hadrons at \ac{LHC} we show
in \fig{chap3:fig:survivors50} (top) the region from which partons escape from the medium by 
plotting the distribution of production points ($x_0$,~$y_0$) for partons fragmenting into 
high-energy hadrons with $\pt^{\rm hadron}>50~\gev$ together with the corresponding 
in-medium path-length distributions (bottom). The chosen values of the scale correspond to 
the delimiting cases shown in \fig{chap3:fig:raalhcomp} of low and high transport coefficient. 
Compared to \fig{chap3:fig:survivors} (right) for $\pt^{\rm hadron}>5~\gev$ the 
``thickness'' of the emission surface is larger (about 50\%) reaching almost $2~\fm$.
However, compared to the size of the overlap region, even at \ac{LHC} hadrons 
with $\pt>50~\gev$ are emitted dominantly from the surface.

\enlargethispage{-0.5cm}
To the present knowledge of the theory (\acs{BDMPS-Z-SW}) the dominance of the surface effect 
limits the sensitivity of the $\RAA$ to the density of the medium, mainly for experimentally 
accessible low-$\pt$ range at \ac{RHIC}. This is demonstrated in \fig{chap3:fig:raavsq} where 
we show the dependence of $\RAA$ as a function of the average transport coefficient evaluated 
with \ac{PQM} in the parton-by-parton approach for $0$--$10$\% most central collisions at 
$\snn=5.5~\tev$. For $10~\gev$ hadrons the nuclear modification factor is sensitive to average
medium densities up to about $15~\gev^2/\fm$, but loses its sensitivity for higher values of 
$\av{\hat{q}}$. For  $100~\gev$ hadrons the sensitive regime might widen to average values of
about $50$.~\footnote{Though, when looking at \fig{chap3:fig:raavsq} one should keep in mind 
that systematic errors influence the theoretical determination of $\RAA$ above
densities of $10~\gev^2/\fm$ to about $\pm0.05$ for $100~\gev$ hadrons.}

In most, if not all, cases high-$\pt$ hadrons are leading particles of high-energy jets
carrying, on average, 1/3 of the jet energy (\sect{chap3:frag}).  Therefore, the natural 
extension to leading-hadron spectroscopy is to investigate in-medium modification of well-known
jet properties and hadro-production of particles, which are associated with high-$\pt$ trigger 
(or leading) particles, reported in \chap{chap6}.
\fi

%

\newif\ifdetlayout
\detlayouttrue
\newif\ifdatarates
\dataratestrue
\newif\ifhlt
\hlttrue
\newif\iftracking

\chapter{ALICE experimental capabilities}
\label{chap4}
\acf{ALICE} is a general-purpose experiment whose detectors 
measure and identify hadrons, electrons, photons, and muons 
at the \ac{LHC}. The \ac{ALICE} detectors are optimized for 
the study of heavy-ion collisions up to the highest energy available. 
As such, the detector system has been designed to be capable of measuring 
properties of the bulk (soft hadronic, large cross section, physics) 
and of rare probes (hard, small cross section, physics). In particular,
\ac{ALICE} has to be able to track and identify particles from 
very low, $\sim 100~\mev$, up to fairly high, $\sim 100~\gev$, 
transverse momenta  
in an environment of extreme particle density.~\footnote{In correspondence
with the expectations, \psect{chap2:partmult}, the \ac{ALICE} detectors 
are designed to cope with multiplicities up to 8000 charged 
particles per pseudo-rapidity unit.}

\section{Layout of the detector system}
\label{chap4:detlayout}
\ifdetlayout
The layout of the \ac{ALICE} detector ---as proposed initially 
together with the physics objectives--- is described in the \ac{ALICE} 
technical proposal~\cite{alicetp1} and in two subsequent
addenda~\cite{alicetp2,alicetp3}. 
The individual detector or sub-detector systems are described 
in detail in technical design reports and addenda~\cite{aliceits,alicetpc,
alicetrd,alicetof1,alicetof2,alicehmpid,alicephos,alicemuon1,alicemuon2,
alicefmd,alicepmd1,alicepmd2,alicezdc}. 
The addenda reflect modifications to the original design considerations 
to meet new experimental objectives given by the recent 
results from \ac{RHIC} and latest theoretical developments. 
A summary of the present status of the \ac{ALICE} detectors can 
be found in~\Ref{pprvol1}.

As shown in \fig{chap4:fig:alicedet}, the \ac{ALICE} setup
consists of three major parts:
\begin{itemize}
\item The central barrel contained inside the magnet with an acceptance 
in pseudo-rapidity of  $-0.9 \leq \eta \leq 0.9$ over the full azimuth angle; 
\item the muon spectrometer at the pseudo-rapidity interval
$-4.0\leq\eta\leq-2.4$; 
\item various multiplicity detectors at $-3.4 \leq \eta \leq 5.1$. 
\end{itemize}

\begin{figure}[htb]
\begin{center}
\ifarxiv
\includegraphics[width=15cm]{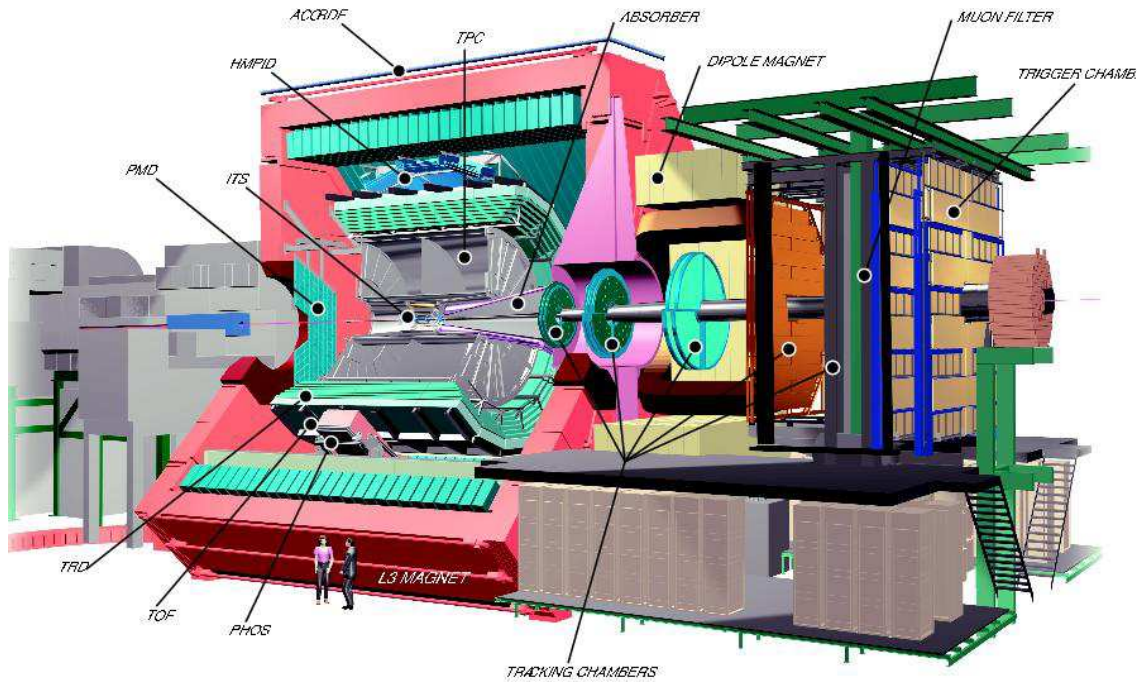}
\else
\includegraphics[width=15cm]{ALICE_layout}
\fi
\end{center}
\vspace{-0.3cm}
\caption[xxx]{The \ac{ALICE} experiment in its final layout. For the sake of 
visibility the \ac{HMPID} detector is shown in the 12 o'clock position 
instead of the 2 o'clock position at which it is actually located.}
\label{chap4:fig:alicedet}
\end{figure}

Hadrons, electrons and photons are detected and identified by a complex 
system of detectors placed in a homogeneous magnetic field of $0.5$~T
surrounding the central barrel at $-0.9 \leq \eta \leq 0.9$.
Charged particle tracking relies on a set of high-granularity detectors: 
an \ac{ITS} consisting of six layers of silicon detectors, 
a large-volume \ac{TPC}, and a high-granularity \ac{TRD}. Particle
identification is performed by measuring the energy loss via
ionization (d$E$/d$x$) in the tracking detectors, the transition 
radiation  in the \ac{TRD}  and the time of flight with a high-resolution 
\ac{TOF} detector. Two smaller single-arm detectors complete the particle 
identification at mid-rapidity via the detection of Cherenkov radiation 
with an \ac{HMPID}, and photons in the \ac{PHOS} using an electromagnetic 
calorimeter based on scintillating crystals. 

The detection and identification of muons or muon pairs from the
decay of heavy quarkonia at  $-4.0 \leq \eta\leq -2.4$ is performed 
with a dedicated spectrometer including a warm dipole magnet.

Last but not least, additional detectors located at large rapidities 
are used to characterize the event and to provide fast interaction 
triggers. Multiplicity detection by the \acs{FMD}, \acs{V0} and 
\acs{T0} detectors covers a wide acceptance of $-3.4 \leq \eta \leq 5.1$ 
for the measurement of charged particles and triggering, supported
by the \ac{PMD} at $2.3 \leq \eta \leq 3.5$ for photon multiplicity
measurement. Finally the \acp{ZDC} counts the spectator nucleons in
heavy-ion collisions close to the beam rapidity.

For convenience, we will give a brief description of the some of the 
\ac{ALICE} detectors in the following sections; for details we refer 
to the respective technical design reports. 

\subsection{Inner Tracking System}
\label{chap4:its}
The \acf{ITS}~\cite{aliceits} is designed and optimized 
for the reconstruction of secondary vertices from hyperon 
and open-charm (or open-beauty) meson decays, and, thus, 
precise tracking and identification of low-\pt\ particles. 
The detector consists of six cylindrical layers of high 
resolution silicon detectors, located at the innermost radius 
of $4~\cm$ to the outermost radius of $44~\cm$. 
It covers the rapidity range of $|\eta|<0.9$ for all vertices located 
within the length of the interaction diamond ($\pm 1\sigma$), 
\ie~ about $10.6~\cm$ along the beam direction. 

\pagebreak
To cope with the high particle density, up to 80 particles per
$\cm^2$, and to achieve the required impact-parameter resolution
of $100~\mu$, pixel detectors have been chosen for the 
innermost two layers, and silicon drift detectors for the following 
two layers. The outer two layers, where the track densities are 
below one particle per $\cm^2$, are equipped with double-sided 
silicon micro-strip detectors. 
The two layers of pixel detectors have about 10$^7$ channels with
one-bit information (signal above threshold). 
The silicon drift detectors contain about $1.4\cdot10^5$ channels 
of 256-deep arrays of digitized 10-bit amplitudes, which are 
compressed in the front-end electronics. Finally, the strip detectors 
have $2.6\cdot10^6$ channels of digitized amplitude information. 

\subsection{Time Projection Chamber}
\label{chap4:tpc}
The \acf{TPC}~\cite{alicetpc} is the main tracking device of the 
central barrel. Together with the other barrel detectors, its task 
is to provide charged-particle momentum measurements and particle 
identification via energy loss measurements (d$E$/d$x$). 
The \ac{TPC} acceptance covers the pseudo-rapidity region of 
$|\eta| < 0.9$; and up to $\abs{\eta} \sim 1.5$ for tracks with 
reduced track length and momentum resolution. In $\pt$ up to $100~\gev$ 
can be measured with good momentum resolution. 

The \ac{TPC} is cylindrical in shape and has an inner radius of 
about $85~\cm$, which is given by the maximum acceptable hit density, 
and an outer radius of $250~\cm$ defined by the length required 
for a d$E$/d$x$ resolution of $<$10\%. Its overall length
along the beam direction is $500~\cm$.
The detector is made of a large field cage, which is divided by a
a thin high voltage electrode in the center, providing an uniform 
electric drift towards the end-caps. 
It is filled with $88~{\rm m}^3$ of Ne/CO$_{2}$ (90\%/10\%), which
is needed for the transport of the primary electrons over $2.5~\m$ 
on either side of the central electrode to the end-plates.~\footnote{The 
choice of the gas mixture is still under discussion~\cite{veenhof2003}, 
and may also include N$_2$.} 
The drift gas is optimized for drift speed, low diffusion, low
radiation length and hence multiple scattering, small space-charge
effect and aging properties.

Multi-wire proportional chambers with cathode pad readout are mounted 
into 18 trapezoidal sectors of each end-plate, leading to
$2\times18$ trapezoidal sectors, each covering $20^{\circ}$ in azimuth. 
Due to the  radial dependence of the track density, the readout is 
segmented radially into two type of readout chambers with slightly 
different wire geometry adapted to the varying pad sizes mentioned below.
The inactive areas between neighboring inner chambers are aligned with 
those between neighboring outer chambers, optimizing the momentum
precision for detected high-momentum tracks, but creating cracks 
in the acceptance, as in about 10\% of the azimuthal angle the detector 
is not sensitive. The readout chambers are made of standard wire planes, 
\ie~they consist of a grid of anode wires above the pad plane, a cathode 
wire plane, and a gating grid. Each sector is divided into 6 sub-sectors, 
sometimes called `patches', four in the outer sector and two in the inner 
sector. In total, there are $2\times 18 \times 6=216$ sub-sectors to be
readout. 

To keep the occupancy as low as possible and to ensure the necessary 
d$E$/d$x$ and position resolution, there are about 560\,000 readout pads
of three different sizes: $4 \times 7.5~\mm^2$ in the inner chambers,
$6 \times 10~\mm^2$ and $6 \times 15~\mm^2$ in the outer chambers. 
The pads are sampled with a frequency of about $5.66~\mhz$, dividing 
the drift time into about 500 time-bins, corresponding to a
total drift time of about $88~\musec$.
Thus, during the drift time about  $3\cdot10^8$ 10-bit amplitudes 
are produced, which subsequently are processed by the \ac{TPC} 
front-end electronics~\cite{thesisroland}.

Normally the gating grid protects the readout chambers for electrons 
coming from the drift volume. The grid only opens after receiving 
a particular trigger signal (\acs{L1}, see \sect{chap4:trigsystem}). 
This helps to prevent the buildup of space charge from positive ions 
drifting back from the multiplication region for non-triggered 
interactions and background.

\subsection{Particle identification detectors}
For a large part of the phase space the identification of particles 
is obtained by a combination of d$E$/d$x$ in the \ac{ITS} and \ac{TPC}, 
and time of flight information from the \acf{TOF} 
detector~\cite{alicetof1,alicetof2}. 

Electron identification above $1~\gev$ is provided by the 
\acf{TRD}~\cite{alicetrd},
because for momenta greater than $1~\gev$ the pion rejection 
capability through energy loss measurement in the \ac{TPC} is 
no longer sufficient. 
For quality of electron identification, the \ac{TRD} consists of six
individual layers. Matching the azimuthal segmentation of the 
\ac{TPC}, there are 18 sectors. Along the beam direction there is 
a 5-fold segmentation. Thus, there are $18 \times 5 \times 6 = 540$ 
detector modules. Each module consists of a radiator of $4.8~\cm$
thickness and a multi-wire proportional readout chamber including 
its front-end electronics. 
Each chamber has about 2000 pads; the pads have a typical area 
of $6-7~\cm^2$ and, in total, cover an active area of about $736~m^2$ 
with about $10^6$ readout channels.
The gas mixture in the readout chambers is Xe/CO$_2$ (85\%/15\%).
Each readout chamber consists of a drift region of $3.0~\cm$;
the drift time is $2.0~\musec$.
The induced signal at the cathode pad plane is sampled in 20 time-bins
spaced $1.5~\mm$ or $100~\ns$ apart.
In conjunction with \ac{ITS} and \ac{TPC}, the \ac{TRD} will
allow to measure, the production of light and heavy meson 
resonances in the di-electron channel, as well as to study the 
di-lepton continuum. 
The identification of high-momentum hadrons is provided by 
the \acf{HMPID} detector~\cite{alicehmpid}, which is based on 
proximity-focusing \ac{RICH} counters. The detector covers 5\% 
of the acceptance of the central detectors, and extents the 
identification of hadrons to \mbox{$\pt \approx  5~\gev$}.

Electromagnetic particles are measured by the \acf{PHOS}~\cite{alicephos},
which is a high-resolution electromagnetic spectrometer with limited 
acceptance at central rapidity. It provides photon identification as well 
as neutral meson identification through the two-photon decay channel. 
The detector is located on the bottom of the \ac{ALICE}
setup, and is built from scintillating lead-tungstate crystals 
coupled with photo-detectors. It covers approximately a quarter of a 
unit in pseudo-rapidity, $-0.12\leq \eta \leq 0.12$, and $100^\circ$ 
in azimuthal angle.

It has been proposed~\cite{aliceemcalpropos} 
to extend the electromagnetic calorimeter coverage of \ac{ALICE}
by a large lead-scintillator sampling \acf{EMCAL}, which will be located between the 
space frame and the magnetic coils adjacent to \ac{HMPID} and opposite to \ac{PHOS}. 
The detector is foreseen to have a central acceptance in pseudo-rapidity of $\abs{\eta}\le0.7$ 
with a coverage of $120^\circ$ in azimuth and is segmented into $\sim20000$ towers
with a resolution of $\Delta E\sim 15\%/\sqrt{E}$.
It will be optimized for the detection of high-$\pt$ photons, neutral pions and 
electrons, and, together with the barrel tracking detectors, 
will improve the jet energy resolution. 

\subsection{Muon spectrometer}
The muon spectrometer~\cite{alicemuon1,alicemuon2} is designed to
detect muons in the polar angle range \mbox{$2^{\circ}-9^{\circ}$}
($-4.0\leq \eta \leq -2.5$). It allows to study vector resonances 
via the $\mu^+\mu^-$ decay channel.
The spectrometer consists of the following components: 
passive front absorber to absorb hadrons and photons from the 
interaction vertex; high-granularity tracking system of 10 
detection planes; large dipole magnet; passive muon-filter wall 
followed by four planes of trigger chambers; inner beam shield 
surrounding the beam pipe to protect the chambers from
particles produced at large rapidities.

The design of the tracking system is driven by two main
requirements: a spatial resolution of about $100~\mum$ and the
capability to operate in an high particle-multiplicity environment.
The requirements can be fulfilled by the use of cathode pad chambers,
which are arranged in five stations: two are placed before, one
inside, and two after the dipole magnet. Each station contains
two chamber planes, and each chamber contains two cathode planes.
Together they provide the two-dimensional hit information. 
To keep the occupancy at a 5$\%$ level, a large segmentation of the 
readout pads has been chosen. For instance, pads as small as 
$4.2\times 6~\mm^2$ are needed for the region of the first station 
close to the beam pipe, where the highest multiplicity is expected. 
Since the hit density decreases with the distance from the beam, larger 
pads can be used at larger radii. Therefore, the total number of 
channels can be kept to about $10^6$.

\subsection{Multiplicity detectors}
Several detector systems placed outside the central barrel
will measure global event characteristics such as the event
reaction plane, multiplicity of charged particles and precise 
time and vertex position of the collision. Their combined 
information can be used to derive the interaction trigger 
signal. 

The \acf{FMD}~\cite{alicefmd} consists of five silicon-strip ring 
counters placed on both sides of the interaction point, 
covering the pseudo-rapidity range $-3.4 < \eta < -1.7$ 
and $1.7 < \eta < 5.1$ for charged-particle multiplicity measurement. 
It has in total about $5\cdot10^4$ channels. The readout time of 
the system is too long to allow the detector to participate in 
the generation of the low-level trigger information.

The \acs{V0} detector~\cite{alicefmd} is made of two scintillator 
arrays located asymmetrically on each side of the interaction 
point. 
It rejects beam--gas interactions 
by the time difference between the two arrays, and measures 
the charged-particle multiplicity using the signal amplitude. 
The \acs{V0} information is used to generate the interaction trigger 
input and to locate the event vertex.

The \acs{T0} detector~\cite{alicefmd} consists of two arrays of 12 
Cherenkov counters each, read out by fine-mesh photo-multiplier 
tubes. The counters provide the event time with a precision of 
$50~\ps$. The arrays are placed asymmetrically on both sides of 
the interaction point. 
The detector is designed to provide the start-time signal for the 
\ac{TOF} detector, discriminate against beam--gas interactions and 
sample particle multiplicity; in addition it provides input for the 
interaction trigger decision.

The \acf{PMD}~\cite{alicepmd1,alicepmd2} is located at $360~\cm$ 
from the interaction point behind the \ac{TPC}, on the opposite 
side of the muon spectrometer, and covers the region $2.3 < \eta < 3.5$. 
It has about $2.2\cdot10^5$ readout channels; its electronics 
is similar to that of the muon tracking chambers.
The \acf{PMD} is able to measure the ratio of photons to charged 
particles, the transverse energy of neutral particles, the elliptic 
flow and the event reaction plane.

Spectator nucleons are detected by means of the \acfp{ZDC}~\cite{alicezdc}, 
which are placed at $116\m$ from the interaction point, on both sides 
of the intersection. The \acp{ZDC} cannot provide an interaction 
trigger input in time, since they are located far inside the tunnel. 
But, using three levels of discrimination they participate in the 
(later) trigger generation for different centrality classes.
\fi

\section{Data volume, rate and acquisition}
\label{chap4:datavol}
\ifdatarates
The data volume and data rate, which is produced by the \ac{ALICE} 
detectors and their respective front-end electronics, depend on both, 
the event rate and the event data volume. 
The event rate is determined by the luminosity of the beam in the collider, 
while the event data volume is defined by the granularity of the detectors 
and the particle multiplicity or rather the occupancy in the various 
detectors. 

\subsection{Event rate}
\label{chap4:eventrate}
The maximum usable luminosity is limited by \ac{LHC} accelerator complex,
by the number of participating experiments and by the dead times of the 
detectors. 

Taking $7.8~\barn$ for the total (geometrical) cross section,
see~\psect{app:glauber}, the event rate for \PbPb\ collisions at the 
\ac{LHC} maximum (initial) luminosity of $\Lumi_0 = 10^{27}~\lum $ will 
be about 8000 minimum-bias collisions per second. This low interaction 
rate is crucial for the design of the experiment, since it allows to use slow, 
but high-granularity detectors, such as the time-projection chamber and 
the silicon drift detectors. The time-averaged luminosity for one participating 
experiment is $0.44\,\,\Lumi_0$~\cite{brandt2000}. Thus, throughout the thesis 
we will use $\Lumi = 5\cdot10^{26}~\lum$ for \PbPb\ collisions, yielding
a minimum-bias rate of $4\khz$.

However, the maximum usable luminosity is limited by the readout of the 
detectors. In particular, at the above event rates the \ac{TPC}, 
which is the slowest detector with an readout time of $88\,\musec$, 
has a significant double event fraction. Additional collisions 
may occur during its readout causing several superimposed events. 
These pile-up events are typically displaced in the time direction, 
but contribute to the local track density and the detector occupancy, 
and consequently lead to an increase of data volume and at the same 
time to a decrease of tracking performance.
The average fraction of double \PbPb\ events during \ac{TPC} readout 
is given by $1-\exp(-2\tau\cdot f) = 0.5$,
where $\tau=88~\musec$ is the drift time and $f=8~\khz$ the interaction 
frequency.~\footnote{We have taken $2\,\tau$, because after opening of 
the gating grid a single \ac{TPC} event may contain displaced
events occurring during the drift time before and after the trigger.}
This effect is specific to the \ac{TPC} as all other sub-detectors
have drift or integration times of up to about $5~\musec$. Restricting 
the double event fraction to below 30\%, we end up with the past--future
protected (`clean') minimum-bias rate of $2~\khz$.

The situation is very different in the case of \pp\ running at nominal
\ac{LHC} \cms~energy of $\sqrt{s}=14~\tev$. The maximum machine 
luminosity, which \ac{ALICE} can tolerate, is about three orders of 
magnitude below the design value for the other experiments, 
$\Lumi^{\rm pp}_{\rm max}=5\cdot10^{30}~\lum$~\cite{pprvol1}. 
At this luminosity the interaction rate is amounts to 
$350~\khz$ assuming that the total \pp\ cross section is $70~\mbarn$. 
Hence, on average the number of pile-up events in the \ac{TPC} rises 
to $60\times 1/2$ events; $97$\% of the data volume corresponds 
to unusable partial events. However, the charged particle density 
in \pp\ collisions is expected to be about $10$ particles per unit of
pseudo-rapidity at mid-pseudo-rapidity (see \pfig{chap2:fig:ncharge}),
resulting in a total of about $2\times30\times 10=600$ ($900$) tracks 
within the (extended) \ac{TPC} acceptance. Clearly, tracking in such a 
pile-up is feasible, since the occupancy is more than an order of 
magnitude below the design value of the \ac{TPC}. 

Concerning event rates involving the operation of the \ac{TPC},
the maximum possible event rate for both minimum-bias \PbPb\ and \pp\ 
interactions is limited by the maximum \ac{TPC} gating frequency 
to approximately $1~\khz$~\cite{alicetpc}.

For total rate estimates, all \ac{LHC} experiments have agreed to use 
an effective time per year of $10^{7}~\s$ for \pp\ and $10^{6}~\s$ for 
heavy-ion operation, since the \ac{LHC} is expected to run essentially 
in the same yearly mode as the \ac{SPS} (starting with several months of 
\pp\ running followed by several weeks of heavy-ion collisions at the end 
of each year).

\subsection{Event data volume}
\label{chap4:eventdatavol}
The event sizes essentially scale linearly with the charged-particle 
multiplicity and the resulting occupancy in the detectors. Furthermore 
---although almost too trivial to mention--- they strongly 
depend on the way the detector information is coded.

Given the amount of readout channels, $3\cdot10^8$, the by far 
highest amount of data is produced by the \ac{TPC} detector. 
Simulations indicate that the average occupancy in the \ac{TPC}
will be about 25\% for the highest-multiplicity assumption
of $\dncde=8000$ taken for the design of the detectors~\cite{alicetpc}.
Multiplying the number of channels with the 10-bit \ac{ADC} dynamic 
range leads to an event size of $375~\mbyte$, which is to be processed 
by the front-end electronics. 
By logarithmically compressing the \ac{ADC} values 
from $10~\bit$ to $8~\bit$, the event size will be reduced to about 
$300~\mbyte$.~\footnote{The \ac{ADC} conversion gain is typically 
chosen that $\sigma_{\mathrm{noise}}$ corresponds to one count. 
The relative accuracy increases with increasing \ac{ADC} 
value, but it is not needed for the upper part of the dynamic range. 
Therefore, the \ac{ADC} values can be non-linearly compressed 
from $10~\bit$ to $8~\bit$ leading to a constant relative accuracy 
over the whole dynamic range.} 
In addition, a 45$^{\circ}$ cone is cut out of the data, since it is 
problematic to resolve individual tracks that have a low \pt and 
cross the \ac{TPC} volume under small angles relative to the beam axis. 
This rejects all particles, which are not in the geometrical acceptance 
of the outer detectors and reduces the data size further by about 40\%. 
Finally, after zero-suppression and run-length encoding the raw event 
size is reduced to $75~\mbyte$, while the event size for all detectors 
is expected to be about $86.5~\mbyte$~\cite{alicehlt}. If the experiment 
is triggering on the 10\% most central, past-future protected, \PbPb\ 
collisions, corresponding to an interaction rate of $200~\hz$,
the data rate produced by the front-end will amount to $17~\gbyteps$
of which the \ac{TPC} creates about $14.5~\gbyteps$ alone. Since on 
average minimum-bias events have a multiplicity of about 20\% compared
to central events, the minimum-bias rate at the maximum \ac{TPC} 
rate of $1~\khz$ amounts to about $21~\gbyteps$.

Regarding \pp\ interactions, the estimated single, minimum-bias 
event size of the \ac{TPC} is approximately $60~\kbyte$ on average. 
Due to the additional data of the 60 half-complete events the total 
volume of the pile-up increases the \ac{TPC} event size to the order 
of $2.5~\mbyte$. The data volume created by the front-end of the 
other detectors in the experiment is negligible in that case. 
Thus, running at the foreseen maximum \ac{TPC} rate of $1~\khz$ will
lead to a total data rate of $2.5~\gbyteps$ for \pp\ interactions
at the maximum tolerable luminosity of $5\cdot10^{30}~\lum$.
It is interesting to note that for a very low luminosity of $10^{29}~\lum$, 
which might be achievable at the start of the \ac{LHC}, essentially
no pile-up will occur. In that case the rate would drop down to the 
order of a few $~\mbyteps$.

\begin{table}[htb]
\begin{center}
\begin{tabular}{l|ccccc}
\hline
\hline
Collision type & Luminosity &  Event rate & Event size & Date rate  & Prob.~rate \\
               & [\lum]     & [\hz]       & [\mbyte]   & [\gbyteps] & [\gbyteps]\\
\hline
Minimum-bias \pp\        & $5\cdot10^{30}$ & $1000$ & $\hide{0}2.5$ & $\hide{0}2.5$ &     \\
Minimum-bias \PbPb\      & $5\cdot10^{26}$ & $1000$        & $21.6$ & $21.1$ & $7.0$ \\
$0$--$10$\% central \PbPb\  & $5\cdot10^{26}$ & $\hide{0}200$ & $86.5$ & $16.9$ & $5.6$ \\
\hline
\hline
\end {tabular}
\end{center}
\vspace{-0.4cm}
\caption[xxx]{Expected \acs{ALICE} event and data rates for different \acs{LHC} runs,
limited in rate and dominated in volume by the participating \acs{TPC}. The event
sizes and correspondingly the rates are without compression. The probable
data rate corresponds to $\dncde=2500$. Central events are past--future protected, 
without protection the numbers increase by a factor of 2.}
\label{chap4:tab:datarates}
\end{table}

\Tab{chap4:tab:datarates} summarizes the expected event and data rates for 
different interactions with participating \ac{TPC} in the readout. 
For \PbPb\ interactions the result strongly
depends on the expected multiplicity. Predictions for the multiplicity in 
central \PbPb~collisions at the \ac{LHC} range at present from 2000 to 6000 
charged particles per pseudo-rapidity unit at mid-pseudo-rapidity, 
while most extrapolations from \ac{RHIC} data favor values of 
$2000$--$3000$~(see \psect{chap2:partmult}). Thus, the `probable' 
value for the particle density (\ac{RHIC} extrapolation) corresponds roughly 
to one-third of the worst-case assumption leading to rates of the
order of $5-7~\gbyteps$. The event sizes stated in this section are 
without further compression. It has been shown~\cite{berger02} that a 
compression factor of about 2 (typically 60\%) can be obtained for real and 
simulated \ac{TPC} data using entropy encoding methods~\cite{shannon1948}. 
Several additional steps of advanced data-reduction methods are envisaged 
for the \ac{TPC}~\cite{thesisvestbo}.

It should be mentioned that other detectors, \eg~the muon spectrometer,
can record data at a much higher rate (roughly $2~\khz$). Where it makes
sense to improve the statistics for specific trigger channels, groups of
detectors might be read out independently.

\subsection{Trigger system}
\label{chap4:trigsystem}
As already indicated in previous sections, the \ac{ALICE} experiment will
operate in different running modes with significantly different characteristics. 
The trigger system~\cite{alicehlt} is responsible for the selection of different 
types of events and controls the readout of the respective detectors when 
certain criteria are met. The system operates in three different levels: 
\ac{L0}, \ac{L1} and \ac{L2}, which have different latencies. Each level 
corresponds to criteria imposed from different detectors. The 
selection criteria get tighter as the trigger level increases. 
Consequently, the rates, at which each trigger level is operated, 
decrease at higher levels.

\ac{L0} and \ac{L1} are fixed-latency triggers. The reason for their 
separation is that in some detectors the front-end electronics requires 
a strobe very early, and, therefore, a first trigger decision must be 
delivered $1.2~\musec$ after the collision has taken place.
The \ac{L0} latency is estimated by the expected transmission time in 
the cables and limited by the dimensions of the detector. In some cases
a triggering detector will not be able to send its input in time. Thus,
every information which can be gathered and transmitted in $1.2~\musec$
is used to make the \ac{L0} decision, while those detectors, which
take longer, contribute to the \ac{L1} trigger, which arrives at the 
detectors $5.3~\musec$ after the \ac{L0} ($6.5~\musec$ after the collision
time). Its latency is given by the expected time \ac{ZDC} to issue a 
trigger signal, including a safety margin of 20\%.

The main message of the \ac{L0} trigger signal is that an interaction 
has taken place. The trigger input is mainly based on the information 
from the \acs{T0} and \acs{V0} counters, but also other detectors,
like \ac{PHOS}, the \ac{EMCAL} and the (independent) pre-\ac{TRD} and
muon triggers, deliver input.

The \ac{L0} trigger  ensures the following criteria:
\begin{itemize}
\item The interaction vertex must be close the the nominal collision point;
\item the forward-backward track distribution of tracks should be consistent with
a \\colliding beam--beam interaction; 
\item the measured multiplicity must exceed a given threshold.
\end{itemize}
No strong centrality decision is taken at this level, while at \ac{L1} 
centrality requirements based on the \ac{ZDC} information can be fulfilled.
At \ac{L1} the fast detectors and pre-triggers (mentioned above) deliver 
more concrete information about the triggered physics signal. Furthermore,
at this time all detectors are strobed. In particular, the \ac{TPC} gate 
is opened, which leads to the restriction that the \ac{L1} trigger 
can operate at a maximum frequency of $1~\khz$, if the \ac{TPC} is 
to participate in the readout. 

The third step, the \ac{L2} decision, has a variable latency. It is mainly
used to wait for the fulfillment of the past-future protection condition. 
Three groups of detectors exists: 
\begin{itemize}
\item Triggering detectors, which need no protection, since, by
design, they must be able to respond to each bunch crossing;
\item detectors with a required protection time less than
$\pm10~\musec$;
\item the \ac{TPC} with a protection interval of $\pm88~\musec$.
\end{itemize}

During the \ac{L2} decision taking time more selective algorithms 
are applied on the data extracted from the different trigger and detectors.
Based on their result an event will be accepted, \ac{L2a}, or rejected,
\ac{L2r}. Since the selection algorithms can differ in processing time,
the latency of the \ac{L2} trigger is not fixed, but has an upper bound as
defined by the \ac{TPC} drift time. After the \ac{L2} trigger, the data 
of the participating detectors are read out from the respective front-end
electronics and fed into the \ac{DAQ} and \ac{HLT} systems. 

\subsection{Data Acquisition system}
\label{chap4:daq}
The \acf{DAQ} system~\cite{alicehlt} collects the data from
the sub-detectors and assembles the sub-event data blocks into 
a full event before the complete event is sent to the mass storage. 
Its architecture is based on conventional \acsp{PC} connected by 
a commodity network, most likely \acs{TCP} over Gigabit Ethernet. 
The data transfer from the front-end electronics of the detectors 
is initiated by the \ac{L2a} trigger. Following, the data are 
transferred in parallel from the sub-detectors over optical fibers, 
so called \acp{DDL}, into the \acp{LDC}, where the sub-event building 
takes place. The sub-events built in the \acp{LDC} are then sent to 
a single \ac{GDC}, where the full event is assembled. The event building 
is managed by the \ac{EBDS} running on all \acs{DAQ} machines.
Its main task is to determine the \ac{GDC} destination for a particular 
event. The \ac{EBDS} communicates its decision to the \acp{LDC}. The fully 
assembled events are finally shipped to the mass storage system and from 
there to the permanent storage system for archiving and further offline 
analysis.

The \ac{DAQ} system is designed to be flexible in order to meet the 
requirements for the different data taking scenarios. Since \pp\ 
interactions produce a data rate of about 15\% compared to \PbPb\ 
interactions (for the worst-case multiplicity), the requirements on the 
system are defined by the expected data rate for the heavy-ion mode. 
In the heavy-ion mode, two main types of events have to be handled. 
The first type consists of central \PbPb\ events at a relatively low 
input rate, but with a large event size. The second type consists of 
events containing a muon pair reported by the trigger, which is read 
out with a reduced detector subset, including the muon spectrometer. 
Much higher event rates at quite small event sizes have to be handled 
in the latter case, of up to $2~\khz$. 

In the \ac{ALICE} proposal~\cite{alicetp1}, the collaboration
estimated the bandwidth of $1.25~\gbyteps$ to mass storage to
provide adequate physics statistics. However, depending on the 
multiplicity, the expected data rate of the \ac{ALICE} detectors 
exceeds the foreseen bandwidth by a factor of 4 (probable case) 
to 12 (worst case) (see \tab{chap4:tab:datarates}). Since the 
proposal several physics objectives have been refined, regarding 
rare processes, where higher statistics (of an order of magnitude)
is always welcome, and, since the bandwidth cannot be increased 
(also because of taping costs), the \ac{HLT} system has been put 
forward in a serious of notes~\cite{roehrich2001,frankenfeld2001,
bramm2002,lindenstruth2003,lindenstruth2004} and is now being 
constructed~\cite{alicehlt}.

Its task is to reduce the data rate to an acceptable level in terms 
of \ac{DAQ} bandwidth and mass storage costs, and, at the same time, 
provide the necessary event statistics. The idea is accomplished by 
processing of the data online, almost in real-time, allowing partial 
or full event reconstruction to select interesting events or sub-events, 
and/or to compress the data efficiently using data compression techniques. 
The same strategy has been proven to efficiently increase statistics
while decreasing bandwidth and taping costs in the online system
of the \acs{STAR} experiment~\cite{adler2002c}. To process the complete 
event online at data rates of $5-20~\gbyteps$ requires a massive 
parallel system. The functionality and architecture of the \ac{HLT} 
system are topics of the next section.

Logically, the trigger is located between \ac{L2} and event building, 
and, thus, in the data flow between the front-end electronics of the 
detectors and the event building of the \ac{DAQ} system. 
Technically, the data is duplicated on the way to the \ac{DAQ} system
(most likely via optical splitters) and fed into the \ac{HLT} 
system, see \fig{chap4:fig:hltflowscheme}. While waiting for the trigger 
decision of the \ac{HLT}, the \ac{DAQ} may build the complete event, 
which it will send to mass storage on \acs{HLT} accept, or disregard 
on \ac{HLT} reject.
\fi

\section{High-Level Trigger system}
\label{chap4:hlt}
\ifhlt
The \acf{HLT} system~\cite{alicehlt} is designed to reduce the recorded 
data rate below the limit of the \ac{DAQ} and mass storage 
bandwidth, while preserving the `interesting' physics content of the 
readout data, and, therefore, increasing the event statistics for these
observables. It is the only trigger system in \ac{ALICE}, which can base 
its decision on the complete event information. Its latency is in principle 
variable, and only limited by its buffer capacity and that of the 
event-building system.

In general, data reduction can be accomplished by either reducing the
event rate or by reducing the event size (or both). The first case 
implies that only a fraction of the available events are sent to mass 
storage. Exactly that had to done without any \ac{HLT} being present, 
since the readout rate coming from the detectors would have to be 
decreased in order to meet the foreseen bandwidth to mass storage. 
However, by introducing the \ac{HLT} data is analyzed online 
and events can be selected on the basis of physics observables other 
than the hardware triggers deliver. 
In the latter case, \ac{ROI} are selected and recorded only and/or data 
compression by modeling techniques are used to reduce the event size itself,
and, thus, increase the possible event rate being sent to mass storage. 
In combination with the fast trigger systems (\ac{TRD} pre-, muon trigger, 
\ac{PHOS} and \ac{EMCAL}) it is therefore possible to selectively read out 
the \ac{TPC} in the region where the respective hardware trigger found an 
interesting candidate. In the \pp\ mode the main application is the online 
removal of the additional pile-up events, and, thus reducing the data rate 
by more than an order of magnitude. In both cases online processing is needed, 
requiring pattern recognition for the reconstruction of the event or at 
least parts of it. In the following we differentiate between selection mode
and data compression mode.

\subsection{Selection mode}
There are two \ac{QCD} physics domains addressed by 
\ac{ALICE}, which can be separated into `mostly soft' and 
`predominantly hard'  with relatively large and small cross sections, 
respectively. It turns out that analysis of observables related to 
soft physics requires modest event statistics of a few $10^6$ \PbPb\ 
and about $10^8$ \pp\ collisions, whereas systematic analysis of hard
signals calls for an additional one or two orders of magnitude, both
in \PbPb\ and \pp. For example, inclusive production of jets
with total transverse energy of more than $150~\gev$, or of the weaker states
in the bottonium family, is expected to occur (within the ALICE
tracking acceptance) about once every 10$^{4}$ central \PbPb\
collisions. Clearly, it is the latter sector of \ac{ALICE} physics where 
the online selection will be required. 

\enlargethispage{0.5cm}
The \ac{HLT} selection mode can be divided into two subclasses: Complete 
event selection or rejection (trigger), and \ac{ROI} readout. The latter
can be regarded as filtering the unwanted information from the event.
Both are based on the online identification of pre-defined physics 
characteristics. They have to be known and specified in advance; 
by studying Monte Carlo and recorded events to train and control 
the operation of trigger/filter. Depending on the topology of the signals, 
either full or partial event reconstruction is needed.

Hard probes provide to a large degree the most topologically 
distinct tracking signatures in the \ac{TPC}. Therefore, most of the online 
trigger algorithms, investigated so far, are based on online tracking of the 
\ac{TPC} data. Further refinement and support will result from using the early 
time information  of the dedicated hardware trigger systems. 
The different feasible trigger modes envisaged to date are described in 
detail in~\Ref{alicehlt}. We will give a brief summary in the following.

\subsubsection{Open charm trigger}
The measurement of open charm, \ie~of $D$-mesons, in heavy-ion
collisions provide a probe, which is sensitive to the collision 
dynamics at both short and long time scales. 
This observable is of main interest at \ac{LHC} 
energies and its detection and systematic analysis is one of the main 
goals of the \ac{ALICE} experiment~\cite{dainese2003}.
It is estimated~\cite{paic00} that for its analysis  $2\cdot10^7$ 
events are needed ($20~\hz$ of central \PbPb\ for $10^6$ seconds). 
If all events are written to tape, 80\% (worst case) and 30\%  
(probable case) of the available \ac{DAQ} bandwidth would be needed
for open charm alone.  
Simulations concentrating on $D^0\rightarrow K\pi$ mode show~\cite{thesisdainese} 
that a signal--to--event of about 1/1000 and a background--to--event of 1/100 
should be obtainable in \ac{ALICE}. The strategy is to detect $D^0$-mesons via 
their characteristic weak decay topology into pions and kaons and to compute 
the invariant mass of tracks originating from displaced secondary vertices using 
a sequence of kinematical and impact parameter cuts. 
The foreseen \ac{HLT} event-selection 
strategy proceeds in two steps: Firstly, a momentum filter, applied as it is
done in offline, reduces the data volume by a factor of about four.
Secondly, an impact-parameter analysis (with very relaxed parameters 
compared to the offline selection cuts) rejects events with no obvious 
D$^0$ candidate, reducing the data rate further (the concrete reduction factor
is not known). It is expected ---although proper simulations are 
outstanding~\footnote{However, in the case the most probable multiplicity
case comes true, it will be questionable, whether the complicated analysis
of the open charm detection should be performed online at all.}--- 
that \ac{HLT} can reduce the rate for the open charm program 
by a factor of 5--10, thus increasing statistics and at the same time releasing 
\ac{DAQ} bandwidth.

\subsubsection{Di-electron trigger}
The yields of $J/\psi$ and $\Upsilon$ production will be important to 
measure, since, for example, at \acs{SPS}, $J/\psi$ suppression has played 
a major role in the discovery of deconfined matter~\cite{satz2002}.~\footnote{Currently, 
there are first hints for $\Upsilon$ production at \acs{RHIC}~\cite{thesistorsten}.}
The bound systems will be reconstructed by their leptonic decay into $e^+e^-$.
The pair will be tracked through the combined barrel, \ac{ITS}, \ac{TPC} and \ac{TRD}. 
The hardware trigger of the \ac{TRD} is constructed to trigger
on high-$\pt$ tracks by online reconstruction of particle trajectories 
in the \ac{TRD} chambers and on the electron candidates by measuring 
the total energy loss and the depth profile of the deposited energy. 
However, the true quarkonium trigger rate is small, \ie~the signal rate 
for $\Upsilon$ is about $10^{-2}~\hz$, and the trigger is dominated by 
background. 
The \ac{HLT} is able to reject background events by using the complete
information of the barrel detectors online. Firstly, \ac{TRD} tracklets 
will be combined with \ac{TPC} and \ac{ITS} tracks. The fit of the combined
track allows the momenta of the candidate pair to be determined more precise 
than by the \ac{TRD} alone, and, thus, the \ac{HLT} will reject secondary 
electrons by sharpening the momentum cut. Secondly, the \ac{HLT} can improve
the identification of the candidate particles using the additional d$E$/d$x$ 
information of the \ac{TPC}. Hence, the background from misidentified pions 
can be reduced. Simulations indicate that event-rate reduction of a factor 
of $10$ or even more is within~reach.

\subsubsection{Di-muon trigger}
\label{chap4:muontrigger}
The measurements of the muon spectrometer are complementary to the quarkonia measurements
involving the \acs{TRD}. The spectrometer is designed to detect vector resonances via the
$\mu^+\mu^-$ decay channel. It will run at the highest possible 
rate in order to record muons with the lowest possible dead time. The spectrometer
is built together with an hardware trigger system, which consists of four 
\acfp{RPC} planes arranged in two stations, one meter apart from each other,
placed behind the muon-filter wall. 
The trigger detector participates also in the \ac{L0} decision, since 
the muon tracking stations require the \ac{L0} signal for track-and-hold.
The task of the muon trigger system is to select events containing a 
muon pair candidate coming from the decay of $J/\psi$ or $\Upsilon$ 
resonances. The background is dominated from low-$\pt$ muons from pion
or kaon decays. The \ac{L0} fires if at least two tracks with opposite
charge above a pre-defined $\pt$ threshold are found.
The value of the threshold typically is set to either select low-\pt\ ($>1~\gev$) 
or high-\pt\ ($>2~\gev$) muons from the $J/\psi$ and $\Upsilon$ resonances, 
respectively. 
However, the coarse-grained segmentation of the dedicated trigger chambers 
do not allow a sharp $p_t$-cut, resulting in a rather large background 
rate at \ac{L0}. The \pt resolution can be improved by performing an additional 
tracking step within the \ac{HLT} system using information from the slower, but
more accurate tracking chambers. Thus, a far better trigger selectivity can be 
achieved. The expected background rejection factor by inclusion of the \ac{HLT} 
algorithm is 5 for the low-\pt\ and 100 for the high-\pt\ threshold~\cite{manso02}.

\subsubsection{Jet trigger}
In the light of the previous chapter, the study of jets as probes of the strongly 
interacting \ac{QCD} matter will be most interesting at \ac{LHC} energies.
However, quite a high number of collected events are required in order to provide 
the necessary statistics at high jet energy. Estimations based on Glauber scaling 
from \pp\ collisions indicate that about one jet with  $\et>100~\gev$
is produced per second in minimum-bias~\PbPb~collisions (see next chapter, 
\pfig{chap5:fig:jetrate}). Depending on the setup of the 
\ac{L1} triggering detectors the \ac{HLT} will be used to either verify the 
\ac{L1} hypothesis or to solely inspect events at \ac{L2}. 
Trigger simulations, reported in \psect{chap5:recontriggerrates}, show that data rates 
in \pp\ and in \PbPb\ can be reduced by about a factor of $50$ ($100$ for \pp), 
while keeping $1/10$ ($1/5$ for \pp) of the events where $\et>50~\gev$ and 
slightly more than half of the events with $\et>100~\gev$. 
Assuming 100\% efficient hardware triggers at \ac{L1}, which
reduce the event rate to a rate that can be tolerated by the \acs{TPC}, \ie~below
$1~\khz$, at total of about $3\cdot 10^{7}$ minimum-bias \PbPb~events per \ac{ALICE} year will be 
recorded, which contain about $1\cdot 10^{6}$ events with $\et>50~\gev$. Without the help 
of hardware triggers the rates will be limited by the maximum inspection rate of the \ac{TPC},
and by a factor of $1/4$ lower.~\footnote{A word about timing seems appropriate. The online
jet trigger currently uses the same algorithm as offline including the same settings. This strategy 
is preferable, since it reduces additional biases in the triggered data sample. With the settings 
listed in \ptab{app:tab:coneparams}, the run time per event of the complete jet finder including 
all data handling and disk accesses is less than $50~\ms$ for central \PbPb\ on a standard Pentium III, 
$800~\mhz$. Since we use different parameters in \pp, which increase the precision of the
found jets, the run time is the same for \pp\ and central \PbPb. However, for the same
settings as in \PbPb\ the run time in \pp\ (without pile-up) reduces to $10~\ms$ per event.}

\subsubsection{Pile-up removal}
In the case of \pp\ running, the foreseen running luminosity of
$5\cdot 10^{30}~\lum$ will lead to about 60 superimposed, half-complete 
events within the \ac{TPC} frame, \ie~97\% overhead in the data stream.
The additional piled-up events are displaced along the beam axis
and will be disregarded during offline analysis.
Using \ac{HLT} to reconstruct the \ac{TPC} tracks online, the tracks 
corresponding to the original (triggered) event are identified,
while the tracks belonging to the pile-up events are disregarded 
from the readout data stream. Although average spacing in time 
is about $3~\musec$, \ie~about $9~\cm$ in the \ac{TPC}, the events are 
still not necessarily ordered in drift distance because of the variation 
of the primary vertex position, which will randomize distances. 
This influences the online capability to identify the sub-event that 
belongs to the trigger event and sets a limit to the number of pile-up
events, which can be handled.
Simulations have revealed that an overall event size reduction of
1/10 can be achieved while retaining an efficiency of more than 95\%
for the primary tracks of the event. The pile-up removal capabilities
of the \ac{HLT} in the \pp\ mode are even more important, since
to be triggered physics signals must be identified to belong to
the trigger event and not to the pile-up.

\subsection{Data compression mode}
The option to compress the readout data online allows to improve 
the physics capabilities of the experiment without performing 
selective readout. In principle, the full event rate could be 
written to mass storage, if a compression factor of about  5--10 
can be achieved. 

We concentrate again on the \ac{TPC} detector, since it produces more 
than 85\% of the total event size alone. It has been shown~\cite{berger02},
that \ac{TPC} data by means of loss-less compression techniques can be
compressed, at most, by a factor of two.
However, the most efficient data compression is expected by using
data-modeling compression methods, which are highly adapted to the 
underlying \ac{TPC} data. Such compression algorithms exploit the fact 
that the relevant information is contained in the reconstructed cluster 
centroids and the track charge depositions, rather in the \ac{ADC}
values, from which they are inferred. Thus, the parameters of the 
cluster can be stored as deviations from a model. If the model is well 
adapted to the data (clusters), the resulting bit-rate needed to code 
the data, will be small. Since the clusters in the \ac{TPC} critically 
depend on the track parameters, the reconstructed tracks and clusters 
can be used to build an efficient model of the data. In contrast to the 
loss-less coding algorithms mentioned above, this technique does not 
keep the original data unmodified, since clusters are stored rather than 
the \ac{ADC} values, which make the clusters. Recent 
studies~\cite{thesisvestbo} seem to indicate that compression factors of 
$5$--$10$ could be achieved using such a compression scheme.
However, any data compression method, which is not loss-less on the binary 
level, must be performed with caution to assure the validity of the measured 
physical observables. Clearly, data-modeling techniques are not to 
be used in the first years of the experiment, but might be an interesting
option for later on. 

\subsection{Architecture}
\label{chap4:hltarch}
The design of the \ac{HLT} system is driven by the large amount of 
data it is foreseen to process, the large uncertainty in the expected rate 
and the complexity of the processing task, which requires a massive parallel, 
flexible and extensible computing system. 

The \ac{HLT} system is therefore planned to consist of a large \acs{PC} 
cluster farm with several hundred (up to a thousand) separate nodes. 
Its architecture is mainly determined by two constraints. 
Firstly, the data flow has an inherent granularity and parallelism, which
is given by the readout segmentation of the detectors.
Secondly, the system is responsible for issuing a trigger decision based 
on information derived from a partial or complete event reconstruction. 
Therefore, the reconstructed data finally has to be collected at
a global layer, in which the final trigger algorithms are 
implemented.~\footnote{This is even holds, if the system is running in 
compression only mode. In that case the trigger decision is always positive,
but the compressed event has to be sent to \ac{DAQ} in any case via
dedicated, the \ac{DAQ} interfacing, nodes, which most probably will 
coincide with the nodes at the global layer.} 
Both requirements demand a hierarchical tree-like topology with an
high degree of inter-connectivity.

The foreseen data flow architecture is shown in \fig{chap4:fig:hltflowscheme}.
The data is duplicated on the way into the \ac{DAQ} system and enters 
the receiving nodes of the \ac{HLT} system. These \ac{FEP} constitute the first
layer of the \ac{HLT} system. Each \ac{DDL} is mounted on a \ac{HLT-RORC}, 
which is a custom designed \acs{PCI} card hosted by every \ac{FEP}. 
Several \acp{HLT-RORC} may be placed in one \ac{FEP}, depending on the bandwidth 
and processing requirements. Every \ac{HLT-RORC} is equipped with additional
co-processor functionality for designated pre-processing pattern-recognition
steps of the data in order to take load off the \acp{FEP}.
The total number of \acp{HLT-RORC} is defined by the readout granularity of the
detectors and corresponds to the total number of \acp{D-RORC}. For the \ac{TPC} 
detector, which as the main data source in the experiment is the biggest contributor, 
the readout is divided into its respective 216 sub-sectors, as mentioned 
in \sect{chap4:tpc}. Every sub-sector is read out by a single \ac{DDL}, and, 
thus, there $216$ fibers are needed for the \ac{TPC} alone. In total taking into 
account all detectors there are about $400$ \acp{DDL}.

\begin{figure}[htb]
\begin{center}
\includegraphics[width=15cm]{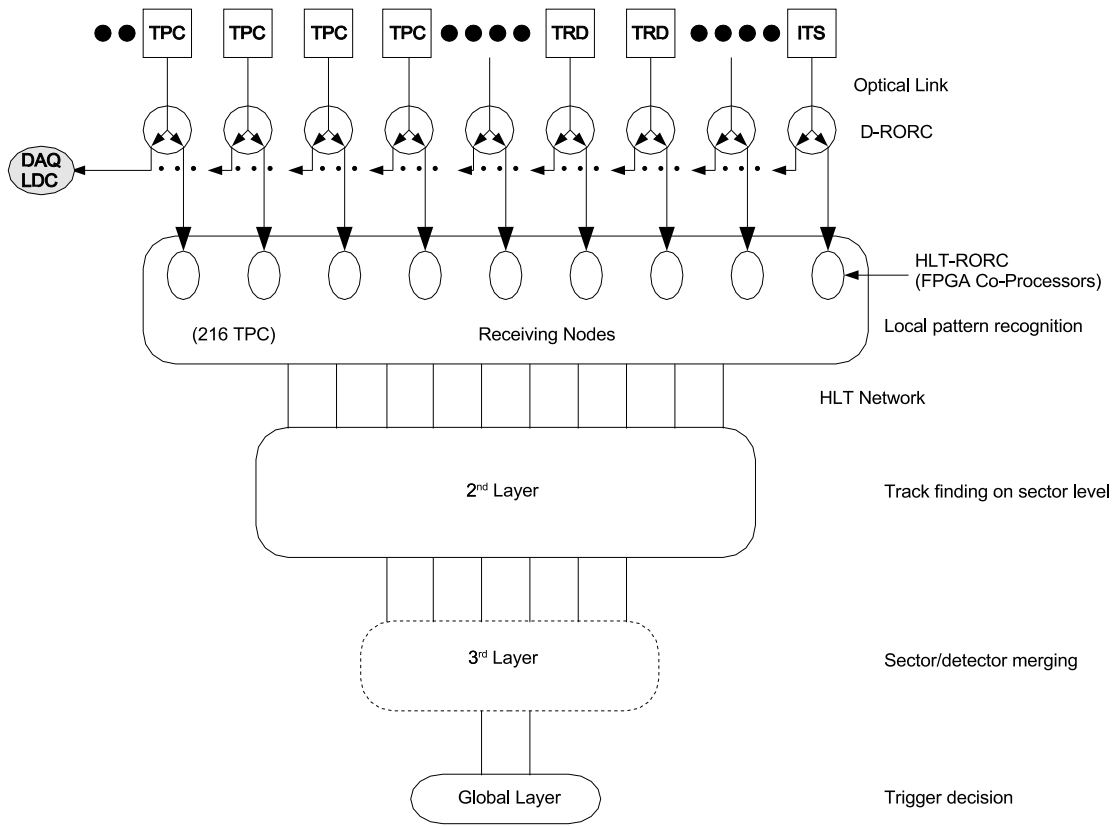}
\end{center}
\vspace{-0.3cm}
\caption[xxx]{Data flow architecture of the \ac{HLT} system. The detector raw
data is duplicated and received by \ac{DAQ} and \ac{HLT}. The architecture 
is inherent hierarchical, adapted to the parallelism of the data readout
and the various tasks of the pattern recognition.}
\label{chap4:fig:hltflowscheme}
\end{figure}

The first pattern recognition step happens while reading the data into the system. 
It is performed by the \ac{FPGA} co-processor hosted on every \ac{HLT-RORC}. 
In the case of the \ac{TPC} local pattern recognition tasks, \ie~cluster 
finding~\cite{grastveit2003} and/or Hough transformation~\cite{thesisalt}
might be done. The processed data is transferred via the \acs{PCI} into the 
main memory of the \ac{FEP}, where further analysis takes place. From then 
on the data is an integral part of the \ac{HLT} communication framework described 
below. It will ---transparently to the \ac{FEP} and only on request--- 
ship the data to a node of the next processing layer. At every layer there are 
depending on the type of data and on the processing task to be done as many nodes 
as necessary to stay within the latency budget. Output data produced by a node 
of a certain layer is shipped within the communication framework to the next 
layer until the final stage has been reached.
In this way, the processing hierarchy follows a tree-like structure, where 
successively larger fractions of an event are processed and merged.
At the second layer, the processing typically includes track finding within the 
\ac{TPC} sectors. At the global level, finally, the necessary fractions have been 
collected and merged into the reconstructed event, \ie~ tracks from the different 
sub-sectors are merged and fitted and might be combined with tracks from \ac{ITS} 
and \ac{TRD}. Thus, the complete event (or necessary parts of it) are analyzed by 
the selection algorithms. The final trigger decision for the event is taken based 
on the output of the selection algorithms. The decision together with the 
corresponding data (if any) is transmitted to the \ac{DAQ} system. 
The interface between \ac{DAQ} and \ac{HLT} is given by a number of \acp{DDL} 
between a set of \ac{HLT} event merger nodes at the global level and a number of 
\ac{DAQ} \acp{LDC}. 

An essential part of the \ac{HLT} system is interprocess communication 
and data transport within the system. A generic communication framework has been
developed~\cite{thesistimm} with emphasis on efficiency in terms of 
\acs{CPU} power, flexibility in terms of network topology and fault 
tolerance regarding failing nodes.
\enlargethispage{0.8cm}
The framework implements an interface between different analysis steps 
(also between different layers) by defining data producers and data consumers. 
For efficiency data is not communicated between different processes, but rather 
a descriptor of the data including a reference to the actual data in shared memory 
is sent. It is therefore ensured that data stays in memory as long as possible, 
avoiding unnecessary copying (within a single node and over network).
The framework basically consists of a number of independent software
components, which can be connected together in an arbitrary fashion. 
The generic interface allows the processing modules to have a
common interface which is independent of the underlying transport interface. 
\fi

\iftracking
\section{Tracking performance}
\label{chap4:tracking}

\fi

%

\newif\ifcross
\crosstrue
\newif\ifreconpp
\reconpptrue
\newif\ifreconpbpb
\reconpbpbtrue
\newif\ifreconspectrum
\reconspectrumtrue
\newif\ifbacktoback
\backtobacktrue

\chapter{Jets in ALICE}
\label{chap5}
It is our aim to investigate the potential of the \ac{ALICE} detectors 
for the measurement of high-energy jets. We start by estimating inclusive
single-jet rates in \sect{chap5:jetrates}. In \sect{chap5:jetreconpp}
and \sect{chap5:jetreconpbpb} we concentrate on the quality of the jet 
reconstruction at fixed jet energy for various detector settings in \pp\ and 
\PbPb~collisions, respectively. In \sect{chap5:reconjetspectrumpp} we present the 
complete jet spectrum reconstructed by means of offline and online tracking 
algorithms in \pp~collisions and discuss the \ac{HLT} trigger performance. 
Finally, in \sect{chap5:b2bjetrates} we estimate the yield of back-to-back 
jet and photon--jet production in the central \ac{ALICE} acceptance.

\section{Expected single-inclusive jet rates}
\label{chap5:jetrates}
\ifcross
We repeatedly have mentioned the large hard scattering cross section as 
compared to the geometrical cross section at \ac{LHC} energies, see 
\psect{chap2:highQ}, which is quantified in the following.

\subsection{Partonic and hadronic cross sections} 
\label{chap5:phcrosssections}
Using the \acs{EKS} program~\cite{ellis1992} we compute the expected inclusive 
single-jet partonic cross section at mid-pseudo-rapidity for \pp\ collisions at 
\cms\ energies of $1.8$, $5.5$ and $14~\tev$.
As reported in~\psect{chap3:inclusivexsec}, the \ac{NLO} calculation performed 
at parton level agrees with the measured, and corrected, cross section at 
highest \acs{Tevatron} energies of up to $2~\tev$. 
The values chosen here for the parameters of \acs{EKS} mostly correspond to the 
values introduced there, \ie~using a cone of $R=0.7$, a parton separation value 
of $R_{\rm sep}=1.3$ and all scales at half of the highest jet energy found in 
the event. Taking \acs{CTEQ}~5L \acp{PDF} is safe, since the obtained cross 
sections for jet energies between $30\le\Etj\le250~\gev$ at mid-rapidity are not 
sensitive to that choice. For \acs{EKS} in general, systematic errors are estimated 
to be about 20\%~\cite{soper1997}. Note that $x<0.1$, since the discussion is limited to 
(low) jet energies of less than $250~\gev$, and, thus the large uncertainty in the 
gluon distribution at high-$x$ does not play a role~\cite{martin2004}. 

\begin{figure}[htb]
\begin{center}
\includegraphics[width=12cm]{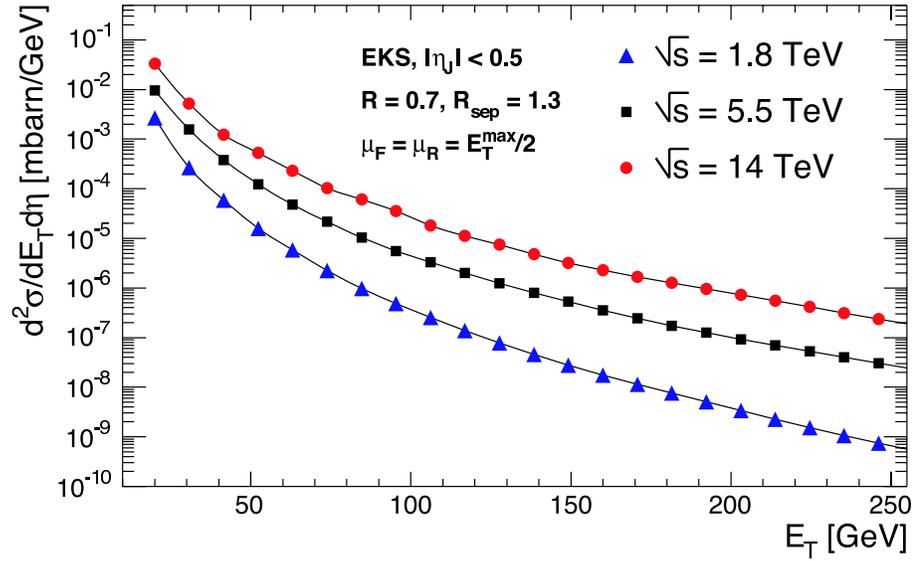}
\end{center}
\vspace{-0.5cm}
\caption[xxx]{Inclusive, partonic single-jet cross section at mid-pseudo-rapidity 
for \pp\ collisions at \cms\ energies of $1.8$, $5.5$ and $14~\tev$ 
calculated with the \acs{EKS} program at \acs{NLO} with the set of parameters 
as reported in the figure. Compare to \pfig{chap3:fig:cdfcross}.}
\label{chap5:fig:spectraeks}
\end{figure}

\begin{figure}[htb!]
\vspace{1.2cm}
\begin{center}
\includegraphics[width=12cm]{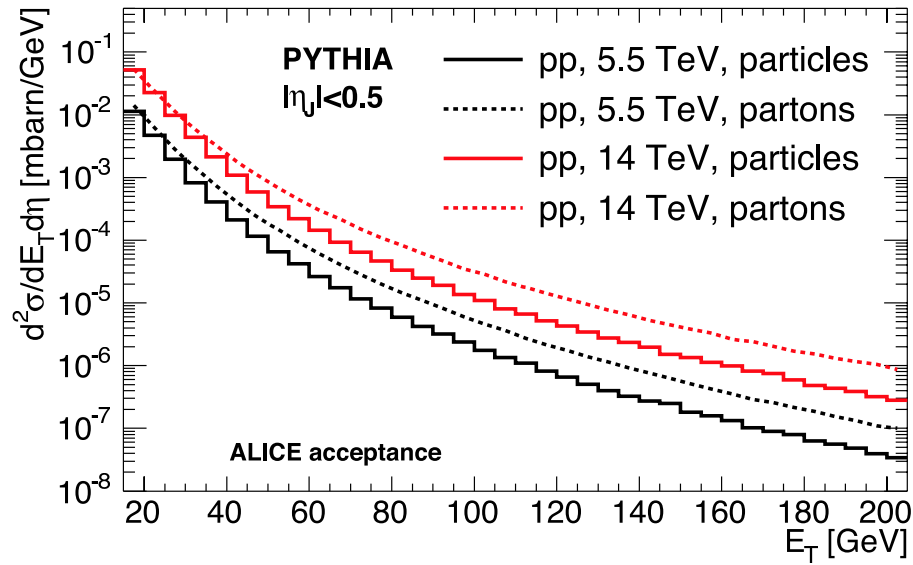}
\end{center}
\vspace{-0.5cm}
\caption[xxx]{Inclusive single-jet cross section at mid-pseudo-rapidity for \pp\ 
interactions at the \acs{LHC} \cms\ energies of $5.5$ and $14~\tev$ simulated 
with \acs{PYTHIA} at \acs{LO}. The continuous lines show the jet spectrum,
identified with the \acs{UA1} cone finder for $R=1.0$ using particles in the 
central \acs{ALICE} acceptance ($\abs{\eta}<1$). The dashed lines show the 
spectrum of the outgoing partons from the hard scattering.}
\label{chap5:fig:spectrapythia}
\end{figure}

As can be seen in~\fig{chap5:fig:spectraeks}, above $50~\gev$ the cross sections
at \ac{LHC} are predicted to be more than an order of magnitude higher than those 
at high \acs{Tevatron} energies. With \cms~energy scaling of dimensionless 
cross-section ratios at the same value of $\xt$
\begin{equation}
\label{chap5:eq:qcdscaling}
R\left(\xt=\frac{2\et}{\sqrt{s}}=\frac{2\et^*}{\sqrt{s^*}}\right) = 
\frac{\et^3 \,\frac{\dd\sigma}{\dd\et} (\et,\sqrt{s})}
{{\et^*}^3 \,\frac{\dd\sigma}{\dd\et} (\et^*,\sqrt{s^*})} \approx 1\;,
\end{equation}
the value of the cross section may be estimated for higher \cms\ energies at 
corresponding $\xt$ according to
\begin{equation}
\label{chap5:eq:etscaling}
\frac{\dd\sigma}{\dd\et} \left(\frac{\sqrt{s^*}}{\sqrt{s}}\,\et,\sqrt{s}\right) 
\approx \left(\frac{\sqrt{s}}{\sqrt{s^*}} \right)^ 3 
\, \frac{\dd\sigma}{\dd\et} (\et,\sqrt{s^*})\;.
\end{equation}

The scaling of the ratio, \eq{chap5:eq:qcdscaling}, exactly holds in the parton model.
For $\sqrt{s}=630$ and $\sqrt{s^*}=1800~\gev$, as reported by \acs{D0}~\cite{abbott2000} 
and \ac{CDF}~\cite{affolder2001}, the scaling is approximately true ($R~\sim 1.5$ at 
$\xt>0.15$), but uncertainties are large (of up to $50$\%); power-like corrections 
breaking the simple energy scaling must be introduced~\cite{mangano2001}.

Nevertheless, applying \eq{chap5:eq:etscaling} with $\dd\sigma/\dd\et \sim \et^{-n}$
gives a value of $\left({\sqrt{s^*}} / {\sqrt{s}}\right)^{n-3}$ for the increase 
of the cross-section ratio at the same value of $\et$.
Taking $n=6$ yields a factor of $30$ at ${\sqrt{s^*}} / {\sqrt{s}}=3$ for the 
extrapolation from $\sqrt{s}=1.8$ to $5.5~\tev$ and about $500$ at 
${\sqrt{s^*}} / {\sqrt{s}}=7.8$ for the extrapolation to $14~\tev$.

Turning from partons to particles, in \fig{chap5:fig:spectrapythia} 
we show the corresponding inclusive single-jet cross section at 
mid-pseudo-rapidity for \pp\ interactions at the \ac{LHC} \cms\ 
energies of $5.5$ and $14~\tev$ simulated with \acs{PYTHIA}. The
general settings of the simulation are described in~\psect{app:pythia}.
The cone finder applied on the particle level with radius of $R=1$ 
is a variant of \acs{UA1} cone finder~\cite{arnison1983,albajar1988} 
using the Snowmass accord, \peq{chap3:eq:snowmass}. 
All particles within the central \ac{ALICE} acceptance are taken into account, 
$-1<\eta<1$; no detector response is included. To be accepted in the calculation 
of the cross section, the jet axis is required to be within the interval 
$-0.5<\eta_{J}<0.5$ and, thus, the cross section is averaged over the central 
region. The dashed lines show the corresponding partonic jet-spectrum 
without \ac{ISR} and \ac{FSR},~\ie the outgoing partons from the hard scattering 
as calculated by \acs{PYTHIA} at \ac{LO}.~\footnote{Note that taking the ratio of 
the \acs{EKS} cross sections at \ac{NLO} shown in \fig{chap5:fig:spectraeks} 
to the partonic \acs{PYTHIA} cross sections at \ac{LO} leads to a factor 
of $K_{\rm NLO}\approx 1.5$.} At very low jet energy, the cross section measured at 
particle level almost agrees with the partonic cross section, but, clearly, for higher 
energy the influence of the detector acceptance reduces the 
measurable fraction of the (partonic) cross section to about $20$\%.

\subsection{Yields at L1 or L2 inspection rate} 
\label{chap5:intyields}
We are interested in the integrated jet yield at mid-pseudo-rapidity 
that can be measured with \ac{ALICE} within a running year. At first, one 
one may estimate the number of produced jets per second, $N(\et^{\rm min})$, 
which at least contain the minimum transverse energy, $\et^{\rm min}$, 
\begin{equation}
\label{chap5:eq:jetrate}
N(\et^{\rm min}) = \Lumi \, \int_{\et^{\rm min}}^{\infty}\,
\frac{\dd\sigma}{\dd\et} \,\dd\et \;.
\end{equation}
The average luminosities reported in \psect{chap4:eventrate} amount to 
$\Lumi=0.5~\mbarn^{-1}\s^{-1}$ for \PbPb\ and 
$\Lumi=5\cdot10^{3}~\mbarn^{-1}\s^{-1}$ for \pp.~\footnote{However, one should keep in mind
that for \PbPb\ the average may be lower by about a factor of 2, in case three experiments
will run.}
In \pp\ at the two \cms\ energies $\frac{\dd\sigma}{\dd\et}$ is simply given by 
the (hadronic) cross section calculated in \acs{PYTHIA}. For the extrapolation from 
\pp\ to \PbPb\ at $\snn=5.5~\tev$ we scale $\dd\sigma/\dd\et$ according to binary 
scaling in the Glauber framework, \peq{app:eq:binhcs}, 
with $\sigma^{\rm geo}_{\rm PbPb}=7.8~\barn$ and 
$\av{T_{\rm AB}}=23.3~\mbarn^{-1}$ for $0$--$10$\% central and 
$\av{T_{\rm AB}}=5.5~\mbarn^{-1}$ for minimum-bias collisions, respectively.

\begin{figure}[htb]
\begin{center}
\includegraphics[width=12cm]{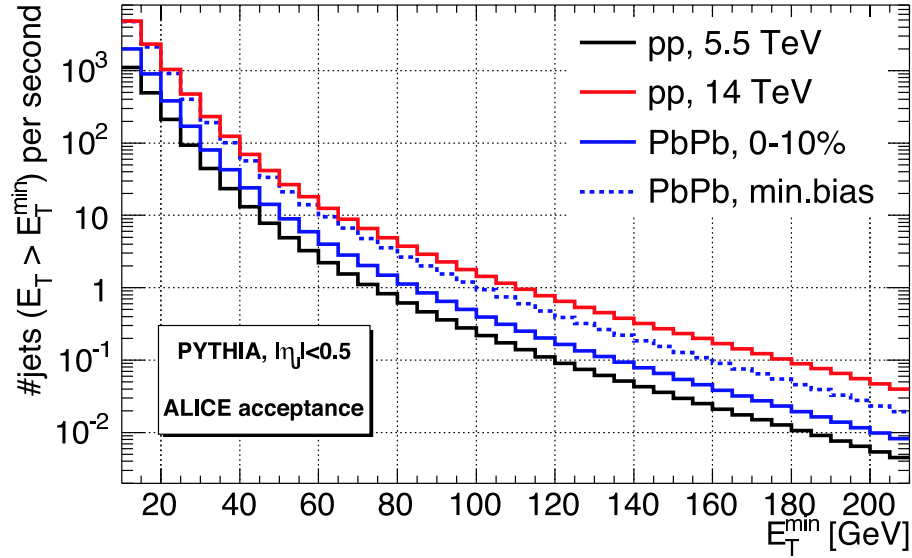}
\end{center}
\vspace{-0.3cm}
\caption[xxx]{Jet yield per second, \eq{chap5:eq:jetrate}, for jets with 
$\et>\et^{\rm min}$ at mid-pseudo-rapidity in minimum-bias \pp~collisions 
at $5.5$ and $14~\tev$, as well as the binary scaled extrapolation
for central and minimum-bias \PbPb~collisions at $\snn=5.5~\tev$. 
The jet sample corresponds to the spectrum shown in \fig{chap5:fig:spectrapythia}.}
\label{chap5:fig:jetrate}
\end{figure}

\ifallpages
\pagebreak
\fi
The integrated jet yield per second, \eq{chap5:eq:jetrate}, 
is shown in \fig{chap5:fig:jetrate} as a function of the minimum energy for 
minimum-bias \pp~collisions at $\sqrt{s}=5.5$ and $14~\tev$, as well as for central 
and minimum-bias \PbPb~collisions at $\snn=5.5~\tev$. Thus, at $\et^{\rm min}=100~\gev$, 
on average, one jet per second will be produced in minimum-bias \PbPb~collisions at 
$\snn=5.5~\tev$  and about four in ten seconds for $0$--$10$\% most central collisions; 
the rate in \pp\ at $\sqrt{s}=5.5~\tev$  is about one jet in every two seconds and 
about one jet per second at $14~\tev$.

It is important to note that for a given centrality the estimated rates are much lower 
than the \ac{L1} trigger rate ($100~\khz)$ and for $\et\ge30~\gev$ lower than the maximum 
gating frequency of the \ac{TPC} ($1~\khz)$. 
Therefore, the total yield per year can be estimated by
\begin{equation}
\label{chap5:eq:jetyield1}
Y^{\rm year}_{\rm L1}(\et^{\rm min}) = \epsilon_{\rm L1} \, t_{\rm run} \, N(\et^{\rm min})\;,
\end{equation}
where $t_{\rm run}=10^{7}~\s$ for \pp\ and $10^{6}~\s$ for \PbPb\ denotes the effective 
running time per year and $\epsilon_{\rm L1}$ is the jet-detection 
efficiency of the \ac{TRD}, \ac{PHOS} and \ac{EMCAL} (pre-) triggering 
complex at \ac{L1}. Of course, in reality the efficiency will depend on 
$\et^{\rm min}$. Assuming an optimal trigger, $100$\% efficiency for the signal,
$\epsilon_{\rm L1}=1$, and low accidental trigger rate compared to $1~\khz$, 
we end up with the total yields reported for $\et^{\rm min}=50$, $100$, $150$ 
and $200~\gev$ in \tab{chap5:tab:jetyield1}.

\begin{table}[htb]
\begin{center}
\begin{tabular}{l|cccc}
\hline
\hline
Collision type & \pp, $5.5~\tev$  & \pp, $14~\tev$ & \PbPb, minimum-bias  & \PbPb, $0$--$10$\% \\
\hline
$Y^{\rm year}_{\rm L1}(50~\gev)$  & $4.9\cdot10^{7}$ &  $2.3\cdot10^{8}$ &  $2.1\cdot10^{7}$ & $8.9\cdot10^{6}$ \\
$Y^{\rm year}_{\rm L1}(100~\gev)$ & $2.2\cdot10^{6}$ &  $1.4\cdot10^{7}$ &  $9.4\cdot10^{5}$ & $4.0\cdot10^{5}$ \\
$Y^{\rm year}_{\rm L1}(150~\gev)$ & $3.0\cdot10^{5}$ &  $2.3\cdot10^{6}$ &  $1.3\cdot10^{5}$ & $5.4\cdot10^{4}$ \\
$Y^{\rm year}_{\rm L1}(200~\gev)$ & $5.4\cdot10^{4}$ &  $4.7\cdot10^{5}$ &  $2.3\cdot10^{4}$ & $9.8\cdot10^{3}$ \\
\hline
\hline
\end{tabular}
\end{center}
\vspace{-0.4cm}
\caption[xxx]{Accumulated jet yield per \acs{ALICE} run year, \eq{chap5:eq:jetyield1}, 
at mid-pseudo-rapidity for optimum hardware triggers at \ac{L1} exploiting the
production rate.}
\label{chap5:tab:jetyield1}
\end{table}

The rates estimated above are production rates, \ie~relative to the 
minimum-bias collision rate, $~\Lumi\,\sigma^{\rm tot}$, which can only be exploited
by fast dedicated triggers in hardware. As discussed in \psect{chap4:hlt},
the \ac{HLT} system might be used in addition to the hardware triggers or stand-alone
to either verify the \ac{L1} hypothesis or to online search for jets using the 
detector information of the complete event.

\ifallpages
\pagebreak
\fi
Let's assume for a moment that no other hardware triggers (except from centrality 
detection) are available. In this case the inspection rate is limited to the maximum 
gating frequency of $1~\khz$, or lower for past--future protection in central 
\PbPb~collisions. Taking for $\sigma^{\rm tot}$ the value of the \pp\ inelastic cross 
section of $70~\mbarn$ and $79~\mbarn$~\cite{berardi2004} and for \PbPb\ the geometrical 
cross section, $\sigma^{\rm geo}_{\rm PbPb}=7.8~\barn$, we compute the ratio of maximum 
\ac{TPC} inspection rate at \ac{L2} over the collision rate, $r_{\rm TPC}$. 
It is $r_{\rm TPC}=1/350$ and $r_{\rm TPC}=1/400$ for \pp\ at $\sqrt{s}=5.5~\tev$ and 
$14~\tev$, whereas it is $r_{\rm TPC}=1/4$ for \PbPb\ collisions at $\snn=5.5~\tev$. 
For the case, the \ac{HLT} runs without the help of hardware triggers, 
we define the expected integrated yield per \ac{ALICE} run year as
\begin{equation}
\label{chap5:eq:jetyield2}
Y^{\rm run}_{\rm HLT}(\et^{\rm min}) = \epsilon_{\rm HLT} \, r_{\rm TPC} 
\, t_{\rm run} \, N(\et^{\rm min})\;,
\end{equation}
where $\epsilon_{\rm HLT}$ denotes the efficiency of the \ac{HLT} jet finder, which 
depends on $\et^{\rm min}$. 

Assuming $\epsilon_{\rm HLT}=1$ the integrated yield is shown 
in \fig{chap5:fig:jetyield2} for minimum-bias \pp~collisions at $\sqrt{s}=5.5$ and $14~\tev$, 
as well as for central and minimum-bias \PbPb~collisions at $\snn=5.5~\tev$.~\footnote{For 
past--future protected, $0$--$10$\% central \PbPb~collision the yield will be by a factor of 2 lower.}.
The total integrated yield for $\et^{\rm min}=100~\gev$ in minimum-bias \PbPb~collision amounts to about 
$2\cdot10^5$; in $0$--$10$\% central collisions it is about an order of magnitude lower, on the level
of $10^4$ events per year and about the same for the reference measurements in \pp. These numbers are 
at the the statistical limit needed for the analysis of jet properties at high-$z$.~\footnote{In about
$10^4$ events one may expect about 100 events at high $z=\pt/\Ptj$ (see \pfig{chap3:fig:cdfpartfrac}).}
\Tab{chap5:tab:jetyield2} summarizes the expected total jet yield for $\et^{\rm min}=50$, $100$, 
$150$ and $200~\gev$ in the case the \ac{HLT} inspects the \ac{TPC} without \ac{L1} triggers.

\begin{figure}[hbt]
\begin{center}
\includegraphics[width=12cm]{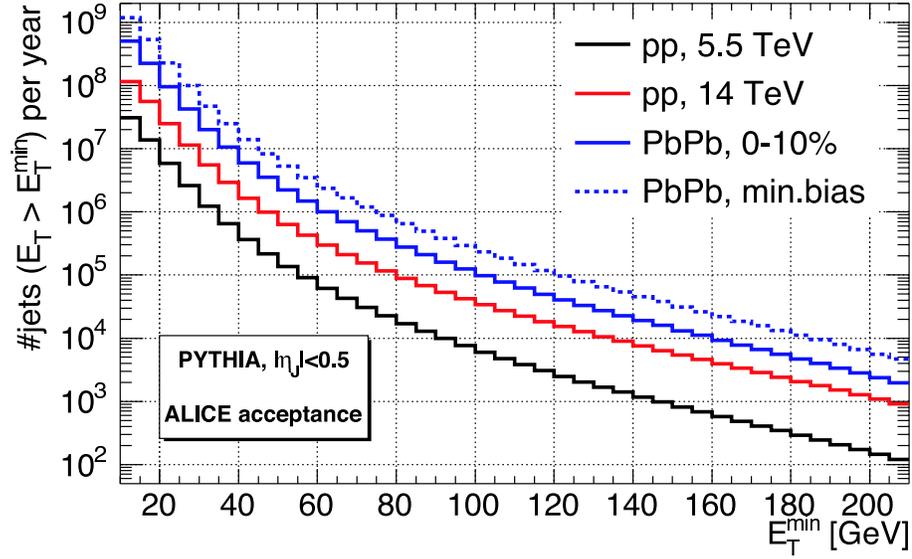}
\end{center}
\vspace{-0.4cm}
\caption[xxx]{Total jet yield per \acs{ALICE} run year, \eq{chap5:eq:jetyield2}
with $\epsilon_{\rm HLT}=1$ limited by the \acs{TPC} inspection rate, for jets with 
$\et>\et^{\rm min}$ at mid-pseudo-rapidity in minimum-bias \pp~collisions at $5.5$ and $14~\tev$, 
as well as the binary scaled extrapolation for central and minimum-bias \PbPb~collisions at 
$\snn=5.5~\tev$. The corresponding jet production rate is shown in \fig{chap5:fig:jetrate}.}
\label{chap5:fig:jetyield2}
\end{figure}

\begin{table}[htb!]
\vspace{0.8cm}
\begin{center}
\begin{tabular}{l|cccc}
\hline
\hline
Collision type & \pp, $5.5~\tev$ & \pp, $14~\tev$ & \PbPb, minimum-bias  & \PbPb, $0$--$10$\% \\
\hline
$Y^{\rm year}_{\rm HLT}(50~\gev)$  & $1.4\cdot10^{5}$ & $6.4\cdot10^{5}$ & $5.3\cdot10^{6}$ & $2.2\cdot10^{6}$ \\
$Y^{\rm year}_{\rm HLT}(100~\gev)$ & $6.1\cdot10^{3}$ & $3.4\cdot10^{4}$ & $2.3\cdot10^{5}$ & $9.8\cdot10^{5}$ \\
$Y^{\rm year}_{\rm HLT}(150~\gev)$ & $8.3\cdot10^{3}$ & $5.6\cdot10^{3}$ & $3.2\cdot10^{4}$ & $1.3\cdot10^{4}$ \\
$Y^{\rm year}_{\rm HLT}(200~\gev)$ & $1.5\cdot10^{3}$ & $1.1\cdot10^{3}$ & $5.8\cdot10^{3}$ & $2.4\cdot10^{2}$ \\
\hline
\hline
\end{tabular}
\end{center}
\vspace{-0.4cm}
\caption[xxx]{Accumulated jet yield per \acs{ALICE} run year, \eq{chap5:eq:jetyield2}
with $\epsilon_{\rm HLT}=1$, at mid-pseudo-rapidity for the case that the \ac{HLT} runs
without dedicated hardware triggers at the maximum \ac{TPC} inspection rate of $1~\khz$ and
at $200~\hz$ in central \PbPb\ collisions.}
\label{chap5:tab:jetyield2}
\end{table}

In practice, of course, \ac{ALICE} will run with a mix of hardware triggers at \ac{L1}
and further improve the \stn~ratio by \ac{HLT} inspection of the triggered events.
In this case, the yield will be given by a combination of \eq{chap5:eq:jetyield1}
and \eq{chap5:eq:jetyield2},
\begin{equation}
\label{chap5:eq:jetyield3}
Y^{\rm run}(\et^{\rm min}) = \epsilon_{\rm HLT} \, \epsilon_{\rm L1} \, t_{\rm run} 
\, N(\et^{\rm min})\;,
\end{equation}
and, thus, in the best case, will correspond to \eq{chap5:eq:jetyield1}.

\pagebreak
The total rate of accepted events per second,
\begin{equation}
\label{chap5:eq:eventrate}
N_{\rm acc} =  f_{\rm HLT} \, f_{\rm L1} \,\Lumi \, \sigma^{\rm tot}\;,
\end{equation}
is determined by the fraction of accepted events of the involved triggers,
$f_{\rm i}$. At \ac{L1} the efficiency should be as high as possible, while
the rejection of pure background should only reduce the event rate below the 
\ac{TPC} readout rate, \ie~around $1/100$ for \pp\ and about $1/5$ for \PbPb. 
The \ac{HLT} system then must verify the trigger hypothesis and reduce the 
rate of accepted events below an affordable limit, which will be discussed in
\sect{chap5:reconjetspectrum}.
\fi

\section{Jet reconstruction in pp for fixed energy}
\label{chap5:jetreconpp}
\ifreconpp
It is instructive to at first start with the jet reconstruction capabilities
of \ac{ALICE} in \pp~collisions. We generate samples of jets with $\Etj=50$, $100$, 
$150$, $200$ and $250~\gev$ at $\sqrt{s}=5.5~\tev$ using \acs{PYTHIA}
(see \psect{app:pythia} for parameter settings); every sample contains about 
3000 events, and every event contains at least one triggered jet in $\abs{\eta}<0.5$ 
($R=1$) within $\pm 2$\% of the required jet energy $\Etj$ (also denoted as $\et^{\rm mc}$). 

\subsection{Ideal detector response} 
\label{chap5:idealdetresponse}
For jet reconstruction under ideal conditions we distinguish three types of detectors:
\begin{itemize}
\item Ideal detector keeping all particles except neutrinos ({\em ideal detector});
\item Ideal charged-particle tracking and electromagnetic calorimeters detecting all 
charged particles as well as photon radiation and neutral pions ({\em ideal barrel+em});
\item Ideal charged-particle tracking detecting all charged particles ({\em ideal barrel}).
\end{itemize}
Without simulating the detailed detector response the detectable particle types, which correspond 
to the different scenarios, are taken from the Monte Carlo. For a given detector 
type, these particles must furthermore pass geometrical ($-0.9\le\eta\le0.9$) and kinematical 
($0.5~\gev\le\pt\le100~\gev$) cuts. These cuts anticipate that lower momenta in pile-up \pp\ 
or in \PbPb\ will not be efficiently reconstructed, while for higher momenta the 
$\pt$-resolution of the barrel tracking detectors will be severely degraded or not measurable.
The selected particles are then grouped into jets 
using the developed cone finder algorithm (see \psect{app:conefinder}) with a cone size 
of $R=1$. The jets reconstructed that way are subject to be discussed in the following.

\enlargethispage{1cm}
\Fig{chap5:fig:ppetmeanresideal} shows the average fraction of reconstructed jet energy, 
$\av{\et^{\rm rec}/\et^{\rm mc}}$, and the reconstructed energy resolution, 
$\sigma(\et^{\rm rec})/\av{\et^{\rm rec}}$, as a function of the jet-trigger energy for
the three different cases of ideal particle reconstruction and jet finding. It is obvious
that for an ideal detector the average energy fraction is very close to one and
the resolution is better than $5$\% decreasing to $1$\% with increasing jet energy. Using
only charged-particle tracking of the barrel detectors, independently of the jet energy
a mean of slightly less than $60$\% with a constant resolution of about $30$\% is obtained,
which is close to the value of $\sim 0.6$ realized in 
nature~(isospin conservation plus violating decays).
Using information provided by ideal electromagnetic calorimeters the mean increases to about 80\%
with about $20$\% resolution.~\footnote{Depending on the knowledge of the hadronic response in the 
calorimeter for other neutral particles (such as neutrons) the mean might increase by up to $10$\%.} 
It is interesting to note that due to large fluctuations in the ratio of charge-to-neutral or 
electromagnetic-to-neutral particles the resolution is constant and, thus, not decreasing with jet 
energy.

\begin{figure}[htb]
\begin{center}
\subfigure[Mean energy fraction (ideal)]{
\label{chap5:fig:ppetmeanideal}
\includegraphics[width=7cm]{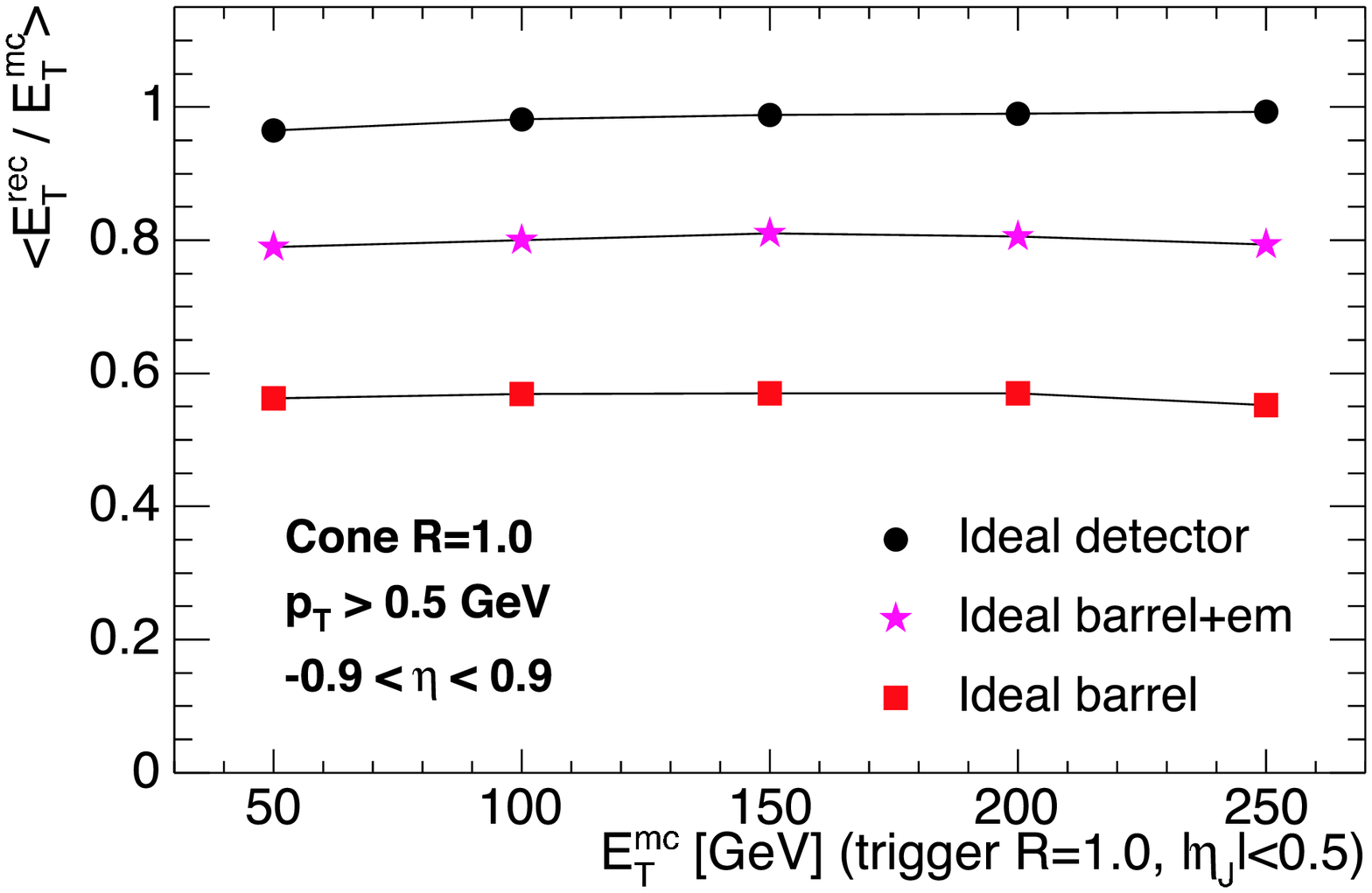}}
\hspace{0.5cm}
\subfigure[Energy resolution (ideal)]{
\label{chap5:fig:ppetresideal}
\includegraphics[width=7cm]{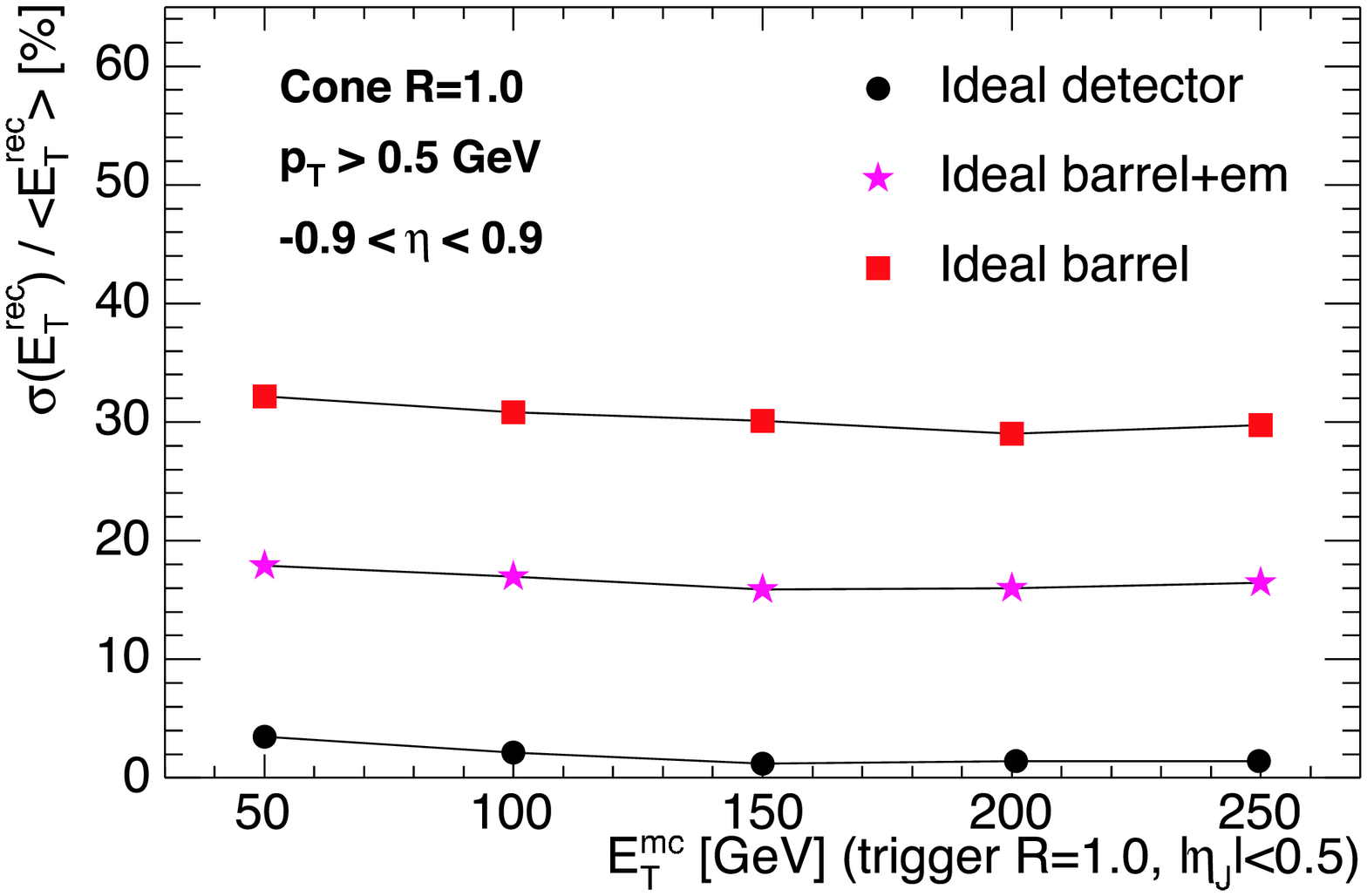}}
\end{center}
\vspace{-0.5cm}
\caption[xxx]{Average fraction of reconstructed jet energy, 
$\av{\et^{\rm rec}/\et^{\rm mc}}$~\subref{chap5:fig:ppetmeanideal}, 
and reconstructed energy resolution, 
$\sigma(\et^{\rm rec})/\av{\et^{\rm rec}}$~\subref{chap5:fig:ppetresideal}, 
both, as a function of the jet-trigger energy (Monte Carlo) for the different 
ideal cases. Further details are given in the text.}
\label{chap5:fig:ppetmeanresideal}
\end{figure}

\begin{figure}[htb!]
\vspace{0.3cm}
\begin{center}
\includegraphics[width=10cm]{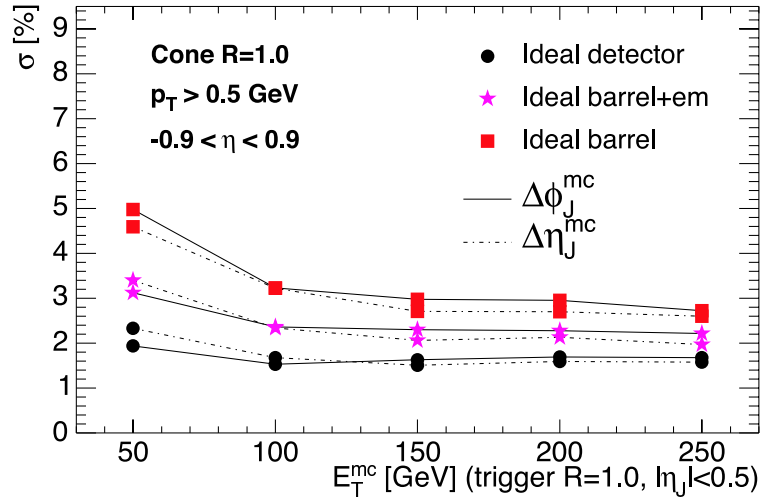}
\end{center}
\vspace{-0.3cm}
\caption[xxx]{Spatial resolution of the reconstructed jets, $\sigma(\Delta \phi^{\rm mc}_{J})$ and 
$\sigma(\Delta \eta^{\rm mc}_{J})$, both, as a function of the jet-trigger energy (Monte Carlo) 
for the different ideal cases. The spatial differences are measured relative to the direction
of the triggered jet, $\Delta \phi^{\rm mc}_{J}=\phi^{\rm rec}_{J}-\phi^{\rm mc}_{J}$ and 
$\Delta \eta^{\rm mc}_{J}=\eta^{\rm rec}_{J}-\eta^{\rm mc}_{J}$.}
\label{chap5:fig:ppspaceresideal}
\end{figure}

As outlined in \psect{chap3:jetpp} it might be important to reconstruct the jet axis 
accurately enough to resolve the direction of the primary parton from which the jet originates.
We define the measured jet axis relative to the triggered jet (which ideally corresponds 
with the parton direction) and compute the spatial distributions of 
$\Delta \phi^{\rm mc}_{J}=\phi^{\rm rec}_{J}-\phi^{\rm mc}_{J}$ and 
$\Delta \eta^{\rm mc}_{J}=\eta^{\rm rec}_{J}-\eta^{\rm mc}_{J}$ for every input energy 
$\et^{\rm mc}$ and detector type.~\footnote{As usual, $\phi$ and thus,
$\Delta \phi$, $\sigma(\phi)$ or its root mean square (rms) are given in radians.} 
The extracted widths of the Gaussians, which correspond to the spatial resolution of the 
reconstructed jets, are plotted in \fig{chap5:fig:ppspaceresideal} as a function of 
$\et^{\rm mc}$. As can be seen, $\sigma(\Delta \phi^{\rm mc}_{J})$ and 
$\sigma(\Delta \eta^{\rm mc}_{J})$ slightly decrease with jet energy. 
In the ideal case, the resolution is limited by the difference in the definition of the jet 
finding and jet trigger algorithm and, in general, by intrinsic effects on the parton level. 
However, most notably, even in the case of charged-particle tracking only, the spatial resolution 
is better than $5$\% at lowest increasing to about $3$\% at highest input energy.~\footnote{Note
that we specify the spatial resolutions, $\sigma$, in percent points rather than in absolute numbers, 
even though no ratio is taken for their computation.}

\subsection{Simulated detector response} 
\label{chap5:simdetresponse}
In the following we estimate the influence of the detector response and of the 
performance of the different particle-reconstruction methods on jet reconstruction..
The expected detector response may be included in the simulation using the
\acs{GEANT3} interface of \acs{ALIROOT}~\cite{pprvol1}. Based on the assembled
hit information the foreseen tracking algorithms can be applied. For the purpose 
of the thesis we distinguish three cases:
\begin{itemize}
\item Charged-particle tracking in the barrel with the offline code using 
the combined tracking information of \ac{ITS}, \ac{TPC} and \ac{TRD}
({\em offline barrel});
\item charged-particle tracking in the \ac{TPC} with the \ac{HLT} online code,
which is based on cluster finding and track follower ({\em tracker});
\item charged-particle tracking in the \ac{TPC} with the \ac{HLT} online code,
which uses the improved stand-alone Hough transform ({\em hough}).
\end{itemize}

The different methods and in particular their tracking performance have been briefly discussed 
in \chap{chap4}. As for the ideal cases described above, the tracks (particles) are required 
to furthermore pass geometrical ($-0.9\le\eta\le0.9$) and kinematical ($0.5~\gev\le\pt\le100~\gev$) 
selection cuts, before they are grouped into jets using the developed cone finder algorithm (see 
\psect{app:conefinder}) with a cone size of $R=1$. The reconstructed jets, which are based
on charged-particle tracking including simulated detector response, will be discussed 
in the following and compared to the ideal cases discussed above.

\begin{figure}[htb]
\begin{center}
\subfigure[Mean energy fraction (tracked)]{
\label{chap5:fig:ppetmeantracked}
\includegraphics[width=7cm]{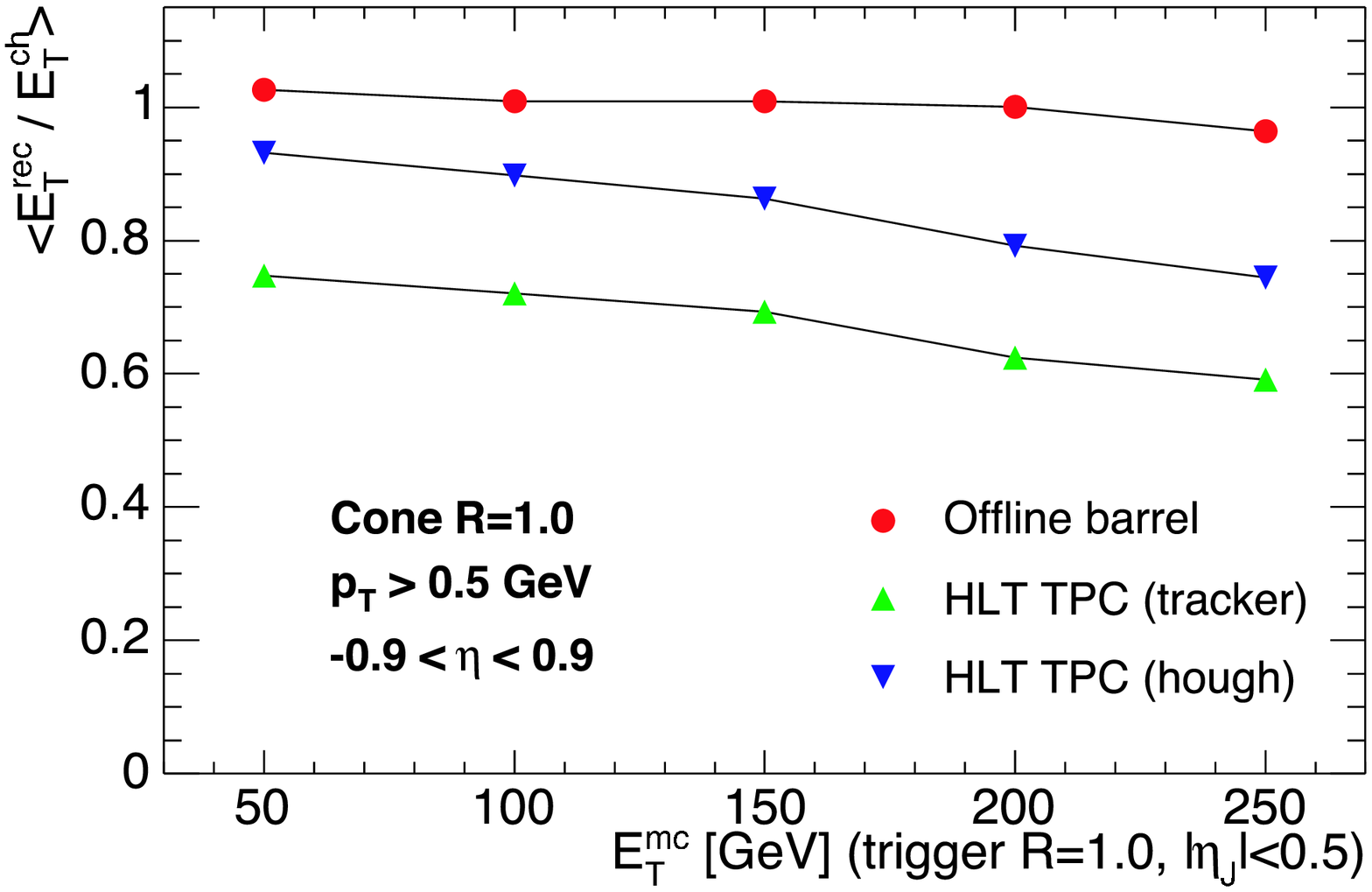}}
\hspace{0.5cm}
\subfigure[Energy resolution (tracked)]{
\label{chap5:fig:ppetrestracked}
\includegraphics[width=7cm]{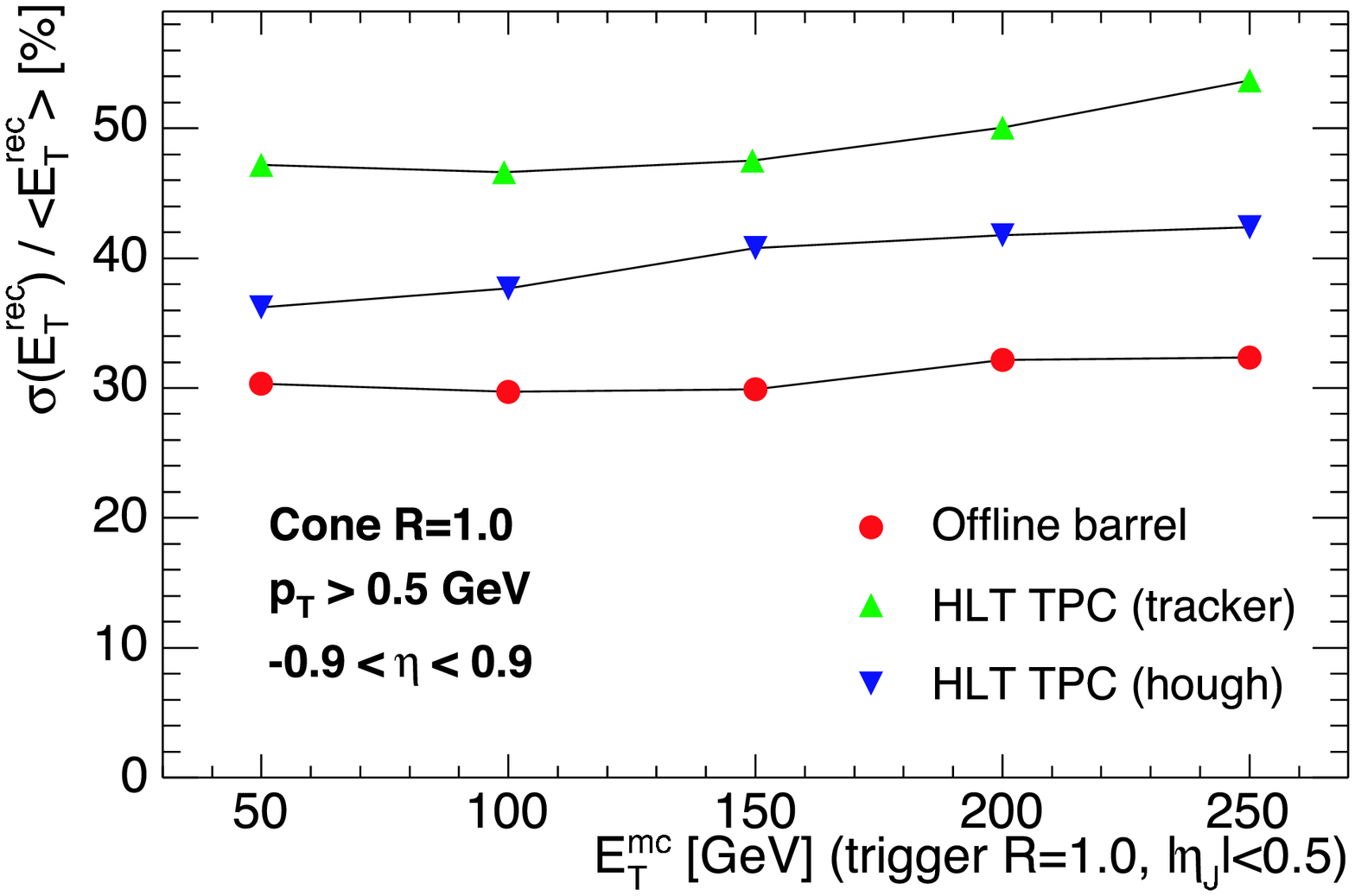}}
\end{center}
\vspace{-0.5cm}
\caption[xxx]{Average fraction of reconstructed charged-jet energy, 
$\av{\et^{\rm rec}/\et^{\rm ch}}$~\subref{chap5:fig:ppetmeantracked}, 
and reconstructed energy resolution, 
$\sigma(\et^{\rm rec})/\av{\et^{\rm rec}}$~\subref{chap5:fig:ppetrestracked}, 
both, as a function of the jet-trigger energy (Monte Carlo) for the different 
tracking cases. The corresponding ideal case (ideal barrel) is shown in 
\fig{chap5:fig:ppetmeanresideal}.}
\label{chap5:fig:ppetmeanrestracked}
\end{figure}

\Fig{chap5:fig:ppetmeanrestracked} shows the average fraction of reconstructed charged-jet 
energy, $\av{\et^{\rm rec}/\et^{\rm ch}}$, and the reconstructed energy resolution, 
$\sigma(\et^{\rm rec})/\av{\et^{\rm rec}}$, as a function of the jet-trigger energy for
the three different cases of charged-particle reconstruction and jet finding. Opposed
to \fig{chap5:fig:ppetmeanresideal}, where we take $\et^{\rm mc}$ of the trigger jet
for the normalization, we normalize to the reconstructible charged jet energy 
(ideal barrel), $\et^{\rm ch}$. Thus, for offline reconstruction the average 
fraction of charged-jet energy is very close to the optimum, and, therefore, 
about $60$\% of the total jet energy are, on average, reconstructed. 
The corresponding resolution is about $30$\%, independent of $\et^{\rm mc}$, and again
close to optimal. The online charged-jet energy fraction is decreasing with the trigger-jet
energy to about $80$\% for the hough and $60$\% for the tracker case. In the regime, which
due to statistic reasons is most interesting for \ac{ALICE}, $50$ to $150~\gev$, the 
average reaches $90$\% for hough and about $75$\% for the tracker case. 
The resolution is slightly worse compared to offline, amounting to about $35$--$40$\% for hough
and about $45$--$55$\% for the tracker, which is due to the compromise of tracking efficiency 
versus running time of the online code. It might be improved once the \ac{HLT} includes combined
tracking using the \ac{ITS} and \ac{TRD} detectors in addition to the \ac{TPC}.

\begin{figure}[htb]
\begin{center}
\subfigure[Spatial resolution in $\phi$ (tracked)]{
\label{chap5:fig:ppresphitracked}
\includegraphics[width=7cm]{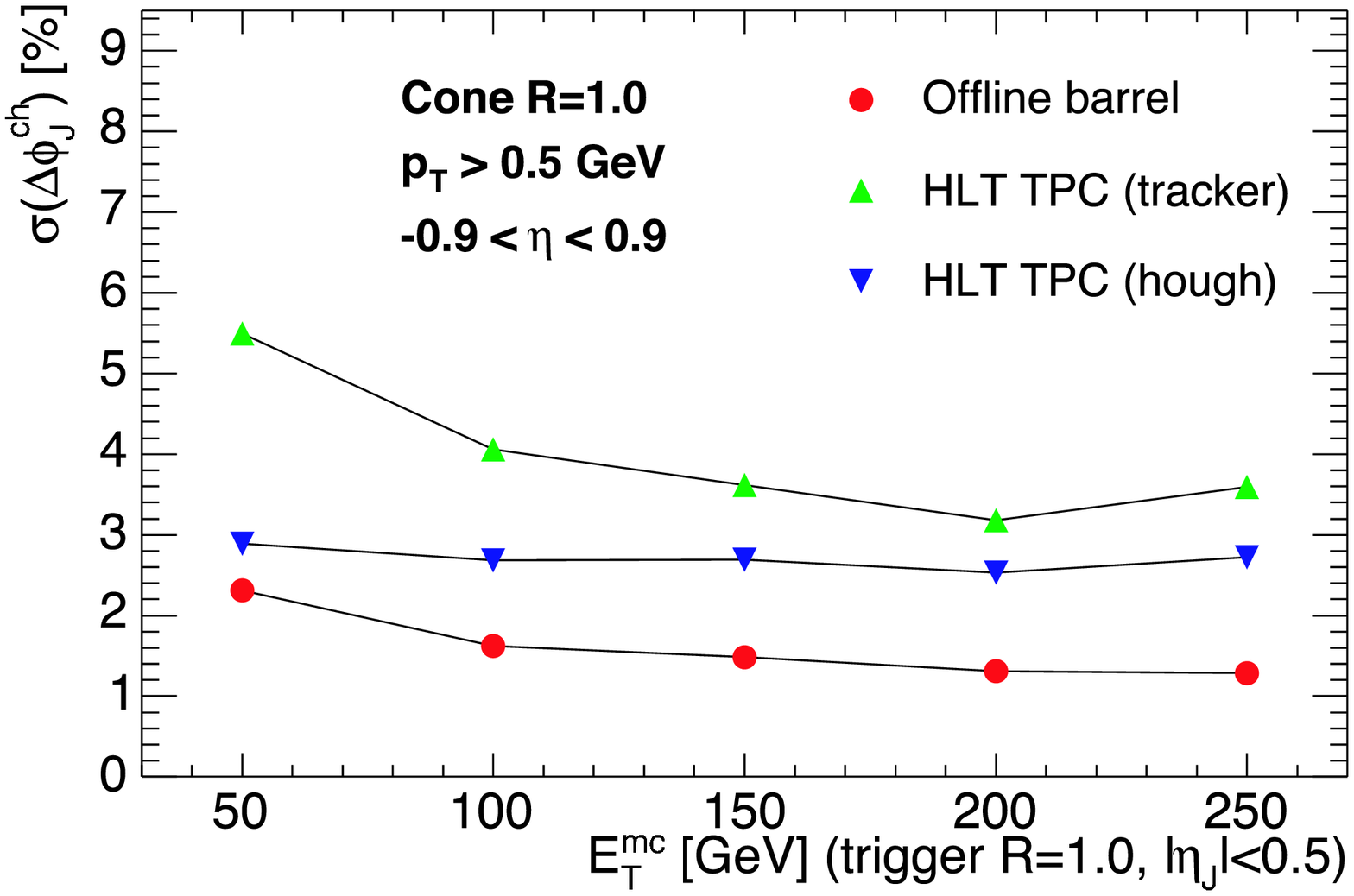}}
\hspace{0.5cm}
\subfigure[Spatial resolution in $\eta$ (tracked)]{
\label{chap5:fig:ppresetatracked}
\includegraphics[width=7cm]{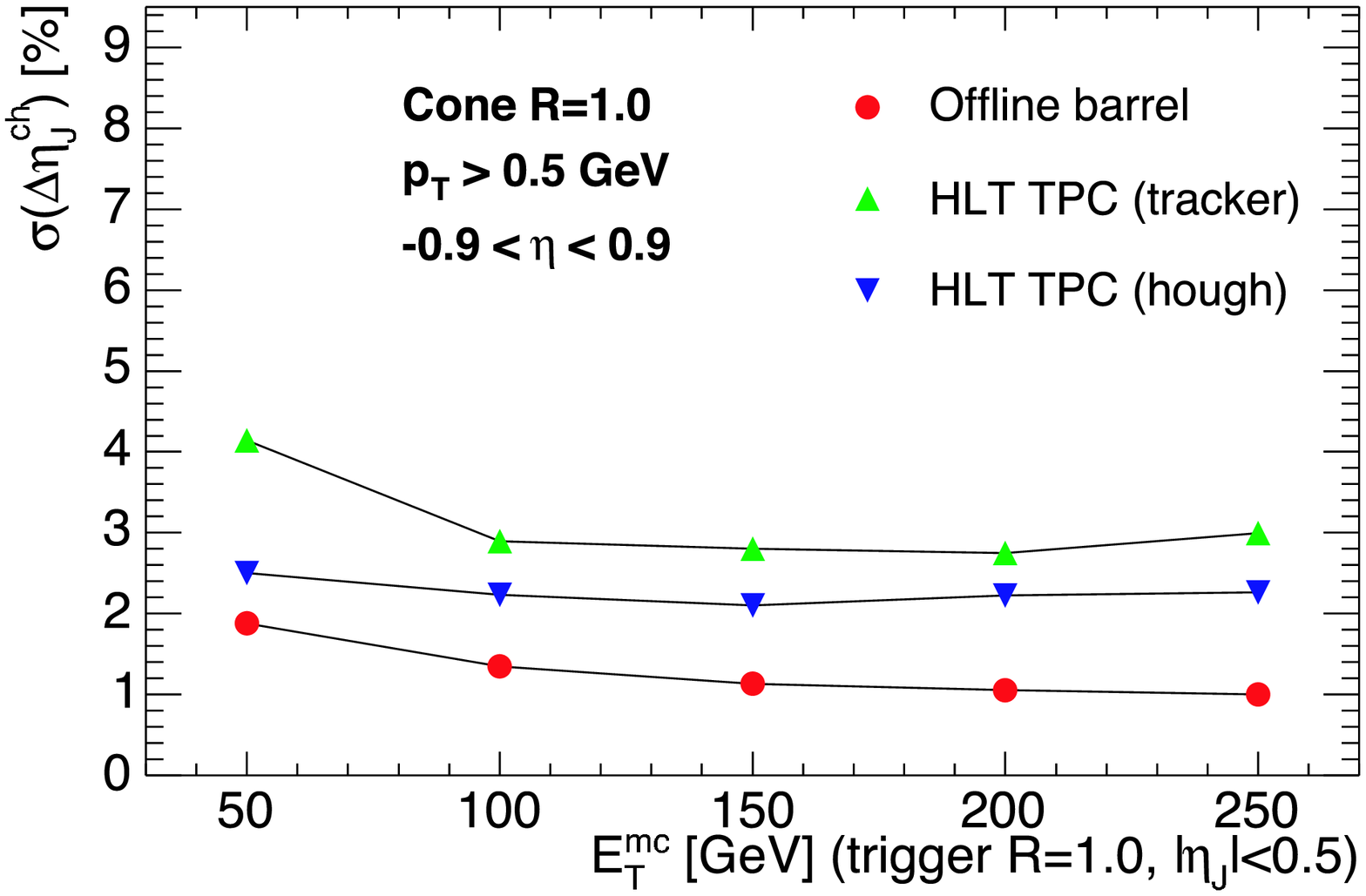}}
\end{center}
\vspace{-0.5cm}
\caption[xxx]{Spatial resolution of the reconstructed jets, 
$\sigma(\Delta \phi^{\rm ch}_{J})$~\subref{chap5:fig:ppresphitracked} 
and $\sigma(\Delta \eta^{\rm ch}_{J})$~\subref{chap5:fig:ppresetatracked}, 
both, as a function of the jet-trigger energy (Monte Carlo) for the different 
tracking cases. The spatial differences are measured relative to the charged jet
in the ideal barrel case, $\Delta \phi^{\rm ch}_{J}=\phi^{\rm rec}_{J}-\phi^{\rm ch}_{J}$ 
and $\Delta \eta^{\rm ch}_{J}=\eta^{\rm rec}_{J}-\eta^{\rm ch}_{J}$. The corresponding 
ideal case (ideal barrel) is shown in \fig{chap5:fig:ppspaceresideal}.}
\label{chap5:fig:ppspacerestracked}
\end{figure}

For the estimation of the space resolution we define the measured jet axis relative to the 
charged jet direction (ideal barrel), 
$\Delta \phi^{\rm ch}_{J}=\phi^{\rm rec}_{J}-\phi^{\rm ch}_{J}$ 
and $\Delta \eta^{\rm ch}_{J}=\eta^{\rm rec}_{J}-\eta^{\rm mc}_{J}$. 
The extracted widths are shown in \fig{chap5:fig:ppspacerestracked} as a 
function of $\et^{\rm mc}$. As expected, the offline combined tracking yields the best resolution of about 
$2$\% (increasing to $1$\%), whereas the online Hough track-finder has a constant resolution of $3$\% 
in $\phi$ and $2.5$\% in $\eta$ direction. At low Monte Carlo input there is an noticeably difference 
of almost a factor two between the online tracking based on cluster finding in the \ac{TPC} and the 
others, which reduces with increasing input energies.

\begin{figure}[htb]
\begin{center}
\subfigure[Spatial resolution in $\phi$ (tracked)]{
\label{chap5:fig:ppresphitrackedvstrigger}
\includegraphics[width=7cm]{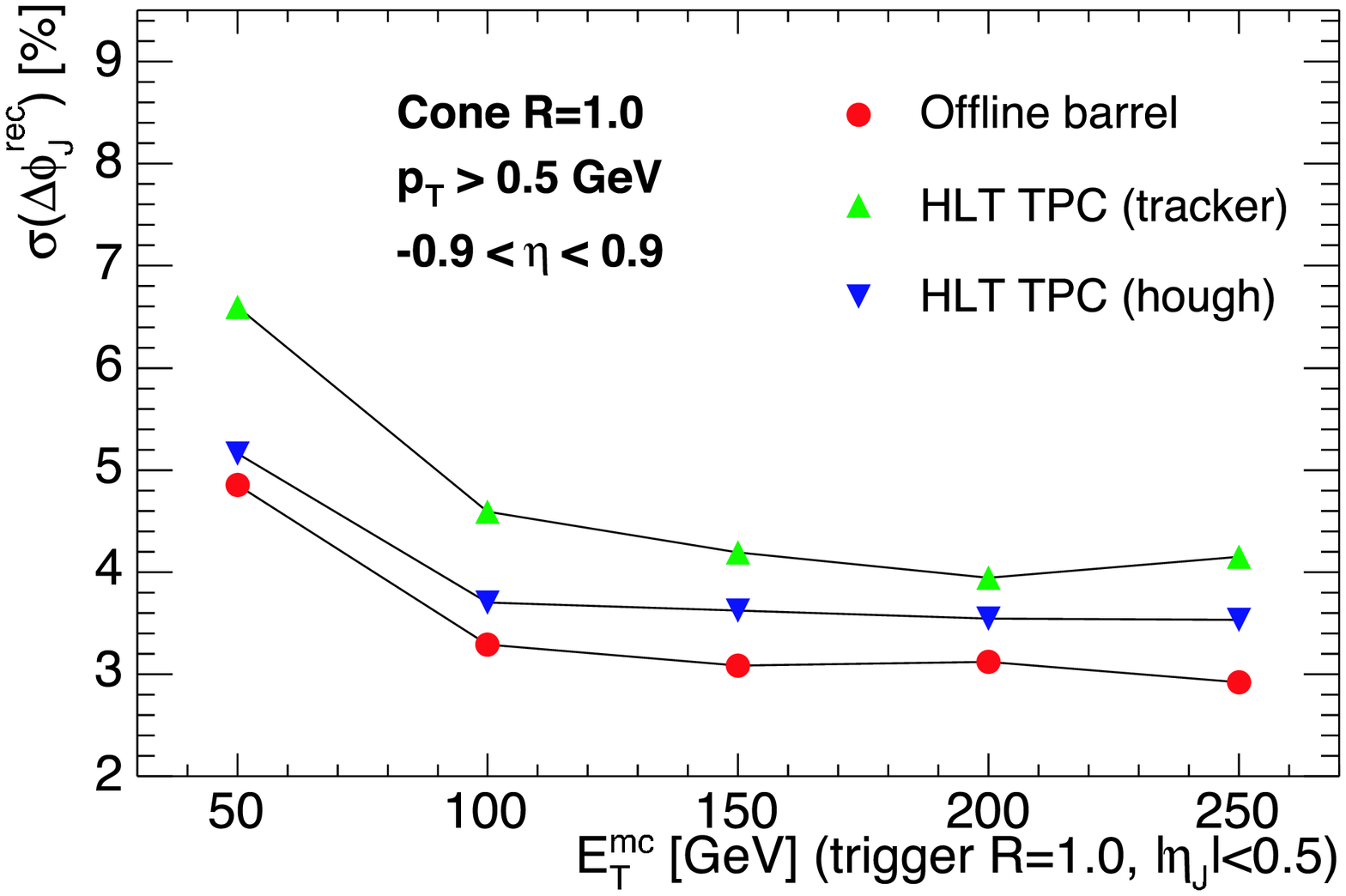}}
\hspace{0.5cm}
\subfigure[Spatial resolution in $\eta$ (tracked)]{
\label{chap5:fig:ppresetatrackedvstrigger}
\includegraphics[width=7cm]{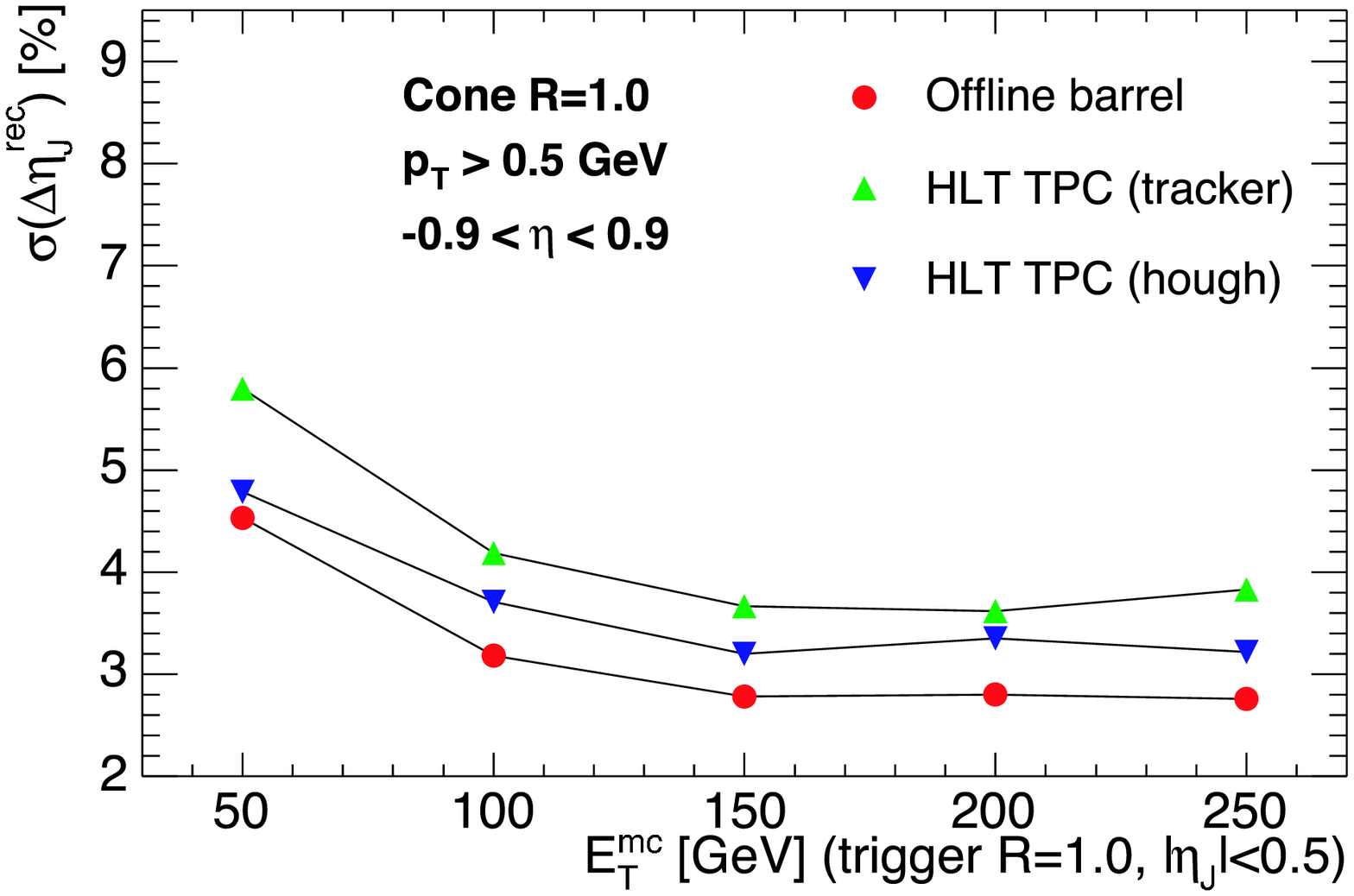}}
\end{center}
\vspace{-0.5cm}
\caption[xxx]{Spatial resolution of the reconstructed jets, 
$\sigma(\Delta \phi^{\rm rec}_{J})$~\subref{chap5:fig:ppresphitrackedvstrigger} 
and $\sigma(\Delta \eta^{\rm rec}_{J})$~\subref{chap5:fig:ppresetatrackedvstrigger}, 
both, as a function of the jet-trigger energy (Monte Carlo) for the different 
tracking cases. The spatial differences are measured relative to the trigger jet, 
$\Delta \phi^{\rm rec}_{J}=\phi^{\rm rec}_{J}-\phi^{\rm mc}_{J}$ 
and $\Delta \eta^{\rm rec}_{J}=\eta^{\rm rec}_{J}-\eta^{\rm mc}_{J}$.}
\label{chap5:fig:ppspacerestrackedvstrigger}
\end{figure}

The resolution, $\sigma(\phi_{J}^{\rm rec})$ and $\sigma(\eta_{J}^{\rm rec})$, 
relative to the trigger jet instead to the charged measured jet could be estimated from 
\fig{chap5:fig:ppspaceresideal} and \fig{chap5:fig:ppspacerestracked} according to 
\[\sigma(\phi_{J}^{\rm rec})=\sqrt{\sigma(\phi_{J}^{\rm mc})^2+\sigma(\phi_{J}^{\rm ch})^2}\] 
and similar for $\sigma(\eta_{J}^{\rm rec})$. 

\pagebreak
For convenience, the resolution is computed directly by comparing the reconstructed
jets with the input jets shown in \fig{chap5:fig:ppspacerestrackedvstrigger}.
We end up with a jet-reconstruction resolution of about $5$--$7$\% at low and $3$--$4$\% at high jet 
energies. However, it is very likely that the resolution in \pp~collisions will degrade  
once the additional pile-up events will be taken into account in the simulation.
\fi

\section{Jet reconstruction in Pb--Pb for fixed energy}
\label{chap5:jetreconpbpb}
\ifreconpbpb
Jet reconstruction in \PbPb~collisions at \ac{LHC} energies may be affected by the 
bulk of soft and semi-hard particles, which make up the background, \ie~the
`underlying event' in terms of \pp\ physics. As outlined in \psect{chap2:partmult}
the expected multiplicity and, thus, also the `hardness' of the background
are at present the main unknown. For the estimation of the jet reconstruction 
capabilities of \ac{ALICE} in central \PbPb~collisions we chose \acs{HIJING} 
for the generation of background events, in which \acs{PYTHIA} events containing
jet signals are implanted. \acs{HIJING},  mainly used with default options (\psect{app:hijing}), 
generates a rather soft background of about $6000$ charged particles, $2000$ for 
$\pt>0.5~\gev$, in $\abs{\eta}<0.5$. Corresponding to the sample of signal events 
discussed above for \pp, we generate 3000 central ($0$--$10$\%) background events in which 
to embed the jets at different energies. 

\begin{figure}[htb]
\begin{center}
\includegraphics[width=12cm]{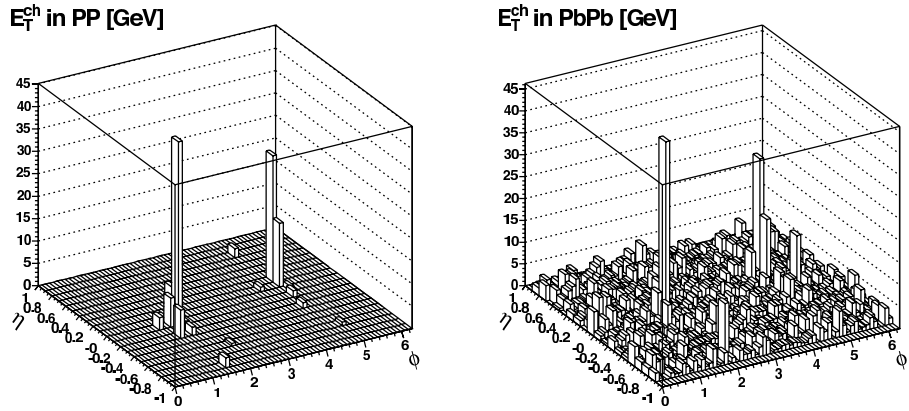}
\end{center}
\vspace{-0.4cm}
\caption[xxx]{Charged tower configuration in the $\eta$-$\phi$ plane for a $100~\gev$ jet 
in \pp~(left) and for the same jet embedded in a central \PbPb~event (right). In both
cases a particle $\pt$-cut of $0.5~\gev$ has been applied.}
\label{chap5:fig:embedded}
\end{figure}

\begin{figure}[htb!]
\vspace{0.5cm}
\begin{center}
\includegraphics[width=8cm]{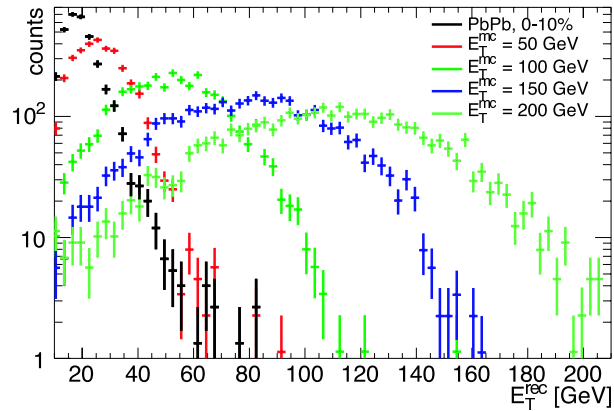}
\end{center}
\vspace{-0.4cm}
\caption[xxx]{Distribution of reconstructed (charged) jets found in the signal classes 
compared to pure background ($0$--$10$\% \PbPb, \acs{HIJING}) as a function of the 
reconstructed energy for the ideal barrel. $R=0.3$ and $\pt\ge2~\gev$ are used.}
\label{chap5:fig:mixedillustration}
\end{figure}

Since, so far, no detector response is simulated, the mixed event contains all particles 
of the signal and as well of the background event, both, above a $\pt$-cut and within 
the central \ac{ALICE} acceptance of $-0.9<\eta<0.9$. The initial tower configuration of 
the cone finder (\psect{app:conefinder}) is illustrated in \fig{chap5:fig:embedded} for a 
$100~\gev$ jet in the case of the ideal barrel. Clearly, in the chosen example the signal 
sticks out of the background (\cf \pfig{chap3:fig:cdfjets}). 

However, for illustration of the situation in the ideal barrel case, we show in 
\fig{chap5:fig:mixedillustration} the distribution of reconstructed (charged) jets 
corresponding to the signal classes and pure background as a function of the 
reconstructed energy. 
To anticipate the conditions for mixed events 
(see below) we use $R=0.3$ and $\pt\ge2~\gev$ for all classes. 
In the case of the ideal barrel, the distribution arising from pure background partially 
overlays the distribution arising from jets of about $50~\gev$.
Therefore, one qualitatively may expect that the recognition and reconstruction 
of these, rather low, energy jets within the underlying heavy-ion background will be 
degraded compared to pure \pp. In the following, we will quantify this observation,
mainly focusing on the case of ideal barrel tracking (charged particles). 

\subsection{Background fluctuations}
A single jet with a cone of $R=0.7$ covers $14$\% of the \ac{ALICE} central
acceptance. Assuming only about 5000 charged particles with on average 
$0.5~\gev$ within the acceptance, the expected energy inside the cone amounts
on average to about $350~\gev$ arising from uncorrelated background alone. 

We start by comparing the average energy content,
\begin{equation*}
\et^{\rm jet}(r) = \frac{1}{\rm N_{\rm jets}} \sum_{\rm jets} \Etj(0,r)\;,
\end{equation*}
inside cones of real jets centered at jet axes, \cf~\eq{chap3:eq:intjetshape}, 
with the average content of cones centered at randomly chosen axes in background events,
\begin{equation*}
\et^{\rm bg}(r) = \frac{1}{\rm N_{\rm axes}} \sum_{\rm rand.axes} \et^{\rm bg}(0,r)\;.
\end{equation*}

\begin{figure}[htb]
\begin{center}
\subfigure[Average background]{
\label{chap5:fig:etbackcomp}
\includegraphics[width=7cm]{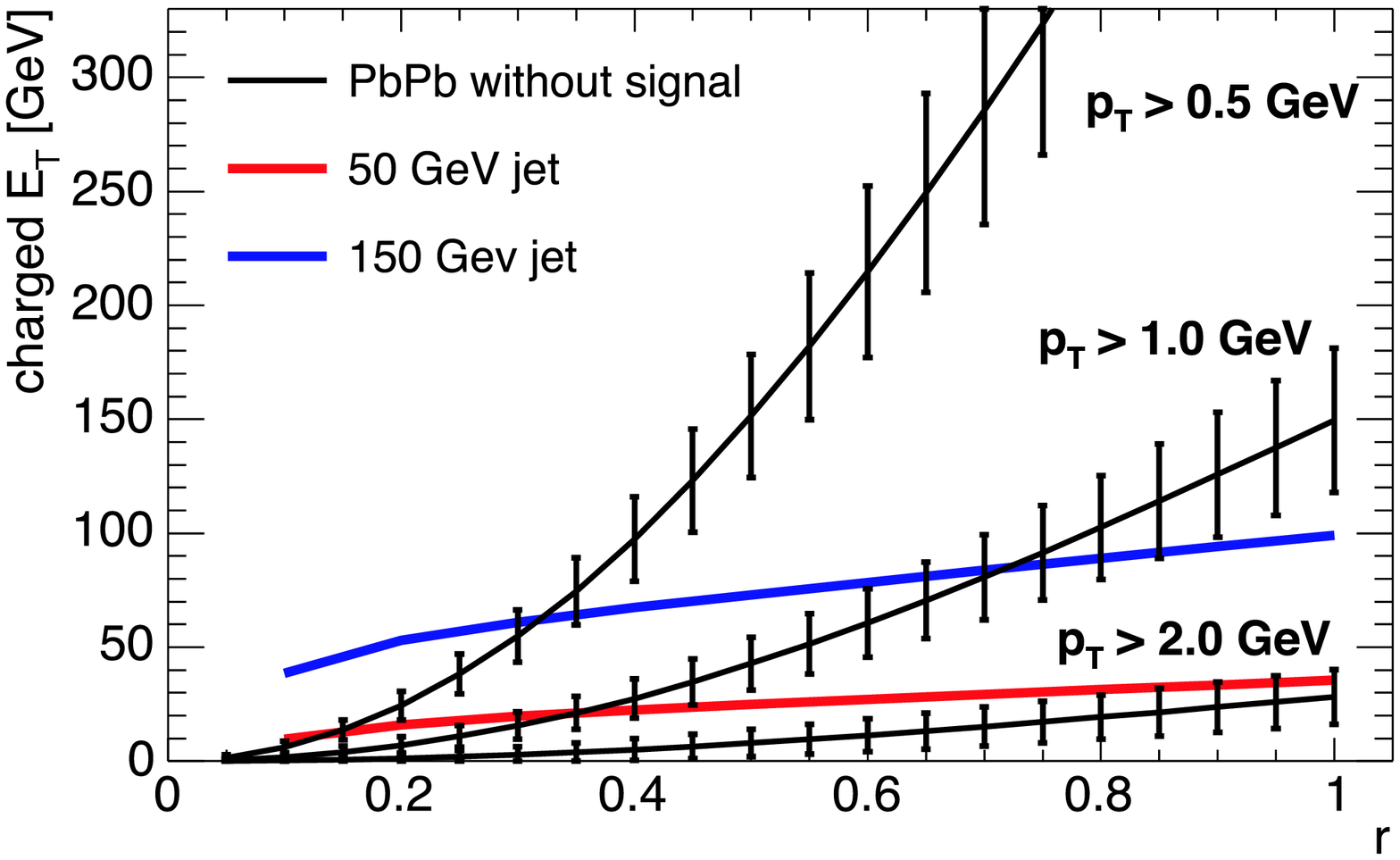}}
\hspace{0.5cm}
\subfigure[Root mean squared background]{
\label{chap5:fig:rmsbackcomp}
\includegraphics[width=7cm]{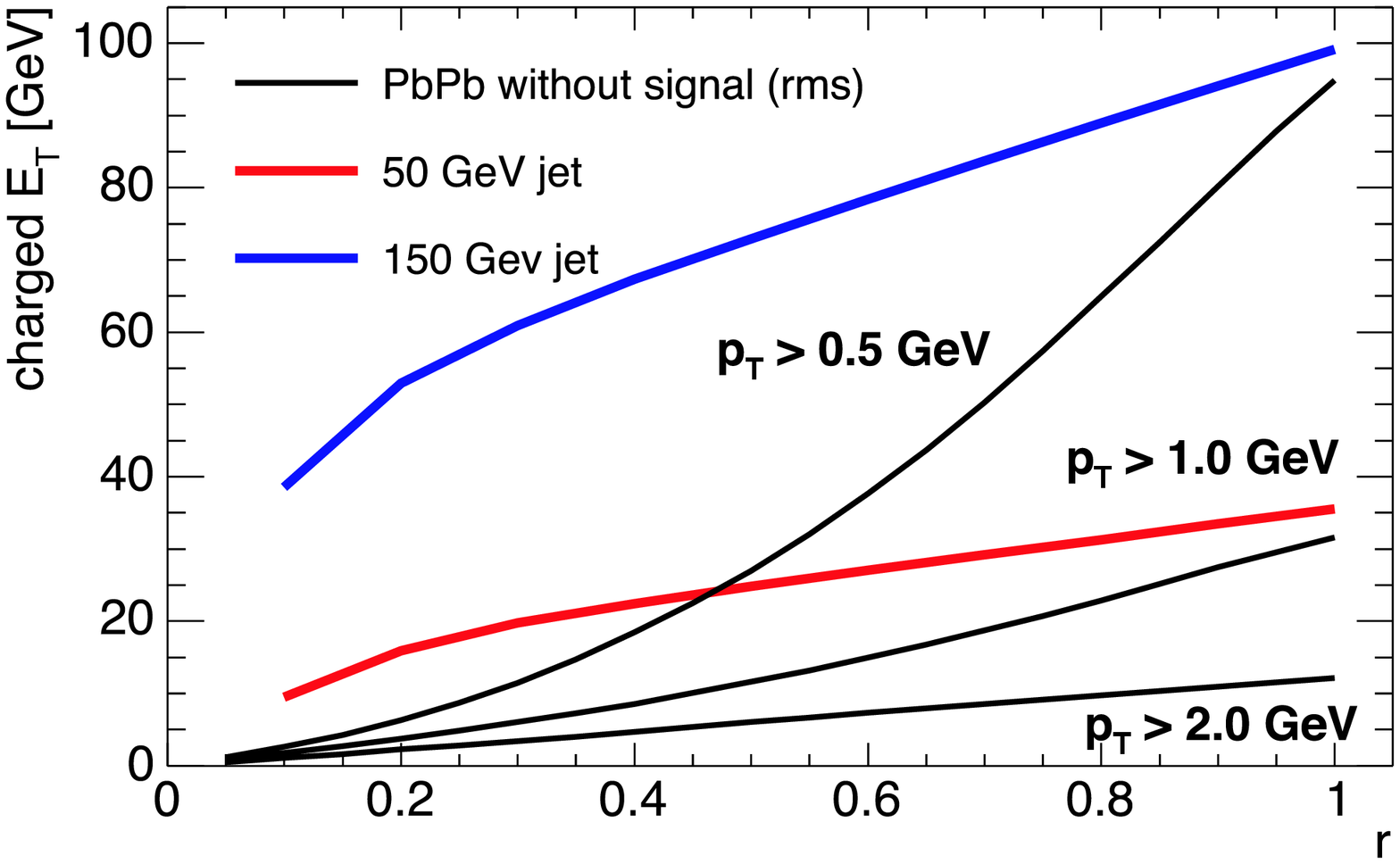}}
\end{center}
\vspace{-0.5cm}
\caption[xxx]{Average transverse energy content of $50$ and $150~\gev$ jets, 
$\et^{\rm jet}(r)$, compared to the average~\subref{chap5:fig:etbackcomp} and 
the rms~\subref{chap5:fig:rmsbackcomp} content of randomly chosen cones in the 
background ($0$--$10$\% \PbPb, \acs{HIJING}) as a function of $r$ in the case of
ideal barrel. The background is suppressed by $\pt$-cuts of $0.5$, $1$ and $2~\gev$.}
\label{chap5:fig:etrmsbackcomp}
\end{figure}

\Fig{chap5:fig:etrmsbackcomp} shows the average energy content of $50$ and 
$150~\gev$ jets compared with the mean and rms of the background as a function 
of the cone size, $r=\sqrt{(\phi-\phi_{\rm c})^2+(\eta-\eta_{\rm c})^2}$. 
The expected behavior of $\et^{\rm bg}(r)$ to be proportional to $r^2$ and 
its rms, $\Delta \et^{\rm bg}(r)$, to be proportional to $r$ are approximately 
true. The $\pt$-cuts of $0.5$, $1$ and  $2~\gev$ have been applied to suppress 
the uncorrelated background. It is obvious that in \PbPb~at $\snn=5.5~\tev$ one can not 
apply a cone finder with $R=0.7$ or higher as it is typically done in \pp, since for 
these radii the energy in the jet cone is dominated by the background; furthermore the 
fluctuations of the background are comparable to the jet energy. Instead, one must reduce 
the cone size to about $R=0.3$, and, in addition, apply a particle $\pt$-cut of $1$ or 
$2~\gev$ to resolve jet energies of the order of $50~\gev$ and below. However, the 
$\pt$-cut is less effective as one would think, since the reduction of the number of low-$\pt$ 
particles increases the mean and the rms inside the cone (assuming uncorrelated particle
production).

\begin{figure}[htb]
\begin{center}
\includegraphics[width=10cm]{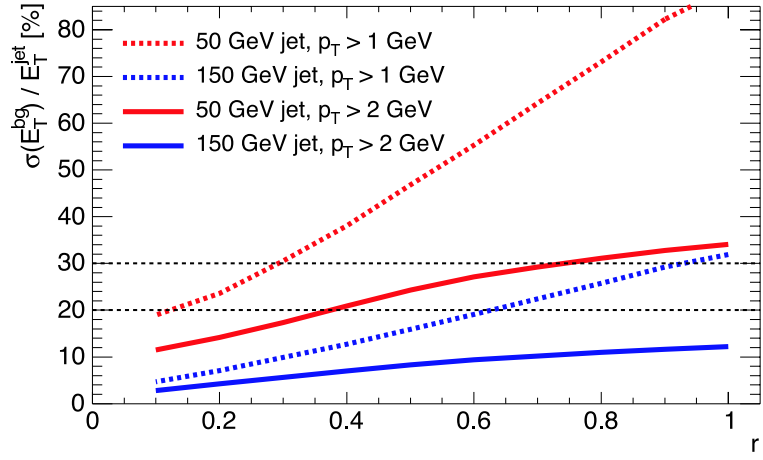}
\end{center}
\vspace{-0.5cm}
\caption[xxx]{The ratio of background fluctuations over jet-signal energy, 
$\sigma(\et^{\rm bg})/\Etj$, for $50$ and $150~\gev$ jets as a function of $r$ 
for the case of the ideal barrel. The $\pt$-cuts of $1$ and $2~\gev$ are applied 
only in the background events ($0$--$10$\% \PbPb, \acs{HIJING}).}
\label{chap5:fig:rmsoveretbackground}
\end{figure}

The ratio of the background fluctuations over the jet-signal energy, 
$\sigma(\et^{\rm bg})/\Etj$, for the $50$ and $150~\gev$ jets as a function 
of $r$ is shown in \fig{chap5:fig:rmsoveretbackground}. 
Since the jet energy resolution in \PbPb\ is roughly given by the resolution 
of the jet signal in \pp\ and the fluctuations of the background,
\[\sigma(\et^{\rm rec})=\sqrt{\sigma(\Etj)^2+\sigma(\et^{\rm bg})^2}\;,\]
one should aim for $\sigma(\et^{\rm bg})/\Etj\le\sigma(\Etj)/\Etj$.
From \fig{chap5:fig:ppetresideal} we know that the latter is about $30$\% for 
ideal barrel tracking (and $20$\% if ideal electromagnetic calorimetry is included).
Thus, in \PbPb\ jet reconstruction at $R=0.3$ with $\pt$-cut of $2~\gev$ seems to 
be preferable, at least, for a first pass to identify jets (\eg~for the jet
trigger in the \ac{HLT} system). In a second pass for a refined analysis one 
might increase the cone size and decrease the $\pt$-cut.

\subsection{Out-of-cone fluctuations}
So far, we did not consider the effect of the reduced cone size and the $\pt$-cut 
on the jet signal. Since we are aiming for jet production at mid-pseudo-rapidity, 
both restrictions are correlated and deteriorate the measured signal. 
Though it has been shown that for $\Etj\ge50~\gev$ on average $80$\% of the charged 
energy is contained within a cone radius of about 0.3~(\cf \fig{chap3:fig:intjetshape}),
on jet-by-jet basis particles produced outside of the reduced cone may significantly 
degrade the jet-energy resolution.

\begin{figure}[htb]
\begin{center}
\ifarxiv
\includegraphics[width=12.5cm]{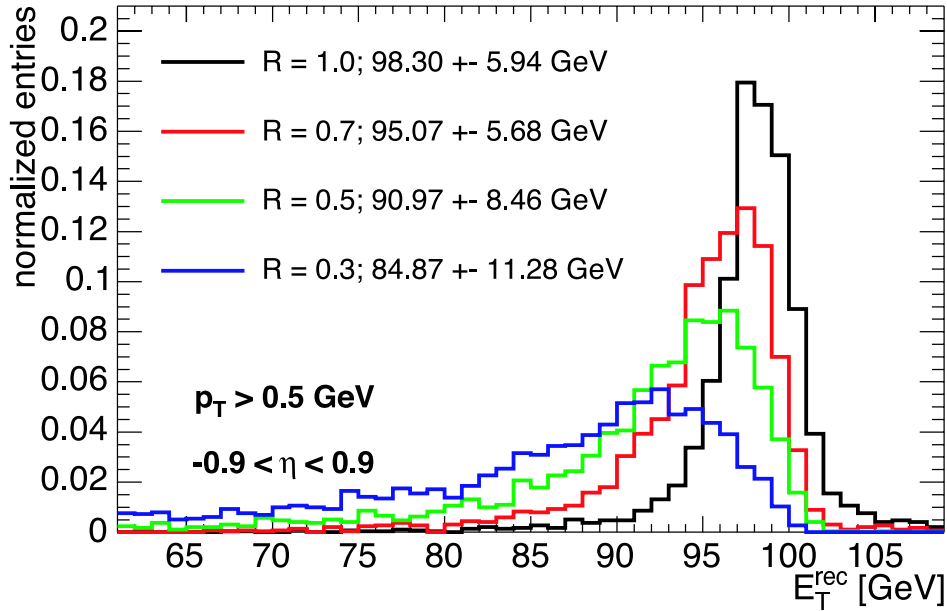}
\else
\includegraphics[width=13.2cm]{cOutOfConeFlucPP}
\fi
\end{center}
\vspace{-0.5cm}
\caption[xxx]{Out-of-cone fluctuations for fixed $\et^{\rm mc} = 100~\gev$ and cone radii
of $R=0.3$, $0.5$, $0.7$ and $1.0$ in the case of an ideal detector. The 
mean and rms values of the $\et^{\rm rec}$ distribution are shown in the legend.}
\label{chap5:fig:outofconeflucpp}
\end{figure}

\begin{figure}[htb!]
\vspace{1cm}
\begin{center}
\ifarxiv
\includegraphics[width=12.5cm]{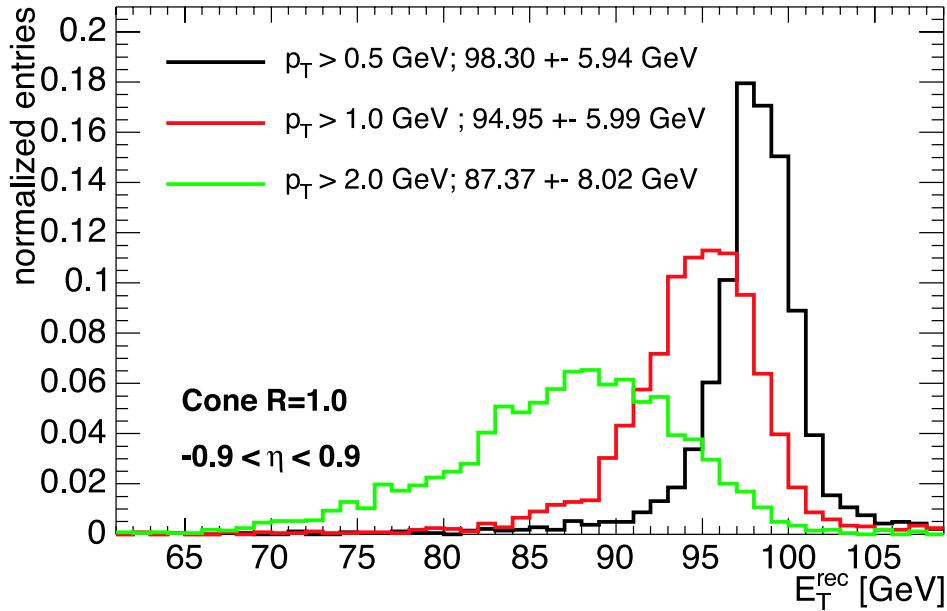}
\else
\includegraphics[width=13.2cm]{cOutOfConeFlucPtPP}
\fi
\end{center}
\vspace{-0.5cm}
\caption[xxx]{Effect of the $\pt$-cut on the reconstructed energy distribution for 
fixed $\et^{\rm mc} = 100~\gev$ and $R=1.0$ in the case of an ideal detector. The 
mean and rms values of the $\et^{\rm rec}$ distribution are shown in the legend.}
\label{chap5:fig:outofconeflucpppt}
\end{figure}

\enlargethispage{1cm}
The out-of-cone fluctuations are simplest illustrated by plotting the reconstructed 
transverse energy distribution, $\et^{\rm rec}$, for fixed input energy using different 
radii of the cone finder. As shown in \fig{chap5:fig:outofconeflucpp} for 
$\et^{\rm mc} = 100~\gev$ the mean reconstructed values are reduced by up to $15$\%; 
furthermore the distribution is slightly broadened and contains a very long tail 
to smaller reconstructed energies reaching even zero. Instead, we observe in 
\fig{chap5:fig:outofconeflucpppt}, where for fixed cone radius of $R=1$ the $\pt$-cut 
varied that the shape of the distribution is preserved, while the mean energy 
reduces to about the same amount as the out-of-cone fluctuations. Since the shape of 
the distribution remains Gaussian, the energy resolution is less affected than for 
reduced cone sizes. It is questionable, whether one should take out-of-cone fluctuations 
into account for determination of the quality of the jet reconstruction. High-energy jets, 
which strongly fluctuate to the left, will be overwhelmed by the more likely produced 
jets at lower energies, and, thus, will simply not be detected; a fact, which contributes 
to jet detection (in-)efficiency, rather than to the quality of the jet reconstruction 
algorithm itself.

\begin{figure}[htb]
\begin{center}
\subfigure[Mean energy fraction (ideal)]{
\label{chap5:fig:ppetmeanidealcut}
\includegraphics[width=7cm]{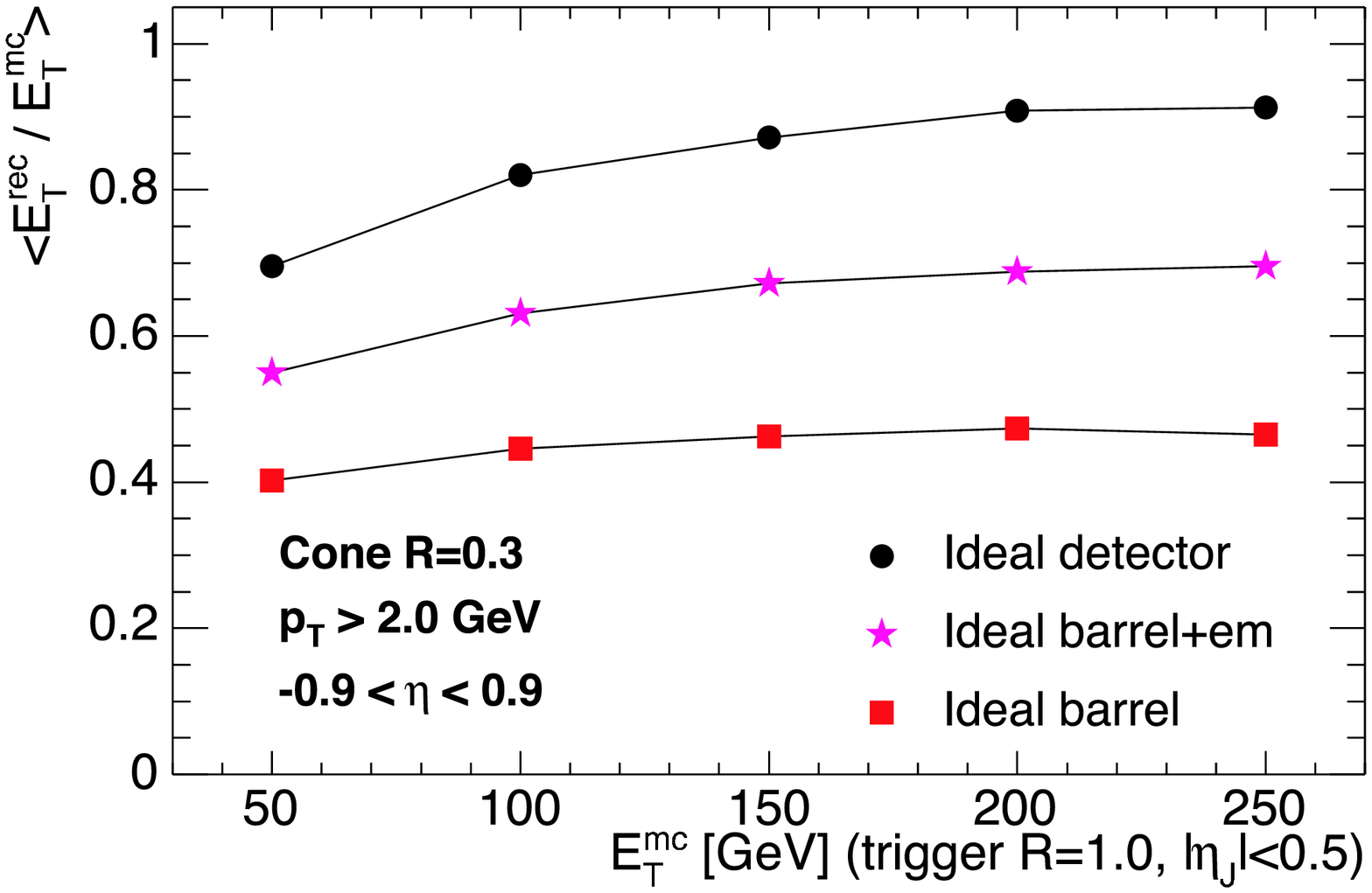}}
\hspace{0.5cm}
\subfigure[Energy resolution (ideal)]{
\label{chap5:fig:ppetresidealcut}
\includegraphics[width=7cm]{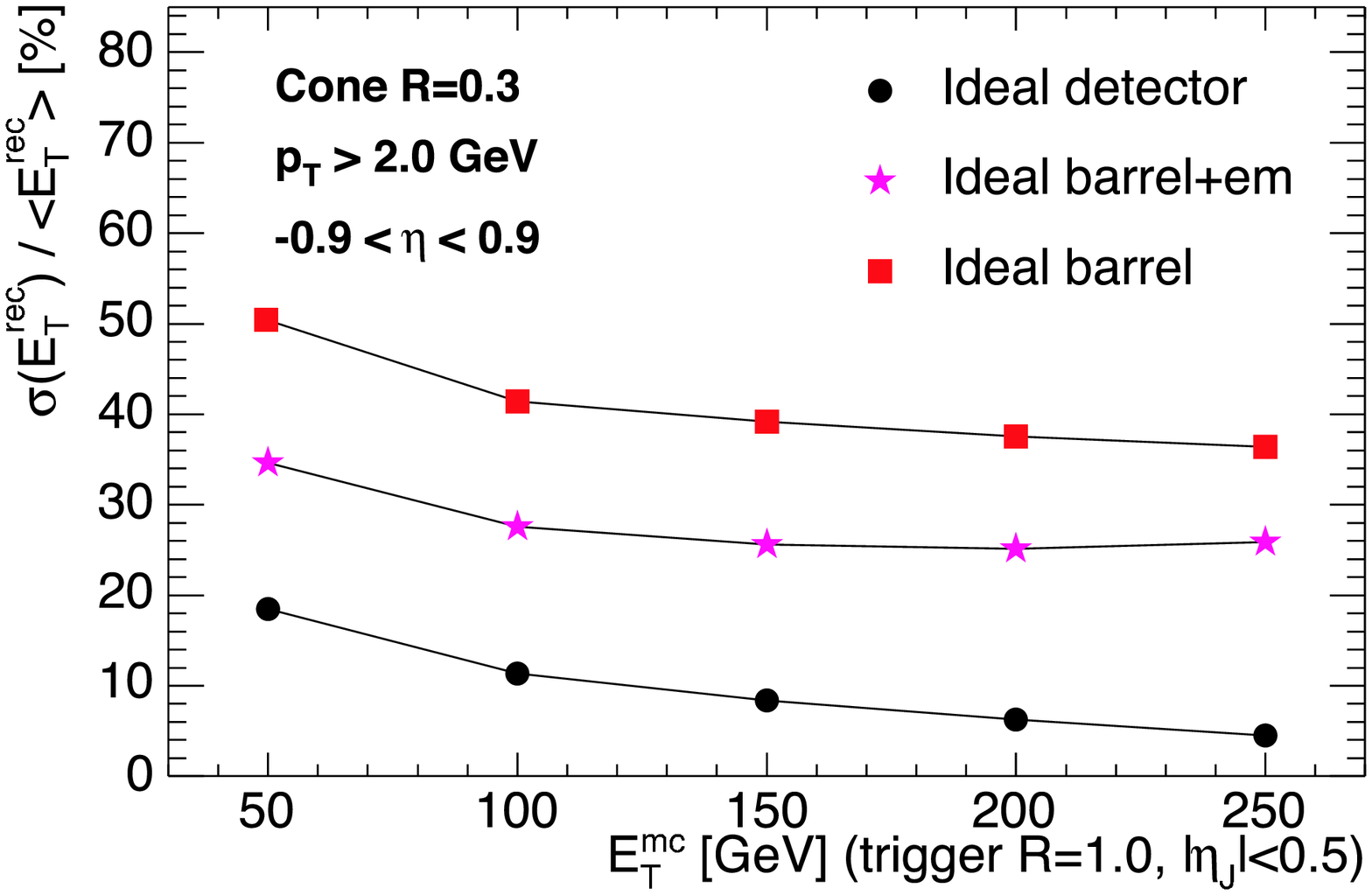}}
\hfill
\subfigure[Mean energy fraction (tracked)]{
\label{chap5:fig:ppetmeantrackedcut}
\includegraphics[width=7cm]{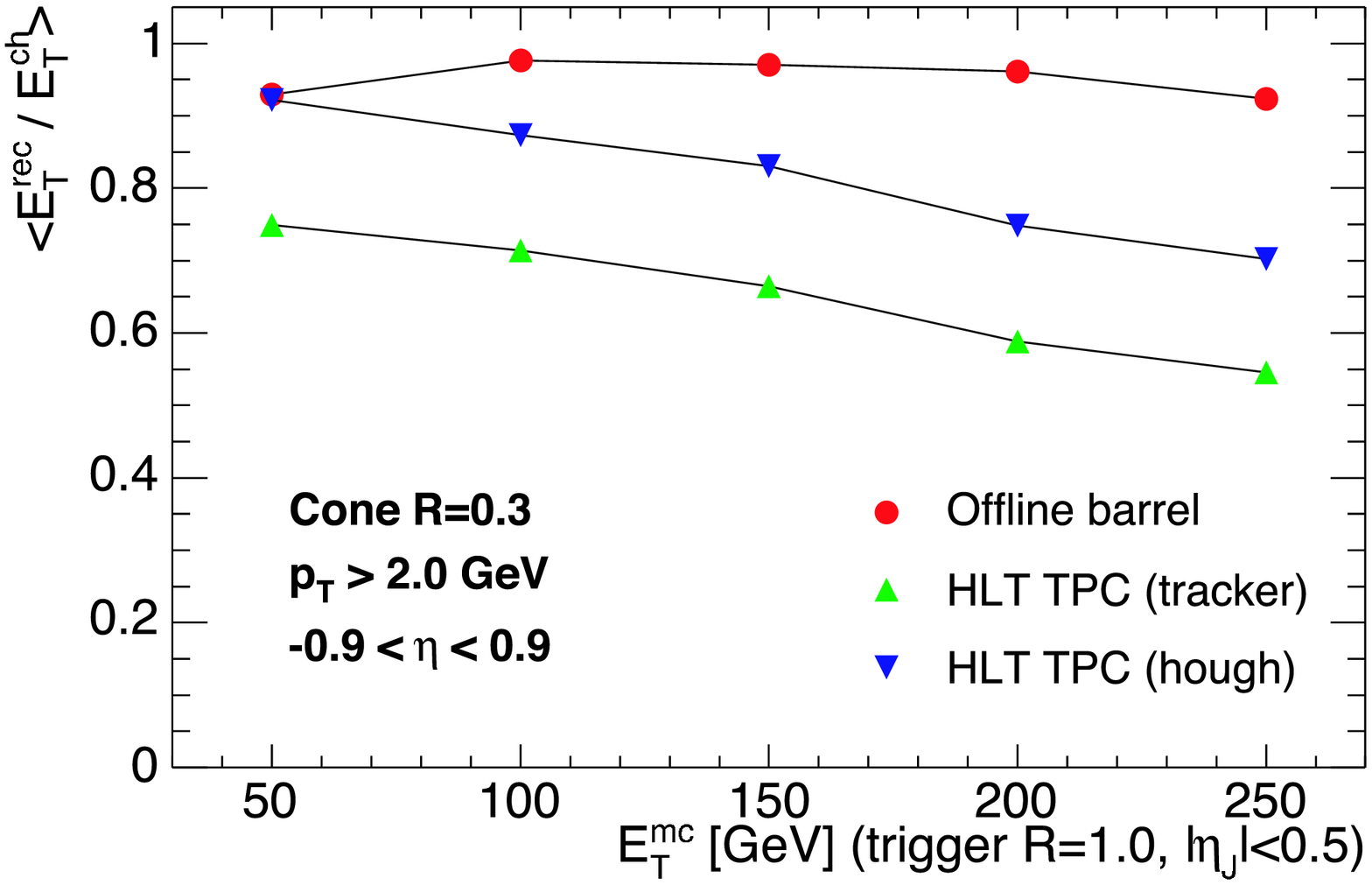}}
\hspace{0.5cm}
\subfigure[Energy resolution (tracked)]{
\label{chap5:fig:ppetrestrackedcut}
\includegraphics[width=7cm]{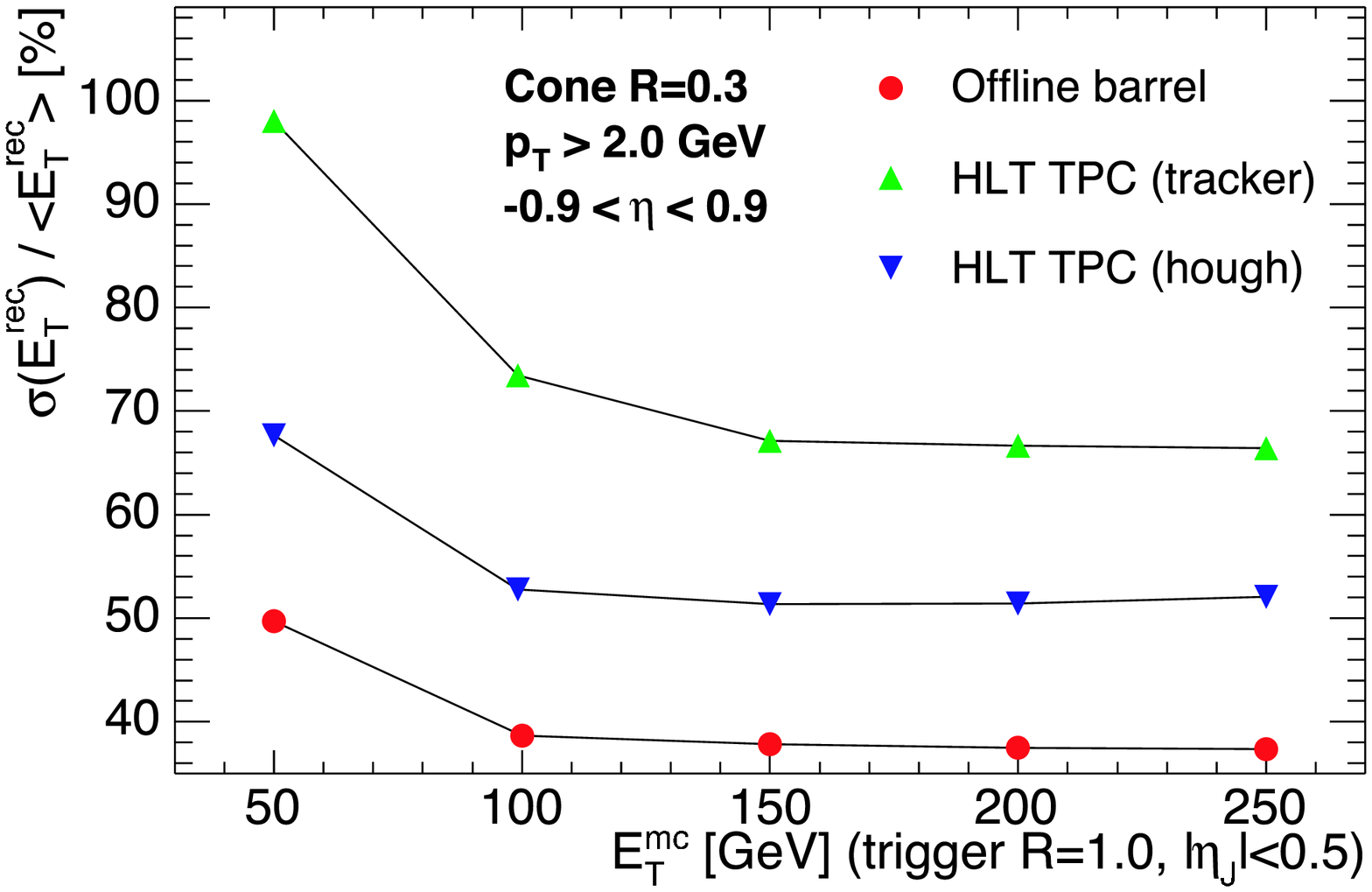}}
\end{center}
\vspace{-0.5cm}
\caption[xxx]{Average fraction of reconstructed jet energy, 
$\av{\et^{\rm rec}/\et^{\rm mc}}$~\subref{chap5:fig:ppetmeanidealcut}, 
and corresponding energy resolution, 
$\sigma(\et^{\rm rec})/\av{\et^{\rm rec}}$~\subref{chap5:fig:ppetresidealcut}, 
for ideal cases, as well as,
average fraction of reconstructed charged-jet energy, 
$\av{\et^{\rm rec}/\et^{\rm ch}}$~\subref{chap5:fig:ppetmeantrackedcut}, 
and corresponding energy resolution, 
$\sigma(\et^{\rm rec})/\av{\et^{\rm rec}}$~\subref{chap5:fig:ppetrestrackedcut}, 
for different tracking cases as a function of the jet-trigger energy (Monte Carlo). 
Opposed to \fig{chap5:fig:ppetmeanresideal} and \fig{chap5:fig:ppetmeanrestracked} 
a cone size of $R=0.3$ and a $\pt$-cut of $2~\gev$ are used. Signal events only 
without embedding into background are taken.}
\label{chap5:fig:ppetmeanrescut}
\end{figure}

Nevertheless, since in practice jet finding and quality of jet reconstruction are not 
independent and since one must understand the performance and introduced biases of the online 
trigger, it is worth to study the impact of the fluctuations in \pp\ at first. Presented in 
\fig{chap5:fig:ppetmeanrescut} is the combined effect as a function of $\et^{\rm mc}$ by applying 
both restrictions, $R=0.3$ and $\pt\ge2~\gev$, to the samples of signal events (\eg~\pp\ without 
mixing). The result is quite striking. Even in the case of an ideal detector the mean of the
reconstructed jet-energy fraction drops to about $70$\% at $50~\gev$, though, as expected, 
increasing with increasing input energy to the level of $95$\%. In the case of ideal barrel tracking 
the mean reconstructed fraction drops by about $20$\% to the level of $40$\% 
(\cf~\fig{chap5:fig:ppetmeanresideal}). 
The resolution decreases to about $50$\% for $50~\gev$ and about $40$\% for the higher energies. 
Compared to ideal barrel, offline tracking is only slightly affected, but still the overall 
mean reduces to $37$\% with a resolution of $50$\% at $\et^{\rm mc}=50~\gev$. The impact of the
additional cuts on the online tracking is more significant (\cf~\fig{chap5:fig:ppetmeanrestracked}). 
The mean reconstructed energy fraction yields around $35$\% ($30$\%) at low energies decreasing 
to $30$\% ($25$\%), whereas the resolution is about $50$\% ($60$\%) for the hough (tracker) case. 
At $\et^{\rm mc}=50~\gev$ the online resolution degrades to $70$\% ($100$\%), which might impose 
a problem for the \ac{HLT} trigger since jets at that and lower energies are abundantly produced. 

\begin{figure}[htb]
\begin{center}
\subfigure[Spatial resolution in $\phi$ (tracked)]{
\label{chap5:fig:ppresphitrackedvstriggercut}
\includegraphics[width=7cm]{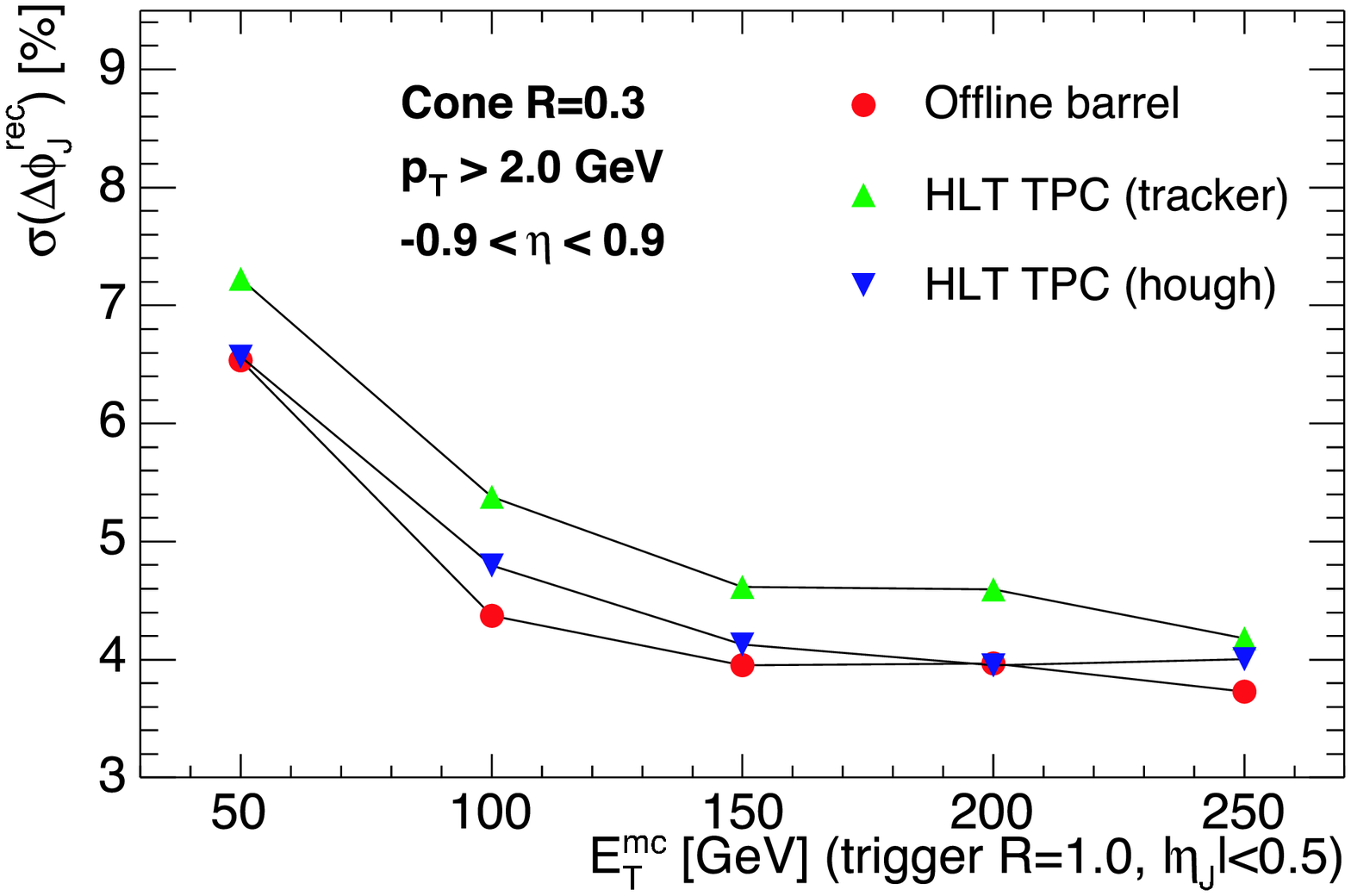}}
\hspace{0.5cm}
\subfigure[Spatial resolution in $\eta$ (tracked)]{
\label{chap5:fig:ppresetatrackedvstriggercut}
\includegraphics[width=7cm]{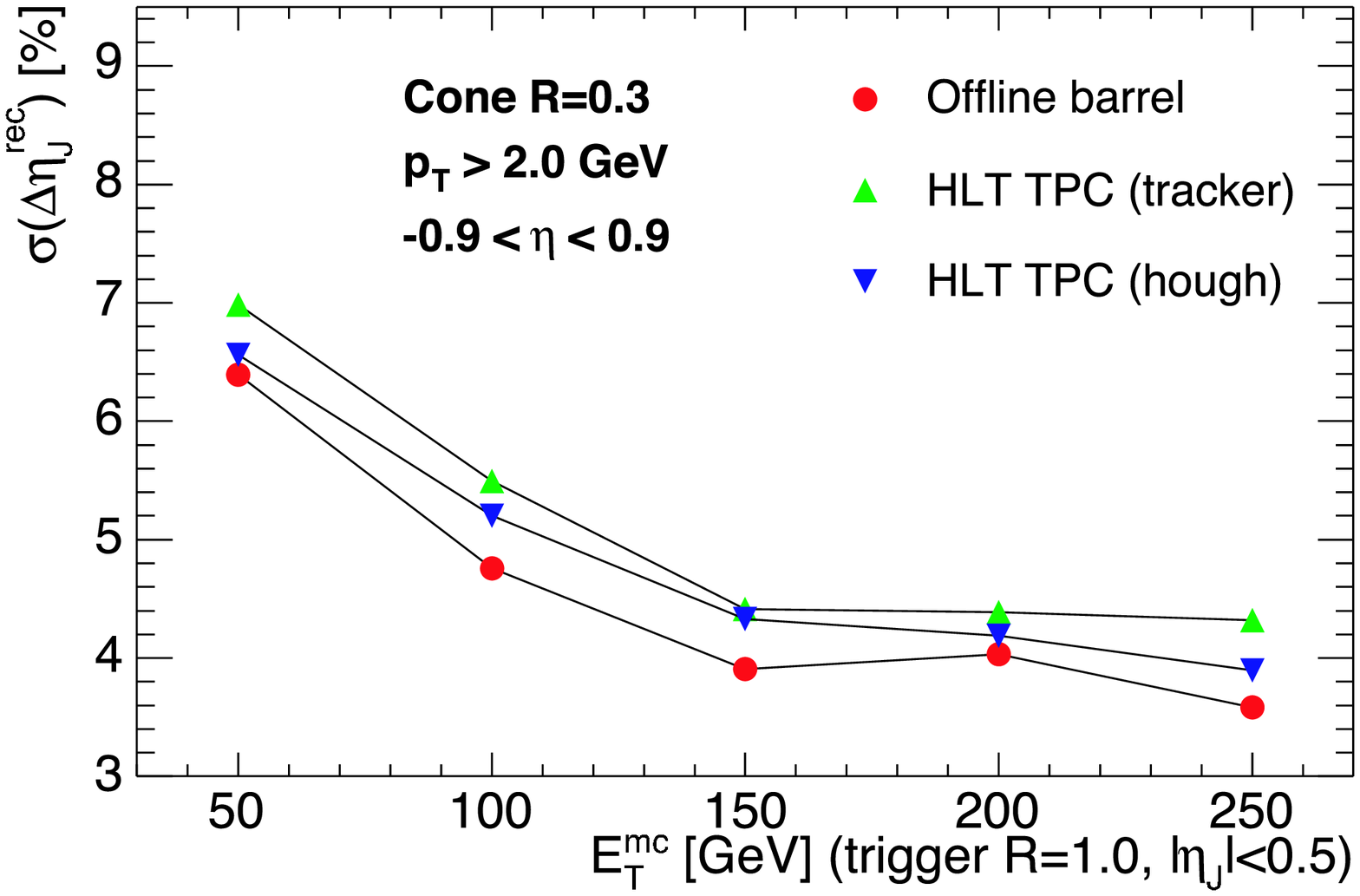}}
\end{center}
\vspace{-0.5cm}
\caption[xxx]{Spatial resolution of the reconstructed jets in signal events only, 
$\sigma(\Delta \phi^{\rm rec}_{J})$~\subref{chap5:fig:ppresphitrackedvstriggercut} 
and $\sigma(\Delta \eta^{\rm rec}_{J})$~\subref{chap5:fig:ppresetatrackedvstriggercut}, 
relative to the trigger jet a function of $\et{\rm mc}$. 
Opposed to \fig{chap5:fig:ppspacerestrackedvstrigger} a cone size of $R=0.3$ and a $\pt$-cut 
of $2~\gev$ are used.}
\label{chap5:fig:ppspacerestrackedvstriggercut}
\end{figure}

In \fig{chap5:fig:ppspacerestrackedvstriggercut} the spatial resolution, $\Delta \phi^{\rm rec}_{J}$ 
and $\Delta \eta^{\rm rec}_{J}$, are shown for the signal events (\eg~\pp\ without mixing) in the 
different tracking cases. Since high-energy jets typically contain at least $8$--$10$ charged
particles at high-$\pt$, the spatial resolution only little worsens compared to unconstrained jet 
reconstruction (\cf \fig{chap5:fig:ppspacerestrackedvstrigger}). The biggest change is again 
observed at lowest jet energies. There the resolution is about $7$\%, but increases to about $4$\% 
at higher energies. 

\subsection{Ideal detector response}
The findings of the last section refer to pure jet events, \ie~\pp\ only. 
In this section, we repeat the analysis for signal events, which are embedded 
into $0$--$10$\% central \PbPb\ events. Since for mixed events no realistic 
detector response is yet available, we use simulations at the Monte Carlo level 
for the different ideal scenarios defined in \sect{chap5:idealdetresponse}.

\begin{figure}[htb]
\begin{center}
\subfigure[Mean energy fraction (ideal)]{
\label{chap5:fig:mixedetmeanideal}
\includegraphics[width=7cm]{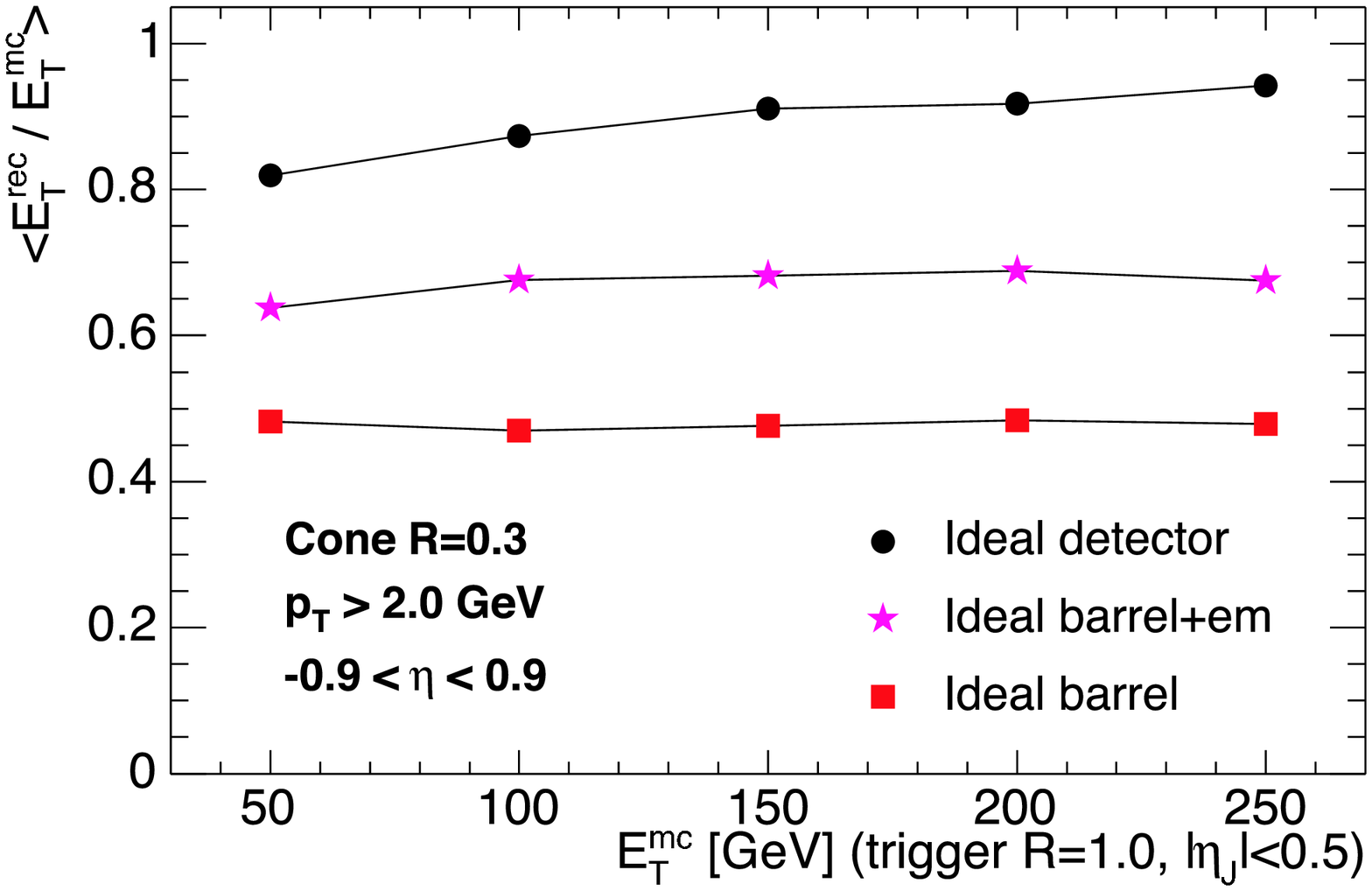}}
\hspace{0.5cm}
\subfigure[Energy resolution (ideal)]{
\label{chap5:fig:mixedetresideal}
\includegraphics[width=7cm]{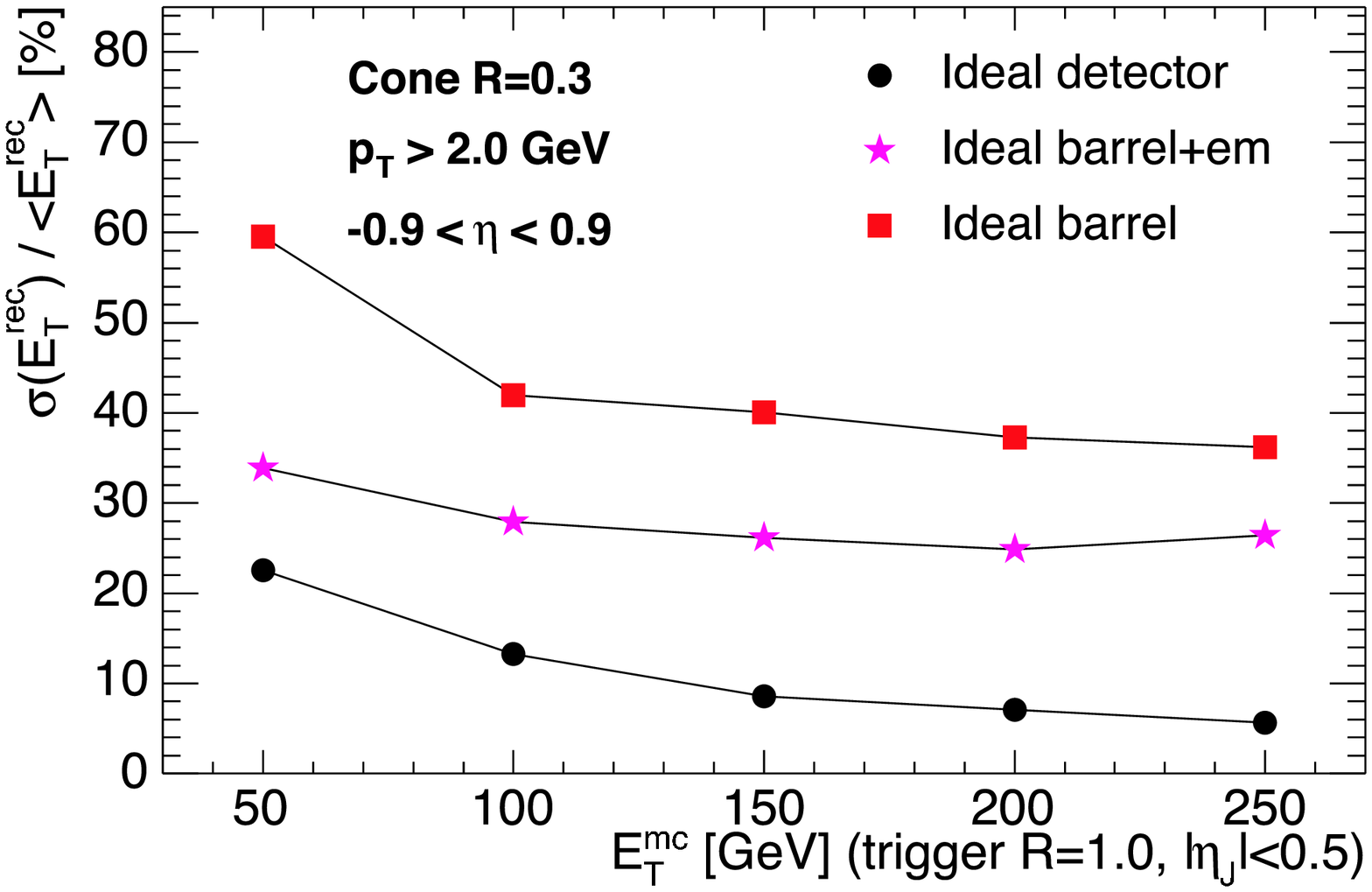}}
\end{center}
\vspace{-0.5cm}
\caption[xxx]{Average fraction of reconstructed jet energy, 
$\av{\et^{\rm rec}/\et^{\rm mc}}$~\subref{chap5:fig:mixedetmeanideal}, 
and reconstructed energy resolution, 
$\sigma(\et^{\rm rec})/\av{\et^{\rm rec}}$~\subref{chap5:fig:mixedetresideal}, 
both, as a function of the jet-trigger energy (Monte Carlo) for the different 
ideal cases. The signal jets are embedded into background ($0$--$10$\% \PbPb, 
\acs{HIJING}); $R=0.3$ and $\pt\ge2~\gev$ are used.}
\label{chap5:fig:mixedetmeanresideal}
\end{figure}

The average fraction of reconstructed jet energy, $\av{\et^{\rm rec}/\et^{\rm mc}}$,
and the energy resolution, $\sigma(\et^{\rm rec})/\av{\et^{\rm rec}}$, are shown in 
\fig{chap5:fig:mixedetmeanresideal} as a function of the jet trigger energy, 
$\et^{\rm mc}$. In order to suppress the background, we $R=0.3$ and $\pt\ge2~\gev$
are used, as explained in the last section. However, compared to \pp\ with the same 
cuts~(\cf~\fig{chap5:fig:ppetmeanidealcut}) the average mean for the embedded jets 
still is lifted by the remaining fraction of the underlying event. As expected, the effect 
is strongest for lowest jet energies; the higher the energy the less the impact of the 
underlying background fluctuations. Due to the additional fluctuations induced by
the underlying event the resolution for the $50~\gev$ jets further degrades (at most $10$\%
for the ideal barrel), whereas it is almost the same in all other cases.

\begin{figure}[htb!]
\begin{center}
\subfigure[Ideal detector]{
\label{chap5:fig:mixedtrueetidealall}
\includegraphics[width=7cm]{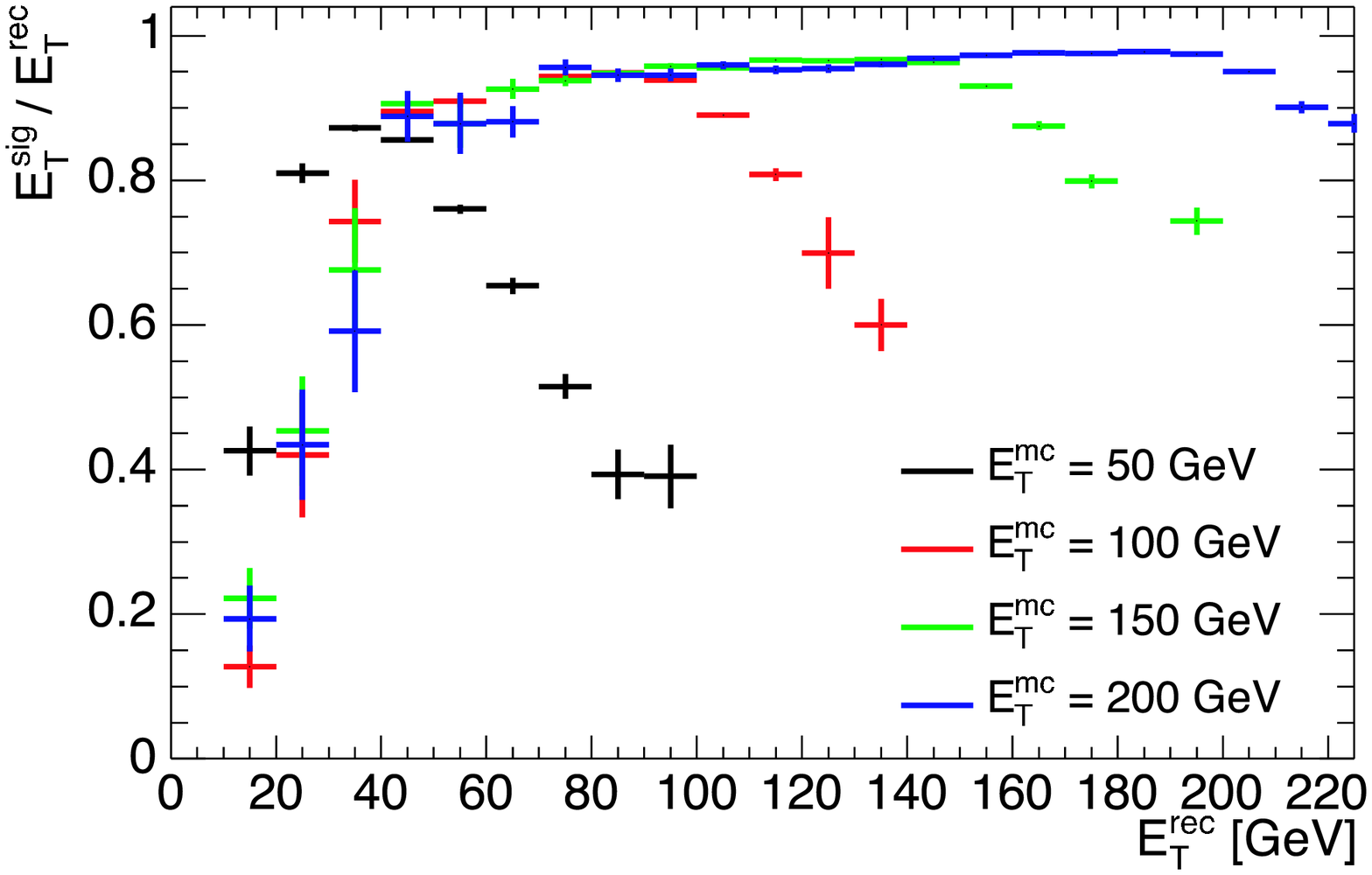}}
\hspace{0.5cm}
\subfigure[Ideal barrel]{
\label{chap5:fig:mixedtrueetidealcharged}
\includegraphics[width=7cm]{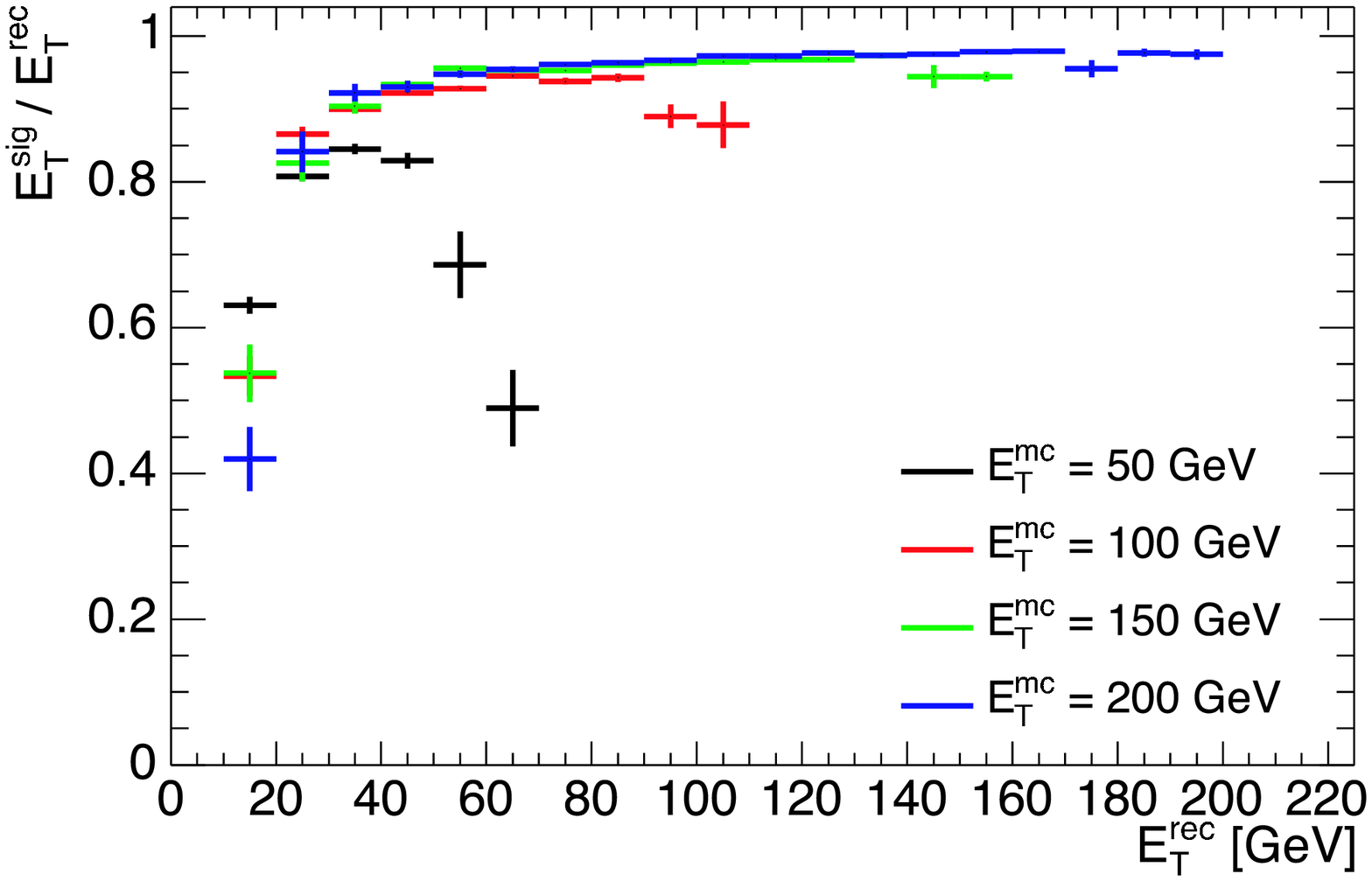}}
\end{center}
\vspace{-0.5cm}
\caption[xxx]{Distributions of fractions of the jet-signal energy relative to total reconstructed 
energy for the ideal case~\subref{chap5:fig:mixedtrueetidealall} and the ideal barrel
\subref{chap5:fig:mixedtrueetidealcharged} as a function of the reconstructed energy, 
$\et^{\rm rec}$. The signal jets are embedded into background ($0$--$10$\% \PbPb, 
\acs{HIJING}); $R=0.3$ and $\pt\ge2~\gev$ are used.}
\label{chap5:fig:mixedtrueetideal}
\end{figure}

The distribution of energy fraction, which originates from the jet in $R=0.3$, relative to 
the contribution of the energy due to background inside the jet cone is conveniently reported 
in \fig{chap5:fig:mixedtrueetideal}. Shown for the ideal detector and ideal barrel cases are the 
fraction of signal energy over the reconstructed energy as a function of the reconstructed energy. 
\enlargethispage{0.5cm}
Since the mean of the reconstructible fraction from the analysis in \pp\ is known, on average, 
the reconstructed jet energy arises to about $85$\% for $50~\gev$ and about $90$--$95$\% 
for the higher jet energies due to the signal, almost independent of the detector scenario.
Opposed to the ideal barrel the contribution of the uncorrelated background to the reconstructed 
jet energy is clearly apparent for the ideal detector, since neither the jets from the signal, nor 
the jets from the background are biased to extreme fragmentation into charged particles, only. 
Note that in the scope of the thesis no attempt has been undertaken to correct for the remaining 
contribution of the background.~\footnote{A simple approach might be to increase the minimum 
proto-jet energy to the level of the average content in the cone due to the remaining background.}

\begin{figure}[htb]
\begin{center}
\includegraphics[width=10cm]{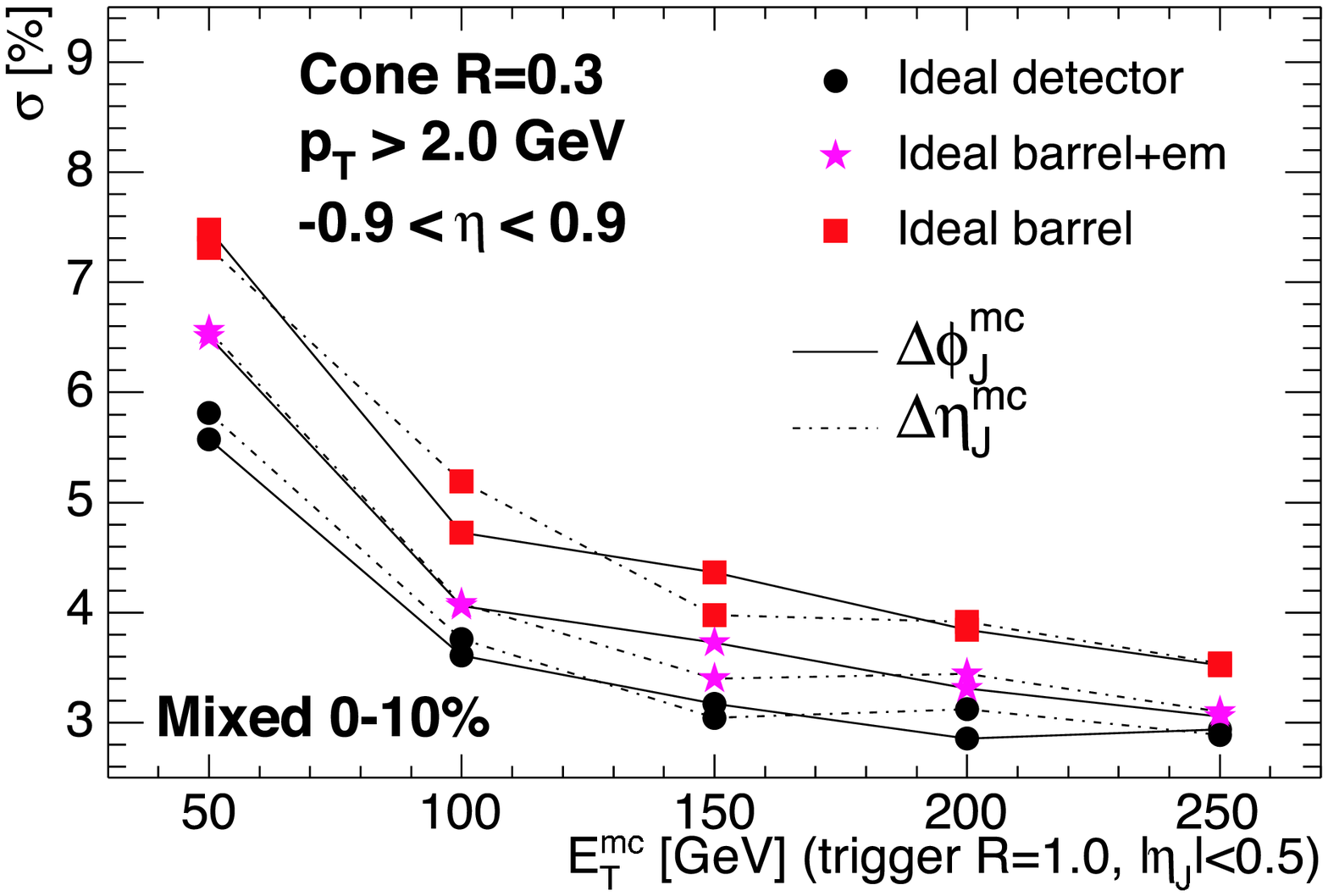}
\end{center}
\vspace{-0.3cm}
\caption[xxx]{Spatial resolution of the reconstructed jets, $\sigma(\Delta \phi^{\rm mc}_{J})$ and 
$\sigma(\Delta \eta^{\rm mc}_{J})$, both, as a function of the jet-trigger energy (Monte Carlo) 
for the different ideal cases. The spatial differences are measured relative to the direction
of the triggered jet, $\Delta \phi^{\rm mc}_{J}=\phi^{\rm rec}_{J}-\phi^{\rm mc}_{J}$ and 
$\Delta \eta^{\rm mc}_{J}=\eta^{\rm rec}_{J}-\eta^{\rm mc}_{J}$. The signal jets are embedded 
into background ($0$--$10$\% \PbPb, \acs{HIJING}); $R=0.3$ and $\pt\ge2~\gev$ are used.}
\label{chap5:fig:mixedspaceresideal}
\end{figure}

Finally, the spatial resolution of the reconstructed jets in $\phi$-direction, 
$\sigma(\Delta \phi^{\rm mc}_{J})$, and in $\eta$-direction, $\sigma(\Delta \eta^{\rm mc}_{J})$, 
as a function of the jet-trigger energy, $\et^{\rm mc}$, for the different ideal cases is 
shown in \fig{chap5:fig:mixedspaceresideal}. The spatial differences are 
measured relative to the direction of the triggered jet, 
$\Delta \phi^{\rm mc}_{J}=\phi^{\rm rec}_{J}-\phi^{\rm mc}_{J}$ and 
$\Delta \eta^{\rm mc}_{J}=\eta^{\rm rec}_{J}-\eta^{\rm mc}_{J}$. At lowest energy, the mean 
values are found to amount to about $0.005$ indicating a moderate bias by the background. 
The resolution for the charged barrel is about $7$--$8$\%, which is $2$\% worse compared 
to the resolution obtained in \pp\ with the same constraints on the reconstruction 
(\cf~\fig{chap5:fig:ppspacerestrackedvstriggercut}). As expected for higher energies the 
additional effect of the medium seems not to be apparent.

\subsection{Simulated detector response}
It is probably reasonable to assume that high-$\pt$ charged-particle tracking is almost not 
affected by the underlying soft event. From the analysis of pure signal (\pp) events, we know 
that the offline reconstruction (charged barrel) almost reaches values close to the optimum 
obtained by the ideal barrel case. In addition, we have shown that the jet-reconstruction 
performance is dominated by out-of-cone and background fluctuations. 
Therefore, we expect to reach a spatial resolution of around $10$\% with a mean and width of 
about $50$\% of the jet energy once the complete detector response is included.~\footnote{There 
was a large production of simulated events containing signal jets embedded into \acs{HIJING} 
from spring to autumn 2004. However at writing of the thesis the data was not available for 
distributed analysis.} The inclusion of the \ac{EMCAL} is expected to improve the mean by 
about $10$ percent points and the resolution by about $20$ percent points, 
an expectation which is in accordance with~\Ref{blyth2004}.
\fi

\section{Reconstructed jet spectra and trigger rates}
\label{chap5:reconjetspectrum}
\ifreconspectrum
After the reconstruction capabilities of the \ac{ILCA} cone finder have been characterized 
for different detector scenarios at fixed input jet energies, we will discuss its performance 
for the realistic spectrum.

\subsection{Single-inclusive jet spectra in pp}
\label{chap5:reconjetspectrumpp}
We generate the input jet spectrum with \acs{PYTHIA} at $\sqrt{s}=5.5~\tev$ using 10000 events per 
$\pt^{\rm hard}$ interval in the way described in \psect{app:pythia}. Therefore, the distribution 
of jets given to the simulation framework is distributed according to the cross section shown 
in \fig{chap5:fig:spectrapythia}. 

\begin{figure}[htb!p]
\begin{center}
\subfigure[Reconstructed spectrum (ideal)]{
\label{chap5:fig:spectracomparisonm}
\includegraphics[width=12cm]{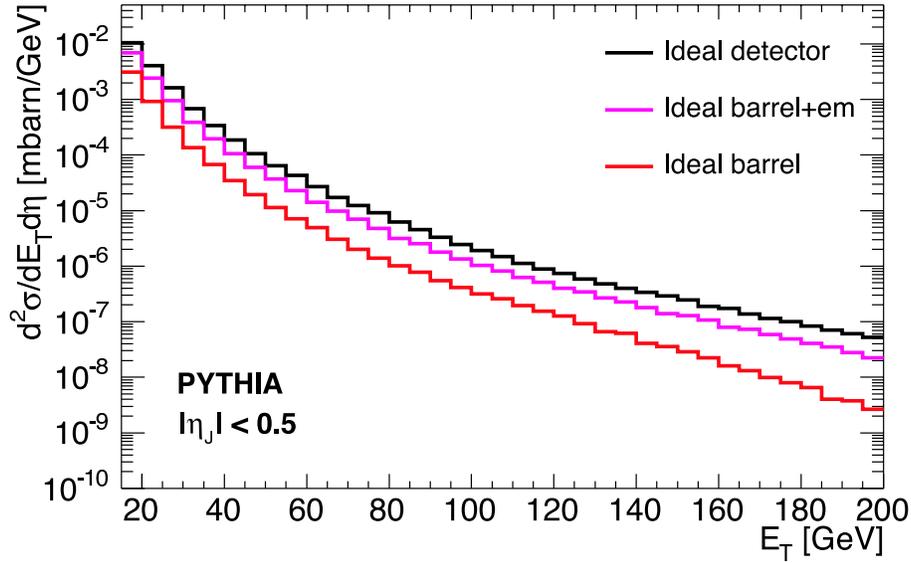}}
\vfill
\vspace{0.5cm}
\subfigure[Reconstructed spectrum (tracked)]{
\label{chap5:fig:spectracomparisont}
\includegraphics[width=12cm]{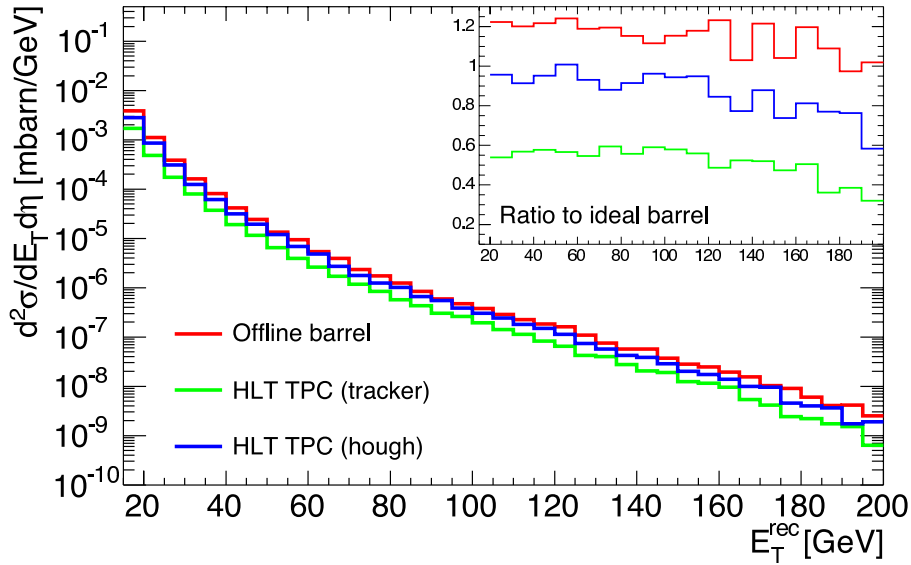}}
\end{center}
\vspace{-0.5cm}
\caption[xxx]{Inclusive single-jet cross section at mid-pseudo-rapidity for 
ideal~\subref{chap5:fig:spectracomparisonm} and charged-particle tracking 
cases~\subref{chap5:fig:spectracomparisont} as a function of the reconstructed
jet energy in \pp\ collisions at $\sqrt{s}=5.5~\tev$. The transverse energy is
not corrected for introduced biases or inefficiencies of the tracking. The inset 
shows the ratio of the spectrum obtained by the different tracking methods to 
the spectrum obtained by the ideal barrel case. All cases are for a cone of $R=0.7$ 
and $\pt>0.5~\gev$. The input distribution is the same as in \fig{chap5:fig:spectrapythia}
and the corresponding cross section agrees with the cross section deduced in the ideal 
detector case.}
\label{chap5:fig:spectracomparisonmt}
\end{figure}

As explained in the previous sections, in the ideal detector cases we just keep the detectable 
particle types from the Monte Carlo, which correspond to the different scenarios (ideal detector, 
ideal barrel plus electromagnetic calorimeter and ideal barrel). The spectrum or rather the 
deduced cross section, reconstructed  with the developed \ac{ILCA} cone finder using $R=0.7$ 
and \mbox{$0.5<\pt<100~\gev$}, is shown in \fig{chap5:fig:spectracomparisonm} for the ideal cases
as a function of the reconstructed transverse jet energy, $\et^{\rm rec}$, averaged over 
$\abs{\eta}<0.5$. The reconstructed energies are not corrected for introduced biases by the different 
detector types. Especially for the higher energies one notices that the cross section estimated merely 
based on the ideal barrel detectors underpredicts the real cross section by about one order of magnitude. 
However, on average, one may correct the obtained cross section by dividing the reconstructed 
energy with the reconstructible fraction, $0.60$. The corrected distribution corresponds to the 
spectrum found with the ideal detector within a few percent. As expected from the discussion in 
the previous sections the inclusion of the \ac{EMCAL} improves the resolution of the jet energy 
and, therefore, the determination of the cross section. Without correction, in this case  
the reconstructed cross section is merely little below the ideal measurement (and the input).
But more importantly, it not only allows one to improve the average and jet-by-jet resolution, 
but also to trigger on neutral particles in the fragmentation of jets.

Since we are dealing with \pp\ events, we include the detailed detector response of the barrel 
detectors in the simulation and apply the tracking algorithms onto the digitized hit information 
in the \acs{ALIROOT} framework (offline barrel, \ac{HLT} tracker, \ac{HLT} hough).
The reconstructed tracks with \mbox{$0.5<\pt<100~\gev$} are grouped into jets using the \ac{ILCA} 
cone finder for a cone size of $R=0.7$. The spectrum or rather the deduced cross section, which 
results for the different cases, averaged over $\abs{\eta}<0.5$, is shown in 
\fig{chap5:fig:spectracomparisont}. Since charged-particle tracking, at best, can reproduce the result 
of the ideal barrel, we display the ratio of the spectra obtained based on tracking algorithms to the 
spectrum obtained based on the ideal barrel case in the inset of the same figure. The reconstructed
spectrum based on offline barrel tracking using the combined information of \ac{ITS}, \ac{TPC} and 
\ac{TRD} is reasonably close to the optimum, within $20\%$ over the whole $\et$-range. However, it mainly 
overestimates low jet energies. This seems to be in contrast to the findings of \sect{chap5:jetreconpbpb}, 
where for fixed input energy the offline tracker reaches the optimal values. However, looking again 
at \fig{chap5:fig:ppetmeantracked} the precise values of the mean fraction for jets below 
$150~\gev$ are slightly above one. At $50~\gev$ input the fraction is about $1.035$, leading to an 
average increase (shift to the right) of the distribution by approximately $1.035^{6}=1.23$. Since 
the spectrum is dominated by jets at low energies this might explain the apparent discrepancy to the 
ideal barrel. The distribution obtained by the \ac{HLT} Hough-track finder is close to the optimum for 
energies up to $100~\gev$. Though, at higher energies it differs by $20$--$40$\% due the decreasing 
jet-energy resolution. The spectrum deduced in the case of the \ac{HLT} tracker underestimates the 
ideal case by about $40$--$60$\%. Since that is the case throughout the probed jet-energy range, 
one might apply a constant correction. 
However, for the quality of the trigger it is more important to understand and control potential 
biases than to optimize the performance of the jet recognition.

\subsection{Trigger rates in pp and Pb--Pb}
\label{chap5:recontriggerrates}
In the following we will introduce a very simple algorithm, 
which is supposed to run on the trigger nodes of the \ac{HLT} system.
It is supposed to trigger, if the online version of the jet finder, \ie~\ac{ILCA} with the 
same settings as before, finds a jet in the event with more than $m~\gev$ in the cone of $R=0.7$. 
The reconstructed energy obtained by the trigger is not corrected for biases. Rather, we adapt the 
value of $m$ to the particular running conditions (\eg~available detectors in the event, size 
of $R$ and $\pt$-cut). Therefore, in the case of the ideal barrel, $m$ sets the lower limit of 
the charged energy, which the triggered jet is required to have. More generally, the trigger 
accepts all events, where it finds at least one jet with
\begin{equation*}
\et^{\rm rec}> m\;.
\end{equation*}
Note that the definition of a jet ultimatly is linked to the algorithm used to find it~(\cf 
\psect{chap3:jetdefintion}). Thus, running the same algorithm online as well as offline is preferable 
in order to minimize additional bias.

In order to evaluate the trigger algorithm, for every event in the simulation the jet is recorded, 
as originally defined by the \acs{UA1} cone finder in \acs{PYTHIA}. In that way, we are able to 
bookmark original jets, which did or did not trigger.~\footnote{Note that in previous sections 
these original jets were called `trigger jets', since they also trigger the \acs{PYTHIA} event 
generator to accept the particular event.} 

\begin{figure}[htb]
\begin{center}
\subfigure[Spectrum of triggered jets in \pp~(ideal)]{
\label{chap5:fig:pptriggerspectrummonte}
\includegraphics[width=7cm]{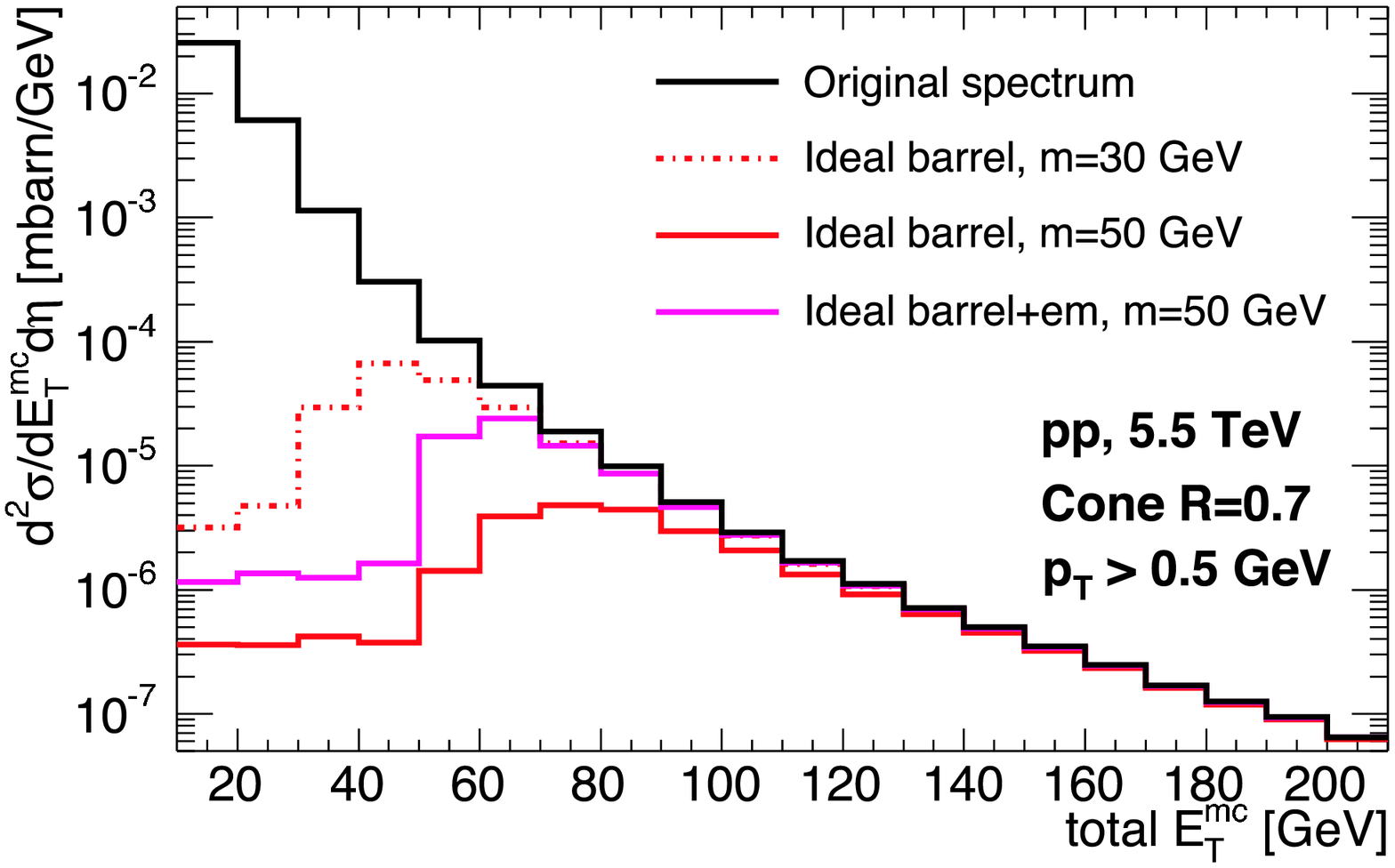}}
\hspace{0.5cm}
\subfigure[Rate of triggered jets in \pp~(tracked)]{
\label{chap5:fig:pptriggerratetracked}
\includegraphics[width=7cm]{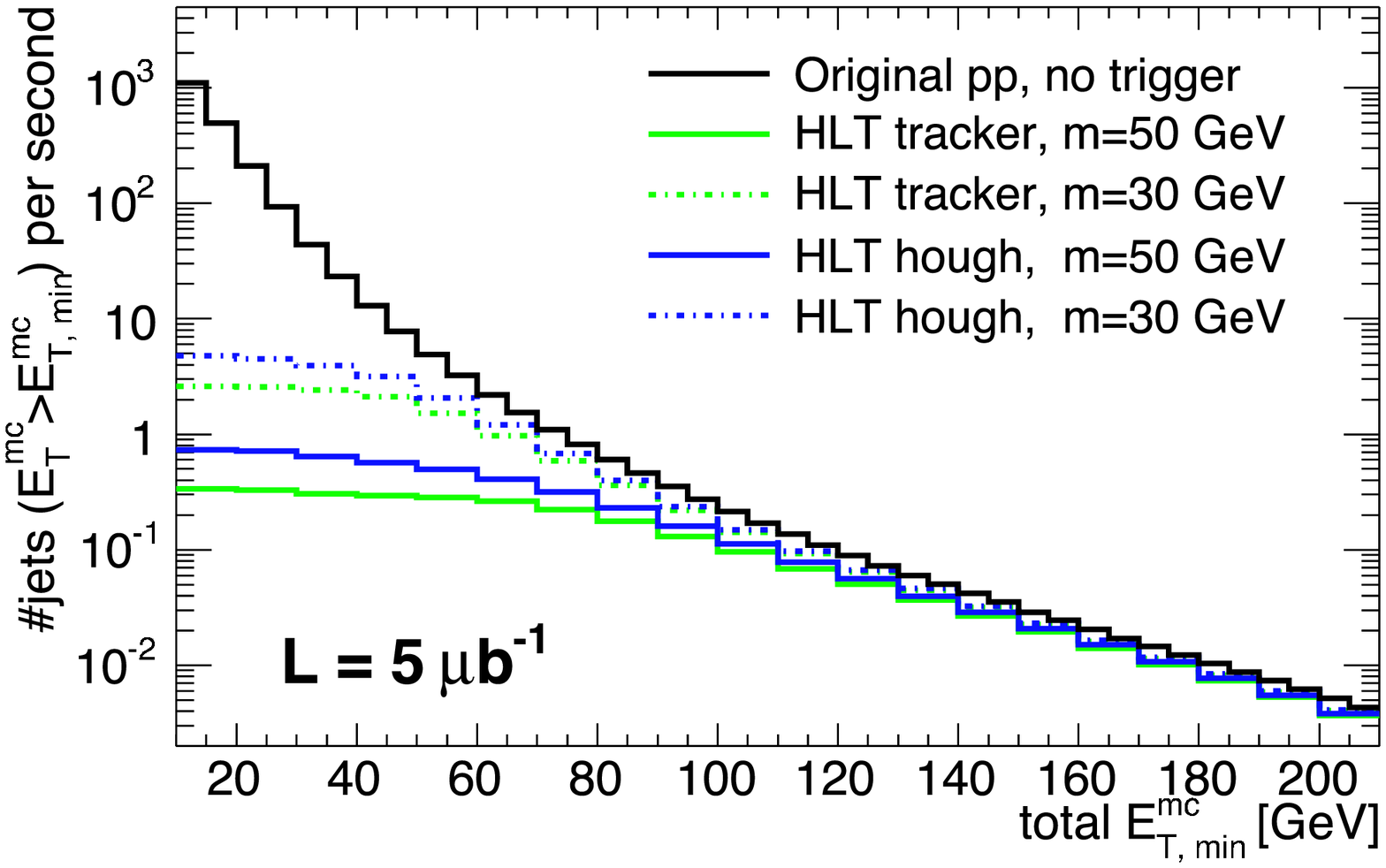}}
\end{center}
\vspace{-0.5cm}
\caption[xxx]{\subref{chap5:fig:pptriggerspectrummonte}~Original input spectrum compared to the spectrum 
of triggered jets obtained for different ideal detector cases and values of the required energy 
in the cone~($m$). The spectra are normalized to the inclusive single-jet cross section at 
mid-pseudo-rapidity in \pp\ at $\sqrt{s}=5.5~\tev$.
\subref{chap5:fig:pptriggerratetracked}~Rate of triggered jets with 
$\et^{\rm mc}>E_{\rm T,\,min}^{\rm mc}$ for the \acs{HLT} cases at different values of the required 
energy~($m$) compared to the production rate at full \acs{L1} rate. In both figures $R=0.7$ and 
$\pt>0.5~\gev$ are used; opposed to \fig{chap5:fig:spectracomparisonmt} results are given as a function 
of the total $\et$ of the Monte Carlo trigger jets.}
\label{chap5:fig:pptriggerresults}
\end{figure}

\begin{figure}[htb!]
\vspace{0.5cm}
\begin{center}
\subfigure[Trigger efficiency in \pp~(ideal)]{
\label{chap5:fig:pptriggereffmonte}
\includegraphics[width=7cm]{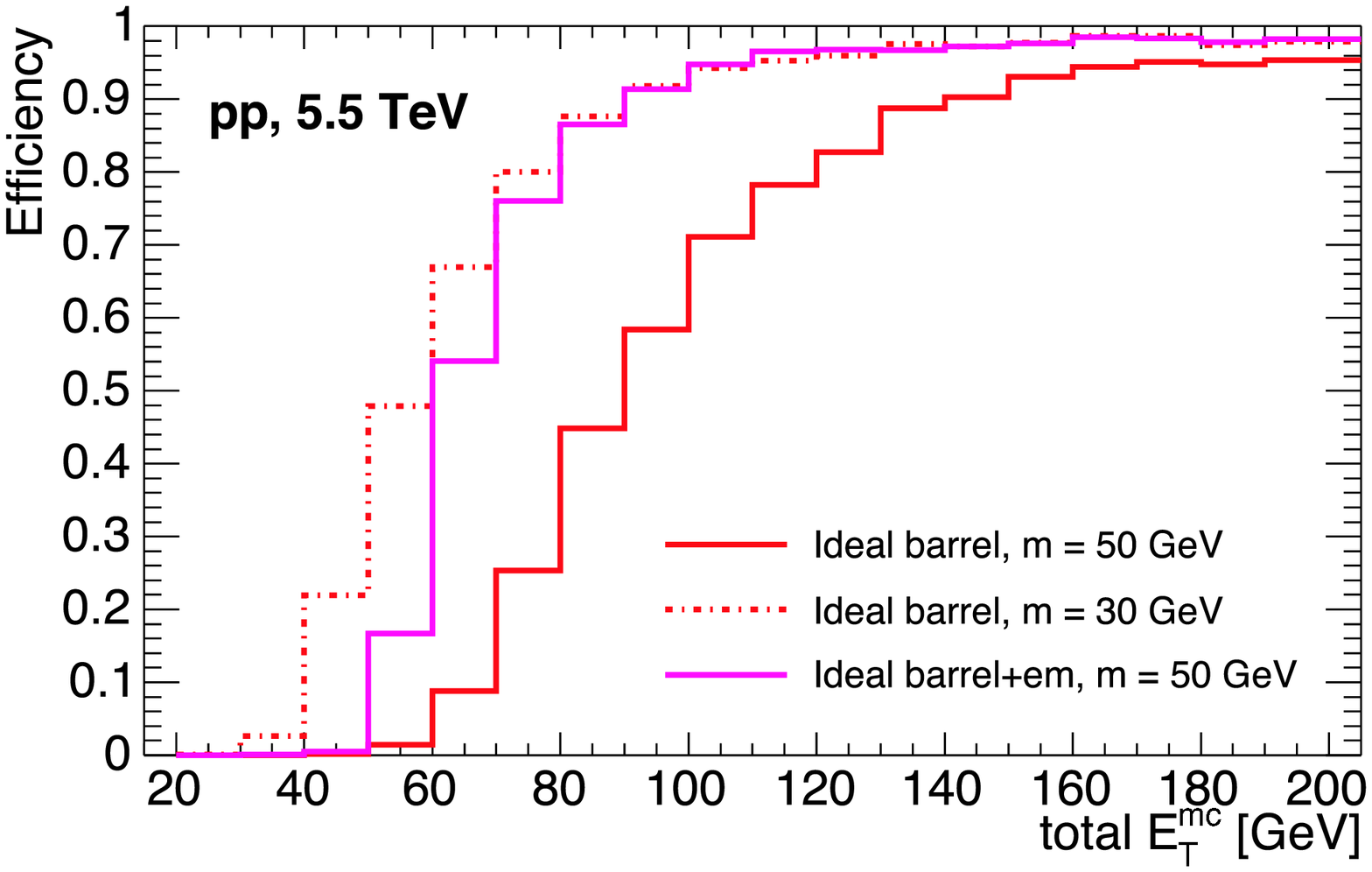}}
\hspace{0.5cm}
\subfigure[Trigger efficiency in \pp~(tracked)]{
\label{chap5:fig:pptriggerefftracks}
\includegraphics[width=7cm]{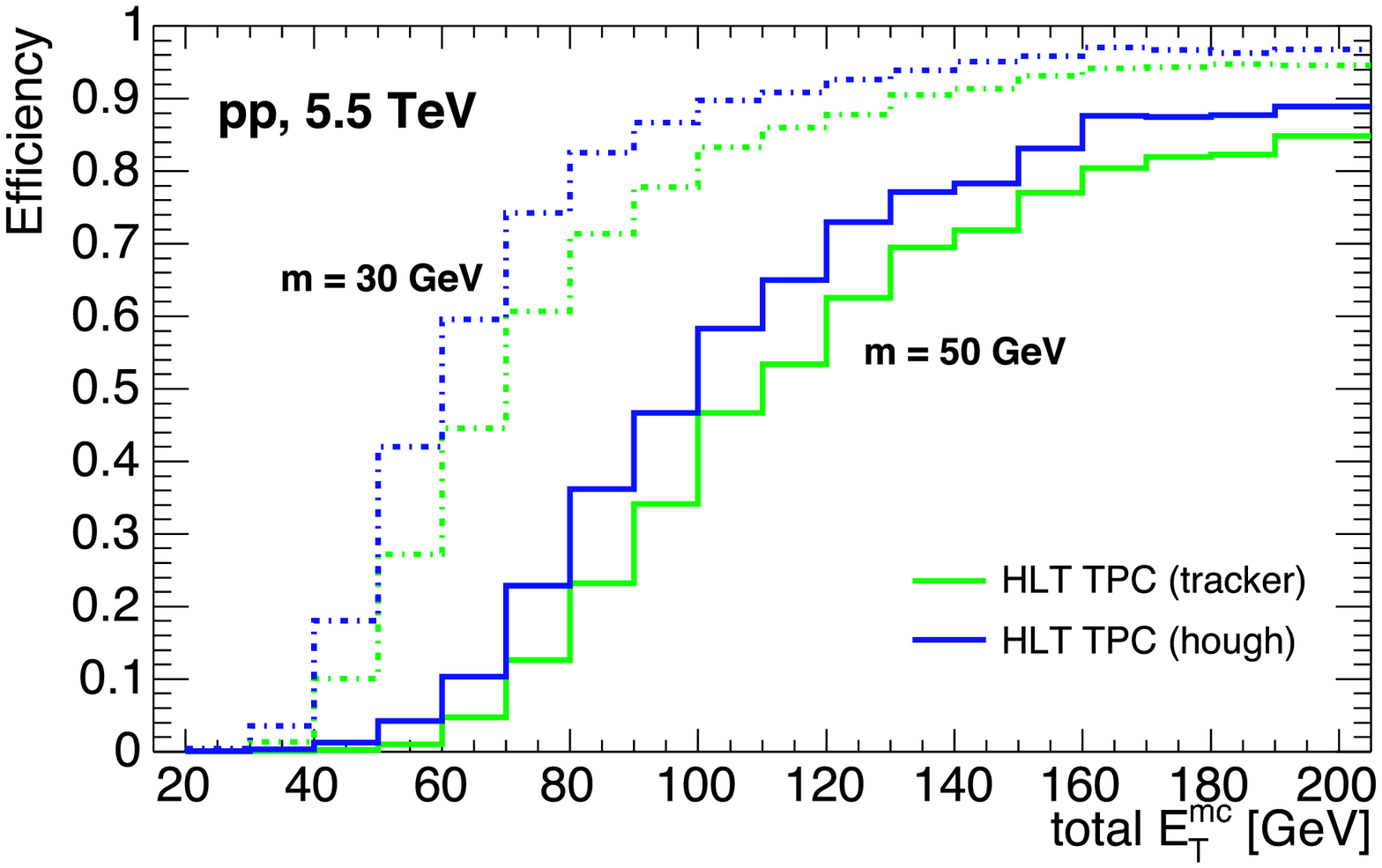}}
\end{center}
\vspace{-0.5cm}
\caption[xxx]{Trigger efficiency for ideal~\subref{chap5:fig:pptriggereffmonte} and 
tracking~\subref{chap5:fig:pptriggerefftracks} cases as a function of the total 
$\et^{\rm mc}$ given by the Monte Carlo jets at different values of the required 
energy~($m$). See \fig{chap5:fig:pptriggerresults} for further details.}
\label{chap5:fig:pptriggereffmt}
\end{figure}

In \fig{chap5:fig:pptriggerspectrummonte} the original input spectrum is compared to the triggered 
spectrum of jets, depending on different ideal detector cases and values of the required energy 
in the cone. The shown spectra are normalized to the inclusive single-jet cross section at 
mid-pseudo-rapidity in \pp\ at $\sqrt{s}=5.5~\tev$. As expected, the cut on the 
energy in the cone suppresses low energy jets. Since the energy resolution in case of an ideal 
\ac{EMCAL} is significantly higher, the choice of $m=50~\gev$  leads to a sharper cut than for the 
ideal barrel only. This is quantified in \fig{chap5:fig:pptriggereffmonte}, where the efficiency 
of the trigger is plotted for the same settings. The efficiency is defined by the fraction of triggered 
jets to total jets (entering the simulation) as a function of the jet energy, \ie~the ratio 
of the triggered to the input spectrum in \fig{chap5:fig:pptriggerspectrummonte}. 
Due to the increased resolution induced by the \ac{EMCAL}, the rise of the efficiency occurs around the 
value given by $m$, and, furthermore, is relatively steep. In \fig{chap5:fig:pptriggerratetracked},
we show the rate of triggered jets with $\et^{\rm mc}>E_{\rm T,\,min}^{\rm mc}$ for the \ac{HLT} tracking 
case compared to different values of the required energy. The accepted number of events per second, 
$N_{\rm acc}$, \eq{chap5:eq:eventrate}, corresponds to the rate at $E_{\rm T,\,min}^{\rm mc}=0$. 
For the calculation of the rate it is assumed that jet events enter the system at full \ac{L1} 
production rate (labeled `Original pp, no trigger', \ie~$m=0$, \cf~\fig{chap5:fig:jetrate}). 
Thus, for \pp\ with the chosen values of $m$, the trigger stays well below an event rate of $10~\hz$,
if a value of $m=30~\gev$ is used. Obviously, one could even afford to use a smaller value of $m$. This 
is opposed to \PbPb, see below. The corresponding efficiencies of the two \ac{HLT} cases are shown in 
\fig{chap5:fig:pptriggerefftracks}. In \tab{chap5:tab:totalyieldppwithtracks}, we report the 
resulting yearly yields in \pp~at $\sqrt{s}=5.5~\tev$, compared to rates, one would obtain with the 
trigger based on the ideal barrel, and compared to the total number of produced jets at \ac{L1} (taken from
\tab{chap5:tab:jetyield1}). In addition, we report the total number of accepted events per year. Since 
there is a lower cut on the jets in the simulation (\cf to the first $\pt^{\rm hard}$-interval), $N_{\rm acc}$ 
is evaluated at $E_{\rm T,\,min}^{\rm mc}=15~\gev$ and, therefore, not precise. Note that the triggered

\begin{table}[htb]
\begin{center}
\begin{tabular}{l|cc|cc|cc|c}
\hline
\hline
Case &\multicolumn{2}{c|}{Ideal barrel} &\multicolumn{2}{c|}{\acs{HLT} tracker} &\multicolumn{2}{c|}{\acs{HLT} hough} & Total \acs{L1}\\
$m$ [$\gev$] & $30$ & $50$ & $30$ & $50$ & $30$ & $50$ & $0$ \\
\hline
$Y(50~\gev) $ & $2.3\cdot10^{7}$ & $4.8\cdot10^{6}$ & $1.5\cdot10^{7}$ & $2.8\cdot10^{6}$ & $2.1\cdot10^{7}$ & $4.8\cdot10^{6}$ & $4.9\cdot10^{7}$ \\
$Y(100~\gev)$ & $1.6\cdot10^{6}$ & $1.3\cdot10^{6}$ & $1.4\cdot10^{6}$ & $9.6\cdot10^{5}$ & $1.5\cdot10^{6}$ & $1.1\cdot10^{6}$ & $2.2\cdot10^{6}$ \\
$Y(150~\gev)$ & $2.4\cdot10^{5}$ & $2.3\cdot10^{5}$ & $1.9\cdot10^{5}$ & $2.3\cdot10^{5}$ & $2.3\cdot10^{5}$ & $2.1\cdot10^{5}$ & $3.0\cdot10^{5}$\\
$Y(200~\gev)$ & $4.2\cdot10^{4}$ & $4.1\cdot10^{4}$ & $4.0\cdot10^{4}$ & $3.7\cdot10^{4}$ & $4.1\cdot10^{4}$ & $3.8\cdot10^{4}$ & $5.4\cdot10^{4}$ \\
\hline
$N_{\rm acc}$ & $4.4\cdot10^{7}$ & $5.1\cdot10^{6}$ & $2.6\cdot10^{7}$ & $3.4\cdot10^{6}$ & $4.8\cdot10^{7}$ &  $7.3\cdot10^{6}$ & $5.0\cdot10^{9}$\\ 
\hline
\hline
\end{tabular}
\end{center}
\vspace{-0.4cm}
\caption[xxx]{Jet yield per \acs{ALICE} run year in \pp~at $\sqrt{s}=5.5~\tev$ corresponding to the rates shown in 
\fig{chap5:fig:pptriggerratetracked} compared to the rates obtained for the ideal barrel and the total 
produced jets at full \acs{L1} rate. $N_{\rm acc}$ is evaluated at \mbox{$E_{\rm T,\,min}^{\rm mc}=15~\gev$}.}
\label{chap5:tab:totalyieldppwithtracks}
\end{table}

Coming to central \PbPb~collisions, the situation for the trigger changes. We embed the \pp\ spectrum, 
or rather the corresponding \pp\ events, generated with \acs{PYTHIA} as mentioned above in $0$--$10$\% 
central \acs{HIJING} events with settings listed in \psect{app:hijing}.~\footnote{In total, 2500 
different background events are used, \ie~four signal events share the same background per 
$\pt^{\rm hard}$ interval, and,therefore, will be used in a total of $4\times16$ times.} 
In that way, the performance of the trigger algorithm in \PbPb\ can be evaluated as before, 
since the signal jets may be identified with the bookmarked original jets from \acs{PYTHIA}. 
As usual in \PbPb, $R=0.3$ and a cut of $\pt>2~\gev$ are used, also for the trigger. Since no 
realistic detector response is available, different ideal cases are evaluated.

\Fig{chap5:fig:pbpbtriggerratemonte} shows the trigger rate of signal jets with 
$\et^{\rm mc}>E_{\rm T,\,min}^{\rm mc}$ as a function of $E_{\rm T,\,min}^{\rm mc}$ for 
different choices of $m$. 
For the calculation of the rates it is assumed that jet events enter the system at full \ac{L1} 
minimum-bias production rate (labeled as `PbPb, minimum-bias', \ie~$m=0$, \cf~\fig{chap5:fig:jetrate}). 
Note that the background is still approximated by $0$--$10$\% \acs{HIJING}, not adjusted to 
minimum-bias.~\footnote{For this and other reasons mentioned below, the obtained event rates, 
triggered in total, are a rather crude, but upper limit, estimation.}
The total number of triggered events per second corresponds to $E_{\rm T,\,min}^{\rm rec}=0$. 
It is indicated as thin horizontal lines for the different settings and values of $m$.
Thus, the trigger for $m=30~\gev$ does not significantly reduce the rate arising
from the \acs{HIJING} background. The reason is that the background itself contains jet signatures 
and correlations (\acs{HIJING} quenched), which even with the proposed cone finder settings
$R=0.3$ and $\pt>2~\gev$ lead to reconstructed jets of $50~\gev$, and even more. 
In \fig{chap5:fig:fractionofaccbackground} the fraction of accepted background is estimated
as a function of $m$ for the ideal cases. In the case of the ideal barrel, for $m=30~\gev$ 
about every tenth event will trigger, even if no signal is embedded. However, for $m=50~\gev$ 
this accidental trigger rate is already suppressed by one order of magnitude. 
In \fig{chap5:fig:pbpbtriggereffmonte}, the efficiency, defined as the ratio of the number of 
triggered signal jets to the total number of signal jets, is plotted as a function 
of $E_{\rm T,\,min}^{\rm mc}$. \Fig{chap5:fig:pbpbtriggersbmonte} shows the corresponding
significance of the trigger, defined as the ratio of the number of triggered signal jets to total 
number of triggered events. 

\begin{figure}[htb]
\begin{center}
\subfigure[Trigger rate in minimum-bias \PbPb~(ideal)]{
\label{chap5:fig:pbpbtriggerratemonte}
\includegraphics[width=7cm]{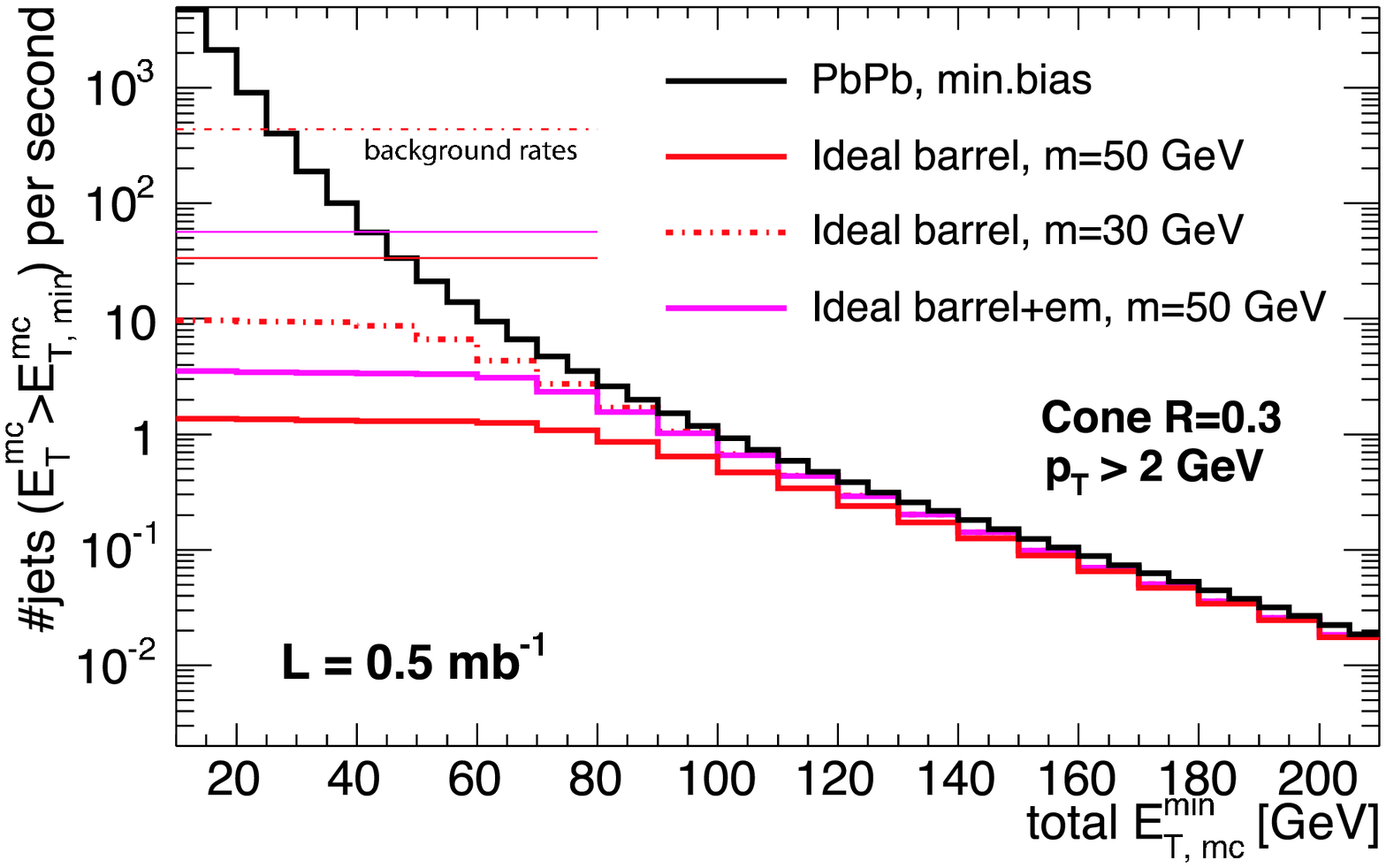}}
\hspace{0.5cm}
\subfigure[Fraction of accepted background (ideal)]{
\label{chap5:fig:fractionofaccbackground}
\includegraphics[width=7cm]{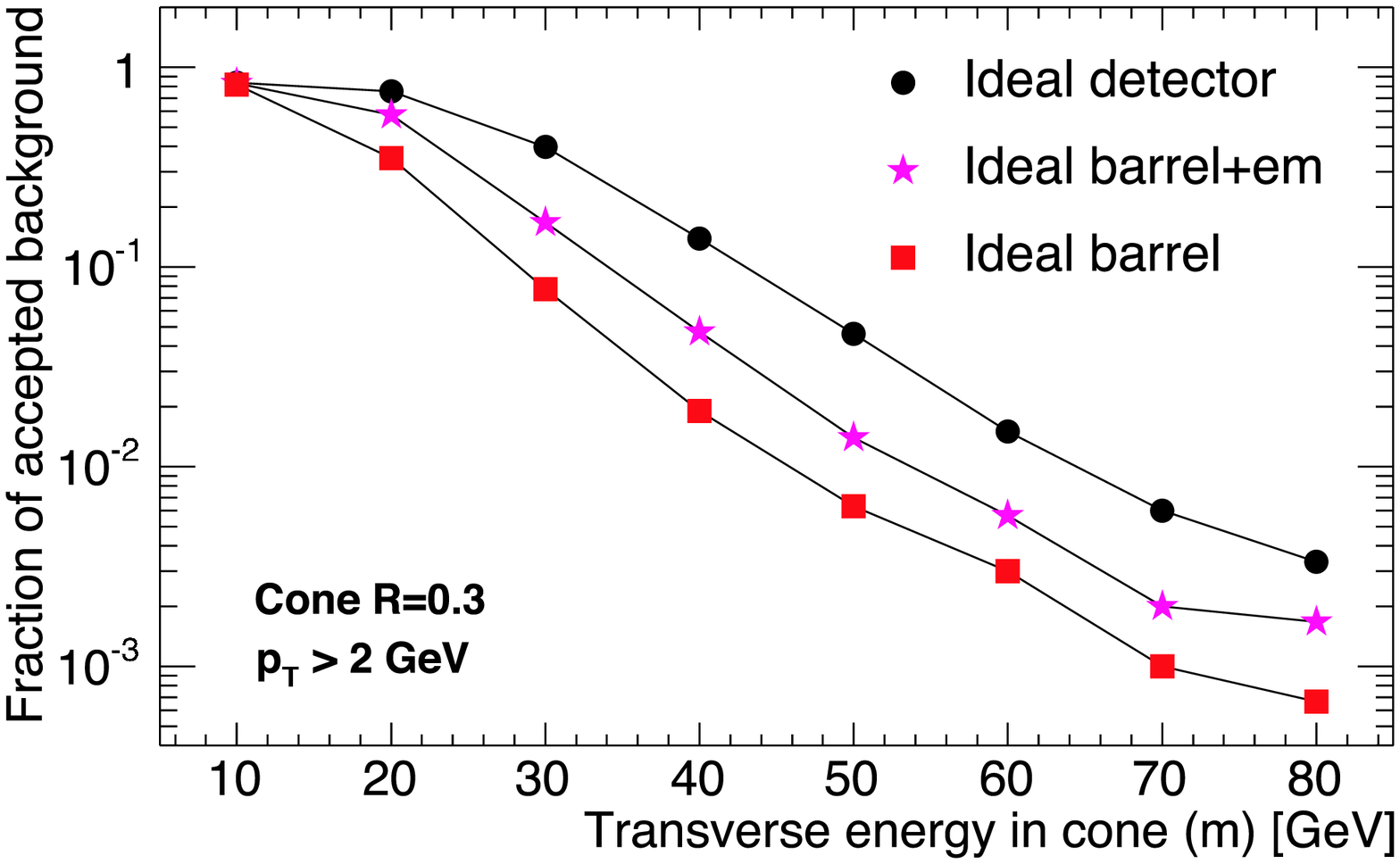}}
\end{center}
\vspace{-0.5cm}
\caption[xxx]{\subref{chap5:fig:pbpbtriggerratemonte}~Trigger rate of signal jets with 
$\et^{\rm mc}>E_{\rm T,\,min}^{\rm mc}$ for ideal cases at different values of the 
required energy~($m$). The minimum-bias input contains the original \pp\ spectrum 
embedded into background ($0$--$10$\% \PbPb, \acs{HIJING}). The horizontal lines 
show the total rate of triggered events per second.
\subref{chap5:fig:fractionofaccbackground}~Fraction of accepted events in pure background 
($0$--$10$\% \PbPb, \acs{HIJING}) for ideal cases as a function of the required energy~($m$). 
In both figures $R=0.3$ and $\pt\ge2~\gev$ are used.}
\label{chap5:fig:pbpbtriggerrate}
\end{figure}

\begin{figure}[htb!]
\vspace{0.5cm}
\begin{center}
\subfigure[Trigger efficiency in central \PbPb~(ideal)]{
\label{chap5:fig:pbpbtriggereffmonte}
\includegraphics[width=7cm]{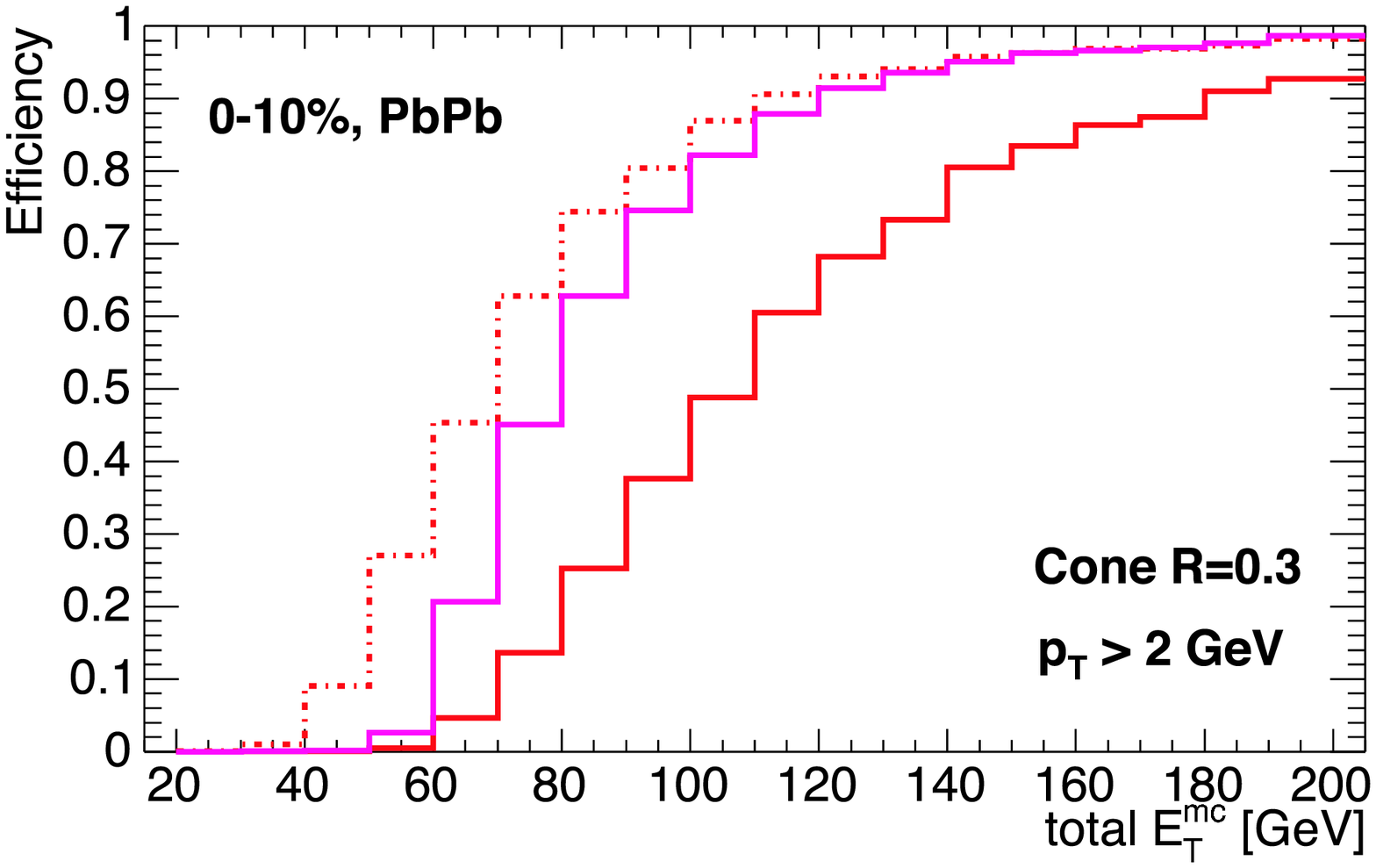}}
\hspace{0.5cm}
\subfigure[Significance in central \PbPb~(ideal)]{
\label{chap5:fig:pbpbtriggersbmonte}
\includegraphics[width=7cm]{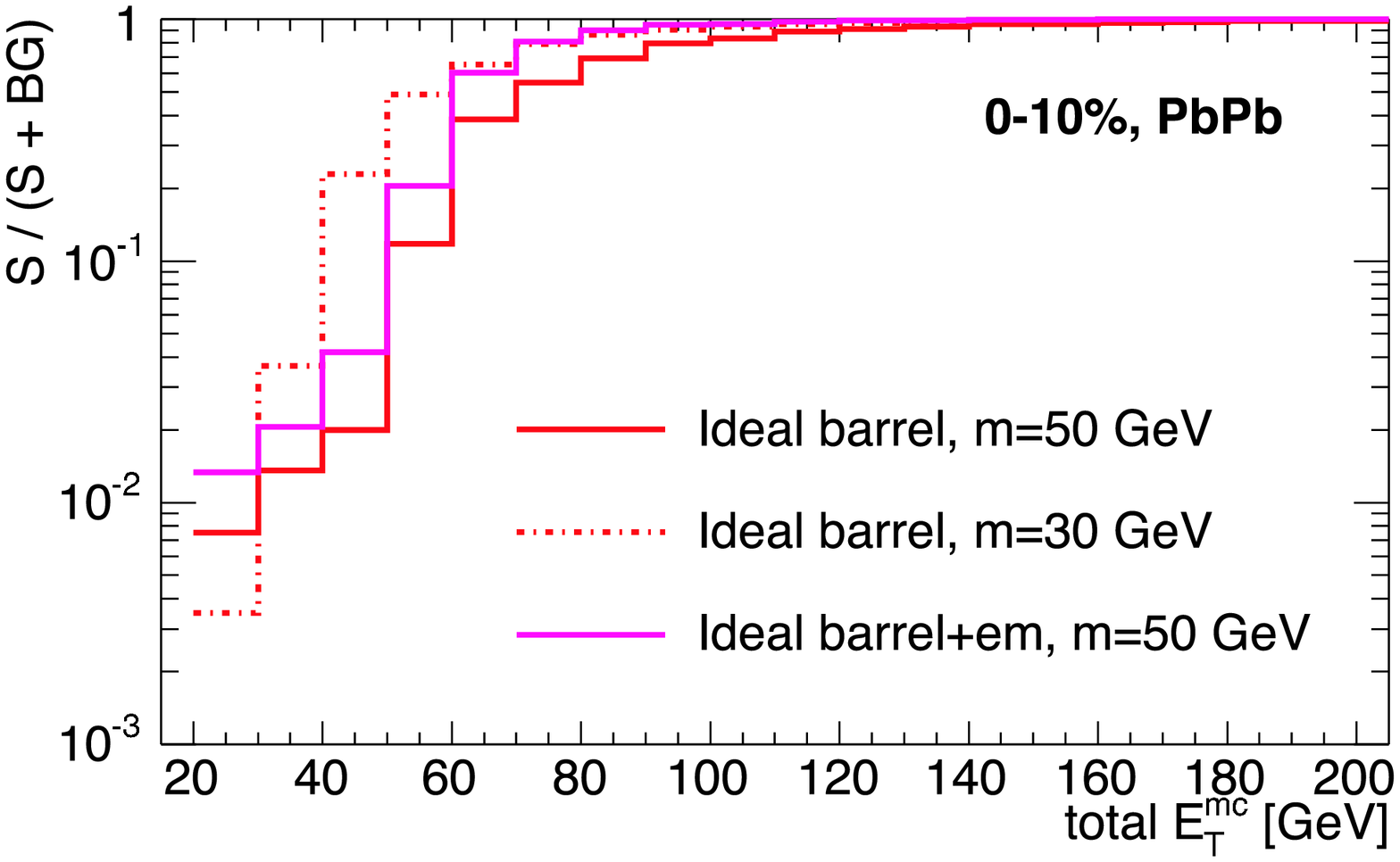}}
\end{center}
\vspace{-0.5cm}
\caption[xxx]{Trigger efficiency~\subref{chap5:fig:pbpbtriggereffmonte} and 
significance~\subref{chap5:fig:pbpbtriggersbmonte} for ideal cases as a function 
of the total $\et$ given by the Monte Carlo jets at different values of the 
required energy~($m$). The \pp\ spectrum is embedded into background ($0$--$10$\% \PbPb, 
\acs{HIJING}); $R=0.3$ and $\pt\ge2~\gev$ are used.}
\label{chap5:fig:pbpbtriggermonte}
\end{figure}

Both figures confirm that, while the lower value of $m$ generally improves the efficiency, 
it reduces the significance mainly at low jet energies leading to a high number of accepted events. 
Thus, for \PbPb~the trigger threshold must be increased to about $m=50~\gev$, in order to 
acquire event rates, which can be handled by the \ac{DAQ} system. 
For $m=50~\gev$, \ac{ALICE} should preferably run in the barrel+em case, which greatly improves the 
efficiency and \stn\ ratio. Finally, we report the resulting yearly yields in \tab{chap5:tab:totalyieldpbpb} 
for minimum-bias \PbPb~at $\snn=5.5~\tev$ compared to total number of produced jets 
at full \ac{L1} rate (taken from \tab{chap5:tab:jetyield1}). The corresponding yields for central 
events may approximately deduced by dividing the minimum-bias yields by a factor of $2.5$. In addition, we 
report the total accepted number of events per year evaluated at $E_{\rm T,\,min}^{\rm mc}=15~\gev$.~\footnote{As
mentioned above, the accepted number of events may be overestimated, since we use $0$--$10$ central \acs{HIJING}
but scale our expectation to minimum-bias \PbPb~collisions.}

\begin{table}[htb]
\begin{center}
\begin{tabular}{l|cc|c|c}
\hline
\hline
Case &\multicolumn{2}{c|}{Ideal barrel} & Ideal barrel+em & Total \acs{L1}\\
$m$ [$\gev$] & $30$ & $50$ & $50$ & $0$ \\
\hline
$Y(50~\gev) $ & $6.6\cdot10^{6}$ & $1.3\cdot10^{6}$ & $3.3\cdot10^{6}$ & $2.1\cdot10^{7}$ \\
$Y(100~\gev)$ & $6.7\cdot10^{5}$ & $4.7\cdot10^{5}$ & $6.6\cdot10^{5}$ & $9.4\cdot10^{5}$ \\
$Y(150~\gev)$ & $9.8\cdot10^{4}$ & $8.9\cdot10^{4}$ & $9.9\cdot10^{4}$ & $1.3\cdot10^{5}$ \\
$Y(200~\gev)$ & $1.8\cdot10^{4}$ & $1.7\cdot10^{4}$ & $1.8\cdot10^{3}$ & $2.3\cdot10^{4}$ \\
\hline
$N_{\rm acc}$  & $4.4\cdot10^{8}$ & $3.4\cdot10^{7}$ & $5.7\cdot10^{7}$ & $2.1\cdot10^{9}$ \\
\hline
\hline
\end{tabular}
\end{center}
\vspace{-0.4cm}
\caption[xxx]{Jet yield and total number of triggered events per \acs{ALICE} run year in minimum-bias \PbPb~at 
$\snn=5.5~\tev$ corresponding to the rates shown in \fig{chap5:fig:pbpbtriggerratemonte} compared 
to the production at full \acs{L1} rate. $N_{\rm acc}$ is evaluated at \mbox{$E_{\rm T,\,min}^{\rm mc}=15~\gev$}.}
\label{chap5:tab:totalyieldpbpb}
\end{table}

In summary, in \pp\ for $m=30~\gev$ and in \PbPb\ for $m=50~\gev$ the rate may be reduced by about a
factor of $50$ for \PbPb\ ($100$ for \pp), while keeping $1/10$~($1/5$ for \pp) of the events with 
$E_{\rm T,\,min}^{\rm mc}=50~\gev$ 
and more than half of the events with $E_{\rm T,\,min}^{\rm mc}=100~\gev$. Running in the case of the ideal 
barrel will provide enough statistics at high energies. However, one should keep in mind that without an 
\ac{EMCAL} a bias towards charged-particle fragmentation will be introduced in the recorded data sample,
and ---throughout the thesis--- full $2\pi$ coverage in $\phi$ is assumed for the \ac{EMCAL}. 
Furthermore, without hardware triggering at \acs{L1} (neglecting the possibility of the \ac{TRD}) the yields 
will drop by about a factor of $350$ in \pp\ and a factor of $4$ in \PbPb~(see \sect{chap5:intyields}).
\fi

\section{Expected back-to-back jet rates}
\label{chap5:b2bjetrates}
\ifbacktoback
It might be useful to estimate the cross section of jet pairs which emerge back-to-back
in azimuth in the \ac{ALICE} central acceptance. Typically one computes the dijet mass 
spectrum, $\dd^3\sigma/\dd M_{\rm jj}/\dd\eta_{1}\dd\eta_{2}$, where the dijet mass in the 
\cms\ system is defined according to
\[
M_{\rm jj}^2=2 \et^{\rm jet1} \, \et^{\rm jet2} 
\left( \cosh(\Delta \eta^{\rm 1\,2}_{J}) - \cos(\Delta \phi^{\rm 1\,2}_{J}) \right)
\]
assuming the jets are massless~\cite{abbott1998}. In our case we are interested in back-to-back jets, 
where $\Delta \phi^{\rm 1\,2}_{J}\approx \pi$. To compute the corresponding spectrum we generate 
pairs of jets using \acs{PYTHIA} in the way described in \psect{app:pythia}. Since we need to be 
able to reconstruct both jets (of the dijet) in the central barrel their jet axes must be in the 
interval of $-0.5<\eta^{\rm jet1}_{J},\,\eta^{\rm jet2}_{J}<0.5$ to be accepted. Furthermore, 
ensuring that the jets are pointing back-to-back in azimuth we require 
\mbox{$\frac{5}{6}\pi\le\abs{\Delta \phi_{J}^{\rm 1\,2}}\le\frac{7}{6}\pi$}.
In addition, we might apply a restriction to the magnitude of their relative transverse
energy
\[
\et^{\rm jet2} \ge f \, \et^{\rm jet1}\;,
\] 
where $0\le f \le 1$ and $\et^{\rm jet2} \ge 10~\gev$ assuming $\et^{\rm jet1}$ is the 
energy of the leading jet.

\begin{figure}[htb]
\begin{center}
\includegraphics[width=12cm]{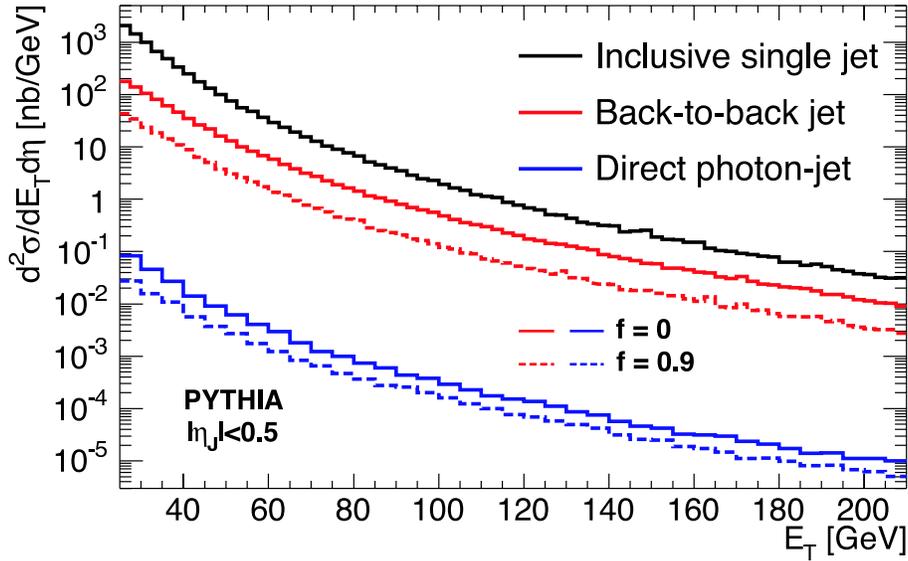}
\end{center}
\vspace{-0.3cm}
\caption[xxx]{Back-to-back jet and direct photon--jet cross sections for $f=0$ and $f=0.9$
at mid-pseudo-rapidity in \pp\ interactions at $5.5~\tev$ simulated with \acs{PYTHIA} 
compared to the inclusive single-jet cross section at the same energy. 
In all cases the jets are identified with the \acs{UA1} cone finder for $R=1.0$ 
using particles in the central \acs{ALICE} acceptance ($\abs{\eta}<1$). 
For further details see the text describing the figure.}
\label{chap5:fig:backtoback}
\end{figure}

In \fig{chap5:fig:backtoback} we report the cross section, $\dd^2\sigma/\dd\et\,\dd\eta$, 
for jet pairs fulfilling the back-to-back conditions compared to the single-inclusive
spectrum. The jets are reconstructed with the \acs{UA1} cone finder for $R=1.0$ using particles 
without detector response in the central \ac{ALICE} acceptance ($\abs{\eta}<1$). The single-jet 
spectrum is the same as shown in \fig{chap5:fig:spectrapythia}, but note the change of 
the scale to $\nb$.~\footnote{However, since we require $\et^{\rm jet2} \ge 10~\gev$
we must omit the first $\pt^{\rm hard}$ interval of $5~\gev<\pt^{\rm hard}<15~\gev$, where 
the trigger conditions get increasingly more difficult to fulfill. Instead we present the spectrum 
beginning at $25~\gev$ rather than at $15~\gev$ as in \fig{chap5:fig:spectrapythia}.}
Compared to the single-inclusive the back-to-back jet cross section for $f=0$ is about an order 
of magnitude smaller at low energies, while for $\et>100~\gev$ it is about a factor of $5$ smaller. 
Taking $f=0.9$ dramatically reduces the cross section up to a factor of $50$ at low energies, 
while at $50~\gev$ the difference is about a factor of $20$ gradually reducing to about 
difference of a factor of $10$ at highest energies. 

For completeness we also show in \fig{chap5:fig:backtoback} the direct photon--jet cross section
for $f=0$ and $f=0.9$, respectively. Here, the leading jet is replaced by prompt photons 
(mainly $\quark{q}\qubar{q} \rightarrow \quark{g}\quark{\gamma}$ and 
$\quark{q}\quark{g} \rightarrow \quark{q}\quark{\gamma}$).~\footnote{Note that we do not intend
to go into the discussion of prompt versus bremsstrahlung production~\cite{arleo2004}.}
The prompt photon together with reconstructed hadronic jet from the fragmentation of the quark or 
gluon is required to fulfill the same conditions as above. Due to the electromagnetic origin
of the prompt photon the cross section is suppressed by more than three orders of magnitude.
We see that the requirement of $f=0.9$ reduces the cross section by an additional factor of about 
$4$ at low energies to about $2$ at high energies.

Finally in \tab{chap5:tab:totalyieldback2back} and \tab{chap5:tab:totalyieldgammajet} we report the
yearly yield obtained at \ac{L1} for back-to-back jet and photon--jet production in \pp~collisions at 
$\sqrt{s}=5.5~\tev$ and in minimum-bias \PbPb~collisions at $\snn=5.5~\tev$ according to 
\eq{chap5:eq:jetyield1} as explained in \sect{chap5:intyields}. 
Back-to-back jet or photon--jet correlation might play an important role in the understanding of 
final partonic state effects in \PbPb~collisions. However, we see that mainly for the interesting
case of photon--jet correlations, where the quark or gluons jet contains almost the same energy
as the prompt photon in the \ac{ALICE} acceptance, we run into the statistical limit at energies
beyond $25~\gev$. 

\begin{table}[htbp]
\begin{center}
\begin{tabular}{l|cc | cc}
\hline
\hline
Collision type & \multicolumn{2}{c|}{\pp, $5.5~\tev$} & \multicolumn{2}{c}{\PbPb, minimum-bias} \\
Fraction of $\et$ ($f$) & $0$ & $0.9$ & $0$ & $0.9$ \\
\hline
$Y_{\rm L1}^{\rm year}(50~\gev)$  & $3.8\cdot10^{6}$ & $9.4\cdot10^{5}$ & $1.6\cdot10^{6}$ & $4.1\cdot10^{5}$ \\
$Y_{\rm L1}^{\rm year}(100~\gev)$ & $2.3\cdot10^{5}$ & $6.2\cdot10^{4}$ & $9.9\cdot10^{4}$ & $2.7\cdot10^{4}$ \\
$Y_{\rm L1}^{\rm year}(150~\gev)$ & $3.6\cdot10^{4}$ & $1.1\cdot10^{4}$ & $1.5\cdot10^{4}$ & $4.6\cdot10^{3}$ \\ 
$Y_{\rm L1}^{\rm year}(200~\gev)$ & $6.7\cdot10^{3}$ & $2.1\cdot10^{3}$ & $2.9\cdot10^{3}$ & $8.9\cdot10^{3}$ \\
\hline
\hline
\end{tabular}
\end{center}
\vspace{-0.4cm}
\caption[xxx]{Back-to-back jet yield per \acs{ALICE} run year in \pp~collisions at $\sqrt{s}=5.5~\tev$ and in 
minimum-bias \PbPb~collisions at $\snn=5.5~\tev$, both at full \acs{L1} rate.}
\label{chap5:tab:totalyieldback2back}
\end{table}

\begin{table}[htbp]
\begin{center}
\begin{tabular}{l|cc | cc}
\hline
\hline
Collision type & \multicolumn{2}{c|}{\pp, $5.5~\tev$} & \multicolumn{2}{c}{\PbPb, minimum-bias} \\
Fraction of $\et$ ($f$) & $0$ & $0.9$ & $0$ & $0.9$ \\
\hline
$Y_{\rm L1}^{\rm year}(25~\gev)$  & $1.0\cdot10^{4}$ &  $3.6\cdot10^{3}$ &  $4.3\cdot10^{3}$ & $1.6\cdot10^{3}$ \\
$Y_{\rm L1}^{\rm year}(50~\gev)$  & $1.1\cdot10^{3}$ &  $4.8\cdot10^{2}$ &  $4.5\cdot10^{2}$ & $2.1\cdot10^{2}$ \\
$Y_{\rm L1}^{\rm year}(100~\gev)$ & $8.4\cdot10^{1}$ &  $4.5\cdot10^{1}$ &  $3.6\cdot10^{1}$ & $2.0\cdot10^{1}$ \\
$Y_{\rm L1}^{\rm year}(150~\gev)$ & $1.6\cdot10^{1}$ &  $9.0\cdot10^{0}$ &  $7.0\cdot10^{0}$ & $4.0\cdot10^{0}$ \\
\hline
\hline
\end{tabular}
\end{center}
\vspace{-0.4cm}
\caption[xxx]{Direct photon--jet yield per \acs{ALICE} run year in \pp~collisions at $\sqrt{s}=5.5~\tev$ and in 
minimum-bias \PbPb~collisions at $\snn=5.5~\tev$, both at full \acs{L1} rate.}
\label{chap5:tab:totalyieldgammajet}
\end{table}
\fi

%

\newif\ifinclusivecorr
\inclusivecorrtrue
\newif\ifleadingjets
\leadingjetstrue

\chapter{Jet quenching in ALICE}
\label{chap6}
The preceding discussion has outlined the research interest in jet physics 
in heavy-ion collisions at the \ac{LHC} and the potential of the central 
\ac{ALICE} detectors for jet identification and reconstruction. 
Qualitatively, we expect in these collisions hard scattered partons to be 
used as tomographic probes of the produced partonic matter and, thus, the 
properties of produced hadronic jets to be modified. 

In the present chapter, we will give an outline of the perspectives to study
jet modification at \ac{LHC} conditions. At first, we introduce 
in \sect{chap6:hadronizationvstherm} two qualitative regimes of jet energy, 
where in-medium effects may be studied. In the subsequent sections we discuss
the potential for jet-quenching measurements in \ac{ALICE}, briefly touching 
the low energy regime in \sect{chap6:inclcorr}, while focusing in 
\sect{chap6:leadingjets} onto jets accessible only at \ac{LHC}.

\section{Low and high energy regimes}
\label{chap6:hadronizationvstherm}
In order to separate qualitative different manifestations of in-medium effects,
one may determine on the basis of the measured hadronic final state, 
whether interaction processes occur in the medium, and, 
whether they manifest in terms of partonic or hadronic degrees of freedom. 
Following~\Ref{wiedemann2004} we consider a parton of high $\et$ produced in a hard 
collision. If the parton escapes into the vacuum, it will reduce its initial virtuality, 
$Q$, by perturbative parton splitting, until after some time, $\propto 1/Q_{\rm hadr}$, 
it is degraded to the hadronic scale of about $Q_{\rm hadr}\approx 1~\gev$. 
Numerical estimates for the time scale of hadronization vary 
significantly~\cite{wang2003,kopeliovich1990}, but due to the Lorentz boost to the 
laboratory frame the scale is proportional to the energy,
\begin{equation}
L_{\rm hadr} \simeq \const/Q_{\rm hadr}^2\, \et\;,
\end{equation}
where $\const=2$ to account for multiple parton branching.
If instead the hard parton escapes into an infinitely extended \ac{QGP},
the initial perturbative parton splitting is more efficient because of 
medium-induced gluon radiation. Since the parton cannot hadronize in the 
dense medium, after some time its partonic fragments will no longer be 
distinguished from the heat bath: it is thermalized. To estimate the corresponding 
time scale, $L_{\rm therm}$, one may require that the hard parton has lost all its energy 
through medium-induced gluon radiation. According to the \acs{BDMPS-Z} energy loss 
formula, \peq{chap3:eq:avdE}, the partonic thermalization length is given by
\begin{equation}
L_{\rm therm} \simeq \sqrt{4/(\as\,C_{\rm R}\,\hat{q})} \, \sqrt{\et}\;.
\end{equation}

These simplified estimates illustrate that for high $\et$ perturbative mechanisms may 
indeed remain undisturbed by hadronization over a significant time scale.
Depending on its in-medium path length, $L_{\rm med}$, the hard parton 
will either be absorbed, \mbox{$L_{\rm therm} < L_{\rm med} < L_{\rm hadr}$}, 
or it has a sufficiently large transverse energy to suffer only 
the onset of thermalization processes, \mbox{$L_{\rm med} < L_{\rm therm}
< L_{\rm hadr}$}. It is the latter case, in which the parton appears 
as a medium-modified jet. For lower transverse energies, there is not only 
a competition between the hadronization and the thermalization mechanism,
\mbox{$L_{\rm hadr} \sim L_{\rm therm}$}, but also the possibility that the 
medium interferes with the dynamics of hadronization, \mbox{$L_{\rm hadr} 
\sim L_{\rm med}$}. 

Taking $\as\,C_{\rm R}=1$ and $\hat{q}=1~\gev^2/\fm$, the condition 
\mbox{$L_{\rm med} < L_{\rm therm} < L_{\rm hadr}$} is fulfilled for partons with 
$\et\gsim 10~\gev$, whereas already at $\hat{q}=5~\gev^2/\fm$ the conditions requires 
$\et\gsim20~\gev$ (for $L_{\rm med}=2~\fm$). Thus, at \ac{LHC} opposed to \ac{RHIC},
for high transverse energies there might be the chance ---depending on the concrete 
value of the transport coefficient--- 
to study the evolution of out-of-equilibrium partons mostly undisturbed by hadronization. 

For these jets, the interaction with the medium is expected to manifest 
in the modification of jet properties deviating from known fragmentation processes in 
vacuum. Calculations predict that the energy lost by the parton remains inside the jet 
cone, although redistributed in transverse phase space~\cite{wiedemann2000c,salgado2003}. 
The corresponding jet-production cross section is expected to follow binary scaling. 
However, the jet shape is claimed to broaden and the jet multiplicity to soften and 
increase~\cite{salgado2003b}. Ideally, one should reconstruct the hadronic energy, which
for outstanding high-energy jets may be associated with the energy of the parent parton.
For these jets, we will compare differences of known properties to \pp\ and peripheral 
\PbPb~collisions. Varying the jet energy, may characterize parton interactions 
in dense colored matter over the widest possible energy range. These measurements 
require the full reconstruction power of the central barrel, 
ideally along with the \ac{EMCAL}.

\section{Inclusive leading-particle and jet-like correlations}
\label{chap6:inclcorr}
\ifinclusivecorr
We restrict the discussion to low energies, $5~\gev<\et<30~\gev$, of the jet spectrum in \PbPb\ collisions, 
where there is some overlap with mini-jet and jet production at ongoing and future \ac{RHIC} measurements.

The main problem of jet identification in heavy-ion collisions is the complexity of the underlying 
high-multiplicity background, as we have seen in the previous chapter. At these energies, jets cannot 
be reconstructed as identified objects, neither at \ac{RHIC}, nor at \ac{LHC}. 
Therefore, in this range one is limited to discerning \ac{QCD} medium effects from intermediate to 
high-$\pt$ inclusive spectra and angle or energy correlation studies. 
This restricts what can be unambiguously learned about jet quenching, since, for example, no primary parton 
momentum can be deduced. 

However, such measurements may provide the best means to directly compare the matter produced at 
\ac{RHIC} with that produced at the \ac{LHC}. As outlined above, the physics issues in the low energy 
regime, common to both colliders, may be distinctly different from the higher energy regime, uniquely 
accessible at \ac{LHC}. It is expected and supported by \ac{RHIC} measurements that in-medium 
modifications of the jet structure will be stronger at low jet energies 
effects. Therefore, it will be interesting to quantify changes in the produced matter from \ac{RHIC} to 
\ac{LHC}. 

\begin{figure}[htb]
\begin{center}
\subfigure[Particle density (ideal barrel)]{
\label{chap6:fig:partcorr}
\includegraphics[width=7.2cm, height=4.5cm]{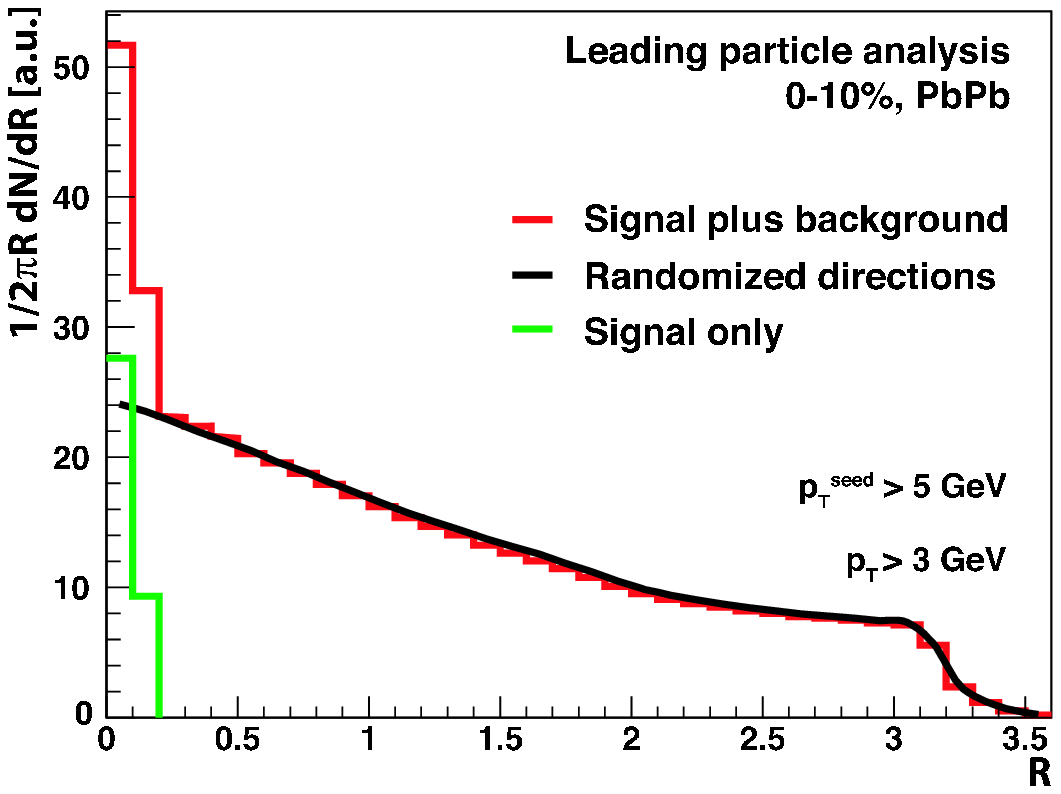}}
\hspace{0.1cm}
\subfigure[Fragmentation estimation (ideal barrel)]{
\label{chap6:fig:leadfrag}
\includegraphics[width=7.2cm, height=4.5cm]{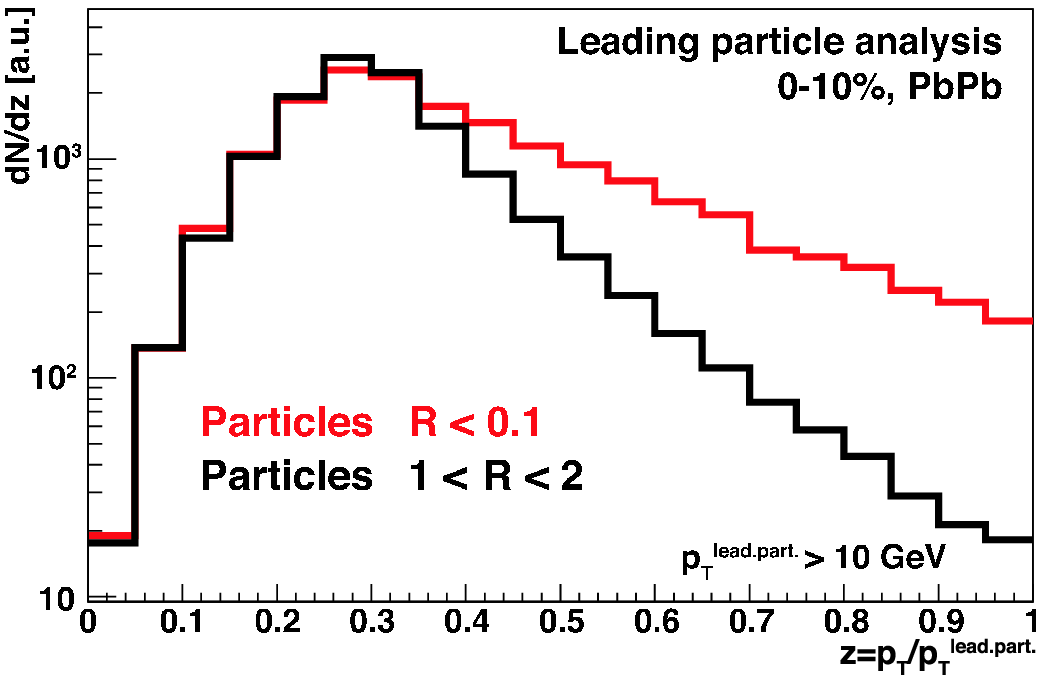}}
\end{center}
\vspace{-0.5cm}
\caption[xxx]{\subref{chap6:fig:partcorr}~Particle-density distribution, $(2\pi R)^{-1} \dd N/\dd R$, 
for correlated and uncorrelated particles as a function of $R$ obtained with 
$\pt^{\rm seed} = 5~\gev$ and a $\pt$-cut of $3~\gev$ in $0$--$10$\% central \PbPb~collisions. 
\subref{chap6:fig:leadfrag}~Estimation of the fragmentation function of correlated 
particles compared to the uncorrelated background in $0$--$10$\% central \PbPb~collisions.
Both figures are adapted from \Ref{cormier2004}.}
\label{chap6:fig:corrfrag}
\end{figure}

\enlargethispage{0.5cm}
To study correlations with leading particles, an algorithm similar to the one used for the 
\acs{CDF} charged-jet analysis is used~\cite{affolder2001b}. Central \PbPb~collisions are simulated
using \acs{HIJING} as explained in \psect{app:hijing}. The discussion is limited to the case of the 
ideal barrel.

All charged particles with $\pt>\pt^{\rm seed}$ are initially regarded as leading-particle 
candidates, $P_i$, ordered according to their transverse momentum, \ie~$P_0>P_1>\cdots>P_n$. 
The algorithm starts with the highest candidate, $P_0$, and records the distances $R$ in the $\eta\times\phi$ 
space between all particles (above a certain $\pt$-threshold) and $P_0$. If another candidate is found 
within a distance $R < R_{\rm sep}=1.3$ it is eliminated from the list of candidates. The procedure continues 
with the next candidate on the list until no candidate is left. The algorithm is a natural extension of 
the cone algorithms (see \psect{chap3:jetalgorithms}) to be used for inclusive studies in the low jet-energy 
region for heavy-ions collisions. 

To visualize possible angular correlations, one may plot the density, $(2\pi R)^{-1} \dd N/\dd R$, 
as a function of the distance, $R$ shown in \fig{chap6:fig:partcorr}. The distribution is obtained for 
$\pt^{\rm seed} = 5~\gev$ with a $\pt$-cut on all particles of $3~\gev$. After subtracting the 
corresponding distribution for randomized leading-particle directions, a clear near-side correlation 
signal is visible for $R<0.3$, which results from the mini-jets present in the \acs{HIJING} simulation 

The distribution of the ratio given by the transverse momentum of the correlated particles 
to the transverse momentum of the leading-particle, $z=\pt / \pt^{\rm lead.part.}$, is an 
estimator of the jet \ac{FF}. It is shown in \fig{chap6:fig:leadfrag} for particles 
in the near-side correlation with $R<0.1$, compared to the background distribution obtained from 
uncorrelated particles with $1<R<2$. The signal dominates by a large factor at higher $z$-values. 
The \ac{FF} approximated by the correlated particles suffers from the fact that the estimation 
of the jet energy by $\pt^{\rm lead.part.}$ is rather poor, smearing out the true 
jet \ac{FF}. However, since in-medium modification of the \ac{FF} will be strong for low energy 
jets, it is reasonable to  expect that the quenching effect should be observed in the measured 
leading-particle distribution, for example by comparing central versus peripheral \PbPb\ collisions. 

This section limited in scope is based on~\Refs{accardi2003,cormier2004}, where the reader may find 
further details and observables, which are partially covered in the next section, at high $\et$. 
\fi

\section{Identified jets at high energy}
\label{chap6:leadingjets}
\ifleadingjets
We will now focus on studies with identified jets for $\et\gsim50~\gev$.
In combination with results at lower energies from correlation methods, mentioned above, 
these measurements might complete the picture of medium-induced parton-energy-loss 
phenomena. It is our aim to deduce medium effects based merely on the comparison 
with measurements in \pp.

The unquenched spectrum representing the \pp\ reference is generated as in 
\psect{chap5:reconjetspectrum}, using 10000 \acs{PYTHIA} events at $\sqrt{s}=5.5~\tev$ per 
$\pt^{\rm hard}$ interval. 
Quenched signal jets are simulated in the same way, however, by a modified version of \acs{PYTHIA}, 
described in \psect{app:mcqm}.
It introduces parton energy loss via final-state gluon radiation, in a rather ad-hoc way:
Before the partons originating from the hard $2$-to-$2$ process (and the gluons 
originating from \ac{ISR}) are subject to fragment, they are replaced according 
\[
{\rm parton}_i(E) \rightarrow {\rm parton_i}(E-\Delta E) + n(\Delta E)\,{\rm gluon}(\Delta E/n(\Delta E))
\]
conserving energy and momentum. We distinguish between two toy models, 
where $\Delta E$ and $n(\Delta E)$ are given as follows:
\begin{itemize}
\item {\em fixed} energy loss: $\Delta E/E=0.2$ and $n=1$ independent of parton type and parton
production point in the system (geometry);
\item {\em variable} energy loss: $\Delta E$ given by \ac{PQM} for the non-reweighted case,
dependent on medium density, parton type and parton production point in the system (geometry).
In the following discussion, we will compare jet samples prepared for $k=5\cdot10^{5}$, $5\cdot10^{6}$ 
and $1\cdot10^{7}~\fm$, which for $0$--$10$\% central \PbPb~collisions lead to $\av{\hat{q}}=1.2$, 
$12$ and $24~\gev^2/\fm$.
The choice of the first value implies only a very little modification of the embedded 
quenched jets compared to the embedding with the standard \acs{PYTHIA}. 
The second corresponds 
to the value of $k$ found to describe the $\RAA$ at \ac{RHIC}, whereas the third is a conservative 
choice below the extrapolation from \ac{RHIC} to \ac{LHC} (in fact, a factor of $5$ lower, see 
\psect{chap3:pqm}). The number of radiated gluons, $1\le n(\Delta E)\le6$, is determined by the 
condition that each gluon must have less energy than the quenched parton from which it was 
radiated away.~\footnote{There is another quenching procedure~\cite{lokhtin2004}, which generates
medium-modified jets in \acs{PYTHIA}, simulating medium-induced rescattering and energy loss of 
hard partons in an expanding \ac{QGP}. For typical settings, this model, however, introduces 
about $40$ additional gluons.}
\end{itemize}

The fixed model with the chosen parameters applied to leading-hadron analysis, via \eq{chap3:eq:raaapprox}, 
would approximately yield $\RAA\sim0.25$ (for $n(\pt)\sim 7$), which is comparable to the value obtained
by \ac{PQM} for $\av{\hat{q}}=12~\gev^2/\fm$ (\cf~\pfig{chap3:fig:raalhcomp}). Nevertheless, both models are 
conceptually very different. In the fixed model all initial partons are quenched in the same way regardless 
of their identity and production point in the collision, \ie~without surface effect. Instead \ac{PQM} 
introduces the \acs{BDMPS-Z-SW} framework and the geometry of the collision, and, thus, partons are treated 
based on their identity and depending on their path trough the medium. In the latter case, there will be
jets, which suffer no or almost no quenching, and jets, which suffer maximum quenching. However, due to 
the trigger bias, one is predominantly sensitive to jets, emerging close to the surface.

The sample of jets prepared with the modified \acs{PYTHIA} version is then embedded into $0$--$10$\% 
central \acs{HIJING} events.~\footnote{As before, a total of 2500 background events is used.} 
These events represent the quenched spectrum in \PbPb\ and will be discussed in the following.
As in the previous section the analysis is mostly restricted to the case of the ideal barrel, 
\ie~charged particles in the central \ac{ALICE} acceptance taken from Monte Carlo 
without simulated detector response. To cope with the soft background in \PbPb\ we use the cone finder,
\acs{ILCA}, with $R=0.3$ and a cut of $\pt>2~\gev$, also for the reference measurement in \pp. 
The number of jets with highest initial energies ($\sim200~\gev$) in the sample roughly 
corresponds to the number within the integrated yield of one \ac{ALICE} year in \PbPb\ 
($\sim10^{4}$), assuming we can fully exploit the maximum \ac{L1} jet production rate. 
For lower jet energies, weights reflecting their increase in cross section are introduced 
to account for their predominance in the sample. Statistical errors for these events are therefore
not representative.

In the following it is our aim to discuss simple, model-independent observables, which might be 
sensitive to in-medium modification of jets, but insensitive to soft particles from the underlying 
background event, and to compare the different quenching scenarios.~\footnote{See also the discussion
in ~\Ref{morsch2005}, which disentangles various effects.} 
For the scope of the thesis, the analysis will be restricted to jets with highest energy
in the event. We know from the previous chapter that the energy resolution in \PbPb\ in the case 
of the ideal barrel alone is limited to a mean and width of about $50$\%. Therefore, we will 
introduce for all observables a lower bound on the energy of jets, which are included in the analysis, 
\mbox{$\et^{\rm rec}> E_{\rm T,\,min}^{\rm rec}$}. 
This way it is ensured that the true energy of the jets in the sample cannot be lower than 
$E_{\rm T,\,min}^{\rm rec}$, however the jets still may contain contributions from the 
underlying background. Due to the steeply falling jet-production spectrum, the analysis will be 
dominated by jets, whose true energy leads, on average, to about $E_{\rm T,\,min}^{\rm rec}$.

\subsection{Longitudinal and transverse momentum distributions}
The physics signature of medium-induced gluon radiation generally is believed to be visible in 
the modification of the jet \acf{FF} as measured through the longitudinal and transverse momentum 
distributions of associated hadrons within the jet. 
The momenta parallel to the jet axis, 
$p_{\rm L} = p_{\rm hadron} \, \cos(\theta_{\rm jet},\theta_{\rm hadron})$, 
are expected to be reduced~(jet quenching), while the momenta in the transverse direction, 
$j_{\rm T} = p_{\rm hadron} \, \sin(\theta_{\rm jet},\theta_{\rm hadron})$, to be 
increased~(transverse heating). 

\begin{figure}[htb!]
\begin{center}
\subfigure[Longitudinal for $\et^{\rm rec}>30~\gev$ (ideal barrel)]{
\label{chap6:fig:jetfraglongidealb}
\includegraphics[width=12.5cm]{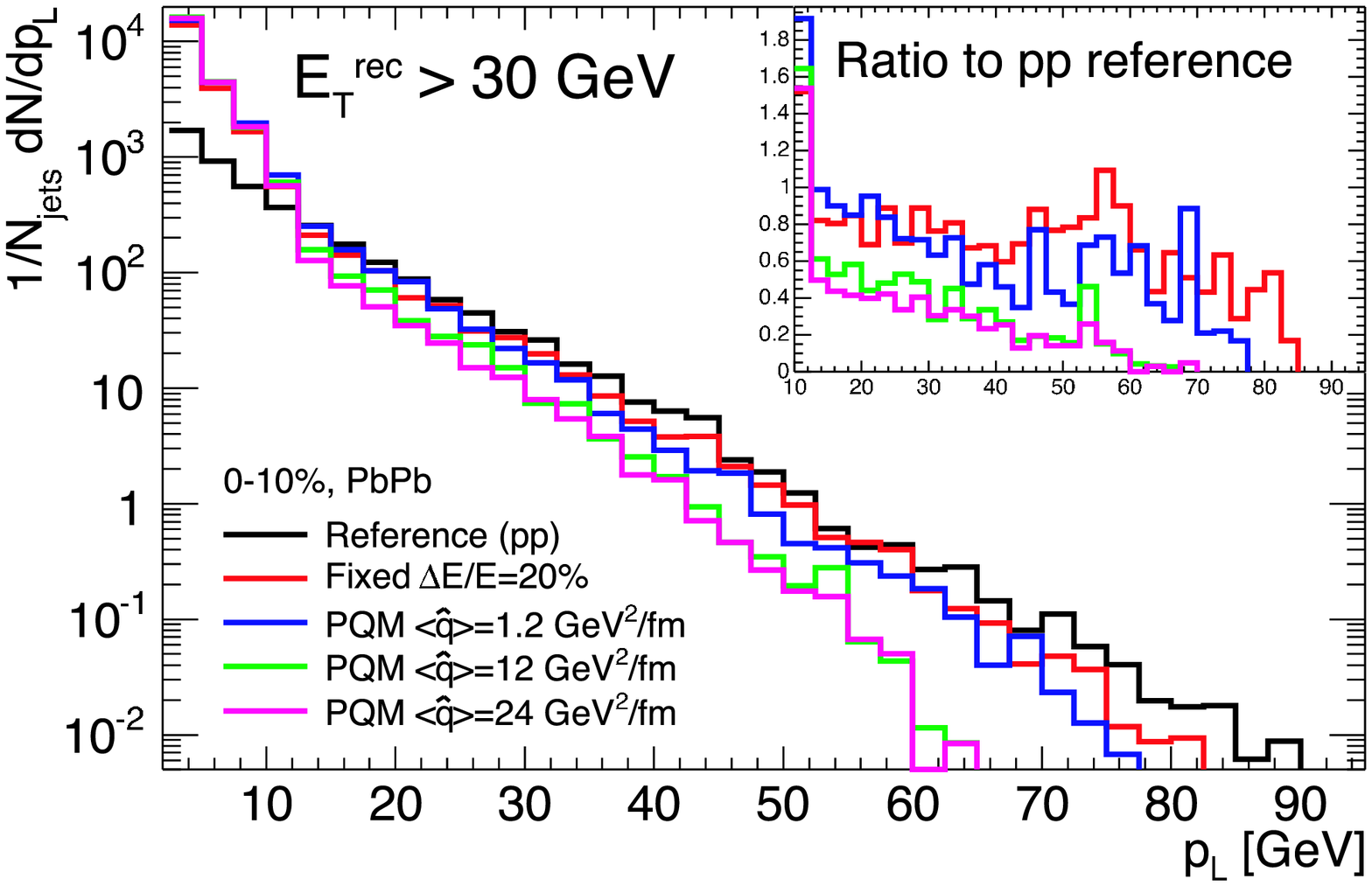}}
\vfill
\vspace{0.5cm}
\subfigure[Longitudinal for $\et^{\rm rec}>80~\gev$ (ideal barrel)]{
\label{chap6:fig:jetfraglongidealb2}
\includegraphics[width=12.5cm]{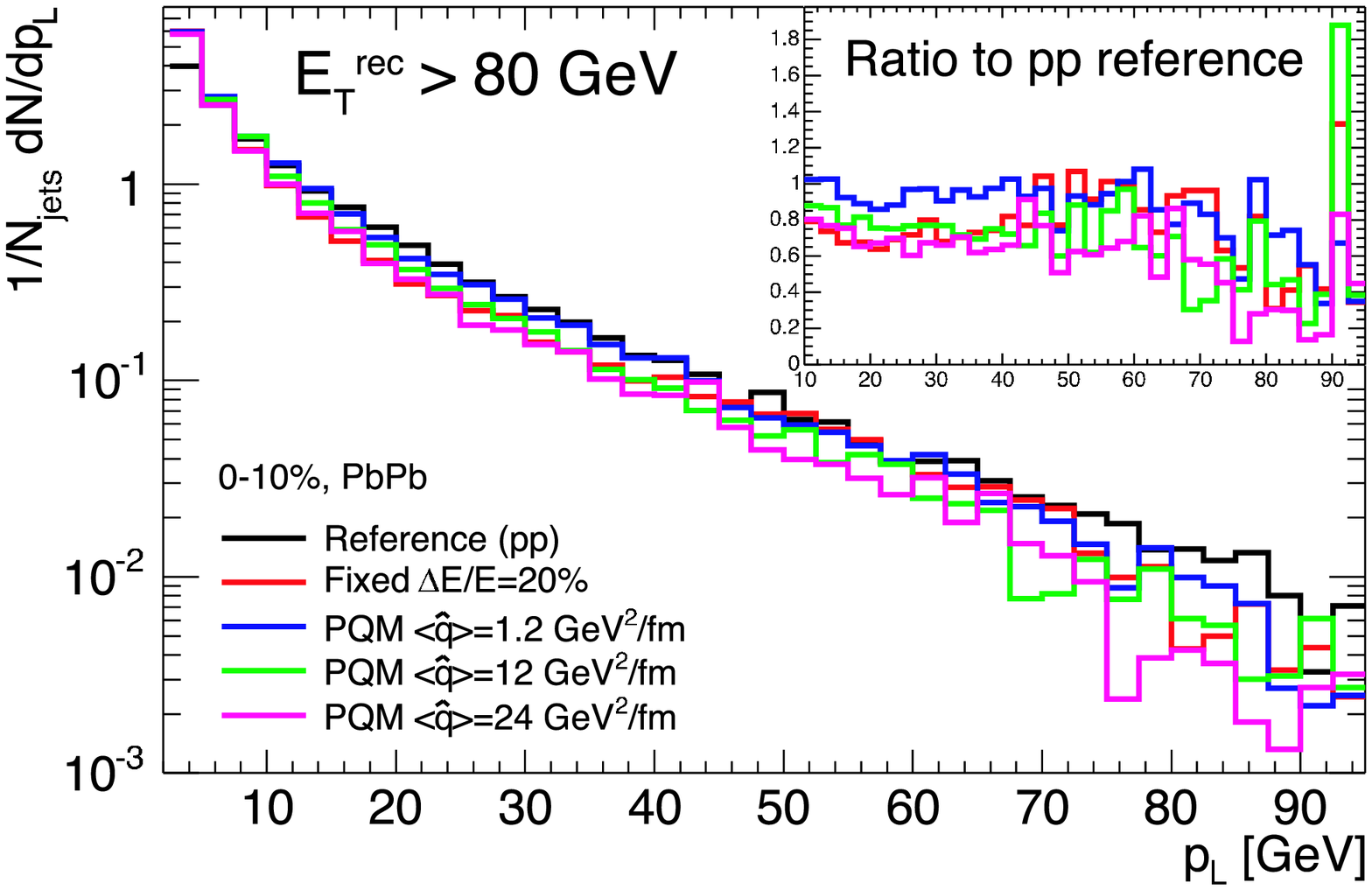}}
\end{center}
\vspace{-0.5cm}
\caption[xxx]{Longitudinal momentum distributions, with respect to the jet axis,
for charged particles in jets with $\et^{\rm rec}>30~\gev$~\subref{chap6:fig:jetfraglongidealb} 
and $\et^{\rm rec}>80~\gev$~\subref{chap6:fig:jetfraglongidealb2}
for different quenching scenarios in $0$--$10$\% central \PbPb\ compared 
to \pp\ collisions. The inset shows the ratio of the distribution 
for the different \PbPb~cases to the distribution obtained in \pp.
In both figures the jets are identified with the \acs{ILCA} cone finder 
using $R=0.3$ and $\pt>2~\gev$ in case of the ideal barrel.}
\label{chap6:fig:jetfragidealb}
\end{figure}

In \fig{chap6:fig:jetfraglongidealb} we show the longitudinal distribution 
for different quenching scenarios in $0$--$10$\% central \PbPb\ compared 
to \pp\ collisions taking into account all reconstructed jets where 
$\et^{\rm rec}>30~\gev$. The expected behavior is clearly visible: higher medium 
density leads to stronger suppression of the momenta along the jet axis and 
enhancing of smaller momenta. At $p_{\rm L}\gsim25~\gev$ the effect of the 
medium-induce radiation becomes apparent for the two dense \ac{PQM} cases.
However, at low $p_{\rm L}\lsim 10~\gev$, the strong change of the longitudinal 
distribution is predominantly induced by the remaining hadrons of the underlying 
background, since in this region all quenched case agree. It is obvious that jet 
quenching effects will be measurable in the ratio of the longitudinal distribution
obtained for jets in \PbPb\ with respect to \pp.  
However, given the statistics we have at hand, the quantitative distinction 
between, for example, the two dense \ac{PQM} scenarios seems to be impossible. 

\begin{figure}[htb!]
\begin{center}
\subfigure[Transverse for $\et^{\rm rec}>30~\gev$ (ideal barrel)]{
\label{chap6:fig:jetfragtransidealb}
\includegraphics[width=10cm]{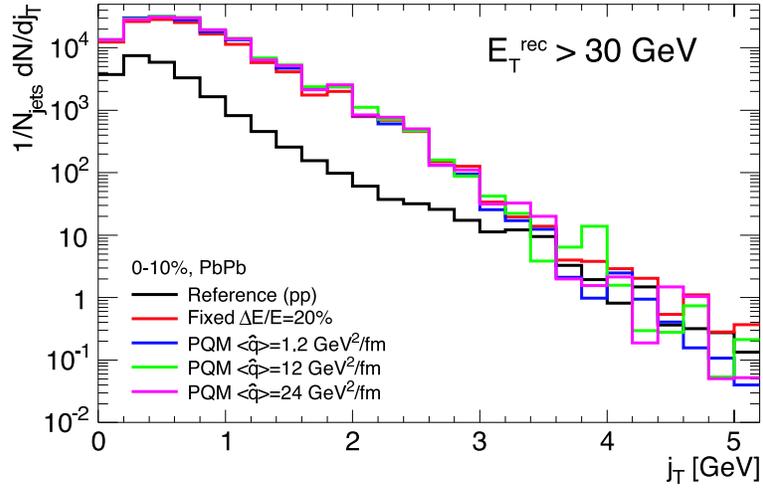}}
\subfigure[Transverse for $\et^{\rm rec}>80~\gev$ (ideal barrel)]{
\label{chap6:fig:jetfragtransidealb2}
\includegraphics[width=10cm]{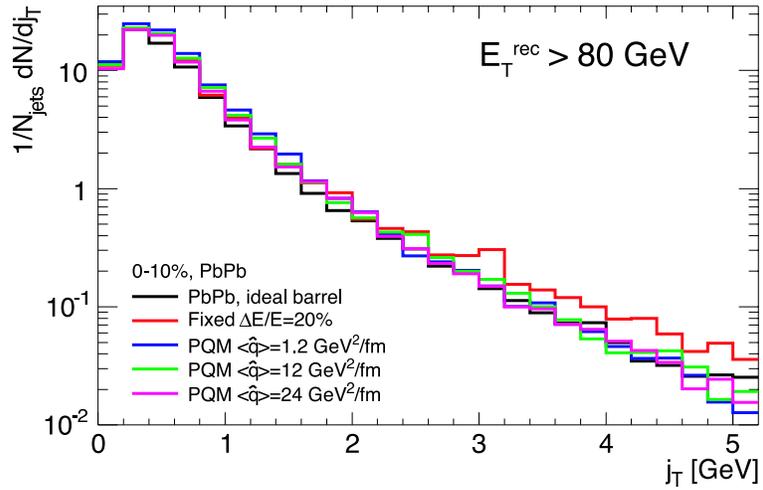}}
\end{center}
\vspace{-0.5cm}
\caption[xxx]{Transverse momentum distributions, with respect to the jet axis,
for charged particles in jets with $\et^{\rm rec}>30~\gev$~\subref{chap6:fig:jetfraglongidealb} 
and $\et^{\rm rec}>80~\gev$~\subref{chap6:fig:jetfraglongidealb2}
for different quenching scenarios in $0$--$10$\% central \PbPb\ compared 
to \pp\ collisions. In both figures the jets are identified with the 
\acs{ILCA} cone finder using $R=0.3$ and $\pt>2~\gev$ in case of the 
ideal barrel.}
\label{chap6:fig:jetfragtidealb}
\end{figure}

To reduce the contribution of the background, we increase the cut 
on the jet energy to $\et^{\rm rec}>80~\gev$ as shown in \fig{chap6:fig:jetfraglongidealb2}. 
For low momenta the shape of the quenched distributions now turn out to be quite similar to the 
\pp\ reference. Measuring the excess of low $p_{\rm L}$ with respect to \pp\ will be
challenging, but of main interest for the comparison with theory. Looking at the ratio of the \PbPb\ 
cases with respect to the \pp\ reference, the difference in the strength of the medium-induced 
radiation becomes seemingly apparent for the two dense cases of \ac{PQM}, and, by eye, both may be 
distinguished. Although, for quantitative studies, high statistics will be needed at higher $p_{L}$. 
For jets, merely reconstructed with the ideal barrel, energies of $\et^{\rm rec}>80~\gev$ might be 
at the statistical limit, since the jets, which contribute to the analysis, mainly arise from jets 
with true energies of about $\et\gsim 120~\gev$. As a consequence of the trigger bias and due to the 
surface effect, a considerable fraction of jets does not (or almost not) suffer from the quenching. 
This is opposed to the fixed-energy loss model, where jets from all depths are equally quenched. 
It, therefore, will be fruitful to study the dependence of the $p_{\rm L}$ distribution with 
collision centrality and ---if statistics allows--- with respect of the jet direction to the 
reaction plane.

The momentum distribution transverse to the jet axis is plotted in 
\fig{chap6:fig:jetfragtransidealb} for $\et^{\rm rec}>30~\gev$. 
Strong broadening of the transverse distribution compared to the vacuum case is observed. 
However, the observed modification originates from the mixing with the \acs{HIJING} event
rather from the interactions of the primary parton with the dense medium, since the
modified \acs{PYTHIA} version does not include such effects.~\footnote{It has been suggested
to introduce the medium effect of the broadening, $\kt~\sim\sqrt{\hat{q}\,L}$, into the
quenching routine. However, we abandon the idea, since neither it has been known how to 
distribute the effect among the radiated gluons, nor how to deal (within \acs{PYTHIA}) with 
the large momentum transfer, the broadening would imply.}
This is confirmed by \fig{chap6:fig:jetfragtransidealb2} where in the analysis only 
jets with $\et^{\rm rec}>80~\gev$ are contributing. Since all models agree with the
\pp\ reference, it seems possible that one can exclusively measure transverse broadening 
for these jets, if it exists in nature.

Since the $j_{\rm T}$ distribution is expected to significantly broaden in central 
\PbPb~collisions~\cite{salgado2003b}, and since its average has been measured in numerous collider 
experiments and found to be relatively insensitive to the collision energy~\cite{angelis1980}, 
the average particle-momentum associated in a jet transverse to the jet axis,
$\av{j_{\rm T}}$, might be a key observable at the \ac{LHC}. 
For our toy models we find $\av{j_{\rm T}}=0.7~\gev$ for $\et^{\rm rec}>30~\gev$ 
and $\av{j_{\rm T}}=0.63~\gev$ for $\et^{\rm rec}>80~\gev$ compared to 
$\av{j_{\rm T}}=0.51~\gev$ and $\av{j_{\rm T}}=0.61~\gev$ in the vacuum.~\footnote{The fixed 
energy-loss model for $\et^{\rm rec}>80~\gev$ systematically leads to higher 
values of $j_{\rm T}$, resulting in $\av{j_{\rm T}}=0.66~\gev$. The effect is verified to be
even visible, with about $1.5\sigma$, in the quenched spectrum without mixing with the \acs{HIJING}
background. The reason is not quite clear, but since we find the same behavior without mixing, we may 
positively conclude that the transverse momentum distribution for jets with high energy ($\et\gsim120~\gev$)
can be measured with little influence of the underlying event.}

It will be interesting to study 
$\av{j_{\rm T}}$ as a function of the minimum jet energy, $\et^{\rm rec}>E_{\rm T,\,min}^{\rm rec}$, 
and to link the observed distributions in reconstructed jets with those deduced for jets in the 
low energy range from correlations studies of leading particles, also for varying centrality.

\subsection{Leading-particle fragmentation}
The modification of the jet \ac{FF} must be reflected in the quenching of the leading-particle
transverse momentum, because prior to hadronization the primary parton has lost a considerable 
fraction of its energy into gluon radiation while traversing the dense medium.~\footnote{Note that 
for jets at central rapidities the transverse momentum of the leading particle in the \cms~system 
roughly corresponds to the longitudinal momentum along the jet axis.}
Since it is expected that the radiated gluons will subsequently fragment into hadrons inside the 
jet cone, the total energy of the jet and its spatial distribution are only moderately 
modified. It has been shown~\cite{salgado2003b} that the medium-induced broadening of the jet 
reduces the energy inside $R=0.3$ by $\sim 15$\% and by $\sim7$\% for jets with $\et=50$ and 
$\et=100~\gev$, respectively, and already at $R=0.7$ the effect reduces to about $2$\%. 
However, one must be cautious about these findings, since this prediction has been calculated 
assuming a rather low value of the gluon density at \ac{LHC} conditions. Furthermore, 
the calculation has merely been performed at the parton level. 

Nevertheless, assuming the energy remains within the cone and assuming that the estimation of $\et$ 
with the reconstructed (charged) energy in the reduced size of the cone is accurate enough to not 
completely mask the effect, the fraction \mbox{$z=\pt^{\rm lead.part.}/\et$} might be sensitive to 
the in-medium modification of the jet \ac{FF}.

\begin{figure}[htb]
\begin{center}
\subfigure[Leading-particle frag. (ideal barrel)]{
\label{chap6:fig:jetfraglead50to350idealb}
\includegraphics[width=7.2cm]{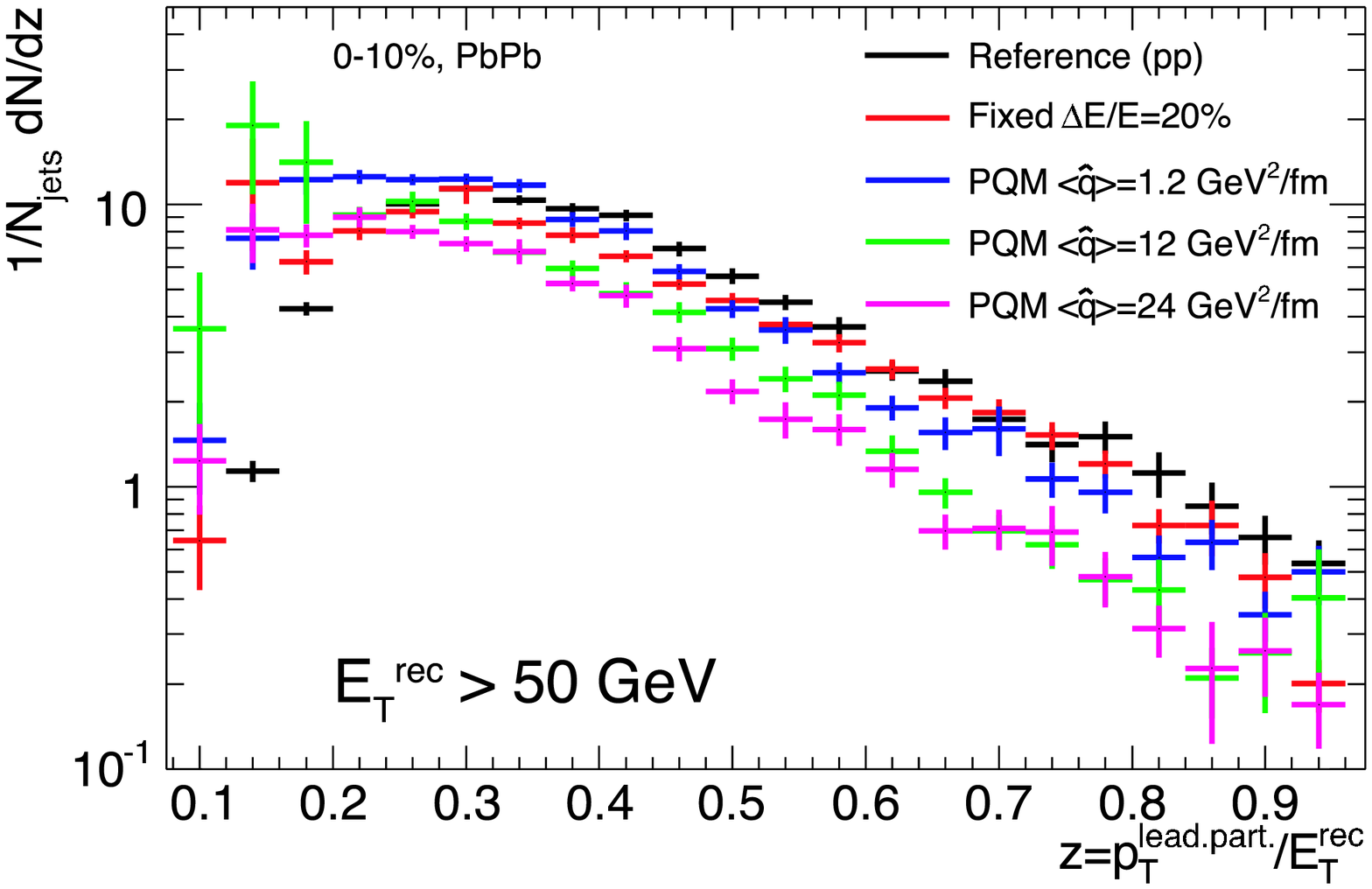}}
\hspace{0.1cm}
\subfigure[Fraction for $\et^{\rm rec}>E_{\rm T,\, min}^{\rm rec}$ (ideal barrel)]{
\label{chap6:fig:jetfragleadminetidealb}
\includegraphics[width=7.2cm]{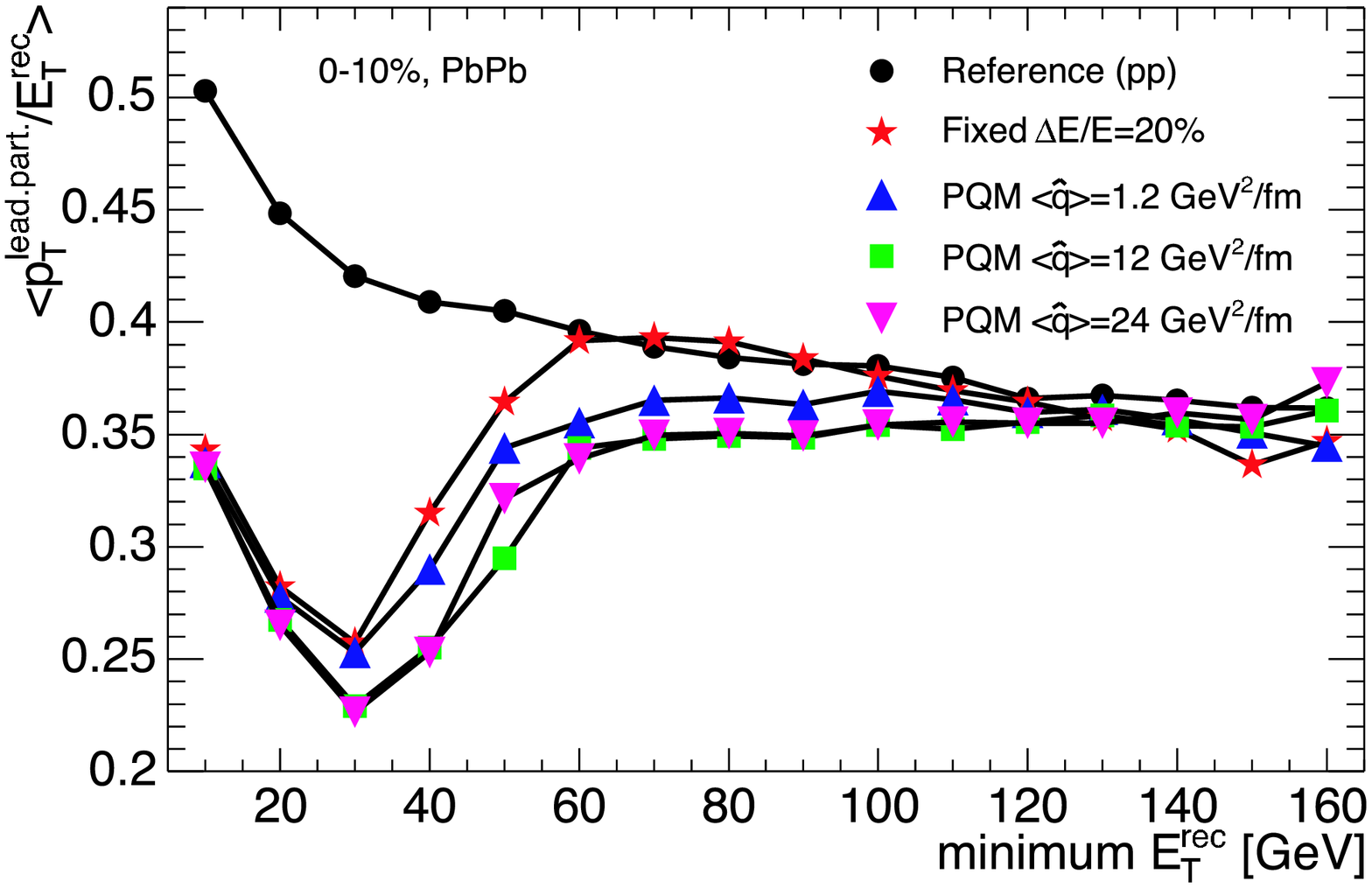}}
\subfigure[Leading-particle frag. (ideal barrel+em)]{
\label{chap6:fig:jetfraglead50to350idealbem}
\includegraphics[width=7.2cm]{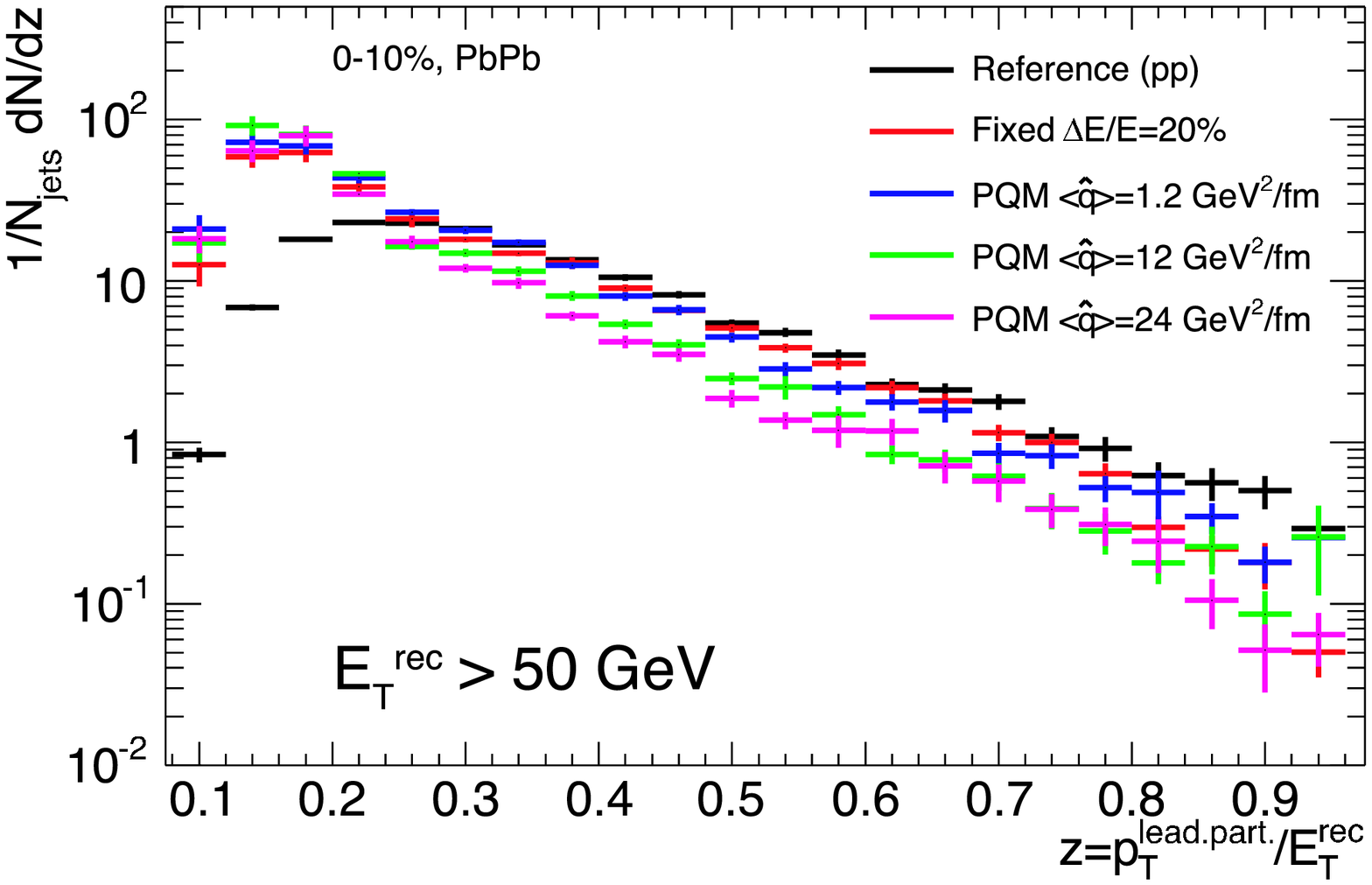}}
\hspace{0.1cm}
\subfigure[Fraction for $\et^{\rm rec}>E_{\rm T,\, min}^{\rm rec}$ (ideal barrel+em)]{
\label{chap6:fig:jetfragleadminetidealbem}
\includegraphics[width=7.2cm]{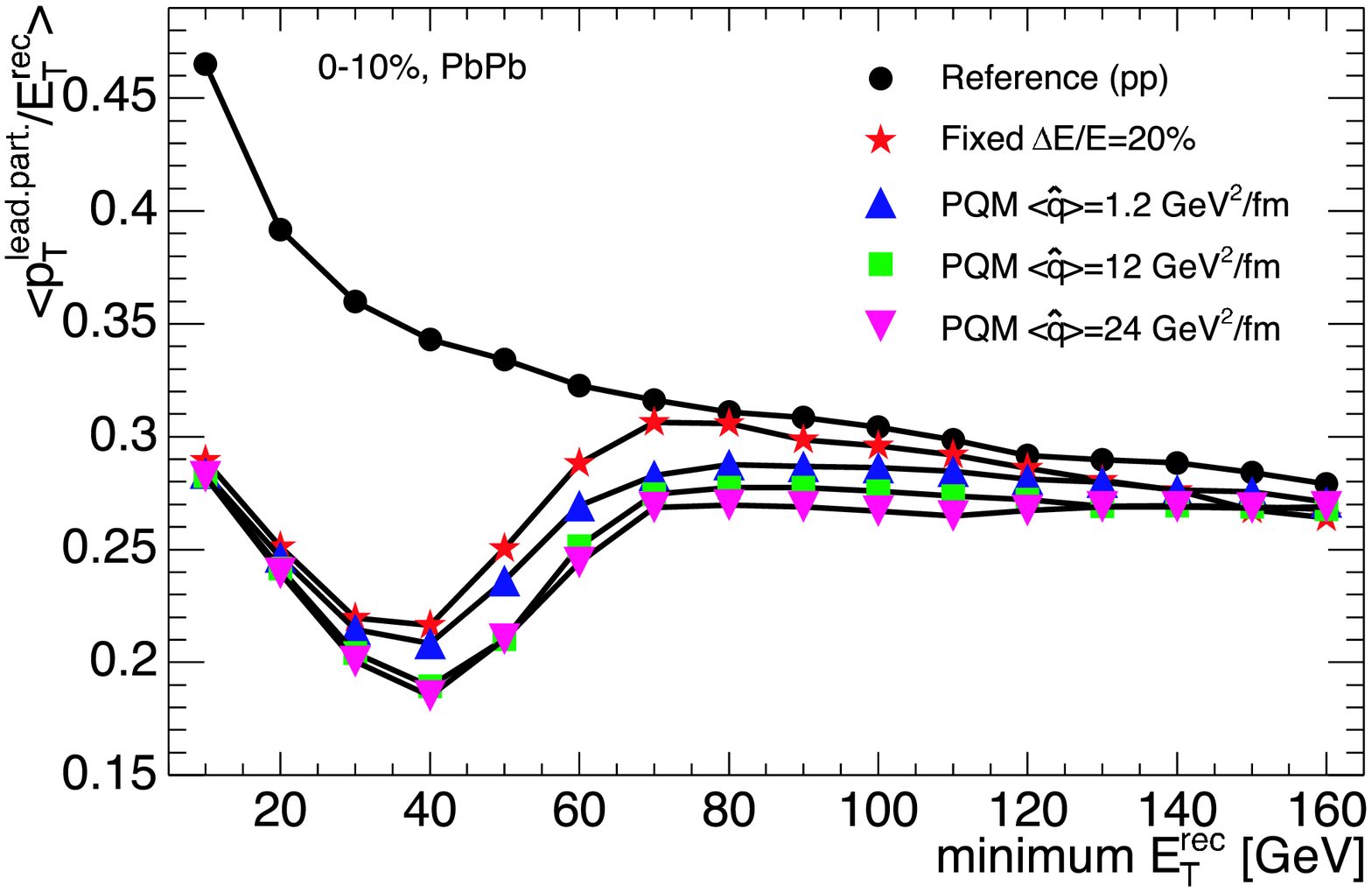}}
\end{center}
\vspace{-0.5cm}
\caption[xxx]{Leading-particle fragmentation in jets with $\et^{\rm rec}>50~\gev$ 
in the case of the ideal barrel~\subref{chap6:fig:jetfraglead50to350idealb}
and of ideal barrel+em~\subref{chap6:fig:jetfraglead50to350idealbem}
for different quenching scenarios in $0$--$10$\% central \PbPb\ compared to \pp\ collisions.
Fraction of jet energy carried by the leading particles
in jets with $\et^{\rm rec} > E^{\rm rec}_{\rm T,\,min}$ as a function 
of $E^{\rm rec}_{\rm T,\,min}$ 
in the case of the ideal barrel~\subref{chap6:fig:jetfragleadminetidealb}
and of ideal barrel+em~\subref{chap6:fig:jetfragleadminetidealbem} 
for different quenching scenarios in $0$--$10$\% central \PbPb\ compared to \pp\ collisions.
In all figures the jets are identified with the \acs{ILCA} cone finder 
using $R=0.3$ and $\pt>2~\gev$.}
\label{chap6:fig:jetfragleadideal}
\end{figure}

\Fig{chap6:fig:jetfraglead50to350idealb} shows the fragmentation function for leading particles, 
$\dd N/\dd z$ for \mbox{$z=\pt^{\rm lead.part.}/\et^{\rm rec}$}, in jets where 
$\et^{\rm rec}>50~\gev$. Indeed, the expected behavior is visible, higher medium density leads to 
stronger suppression of the leading particles, mainly at large values of $z$. However, there is
the indication of a contribution by the background, at low $z$, shifting the mean of distribution 
to the left. As before, we can study the artificial contribution of the heavy-ion background by 
variation of $E^{\rm rec}_{\rm T,\,min}$.

In vacuum the average fraction of the jet energy carried by the leading particle is known, 
$\av{z}\simeq0.3$ for $\et\gsim 100~\gev$ (for $R=0.7$ and $\pt>0.5~\gev$, see \psect{chap3:frag}). 
Since the average value most likely will change according to the modification of the distribution, 
we compute $\av{z}$ for jets with $\et^{\rm rec} > E^{\rm rec}_{\rm T,\,min}$ as a function 
of $E^{\rm rec}_{\rm T,\,min}$ shown in \fig{chap6:fig:jetfragleadminetidealb}.
At low $E^{\rm rec}_{\rm T,\,min}\lsim50~\gev$ all quenching models agree and differ strongly from 
the reference measurement. In this region, the value of $\av{z}$ is determined by properties of the 
background, resulting in artificial jets, whose leading particle does not arise from jet fragmentation.
Instead, at high $E^{\rm rec}_{\rm T,\,min}$ the reconstructed jets are biased to extreme 
fragmentation and $\av{z}$ reaches the magnitude of the \pp\ reference. However, 
there is indication that within $50~\gev<E^{\rm rec}_{\rm T,\,min}<100~\gev$ different toy 
models may be discriminated by the different values of $\av{z}$, but the effect is smeared by the 
poor energy resolution. The situation slightly improves, when the information of the \ac{EMCAL} is added. 
The fragmentation of leading particles for $E^{\rm rec}_{\rm T,\,min}\lsim50~\gev$ in the case of the 
ideal barrel+em is shown in \fig{chap6:fig:jetfraglead50to350idealbem} and 
$\av{z}$ as a function of $E^{\rm rec}_{\rm T,\,min}$ in \fig{chap6:fig:jetfragleadminetidealbem}.
Due the improved energy resolution and to the increase in statistics for higher $\et^{\rm rec}$ different 
models may, by eye, be distinguished with only little influence of the background over the range of 
$60~\gev<E^{\rm rec}_{\rm T,\,min}<120~\gev$. Again, the two dense \ac{PQM} scenarios cannot
be distinguished.

As for the longitudinal and transverse fragmentation it will be interesting to study the dependence of 
the leading-particle fragmentation and its average on centrality and jet direction with respect to the 
event plane. Generally, one might prefer to avoid normalizing by the reconstructed 
jet energy, since it introduces various additional biases. Instead, one could analyze the raw $\pt$ 
distribution for leading particles, in the same way as done above for all associated particles 
in the jet.

\subsection{Particle multiplicity and momentum}
The modifications of the jet \ac{FF} we discussed so far are connected to the expectation that due 
to the medium-induced gluon radiation, many rather, soft hadrons are produced in the fragmentation
of the primary parton. Therefore, related and simple observables are the average number of particles in
the cone and their average transverse momentum. 

\begin{figure}[htb]
\begin{center}
\subfigure[Average number (ideal barrel)]{
\label{chap6:fig:jetnumidealb}
\includegraphics[width=7.2cm]{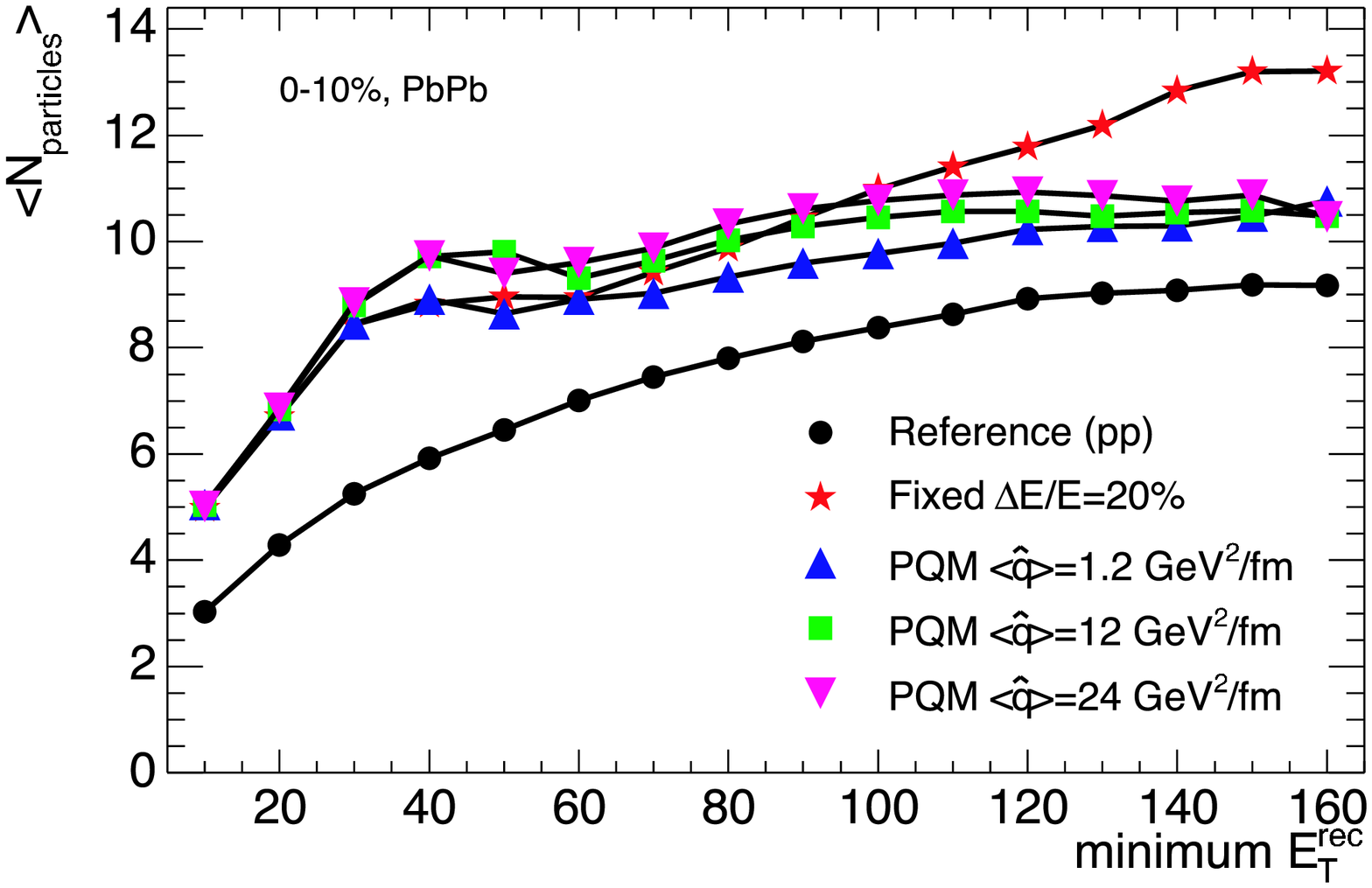}}
\hspace{0.1cm}
\subfigure[Average particle-$\pt$ (ideal barrel)]{
\label{chap6:fig:jetmeanidealb}
\includegraphics[width=7.2cm]{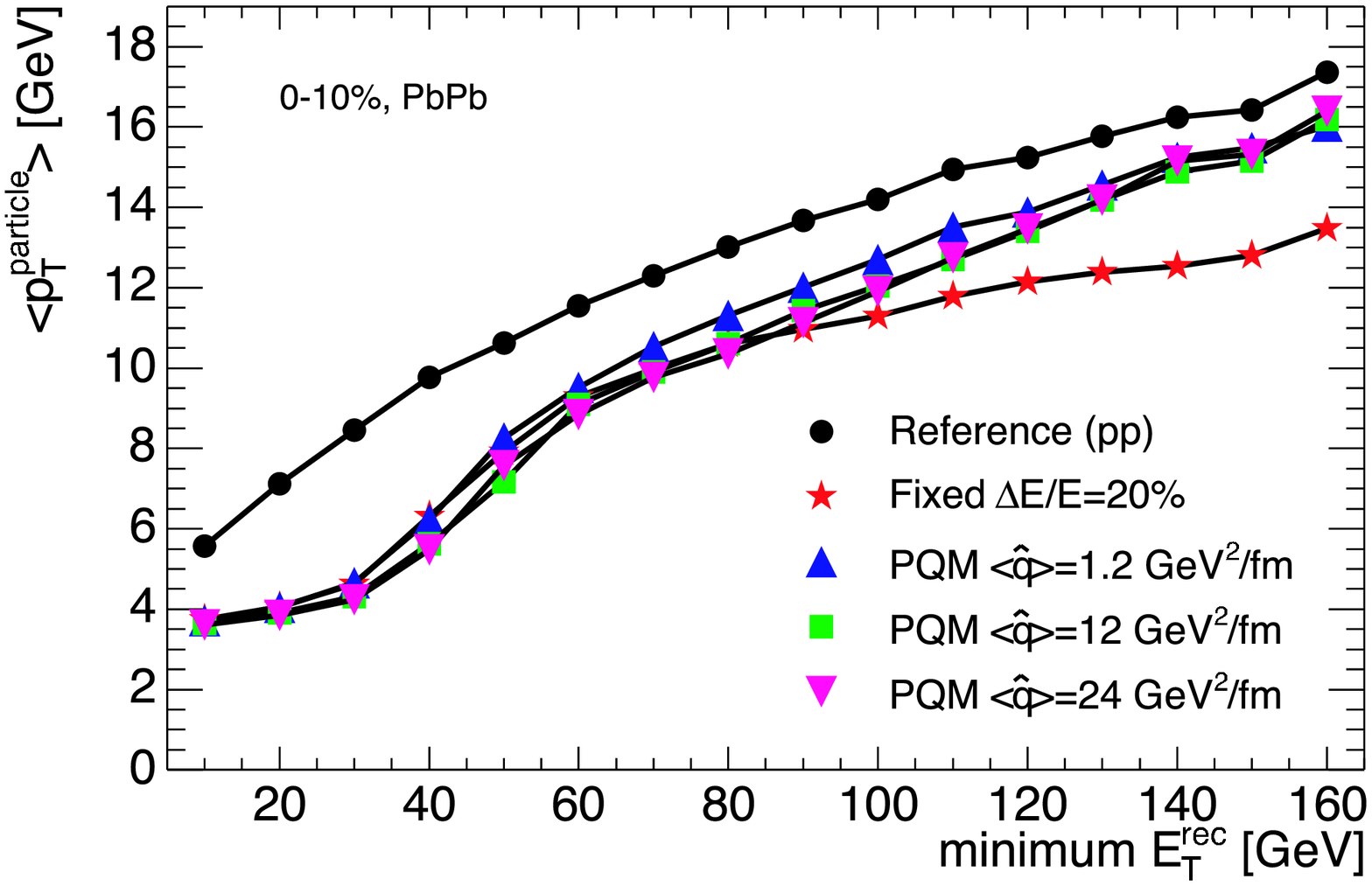}}
\subfigure[Average number (ideal barrel+em)]{
\label{chap6:fig:jetnumidealbem}
\includegraphics[width=7.2cm]{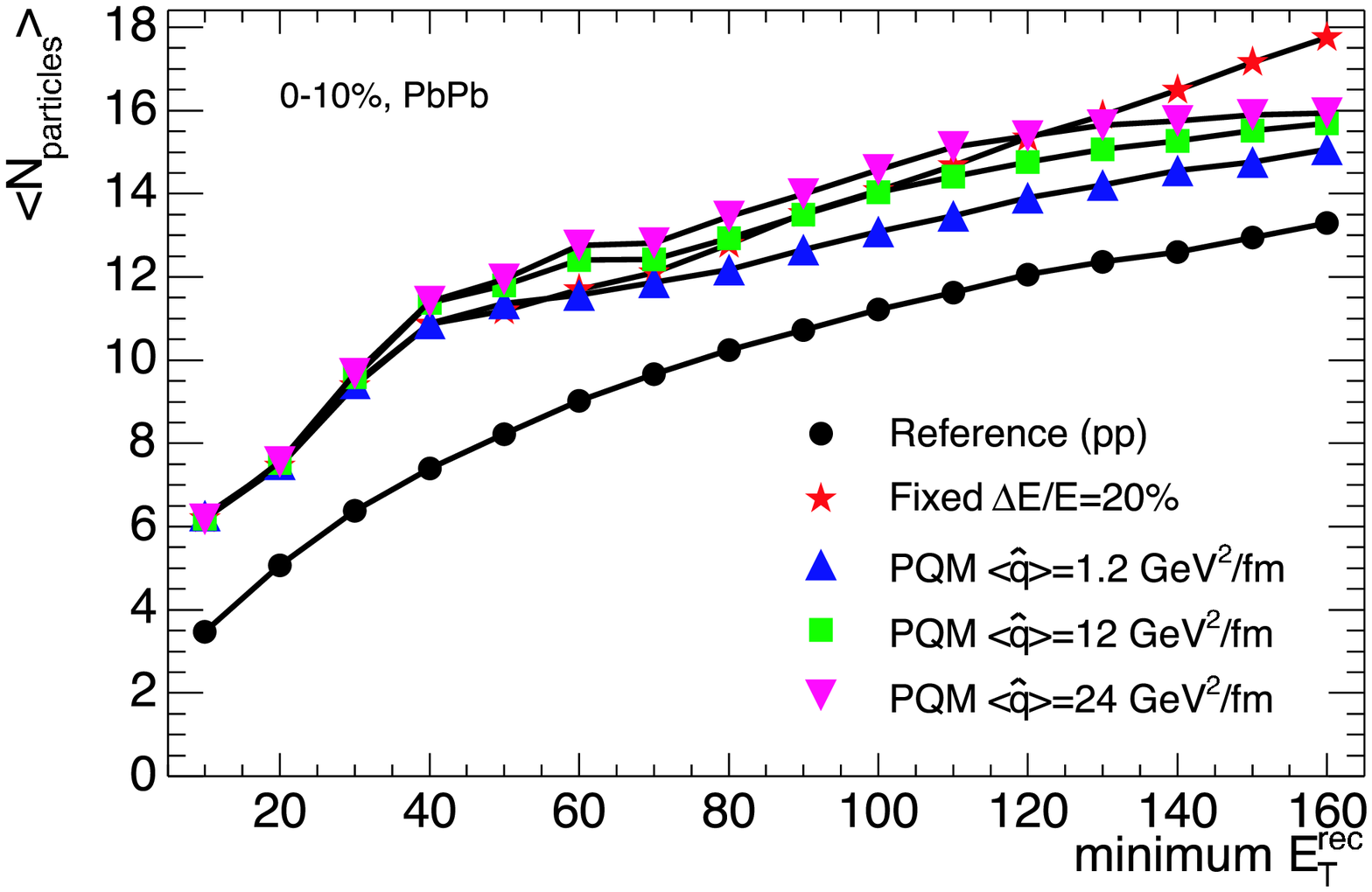}}
\hspace{0.1cm}
\subfigure[Average particle-$\pt$ (ideal barrel+em)]{
\label{chap6:fig:jetmeanidealbem}
\includegraphics[width=7.2cm]{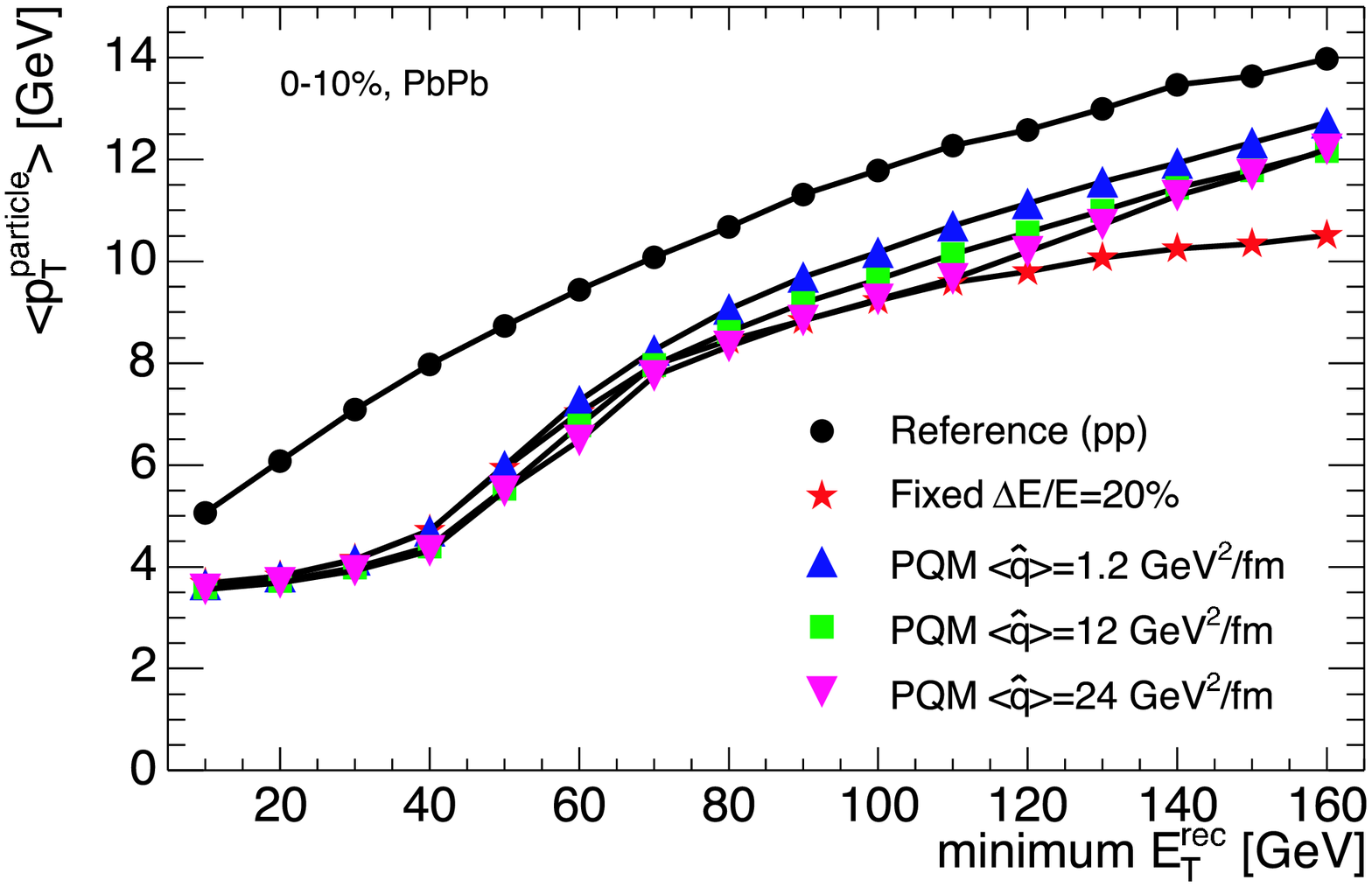}}
\end{center}
\vspace{-0.5cm}
\caption[xxx]{Average number of particles (left) and mean particle-$\pt$ (right) in jets with 
$\et^{\rm rec} > E^{\rm rec}_{\rm T,\,min}$ as a function of $E^{\rm rec}_{\rm T,\,min}$ 
in the case of the ideal barrel (top) and ideal barrel+em (bottom) for different quenching 
scenarios in $0$--$10$\%  central \PbPb\ compared to \pp\ collisions. 
In all figures the jets are identified with the \acs{ILCA} cone finder using $R=0.3$ and 
$\pt>2~\gev$.}
\label{chap6:fig:jetnummeanideal}
\end{figure}

Both are shown in \fig{chap6:fig:jetnummeanideal} in jets with $\et^{\rm rec} > E^{\rm rec}_{\rm T,\,min}$ 
as a function of $E^{\rm rec}_{\rm T,\,min}$ compared for the case of the ideal barrel and the ideal 
barrel+em. The general picture is as before: at low $E^{\rm rec}_{\rm T,\,min}$ the background masks 
the difference in the toy models, whereas starting from about $E^{\rm rec}_{\rm T,\,min}\gsim50~\gev$ 
the properties of the identified jets and their different quenching histories may be 
distinguished. The effect is clearest seen when the $\et$-resolution is improved by the 
\ac{EMCAL}, however using only the charged barrel provides enough information to separate
the fixed quenching scenario from the \ac{PQM} cases. As indicated already in previous plots,
the fixed-loss model strongly deviates at high $E^{\rm rec}_{\rm T,\,min}$ from the other
models.

\subsection{Integrated jet shapes}
Since it is expected that jet quenching leaves a negligible signature in the integrated, calorimetric 
energy content, it will be important to measure a change in the contribution of soft hadrons to the 
integrated shape. 

\begin{figure}[htb]
\begin{center}
\subfigure[$\psi(r)$ for $\et^{\rm rec}>80~\gev$ (ideal barrel)]{
\label{chap6:fig:jetshape50to350idealb}
\includegraphics[width=7.2cm]{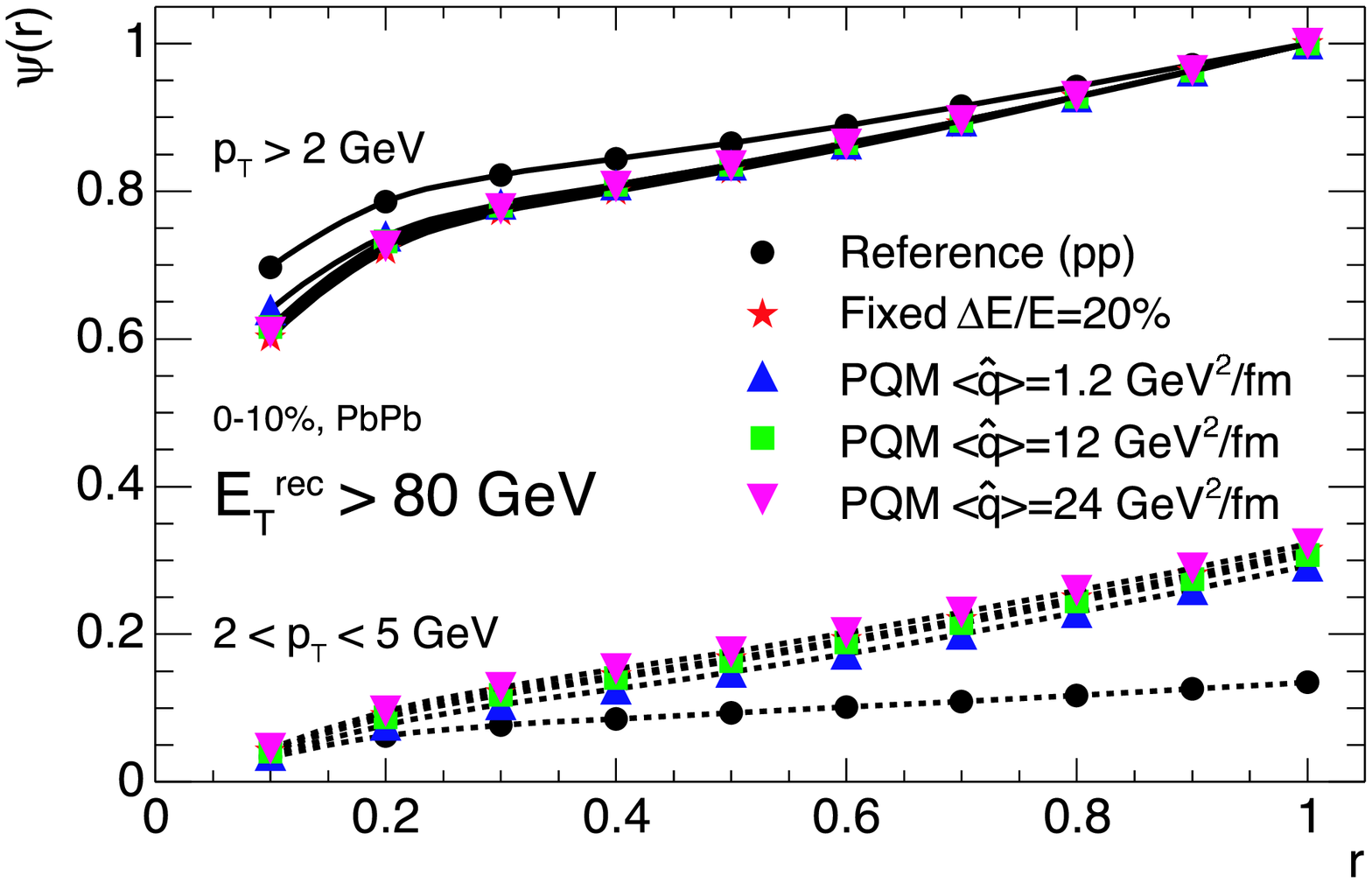}}
\hspace{0.1cm}
\subfigure[$\psi(0.3)$ for $\et^{\rm rec}>E_{\rm T,\, min}^{\rm rec}$ (ideal barrel)]{
\label{chap6:fig:jetshapeminetidealb}
\includegraphics[width=7.2cm]{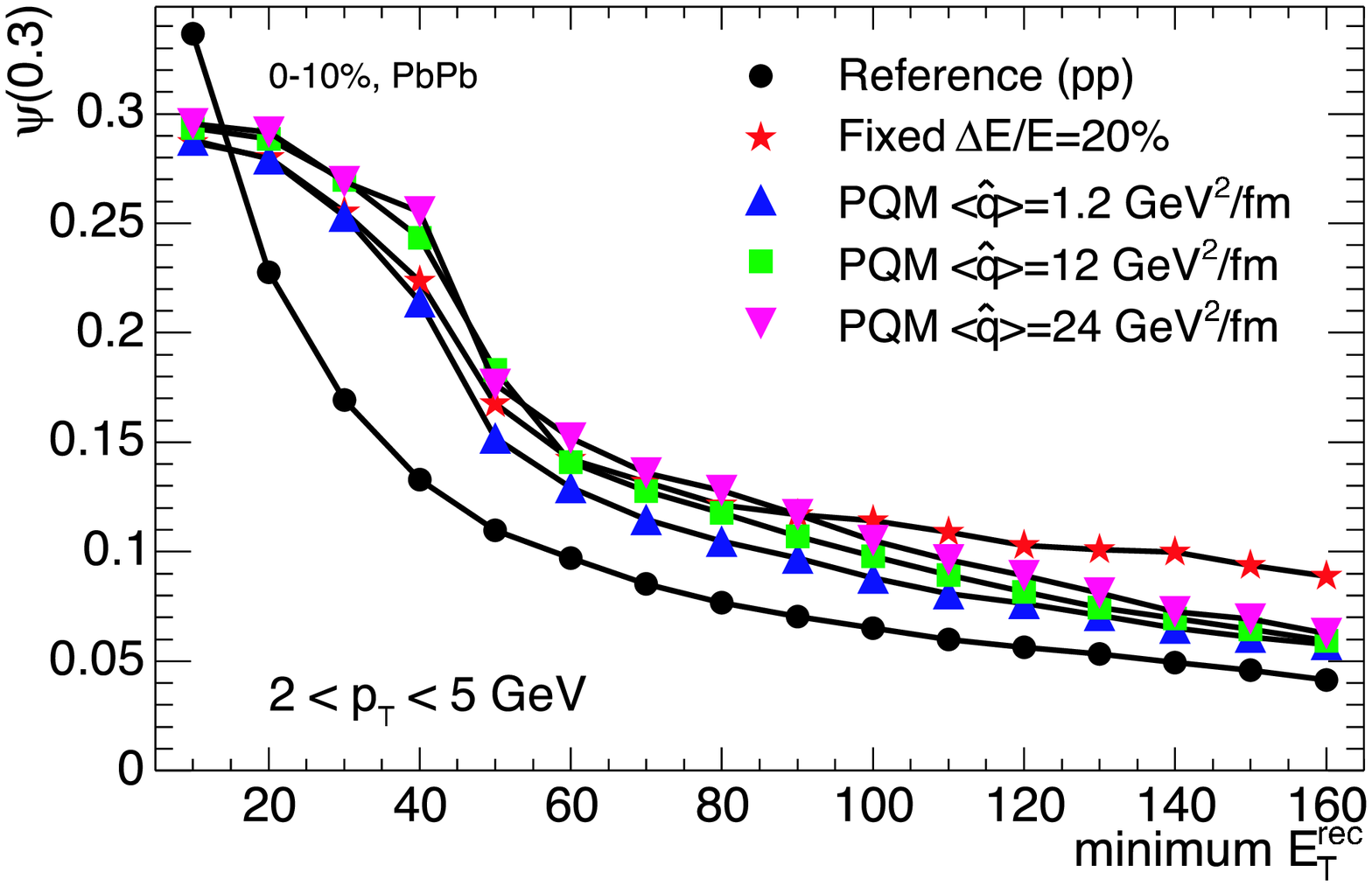}}
\end{center}
\vspace{-0.5cm}
\caption[xxx]{\subref{chap6:fig:jetshape50to350idealb}~Integrated jet shapes, $\psi(r)$
and $\psi(r)$ where $\pt\le5~\gev$, both normalized at $r=1$, of jets with $\et^{\rm rec}>80~\gev$ 
for different quenching scenarios in $0$--$10$\% central \PbPb\ compared to \pp\ collisions.
\subref{chap6:fig:jetshapeminetidealb}~Integrated jet shape $\psi(0.3)$ where
$\pt\le5~\gev$, normalized at $r=1$, for jets with $\et^{\rm rec} > E^{\rm rec}_{\rm T,\,min}$
as a function of $E^{\rm rec}_{\rm T,\,min}$ for different quenching scenarios in $0$--$10$\% 
central \PbPb\ compared to \pp\ collisions. In both figures the jets are identified 
with the \acs{ILCA} cone finder using $R=0.3$ and $\pt>2~\gev$ in the ideal barrel.}
\label{chap6:fig:jetshapesidealb}
\end{figure}
\fi

In \fig{chap6:fig:jetshape50to350idealb} we show the integrated jet shapes, $\psi(r)$, 
and $\psi(r)$ for particles with \mbox{$\pt\le5~\gev$}, in jets with $\et^{\rm rec}>80~\gev$
as a function of $r$. The jets have been defined only for $R=0.3$, however, 
we still normalize at $r=1$. A significant fraction of the shape arises due to the soft 
background, which ---mainly at large $r$--- stimulates a contribution of soft particles 
seen, for example, in the linear rise of the shape for particles with $\pt\le5~\gev$. Obviously,
corrections could be applied. For example, one may substract the contribution of the background 
evaluated for random jet axis before normalizing the shape at $r=1$ (or at $r\gsim 0.3$). 
Once the contribution of the background is under control it will be interesting to compare 
the shapes measured in \PbPb\ to the measured values in the vacuum, 
\ie~in the same way as in \pfig{chap3:fig:intjetshapes}.

For the moment, we do not apply any correction, but rather evaluate the shape at fixed 
\mbox{$r_0=0.3$}, the value for which the jets are defined. 
In \fig{chap6:fig:jetshapeminetidealb} \mbox{$\psi(0.3)$} is shown restricted to soft particles 
with $\pt\le5~\gev$ for jets with $\et^{\rm rec} > E^{\rm rec}_{\rm T,\,min}$ as a function of 
$E^{\rm rec}_{\rm T,\,min}$. The picture is the same as before: Below  
$E^{\rm rec}_{\rm T,\,min}=50~\gev$ the contribution of soft particles in the background contributes, 
for higher values of $E^{\rm rec}_{\rm T,\,min}$, the toy models lead to slightly different values of 
the shape, and reconcile at highest values of $E^{\rm rec}_{\rm T,\,min}$. In correspondance with the 
findings above, the fixed energy-loss model deviates from the \ac{PQM} models. 

%

\chapter{Summary}
\label{chap7}
In this work, we study the performance of the \ac{ALICE} detector 
for the measurement of high-energy jets at mid-rapidity 
in ultra-relativistic \AAex\ collisions at $\snn=5.5~\tev$ and 
their potential for the characterization of the partonic matter 
created in these collisions.

\medskip
In our approach, jets at $\Etj\gsim30~\gev$ are reconstructed with a cone jet finder, as 
typically done for jet measurements in hadronic collisions. However, the presence of numerous
mini-jets in the heavy-ion environment makes it necessary to reduce the cone size, the radius 
in the plane spanned by pseudo-rapidity and azimuth, from its nominal value of $R\sim0.7$ 
used at hadron colliders to $R=0.3$. In addition, the high-particle multiplicity density 
of the soft bulk at mid-rapidity requires the introduction of a cut in transverse momentum 
for charged hadrons with $\pt<2~\gev$. 

In central lead--lead (\PbPb) collisions, jets of about $50~\gev$ and higher will be measurable 
with \ac{ALICE}, but intrinsic fluctuations in the jet fragmentation, out-of-cone fluctuations 
and the remaining underlying mini-jet background limit the energy resolution. This is even valid for 
an ideal detector and for jets with far higher energy. Without the \ac{EMCAL}, the resolution is 
mainly dominated by intrinsic fluctuations in the ratio of charged-to-neutral particles in the jet 
fragmentation. The mean reconstructed fraction amounts to $50$\% (and a width of about~$50$\%). 
Including the \ac{EMCAL}, the mean fraction increases to about $60$\% (and a width of~$30$\%). 
In both cases, however, the spatial direction of the jet axis can be reconstructed with a 
resolution (in $\eta$ and $\phi$) of better than $0.01$, at $50~\gev$, which is enough for 
matching with the direction of the initial parton. The contribution of the background to the 
energy content within the reduced cone is sizeable, about $15\%$ for $50~\gev$, even with 
the proposed cuts.

It will be increasingly difficult to resolve the signal from the background for energies 
lower than $\sim50~\gev$. In this regime, where jet rates are considerably large, inclusive 
measurements will be applied. The jet rate in minimum-bias \PbPb\ for jets produced within 
the central \ac{ALICE} acceptance drops from one jet per event with \mbox{$\Etj\ge20~\gev$} 
to about one jet per $1000$ events with \mbox{$\Etj\ge100~\gev$}. Therefore, for jets with
more that $50~\gev$ triggering becomes relevant. 

A jet finder running online in the \acf{HLT} ---using the same algorithm and parameters 
as for the offline analysis, which is within the time budget--- 
will suppress the recorded data rate by a factor of $50$ for minimum-bias \PbPb\ collisions,
while keeping $1/10$ of the jets with $\Etj\ge50~\gev$ and $1/2$ with $\Etj\ge100~\gev$.
Even if the \ac{HLT} system inspects minimum-bias \PbPb\ events without further help of 
hardware triggers at \ac{L1}, in total $10^{7}$ events will be recorded in one \ac{ALICE} 
year at $\Lumi=0.5\,\mbarn^{-1}\s^{-1}$, containing about $4\cdot10^{5}$ jets with 
$\Etj\ge50~\gev$ and about $10^{5}$  with $\Etj\ge100~\gev$.
The \ac{HLT} system without further help by hardware triggers is able to provide the 
necessary statistics. However, for minimization of the trigger bias and for increase 
in statistics (even if moderate), the \ac{HLT} should be supported by hardware triggers 
at \ac{L1}, as for example foreseen by the \ac{EMCAL}.


\pagebreak
For the study of the sensitivity of high-energy jets to in-medium effects, quenched 
jets are embedded into central \PbPb\ collisions at $\snn=5.5~\tev$. The jets are prepared 
for medium densities of $\av{\hat{q}}=1.2$, $12$ and $24~\gev^2/\fm$ using a modified 
\acs{PYTHIA} version, which includes partonic energy loss in the \acs{BDMPS-Z-SW} 
framework together with a realistic description of the collision geometry.

Our analysis addresses the high-energy regime, where jets may be reconstructed as distinct 
objects, $\Etj\ge50~\gev$. Still, the comparison with the \pp\ reference reveals that the 
contamination of the jet cone by uncorrelated particles from the underlying background 
may severely influence measured jet properties, even up to jet energies of $100~\gev$. 
Since low-energy jets are predominantly produced, properties of the background mask and, 
even worse, partially mimic the effect of the medium. However, with increasing energy 
\mbox{threshold} for the reconstructed jets in the analysis, the influence of the 
background reduces and most observables show moderate sensitivity, at least.

Even within the limited approach of the quenching procedure, there are a few 
observables that deserve to be mentioned explicitly. 
The longitudinal particle-momentum distribution of associated particles 
along the jet axis is strongly suppressed, measurable for $p_{\rm L}\gsim 15~\gev$,
for example, in the ratio with respect to the measurement in \pp. The distribution of 
the fraction of energy that the leading particle carries changes with respect to the 
reference. It is shifted towards smaller values, and suppressed at intermediate to high 
values, $z\gsim0.5$. Both distributions have in common that they rely on 
high-transverse-momentum particles, which are comparatively rare in the underlying background. 
Instead, the integrated jet shape, by construction, suffers from the soft contribution it 
receives, since it accounts for all particles in the cone, smearing out the effect of the
medium.

The measurement in the charged-particle sector together with the increase of the jet energy 
threshold proves sufficient to distinguish between the case of lowest and highest medium density. 
Thus, based on our quenching studies, the detection of in-medium modification in the fragmentation
of high-energy jets seems to be possible with the charged-particle tracking detectors, alone.

A word on the quenching procedure seems to be appropriate. 
Because of the reasoning behind the modified \acs{PYTHIA} version, relative energy losses of
more than $80$\% induce the emission of six additional radiated gluons, which share the lost 
energy of the initial parton at equal parts. 
Owing to the surface effect, present for both dense cases of the medium, modified jets 
appear almost similar. Thus, `what we see is what we put in'.
The limited sensitivity expresses the conceptual difficulties and the lack of theoretical 
predictiveness within the toy model and, in general, within the \acs{BDMPS-Z-SW} framework 
at finite parton energies and for high medium densities. 
A consistent treatment of medium-induced gluon radiation and the impact on hadronic properties 
probably requires a Monte Carlo implementation of the medium-modified parton shower.

According to the \ac{PQM} calculation, high-energy jets with $\Etj\gsim 150~\gev$ might be 
almost extinct at the \ac{LHC}. 
In this case, jet quenching will presumably be detected by calorimetry 
via apparent reduction of cross section, similar to leading-hadron spectroscopy. 
\ac{ALICE} including the \ac{EMCAL} would be well prepared for such a scenario, too.

\else
\input{chaps}
\fi

\ifappendix
\begin{appendix}
%

\chapter{Appendix}
\label{app}

\section{Glauber calculation}
\label{app:glauber}
In the following we summarize the essential features of the Glauber 
formalism~\cite{glauber1970,bialas1976,shukla2003,denterria2003}, 
which we use in \ac{PQM} and for the estimations of $\Ncoll$ and $\Nhard$.
The values, which we use throughout the thesis, can be found
~\ptab{chap5:tab:glauber}.
\subsubsection{Thickness function}
The nuclear density profile for a nucleus A in the transverse 
plane, perpendicular to the beam axis $z$,
\begin{equation}
\label{app:eq:ta}
T_{\rm A}(\vec{s})=\int \dd z \, \rho_{\rm A}(z,\vec{s})\;,
\end{equation}
is known as the nuclear thickness function of nucleus A. We use the Wood-Saxon 
nuclear density $\rho_{\rm A}\equiv\rho_{\rm A}^{\rm WS}(z,\vec{r})$ and renormalize 
$\int \dd\vec{s} \, T_{\rm A}(\vec{s})=A$. The parameters of the Wood-Saxon profile 
for the different nuclei, \ie~gold and lead, can be found in~\cite{atomdata}.
\subsubsection{Overlap function}
For a collision of nucleus A with nucleus B the nuclear overlap 
function, $T_{\rm AB}(\vec{b})$, at impact parameter $\vec{b}$
is defined as
\begin{equation}
\label{app:eq:tab}
T_{\rm AB}(\vec{b})=\int \dd\vec{s} \, 
T_{\rm A}(\vec{s})T_{\rm B}(\vec{b}-\vec{s})\;.
\end{equation}
Due to the azimuthal symmetry of the Wood-Saxon profile the integration is 
conveniently performed using $\dd\vec{s}=2\pi\,s\dd s$; yielding  $A\,B$
for the integral over all impact parameters.
\subsubsection{Cross section}
In the multiple-scattering approximation of the Glauber formalism the
inclusive, inelastic cross-section, $\sigma_{\rm A B},$ for the collision of A 
and B can be derived, leading to
\begin{equation}
\label{app:eq:tcs}
\sigma_{\rm AB}=\int \dd\vec{b} \, \left[ 1 - \exp\left( -\sigma_{\rm NN}\, 
T_{\rm AB}(\vec{b})\right) \right]\;,
\end{equation}
where $\sigma_{\rm NN}$ denotes the \NNex~cross section. We assume that the
interaction probability is solely given by the \pp\ cross section. At \ac{RHIC}
$\sigma_{\rm NN}=40~\mbarn$ is usually used. At \ac{LHC} expectations vary; 
we use $59~\mbarn$ ($67~\mbarn$), which corresponds to the expected inelastic, 
but non-diffractive cross section for $\sqrt{s}=5.5~\tev$ 
($14~\tev$)~\cite{berardi2004}.
%
Using \eq{app:eq:tcs} gives the total (geometrical) 
cross section of $\sigma^{\rm geo}_{\rm PbPb}=7.8~\barn$. Even though, 
\eq{app:eq:tcs} describes a total cross section, to first order it also determines 
the cross section of a single hard process, 
\begin{equation}
\label{app:eq:hcs}
\sigma^{\rm hard}_{\rm AB} \approx  \int \dd\vec{b} \, \sigma^{\rm hard}_{\rm NN}
\, T_{\rm AB}(\vec{b})\;,
\end{equation}
where $\sigma_{\rm NN}^{\rm hard} \ll \sigma_{\rm NN}$ is the cross section
of the hard process in \NN~interactions.
\subsubsection{Binary collisions}
For a given impact parameter the average hard scattering yield can be obtained 
by integrating the probability of the occurrence of an hard process in the 
interaction of a nucleon of one nucleus multiplied with the interaction probability 
along its straight trajectory (in $z$) within the other nucleus,
\begin{equation}
\label{app:eq:nhard}
N^{\rm hard}_{\rm AB}(b) = \sigma^{\rm hard}_{\rm NN} \int \rho_{\rm A}
(z^\prime,\vec{s}) \, \rho_{\rm B}(z^{\prime\prime},\vec{b}-\vec{s})
\, \dd\vec{s} \, \dd z^\prime \, \dd z^{\prime\prime}
= \sigma^{\rm hard}_{\rm NN}\, T_{\rm AB}(\vec{b})\;.
\end{equation}
In the same way, one obtains the average number of inelastic, binary \NN~collisions,
\begin{equation}
\label{app:eq:ncoll}
N^{\rm coll}_{\rm AB}(b) 
=  \sigma_{\rm NN}\, T_{\rm AB}(\vec{b})\;.
\end{equation}
\subsubsection{Yields in centrality classes}
The above formulas are valid at a given impact parameter. The fraction of 
the (geometrical) cross section for the centrality selection $C_1$--$C_2$, 
corresponding to the impact parameter range $b_1 < b < b_2$, is given by  
\begin{equation}
\label{app:eq:fgeo}
f^{\rm geo}_{\rm AB}(b_1,b_2) = \int_{b_1 < b < b_2} \dd\vec{b} 
\, P(\vec{b})
\end{equation}
where the probability distribution of impact parameters reads
\begin{equation}
\label{app:eq:p}
P(\vec{b}) = \left[ 1 - \exp\left( -\sigma_{\rm NN}\, 
T_{\rm AB}(\vec{b})\right) \right] \,/\, \sigma^{\rm geo}_{\rm AB}\;.
\end{equation}
For the calculation of the yields in the centrality class, $C_1$--$C_2$,
one needs to take into account the conditional probability that a collision in the given 
centrality range takes place. For the average nuclear overlap function in the 
given centrality class we obtain
\begin{equation}
\label{app:eq:avtab}
\av{T_{\rm AB}}(b_1,b_2) = \int_{b_1 < b < b_2} \dd\vec{b} \, P(\vec{b}) 
\, T_{\rm AB}(\vec{b}) \,/\, f^{\rm geo}_{\rm AB}(b_1,b_2) \;.
\end{equation}
Using \eq{app:eq:nhard} and \eq{app:eq:ncoll} leads
to the yields,
\begin{equation}
\label{app:eq:avnhard}
\av{N^{\rm hard}_{\rm AB}}(b_1,b_2) = \av{T_{\rm AB}}(b_1,b_2) \, \sigma^{\rm hard}_{\rm NN}
\end{equation}
for the average number of hard processes and
\begin{equation}
\label{app:eq:avncoll}
\av{N^{\rm coll}_{\rm AB}}(b_1,b_2) = \av{T_{\rm AB}}(b_1,b_2) \, \sigma_{\rm NN}
\end{equation}
for the average number of binary \NN~interactions in the selected centrality range.
\subsubsection{Binary scaling}
From the last two equations we get binary collision scaling of hard processes according
\begin{equation}
\label{app:eq:binhard}
\av{N^{\rm hard}_{\rm AB}}(b_1,b_2) =  \av{N^{\rm coll}_{\rm AB}}(b_1,b_2) \, 
\sigma^{\rm hard}_{\rm NN} / \sigma_{\rm NN}
\end{equation}
and using \eq{app:eq:hcs} with \eq{app:eq:avtab} we find for the cross section
\begin{equation}
\label{app:eq:binhcs}
\sigma^{\rm hard}_{\rm AB}(b_1,b_2) = \av{T_{\rm AB}}(b_1,b_2)\, 
f^{\rm geo}_{\rm AB}(b_1,b_2)\,\sigma^{\rm geo}_{\rm AB} \sigma^{\rm hard}_{\rm NN} 
\end{equation}
\subsubsection{Participant nucleons}
In the wounded nucleon model~\cite{bialas1976,broniowski2001} the number of
participating nucleons in the overlap region is on average given by
\begin{equation}
N^{\rm part}_{\rm AB}(b) = 2A \, \int \dd\vec{s} \,T_{\rm A}(\vec{s}-\vec{b}) 
\, \left(1-(1-\sigma_{\rm NN} T_{\rm A}(\vec{b}))^{A}\right)
\end{equation}
for symmetric collisions $A=B$. Its average value over a given centrality
class is calculated in the same way as it is done for the nuclear overlap 
in~\eq{app:eq:avtab}.

\begin{table}[htb]
\begin{center}
\begin{tabular}{rccccc}
  $f^{\rm geo}_{\rm AB}$ [\%] & $b_{\rm min}$ [$\fm$] & $b_{\rm max}$ [$\fm$] 
& $\av{\Ncoll}$ & $\av{\Npart}$ & $\av{T_{\rm AB}}$ [$\mbarn^{-1}$] \\
\hline
\hide{00}0--5 & \hide{0}0   & 3.7  & 1550       & 369        & 26.4 \\
\hide{0}5--10 & \hide{0}3.7 & 5.1  & 1183       & 306        & 20.2 \\
\hide{0}0--10 & \hide{0}0   & 5.1  & 1376       & 339        & 23.3 \\
10--20        & \hide{0}5.1 & 7.2  & 814        & 235        & 13.8 \\
20--30        & \hide{0}7.2 & 8.8  & 474        & 160        & 8.05 \\
30--40        & \hide{0}8.8 & 10.1 & 259        & 105        & 4.40 \\
40--50        & 10.1        & 11.3 & 132        & \hide{0}65 & 2.25 \\
50--60        & 11.3        & 12.3 & \hide{0}58 & \hide{0}36 & 0.98 \\
60--70        & 12.3        & 13.3 & \hide{0}22 & \hide{0}18 & 0.38 \\
70--80        & 13.3        & 14.2 & 7.6        & 7.7        & 0.14 \\
Min.~bias     & \hide{0}0   & 100  & 326        & 100        & 5.52 \\
\hline
\hline
\end {tabular}
\end{center}
\vspace{-0.4cm}
\caption[xxx]{Number of inelastic \NN~collisions, number of wounded nucleons and overlap function
calculated in the optical Glauber model for $\sigma_{\rm NN}=59~\mbarn$, corresponding to
$\sigma^{\rm geo}_{\rm PbPb}=7.8~\barn$, for \PbPb~at $\sqrt{s}=5.5.~\tev$.} 
\label{chap5:tab:glauber}
\end{table}

\section{PYTHIA parameters}
\label{app:pythia}
In \tab{app:tab:pythiaparams}, we report the list of parameters, which 
we use for the creation of jets (signal events) with the \acs{PYTHIA} event 
generator~\cite{mcpythia1987,mcpythia1994,mcpythia2001}, version 6.214, which
is packaged in \acs{ALIROOT} (classes {\tt AliPythia} and {\tt AliGenPythia}). 

The main settings for the generation of the single-inclusive jet spectrum 
involve the lower and upper value of $\pt^{\rm hard}$, the values of the cone 
finder, type \acs{UA1}~\cite{arnison1983,albajar1988}, as well as the 
jet-trigger conditions, which are implemented in the {\tt AliGenPythia} class. 
Typically we require the event to contain at least one jet with 
$\abs{\eta_{\rm J}}\le0.5$ and $\Etj\ge10~\gev$. The ratio of triggered to 
generated jets for a fixed interval of $\pt^{\rm hard}$ together with
the corresponding (hard) cross section as given by \acs{PYTHIA} determine 
the weight, which we apply to every event (or jet of the event) generated for 
the particular interval. The values of the weighted cross section are listed in 
\tab{app:tab:pythiaweights}.

\begin{table}[htb]
\begin{center}
\begin{tabular}{llc}
{Description} & {Parameter} & {Value} \\
\hline
Process types & MSEL & 1\hide{.}  \\
\hline
Minimum/maximum                & CKIN(3)   & see    \\
parton $\pt^{\rm hard}$~[\gev] & CKIN(4)   & \tab{app:tab:pythiaparams} \\
\hline
\acs{CTEQ}~4L parametrization\footnotemark & MSTP(51) & 4032 \\
Proton \acs{PDF}  & MSTP(52) & 2\hide{.} \\
\hline
Switch off resonance decays & MSTP(41) & 1\hide{.} \\
\hline
Switch off multiple interactions  & MSTP(81) & 0\hide{.} \\
& PARP(81) & 0\hide{.} \\
& PARP(82) & 0\hide{.} \\
\hline
Initial/final state radiation on & MSTP(61) & 1\hide{.} \\
& MSTP(71) & 1\hide{.} \\
\hline
Intrinsic $k_{\rm t}$ from Gaussian (zero mean) & MSTP(91) &  1\hide{.} \\
\hspace{0.75cm}width $\sigma$ [$\gev$]               & PARP(91) & 1.\\
\hspace{0.75cm}upper cut-off (at $5\sigma$) [$\gev$] & PARP(93) & 5.\\
\hline
Cone jet finder (pycell) &  & \\
\hspace{0.75cm}$\abs{\eta}$ of the ``detector'' & PARU(51) & 2\hide{.} \\
\hspace{1.5cm}number of cells in $\eta$ & MSTU(51) & 274\\
\hspace{1.5cm}number of cells in $\phi$ & MSTU(52) & 432\\
\hspace{0.75cm}threshold [$\gev$]   & PARU(58) & 0.\\
\hspace{0.75cm}seed [$\gev$]        & PARU(52) & 4.\\
\hspace{0.75cm}min et [$\gev$]      & PARU(53) & 5.\\
\hspace{0.75cm}radius               & PARU(54) & 1.\\
\hspace{0.75cm}Snowmass accord      & MSTU(54) & 2\hide{.}\\
\hline
\hline
\end{tabular}
\end{center}
\vspace{-0.4cm}
\caption[xxx]{\acs{PYTHIA} parameter settings for the generation of jets in \pp\
collisions at $\sqrt{s}=5.5$ and $14~\tev$. Non-specified parameters are left 
to \acs{PYTHIA} 6.214 defaults.}
\label{app:tab:pythiaparams}
\end{table}

\footnotetext{We use the same \ac{PDF}, \ac{CTEQ}~5L, for \pp, as well as for \PbPb. 
The difference to jet production including nuclear effects, \eg~\ac{EKS98}, can
be neglected.} 

\begin{table}[htb]
\begin{center}
\begin{tabular}{ccc}
Min.~$\pt^{\rm hard}$ & Max.~$\pt^{\rm hard}$ & Weight \\
~[$\gev$] & [$\gev$] & [$\mbarn$] \\
\hline
\hide{00}5 & \hide{0}15 & $3.218\cdot10^{-2}$ \\
\hide{0}15 & \hide{0}20 & $6.475\cdot10^{-2}$ \\
\hide{0}20 & \hide{0}24 & $2.406\cdot10^{-2}$ \\
\hide{0}24 & \hide{0}30 & $1.652\cdot10^{-2}$ \\
\hide{0}30 & \hide{0}35 & $6.107\cdot10^{-3}$ \\
\hide{0}35 & \hide{0}42 & $4.066\cdot10^{-3}$ \\
\hide{0}42 & \hide{0}50 & $4.067\cdot10^{-3}$ \\
\hide{0}50 & \hide{0}60 & $1.052\cdot10^{-3}$ \\
\hide{0}60 & \hide{0}72 & $5.063\cdot10^{-4}$ \\
\hide{0}72 & \hide{0}86 & $2.313\cdot10^{-4}$ \\
\hide{0}86 & 104        & $1.164\cdot10^{-4}$ \\
104        & 124        & $4.752\cdot10^{-5}$ \\
124        & 149        & $2.270\cdot10^{-5}$ \\
149        & 179        & $9.457\cdot10^{-6}$ \\
179        & 214        & $3.963\cdot10^{-6}$ \\
214        & 250        & $1.476\cdot10^{-6}$ \\
\hline
\hline
\end{tabular}
\end{center}
\vspace{-0.4cm}
\caption[xxx]{The values of the weight corresponding to the 
$\pt^{\rm hard}$ interval in \acs{PYTHIA}.}
\label{app:tab:pythiaweights}
\end{table}

\pagebreak
\section{HIJING parameters}
\label{app:hijing}
In \tab{app:tab:hijingparams}, we report the list of parameters, which 
we use for the creation of \PbPb\ (background) events with the \acs{HIJING} 
event generator~\cite{mchijing1991,mchijing1994}, version 1.36, which is 
packaged in \acs{ALIROOT} (classes {\tt AliGenHijing} and {\tt THijing}). 
The impact parameters corresponding to the definition of the centrality classes
used in the simulation are listed in \tab{app:tab:hijingimpact}.

\begin{table}[htb]
\begin{center}
\begin{tabular}{llc}
{Description} & {Parameter} & {Value} \\
\hline
Switch on jet quenching          & IHPR2(4)  & 1 \\
(hijing default)                 & IHPR2(50) & 0 \\
\hline
Initial/final state radiation on & IHPR2(2) & 3 \\
\hline
Switch off resonance decays      & IHPR2(12) & 1 \\
\hline
Switch on shadowing              & IHPR2(6) & 1 \\
\hline
Switch off jet trigger           & IHPR2(3) & 0 \\ 
\hline
\hline
\end{tabular}
\end{center}
\vspace{-0.4cm}
\caption[xxx]{\acs{HIJING} parameter settings for the generation of \PbPb\
collisions at $\sqrt{s}=5.5$. Non-specified parameters are left 
to \acs{HIJING} 1.36 defaults.}
\label{app:tab:hijingparams}
\end{table}

\begin{table}[htb]
\begin{center}
\begin{tabular}{ccl}
$b^{\rm min}~[\fm]$ & $b^{\rm max}~[\gev]$ & Name\\
\hline
\hide{0}0.0 & \hide{0}5.0 & kHijing\_cent1 \\
\hide{0}0.0 & \hide{0}2.0 & kHijing\_cent2 \\
\hide{0}5.0 & \hide{0}8.6 & kHijing\_per1\\
8.6  & 11.2 & kHijing\_per2 \\
11.2 & 13.2 & kHijing\_per3 \\
13.2 & 15.0 & kHijing\_per4 \\
15.0 & 100  & kHijing\_per5 \\
\hline
\hline
\end{tabular}
\end{center}
\vspace{-0.4cm}
\caption[xxx]{The impact parameter values for \acs{HIJING} corresponding to the centrality class.}
\label{app:tab:hijingimpact}
\end{table}

\section{Cone finder parameters}
\label{app:conefinder}
Throughout the thesis we mention two cone finders used to reconstruct jets,
\acs{UA1} and \ac{ILCA}. The \acs{UA1} cone finder is merely used for the trigger 
in the generation of the \acs{PYTHIA} signal events with the settings listed in 
\tab{app:tab:pythiaparams}.
The \ac{ILCA} algorithm is described in \psect{chap3:ilca}. The parameters used
in \pp\ and \PbPb\ are listed in \tab{app:tab:coneparams}. We always use the Snowmass 
convention as the recombination scheme for the calculation of the jet variables.

\begin{table}[htb]
\begin{center}
\begin{tabular}{lcc}
Description & \pp\ & \PbPb \\
\hline
Radius $R$                         & $0.7$  & $0.3$  \\
Particle $\pt$-cut [$\gev$]        & $0.5$  & $2.0$  \\
Minimum tower/seed energy [$\gev$] & $0.0$  & $2.0$  \\
Minimum proto-jet energy [$\gev$]  & $0.0$  & $2.0$  \\
Shared fraction $f$ [\%]      & $50$   & $50$   \\         
Exclusion distance $\epsilon$ & $0.01$ & $0.01$ \\
Maximum iterations            & $100$  & $100$  \\
Minimum jet energy [$\gev$]   & $5.0$  & $10.0$ \\
Tower size in $\phi$          & $0.05$  & $0.05$ \\
Tower size in $\eta$          & $0.05$  & $0.05$ \\
\hline
\hline
\end{tabular}
\end{center}
\vspace{-0.4cm}
\caption[xxx]{The settings used for \acs{ILCA} in \pp\ and \PbPb\ mode.}
\label{app:tab:coneparams}
\end{table}

\section{Monte Carlo quenching model}
\label{app:mcqm}
Since there exists no consistent Monte Carlo implementation of the medium-modified 
parton shower, it has been decided to introduce a parton quenching routine, called 
{\tt Quench} (member of {\tt AliPythia} inside \acs{ALIROOT}) into the process of event 
generation.~\footnote{The quenching procedure has been developed by A.~Morsch.}  
In some sense the quenching procedure can be regarded as an ``afterburner'' 
to the generation of partonic jets by \acs{PYTHIA}:

\begin{itemize}
\item At beginning of the generation of quenched jet events, standard \acs{PYTHIA} is used 
with settings explained in \sect{app:pythia}. However, it will stopped after creation 
of the final (partonic) jet system and before the start of final-state 
fragmentation~(switch~MSTJ(1,0)). 
\item Then, the quenching procedure modifies the final jet system 
according to the specified parameters and includes the radiated gluons into the event 
record. 
\item At the end, \acs{PYTHIA} is called again to perform final-state fragmentation
and hadronization (switch MSTJ(1, 1), followed by calling~{\tt pyexec}).
\end{itemize}

The actual quenching is performed in a loop: 
Every parton assigned to a partonic jet (initial parton),
two outgoing from the hard scatter and two from \acs{ISR},
is quenched by 
a factor $1-z$ using light-cone variables in the reference frame of the initial parton, 
$(E + \pz)^{\rm new} = (1-z) \, (E + \pz)^{\rm old}$,
where $z$ is the fractional energy loss to be applied per radiated gluon.
The lost momentum is first balanced by one additional gluon with non-vanishing virtuality, 
$Q> 0$, which subsequently splits into two gluons conserving the lost energy. 
Depending on the number of additional gluons, $n$, requested to be radiated per initial parton,
the fractional energy loss will be applied in $n$ iterations, such that $z$ is given
by $z=1-(1-\epsilon)^{-n}$, where $\epsilon=\Delta E^{\rm jet}/E^{\rm jet}$ is the fractional 
energy loss of the initial parton.

The two parameters of the quenching function, $\epsilon$ and $n$, may either be set to fixed values, 
or may be calculated by \ac{PQM}. \ac{PQM} is introduced and discussed in detail in \psect{chap3:pqm}.
For the fixed quenching case discussed in \chap{chap6}, we set $\epsilon=0.2$ and $n=1$. 

For \ac{PQM} settings the user has to chose a value of $k$ and the collision centrality.~\footnote{In 
principle, one may choose between reweighted, instead of non-reweighted, constraints.}
The fractional energy loss, $\epsilon_i$, for every initial parton, $i$, is calculated by \ac{PQM} on 
the basis a common origin by evaluating the \acs{BDMPS-Z-SW} energy loss along the path in the transverse 
plane determined by the emission angle, $\phi_i$ and the parton type, $t_i$. All necessary information 
are known from \acs{PYTHIA}. The production point is determined randomly from the impact parameter 
distribution of the chose centrality and the corresponding Monte Carlo evaluation of the nuclear overlap 
function. In order to avoid that a radiated gluon acquires more energy than the quenched initial
parton, from which it originates, $n$ is computed according to $n=1 + (\epsilon / (1-\epsilon))$
for $1\le n \le 6$, integer value. The limit to $n\le6$ is rather arbitrary and due to 
technical reasons within \acs{PYTHIA}. It has been adjusted to treat all energy loses, 
$\epsilon>0.8$, on equal footing.

\ifacro
%

\chapter{List of acronyms}

\begin{acronym}[BDMPS-Z-SW]
\acro{ALIROOT}    {ALICE Offline Framework based on ROOT\acroextra{ (\hrefurl{http://aliweb.cern.ch/offline/})}}
\acro{ADC}        {Analog Digital Converter}
\acro{AGS}        {Alternating Gradient Synchrotron}
\acro{ALICE}      {A Large Ion Collider Experiment\acroextra{ (\hrefurl{http://alice.web.cern.ch/Alice/})}}
\acro{BDMPS-Z}    {Baier-Dokshitzer-Mueller-Peign\'e-Schiff--Zakharov}
\acro{BDMPS-Z-SW} {BDMPS-Z--Salgado-Wiedemann}
\acro{BRAHMS}     {BRAHMS\acroextra{ (\hrefurl{http://www.rhic.bnl.gov/brahms/})}}
\acro{D0}         {D0\acroextra{ (\hrefurl{http://www-d0.fnal.gov/})}}
\acro{CDF}        {Collider Detector at Fermilab\acroextra{ (\hrefurl{http://www-cdf.fnal.gov/})}}
\acro{CDF-FF}     {CDF fragmentation function}
\acro{CERN}       {European Organization for Nuclear Research}
\acro{CGC}        {Color Glass Condensate}
\acro{CPU}        {Central Processing Unit}
\acro{CTEQ}       {The Coordinated Theoretical-Experimental Project on QCD\acroextra{ (\hrefurl{http://www.phys.psu.edu/~cteq/})}}
\acro{DAQ}        {Data Acquisation}
\acro{DDL}        {Detector Data Link}
\acro{DGLAP}      {Dokshitzer-Gribov-Lipatov-Altarelli-Parisi}
\acro{DIS}        {Deep Inelastic Scattering}
\acro{DPM}        {Dual Parton Model}
\acro{DPMJET}     {Dual Parton Model JET\acroextra{ (\hrefurl{http://siwaps.physik.uni-siegen.de/kolloquium/dpmjet/})}}
\acro{D-RORC}     {DAQ Readout Receiver Card}
\acro{EBDS}       {Event Building and Distribution System}
\acro{EKS}        {Ellis-Kunszt-Soper jet program\acroextra{ (\hrefurl{http://zebu.uoregon.edu/~soper/EKSJets/jet.html})}}
\acro{EKRS}       {Eskola-Kolhinen-Ruuskanen-Salgado}
\acro{EKRT}       {Eskola-Kolhinen-Ruuskanen-Tuominen saturation model }
\acro{EKS98}      {Eskola-Kolhinen-Salgado  nPDF parametrization}
\acro{EMC}        {European Muon Collaboration}
\acro{EMCAL}      {Electromagnetic Calorimeter}
\acro{FEP}        {Front-End Processors}
\acro{FF}         {Fragmentation Function}
\acro{FMD}        {Forward Multiplicity Detector\acroextra{ (\hrefurl{http://fmd.nbi.dk/})}}
\acro{FPGA}       {Field Programmable Gate Array}
\acro{FSR}        {Final State Radiation}
\acro{GDC}        {Global Data Concentrator}
\acro{GEANT3}     {GEANT3\acroextra{ (\hrefurl{http://wwwasd.web.cern.ch/wwwasd/geant/index.html})}}
\acro{GRV}        {Glück-Reya-Vogt PDF parametrization\acroextra{ (\hrefurl{http://zebu.uoregon.edu/~parton/partonGRV.html})}}
\acro{H1}         {H1\acroextra{ (\hrefurl{http://www.desy.de})}}
\acro{HERA}       {Hadron Elektron Ring Anlage\acroextra{ (\hrefurl{http://www.desy.de})}}
\acro{HERWIG}     {Hadron Emission Reactions With Interfering Gluons\acroextra{ (\hrefurl{http://hepwww.rl.ac.uk/theory/seymour/herwig/})}}
\acro{HIJING}     {Heavy-Ion Jet Interaction Generator\acroextra{ (\hrefurl{http://www-nsdth.lbl.gov/~xnwang/hijing/})}}
\acro{HKM}        {Hirai-Kumano-Miyama nPDF parametrization}
\acro{HMPID}      {High-Momentum Particle Identification Detector}
\acro{HLT}        {High-Level Trigger}
\acro{HLT-RORC}   {HLT Readout Receiver Card}
\acro{ILCA}       {Improved Legacy Cone Algorithm}
\acro{IKF}        {Institut für Kernpysik Frankfurt}
\acro{ISAJET}     {ISAJET \acroextra{ (\hrefurl{http://www.phy.bnl.gov/~isajet/})}}
\acro{ISR}        {Initial State Radiation}
\acro{ITS}        {Inner Tracking System}
\acro{JetClu}     {Jet Clustering algorithm, CDF Run I}
\acro{JETRAD}     {JETRAD \acroextra{ (\hrefurl{http://theory.fnal.gov/people/giele/jetrad.html})}}
\acro{KKP}        {Kniehl-Kramer-Pötter FF parametrization\acroextra{ (\hrefurl{http://www.desy.de/~poetter/kkp.html})}}
\acro{L0}         {Level 0}
\acro{L1}         {Level 1}
\acro{L2}         {Level 2}
\acro{L2a}        {Level 2 accept}
\acro{L2r}        {Level 2 reject}
\acro{LDC}        {Local Data Concentrators}
\acro{LEP}        {Large Electron Positron Collider}
\acro{LHC}        {Large Hadron Collider\acroextra{ (\hrefurl{http://www.cern.ch})}}
\acro{LO}         {Leading Order}
\acro{LPHD}       {Local Parton-Hadron Duality}
\acro{LPM}        {Landau-Pomeranchuk-Migdal}
\acro{MidPoint}   {Jet clustering algorithm with seeds and mid-points}
\acro{MRST}       {Martin-Roberts-Stirling-Thorne PDF parametrization\acroextra{ (\hrefurl{http://zebu.uoregon.edu/~parton/partonMRS.html})}}
\acro{NLO}        {Next-to-Leading Order}
\acro{NNLO}       {Next-to-Next-to-Leading Order}
\acro{nPDF}       {nuclear-modified Parton Distribution Function}
\acro{PC}         {Personal Computer}
\acro{PCI}        {Peripheral Component Interconnect}
\acro{PDF}        {Parton Distribution Function}
\acro{PHENIX}     {PHENIX\acroextra{ (\hrefurl{http://www.phenix.bnl.gov/})}}
\acro{PHOBOS}     {PHOBOS\acroextra{ (\hrefurl{http://www.phobos.bnl.gov/})}}
\acro{PHOS}       {Photon Spectrometer}
\acro{PMD}        {Photon Multiplicity Detector}
\acro{PQM}        {Parton Quenching Model}
\acro{PYTHIA}     {PYTHIA\acroextra{ (\hrefurl{http://www.thep.lu.se/~torbjorn/Pythia.html})}}
\acro{pQCD}       {perturbative Quantum Chromodynamics}
\acro{QCD}        {Quantum Chromodynamics}
\acro{QED}        {Quantum Electrodynamics}
\acro{QGP}        {Quark-Gluon Plasma}
\acro{RHIC}       {Relativistic Heavy Ion Collider\acroextra{ (\hrefurl{http://www.bnl.gov})}}
\acro{RICH}       {Ring Imaging Cherenkov}
\acro{ROI}        {Region of Interest}
\acro{RPC}        {Resistive Plate Chamber}
\acro{sQGP}       {strongly coupled QGP}
\acro{SFM}        {String Fusion Model}
\acro{SHM}        {Statistical Hadronization Model}
\acro{SPS}        {Super Proton Synchrotron\acroextra{ (\hrefurl{http://www.cern.ch})}}
\acro{STAR}       {STAR\acroextra{ (\hrefurl{http://www.star.bnl.gov})}}
\acro{T0}         {T0\acroextra{ (\hrefurl{http://fmd.nbi.dk/})}}
\acro{TCP}        {Transmission Control Protocol}
\acro{THM}        {Thermal Hadronization Model}
\acro{Tevatron}   {Tevatron Collider at Fermilab\acroextra{ (\hrefurl{http://www.fnal.gov/})}}
\acro{TPC}        {Time Projection Chamber}
\acro{TRD}        {Transition-Radiation Detector}
\acro{TOF}        {Time Of Flight}
\acro{UA1}        {Underground Area 1}
\acro{V0}         {Vertex detector, ALICE}
\acro{ZDC}        {Zero-Degree Calorimeter}
\acro{ZEUS}       {ZEUS\acroextra{ (\hrefurl{http://www.desy.de})}}
\end{acronym}

\fi
\end{appendix}
\else
\ifacro
 
\fi
\fi

\ifbib
\bibliographystyle{unsrtnat}
\bibliography{arbeit}
\fi

\ifindex
\IndexPrologue{\chapter*{Index}}
\printindex
\fi

\iffinal
%

\chapter*{Acknowledgements}
\thispagestyle{empty}
Most of all, I wish to express my gratitude to my supervisor, 
Reinhard Stock, for his excellent support and providing ideal
working conditions. I appreciate interesting discussions, 
both, scientific and  social, and his intellectual insight paired 
with warm humanity. I am also very thankful for the financial aid 
to attend schools and conferences and for the possibility to spend 
long periods at \acs{CERN}.

\vspace{0.01cm}\noindent
Furthermore, I would like to thank Dieter Röhrich for supervision 
at the inital stages of the study and the financial backup for 
several conferences over the past years.

\vspace{0.01cm}\noindent
Special thanks go to my mate, Anders Vestb{\o}, for introducing 
me into the project and a great time during my numerous visits to 
Bergen. I appreciate many discussions, scientific, not scientific and 
not scientific at all, and entertaining interactions of all kinds.

\vspace{0.01cm}\noindent
I wish to express my warmest gratitude to Andrea Dainese for the
fruitful and very close collaboration in developing \ac{PQM}. 
Thanks also for very useful comments to the manuscript.

\vspace{0.01cm}\noindent
During my stays at \acs{CERN}, I was mainly guided by Andreas Morsch. 
I am extremely grateful for his teachings and manifold help in preparing 
the Monte Carlo simulations. He also contributed very valuable comments 
to the manuscript.

\vspace{0.01cm}\noindent
I also warmly thank Roland Bramm for a pleasant time shared working 
together during our stays at \acs{CERN} and \acs{IKF}. I remember many 
stimulating discussions connecting various seemingly uncorrelated topics
of science and life.

\vspace{0.01cm}\noindent
Over the past years, I enjoyed stimulating discussions with Guy Pai\'c
and his participation in \ac{PQM}. Many thanks for his comments to the 
manuscript. 

\vspace{0.01cm}\noindent
I wish to thank Cvetan Cheshkov for close collaboration.

\vspace{0.01cm}\noindent
I also owe a debt to Peter Hristov for support and help with \acs{ALIROOT}.

\vspace{0.01cm}\noindent
I wish to thank Yiota Foka for help with the manuscript.

\vspace{0.01cm}\noindent
I wish to express my special gratitude to Werner Amend for his invaluable
technical and organisational support over the past years.

\vspace{0.01cm}\noindent
I am grateful to the folks in the Experimental Nuclear Physics Group 
in Frankfurt and Bergen for providing a nice environment for both, research 
and friendship. In particular, I would like to mention Jens Ivar J{\o}rdre, 
Are Severin Martinsen, Boris Wagner, Dominik Flierl, Harald Appelshäuser,
Jutta Berschin, Peter Dinkelaker, Heidrun Rheinfels, Thorsten Kollegger, 
Marek Gazdzicki, Rainer Renfordt, Thomas Dietel and Wolfgang Sommer. 
In one or the other way, all have contributed to my work. Thank you!

\vspace{0.3cm}\noindent
Finally, I would also like to express my gratitude to my parents. Their 
endless love encouraged my research over the years. 

\vspace{0.7cm}\noindent 
Frankfurt am Main, February 2005\\
Constantinos Albrecht Loizides

\fi

\end{document}